\documentclass[12pt]{ucthesis}

\usepackage{ifthen}

\usepackage{fancyheadings}
\usepackage{fancyvrb}
\usepackage{aastex_hack_mwv}

\usepackage{subfigure}
\usepackage{alphalph}

\usepackage{deluxetable}
\usepackage{natbib}
\bibstyle{aa}

\usepackage{graphicx}
\graphicspath{{../Figures/}}
\usepackage{times}

\setkeys{Gin}{clip=false,keepaspectratio=true,height=\textwidth,width=\textwidth}

\usepackage{xspace}

\let\url\relax
\usepackage{hyperref}
\hypersetup{
            pdftitle={Thesis Notes},
            pdfauthor={W. Michael Wood-Vasey, UC Berkeley}
            bookmarks=true,
            bookmarksnumbered=true,
            colorlinks=false,
            linkcolor=blue
           }

\usepackage{xspace}
\def\sq{\hbox{\rlap{$\sqcap$}$\sqcup$}}
\def\deg{\hbox{$^\circ$}\xspace}
\def\sqdeg{\sq\deg\xspace}
\def\arcmin{\hbox{$^\prime$}\xspace}
\def\arcsec{\hbox{$^{\prime\prime}$}\xspace}

\newcommand{\file}[1]{\textit{#1}}
\newcommand{\code}[1]{\textit{#1}}
\newcommand{\computer}[1]{\textit{#1}}

\newcommand{\OM}      {${\Omega}_M$\xspace}
\newcommand{\OL}      {${\Omega}_{\Lambda}$\xspace}

\def\lesssim{\mathrel{\hbox{\rlap{\hbox{\lower4pt\hbox{$\sim$}}}\hbox{$<$}}}}
\def\gtrsim{\mathrel{\hbox{\rlap{\hbox{\lower4pt\hbox{$\sim$}}}\hbox{$>$}}}}

\graphicspath{{Figures/}}

\begin{document}
\ifx\href\undefined\else\hypersetup{linktocpage=true}\fi 

\title{Rates and Progenitors of Type Ia Supernovae}
\author{William Michael Wood-Vasey}
\degreeyear{2004}
\degreesemester{Fall}
\degree{Doctor of Philosophy}

\numberofmembers{4}
\cochaira{Professor George Smoot}
\cochairb{Professor Saul Perlmutter}
\othermembers{Professor Eugene Commins \\ Professor Geoff Marcy}

\prevdegrees{B.A. (Harvey Mudd College) 1998\\
             M.A. (University of California, Berkeley) 2000}

\field{Physics} 
\campus{Berkeley}
%

\maketitle
\approvalpage
\copyrightpage

\newcommand{\snefound}{        12}
\newcommand{\sneiaused}{     7.75}
\newcommand{\snesimexpected}{   5.67 \pm    0.05}
\newcommand{\snrate}{  3.67\ _{- 1.66}^{+ 1.20}\mathrm{(stat)}\ _{- 0.11}^{+ 0.11}\mathrm{(sys)}~h^3 \times 10^{-4} }
\newcommand{\snrateSNu}{  1.07\ _{- 0.48}^{+ 0.35}\mathrm{(stat)}\ _{- 0.03}^{+ 0.03}\mathrm{(sys)}~h^2 }
\newcommand{\snrateconf}{[  2.01,  4.87]~h^3 \times 10^{-4} }
\newcommand{\snrateSNuconf}{[  0.81,  1.97]~h^2 }


\pagenumbering{arabic}
%

\pdfbookmark{Abstract}{abs}
\begin{abstract}



The remarkable uniformity of Type~Ia
supernovae has allowed astronomers to use them as distance indicators
to measure the properties and expansion history of the Universe.
However, Type~Ia supernovae exhibit
intrinsic variation in both their spectra and observed brightness.
The brightness variations
have been approximately corrected by various methods, but there remain
intrinsic variations that limit the statistical
power of current and future observations of distant supernovae for
cosmological purposes.  There may be systematic effects in this
residual variation that evolve with redshift and thus limit
the cosmological power of SN~Ia luminosity-distance experiments.

To reduce these systematic uncertainties, we need a deeper understanding 
of the observed variations in Type~Ia supernovae.  Toward
this end, the Nearby Supernova Factory has been designed to discover
hundreds of Type~Ia supernovae in a systematic and automated fashion
and study them in detail.  This project will observe these supernovae
spectrophotometrically to provide the homogeneous high-quality data
set necessary to improve the understanding and calibration of these
vital cosmological yardsticks.

From 1998 to 2003, in collaboration with the Near-Earth Asteroid
Tracking group at the Jet Propulsion Laboratory, a systematic and
automated searching program was conceived and executed using the
computing facilities at Lawrence Berkeley National Laboratory and the
National Energy Research Supercomputing Center.  An automated search
had never been attempted on this scale.  A number of planned future
large supernovae projects are predicated on the ability to find
supernovae quickly, reliably, and efficiently in large datasets.

A prototype run of the SNfactory search pipeline conducted from 2002
to 2003 discovered 83~SNe at a final rate of 12~SNe/month.  A large,
homogeneous search of this scale offers an excellent opportunity to
measure the rate of Type~Ia supernovae.  This thesis presents a new
method for analyzing the true sensitivity of a multi-epoch supernova
search and finds a Type~Ia supernova rate from $z\sim0.01$--$0.1$ of
$r_V = \snrate$~SNe~Ia/yr/Mpc$^3$ from a preliminary analysis of a subsample
of the SNfactory prototype search.

Several unusual supernovae were found in the course of the SNfactory
prototype search.  One in particular, SN~2002ic, was the first SN~Ia
to exhibit convincing evidence for a circumstellar medium and offers
valuable insight into the progenitors of Type~Ia supernovae.


\abstractsignature
\end{abstract}

\setcounter{page}{0}


\pagenumbering{roman}

\begin{frontmatter}

\begin{dedication}

\end{dedication}

\setlength{\evensidemargin}{0.0in}

\tableofcontents

\listoffigures

\listoftables

\begin{acknowledgements}

{\bf Personal}

As a dissertation is a rather impersonal and public document, 
I will simply leave a list of thanks for the support and
encouragement of the following 
in the writing of this dissertation and the 
surviving of past 6 years of graduate school. 
To my family, for believing in me and raising me to know no limits.

To Elaine

To Alysia, for 

To Chris, my personal nutrionist and trainer.

I would like to thank Greg Aldering for his mentorship and advice over
these past 6 years.

To INPA for providing a warm and convivial environment for my
graduate studies.

{\bf Professional}

This research has made use of the NASA/IPAC Extragalactic Database (NED) which is operated by the Jet Propulsion Laboratory, California Institute of Technology, under contract with the National Aeronautics and Space Administration.

My graduate research was supported in part by a Graduate Research Fellowship from the National Science Foundation.

\end{acknowledgements}

\end{frontmatter}


\pagenumbering{arabic}

\part{Background and Context}
\label{part:introduction}
%

\chapter{Cosmology with Type Ia Supernovae}

Cosmology was revolutionized at the end of the 20th century by the
remarkable discovery that the expansion of the Universe is
accelerating.  This completely unexpected result sparked a flurry of
theoretical and experimental activity to understand and better
describe this surprising behavior.  Type Ia supernovae (SNe~Ia) were
at the heart of this discovery.

The usefulness of SNe~Ia as cosmological probes extends from the
evolution and history of the dynamics of the Universe to the formation
of the galaxies and stars within.  Serving as cosmological beacons,
SNe~Ia provide reference points in the cosmic fabric through the
characteristic brightness of their explosions.  Their unique nature
makes studies of SN~Ia progenitors vital to understanding stellar
formation and death.  Measurements of the rate of supernovae as a
function of redshift provide valuable clues to star formation in and
evolution of galactic populations.

While this dissertation focuses on the rates and progenitors of
SNe~Ia, the work presented here was done in the context of improving
the cosmological utility of SNe~Ia.  Thus, this introduction begins
with a survey of SNe~Ia and concludes with an overview of how
SNe~Ia can be used to study the expansion history of the Universe and
a discussion of the improvements in our understanding of SNe~Ia 
that are necessary for elucidating the mysteries of dark energy.



\section{Type Ia Supernovae}
\label{sec:sneia}

\subsection{Classification of Supernovae}

Supernovae are classified based on their observed spectral
features and light-curve behavior.  
The spectrum of a supernova
contains a wealth of information about the composition of and
distribution of elements in the exploding star.  The characteristic
broad features of supernovae reflect the spread in velocities inherent
in an expanding atmosphere and are key in confirming that a
new object is indeed a supernova.  The individual elemental lines
provide information about the progenitor system and the new elements
created in the explosion. 
The phenomenological definition of supernova
types derives from spectral features such as hydrogen emission and
silicon absorption.  

\begin{figure}
\plotone{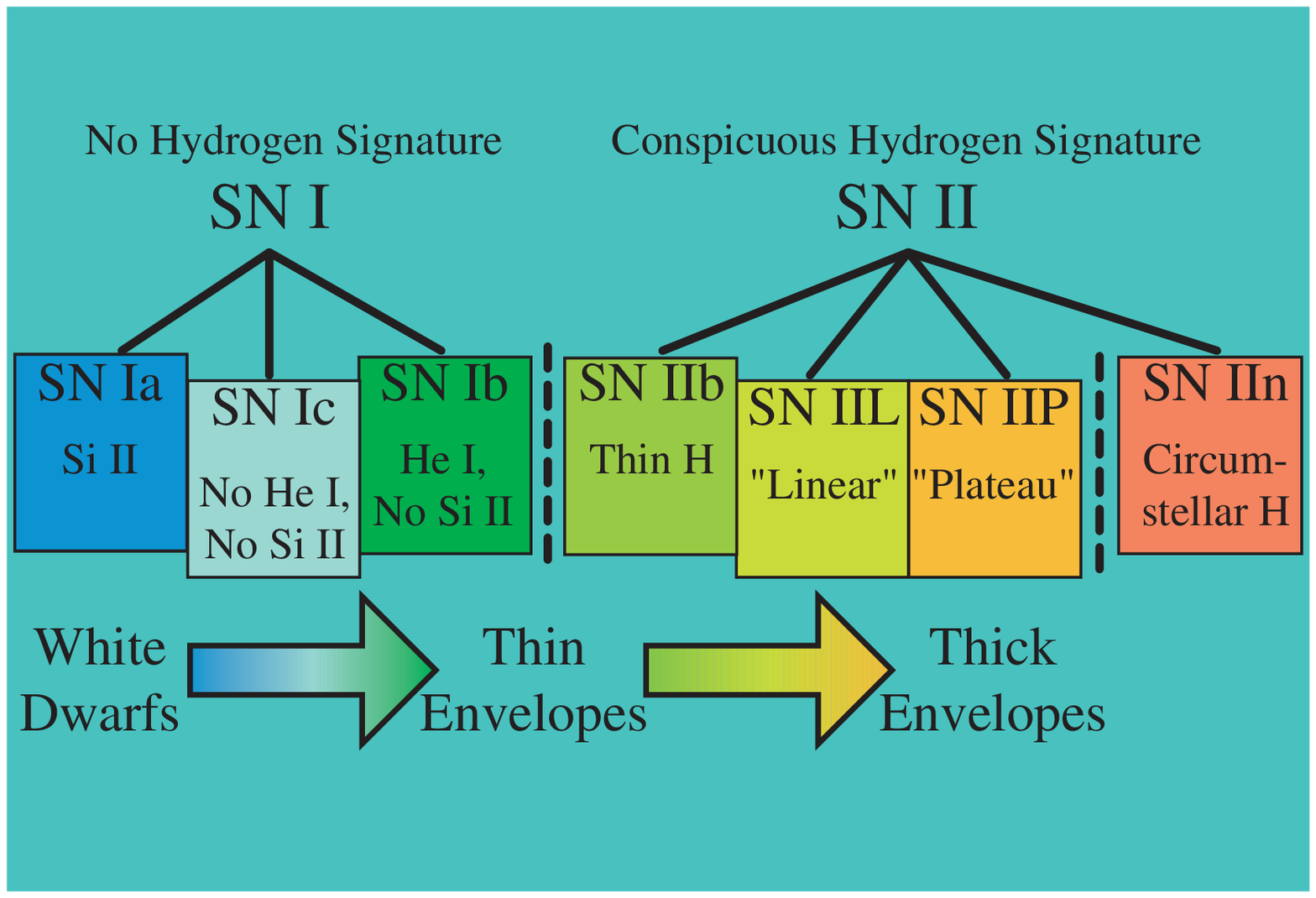}
\caption{Classification of supernova types.  Type~I supernovae exhibit no sign of hydrogen
in their spectra while Type~II do exhibit hydrogen.  Type~I
supernovae are further sub-classified by other spectral features
while Type~II supernovae are further sub-classified by the
shape of their light curves (IIL, IIP) and their spectroscopic features
(IIn).  Type Ia supernovae show the spectroscopic signatures of 
the explosion of a white dwarf while the spectra of other types of
supernovae are dominated by their envelopes.  Type Ib~\&~Ic supernovae have
thin envelopes devoid of hydrogen while Type II supernovae show thick hydrogen envelopes.  But see Chapter~\ref{chp:2002ic} for a Ia-IIn hybrid event that bridges the extremes of this classification scheme.  (Figure courtesy of Rollin Thomas)}
\label{fig:supernovae_types}
\end{figure}

Supernovae lacking hydrogen absorption or emission lines
are classified as
Type~I supernovae while supernovae with hydrogen lines are
classified as Type~II supernovae.  The Type~I supernovae are further
divided into Type~Ia, Type~Ib, and Type~Ic, depending on the presence
or absence of silicon and helium features~\citep{hillebrandt00}.
Type~Ia supernovae show a clear silicon absorption line at 6700~\AA\  
while Type~Ib supernovae show evidence of helium lines.  Type~Ic
supernovae show no traces of silicon or helium, but their spectra resemble
that of Type~Ib at late times.  Type~II supernovae are commonly
divided into four sub-types.  SNe~IIb supernovae are observationally
similar to SNe~Ib, but show evidence for a thin hydrogen envelope at early times.  SNe~IIL have light curves with a steep
``{\bf L}inear'' decline, SNe~IIn exhibit ``{\bf n}arrow'' spectroscopic lines,
and SNe~IIP decline slowly with a long ``{\bf P}lateau'' phase after
maximum-light.  See Fig.~\ref{fig:supernovae_types} for a diagram of the
different supernova classifications.

\subsection{SN~Ia Light Curves}

Of these various types and subtypes of supernovae, SNe~Ia are worthy
of particular attention here because of the homogeneity of their light
curves.
The optical light curve of a SN~Ia is
governed by the decay of the radioactive elements produced in the
explosion.  The dominant radioactive element produced in a SN~Ia
explosion is $^{56}$Ni.  This element decays to $^{56}$Co with a
half-life of around 15 days.  In turn, $^{56}$Co decays to the stable
isotope $^{56}$Fe.  The photons and positrons emitted in these decays
are absorbed by the surrounding material and re-radiated.  This
process leads to the observed rise and fall of a SN~Ia light
curve~\citep{pinto00a,pinto00b}.

\begin{figure}
\begin{center}
\plotone{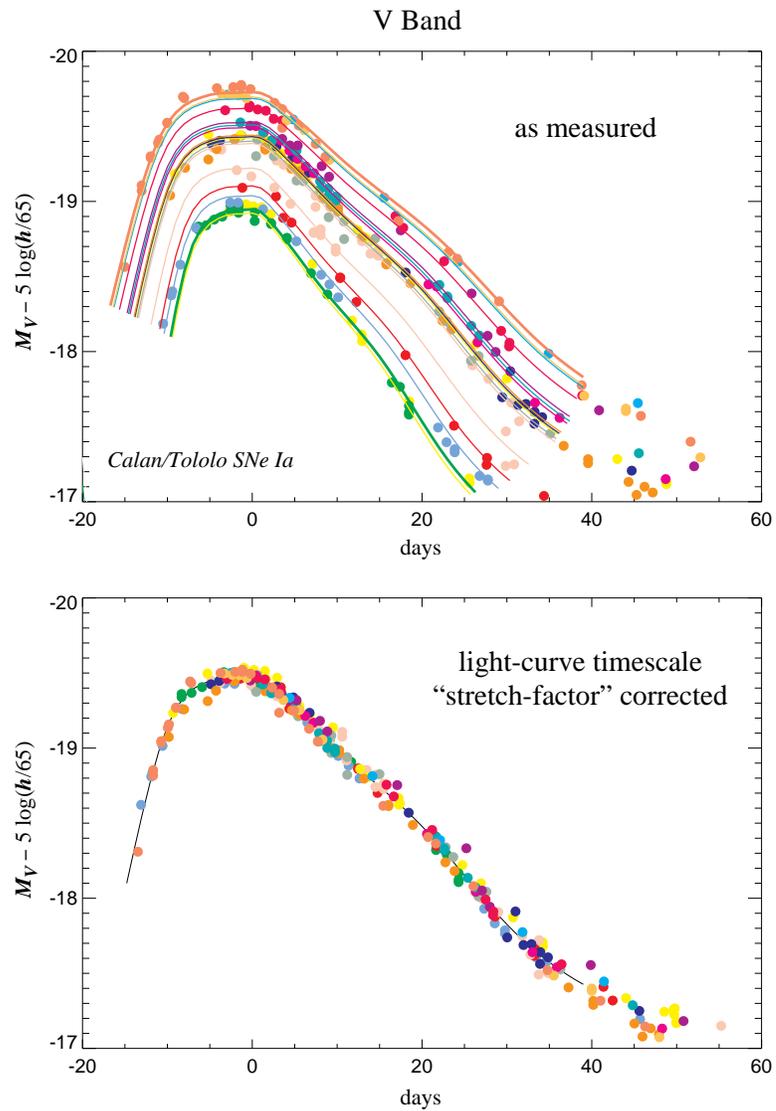}
\end{center}
\caption{Type~Ia supernovae exhibit a remarkable degree of homogeneity
in their light curves.  By fitting for one parameter, the ``stretch''~\citep{perlmutter97a,perlmutter97b,goldhaber01}, the observed variation in Type~Ia SNe can be
reduced to $\sigma_{M} = 0.15$.}
\label{fig:typeIa_homogeneity}
\label{fig:stretch_correction}
\end{figure}

As demonstrated in Fig.~\ref{fig:typeIa_homogeneity}, the light curves
of SNe~Ia are remarkably homogeneous.  
The observed variation (after correcting for redshift) can be well
accounted for by the variation of one parameter.  See
Sec.~\ref{sec:snia_parameterization} for further discussion on this
topic.

\subsection{Progenitor Models}
\label{sec:progenitor_models}

\citet{hillebrandt00} provide a recent summary of the
current understanding of the progenitor models for SNe~Ia, and the
literature reference letter of \citet{wheeler02} gives a guide to the
current literature.  See \citet{livio01} for a review of SN~Ia
progenitor models in the light of available observations.  The
prevailing progenitor model for SNe~Ia is a white dwarf that accretes material from
a companion star until it nears the Chandrasekhar mass
($\sim1.4~M_\sun$) and explodes.

A key clue to the nature of SN~Ia progenitors is the lack of hydrogen in
their observed spectra (but see
\citet{hamuy03b,deng04,wang04,wood-vasey04b}).  Because of
their lack of hydrogen, observed silicon, and characteristic light
curves, SNe~Ia are generally believed to be the result of an explosion
of a degenerate object, namely a white dwarf.  White dwarfs are the
common endpoint in the life of stars of less than three solar masses
($M_\sun$).  After such a star undergoes its red giant phase, it will
lose its outer envelope through winds and other mass-loss events and
finally leave just a carbon and oxygen core.  If the core is less
massive than 1.4~$M_\sun$, it will become a white dwarf (WD) supported by
electron degeneracy pressure.

For this object to explode, it must first become unstable.  
A typical white dwarf will begin at $0.7M_\sun$ but may
accumulate additional mass from a companion star until it
nears the Chandrasekhar limit of $\sim1.4~M_\sun$.  As it approaches
this limit, the interior conditions become unstable to fusion, and 
eventually a runaway fusion reaction begins at one or more points
within the star and rips through the rest of the white dwarf,
causing the explosion and complete destruction of the star.  The
additional mass required is generally significant ($\sim0.7~M_\sun$)
but could reasonably be obtained from another star as discussed in the
following section.  Under the assumption that the capture efficiency for
two separate stars to join in a binary system after formation is
relatively low, the SN~Ia rate is therefore directly related to the initial
binary fraction and initial mass fraction of star formation.

\subsubsection{White Dwarf Accreting from a Companion Star (Single-Degenerate)}
\label{sec:progenitor_models_single_degenerate}

A common scenario for a white dwarf's acquisition of additional mass
is the accretion of material from a less-evolved binary companion in
the red-giant phase.
As the companion star expands, it will eventually overfill the Roche
lobe of the system, causing matter to will flow from its envelope down to the
surface of the white dwarf.  The slowly accumulating hydrogen from
this star will collect onto the surface of the white dwarf, fusing
into helium and eventually carbon and oxygen~\citep{branch95}.  This
model predicts that SNe~Ia should result from very similar progenitor
white dwarfs at the same limiting mass and thus offers an explanation
for the relative homogeneity of SNe~Ia.  However, the
accretion rate necessary to ensure the slow conversion of hydrogen to
heavier elements without explosive burning and mass-loss events
from the white dwarf falls in a relatively narrow
range~\citep{nomoto82a}.

Considerations of stellar structure and evolution allow for four
possible compositions for a white dwarf that nears the Chandrasekhar
mass through accretion: helium; oxygen-neon-magnesium; carbon-oxygen;
and carbon-oxygen with a helium shell 
\citep{branch95,nomoto97,hillebrandt00}.

Helium (He) WDs can undergo helium ignition and explode when their
mass reaches 0.7~$M_\sun$.  However, the central burning in such a 
process is very complete and produces just $^{56}$Ni and unburned
helium~\citep{nomoto77}.  This progenitor model fails to explain
the array of intermediate-mass elements observed in the spectra 
of SNe~Ia.

\citet{gutierrez95} demonstrate that an oxygen-neon-magnesium (O-Ne-Mg) star 
is more likely to collapse down to a neutron star than to
explode.  In addition, these stars are too rare to be major
contributors to the observed number of SNe~Ia~\citep{branch95}.  

By process of elimination, one is left with carbon-oxygen (C-O) WDs and C-O WDs with a helium shell (C-O-He) as
the remaining progenitor candidates.  C-O WDs form naturally in the 
end-of-life stage of intermediate-mass stars of up to $10~M_\sun$.  For
stars in this mass range, carbon fusion is the final stage of nuclear burning that 
balances gravitational contraction with electron degeneracy pressure.  The
outer envelope will be lost through the winds in the giant phase.  The
question then becomes how a white dwarf accretes sufficient additional mass to become
unstable to a run-away fusion reaction.

The leading possibility is the so-called ``single-degenerate'' scenario,
in which a less-evolved binary
companion provides additional mass from its envelope of 
hydrogen or helium.  If the accumulated material is hydrogen, then
the accreted matter may eventually be lost through explosive hydrogen burning on the
surface of the white dwarf.  Depending on the rate of accumulation, which is
affected by the proximity, radius, and winds of the donor star in
addition to possible winds from the white dwarf, this hydrogen burning can lead to the
conversion of hydrogen to helium and either the accumulation or loss
of mass from the surface of the star.  If the accumulated material is
helium, then the white dwarf can lose mass through explosive helium burning.
A well-balanced rate of accretion and
fusion can lead to mass-gain through the accumulation of carbon and the eventual explosion
of the entire star from an initial instability in the interior of the
star.

A C-O WD can accumulate or be created with an additional layer of He.
If the amount of accumulated He becomes great enough, a He flash
can create a significantly off-center explosion that could
tear through the star and result in an
explosion of the entire white dwarf~\citep{nomoto82a}.  However, these explosions are believed to
result in either the complete conversion of the star into $^{56}$Ni if a
detonation wave is launched in the C-O or the burning of the
helium only and a subsequent non-burning explosion of the C-O
core~\citep{nomoto82a,nomoto82b,branch95}.  As with the pure helium
WDs, models of C-O-He WDs do not generate the observed properties of
SNe~Ia.

The problems with the He, O-Ne-Mg, and C-O-He WD progenitor
models outlined above leave the explosion of a pure C-O WD as the leading
candidate for a SN~Ia progenitor.  
The white dwarf is created as a C-O WD, and accreted material fuses on the surface
to become part of the degenerate C-O core.  When this C-O WD nears the Chandrasekhar limit, an explosion starts
as either a deflagration or a detonation at some inner part of the star
and proceeds outward.  Detonation, or super-sonic nuclear burning, burns
material completely before it has a chance to expand and so converts
everything to iron-peak elements; in contrast, deflagration, or 
sub-sonic nuclear burning, burns while
material is expanding and results in slower expansion velocities
insufficient to reproduce observations of peak velocities on the order
of 20,000~km/s~\citep{hillebrandt00}.
In order to explain both the high-velocities and intermediate-mass elements observed in SNe~Ia, a compromise between these two models was proposed:
according to this revised model, a deflagration 
gives the star time to expand and then transforms into a detonation at a
critical transition density, $\rho_\mathrm{DDT}$, resulting in both
intermediate-mass elements from incomplete burning of the already expanded material and high-velocities from the detonation
~\citep{khokhlov91a,khokhlov91b,hoeflich96,iwamoto99}.

\subsubsection{White Dwarf Merger (Double-Degenerate)}
\label{sec:progenitor_models_double_degenerate}

The range of mass accretion rates that allow for
the accumulation of mass onto a white dwarf from a main-sequence or giant
companion is rather narrow.  Concerns about this limited range 
lead to the suggestion that SNe~Ia could be produced instead by
the merger of two white dwarfs~\citep{iben84,webbink84}.  This model avoids
the timing and constrained mass-loss rate problems of the single-degenerate
scenario. In addition, the number of double-degenerate systems should be
relatively large as both stars in a binary system will
eventually evolve into white dwarfs if neither are massive enough to undergo
core collapse.  Eventually, the two stars will spiral inward due to energy
loss through dynamic friction, their earlier respective mass-loss
phases, or gravitational radiation.
As the two white dwarfs near each other, they will become distorted and
eventually merge.  If their combined mass is near to or greater than
the Chandrasekhar limit, then a runaway nuclear reaction will
spread through the merged object and result in an explosion.  

However, it is not immediately clear how this scenario would result in
the observed homogeneity of SNe~Ia.  The combined mass of the white
dwarfs could span a range of values that would result in a variety of
explosion energy and observed light curves.  This variation would be
greater than the peak magnitude dispersion of $\sim0.3$ magnitudes
observed in SNe~Ia~\citep{goldhaber01,knop03} but potentially could
account for outliers in the SN~Ia peak magnitude distribution.

Currently, the most promising model remains the C-O single-degenerate
white dwarf exploding via a deflagration to detonation transition.
But the other mechanisms discussed in the section may also occur and
represent possible contaminants for studies requiring a homogeneous
class of SNe~Ia.

\section{Non-Type~Ia Supernovae}
\label{sec:nonsneia}

All other categories of supernovae (Type~Ib, Ic, and II) are believed
to result from the collapse of a single massive star.  These supernovae can
occur in multiple-star systems, but the fundamental nature of the
explosion derives from the core-collapse and resulting explosive rebound of
the progenitor star.  These events leave behind a collapsed core,
either a neutron star or a black hole, and expel the rest of their mass
out into interstellar space.  In a Type~Ia supernova explosion, in
contrast, the star is completely disrupted and no part of the original
star remains.  This complete destruction of the white dwarf
partially explains why SNe~Ia are intrinsically very
bright while core-collapse supernovae vary significantly in their
brightness due to different initial masses and compositions.  

A core-collapse supernova occurs when the core of a massive star
fuses to iron and can no longer support the star through the release
of energy by fusion.  The star has been steadily burning up the chain
of elements, from carbon and oxygen to silicon and finally to iron.
The last phase takes only seconds and leads to the collapse of the star.
The core is converted into a Chandrasekhar-mass neutron star, and the
infalling material compresses the core and then rebounds with a large
fraction of its initial in-fall kinetic energy.  
This rebound develops into an explosion that blows out all of
the layers above the core.  In the case of more massive stars, the
core rebound is sometimes not enough to let all of the infalling
material escape and a black hole is created.  

As the explosion proceeds through the star, layers of material are
compressed and then re-expanded as the shock passes through them.
This environment, far away from local thermal equilibrium, generates a
wide variety of elements heavier than iron.  As in SNe~Ia, radioactive
$^{56}$Ni is the most common result of this shock-induced fusion and
leads to the observed light curve of the supernova as the
$^{56}$Ni decays to $^{56}$Co and then $^{56}$Fe.  However, in contrast
to the relatively homogeneous SN~Ia population, the variation in composition and opacities in surrounding
material leads to significantly different light curve decline
behaviors in different core-collapse SNe.

The initial mass of the star, the kinetic energy of the explosion,
and the amount of $^{56}$Ni produced are the three dominant factors that
determine the later evolution of the core-collapse supernova
light curve and spectra~\citep{young95}.
This scenario is complicated by rotation of the core, which can
lead to asymmetries in the explosion, including the possible formation
of jets from the poles of the rotation.

Regardless of the details of the explosion, core-collapse SNe are 
direct tracers of the star formation rate as they are the end stage of
short-lived stars.  In contrast, the progenitor systems for SNe~Ia may
take billions of years to evolve.  This difference in progenitor lifetime means
that a study of SN rates offers insight into both the star formation
history of the Universe and the nature of SNe~Ia progenitors.


\section{Supernova Rates}
\label{sec:intro_sn_rates}

Supernovae are very visible indicators of the underlying stellar
population.  The rates of different types of supernovae provide
important clues to the evolution of the star formation rate, initial
mass fraction, galaxy chemical evolution, the evolution of stellar
systems, and the evolution of galaxies.
 
The collapse-induced explosion of massive stars is tightly linked to
the initial mass fraction of the stars formed and to the overall star
formation.  As the most massive stars live the shortest lives,
exploding within $10^7$--$10^8$ years of their formation, these
types of supernovae provide a direct measure of the amount of star
formation activity occurring within a given population.  This
population can be measured as a function of redshift, galaxy age, and
galaxy composition.  These supernovae are the main producers and distributors
of heavy elements and thus are influential in determining
the metallicities of stars in the next cycle of stellar birth.

SNe~Ia, on the other hand, are believed to be the endpoint
in stellar evolution for less-massive stars that form white dwarf
cores in binary systems.  These progenitors take at least a billion years to
evolve through the main-sequence to the white-dwarf stage and yet
more time to accumulate the matter to reach the critical
conditions necessary for the explosion of the stars.  These supernovae
also contribute to the chemical evolution of the galaxy,
in some ways more strongly than core-collapse supernovae,
for SNe~Ia release all of their material to space while
the core-collapse supernovae leave behind a remnant core containing
the bulk of the iron produced by the stars.  However, as the
core-collapse stars are more massive, they may contribute an equal
amount of material to the interstellar medium on a per-supernova
basis, even if the SNe~Ia contribute more as a percentage
of initial mass of the system.  The rates of SNe~Ia versus
core-collapse SNe illuminate the nature of these different evolutionary
pathways and are related to the star formation history of a galaxy
over the past billion years (the rough time scale for a system to evolve
through to a SNe~Ia) to the current star formation rate.  
The evolution of galaxy types and
morphologies is intriguingly tied into this ratio of supernova rates.  As a
relic of a galaxy's history, SNe~Ia also help to trace the previous
evolution and state of a galaxy, even if that galaxy has since merged or been
disrupted.  However, as the progenitor systems of SNe~Ia may wander
in their host galaxy, it is possible that they are preferentially
stripped from galaxies and thus more likely to be found in
intra-cluster space or even outside obvious galaxies.

Studies of the absolute rate of SNe of all types are relevant to a number of questions 
regarding stellar evolution processes.
For example, in the latter stages of life, do all massive stars end in
core-collapse supernovae, or do some lose a large fraction of their
mass and avoid this fate, living out their long retirement years as a
white dwarf or other compact object formed without violent explosion?
The initial mass fraction (IMF) of stars created may vary
significantly with redshift as populations of stars live and die.
Is the IMF strongly correlated with the nature of galaxy formation,
or is it more dependent on more local gas density variations?


Our current limited knowledge of SN rates, particularly nearby SN
rates, is insufficient to provide good constraints on galactic
and stellar populations and evolution.  Even the basic
rate of SNe~Ia is only constrained to $30\%$ at
$z\sim0.01$~\citep{cappellaro99} and $15\%$ at
$z\sim0.5$~\citep{pain02}.  The rate for SNe~Ia at $z\sim0.1$ is only
known to within a factor of 2~\citep{cappellaro99}.
A determination of the local SN~Ia rate to the same precision as the
high-redshift SN~Ia rate would allow for a study of star formation rates 
and offer further clues to the nature of SN~Ia progenitors.

\section{Cosmology with Redshifts and Luminosity Distance}
\label{sec:cosmology}

The homogeneity of SN~Ia light curves can be exploited to make
cosmological measurements by employing the SN~Ia redshifts and
standardized peak magnitudes to determine the relationship between the
scale factor of and luminosity distance in our Universe.  See
\citet{weller02} for a useful summary of deriving cosmological results
from SNe~Ia.  Briefly, the redshift, $z$, is a relative measurement of
the cosmological scale factor, $a$, at the epoch of the supernova,
\begin{equation}
\frac{a_0}{a} = 1+z,
\label{eq:redshift_scale_factor}
\end{equation}
where $a_0$ is the present-day scale factor.  The observed, or apparent, peak magnitude, 
$m$, of a SN~Ia is used to determine its luminosity distance, $d_L$,
defined by
\begin{equation}
f_{\mathrm observed} = \frac{L_{\mathrm source}}{4\pi d_L^2},
\label{eq:luminosity_distance}
\end{equation}
where we consider the bolometric flux $f$ and bolometric luminosity $L$.
Recall that magnitudes are defined as the logarithm of the flux,
\begin{equation}
m = -2.5\log_{10}{f} = -2.5\log_{10}{\frac{L_{\mathrm source}}{4\pi d_L^2}} = -2.5\log_{10}{L_{\mathrm source}} - 5\log_{10}{d_L} -2.5\log_{10}{4\pi},
\end{equation}
modulo constants that defines the distance units and magnitude zeropoint.
By convention, absolute magnitudes are defined 
at a luminosity distance of 10~parsecs (pc).  The difference between
the absolute magnitude, $M$, and apparent magnitude, $m$, of an object is defined as the distance modulus, $\mu=m-M$,
and is related to the luminosity distance as follows:
\begin{eqnarray}
m-M & = &   ( -2.5\log_{10}{L_{\mathrm source}} - 5\log_{10}\frac{d_L}{[{\mathrm pc}]}) - ( -2.5\log_{10}{L_{\mathrm source}} - 5 ) \\
\label{eq:peak_mag_mu}
\mu & = & +5  - 5\log_{10}{d_L}{[{\mathrm pc}]} \\
\mu & = & -25 - 5\log_{10}{d_L}{[{\mathrm Mpc}]} \\
d_L & = & 10^{-(\mu+25)/5}~[\mathrm{Mpc}].
\label{eq:peak_mag_dL}
\end{eqnarray}
Any class of bright astrophysical objects that occur at redshifts
out to $z\sim1$ with a constant $M$ can be used
for cosmological measurements by using
the redshift and Eq.~\ref{eq:peak_mag_dL} of these objects 
to determine the geometry and expansion history of the Universe.

\begin{figure}
\plottwo{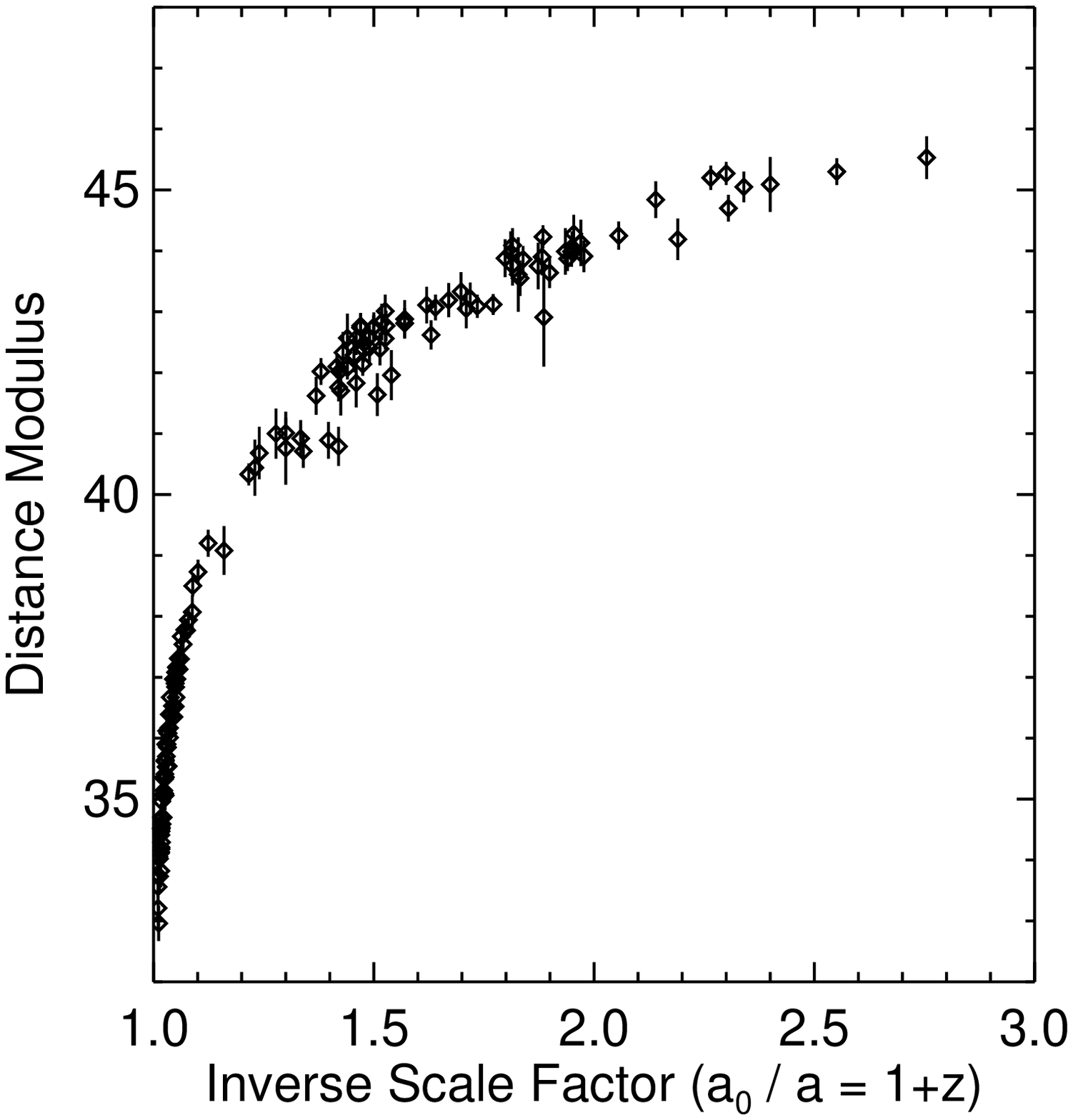}{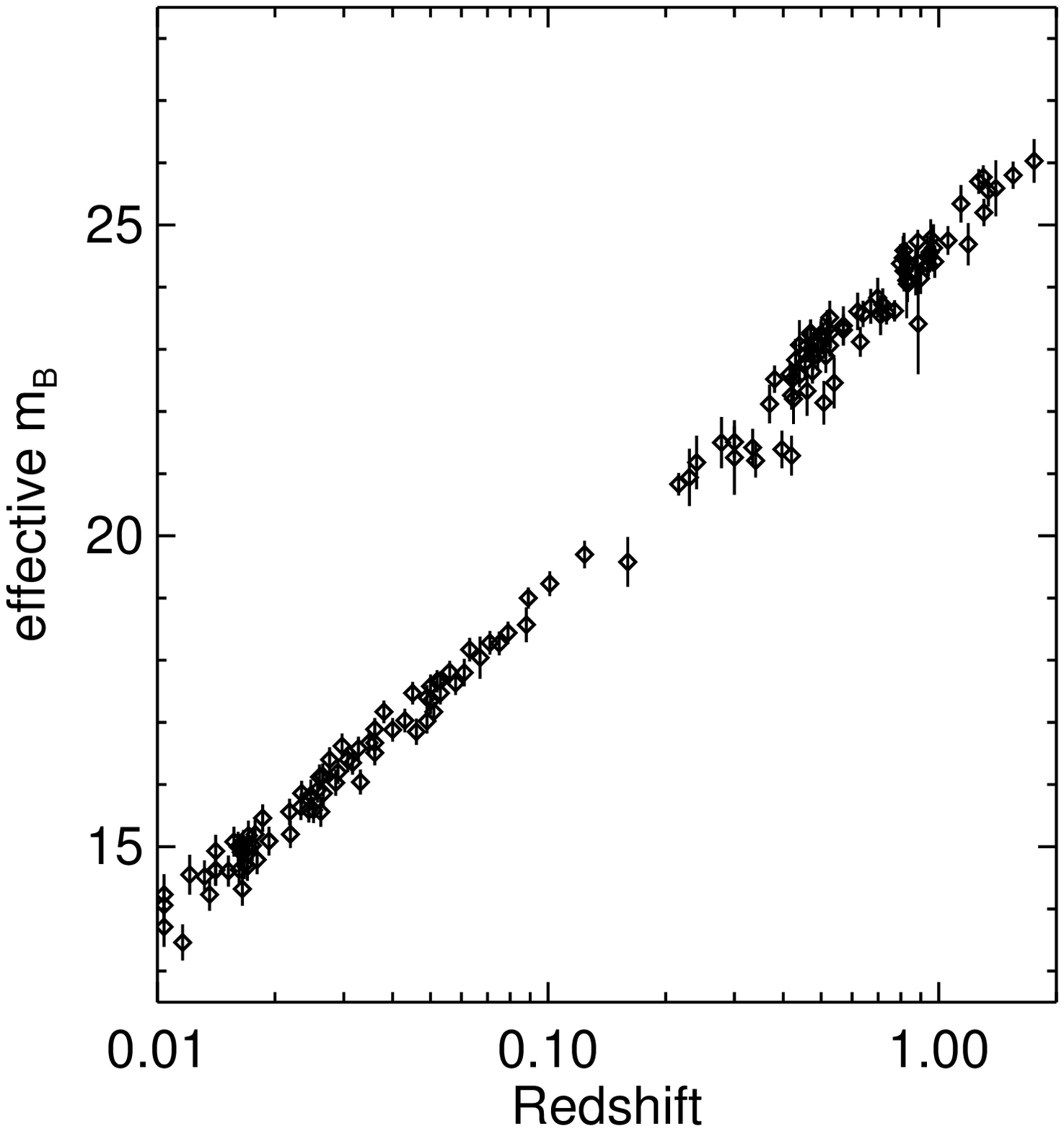}
\caption{The SN~Ia luminosity distance-scale factor relation used to
measure the expansion history of the Universe.  On the left, the luminosity
distance modulus, $\mu=m-M$, versus inverse scale factor.  On the right, the SN~Ia effective
magnitude in the B-band optical filter, $m_B$, versus redshift, shown on a log scale for the redshift.  
This $m_B$ vs. $\log{z}$ format
is the more commonly seen presentation.  Note that the absolute
normalization of the distance modulus is dependent on the assumed
values for $H_0$ and $M$ but is unimportant in the cosmological
measurement of $\Omega_M$ and $\Omega_\Lambda$.  The low-redshift,
($a_0/a\sim1$), region is clearly a critical anchor point for
comparison with higher redshift measurements.  (Data from
\citet{riess04b})}
\label{fig:sne_dL_a0_a}
\end{figure}

The homogeneity of SN~Ia light curves~\citep{branch92,branch93} allows for
relative luminosity distances to be calculated quite accurately
and led to the use of SNe~Ia as
standardizable candles in studies of the geometry and expansion
history of the Universe~\citep{perlmutter97b}.
As an astronomical measurement of $d_L$ involves a
comparison of $m$ and $M$ (Eq.~\ref{eq:peak_mag_dL}), the key benefit
of using SNe~Ia is that they have very similar absolute magnitudes, $M$, at
peak.  Thus, one can construct a plot of $d_L$ versus $a_0/a$ (see
Fig.~\ref{fig:sne_dL_a0_a}) to reveal the expansion history of the
Universe. 
The $H_0$ dependence of $d_L$ and $M$ can be separated out into
\begin{eqnarray}
\mathcal{D}_L & = & H_0 d_L \\
\label{eq:script_dL}
  \mathcal{M} & = & M - 5\log_{10} H_0  + 25 
\label{eq:script_M}
\end{eqnarray}
and thus one can rewrite Eq.~\ref{eq:peak_mag_mu} with no $H_0$ dependence as 
\begin{equation}
m - \mathcal{M} = 5\log_{10}\mathcal{D}_L.
\label{eq:script_peak_mag_dL}
\end{equation}
To fit for the cosmology underlying $\mathcal{D}_L$, one then performs a joint fit from low to high redshift and marginalizes over $\mathcal{M}$.
Alternatively, if one had a well-constrained value for $\mathcal{M}$ 
from a sample of a few hundred nearby SNe~Ia in the smooth Hubble flow,
one could use this value in Eq.~\ref{eq:script_peak_mag_dL}
when fitting higher-redshift supernovae.
Ultimately, the cosmological results are dependent on having a source
with a relatively constant $M$ over a large range of redshifts and are not
specifically dependent on either $M$ or $H_0$.

As Eq.~\ref{eq:peak_mag_dL} shows, it is only the
difference between $m$ and $M$ that enters in the calculation of
$d_L$.  The low observed scatter in $M$ allows for $m-M$ to be a
useful indicator of luminosity distance.  The Hubble constant, $H_0$,
represents a constant offset in the distance scale, but it is not
important in the measurement of the change of $d_L$ with $a$.

After it was demonstrated that SNe~Ia could be discovered in a
reliable manner that allowed for planned and scheduled
follow-up~\citep{perlmutter97b}, two independent teams measured
luminosity distances and redshifts to high-redshift SNe~Ia and
concluded that there was a significant non-matter component of the
total energy density of the Universe that has been causing the
Universe to increase its rate of expansion over the past seven billion
years~\citep{perlmutter98a, garnavich98a,riess98a,perlmutter99}.  This
unknown component became known as ``dark energy,'' and its nature and
source remains the biggest mystery in cosmology and particle
physics today.  For further progress to be made in this important
field, better understanding of the residual variation of SN~Ia light
curves is necessary.

As cosmology has progressed and galaxy clustering
studies~\citep{turner01,allen02,bahcall03,tegmark03} and cosmic
microwave background measurements~\citep{jaffe01,bennett03,spergel03}
agree with the supernova results, we are entering the age of
precision cosmology.  
Further understanding of the history and evolution of the Universe
will come from the next generation of supernova cosmology experiments.

The next great project is measuring the equation
of state of the Universe~\citep{weller02}.
While it may appear most natural to consider the luminosity-distance
vs. redshift measurements in terms of the scale factor $a$, many
models attempting to explain the nature of dark energy are framed in
terms of the equation of state of the dark energy~\citep{linder03a}.
Specifically, these models predict values for the ratio,
$w$, between the pressure, $p$, and density, $\rho$, of the dark energy:
\begin{equation}
w = \frac{p}{\rho}.
\label{eq:equation_of_state}
\end{equation}
Most simply, if the current acceleration of the Universe is caused by
a cosmological constant, then $w=-1$ is constant with time.  For models
where $w$ evolves with the scale factor, $a$, $w$ can be parameterized as
\begin{equation}
w = w_0 + w_a (1-a)
\label{eq:w_a}
\end{equation}
(see \citet{linder03a}).  While Eq.~\ref{eq:w_a} implies a linear
variation with $a$, in general $w$ could be any arbitrary function of
$a$ (see \citet{linder03b}).

Models predicting a constant $w=w_0$, including those invoking a cosmological
constant, are the easiest to test.  Two projects are
presently underway to measure $w_0$ to $\sim10\%$ using SNe~Ia from $0.3 \lesssim z \lesssim 0.8$.  The SuperNova Legacy Survey (SNLS)~\citep{pritchet04} is a part of the
Canada-France-Hawaii Telescope Survey~\citep{cfhls} and will discover
and study 2000 SNe~Ia between 2003 and 2008.  The ESSENCE
project~\citep{essence} is undertaking a similar study using the CTIO
4.0-m telescope to compile a data set of $\sim500$ SNe~Ia by 2006.  Both
projects use 8-m class telescopes for spectroscopic confirmation and
study.  The current scatter of $0.10$--$0.15$~magnitudes in the
luminosity-distance indicator currently used for SNe~Ia and the lack
of well-observed nearby SNe~Ia have become the central limitations in
using SNe~Ia to measure $w$.

The need for improved phenomenological and systematic calibration of
SNe~Ia and the current lack of a high-quality sample of SNe~Ia at low
redshift demand a comprehensive study of nearby SNe~Ia.
Meeting the goals of the SNLS and ESSENCE projects will require a
large sample of nearby supernovae on the same order as the many
hundreds of distant SNe~Ia to be found by each project to enable a
comparison of low- and high-redshift SNe~Ia.

While a sample of nearby SNe~Ia will
provide improved statistical constraints on the
luminosity-distance-redshift Hubble diagram, there is a more pressing need
for an  improved understanding of and controls on systematic errors 
in using SNe~Ia to make cosmological measurements.
It is critical to constrain
possible systematic effects so that any evolution with redshift can be
quantified.
An improved understanding of the astrophysical processes
underlying SNe~Ia and their progenitors will be invaluable in elucidating
the residual variation seen in SN~Ia light curves.

\section{Needed Improvements for SN~Ia Cosmology}
\label{sec:sneia_luminosity_distance}

SNe~Ia exhibit a surprising degree of homogeneity in their absolute
brightness.  They are observed to vary in peak B-band brightness by
only $\sim30\%$~\citep{goldhaber01}.  This variation can be reduced by the use
of a time-scale parameter that characterizes the observed relationship between
light-curve width and brightness for
SNe~Ia~\citep{pskovskii70,pskovskii77b,phillips93,hamuy95,riess95,riess96,perlmutter97a,goldhaber01}
(see Fig.~\ref{fig:typeIa_homogeneity}).
\citet{phillips93} found that SNe~Ia exhibited a dispersion in
their intrinsic brightness that was correlated with their magnitude
decline rate after maximum light, $\delta m_{15}$.  Other
parameterizations of this correlation result in a residual scatter in
SN~Ia maximum B-band luminosity of
$0.15$~magnitudes~\citep{riess96,hamuy96,perlmutter97b,phillips99}.
This $\sim15$\% level of
standardization was sufficient for the original cosmological
measurements that provided clear evidence for an accelerating
Universe.  

The MLCS~\citep{riess95,riess96},
$\Delta m_{15}$~\citep{phillips93,hamuy95,hamuy96}
and stretch~\citep{perlmutter97b,perlmutter99} methods currently used are all
equivalent to the variation of a single parameter.  See
Figure~\ref{fig:stretch_correction} for an example of how the stretch
parameter unifies light curves from a group of SNe~Ia.
The large sample of high-resolution SN~Ia spectra provided by the
SNfactory will hopefully provide additional parameters that will improve
the current standardization of luminosity distances from SNe~Ia.
A detailed set of spectra from a large, well-studied population will
bring improved understanding of both the known first-order and
surmised higher-order variations in SN~Ia light curves.

As efforts continue to measure the expansion history of the Universe
to higher precision and more SNe~Ia are added to the Hubble diagrams
used to fit for the cosmological matter density and dark energy
density, the residual dispersion and unknown systematics of SN~Ia
magnitudes are becoming a significant source of uncertainty.

The dominant source of uncertainty in using SNe~Ia for cosmological 
measurements is in the luminosity distance of the SNe~Ia.
This uncertainty arises both from measurement error of the SN flux
and, more importantly, from variations in the absolute brightness of
SNe~Ia.
The redshift to a given SN host galaxy can be measured very precisely
from galactic emission lines and is only a significant source of
uncertainty at low redshift where peculiar velocities dominate over
the Hubble flow expansion.

It is unknown whether the residual dispersion in SN~Ia light curves
after fitting for the time-scale parameter,
is simply a random distribution
with no correlation with possible observables or whether there is a second
parameter that could reduce this scatter even further.
If the dispersion is random, then very large numbers of supernovae
will be required to reduce the overall uncertainties.  Even then, 
unknown and uncontrolled systematic uncertainties could limit
the accuracy of any such measurement.
While correlations have been observed between selected SN~Ia spectral
features and the $\Delta m_{15}$ time-scale
parameter~\citep{nugent95,mazzali98,riess98b,hatano00}, a detailed
analysis of a large sample of SN~Ia spectra will be necessary to
search for improved spectroscopic indicators
that allow for a deeper understanding and better standardization of
luminosity distances measured with SNe~Ia.
If correlations are found with such secondary observables, then future
high-redshift supernova studies will greatly benefit from the improved
calibration possible from a better understanding of SN~Ia
properties.  The calibration of these vital standardizable candles
needs to be significantly improved to allow the next generation of
large-scale missions, such as the SuperNova Acceleration Probe (SNAP)~\citep{snap04}
and the Joint Dark Energy Mission (JDEM), to explore the dark energy
and expansion history of the Universe.

\part{The Nearby Supernova Factory}
\label{part:snfactory}
\chapter{The Nearby Supernova Factory}
\label{chp:snfactory}

Please see Appendix~\ref{apx:snfactory_summary} for a conference
proceedings article \citep{wood-vasey04a} that provides a 
brief summary of the Nearby Supernova Factory mission and 
the results from the prototype supernova search.

\section{Concept}
\label{sec:snfactory_concept}

In order to fill the vital need for a well-observed sample of nearby
SNe~Ia, the Nearby Supernova Factory (SNfactory)
project~\citep{aldering02b,pecontal03,wood-vasey04a} has been devised
to discover and study in detail $300$~SNe~Ia to better understand and
calibrate these important cosmological tools.  
The SNfactory aims to provide a definitive data set for nearby
SNe~Ia by observing a large and diverse
sample of SNe~Ia to provide a comprehensive sample of the parameter
space of observed SN~Ia variation for comparison with higher-redshift
SNe~Ia.  

Further progress in supernova cosmology requires an
order-of-magnitude increase in the number of well-studied SNe~Ia in
the nearby smooth Hubble flow ($0.03<z<0.08$) to provide a better
understanding of luminosity distance indicators for SNe~Ia and to
anchor the low-redshift end of the SN~Ia Hubble diagram (see
Fig.~\ref{fig:sne_dL_a0_a}).  A two-fold reduction in the scatter of
the SN~Ia luminosity-distance indicator would yield a commensurate
improvement in the constrains on $w_0$ (see Sec.~\ref{sec:cosmology}).
Even without this better calibration of SNe~Ia, the statistical weight
of the 300 SNfactory SNe~Ia alone will provide a factor of two
improvement in both the SNLS~\citep{pritchet04} and
ESSENCE~\citep{garnavich02} measurements of $w_0$ 
and the SNAP~\citep{snap04}
measurement of $w_a$.  The improved calibration of SNe~Ia
that the SNfactory will allow is also a critical contribution needed to
make full use of the large number of SNe~Ia that will be found and
studied with current and upcoming intermediate- and high-redshift
projects such as ESSENCE, SNLS, and SNAP.


The data set acquired by the SNfactory will be a rich source of
information about the SNe~Ia themselves.  From the nature of their progenitors
to their rate of their occurrence, many mysteries remain regarding
these useful cosmological standards.
Detailed, early-time spectra of SNe~Ia will constrain SN~Ia
progenitor models and provide clues to the observed variation in SN~Ia
light curves.
A proper understanding of the efficiency and selection biases in a SN
search is critical to a proper determination of SN rates, for which
the large, automated search of the SNfactory will be an excellent
source.
The SNfactory sample of 300 $z\lesssim0.1$ SNe~Ia will allow
for a determination of the nearby SN~Ia rate to $\sim6$\% to provide
useful constraints on galactic and stellar evolution models.

The first crucial part of the SNfactory project is to discover large
numbers of supernovae in the nearby Hubble flow ($0.03 < z < 0.08$) in
a reliable, continuous fashion.
Discovering hundreds of SNe~Ia in this redshift range requires searching of hundreds of
square degrees of sky every night.  To cover this much area, the
SNfactory uses an automated image processing and subtraction pipeline,
developed as a major focus of the dissertation work presented herein,
to scan images from nightly wide-field asteroid searches. 
This pipeline
has discovered $83$ SNe in its prototype phase of operation.  See
Chapters~\ref{chp:search_pipeline}~\&~\ref{chp:subtractions} for a
description of this successful supernova search project.

The second component of the SNfactory project is spectrophotometric
follow-up of the SNe, using a custom-built instrument dubbed the
SuperNova Integral Field Spectrograph (SNIFS).  This specialized
instrument takes spectrophotometric observations of an object and its
surroundings by using an array of $225$ lenselets fully covering a 
6-arcsecond by 6-arcsecond region.  This array allows for simultaneous
observations of SNe~Ia and their host environment in an automated
fashion.  In parallel with the spectrograph, SNIFS has a multi-color
photometric channel that allows for precise flux-calibration of
the observed spectra by monitoring the atmospheric extinction during
the spectral exposure.
See \citet{lantz03} for a description of the SNIFS instrument.

SNIFS is mounted on the University of Hawaii 2.2-m telescope and
is integrated into the SNfactory search pipeline to automatically
confirm and follow supernovae.  Spectrophotometric observations will
be taken every $3$--$5$ days from well before maximum light through
several months after explosion.  SNIFS is in
the final stages of commissioning and the SNfactory will begin full,
integrated operations in late-2004.

\section{Goals of the Nearby Supernova Factory}
\label{sec:snfactory_goals}

\subsection{New SN~Ia Luminosity Distance Indicator Parameterization}
\label{sec:snia_parameterization}

\begin{figure}
\plotone{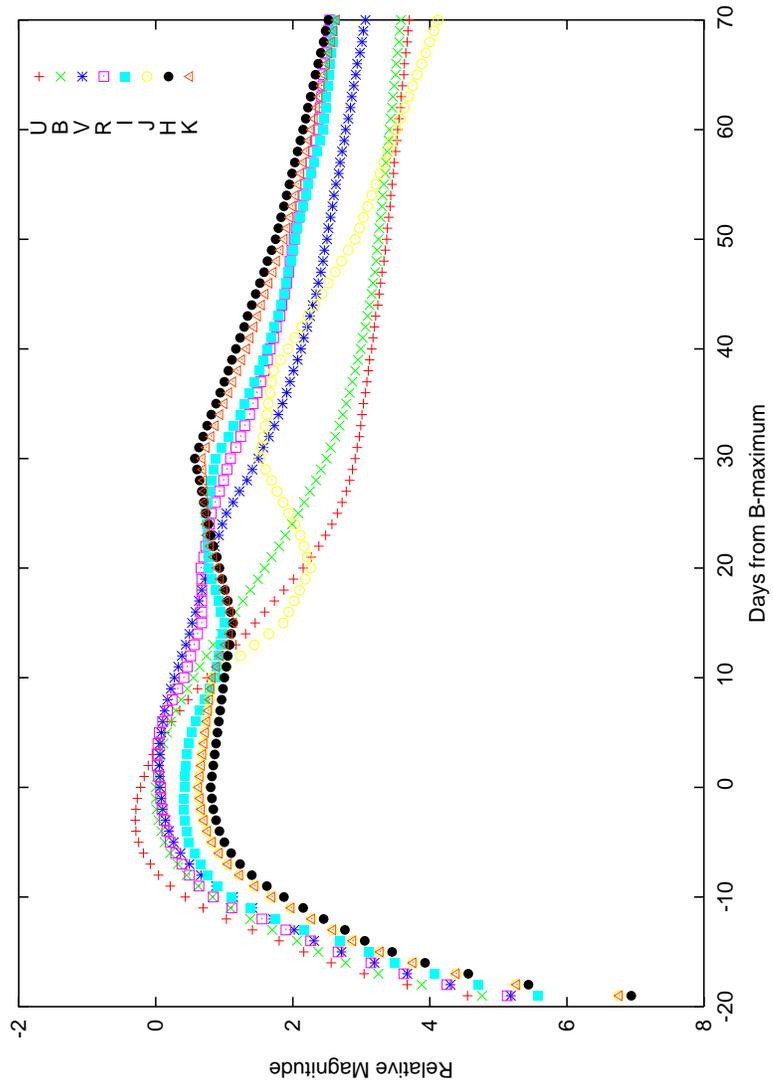}
\caption{Schematic light curves of a SN~Ia in different
standard astronomical filters.  Template courtesy of Peter Nugent.  See \cite{goldhaber01} for a more
complete discussion of SN~Ia light curves and templates.}
\label{fig:Ia_lightcurves}
\end{figure}

Current SNe~Ia calibration techniques use one parameter to describe all of
the observed variation in SNe~Ia light curves (see Sec.~\ref{sec:sneia_luminosity_distance}).
The major physical explanation for the observed variation in SNe~Ia
light curves is differing amounts of $^{56}$Ni created in 
the explosion.
As the light curves of SNe~Ia are determined by the decay of
$^{56}$Ni, variations in the amount
of $^{56}$Ni produced could lead to the observed variations in peak
brightness that follow the width-luminosity relation illustrated in
Fig.~\ref{fig:stretch_correction}.  Models producing different amounts
of $^{56}$Ni exhibit a width-luminosity relation similar to that observed, but they do not
plausibly account for the peculiar SNe~Ia like SN~1991T
(over-luminous, $1.0$~$M_\sun$~$^{56}$Ni~\citep{spyromilio92}) or 
SN~1991bg (under-luminous, 
$0.1$~$M_\sun$~$^{56}$Ni~\citep{filippenko92})~\citep{mazzali01} .


The chemical composition of the progenitor system of a SN~Ia is another physical
mechanism that could lead to differences in observed peak brightness.
The metallicity of the explosion, if not the progenitor itself, can be
studied by an examination of spectral features characteristic of
different elements formed in the explosion.
The quantity of most obvious interest is the ratio of carbon and
oxygen in the progenitor star.  This ratio directly affects the abundances of
elements created in the nuclear burning of the explosion.  Since the
decay of $^{56}$Ni is the primary source for the observed optical
luminosity of the SN, variations in the amount produced of this radioactive
element can have a significant effect on the luminosity of the
SN~Ia.

Indirectly, the light curve can also be affected by hydrodynamic differences
in the exact nature of the detonation/deflagration process that
produced the SN~Ia.  A variety of models and simulations exist
that address these issues, but they await detailed observations to
differentiate between them.

The SNfactory data set will allow for calibration and comparison of
supernova light curve shapes in {\em any} optical color.
With photometric spectroscopy, the SNfactory will be able to
synthesize a light curve at any redshift for any given filter matching
a rest-frame wavelength range from $3500$--$10000$~\AA, producing a
set of high-quality SNe~Ia template light curves for use in comparison
with higher-redshift SNe~Ia.
Reconstructing synthesized light curves from the SNfactory
flux-calibrated spectra at rest-frame wavelengths in the standard
astronomical filters (e.g. $UBVRI$ or $ugriz$) will allow for
immediate recalibration of existing intermediate- and high-redshift
SN~Ia data sets through a new understanding of the family of SN~Ia
light curves.  This improved set of template light curves will allow
for a comprehensive analysis of width-luminosity and color-color
relationships~\citep{wang03} in different filters.

\subsection{Providing a Definitive Set of SN~Ia Spectra}
\subsubsection{K-corrections}

\label{sec:k_corrections}



While the SNfactory data set will allow for
in-depth analyses to address questions regarding the physical nature
of SNe~Ia, an immediate benefit of the SNfactory data will be
an improved phenomenological understanding of SNe~Ia.
In particular, this comprehensive set of SN~Ia spectra will cover the
diversity of the SN~Ia family and will be invaluable in improving
K-corrections for observed magnitudes of SNe~Ia~\citep{kim96,nugent02}.


Observations of SNe~Ia are generally
conducted in several different filters for comparison of their light
curves and absolute magnitudes.  When comparing SNe at different
redshifts, it is important to properly correct the magnitudes in each
filter by the amount of the spectrum shifted into and out of the
filter bandpass as the redshift shifts the spectrum to longer
wavelengths.  These adjustments are known as K-corrections.
For an object with a complicated spectrum such as a
SN~Ia, it is important to track the spectral features as they move
through the different filters.  Correcting appropriately for this
redshift effect requires a detailed knowledge of the spectrum of the
supernova at every epoch.  
Ideally, for every distant supernova, one
would like a spectrum of a well-studied nearby supernova that
is a good match to the distant one.  This like-to-like matching necessitates
having a spectrum of a nearby supernova at every possible epoch for
every different type of supernova.

The SNfactory data set will provide a valuable resource for
both quantifying SN~Ia diversity and establishing the new standard
reference for SNe~Ia spectra.
Most critically, the SNfactory spectra will provide for direct matches
with photometrically observed high-redshift SNe~Ia.  The
K-correction can be done with the high-quality spectrum that is the best
match out of the 300 SNfactory SNe~Ia.  This comprehensive spectral database will bring
deterministic clarity to the current art of
K-corrections~\citep{leibundgut90,hamuy93,kim96,nugent02}.

\subsubsection{Dust extinction correction}

The spectrum of a supernova is affected by the
presence of dust along the line of sight to the 
supernova\footnote{Correction for atmospheric and detector effects is part of the calibration process.}.  Dust
between an observer and an object will both dim and redden
the observed flux from the object.  This dimming can be parameterized
as a function of wavelength~\citep{cardelli89,odonnell94}.  By taking
advantage of this relationship, the brightness of SNe~Ia can be
corrected for extinction by taking the color differences between the
observed supernova spectrum and a reference supernova spectrum from an
extinction-free region.  The latter generally come from elliptical
galaxies, as they are relatively free of dust in comparison to spiral
galaxies.
The number of supernova available for this color-based
extinction correction is currently quite small and more are needed to
minimize the uncertainty in this correction.

The need for well-established intrinsic colors extends beyond just
correction for dust extinction.  Dust is generally smooth in its
absorption variation over wavelength, but obtaining a pure spectrum,
unadulterated by host-galaxy dust, is still necessary to reduce our
reliance on the assumption that the effects of dust are understood.
Dust from the inter-galactic medium and our own galaxy is, of course,
unavoidable, but the former can be observed separately from the SN
while the latter has been well-studied~\citep{schlegel98}.






\subsection{Study of host galaxies}

Studies of the host galaxies of SNe~Ia will help answer questions
about the effects of environment on the formation and explosion of SNe~Ia.
The specialized integral field unit spectrograph built by the
SNfactory will allow for
simultaneous observations of both a supernova and its host galaxy.
Sufficiently large or nearby hosts will be covered by several separate
lenses.  Such spatial coverage will allow for a systematic study of host galaxy
properties and morphology, which is the closest we will ever be able
to come to examining the host environment of a Type Ia supernova.
The actual birthplace of the SN~Ia progenitor is perhaps impossible to truly 
determine as the likely progenitors for SNe~Ia are white dwarfs (see Sec.~\ref{sec:progenitor_models}), 
which can be billions of years old and have thus had the opportunity
to leave their stellar birthplace and wander about their host galaxy.

\section{Operational Strategy}
\label{sec:snfactory_operational}

A coordinated observational program is vital to accomplishing the
ambitious scientific goals of the SNfactory.  The expertise gained
from many successful supernova programs,
in particular a large pilot program organized by the Supernova
Cosmology Project in the spring 1999,
has gone into the design of the SNfactory integrated search and
follow-up pipeline and it is carefully constructed to yield a
homogeneous sample of SNe~Ia suitable for answering the key questions
discussed above.

\subsection{Detection}

\begin{figure}
\begin{center}
\plotone{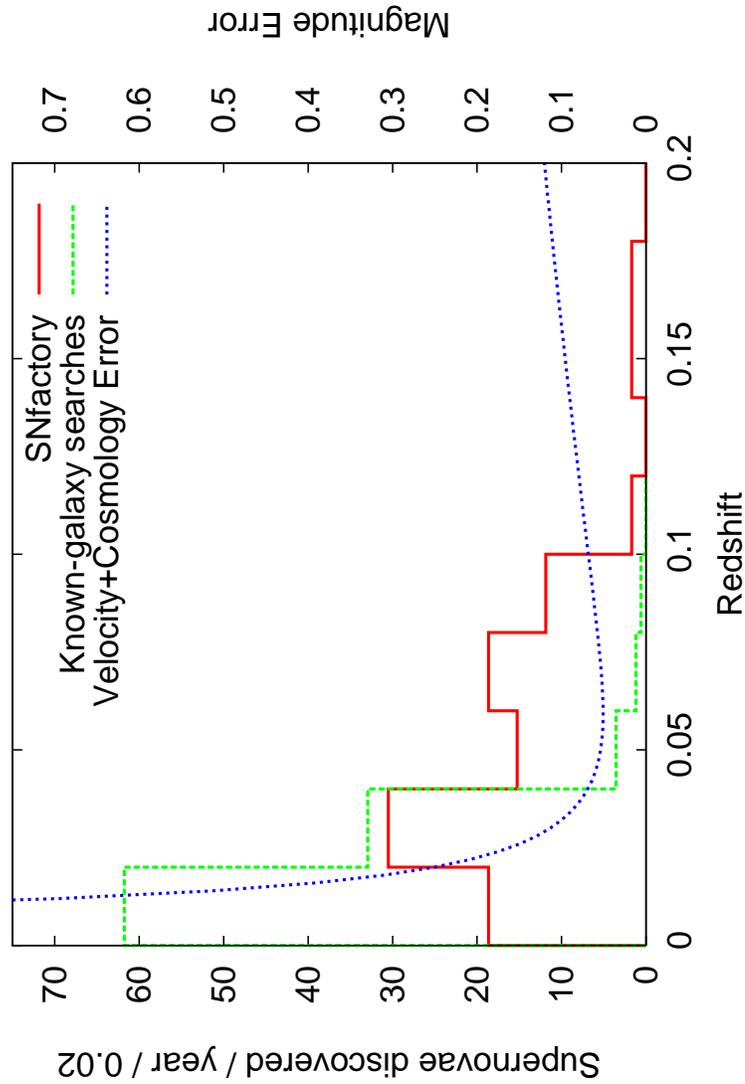}
\end{center}
\caption{The SNfactory is operating in the ``sweet spot'' between
peculiar velocity noise and cosmological uncertainty.  The SNfactory
histogram is the redshift distribution of SNe shown in
Fig.~\ref{fig:redshifts} scaled up to 100~SNe/year.
The known-galaxy search curve has also been scaled to
100~SNe/year.  The peculiar velocity of local galaxies, taken here to be 300~km/s~\citep{landy02}, dominates low-redshift part of the ``Velocity+Cosmology Error'' curve.  The cosmological uncertainty shown here is the difference in distance modulus in an $\Omega_M=0.3$, $\Omega_\Lambda=0.7$ Universe and an $\Omega_M=0.3$, $\Omega_\Lambda=0$ Universe.
}
\label{fig:snf_sweet_spot}
\end{figure}

The SNfactory will focus on SNe~Ia in the smooth linear Hubble flow
($0.03<z<0.08$).  SNe~Ia in this redshift range are sufficiently close
to allow for high signal-to-noise observations but are sufficiently
distant so that the redshift is a good indicator of distance (see Fig.~\ref{fig:snf_sweet_spot}).  In
addition, these SNe~Ia will be close enough so that the non-linear
expansion history of the Universe will not have a large effect on the
luminosity-distance for these supernovae.

The supernovae will be found using data from automated nightly
wide-field asteroid searches and will be studied from several weeks
before through several months after maximum brightness of the
supernovae.  It is necessary to search a large area of the sky every
night, $\sim500~\sqdeg$, to find enough SNe~Ia soon
after their explosion to study 100~SNe~Ia per year.  The volume of space out to a redshift of
$z=0.08$ is limited; on a given search night only $\sim0.01$ SNe~Ia
are expected to be detected before maximum light in any particular
square degree out to a limiting depth of $20.5$~magnitudes.


The SNfactory currently discovers supernovae 
using images from a collaboration with
the Near-Earth Asteroid Tracking (NEAT) group at the Jet Propulsion Laboratory.  In their quest for
asteroids, the NEAT scans the skies every night and looks for objects
that move over the time scale of an hour.  They take three images of a
given field in the sky, spaced fifteen--thirty minutes apart, and
search for objects which move by more than a couple of arcseconds over
this period.  They do this for hundreds of fields every night, covering $500$--$1000$\sqdeg.  This
strategy enables them to find thousands of new asteroids in the main
asteroid belt of our solar system and near-Earth asteroids in
particular.  The SNfactory uses these data and compares the new images
with reference images of the same field from previous years of NEAT
data by subtracting the reference image from the new image and looking
for the objects which remain.
 
The SNfactory search pipeline has been quite productive in a
several-year collaboration with the NEAT group.  In addition, a new arrangement
with the Palomar Consortium became effective in 2004 and is expected
to be equally productive.  Chapter~\ref{chp:search_pipeline} explains
how the SNfactory uses both the NEAT and Palomar Consortium data to
search for supernovae.

\subsection{Study}

\begin{figure}
\begin{center}
\plotone{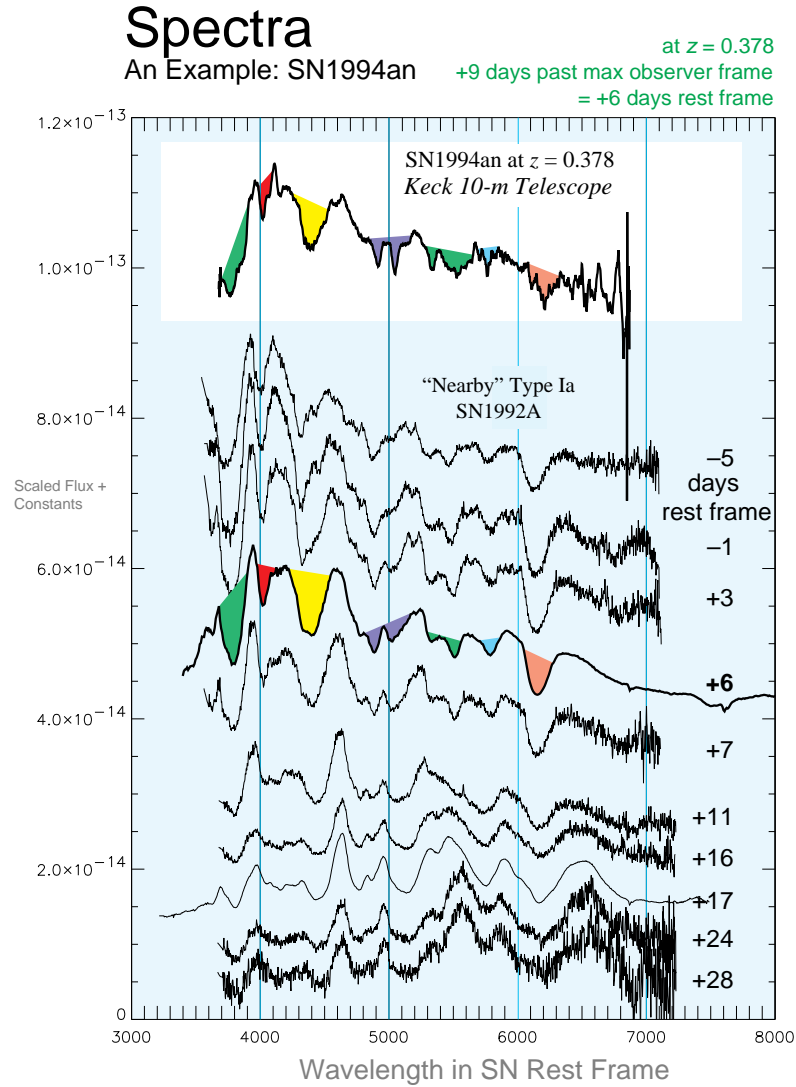}
\end{center}
\caption{Type~Ia supernovae exhibit remarkable uniformity in their
spectral evolution but show differences in specific spectral
features (illustrated here by colored regions) that may provide clues to better calibrate their luminosity
distances.}
\label{fig:spectral_series}
\end{figure}

A systematic study of hundreds of SNe~Ia calls for an automated,
guaranteed-time follow-up strategy.  The goals of the SNfactory
project require $12$--$15$ spectroscopic and photometric observations
of each SN~Ia.  Fig.~\ref{fig:spectral_series} gives an example
of this type of coverage for SN~1992A, although the SNfactory 
coverage will extend to earlier and later epochs in the evolution 
of the SNe~Ia.  The spectroscopic precision required for
spectrophotometric-quality observations necessitates an instrument
specially designed for the purpose.  To enable the comparison and
study of features across the entire observed wavelength regime, the
spectroscopic observations must be flux-calibrated to within one
percent.  This precision allows for photometry to be reconstructed for
any optical filter.  Practically speaking, the precision alignment
required to operate a traditional slit-spectrograph in an automated
mode is quite a challenge.  Few telescopes have the pointing precision
to line-up a target object to within the $0.5$\arcsec necessary for
traditional slit spectroscopy.  An integral field unit (IFU)
spectrograph obviates both static and wavelength-dependent slit loss
by collecting all of the light in a region through an array of imaging
elements covering 99\% of the total area.

SNIFS is a dedicated instrument built specifically
for the SNfactory project.  With two spectroscopic channels
($3000$--$5500$~\AA\ and $5500$--$10000$~\AA) and an integrated
photometric monitoring and guiding camera, SNIFS will provide
flux-calibrated spectroscopy through automated observation.  It
features a $6\arcsec \times 6\arcsec$ field-of-view covered by an
array of $15\times15$ lenselets (with $99\%$ coverage).  This IFU allows
for both simultaneous observation of a supernova and its host galaxy
and for a reduction in the pointing precision required to
$\pm1\arcsec$.  The SNIFS photometric channel features a custom-built
multi-filter that simultaneously monitor five bandpasses to measure
sky absorption and seeing in parallel with the spectroscopic
observations.

As of June, 2004, the SNIFS instrument has just been successfully
commissioned at the University of Hawaii 2.2-m telescope.  The
automated operation and spectrophotometric observations provided by
SNIFS on the UH 2.2-m. will allow the SNfactory to obtain observations
of the hundred supernovae a year required by this ambitious project.

\section{Summary and Plan for Subsequent Chapters}

The current intrinsic dispersion in SN~Ia peak magnitudes and
associated systematic uncertainties will soon become the limiting
factor in our determination of the geometry and expansion history of
the Universe using SNe~Ia.
It is necessary to improve our understanding and knowledge
of SNe~Ia before we can proceed with the next generation of distant
supernova searches.
The integrated detection pipeline and dedicated follow-up resources
of the SNfactory will allow for a detailed study of 300~SNe~Ia
that will provide a critical
foundation for the future of supernova and supernova cosmology
science.

The following three chapters discuss details of the search aspects of
the SNfactory, which was the primary focus of this dissertation work.
Chapter~\ref{chp:search_pipeline} covers the operation of the SNfactory
search pipeline.  Chapter~\ref{chp:subtractions} details the subtraction
software used to detect supernovae.  Chapter~\ref{chp:supernovae_found}
describes the supernovae found in the prototype search of the
SNfactory.  For further discussion of the follow-up aspects of the
SNfactory project, see \citet{aldering02b} and \citet{lantz03}.

\chapter{Search Pipeline Design and Implementation}
\label{chp:search_pipeline}

\section{To catch a supernova}

%

\begin{figure}
\plotone{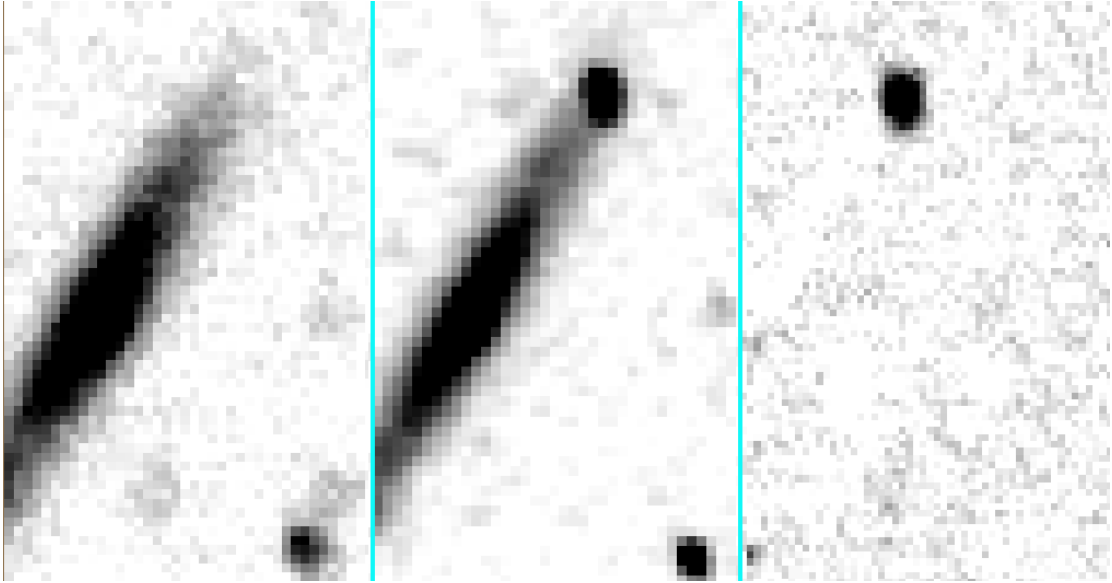}
\caption{(left) A reference image taken on 16 January 2001 (UT) showed 
the galaxy.  (center) A search image taken on 30 June 2001 (UT) showed the
same galaxy with a brighter region.  (right) Subtracting the reference
image from the search image yielded a new object, which was confirmed
to be a Type~Ia supernova.  Note that the galaxy and the star in the
lower-right both disappear in the subtracted image.  The area shown
here represents a small fraction a typical search image.}
\label{fig:sample_subtraction}
\end{figure}

The SNfactory search pipeline discovers supernovae using images
obtained in collaboration with the Palomar Consortium
(Yale/JPL/Caltech) and the Near-Earth Asteroid Tracking (NEAT) group
at the Jet Propulsion Laboratory.  The NEAT group uses the Palomar
1.2-m Samuel Oschin and Haleakala 1.2-m MSSS telescopes in its search
for near-Earth objects while the Palomar Consortium encompasses a
variety of projects using the Samuel Oschin 1.2-m, including searches
for quasars, supernovae, and trans-Neptunian objects.  The NEAT
program scans portions of the sky every night and returns to the same
fields every seven to fourteen days.  The Palomar Consortium uses the
Oschin 1.2-m telescope in drift-scan mode, typically covering a more
limited area of sky than the NEAT program and observing each field
only once per night, repeating the fields a few days later for
the purposes of SNe searching.  The SNfactory uses the repeated
coverage from both types of observations to search for new point
sources.  The search pipeline compares the old (``reference'') and new
(``search'') images of the same field by subtracting the reference
image from the search image and looking for the objects that remain.
These are the candidate supernovae.  Fig.~\ref{fig:sample_subtraction}
shows an example of this process for the discovery of SN~2001dd.



This ostensibly straightforward process of searching via image
subtraction turns out to be more complex than it may at first appear.
The presence of detector artifacts and variations in the image quality
of reference images and search images are some of the most significant
challenges.  The search pipeline uses a sophisticated suite of image
tools, but up to the present time the final step in the vetting
process still requires
human input to separate the good supernovae candidates from the bad.
On the order of one percent of the search images taken are found by
the automated processing and subtraction software to have a
potentially interesting object.  Roughly one percent of those objects
turn out to be supernovae.

From the fall of 2002 through the spring of 2003, a systematic search
for supernovae was carried out with the Palomar 1.2-m Samuel Oschin
telescope.  In 2001, the NEAT group outfitted this telescope with an
automated control system and added a 3-chip 3\sq\deg~field-of-view
(FOV) CCD camera (NEAT12GEN2) at the spherical focal plane at the
primary focus of this Schmidt reflector.  In April of 2003, this
camera was replaced by the QUEST group at Yale with a 112-chip 
9\sq\deg~FOV detector (QUESTII) capable of both drift-scan and
point-and-track observations.  QUESTII became fully operational in
August of 2003.  The SNfactory also processes data from the Haleakala MSSS
1.2-m. telescope with the NEAT4GEN2 detector.  These images are not
currently included in the SNfactory search because of their poor quality and
under-sampled resolution.


\section{Processing Steps toward Supernova Candidates}

The SNfactory pipeline has the ability to automatically process and
search images from all of the aforementioned telescopes through a well-defined
processing framework.
The basic image cleaning routines were adapted from the Supernova
Cosmology Project (SCP) Deeplib C++ framework.
The processing of the NEAT data is described below.
The same steps apply to images from the NEAT4GEN2, NEAT4GEN12, and
QUESTII detectors taken using the telescopes' point-and-track mode.
A special section on the differences involved in the processing the
drift-scan data from the Palomar Consortium follows.
The specific details of the implementation presented here apply to
the SNfactory plans and setup as of the spring of 2004.

The basic search sequence is as follows:
\begin{enumerate}
\ssp
\item Obtain reference images.
\item Reduce reference data.
\item Obtain search images.
\item Reduce search images.
\item Perform subtractions.
\item Perform automated scanning.
\item Perform human scanning.
\item Perform cross-checks.
\item Obtain confirmation image.
\item Obtain confirmation spectrum.
\item Report discovery.
\dsp
\end{enumerate}
The sections that follow discuss the SNfactory's implementation of
these steps.  
In the sections below, the discussion of the processing steps will
diverge a little from the order in which they are presented above. The
discussions of the observation and image reduction steps will be
combined because these processes are the same for both reference and
search images.


\section{Sky Coverage}
\label{sec:sky_coverage}

Using the Oschin and MSSS telescopes, the NEAT group takes many images of the
night sky over a range of right ascension (RA) and
declination (Dec).  All sky fields are covered in at least three
exposures spread over approximately an hour.  This temporal spacing
allows the NEAT group to search for asteroids at the same time
that it enables the SNfactory
search pipeline to eliminate those same asteroids and to minimize
contamination due to cosmic rays.  Ideally, the fields obtained cover
some simple rectangle on the sky to facilitate later searching images
and to provide a complete set of references for future searches in
later years.  In practice, the coverage pattern is more complicated.
A significant challenge for the SNfactory project was transferring all
of these imaging data (up to 60~GB/night) to the computing resources
necessary to process the images.

\section{Data Transfer from Palomar to LBL}
\label{sec:data_transfer}

\begin{figure}
\begin{center}
\scalebox{0.5}{\includegraphics{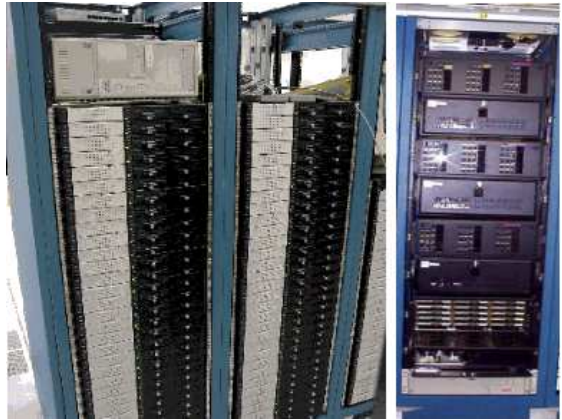}}
\end{center}
\caption{The Parallel Distributed Systems Facility (PDSF) serves several
major research groups.  It schedules jobs across $\sim400$ nodes
with a fair-sharing queuing system.  Images courtesy of NERSC and PDSF.}
\label{fig:pdsf}
\end{figure}

The image data are obtained from the telescope and stored at Lawrence
Berkeley Laboratory (LBL) on the National Energy Research
Supercomputing Center (NERSC) High Performance Storage System (HPSS).
From there, the images are transferred to the NERSC Parallel Distributed Systems
Facility (PDSF) (see Fig.~\ref{fig:pdsf}).  This cluster comprises approximately 200 Dual 1-GHz
Pentium IV PCs with 2 Gigabytes (GB) of memory and 50 GB of scratch disk
space apiece.  Processes are scheduled on the cluster using the 
Sun Grid Engine\footnote{The Sun Grid Engine bears no relation to Grid computing} software package.

As each image is read-out from the telescope, it is saved in a compressed
format to local disk space.  There are currently several hundred
gigabytes of storage at the observatory for this purpose.  This 
storage capacity allows
for the 20--50 GB/night of data in compressed form to be stored with space to buffer
several nights of data in cases of transmission failure.

A high-speed, 45~Megabit-per-second (Mbps) radio internet link has
been established between the Palomar Observatory and the San Diego
Supercomputer Center (SDSC) as part of the High Performance Wireless
Research and Education Network (HPWREN)~\citep{hpwren}. 
This link is used to transfer the images from Palomar to HPSS in near real time.
The 45~Mbps bandwidth comfortably exceeds the data rate from the telescope.
The bandwidth
from SDSC to LBL and NERSC is excellent and several orders of magnitude
greater than necessary to meet the transfer requirements.

The observing plan for the NEAT12GEN2 detector used from 2001--2003
put the telescope through three pointings every four minutes.  There
were 3 CCD detectors on the instrument so each pointing resulted in 3
images.  These images were spaced about 1~degree apart in declination.
Each image was $32$ Megabytes (MB) in its raw form but was compressed
to $\sim16$~MB for transfer.  This produced $144$~MB of data to be
transfered every four minutes, resulting in a data rate of $4.8$~Mbps that used only
one-tenth of the theoretical maximum bandwidth of the radio internet
link and thus left plenty of room for future expansion.  For the
currently operating QUESTII camera, a similar prescription is followed for the
point-and-track images, although the $40$--$60$~GB of data produced
every night generate a higher data rate of $10$~Mbps.  The
drift-scan data obtained through the Palomar Consortium is handled
separately and packaged and stored on HPSS by the Yale QUEST group.

Each night a transfer script is initiated at 1800 local Pacific Time
\footnote{All times given in this chapter will be local Pacific Time.} on the
LBL SNfactory machine at Palomar, Berlioz.  The transfer
script, \code{neat\_to\_hpss.pl}, looks in a known directory for files to transfer.  It is keyed
to images only from that night, so, if there are other files or images
from other nights in that directory, \code{neat\_to\_hpss.pl} ignores them.
The script keeps an updated list
of files to be transferred and a list of files that have already been
transferred.  The transfer to HPSS is accomplished through a
scripted call to \code{ncftp}.  Experiments comparing transfer
rates using \code{scp} and \code{ncftp} to transfer data from Palomar
to HPSS revealed that \code{ncftp} gave better performance by a factor of
two.

With the data rates given above, the transfer process is almost
real-time.  The only delay comes from the few minutes it takes to
compress the images for transfer.  To allow for this delay, the transfer
script currently conservatively waits for ten minutes after the
creation of an image file to make sure the compressed version has been
completely written to disk before adding that file to the list of
image files to transfer. The transfer script for a given night runs
continuously from 1800 to 1755 the next afternoon.  It is then
restarted for the next night at 1800.

At 1000 every morning another script, \code{check\_script.pl}, is run
to verify the transfer of the previous night's images.  This script
compares a list of local image files from the previous night with the
list of files in the appropriate directory on HPSS.  If the file list
names and sizes agree, then an email indicating a successful transfer
is automatically sent to the LBL SNfactory system administrator and
the NEAT collaborators at JPL.  This email includes a list of the
images transferred as well as the images' compressed file sizes.  If
there is a discrepancy between what was transferred and what should
have been transferred, a warning email is sent to the LBL SNfactory
system administrator.  At the same time, however, the check script
automatically flags those images for transfer, causing them to be
resent by the transfer script, which runs throughout the day. The
check script is rerun at 1200 and 1700 to make sure that any images
that were delayed in their compression to disk are included in the
final tally and also to provide confirmation of a complete transfer if
the original run of \code{check\_script.pl} at 1000 did not 
report a successful transfer of the full night of data.

This transfer setup has been working continuously since August 2001.
Minor improvements in handling error conditions such as files of zero
size were in made in September and October of 2001.  After running
completely unattended from January 2002 through May 2003,
\code{neat\_to\_hpss.pl} was slightly adapted to the slightly different
file format of the new QUESTII detector system and has been running
without any need for intervention since.

\section{Data Processing of Palomar Images}
\label{sec:data_processing}

The basic data processing steps are uncompression, conversion to the
standard astronomical FITS format, dark-subtraction, flat-fielding, and
loading into the SNfactory image database.

During the course of a point-and-track observing night at Palomar, the images are
transferred to HPSS after a delay of 10--15~minutes.  Once an hour, a
cron job\footnote{A cron job is a command that
has been scheduled to run at specified times on a given system.  All
PDSF cron jobs for the SNfactory are run from
\code{pdsflx001.nersc.gov}.} is run on the PDSF cluster at NERSC to
fetch the most recent images from HPSS and submit them for processing.
An \code{analysis} table stored in the main SNfactory
PostgreSQL\footnote{\url{http://www.postgresql.org}} database,
currently on \computer{wfbach.lbl.gov}, keeps track of which images
have been processed or submitted for processing.  This database allows
for reductions to be restarted and provides information in the case of
a failed job.  As the NEAT4GEN2, NEAT12GEN2 and point-and-track
QUESTII camera images are reduced in groups according to the dark
calibration image closest in time, the very last group is not
submitted for processing until the morning.  Every day at 1200, a
final cron job runs on the PDSF cluster.  This job is responsible for
finishing the processing of any images not already reduced during the
night.  Any remaining images are downloaded from HPSS and submitted to
the cluster queues for processing.

Grouping the raw images by the closest dark calibration image results in $20$--$30$ data reduction
sets.  These groups are submitted as jobs to the cluster processing queue.
Processing of these jobs takes three to five hours, depending on the load on the
cluster.  After all of the images are reduced, they are saved back to
long-term storage in their processed form.

If all of the processing were done at the same time, it would take
from four to five hours to retrieve and reduce all of the images for a
night.  The retrieval of the files from HPSS generally only takes an
hour of this time as recently saved images that are still in the HPSS
spinning disk cache.  This retrieval delay could be improved by
sending the files first to the cluster and then to HPSS, but such a
strategy raises concerns about security and reliability.  For the
purposes of the SNfactory, \code{ncftp} is preferable to \code{scp}
because the former program permits much faster transfer rates.
However, for reasons of security, running an \code{ncftp} server on
the PDSF cluster machines is not allowed.  Reliability is an important
issue for an automated system and HPSS has a more reliable uptime than
the PDSF cluster; while HPSS is unavailable every Tuesday from
1000--1200 for scheduled maintenance, PDSF experiences less
predictable, unexpected downtime.  Thus, the decision was made to
transfer images originally to HPSS and then download them from HPSS
to PDSF hourly during the night.  This transfer takes less than an
hour and so keeps up during the night with the data rate from the
telescope.  There is, therefore, very little overall delay incurred by having
to download from HPSS.

Each data reduction set is processed by a separate job that runs on
its own CPU in the PDSF cluster.  The job begins by copying the images
to be processed from the cluster central storage to local scratch
space.  Next, the files are decompressed and converted from the NEAT
internal image format (use by the NEAT12GEN2 and NEAT4GEN2 cameras) to
the standard astronomical Flexible Image Transport System (FITS)
format (as defined by NASA~\citep{fits}) as the SNfactory software is
designed to understand this standard format for the processing of
astronomical images.  The QUESTII images come from the telescope as
compressed FITS files.  See Appendix~\ref{apx:compressed_fits} for
a study demonstrating that the pipeline processing is more efficient 
using uncompressed FITS files.

After the images have been converted to standard FITS format, the dark
current from the CCDs is removed from the sky images using dark
images.  These calibration images are taken with the same exposure
time as science images but with the shutter closed.  This type of
exposure measures the amount of signal collected by the CCD from the
background temperature of the device.  It is advisable to remove the
offset created by this detector glow by using dark images from the
same night because dark current can vary over time as well as from
pixel to pixel.  This calibration is particularly important because
the NEAT detectors at both Palomar (NEAT12GEN2) and Maui (NEAT4GEN2)
are thermoelectrically cooled.  This type of cooling allows the
telescope to run unattended since there is no need to continually refill
a nitrogen dewar\footnote{SNIFS uses a closed-system CryoTiger to
maintain cryogenic temperatures, eliminating the need for manual
refills of the dewar.}.  However, thermoelectric cooling also means
that the NEAT CCDs run at a higher temperature than a dewar cooled by
liquid nitrogen and thus a significant number of photo-electrons are
detected by the CCD from the detector itself.  This ``dark-current''
signal needs to be subtracted from any observation of the sky.  The
QUESTII camera is cryogenically cooled and so exhibits a much smaller
dark current, but dark images are still taken and used in the
processing as the dark current is still a noticeable signal that needs
to be removed from the science images.  In practice, the average value
and standard deviation of the dark images from all three detectors remains
relatively constant over a night.

After the dark calibration image is subtracted from each of the other
images in the data set, the next important calibration step is
to account for the pixel-to-pixel variation caused both by variations
in the sensitivity of each pixel and by the different illumination coming
to each pixel through the optics of the telescope.  To correct for
this difference in the effective gain of each pixel, one must construct
a ``flatfield'' image that has an average value of 1 and has the
relative sensitivity for each pixel stored as the value of that
pixel.  The sky images are then divided by this flatfield image to
arrive at images that have an effective equal sensitivity for every
pixel.  The flatfield images are built by taking sets of images of the
sky at different positions and determining the median at each pixel.  The
median process includes an outlier rejection step to recover a more
representative median.  With a sufficiently large set (typically 21
images), this medianing eliminates objects on the image and leaves a
fiducial sky.

For most of the year of 2002, flatfields were built every night from
the individual data sets.  This resulted in flatfields that were most
attuned to that night, but it also ran the risk of not having the best
available flatfields in the event of problems with a particular
night's images.  For a while, the image processing suffered from
flatfields built from too few images, a limitation that led to
residual objects and stars being clearly present in the flatfields.
When images calibrated with these not-so-flat flatfields were used as
references, false objects would appear in the subtractions.  This
difficulty in creating flatfields became quite a problem when it was
realized that on some nights only 70\% of the images were being
successful reduced due to problems in building flatfield images within
a given data set.  Because this situation often developed for small
data sets from which a flatfield could not reliably be built, a
generic flatfield was built from 27 May 2002 images to flatten all of
the data.  

This one-time flatfield generating was successful and led to the idea
of computing generic flatfields to be updated once a month based on
the data from the previous month.  The goal would be to speed up
processing by about twenty minutes per data set (the time it takes to
build the flatfield) and have fewer bad flatfield images.
However, this approach does not use all of the available information from a
given night to build the best flatfield possible.  In the fall of
2002, a compromise was reached: flatfields were generated when
possible from the data set itself, but stock flatfields were used when
there were too few images to generate a flatfield for that set of
data.

Fringing on the QUESTII detectors complicated the process of
generating flatfields.  The QUESTII detectors are both thinner and
more sensitive in the red than the older NEAT detectors and so exhibit
fringing at less than $5\%$ deviation from background in the $z$, $I$,
and RG-610 filters.  The NEAT4GEN2 and NEAT12GEN showed no visible
evidence of fringing.  Although fringing is an additive effect, the
processing pipeline pretends that it is multiplicative for the
purposes of flatfielding the images as described above.
A proper fringing correction is computationally
intensive, and fast processing is a higher priority than precise
photometry.

After being flatfielded, the images are then split up into four
quadrants---one for each of the four different amplifiers.  Each CCD
has four amplifiers to achieve the fast 20-second readout times that
are critical to this type of large-area variable object survey.
Splitting the images into amplifier quadrants allows for the
flexibility to discard bad quadrants when they develop problems.
Because various problems resulting in failed or unreliable amplifiers
have developed several times for both the Haleakala and Palomar
detectors, this splitting scheme has proven quite useful.  There is
some additional areal coverage loss in the subtractions because of the
additional overall edge space in this scheme.  As the images are
dithered (see Appendix~\ref{apx:dither}), a little space around the edges is always lost when the
images are added together.  This areal loss is proportional to the
edge space and so is more expensive for small images, but there is no
practical alternative because dealing with non-rectangular images is
not a viable option.  After the images are split, the aforementioned
quadrants known to be bad are then discarded.  A list of bad
quadrants, including dates, is kept in the reduction scripts and is
referenced to decide which quadrants to eliminate.  For the QUESTII
camera, each of the 112 CCDs has only one amplifier.  There are ten
known bad CCDs in the array, and raw images from these CCDs are not
submitted for processing.

The final step in the basic image calibration is to take the fully
reduced images and move them back to central cluster storage.  A
separate processing job then renames the images to match the SNfactory
canonical name format and registers them with the image database.
This image loading step is done as a separate job because it uses code
written in the Interactive Data Language (IDL), which is a proprietary
fee-for-license language sold by Research Systems Incorporated
(RSI),\footnote{\url{http://www.rsinc.com}} and there is a limit to
the number of simultaneous jobs that are allowed to run under the
SNfactory licensing agreement with RSI.

Once all of the images for a night have been reduced, they are
archived to long-term storage on an HPSS system at NERSC.  This
archiving is done on all of a night's images in one batch so that all
of the images for a given chip or amplifier are saved together in one
tar file using \code{htar}.  Tarring the files together in this way 
results in greatly increased performance in later access to these
images from HPSS as HPSS is designed to handle a few large files much
better than a large number of small files.

The image reduction can be done as early as 1300 every day.  After all
of the images for a night are processed, they are first matched to the
other images from the same night and then matched to historical
reference images from the previous observing season to produce image
sets to submit for subtraction.



\section{QUESTII Image Processing - Some Details}
\label{sec:questii_data_processing}

The steps described above had to be modified slightly to accommodate
the QUESTII camera that was added to the Palomar Oschin 1.2-m
telescope in April, 2003.  This camera can be operated in either of
two modes, drift-scan and point-and-track.  The NEAT group uses the
camera in the standard point-and-track mode, while the QUEST group
uses it in its drift-scan mode.  There are some differences in the
processing of the images taken in each mode.

Drift-scanning is an observing mode in which the telescope is pointed to a
desired declination and hour angle and then left fixed with respect to
the Earth.  The sky thus passes over the CCD as it records.
The readout of the CCD is synchronized to this sky motion by reading
out a row of the CCD and shifting all of the other rows in time with
$15\arcsec$/second.  The QUESTII camera has $0.87\arcsec$ pixels and so
clocks out at $15\arcsec/\mathrm{second} / 0.87\arcsec /\mathrm{pixel}
\approx 17~\mathrm{pixels}/\mathrm{second}$ at a declination of zero degrees.
\footnote{In general, the clock rate is set at $15\arcsec\times\cos{\mathrm{Dec}}$}
In a simplified picture,
this fixed scan results in just one exposure per observing night.  The
QUESTII camera has four separate fingers running parallel along near
constant lines of RA, a setup that results in four images of each part
of the sky observed in a given night.  In practice, there are
generally several scans taken to cover areas in the sky at different
declinations.  In addition, calibration scans are taken to generate
equivalent dark frames and sky flats.

For the QUEST project the sky observations are taken in either of two
filter sets, UBRI and rizz, while the NEAT group has chosen an RG-610
filter, a long-pass filter beginning at $6100$\AA.  The NEAT group has
used unfiltered observations on the previous Palomar camera and at
Haleakala.  The choice of an RG-610 filter for the NEAT observations
with the QUESTII detector would appear non-ideal both for the
purposes of the NEAT group, which selected the filter, and for the
purposes of the SNfactory.  Certainly, if there were no sky, then it
would be in the SNfactory's best interest to conduct the supernova
search unfiltered to maximize the number of discovered supernovae.  At
the same time, however, it is also beneficial to be able to compute
magnitudes on a standard filter system for comparison with other SNe
observations.

The SNfactory discussed the relative merits of filtered vs. unfiltered
observations with the NEAT group and it was determined that the
advantages and disadvantages of using a standard filter balanced out.
The light lost in dark time was balanced out by the extra light
suppressed in bright time.  However, the choice to use a low-pass
filter like the RG-610 is non-ideal as it eliminates a large fraction
of the wavelength range where the Sun is brightest (asteroids are
visible through reflected sunlight) while still allowing contamination
from sky glow.  It does avoid the increased scattered light from the
Moon during bright time, but there is an overall loss of sensitivity
to search for asteroids.  In addition, the long-pass filter results in
fringing on the science images from sky lines in the near-infrared.
For supernova work, the RG-610 filter choice is similarly
inconvenient.  While there are reasons one might choose to use a
filter in supernova searches--one might give up V-band coverage, where
supernova are brighter than in R-band, in order to be on a known
filter system such as a Johnson-Cousins R-band--but there is no such
useful tradeoff in this case.
  
That said, the non-optimal sensitivity caused by this filter choice
should not cause any significant problems with the search because the
improved QE and reduced pixel scale of the new camera yield a net
increased in detection sensitivity of $0.7$ magnitudes over the previous
NEAT12GEN2 camera discussed in Sec.~\ref{sec:data_processing}.

Surprisingly, it was found that the new QUESTII RG610-filtered
observations yielded few residuals when subtracted against the older
unfiltered Palomar images.  There are some residuals at the center of
stars, likely due to differential sampling of the different PSFs, but
there are no obvious cross-filter-related problems.  These good
cross-filter subtractions seem to
indicate that most of the stars used to calculate flux ratios
between images were the same color as the galaxies.  It is also
possible that the galaxies are being slightly over-subtracted to a
degree not clearly visible in the subtractions.  The reliability of
cross-filter subtractions should be investigated further if the
SNfactory decides to use them as a standard part of the search program.

\section{Search Images}

The telescope observing program covers the same fields in the sky as
often as possible while assuring full sky coverage from approximately
$-40$\deg to $+40$\deg in declination (the Haleakala telescope is the
main source for regions with declinations $\lesssim-25$\deg).  In
ideal conditions, the full usable sky pattern can be repeated in as
few as six days, but bad weather and varying moon illuminations make
for a more complicated observing pattern that can average out to as
long as two weeks between initial and repeat coverage.

However, since there is now a multi-year baseline of images to select
from, the SNfactory search pipeline now conducts subtractions against
the full seasons-worth of year-old references for a particular field.
These images are stacked together in a co-addition that spans the
union of the area covered by the reference images. Eventually, a set
of super-reference images should be made that maximizes the use of
information from both good and bad images to produce an optimal
reference tile for each field on the sky.  These premade references would reduce
processing time and allow all of the reference tiles of interest to be
loaded during the season for which they apply.  


Once the search images are taken, they are transferred to LBL and processed as described above.  These
images will then become the reference images for searches when the search
returns again to these fields the following year.
Once the search images are loaded into the database, a matching
program, \code{findmatches.pl}, is run to construct lists of matching fields for the reference
and search nights.  These lists are then used in generating files which describe the
image subtractions to be done to look for supernovae.  This matching
takes about $30$~minutes, comparing all of the images for a night to all
of the images from the previous year and generating an appropriate
list of subtractions to be done.  
As of the summer of 2004, the pipeline performs subtractions against 
reference images taken $360\pm240$~days before to the search images.

\section{Image Differencing: Subtraction}

The SNfactory has adapted the SQL database framework and image
subtraction and scanning IDL code developed by the SCP to the specific
needs of a low-redshift, large-area search.
Chapter~\ref{chp:subtractions} chapter presents a detailed description
of the image subtraction techniques employed and improved upon by the
SNfactory.

\section{Object Identification and Scoring}

After the subtraction is complete, an automated scanning program 
takes the final result of the image subtraction and looks for
remaining objects.  A variety of cuts are applied to the objects found in
the subtraction to eliminate cosmic rays, asteroids, and detector
artifacts such as hot pixels, bad columns, stray light, and an array
of other effects that have been observed over the years.
These cuts are based on ``scores'' that describe features of each candidate
as outlined below.

\subsubsection{Candidate Scores}
\label{sec:scorecuts}

For each object found in the subtracted image, a set of quantitative
scores is calculated by the automated scanning program.
Table~\ref{tab:candidate_scores} shows a list of these scores while
Table~\ref{tab:candidate_scorecuts} shows the thresholds used by the SNfactory.

\begin{table}
\begin{tabular}{ll}
APSIG           &  The signal-to-noise in the candidate aperture (ap.) \\ 
PERINC          &  The percent increase from REF$\rightarrow$NEW in the candidate ap. \\
PCYGSIG         &  Normalized: flux in 2*FWHM ap. - flux in 0.7*FHWM ap. \\
MAXPIXSIG       &  Limit on maximum pixel value (unused) \\
MXY             &  The X-Y moment of the candidate \\
FWX             &  The FWHM of the candidate in X \\
FWY             &  The FWHM of the candidate in Y \\
NEIGHBORDIST    &  Distance to the nearest object in the REF\\
NEIGHBORMAG     &  Magnitude of the nearest object in the REF \\
MAG             &  Magnitude of the candidate \\
THETA           &  Angle between the candidate and nearest object in the REF \\
NEW1SIG         &  Signal-to-noise of candidate in NEW1 \\
NEW2SIG         &  Signal-to-noise of candidate in NEW2 \\
SUB1SIG         &  Signal-to-noise of candidate in SUB1 \\
SUB2SIG         &  Signal-to-noise of candidate in SUB2 \\
SUB2MINSUB1     &  Weighted signal-to-noise difference between SUB1 and SUB2 \\
DSUB1SUB2       &  Difference in pixel coordinates between SUB1 and SUB2 \\
HOLEINREF       &  Signal-to-noise in aperture in the REF \\
BIGAPRATIO      &  Ratio of larger aperture to smaller aperture of candidate \\
OFFSET          &  Correlation with neighbor distance and angle on subtraction \\
RELFWX          &  Candidate FWHM in X divided by NEW image FWHM in X \\
RELFWY          &  Candidate FWHM in Y divided by NEW image FWHM in Y \\
\end{tabular}
\caption{The quantitative scores used by the automated scanning
program to assist in identification of likely supernovae.}
\label{tab:candidate_scores}
\end{table}

\begin{table}
\begin{tabular}{lll}
Score           & Minimum & Maximum \\
APSIG           &  5      & --      \\
PERINC          &  25\%   & --      \\
PCYGSIG         &  0      & --      \\
MAXPIXSIG       &  --     & --      \\
MXY             &  -50    & +50     \\
FWX             &  1.01   & 7.0     \\
FWY             &  1.01   & 7.0     \\
NEIGHBORDIST    &  0.5    & --      \\
NEIGHBORMAG     &  --     & --      \\
MAG             &  --     & --      \\
THETA           &  --     & --      \\
NEW1SIG         &  3.5    & --      \\
NEW2SIG         &  3.5    & --      \\
SUB1SIG         &  3.5    & --      \\
SUB2SIG         &  3.5    & --      \\
SUB2MINSUB1     &  -2.5   & +2.5    \\
DSUB1SUB2       &  --     & 0.5     \\
HOLEINREF       &  -5.0   & --      \\
BIGAPRATIO      &  --     & 5.0     \\
OFFSET          &  --     & --      \\
RELFWX          &  --     & --      \\
RELFWY          &  --     & --      \\
\end{tabular}
\caption{The thresholds for the scores described in Table~\ref{tab:candidate_scores} as set for the Palomar 1.2-m NEAT12GEN2 detector.  Each detector has a separate set of thresholds tuned to eliminate the characteristic false candidates on subtractions from that detector.}
\label{tab:candidate_scorecuts}
\end{table}

These scores are kept as single floating point numbers and can be
restricted by both minimum and maximum limits.  For some scores a
minimum restriction is obvious, as for APSIG and PERINC.  For others,
for example SUB2MINSUB1 and DSUB1SUB2, the minimum possible value of 0
is a good score; for these scores, only a maximum restriction is used.
Table~\ref{tab:candidate_scorecuts} shows the final thresholds used
for the scores in Table~\ref{tab:candidate_scores} for the NEAT12GEN2
detector.  These scores were determined in consultation with the
experienced SNfactory scanners (see Sec.~\ref{sec:human_scanners}) to
eliminate non-supernova objects detected in the subtractions in favor
of the good supernova candidates.  Eventually, the optimal thresholds
for these scores should be determined by analyzing the quantitative
score parameter space occupied by the real supernovae as compared with
the non-supernova objects.

\subsubsection{More on Scores Added for the SNfactory}

To reduce the number of objects to be scanned by a person,
a number of scores were added to the already existing list
from the SCP subtraction code: 
HOLEINREF, BIGAPRATIO, OFFSET, RELFWX, and RELFWY.

HOLEINREF was created to discriminate against depressed regions in the
reference image that would naturally lead to false positive regions in
the subtraction.  This score calculates the signal-to-noise ratio in a
1-pixel aperture at the candidate location.  It is designed more to
eliminate defects and bad pixels than to discard large-scale
depressions.  But, as the large-scale negative regions on the
reference will not lead directly to a point-source-like object on the
subtraction, it is more important to deal with the pixel-level
defects.

RELFWX and RELFWY were designed to ensure that the object shape in the
subtraction is at least roughly consistent with point-sources in the
NEW image.  These scores are particularly helpful in eliminating
artifacts in the subtraction due to bad or elevated columns in the NEW
or REF images.  Similarly, the MXY score detects objects with X-Y
moments that would indicate a significant distortion.  It is not
weighted by the characteristic MXY of the NEW image, although perhaps
such a RELMXY score could be considered in future improvements.

A table is kept in the PostgreSQL database that stores all of these
scores for every object found on every subtraction.  This table allows
the SNfactory to search the parameter space defined by these scores to
distinguish supernova candidates from non-supernova objects and
subtraction or detector artifacts.  Currently the scores are
implemented independently, but it is likely that there are more
complicated combinations of scores that might prove more useful in
discriminating good candidate supernovae.  For example, the HOLEINREF
score could be used in conjunction with the APSIG score.  As the
HOLEINREF score is designed to eliminate objects that result from
negative regions in the REF, one could calculate the contribution that
the negative amount measured by the HOLEINREF score would have on the
APSIG and increase the APSIG cut by that amount.  In this way, an
object with an APSIG of 100 could still be accepted even if its
HOLEINREF score were -5 (currently excluded by the default score
cuts).

HOLEINREF is an important score and is particularly helpful in
quantifying the contribution to the subtracted images from negative
regions in the REF.  Because of the display of the images, human
scanners often have difficulty determining the extent of a negative
region in the REF.  For this reason, there is an inverted grey-scale 
display of the region around a candidate in the REF image shown in
the tiles view (see Fig.\ref{fig:scantng}).

RELFWX and RELFWY were very effective in eliminating objects obviously
inconsistent with a PSF.  Specifically, false objects from unmasked
bad columns typically have a non-symmetric shape elongated along the
column.  Cosmic rays have generally been eliminated by the multiple
NEW images, but low pixels in the REF that are not caught by HOLEINREF
can still persist and result in objects in the subtraction.  However,
these objects, are narrower than the characteristic PSF score and so
can be eliminated by cuts on RELFWX and RELFWY.

\begin{figure}
\plotone{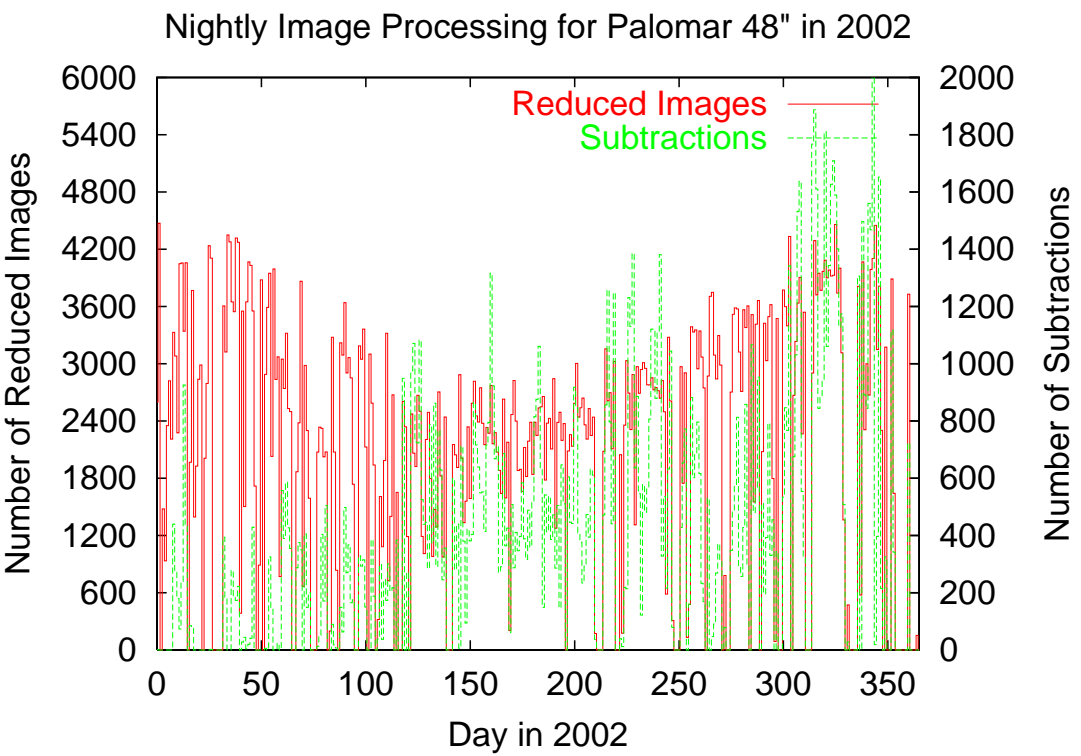}
\caption{The nightly image processing numbers for the Palomar 48"
Oschin telescope for 2002.  The solid (red) line denotes the number of images
(measured in individual CCD amplifiers) and the dotted (green) line
gives the number of subtractions completed for the night. }
\label{fig:oschin_reductions_subtractions_2002}
\end{figure}

\begin{figure}
\plotone{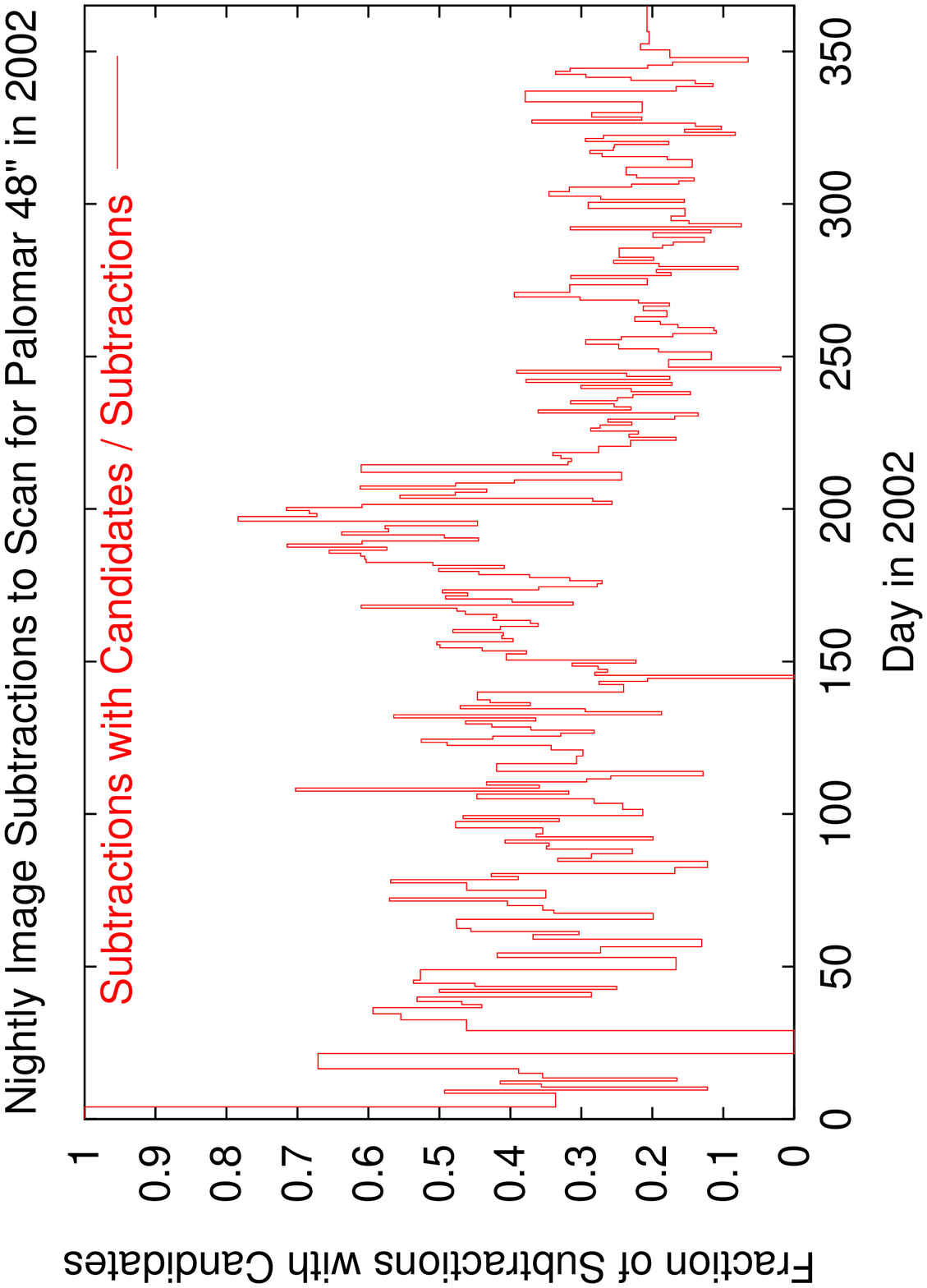}
\caption{The fraction of the subtractions (see
Fig.~\ref{fig:oschin_reductions_subtractions_2002}) that had to be scanned
by human eye in 2002 given the automatic scanning score cuts in use at the time.}
\label{fig:oschin_subtractions_fraction_2002}
\end{figure}

\begin{figure}
\plotone{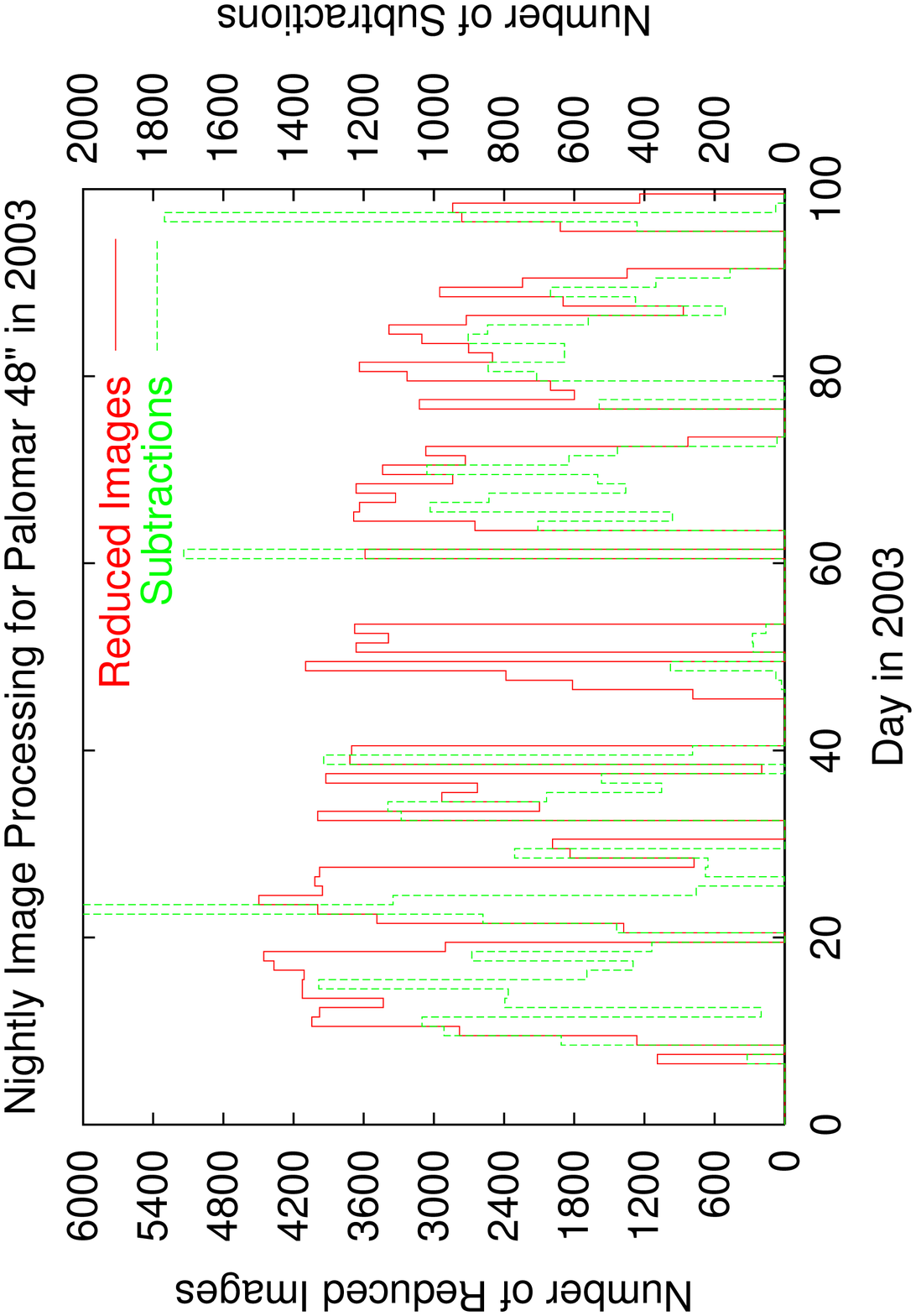}
\caption{The nightly image processing numbers for the Palomar 48"
Oschin telescope for 2003.  The solid (red) line denotes the number of images
(measured in individual CCD amplifiers) and the dotted (green) line
gives the number of subtractions completed for the night. }
\label{fig:oschin_reductions_subtractions_2003}
\end{figure}

\begin{figure}
\plotone{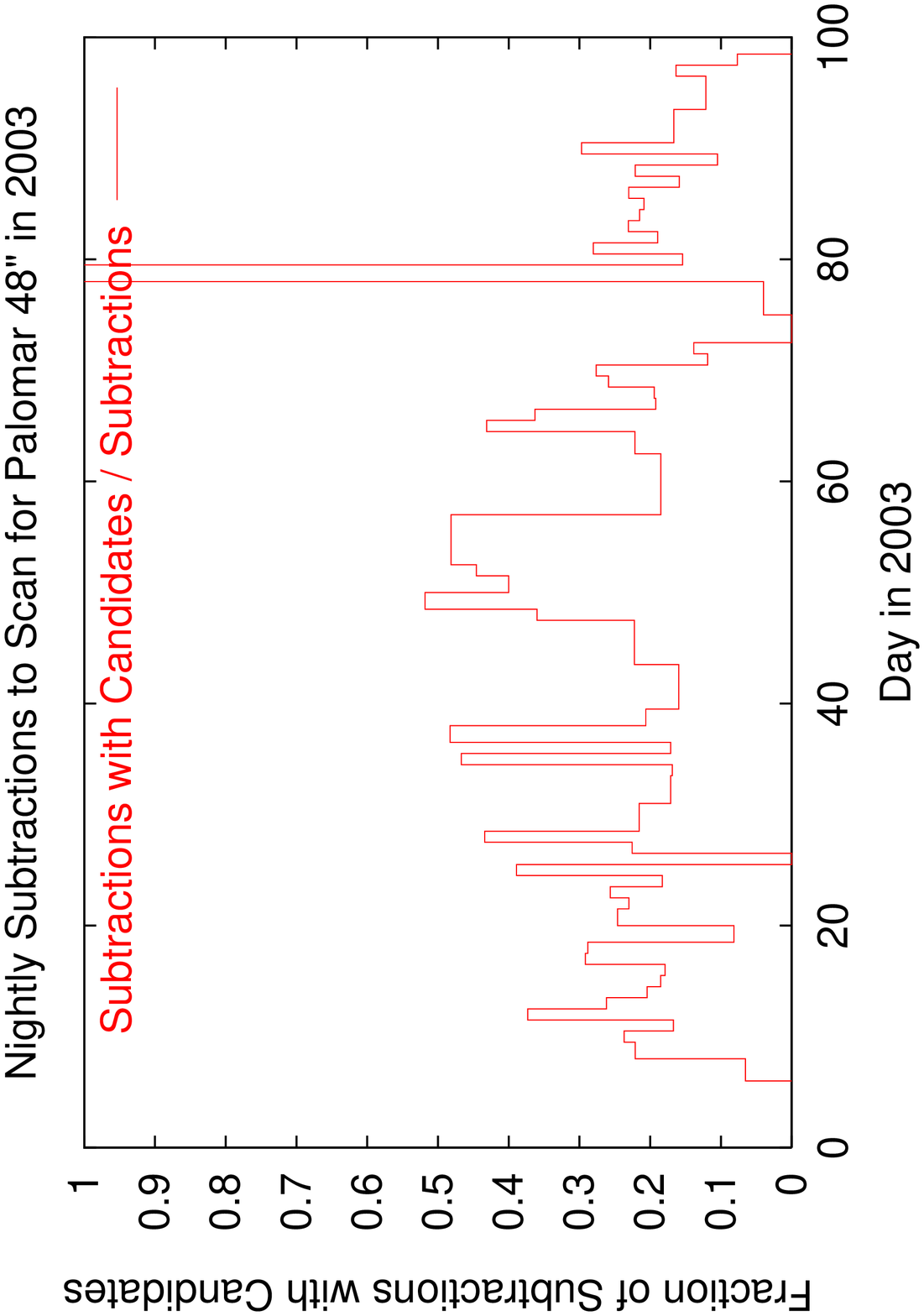}
\caption{The fraction of the subtractions (see
Fig.~\ref{fig:oschin_reductions_subtractions_2003}) that had to be scanned
by human eye in 2003 given the automatic scanning score cuts in use at the time.}
\label{fig:oschin_subtractions_fraction_2003}
\end{figure}

All of the tasks detailed up to this point are fully automated and run
daily with minimal human supervision.
Figs.~\ref{fig:oschin_reductions_subtractions_2002},~\ref{fig:oschin_subtractions_fraction_2002},~\ref{fig:oschin_reductions_subtractions_2003},~and~\ref{fig:oschin_subtractions_fraction_2003}
show the processing statistics for 2002 and 2003.


\section{Human Scanning of Supernova Candidates}
\label{sec:human_scanners}

After the pipeline has automatically processed the search images, run
the subtractions, and classified the new objects found in the
subtractions, a human eye makes the final decision regarding
candidate supernovae.
The automated scanning program passes a list of flagged candidates on to a
human scanner who then looks at the subtracted, search, and reference
images and decides whether or not each computer-flagged candidate
is a real, variable object.

Scanners learn to search for supernovae in the SNfactory data stream
by examining the almost one hundred supernovae already found by the
search pipeline to learn what supernovae look like in the SNfactory
subtractions.
LBL staff scientists and postdoctoral
researches who have already been trained to scan for supernovae for
the high-redshift SCP searches can rapidly adjust to scanning for
nearby supernovae.

The SNfactory search pipeline ran with a very conservative set of cuts
in 2002 and $20$--$40$\% of the successful subtractions for each night
had to be scanned by a person.  During the Spring 2002 academic
semester, a team of six undergraduates scanned the subtracted images.
By the end of the semester, their training and experience enabled them
to keep up with the rate of subtractions when they each worked a few
hours a week.

However, many of the supernovae discovered in Spring 2003 were found
using very restrictive cuts such that only a dozen or so images a
night would have had to be scanned.  These more aggressive cuts were
instituted after the undergraduates stopped working for the semester
and after the number of subtractions attempted and completed was
increased by considering matching reference and search images from
larger date ranges in the database.  It is currently possible to the
meet the scanning burden with $3$--$6$ staff scientists and
postdoctoral researchers working less than an hour per day.

\subsection{Scanning Interface}

\begin{figure}
\subfigure[REF]{
\includegraphics[width=2.5in]{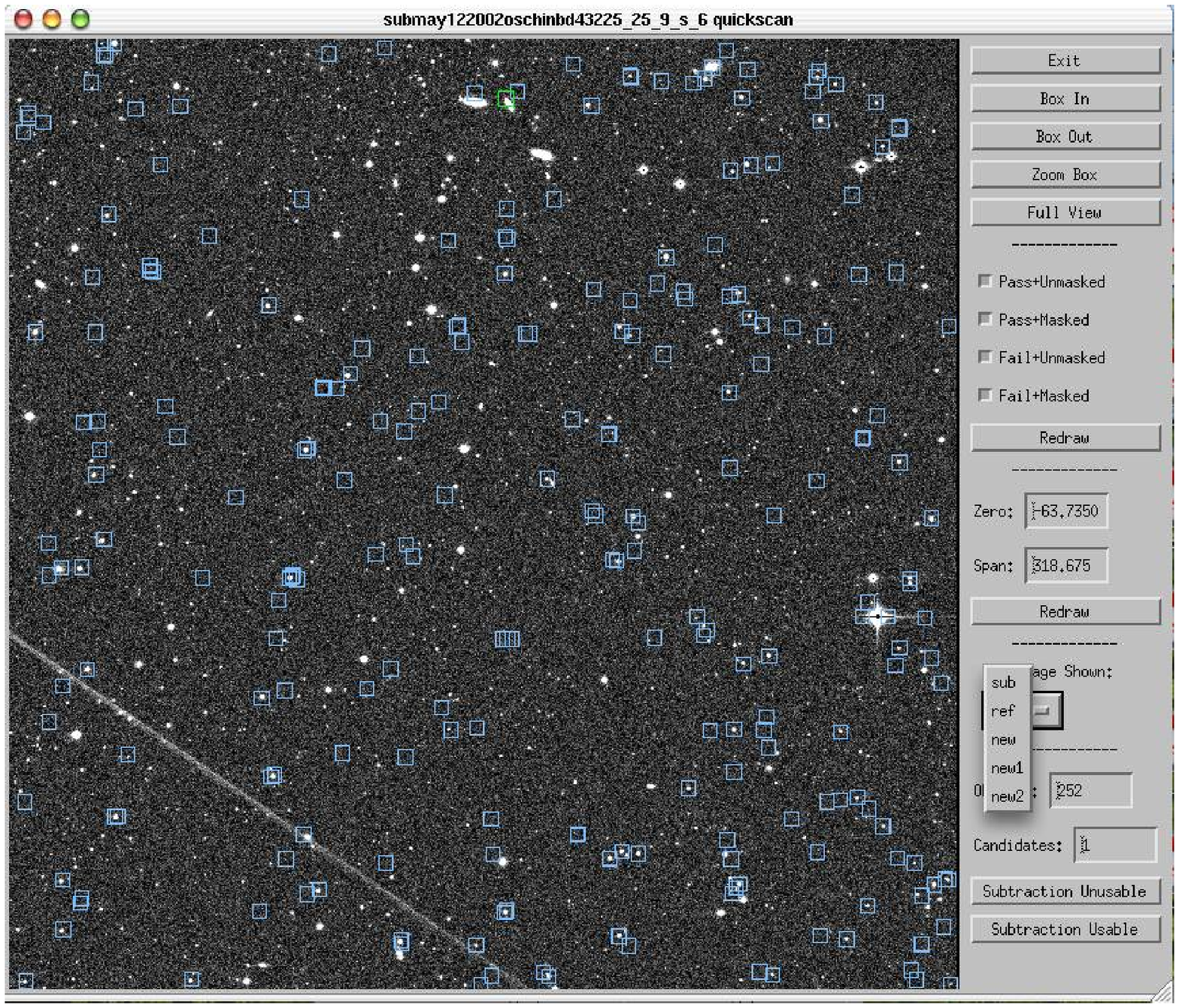}
}
\hspace{0.3in}
\subfigure[NEW]{
\includegraphics[width=2.5in]{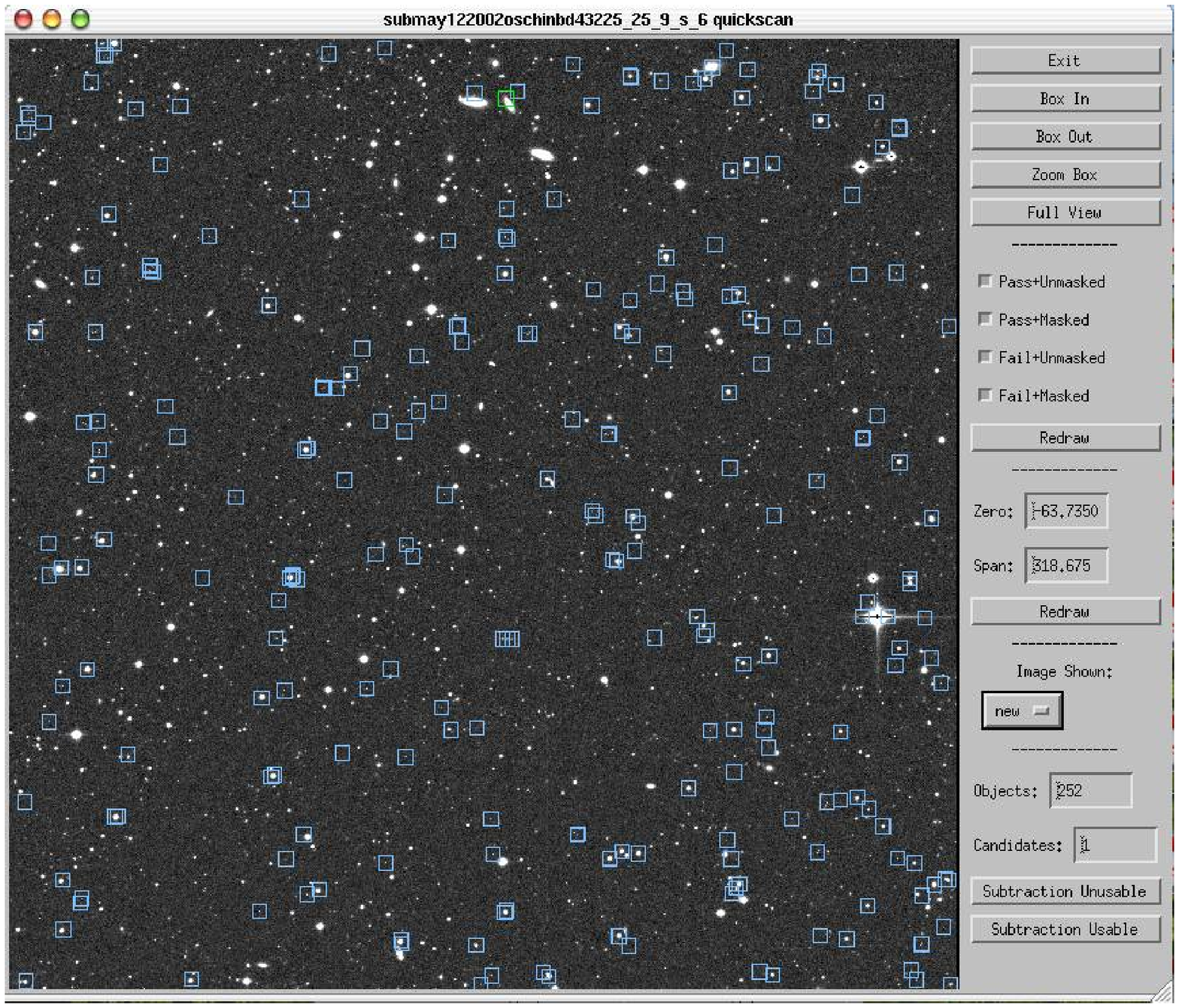}
}
\vspace{0.3in}
\subfigure[SUB]{
\includegraphics[width=2.5in]{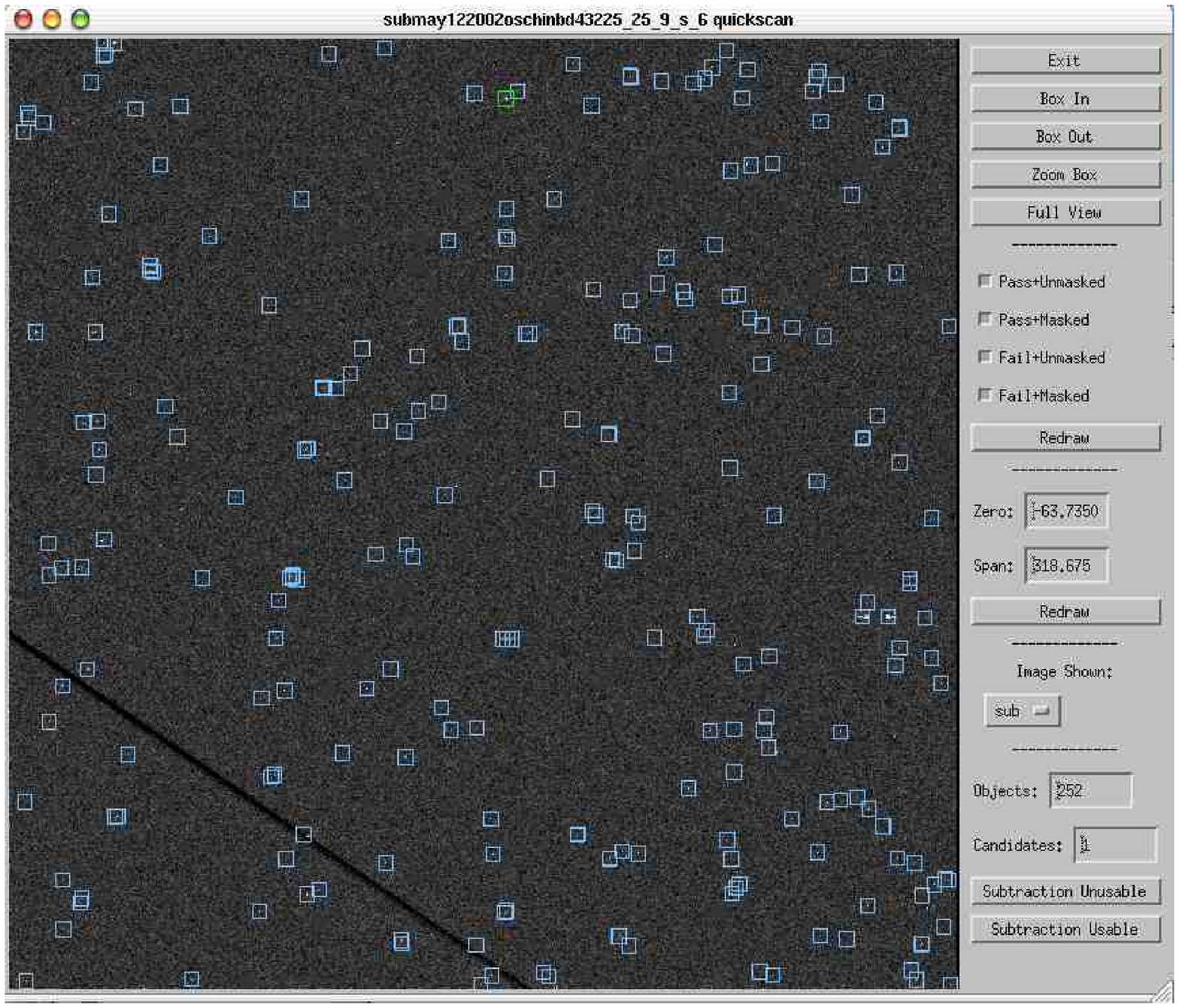}
}
\hspace{0.3in}
\subfigure[Tiles]{
\includegraphics[width=2.5in]{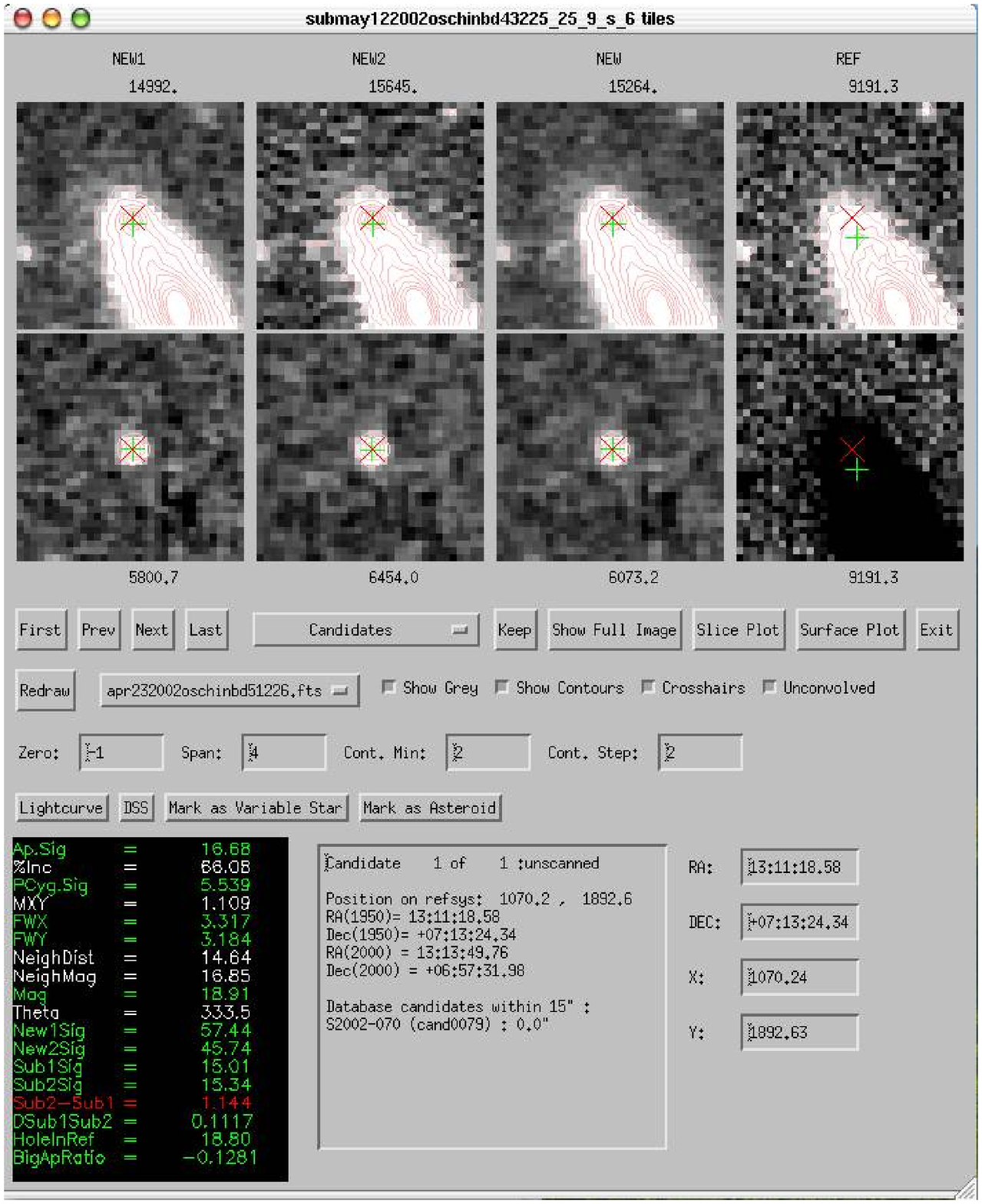}
}
\caption{The scanning software used to check subtractions flagged by
the SNfactory automated system as having candidates of interest.
Shown here are the (a) reference images, (b) new images, (c) subtracted images,
and (d) the close-up ``tiles'' view for SN~2002cx.}

\label{fig:scantng}
\end{figure}

The scanning interface used by the SNfactory search (see
Fig.~\ref{fig:scantng}) represents a mild evolution from the software used by
the SCP.  Each subtraction is first seen whole, with objects and good
candidates marked.  Then the user looks at each candidate individually
in a ``tiles'' view.  A variety of options, in the form of buttons in
the graphical user interface, were added for the SNfactory search to
aid the scanner in classifying candidates.  For example, there is a
button on the tiles view window to generate a lightcurve for the RA
and Dec of the current candidate from all available NEAT images;
another to retrieve an image of that region of the sky from the
Digitized Sky Survey~\citep{dss}; and a number of different fields and
toggles to enhance the display of the candidate.

To check for known causes of variable objects that are not supernovae,
a number of cross-checks are performed:
\ssp
\begin{enumerate}
\item Check for known asteroids (MPEC Minor Object Catalog).
\item Check for classification of object as star/galaxy (POSS/APM catalogs).
\item Check year-old references to spot long-term variable objects
     (variable stars, novae, i.e. not supernovae) on both archived NEAT images
     and DSS/POSS images.
\item Examine automatically generated lightcurve of candidate.
\end{enumerate}
\dsp
Once a determination is made that a candidate is a possible
supernova, the scanner presses the ``Keep'' button and adds 
comments and an assessment of the candidate.  The
candidate is then automatically entered into the database of candidates.
Currently a human supervisor checks the list of new candidates and
submits promising ones for follow-up, but this process can be easily
automated once all of the SNfactory scanners are thoroughly trained.

\section{Confirmation Image}

Promising candidates are automatically submitted to the NEAT target
list for the next night of observation.  The current turn-around time
from the search image being taken to the candidate being submitted for
confirmation is 2~days.  This time lag will be decreased to just
1~day, i.e. the next night after the search image was taken, by later
in 2004 by submitting subtractions during a night as soon as each full
search triplet is available.  Once the potentially confirming data is
transferred from the telescope, the same region of sky is checked to
verify that the variable object is still there.

\subsection{Confirmation Spectrum}

To make a final determination that an object under consideration is
indeed a supernova, it is necessary to take a confirmation spectrum.
For a supernova observed photometrically with a 1-m telescope,
spectroscopic observations can be obtained with a 2- to 3-m class
telescope.  Thus, the UH 2.2-m telescope with the new SNIFS instrument
is a good follow-up resource to study SNe discovered with the 1.2-m
telescopes used by the NEAT project.


Using the spectra of a number of well-studied SNe, templates of
characteristic SNe can be constructed at a variety of times, or
epochs, after explosion.
These spectral templates at different epochs
help determine the type of a SN (Type~I a/b/c, or Type~II
{\bf L}inear/{\bf P}lateau/{\bf n}arrow/{\bf b} + 1987A).  At the simplest level, Type~I and Type~II supernovae are distinguished by the absence or presence of
hydrogen lines in their spectra (See Sec.~\ref{sec:sneia}).  The
sub-classes are determined by both lightcurve behavior and spectral
features.  Spectroscopic and photometric observations will be combined
to determine a type for all of the SNe observed as part of the
SNfactory program.  An early and accurate classification of each SN
candidate will be important in focusing the SNfactory follow-up
resources on the SNe~Ia of interest.

\subsection{Report Discovery}

During the prototype search run, after obtaining a sufficient number
of images of a candidate that revealed it to exhibit behavior
characteristic of a supernova, an IAU circular was submitted to report
the discovery of this candidate as an apparent supernova.

A spectral confirmation was not always possible during the prototype
search.  The SNfactory is grateful to the community for their efforts
to spectroscopically observe and determine the types of the SNe
discovered by the prototype search pipeline.  Now that the SNIFS
instrument is installed and operational on the Hawaii 2.2-m telescope,
supernovae are distributed within the Palomar Consortium and will be
confirmed internally using SNIFS.  Spectroscopic observations of the
candidate will be conducted to confirm that observed candidates
are supernovae and to determine supernova types.





\section{Supernovae found to date by this method}

Eighty-three new supernovae have been found and
accepted by the International Astronomical Union (IAU) using the
techniques described above (see Chapter~\ref{chp:supernovae_found}).  See
Tables~\ref{tab:2002_sne}~\&~\ref{tab:2003_sne} for the IAU
designations for these supernovae.  An additional 17~SNe were found as
part of the search but were first reported in the IAU circulars by
other groups.

\section{Acknowledgements}

This research has made use of the NASA/IPAC Extragalactic Database
(NED), which is operated by the Jet Propulsion Laboratory, California
Institute of Technology, under contract with the National Aeronautics
and Space Administration.

\chapter{Image Differencing to Find Supernovae: Subtractions}
\label{chp:subtractions}

Almost a decade of work has gone into the image differencing, or
subtraction, software used by the SNfactory to search for supernovae.
The same software used by the Supernova Cosmology Project (SCP) to
search for intermediate-redshift ($0.1 \ge z \ge 0.5$) and
high-redshift ($ z\ge 0.5$) supernovae is used by the SNfactory to
search for nearby (z $\le 0.1$) supernovae.  This software, DeepIDL, is
currently IDL-based, with some externally-called C routines, and is
continually being improved and rewritten.  This chapter provides an
outline of the subtraction process and describes some of the specific
changes and improvements that were added to the subtraction software
for the SNfactory automated search pipeline.  See \citet{kim99} for a description
of the full SCP subtraction code used in the original SCP supernova searches.
\footnote{As always, reading the source code
provides the deepest understanding.  The IDL source code is available on the LBL SCP machines under
\file{/home/astro9/deephome/snfactory/idlpro} and
\file{/home/astro9/deephome/snfactory/snidlpro}.  On the PDSF system,
these routines are in
\file{/home/users/wwoodvas/local/deephome/idlpro} and
\file{/home/users/wwoodvas/local/deephome/snidlpro}.}

The subtraction software takes a lists of images for a given search
field and separately registers (aligns) each of the images to a common
reference system.  The images to be used as a reference, REF, are then
added together after they have all been shifted to line up with each
other.  The list of images to be searched is split into two parts,
NEW1 and NEW2, so that there can be two search images to run checks to
make sure a possible candidate isn't an asteroid or cosmic ray.  The
variations in seeing between the co-added images are matched through a
convolution kernel.  The co-added REF image is subtracted from the
NEW=NEW1+NEW2 image.  Objects found in this SUB=NEW-REF image are then
ranked using the SUB image as well as individual subtractions from the
separate NEW1 and NEW2 co-additions: SUB1=NEW1-REF and SUB2=NEW2-REF.
The objects whose scores pass predefined thresholds (see
Sec.~\ref{sec:scorecuts}) are then passed on to be checked.  This
subtraction process consists of many complicated steps but the
framework can be described simply.

\section{Image Database and Lists}

With millions of images in the SNfactory archive, it is important to
have an efficient, well-organized database to keep track of all images
that overlap with a given field on the sky.  The \code{images} table
in SNfactory database stores the minimum and maximum values of RA and
Dec for each image.  These coordinates are always stored in B1950
coordinates for ease of searching.  To match to a given field on the
sky, these minimum and maximum RA and Dec values are compared against
the corners of the field to match and all images that overlap with the
field by at least a specified amount, generally $50$\% of the area of
the reference system image, are returned.  These images are then added
to a list of matching images for a given field on the sky.

\section{Quality Control and Image Registration}

An image quality analysis program,
\code{subng\_reduce\_and\_vet.pro}, is then run on the list of matching images determined above.  This analysis routine requires that 
each image can be successfully processed by the basic image reduction software,
\code{freduceimage2.pro} and that a minimum of 75 stars from the USNO~A1.0
catalog~\citep{usnoa1} are found on the image.  Since the NEAT4GEN2 and NEAT12GEN2 images are
relatively large, $30\arcmin\times30\arcmin$, there are still plenty of
stars on any given field.  Even the smaller QUESTII CCDs
($34\arcmin~\times~8\arcmin$) still typically have $>75$ USNO~A1.0
stars in each frame.

In addition to passing individual quality controls,
\code{subng\_reduce\_and\_vet.pro} requires that images must also be
successfully matched to the reference frame of the subtraction.  The
reference frame for the subtraction is generally chosen to be the
first new search image that passes the initial image-quality cuts.
All of the images for the subtraction must be moved to a common frame
and it reduces the uncertainties introduced by the transformation to
make the common frame one of the images in the subtraction so that the
images can be matched directly to the objects on that image rather
than on an abstract sky space.  The process of matching an image to
another image involves taking the coordinates of all point-sources on
each image, sorting them by brightness, taking the top 300 hundred
objects and producing a transformation between the object lists.  This
preliminary match is then refined using \code{ftranscorrect2.pro}
until the desired transformation tolerance is reached.  The overall
matching between images is handled using the IDL routines called by
\code{ftransimages2.pro}.

Images with good transformations to the reference system image are
kept and stored in a text file to be passed on to the subtraction
software, \code{subng.pro}.  Fig.~\ref{fig:subfile} shows such a
subtraction file for the discovery of SN~2002cx.  Rejecting
images that fail basic image processing and transformation
calculations was a key quality-control device for generating the
thousands of successful subtractions a night necessary for operation
of the SNfactory search pipeline.

\begin{figure}
\begin{verbatim}
REF
apr232002oschinbd51226.fts
apr232002oschinbd54403.fts
apr232002oschinbd61415.fts

NEW1
may122002oschinbd43225.fts
may122002oschinbd50211.fts

NEW2
may122002oschinbd53211.fts
\end{verbatim}
\label{fig:subfile}
\caption{An example of the file list that it sent to the IDL program 
\code{subng.pro} to determine the images to co-add and subtract 
(see Fig.~\ref{fig:idlsubfile}).
This particular file lists the images used for the discovery
subtraction for SN~2002cx.}
\end{figure}

\begin{figure}
\begin{Verbatim}[fontsize=\scriptsize]
subng, $
  '/home/users/wwoodvas/images/20020512p/subdir/submay122002oschinbd43225_25_9_s_6', $
  imgdir='/auto/snfactry2/images/snfactory' , tmpdir='/scratch/wwoodvas/' , $
  snset='S2002' , convmatchalg=6 , dilate = 30, /linear, /noreject , $
  /noupdate , /DONTSHIPTOMASTERSITE , /simplemask 
scantng, 'submay122002oschinbd43225_25_9_s_6', /autoscan, defaults='neat'
shipinterestingtomastersite, 'submay122002oschinbd43225_25_9_s_6'
EXIT
\end{Verbatim}
\caption{An example of a command file used to run the IDL subtraction program \code{subng.pro},
 the automatic scanning program \code{scantng.pro}, and a final program that 
uploads interesting subtractions to be scanned 
\code{shipinterestingtomastersite.pro}.  The final EXIT line exits the IDL session.}
\label{fig:idlsubfile}
\end{figure}

\section{Image Masking}
\label{sec:image_masking}

\begin{figure}
\plottwo{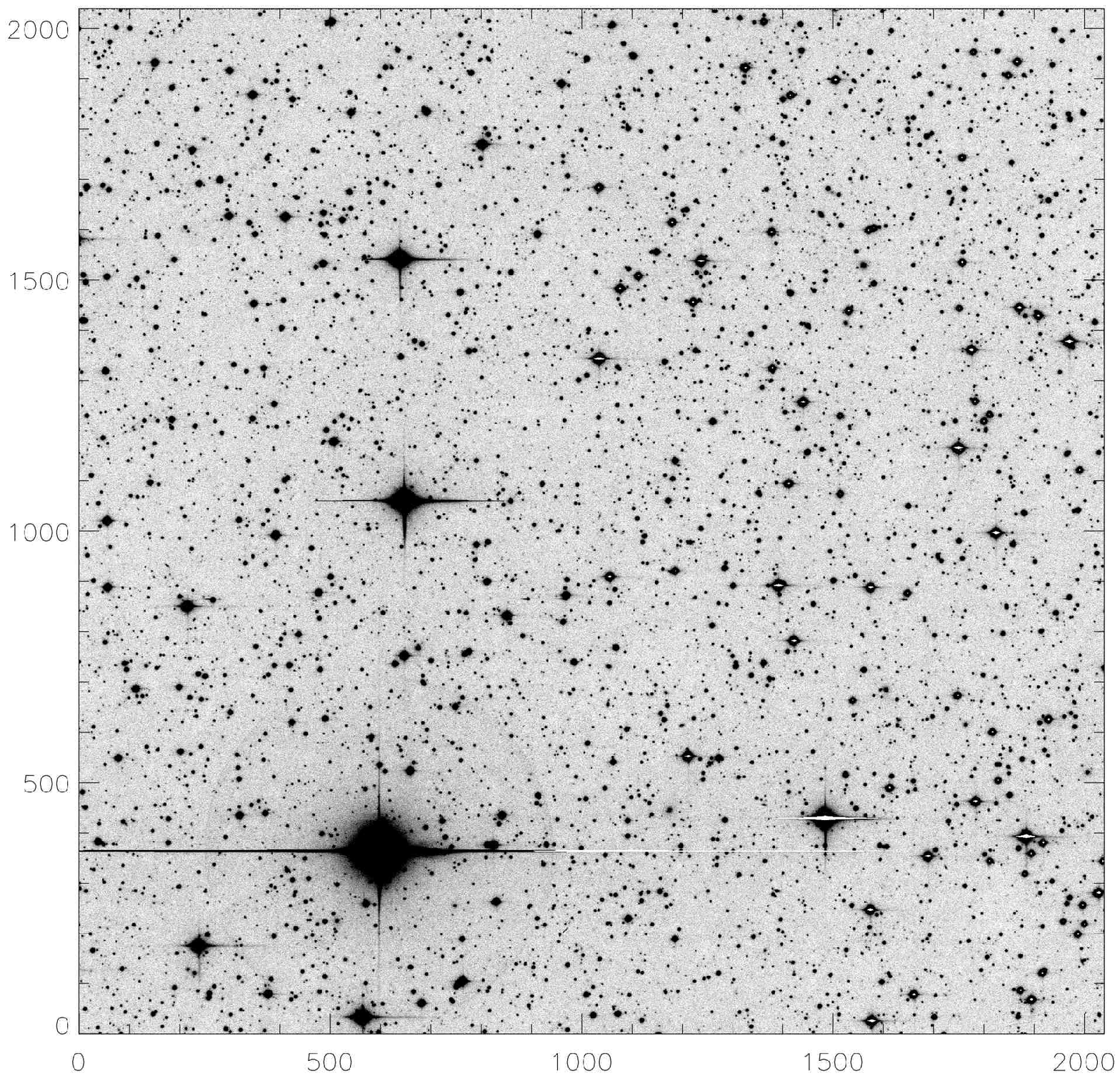}{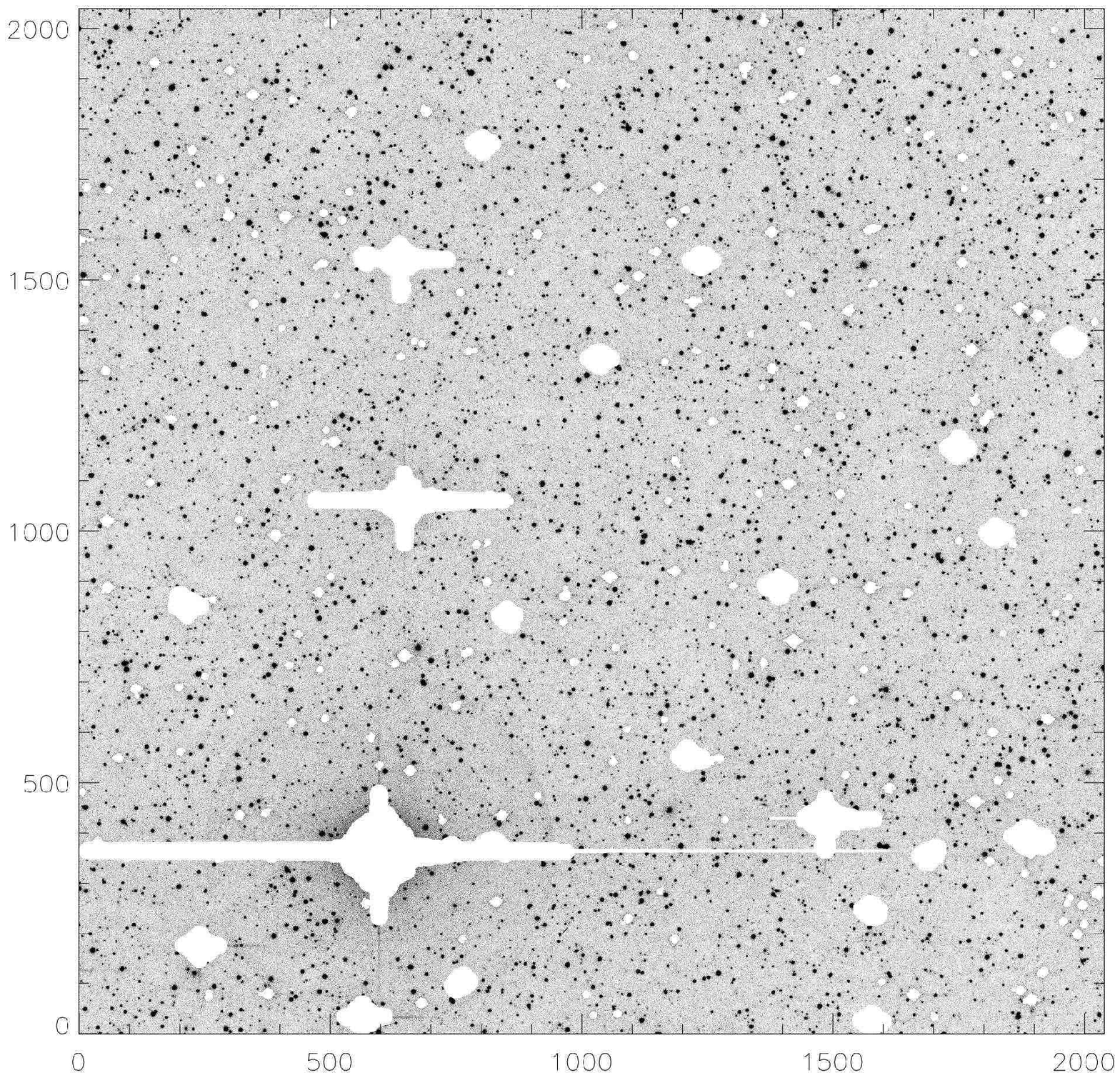}
\label{fig:saturated_stars}
\caption{Saturated stars do not subtract cleanly and need to masked to
avoid spurious candidates in a subtraction.  Note the halos and spikes
around the bright stars in the original image shown on the left.  Also
note that sometimes saturation leads to wrap-around or reflection in
the value from the CCD and turns a saturated positive (black) region
negative (white) (e.g. the star in the lower-right quadrant).  The
masked image on the right shows the result of multiplying the original
image by the mask generated by
\code{makemask.pro}.}
\end{figure}

Even in the best of conditions, images often have non-linear and
non-astrophysical features that need to masked out from the
subtraction to avoid generating spurious candidates.  The two most
common such features are the non-linear response of the CCD as the
signal approaches saturation and blooming spikes from saturated stars.
See Fig.~\ref{fig:saturated_stars} for examples of these problems.

The very brightest saturated stars affect significant portions of the
image, while fainter saturated stars only affect their immediate
surroundings.  Sometimes the center of a saturated star will look
negative, as in the case of some of the stars on the right half of the left-hand image in Fig.~\ref{fig:saturated_stars}.  It is very
important to mask these stars as any negative regions in a reference
image will immediately show up in a subtraction as positive objects.

\subsection{Saturated Stars}
\label{sec:saturated_stars}

Saturation occurs in these images because of the large dynamic range
of the magnitudes of stars in the field and the nature of the CCD
detector.  A CCD is a piece of silicon divided into typically a few
million cells.  Each cell, or pixel, converts incoming photons into
electrons.  When the CCD is read out, the electrons in each row are
shuffled off cell-by-cell and read out by the amplifier.  In the case
of a linear relationship between the number of photons received and
electrons generated, one has a good measure of the original signal.

However, this linearity can only be maintained up to a certain number
of electrons in a given pixel.  Past this point the response becomes
non-linear and the signal is saturated.  Saturation can start to be a
problem for more than just the immediate saturated pixel because
eventually the walls dividing the CCD pixels can no longer contain the
accumulated electrons.  The electrons leak out from the original pixel
and create saturated columns, or readout spikes, extending from the saturated pixel.  The
star in the lower left corner of the image in
Fig.~\ref{fig:saturated_stars} shows this effect clearly.

\subsection{Blooming Spikes, Diffraction Spikes, and Bad Columns}
\label{sec:blooming_spikes}
\label{sec:diffraction_spikes}
\label{sec:bad_columns}

The spikes from saturated stars are a possible source of false
candidates as they extend for hundreds of pixels across the image.
Bright stars generate readout spikes along the readout column and
generate diffraction spikes from the struts, or ``spiders'', holding
up the secondary mirror or prime focus cage.  Because the NEAT (and
many other) detectors are
aligned so that north is perpendicular to a CCD edge and the spiders are
similarly aligned north-south and east-west, the diffraction spikes
appear perpendicular to the readout spikes.  If the readout spikes
weren't there one should see another pair of diffraction spikes.  For
many saturated stars, the readout spikes are brighter, but in
Fig.~\ref{fig:saturated_stars} one can see that the bright star in the
upper-left has relatively symmetric spikes.  These are likely
predominantly due to diffraction spikes.

As one moves away from the star, the counts in the spikes are often no
longer above the saturation level themselves and can only be
determined by tracing out the line from the saturated star or by
looking for contiguous line segments in a given row.

Both approaches to masking are used in the image masking code \code{makemask.pro}.
Often \code{fisofind2.pro} will find the bulk of the blooming spikes near
the star and the object masking from Sec.~\ref{sec:saturated_stars}
will take care of those parts.

The full extension of the spikes is dealt with by specialized code
that scans the image row-by-row and then column-by-column and traces
out line segments.  For each row or column the program finds all
pixels $5\sigma$ above the noise for the image and marks them as a
bad pixel.  It then runs through and creates line segments for
each bad pixel.  These line segments are joined together if
they are within 6 pixels of each other.  Then all line segments that
are at least 10\% of the length of the line are masked.
If more than $30\%$ of a line is masked the whole line
is then marked as masked.

This approach is very successful at masking strong, solid blooming
spikes.  It also handles bad columns from transient or permanent
electronic problems in the CCD.  It does not work as well on fainter
spikes that don't produce solid, contiguous line segments.  These often
come from the optical diffraction spikes which produce fainter,
tapering spikes.  As the number of spurious candidates due
to other sources is decreased, this is starting to become a significant problem and
will be the subject of future investigation.

\section{Co-addition}

The goal of combining images is to maximize the use of all of
the information available in the images while minimizing
the noise.  Once the transformations for all of the
images in a set are calculated, the images are each moved to the
common reference frame so that all of the pixels overlapping with the
reference field match up.  Then statistics can be calculated to determine
which pixels should be rejected as possibly being from cosmic rays or
other irregularities.  Obvious cosmic rays will be separately rejected
by the masking algorithms which detect all saturated objects.  These
masks are passed to the co-addition routine and masked pixels from a
particular image are excluded from the calculation of the new pixel
values in the co-added image.  If all of the images have a given pixel
masked that pixel is masked in a global mask for the subtraction.

The REF co-addition is created as a union of the component images,
while the NEW1 and NEW2 co-additions are created as the intersection
of the respective search images.  This guarantees that all of the
pixels have all three search images to provide information and track
the behavior of objects across the three images.  It is generally
preferable to use references images from the previous year as it can
thus be guaranteed that there will be no supernova light in the
reference image.  No PSF matching is performed in these co-additions
for reasons simplicity and maximizing the signal-to-noise in the pixel
values.  The REF union co-addition uses the relative noise and
zeropoints of the reference images in determining their relative
weights in making up the ``refsum'' image.  The NEW1 and NEW2
intersection co-additions are simple, unweighted sums.  This
unweighted co-addition is appropriate for the NEAT same-night image
search triplets, but a more generalized approach could be
constructed to more optimally deal with search images taken 
with differing exposure times or in varying observing conditions.

\subsection{Union Co-addition}

Each potential reference image is checked to make sure that a
transformation to and from the reference frame is possible and
well-determined.  Images with little or nor overlap with the nominal
reference frame are required to have some path of transformation that
connects to the reference frame.  This allows images that only
overlap by 5--10\% with the reference frame to be used.  However,
in practice a minimum overlap of 50\% is requires since it
is important to ensure that the transformations are all calculated
correctly.  Just one bad transformation can significantly affect the
co-addition because of the chain of transformations used to combine all
of the images.  

The overall co-addition is constructed as the union of all of the
reference images with the intersection with the reference system image.  This
maximizes the use of historical images to ensure that the reference image
is significantly deeper than the search.  This also helps reduce the noise in
the subtraction so that the sensitivity is primarily limited by the depth
of the NEW search images.

\section{Point Spread Function Convolution}

From the co-added REF, NEW1, and NEW2 images, the aggregate
point-spread-function, or ``seeing,'' of each co-added image is
convolved to that of the worst seeing co-added image using a
convolution kernel calculated by \code{subngpsfmatch.pro}.  The co-addition
process itself does not convolve because there is no signal-to-noise
benefit.  However, this means that the PSF on the co-added images is a
combination of the PSF of the constituent images and thus may not
being easily approximated by a Gaussian function.  For the subtraction
steps, the PSF differences between the co-added images need to be
matched so that objects that are unchanged in the epochs of
observations subtract to zero.  If these seeing differences are not
accounted for there will be doughnut-like positive and negative
shapes in the subtractions that would have a net value of zero.  These
doughnuts complicate the search for new objects in the
subtractions.  It is much cleaner to properly match up the PSFs of the
image so that objects which are at equal brightness subtract to zero
and new object show up consistent with a point source according to the
point-spread-function of the co-added image with the worst seeing.

There are a variety of methods available in the current DeepIDL
framework to calculate the convolution kernel.  Currently several
Gaussian functions are used to represent the convolution kernel
between any two of the co-added images.
In the simplest method, the FWHM of a set of good stars is calculated on each image.  The quadrature difference of the Gaussian $\sigma$s corresponding to these FWHMs is then calculated.  This difference, $\Delta\sigma$, is then used as the $\sigma$
for the convolution kernel to be applied to the image with the smaller FWHM seeing.  The method currently used by the SNfactory\footnote{convmatchalg=6}
applies this approach incrementally by only using a given fraction of the calculated $\Delta\sigma$ in each pass.  Four passes with fractions of $[0.5,0.5,0.5,1.0]$ are used.
This process is done separately in the $x$ and $y$ dimensions.
to more fully model the appropriate kernel.  This simple centered-Gaussian approach
has the advantage of speed, but it doesn't do a good job in all cases.
A project for the future is to improve the calculation of these
convolution kernels to yield cleaner subtractions.  This improved
convolution should significantly reduce the number of false objects in
the subtracted images.

\section{Flux Matching of Co-Added Images}

No explicit calculation or adjustment is made to the co-added images
based on relative exposure times.  Instead, the brightness of a
selection of stars on each image is used to calculate the appropriate
flux ratios to normalize each of the images to the same standard.
These stars are picked as the brightest, but not saturated, objects
identified as stars by \code{fcatalog2.pro}.  Applying these ratios to the
co-added images, a meaningful subtraction of REF from NEW1, NEW2, and
NEW1+NEW2 can be realized.  This flux-ratio matching accounts for
atmospheric transmission, sky brightness, exposure time, and other
factors that influence the sensitivity of an image by comparing fluxes
of actual astrophysical objects to generate the desired comparison
subtraction which should then have the correct flux for any candidate
supernova.  The flux ratio is defined with respect to the combined NEW
image, so the flux zeropoint is set by NEW zeropoint.

\section{Actual Subtraction}

All of the above work sets up the final subtraction of the pixel values
of the flux-matched REF image from the flux-matched NEW1 and NEW2
images to generate SUB1 and SUB2 images and an overall SUB image.  The
REF, NEW1, NEW2, and SUB images are the only one saved to disk, as SUB1
and SUB2 can be easily reconstructed from NEW1-REF and NEW2-REF, and
the overall subtraction is the main image of interest.

\section{MAPS Catalog to reject known stars}
\label{sec:mapscatalog}

In a survey with a limiting depth of $20$--$21$ magnitudes, the significant
majority of objects will be stars in our own galaxy.  Imperfect
subtractions result in residuals about these stars.  As these residuals can be
the cause of a large number of false detections, the use of a
catalog of known stars can allow for the rejection of a majority
of false candidates.

The Minnesota Automated Plate Scanner (MAPS) Catalog of the
POSS~I~\citep{apscatalog98} is used to obtain a list of known stars.  This
catalog is from a digitized scan of the first Palomar sky survey
(taken with photographic plates using the main search telescope used
here, the Palomar 1.2-m).  The MAPS project used a neural net
to classify objects as either stars or galaxies in both the
blue and red filters from the POSS~I
survey~\citep{odewahn92,odewahn93,nielsen94,odewahn95}.

To mask known stars for the SNfactory search subtractions, a list was
constructed of objects that were classified as being stars with at
least $80\%$ certainty in both filters.  This list currently contains
$\sim$28 million objects and is stored in a
PostgreSQL\footnote{\url{http://www.postgresql.org}} database, taking up 1.2~GB of disk space on a dual-Pentium III 1~GHz
machine with 1~GB of RAM, currently \computer{lilys.lbl.gov}.  The entries
are stored as integers as described in Table~\ref{tab:apscatalog}.

\begin{table}
\begin{tabular}{lll}
\hline
Field & Format (32-bit Integer) & Scale factor \\
\hline
RA    & Decimal Hours   & $\times 3600 \times 1000$  \\
Dec   & Decimal Degrees & $\times 3600 \times 1000$  \\
Mag   & Magnitude       & $\times 1000$ \\
\hline
\end{tabular}
\caption{
The format of entries stored in the \code{stars\_radec} table in
the \code{apscatalog} database.  This database is used to retrieve a list
of the known stars for a given image and then to generate a mask
appropriate to use to eliminate known stars from being selected as
possible candidates in a subtraction.
}
\label{tab:apscatalog}
\end{table}

This \code{stars\_radec} table is indexed with a BTREE index on RA and
Dec to allow for quick access.  A typical query takes on the order of
a few seconds.

An IDL routine\footnote{In the context of this chapter, an ``IDL routine'' will mean a routine written in IDL rather than a routine provided in the IDL default library.}, \code{makestarmask.pro}, queries the database and
transforms the list of RA and Dec coordinates to pixel positions on
the given image.  The magnitudes returned by the query are then used in
setting the size of the circular mask created for each known star.
Specifically the diameter, $D$, of the circular mask for each star is
calculated using Eq.~\ref{eq:starmask_diameter}.

\begin{eqnarray}
\label{eq:starmask_diameter}
                D = 2^{(\mathrm{mag}_\mathrm{norm}-\mathrm{mag}[k])/3} \times D_\mathrm{norm} \\
  D_\mathrm{norm} = 10~\mathrm{pixels}\ ,\  
\mathrm{mag}_\mathrm{norm} = 15.0~\mathrm{mag}
\end{eqnarray}

This equation works for different telescopes by assuming the telescopes have a
similar pixel sampling as determined by their seeing and pixel scale.
That is, this masking diameter assumes that a star is the same size in pixels on
different images.  $D_\mathrm{norm}$ is controlled by a keyword
argument to the \code{makestarmask.pro} code and so can be easily adjusted if
optimum parameters are determined for a given telescope.

\begin{figure}
\begin{center}
\plottwo{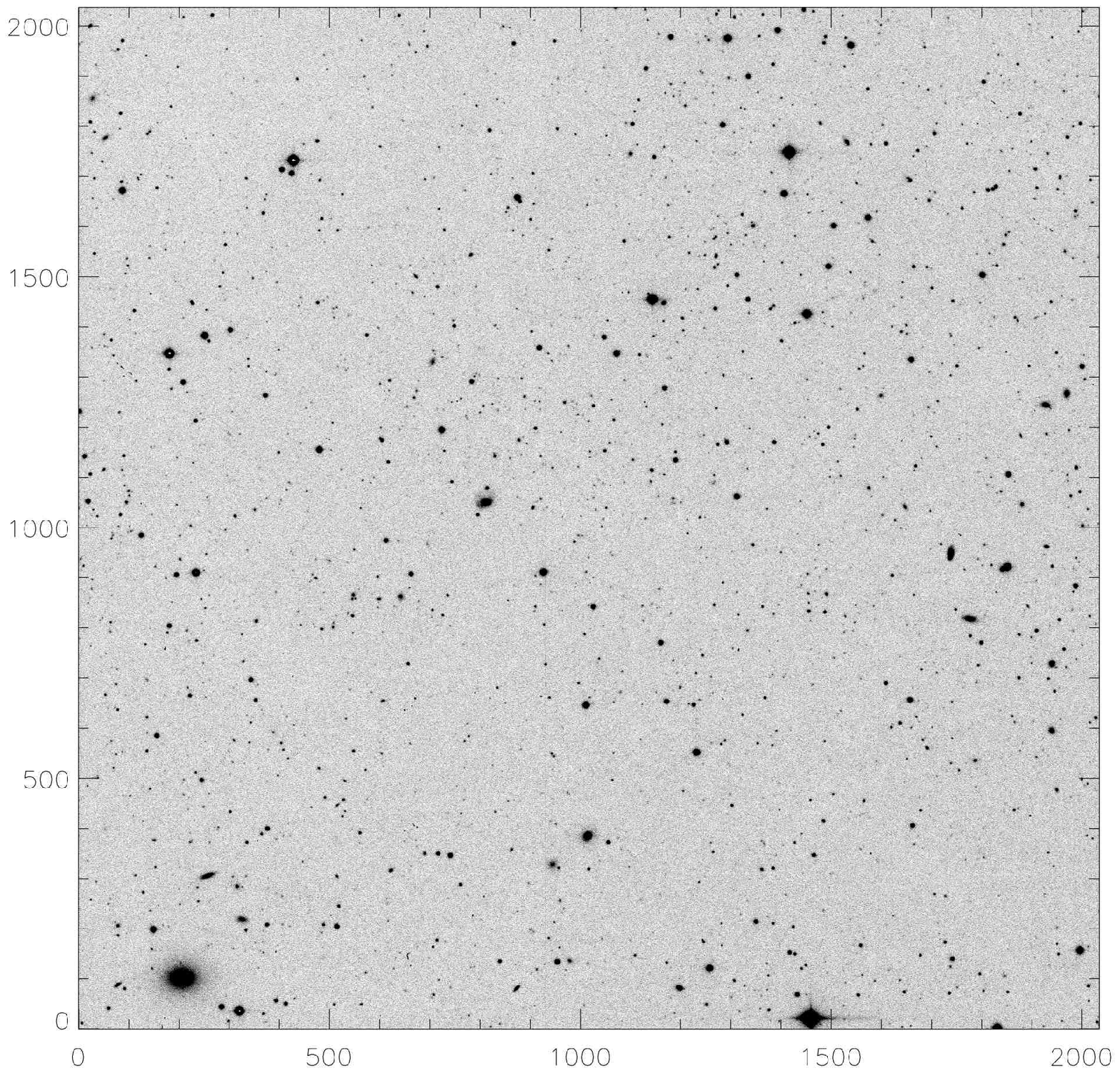}{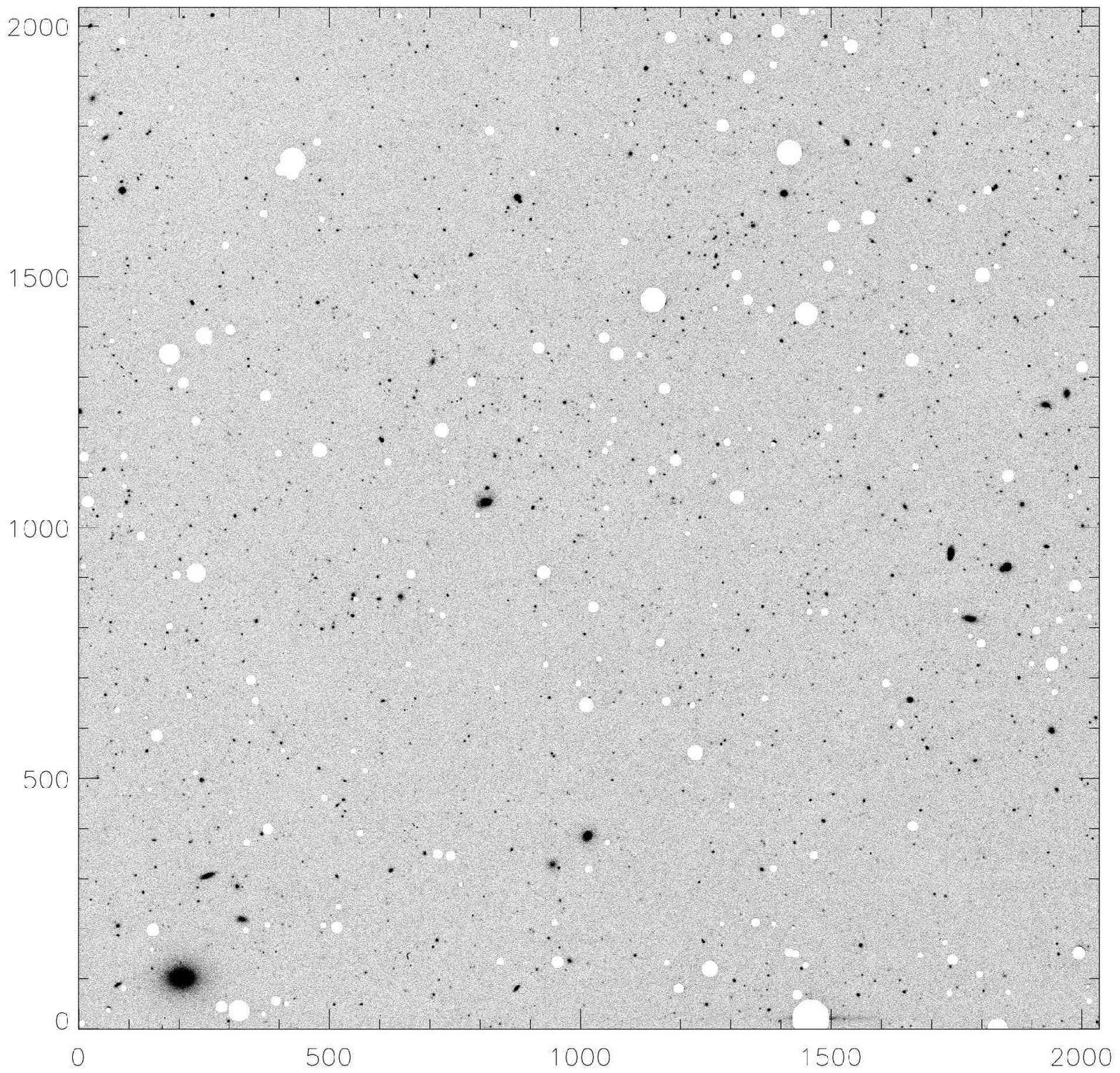}
\end{center}
\caption{(a) NEAT image 'apr82003oschinbc65109.fts'. (b) The same
image with known stars from the APS Catalog~\citep{apscatalog98}
masked out.  The bright object in the lower-left corner is a galaxy.}
\label{fig:starmask}
\end{figure}

See Figure~\ref{fig:starmask} for a sample image and corresponding
star mask.  Notice how the brighter stars are masked out with larger
masking circles.

A \code{starmask.pro} is generated for each subtraction and is applied to
the subtracted image when the image is scanned for objects.  It is not
used, however, in further steps when calculating the fluxes and other
quantities for each candidate.  This \code{starmask.pro} is saved for
future reference during the hand-scanning process.

\section{Future directions}


While the SNfactory automated search pipeline has been very successful,
it can still be improved in both efficiency and sensitivity.
One desirable improvement would be to construct and keep updated a set of
optimum reference images.  The second most obvious step to be improved
is the convolution matching between the REF, NEW1, and NEW2 images.

A set of super-references would use all of the images available in the
NEAT dataset and use the optimum weighting and PSF convolution to
maximize the signal-to-noise of the references.  Doing this for the
tens of thousands of square degrees of sky that the survey covers will
require extensive automated quality control and testing to enabled a
computer to determine the best way to combine every set of images.  It
will also require a well-defined set of reference frames.  Currently,
co-additions to a single reference frame can extend to the edges of
the frame, but the astrometric matching is not as well-constrained
outside of this region.  The geometric scale of both the Palomar 3-CCD
camera, QUESTII 112-CCD camera, and the Haleakala 1-CCD camera are
relatively flat across each CCD so the transformation solution should
be relatively linear and thus hopefully stable to extrapolation and
bootstrapping.  This linearity and stability of these transformations
needs to be verified before building these deep reference images.
These super-references would be updated periodically to use new images
as they become old enough.  This set of optimally-combined images will
provide a useful general-purpose data set for a myriad of scientific
investigations to maximize the value of this Terabyte-scale survey of
the sky.

The simple, yet robust, method currently employed by the SNfactory
subtraction code to calculate convolution kernels to match the
different PSFs between the REF, NEW1, and NEW2 images could be
improved. It currently results in imperfect PSF matching and
characteristic residuals in the subtracted image.  While the score
cuts detailed in Sec.~\ref{sec:scorecuts}, particularly PCYGSIG,
RELFWX, and RELFWY, eliminate the majority of these residuals, a
cleaner subtraction image would result in fewer objects that need to
be checked by a human scanner.  An investigation of adapting the Alard
and Lupton prescription for a flexible convolution
kernel\footnote{http://www.iap.fr/users/alard/package.html}~\citep{alard98,alard00}
was performed, but while some individual subtractions showed
excellent convolution kernel matching, in general, an Alard kernel
parameterization was not found that was robust enough to result in
better subtractions overall.  The framework is there in the
subtraction code\footnote{See \code{subng.pro},
\code{subngpsfmatchalard.pro}, and \code{cvk\_fit\_alard\_kernel.pro}.}
and this work should be pursued further to establish whether or not
the Alard convolution kernel technique should be incorporated into the SNfactory search
pipeline.

The underlying subtraction and image-processing routines
(Chapter~\ref{chp:subtractions}) for the search pipeline should be
completely converted to a non-licensed language.  Having to pay for
IDL licenses places a significant monetary burden on running a large
search on many machines in parallel.  Either additional funding should be obtained
to cover the IDL licensing costs or the current partially
constructed Deeplib C++ framework should be completed to avoid this
significant recurring expense.
Rob Knop is currently working on
developing Deeplib and wrote a new version for the spring and summer
HST searches of 2004.  Adapting to this new version of Deeplib will
involve some changes to the database and auxiliary file structure as
well as changes to DeepIDL to ensure backward compatibility with the
current IDL framework.  Once the new Deeplib reaches a stable release,
it would take approximately one month of a qualified person's time to
get subtractions fully integrated and working in the SNfactory
pipeline using the new Deeplib once it reaches a stable release.

\chapter{Supernovae Found}
\label{chp:supernovae_found}

\section{Introduction}

As of July 11, 2003, the Nearby SNfactory search pipeline had
discovered a total of 99 supernovae.  Eight-three of these were found
by the SNfactory before other supernova searches.  This chapter
presents a brief summary of the search efforts and the
supernovae found.  Sec.~\ref{sec:redshifts} presents an analysis of
redshifts, discovery magnitudes, and other properties of the supernovae found
in this search in the context of a calculation of the rate of nearby
SNe~Ia.

During 2001 and the early part of 2002, a number of supernovae were
found first by other searches and subsequently found by the SNfactory
search pipeline.  For most of 2001, the search wasn't actively running
so the SNfactory was not the first with the discoveries of supernovae
found in this year.  In early 2002, the search pipeline was running
but the subtractions were not being scanned daily, so the SNfactory
discoveries of SN~2002as and SN~2002br were late.  After the IAU
announcement of SN~2001cp, SN~2001dd, SN~2002as, and SN~2002br, each
of these SNe were clearly found in the SNfactory data by the search
pipeline in an analysis of older data.  They would have been found by
the automated search pipeline had it been running when the respective
search images were taken.

In the spring of 2002, the search pipeline discovered its first
original supernova, SN~2002bk.  Eighty-two more
original supernovae were discovered over the next year and it was decided that the search
pipeline had been proven successful and was ready for full operations once
the SNIFS instrument was fully commissioned and operational on the UH 2.2-m.

Sec.~\ref{sec:interesting_sne} details several of the most interesting
objects found as part of the prototype SNfactory search: the unusual
object S2002-028; the Ia-subtype-breaking SN~2002cx; and the first
SN~Ia to exhibit clear evidence for hydrogen and a circumstellar
medium, SN~2002ic.  Sec.~\ref{sec:snfactory_sne} presents discovery 
images and light curves 
of all of the supernovae observed from 2001 to 2003.  Some of
the light curves have had their host galaxies subtracted, others have
not.
These are the rough light curves produced by the automated search
pipeline and so include images from all dates available for that region
of sky.
A simple light-curve based fitting template has been plotted for
a selection of the supernovae to illustrate the ability to distinguish 
supernova type based on the light curve.


\section{Interesting Supernovae}
\label{sec:interesting_sne}

\subsection{S2002-028 - An unusual variable object}

SNfactory candidate S2002-028 was discovered 
on images from 21 March 2002.  Confirmation images were obtained nine
days later on 30 March 2002.  The candidate had risen significantly
between the two epochs.  There was an object present in the reference image
which appeared coincident with the brightening object.  The 
22 February 2002 image did not have the resolution to 
reveal whether or not the already present
object in the 22 February 2002 reference was extended.  If it
were extended the case would have been stronger for this object 
being a supernova.
Otherwise, the possibility of it being a variable star that
flared up in brightness was more of a concern.  The Digital
Sky Survey\footnote{\url{http://stdatu.stsci.edu/cgi-bin/dss\_form}}
plates for the second
generation scan contained the field of S2002-028 for both the ``Red'' and
``Blue'' plates.  An unresolved object was clearly visible in the red plate at
the same location as the supernova.  No
object was apparent in the blue plate at the same location, although
there was an object approximately ten arcseconds to the south which
was not present in the Red plate.  While this blue object was interesting, it was 
difficult to draw any conclusions regarding this nearby object.

\begin{figure}
\plotone{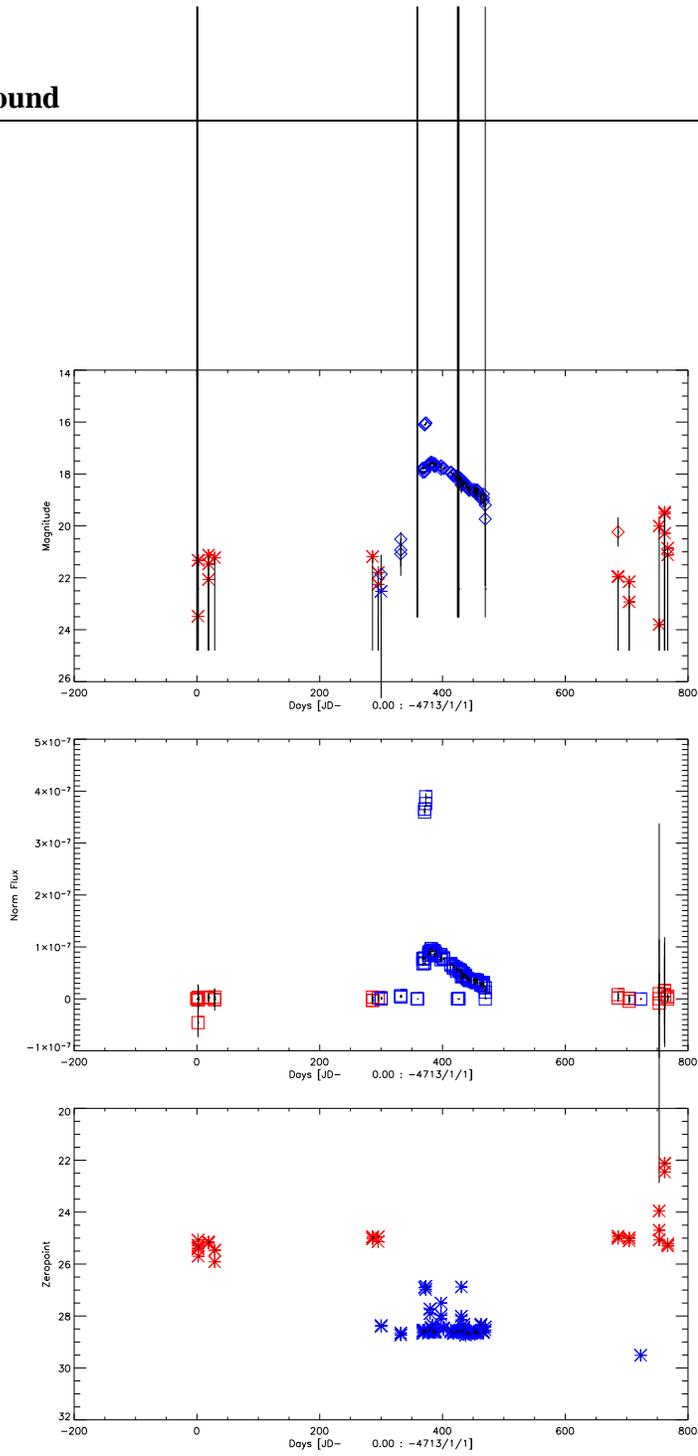}
\caption{The NEAT unfiltered light curve of S2002-028.  Note the very
slow decay relative to a SN~Ia.}
\label{fig:S2002-028_lightcurve}
\end{figure}

After a request to the NEAT team for special follow-up, this object
was observed every other night for over a month with the Samuel Oschin
1.2-m telescope.  Over the next week, the candidate continued to rise slowly,
consistent with a supernova (see Fig.~\ref{fig:S2002-028_lightcurve}).

However, not having spectroscopic confirmation and being hesitant
about releasing a potential dud this early in the prototype search
establish the SNfactory as a source of reliable supernova, this candidate
was not released the candidate to the International Astronomical Union
Circulars (IAUC).

\begin{figure}
\plotone{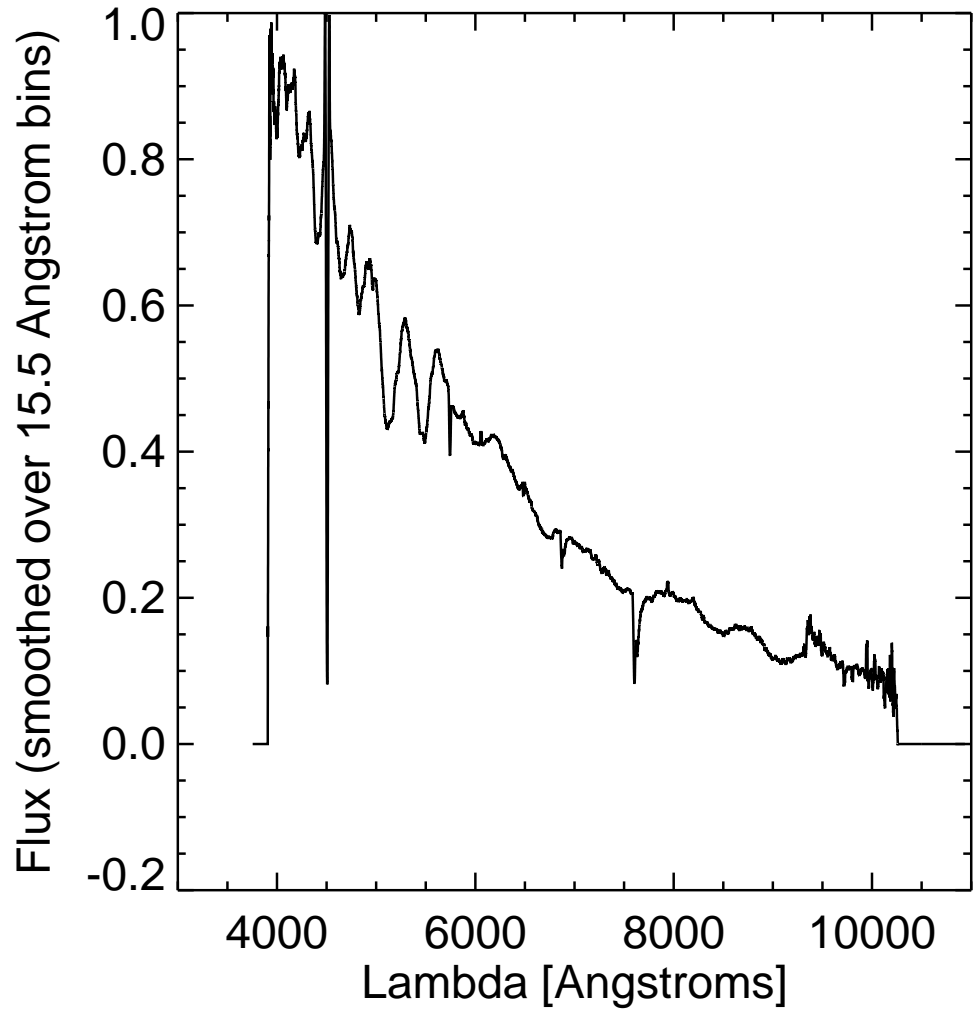}
\caption{A Keck spectrum of S2002-028.  Suggestive broad features
are clearly evident, but no match to a known supernova spectrum can be
made.  The line at $\sim4500$~\AA~is an artifact from combining orders
in the ESI echelle spectrograph.}
\label{fig:S2002-028_keck}
\end{figure}

Over the next several months, S2002-028 decayed slowly in brightness.
Spectroscopic observations were taken with the Keck telescope and an
unusual, unidentified spectrum shown in Fig.~\ref{fig:S2002-028_keck}
was observed.  Consultation with several supernova experts (Peter Nugent,
Rollin Thomas, and Lifan Wang) yielded no
definite identification.  The broad features in the spectrum are
consistent with differential velocity-broadened lines characteristic
of an explosion with a velocity of $\sim6000$~km/s.

\subsection{SN~2002cx}

SN~2002cx~\citep{iauc7902} is a stereotype-breaking SNe~Ia that does
not follow the standard spectroscopic-photometric sequence~\citep{phillips93,hamuy95,riess95,nugent95}.  It is
spectroscopically like SN~1991bg, a sub-luminous event, but
photometrically like SN~1991T, the classic super-luminous
SN~Ia~\citep{li03}.  Fig.~\ref{fig:2002cx_lightcurve} shows the
broad, slow-declining profile of this unusual supernova.  It is
important for the SNfactory to study and understand these unusual
objects to provide insight and of avoid contamination of the
``normal'' SN~Ia sample desired for cosmological work.

\subsection{SN~2002ic}

SN~2002ic~\citep{iauc8019} was discovered in November of 2002 and
turned out to be a remarkable SN~Ia/IIn hybrid exhibiting the clear silicon
absorption feature of SNe~Ia before maximum, but then showing strong
hydrogen emission ~\citep{hamuy03b} characteristic of a SN~IIn.  See
Chapter~\ref{chp:2002ic} for a discussion of this very interesting
supernova.

\section{SNfactory Discoveries}
\label{sec:snfactory_sne}

\begin{figure}
\subfigure[2001cp]{\includegraphics[angle=90,height=2in,width=3in]{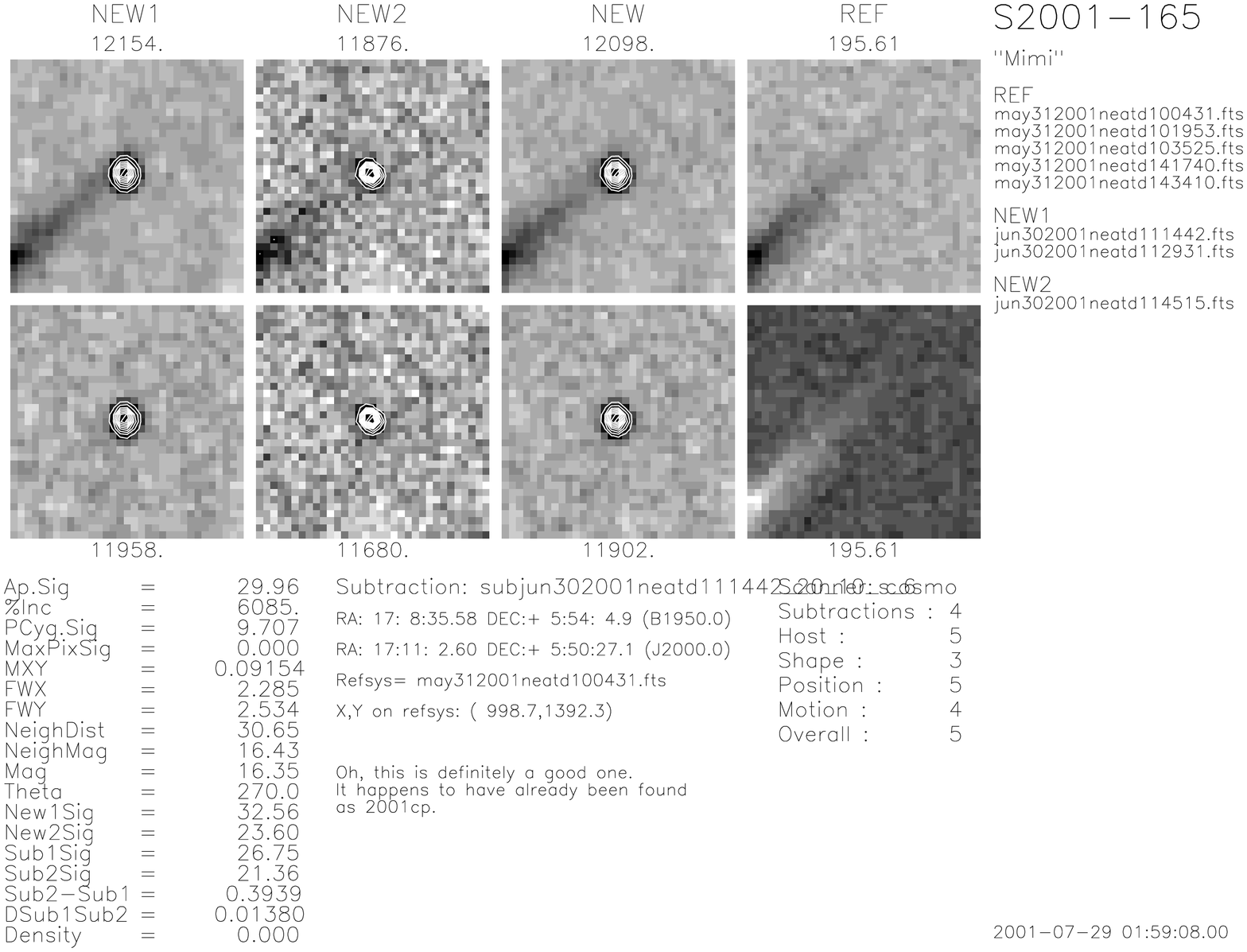}\label{fig:sn2001cp_discovery}}
\hspace{0.3in}
\subfigure[2001cp]{\includegraphics[height=2in]{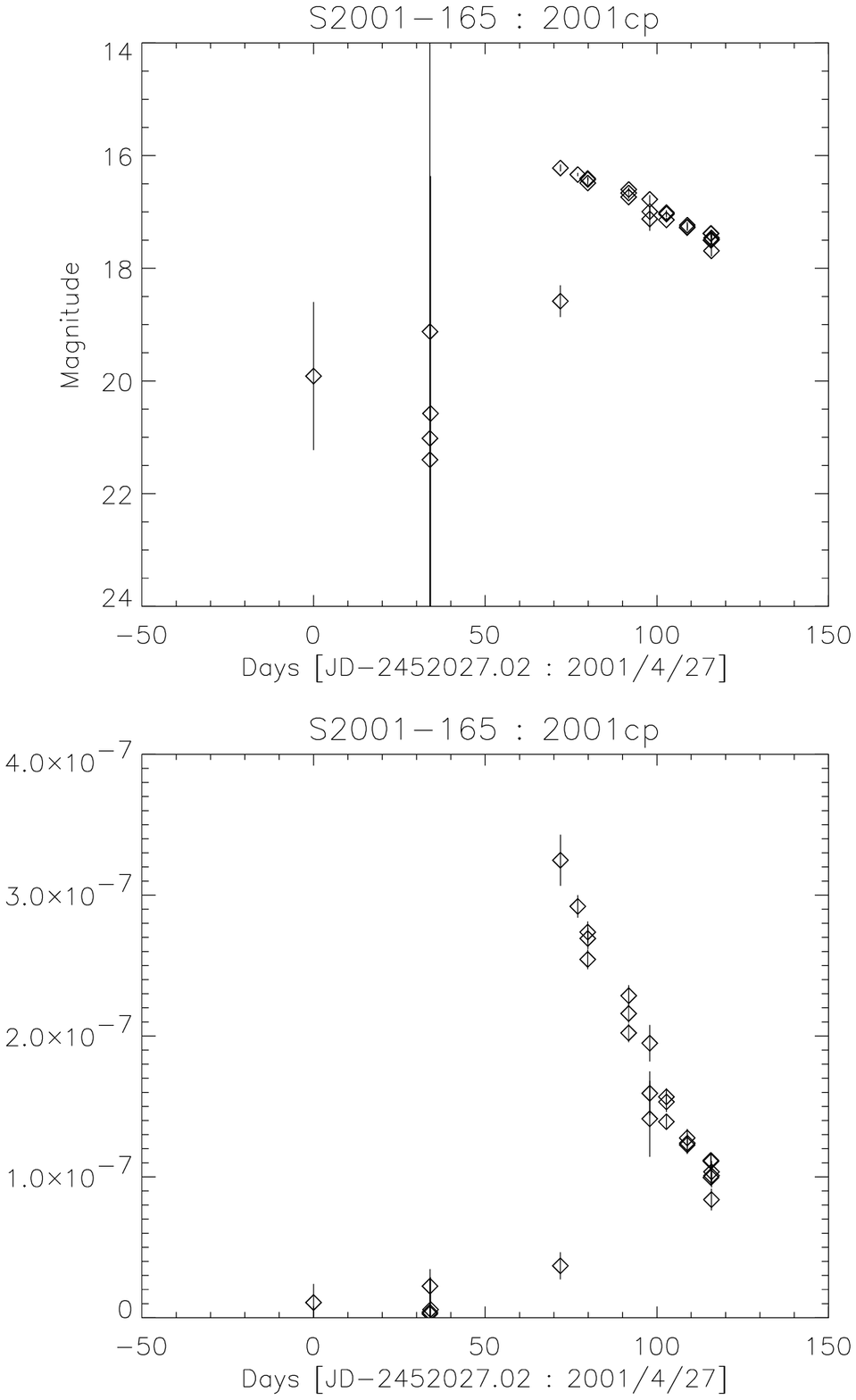}\label{fig:sn2001cp_lightcurve}}
\vspace{0.3in}
\subfigure[2001dd]{\includegraphics[angle=90,height=2in,width=3in]{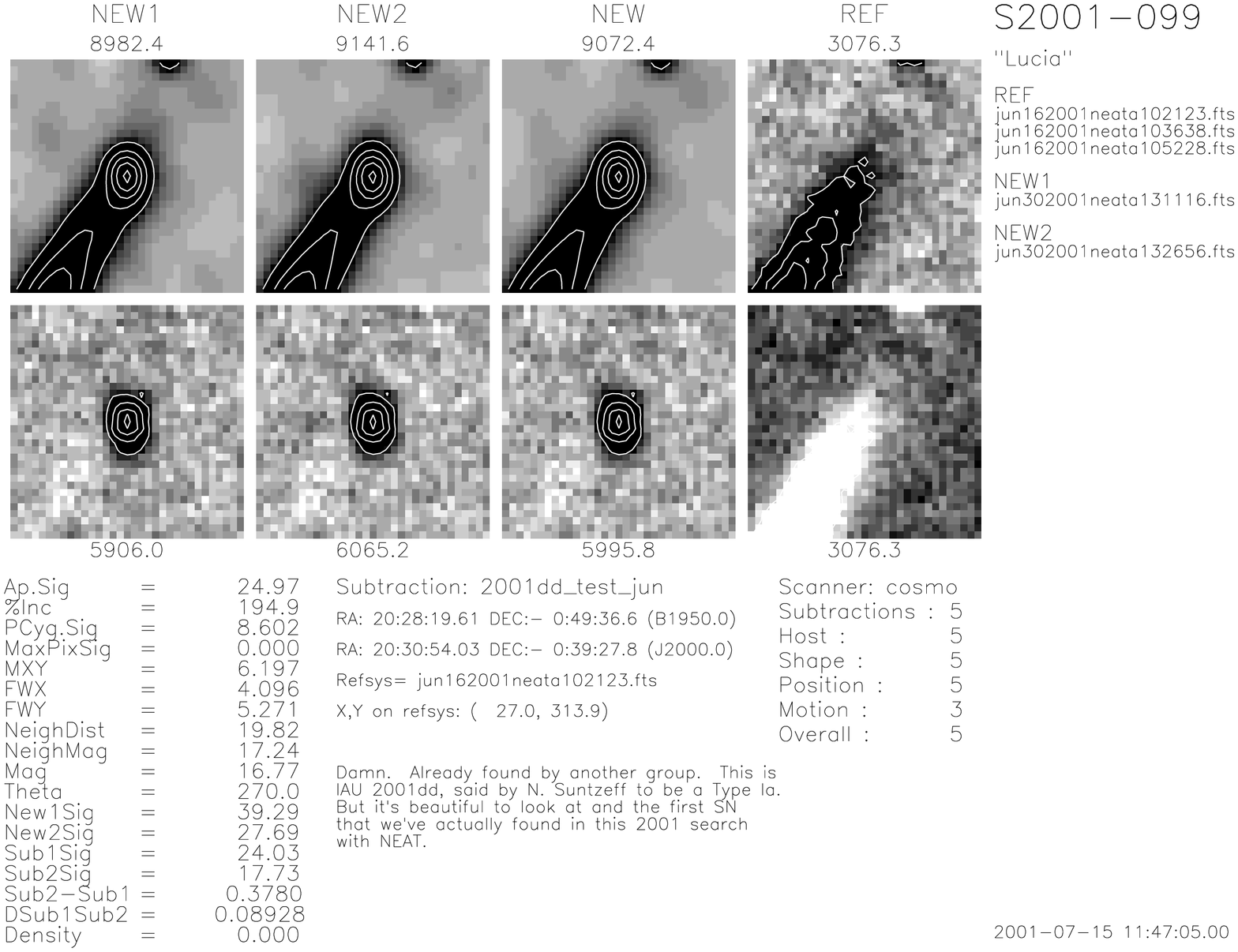}\label{fig:sn2001dd_discovery}}
\hspace{0.3in}
\subfigure[2001dd]{\includegraphics[height=2in]{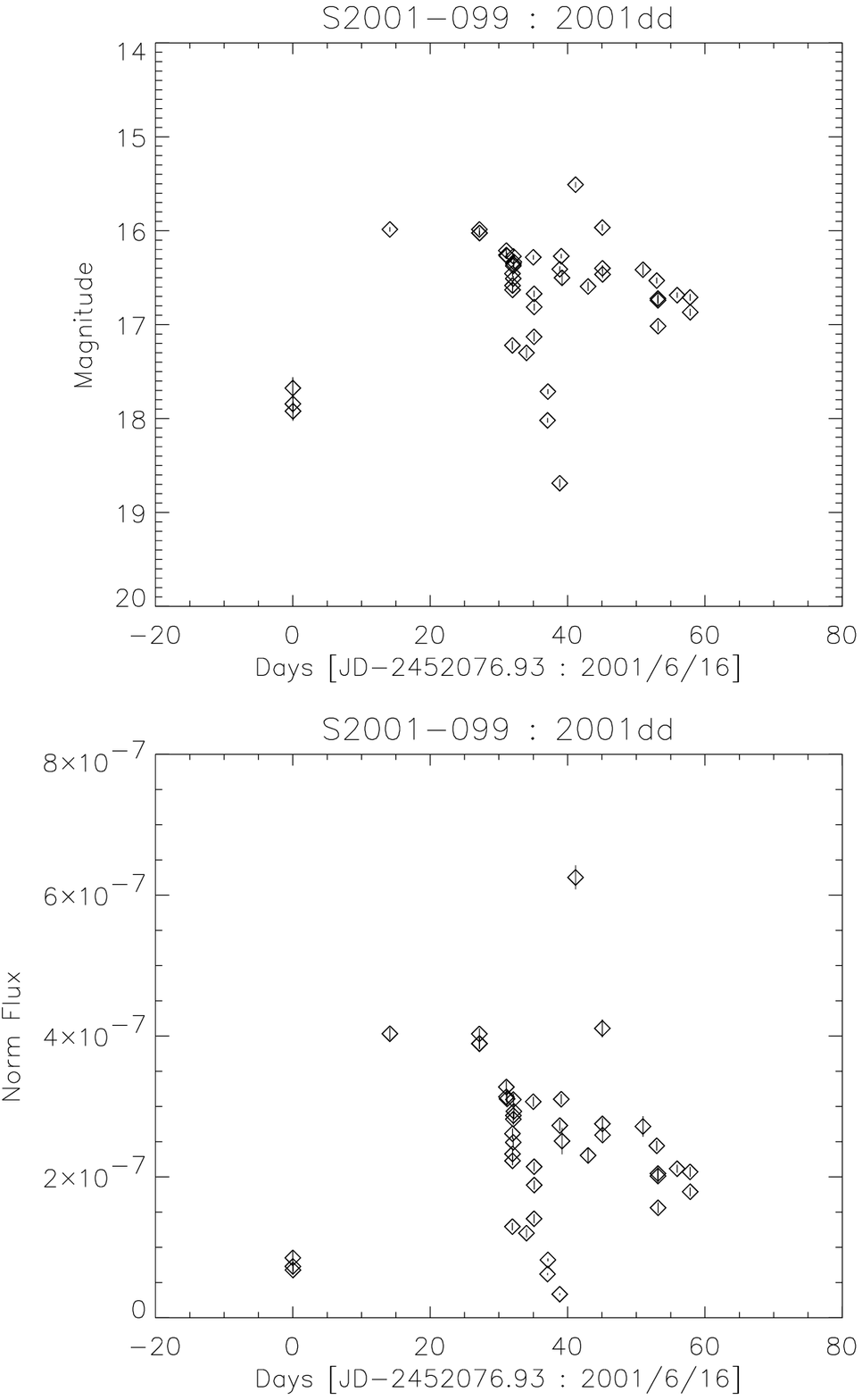}\label{fig:sn2001dd_lightcurve}}
\vspace{0.3in}
\subfigure[2002as]{\includegraphics[angle=90,height=2in,width=3in]{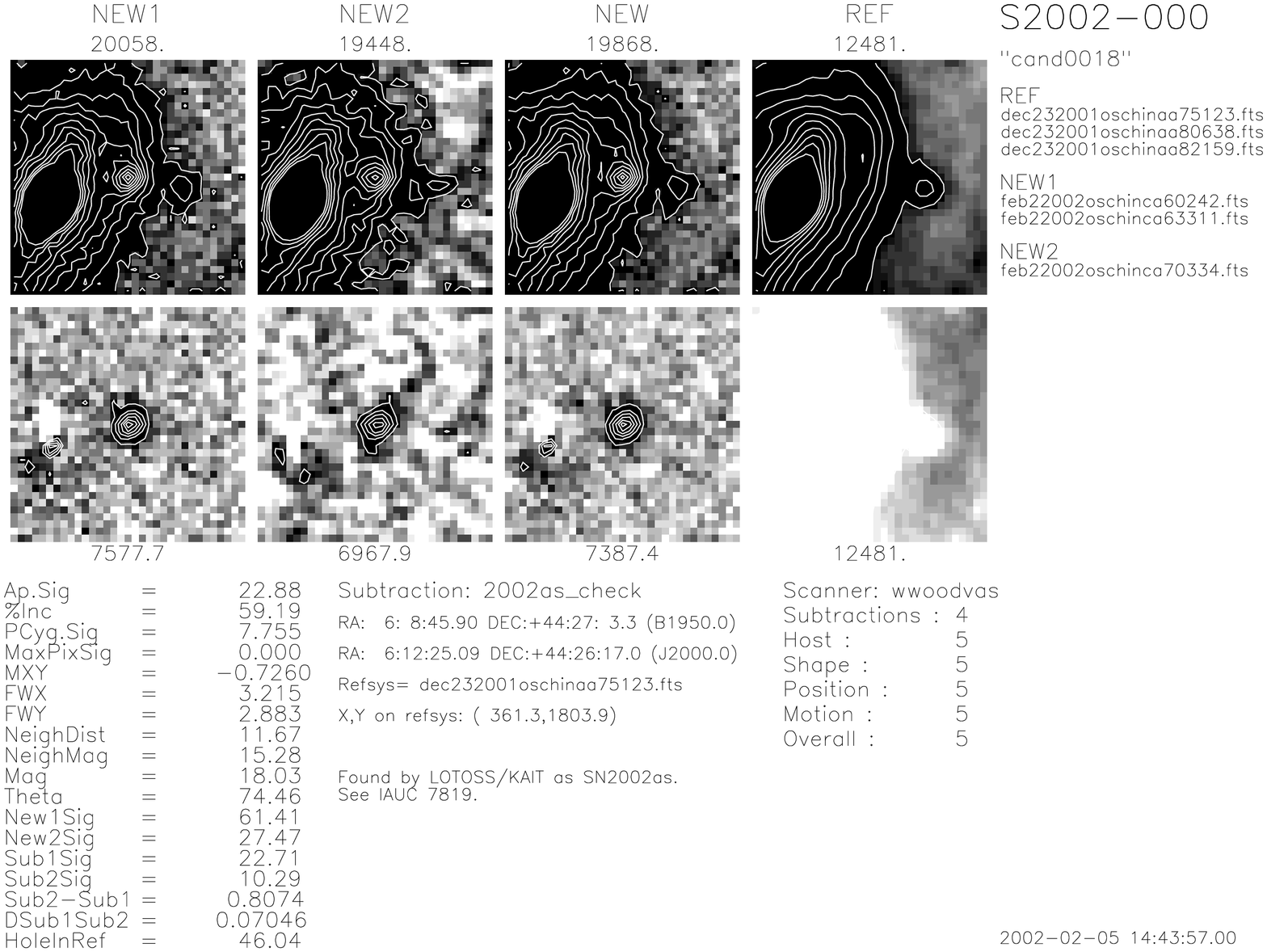}\label{fig:2002as_discovery}}
\hspace{0.3in}
\subfigure[2002as]{\includegraphics[height=2in]{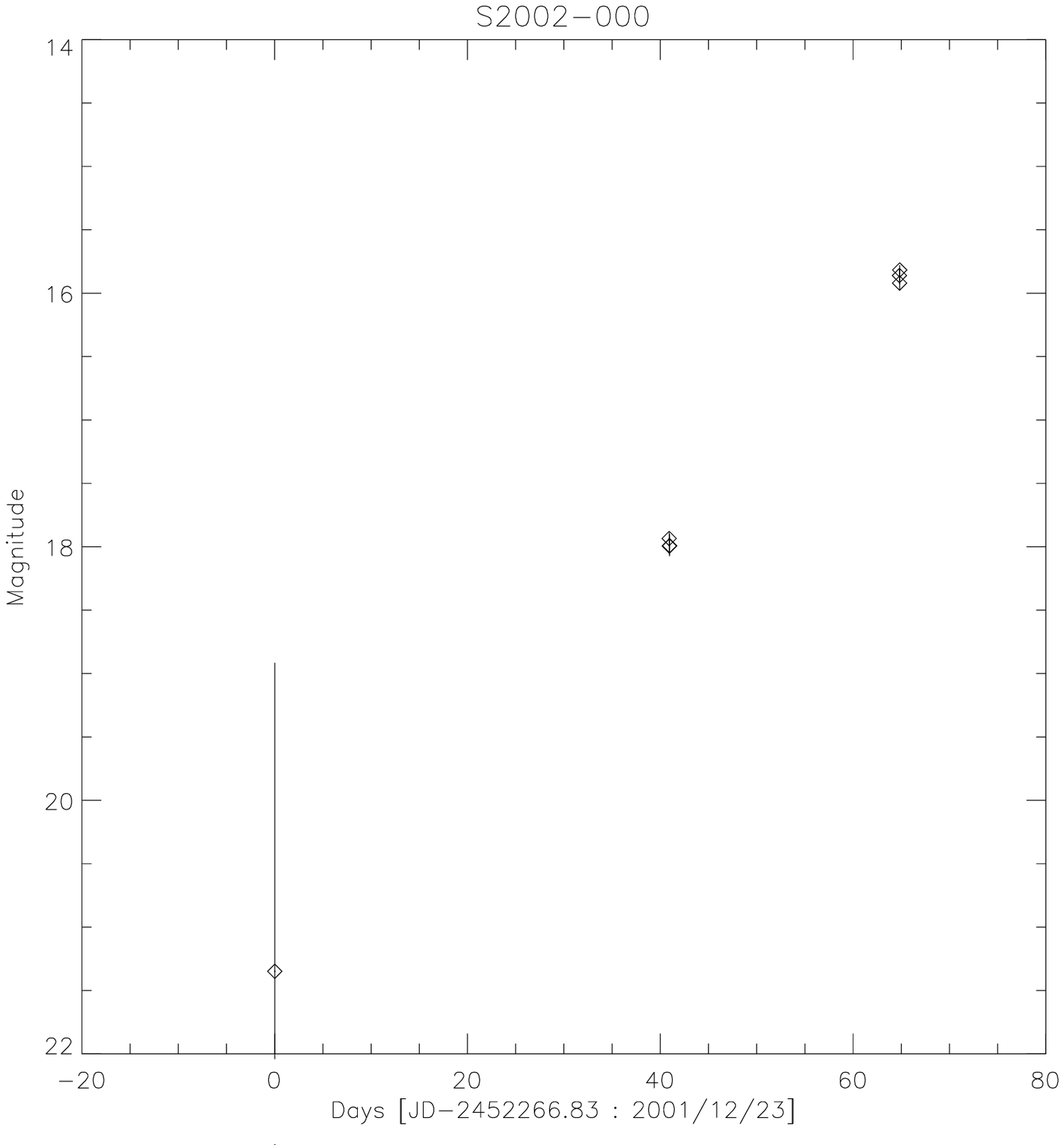}\label{fig:2002as_lightcurve}}
\vspace{0.3in}
\end{figure}

The early shake-down efforts in 2001 yielded two supernovae found after
others had reported them in the IAUCs: SN~2001cp (see
Fig.~\ref{fig:sn2001cp_discovery}) and SN~2001dd
(Fig.~\ref{fig:sn2001dd_discovery}).  Table~\ref{tab:2001_sne} lists
these supernovae along with their discovery magnitudes, redshifts,
and host galaxies.



\begin{deluxetable}{lllll}
\tablewidth{0pc}
\tablecaption{Supernovae in the data from 2001.}
\label{tab:2001_sne}
\tablehead{
\colhead{IAUC Name} & \colhead{SNfactory Name} & \colhead{Discovery Magnitude} & \colhead{Redshift} & \colhead{Comments}
}
\startdata
SN~2001cp  & S2001-165 & 16.35 &  0.02240  & UGC~10738 \\ 
SN~2001dd  & S2001-099 & 16.77 &  0.01969  & UGC~11579 \\
\enddata
\end{deluxetable}

\subsection{A successful prototype search pipeline test: 2002--2003}

The SNfactory discovered $37$ original supernovae in
2002~\citep{wood-vasey03aas1}\footnote{Two SNe from 2002 were reported
in mid-2003~\citep{iauc8149}}.  This was the most ever for the first
year of a supernova search.  The search continued with great success
in 2003 with the discovery of another $46$ original supernovae.  In
2003, the search pipeline discovered its first original supernovae
using the Haleakala 1.2-m telescope (SN~2003bf, SN~2003bn, SN~2003bp,
and SN~2003ey~\citep{iauc8082,iauc8088,iauc8089,iauc8141}).
Supernovae had already been found from with the Haleakala telescope
that were first discovered by others (SN~2001cp, SN~2002hw, and
SN~2002kj~\citep{iauc7645,iauc8014,iauc8053}).  By the time the
Palomar 3-CCD camera was removed in April, the search pipeline had
discovered $41$~supernovae and had reached the target rate of
12~SNe/month.  As two-thirds of the spectroscopically-typed supernovae
discovered by the early SNfactory efforts have been Type~Ia, one can
extrapolate that $12$~SNe/month will yield $8$~SNe~Ia/month $=
96$~SNe~Ia/year $\approx100$~SNe~Ia/year.  This satisfies the design
requirement for the SNfactory automated search pipeline.

\begin{deluxetable}{lllll}
\tablewidth{0pc}
\tablecaption{Supernovae found in 2002.}
\ssp
\tablehead{
\colhead{} & \colhead{} & \colhead{Discovery} & \colhead{} & {} \\
\colhead{IAUC Name} & \colhead{SNfactory Name} & \colhead{Magnitude} & \colhead{Redshift} & {Comments}
}
\startdata
SN~2002as  & S2002-000 &  18.03 &  0.02248  & \\
SN~2002bk  & S2002-006 &  18.51 &    0.057  & \\
SN~2002br  & S2002-058 & -9.404 & 0.034184  & disc. mag. in error \\
SN~2002cq  & S2002-054 &  19.21 &           & \\
SN~2002cx  & S2002-070 &  18.91  &    0.024  & \\
SN~2002cz  & S2002-080 &  18.58 &           & \\
SN~2002da  & S2002-078 &  18.81 &           & \\
SN~2002dg  & S2002-087 &  18.46 &    0.047  & \\
SN~2002dh  & S2002-095 &  15.82 &    0.013  & \\
SN~2002dy  & F2002-074 &  17.78 &  0.03301  & \\
SN~2002ek  & F2002-040 &  19.50 &           & \\
SN~2002el  & F2002-048 &  16.72 &           & \\
SN~2002ep  & F2002-039 &  17.47 &           & \\
SN~2002eq  & F2002-041 &  19.43 &    0.088  & \\
SN~2002ev  & F2002-056 &  18.75 &           & \\
SN~2002ew  & F2002-049 &  17.82 &     0.03  & \\
SN~2002ex  & F2002-054 &  19.12 &    0.038  & \\
SN~2002ez  & F2002-043 &  18.38 &    0.043  & \\
SN~2002fa  & F2002-050 &  19.41 &     0.06  & \\
SN~2002fs  & F2002-059 &  17.44 &    0.038  & \\
SN~2002ft  & F2002-062 &  18.13 &    0.075  & \\
SN~2002fu  & F2002-064 &  19.36 &    0.091  & \\
SN~2002gb  & F2002-058 &  18.89 &    0.074  & \\
SN~2002gd  & F2002-070 &  18.44 &   0.0085  & \\
SN~2002gf  & F2002-071 &  19.19 &    0.086  & \\
SN~2002gg  & F2002-069 &  20.06 &     0.11  & \\
SN~2002gh  & F2002-072 &  19.66 &           & \\
SN~2002gx  & F2002-076 &  18.97 &           & \\
SN~2002gz  & F2002-078 &  18.38 &    0.085  & \\
SN~2002hb  & F2002-081 &  19.13 &     0.09  & \\
SN~2002hf  & F2002-087 &  18.21 &           & \\
SN~2002hj  & F2002-089 &  17.81 &   0.0236  & \\
SN~2002hw  & F2002-141 &  17.24 & 0.017535  & \\
SN~2002ia  & F2002-101 &  19.03 &    0.072  & \\
SN~2002ib  & F2002-097 &  18.24 &   0.0679  & \\
SN~2002ic  & F2002-099 &  18.25 &   0.0664  & \\
SN~2002jf  & F2002-104 &  18.62 &    0.079  & \\
SN~2002jh  & F2002-098 &  18.62 &    0.048  & \\
SN~2002jk  & S2002-081 &  18.46 &           & \\
SN~2002jl  & F2002-106 &  19.50 &    0.064  & \\
SN~2002jp  & S2003-152 &  19.09 & 0.012362  & \\
SN~2002kj  & F2002-140 &  18.51 &           & \\
SN~2002kk  & F2002-046 &  19.42 &           & \\
SN~2002le  & S2002-075 &  19.47 &           & \\
SN~2002lf  & S2002-072 &  19.87 &           & \\
\enddata
\dsp
\label{tab:2002_sne}
\end{deluxetable}

\begin{deluxetable}{lllll}
\tablewidth{0pc}
\tablecaption{Supernovae found in 2003.}
\ssp
\tablehead{
\colhead{} & \colhead{} & \colhead{Discovery} & \colhead{} & {} \\
\colhead{IAUC Name} & \colhead{SNfactory Name} & \colhead{Magnitude} & \colhead{Redshift} & {Comments}
}
\startdata
SN~2003V   & S2003-003 & 18.50 &    0.045  & \\
SN~2003aa  & S2003-113 & 16.30 &  0.01013  & \\
SN~2003ab  & S2003-088 & 18.17 & 0.029414  & \\
SN~2003ae  & S2003-005 & 17.77 &           & \\
SN~2003af  & S2003-011 & 15.60 &     0.02  & \\
SN~2003ap  & S2003-018 & 16.71 &     0.03  & \\
SN~2003av  & S2003-014 & 19.63 &     0.14  & \\
SN~2003aw  & S2003-016 & 17.72 &           & \\
SN~2003ax  & S2003-026 & 18.52 &    0.054  & \\
SN~2003ay  & S2003-025 & 19.16 &    0.073  & \\
SN~2003bf  & S2003-033 & 17.19 &    0.034  & \\
SN~2003bh  & S2003-029 & 19.93 &    0.089  & \\
SN~2003bi  & S2003-030 & 19.59 &           & \\
SN~2003bn  & S2003-073 & 16.75 &    0.013  & \\
SN~2003bo  & S2003-076 & 19.53 &           & \\
SN~2003bp  & S2003-077 & 18.17 & 0.019807  & \\
SN~2003bs  & S2003-080 & 17.73 &     0.05  & \\
SN~2003bt  & S2003-086 & 16.69 & 0.027516  & \\
SN~2003cc  & S2003-112 & 19.50 &           & \\
SN~2003cd  & S2003-111 & 19.98 &           & \\
SN~2003ce  & S2003-094 & 19.49 &           & \\
SN~2003cf  & S2003-115 & 19.01 &           & \\
SN~2003cj  & S2003-138 & 20.04 &           & \\
SN~2003ck  & S2003-129 & 19.26 &           & \\
SN~2003cl  & S2003-147 & 19.51 & 0.168301  & \\
SN~2003cn  & S2003-174 & 18.34 &  0.01811  & \\
SN~2003co  & S2003-120 & 18.60 &   0.0824  & \\
SN~2003cs  & S2003-137 & 19.36 & 0.029583  & \\
SN~2003ct  & S2003-139 & 19.03 & 0.046075  & \\
SN~2003cu  & S2003-145 & 19.91 &           & \\
SN~2003cv  & S2003-156 & 17.62 &    0.028  & \\
SN~2003cw  & S2003-157 & 20.07 &           & \\
SN~2003cx  & S2003-160 & 19.07 &    0.037  & \\
SN~2003cy  & S2003-154 & 18.96 &           & \\
SN~2003cz  & S2003-163 & 20.16 &           & \\
SN~2003dc  & S2003-134 & 19.33 &    0.067  & \\
SN~2003dd  & S2003-133 & 20.08 &           & \\
SN~2003de  & S2003-162 & 17.22 &           & \\
SN~2003df  & S2003-110 & 20.31 &           & \\
SN~2003di  & S2003-116 & 18.00 &           & \\
SN~2003dj  & S2003-126 & 20.01 &           & \\
SN~2003dk  & S2003-150 & 19.19 &           & \\
SN~2003dm  & S2003-167 & 19.73 &           & \\
SN~2003dn  & S2003-164 & 20.02 &           & \\
SN~2003do  & S2003-171 & 19.91 &           & \\
SN~2003dp  & S2003-170 & 19.98 &           & \\
SN~2003dq  & S2003-172 & 18.91 &    0.046  & \\
SN~2003ee  & S2003-149 & 16.47 &    0.021  & \\
SN~2003ef  & S2003-191 & 17.48 &           & \\
SN~2003eo  & S2003-141 & 20.20 &           & \\
SN~2003ex  & S2003-130 & 19.11 &           & \\
SN~2003ey  & S2003-197 & 18.75 &    0.065  & \\
SN~2003gc  & S2003-144 & 18.00 &           & \\
SN~2003gi  & F2003-042 & 19.18 & 0.013006  & \\
SN~2003gt  & F2003-000 & 18.34 & 0.015657  & \\
\enddata
\dsp
\label{tab:2003_sne}
\end{deluxetable}

\section{Current Status of SNfactory search -- 2004}

The SNfactory search has been proven to yield the necessary number of
supernovae to support the SNfactory follow-up program.  Operation has
begun again as of July, 2004.  Data from the QUEST drift-scan program
has been integrated into the search pipeline.  Using the QUESTII
drift-scan data necessitated the incorporation of the more specific
coverage pattern of the drift-scan program and the flexibility to
reconstruct neighboring segments along the scan strip to maximize
overlap in the subtractions.  In addition, more robust matching and
image quality routines were required to deal with the larger number of
marginal quality CCDs on the QUESTII detector.  The current plan to
use the QUEST drift-scan data is to take references a month prior and
then one new image taken on each of two search nights separated by two
or three days.  Subtractions can be run across different red filters
(Johnson R vs. Gunn r) if necessary.  In preliminary tests, these
cross-filter subtractions have posed no serious difficulties.  There
are specific tweaks in the matching code to activate this behavior for
the QUESTII drift-scan subtractions.
 
Improvements to the automatic candidates screening will be helpful in
reducing the human burden in searching for supernovae.  Currently
$1$--$5$\% of the images need to be checked by hand and $1$\% of those
have supernovae.  This represents a few hours of person-time every day
to scan.  This is workable, but a more fully automated system would be
preferable for better consistency and determinism.  Routes toward an
improved system include improvements in the suite of candidate scores
and an investigation of optimizing the PSF-matching required to
compare images taken under different atmospheric and telescopic
conditions.  Searching for SNe using the QUEST drift-scan data has
introduced some benefits and complications in the candidate scoring 
as the different search nights used in NEW1 and NEW2 have, in general,
different PSF shapes.  This night-to-night PSF variation reduces
artifacts due to improper PSF matching because they are eliminated by
the separate requirements in SUB1 and SUB2.  However, the effective
PSF in the overall NEW image can be harder to match to the REF images.
Additionally, over two days between subsequent QUESTII drift-scan search observations, practically everything in the Solar
System moves sufficiently to be eliminated by its apparent motion.
Only true, stationary astrophysical objects should appear similar
enough to be present in the joint subtraction.  With its increased
area and sensitivity, the new QUESTII camera has the potential to
double the SN discovery rate of the SNfactory.


\newpage


\clearpage\pagebreak

\begin{figure}
\subfigure[2002bk]{\includegraphics[angle=90,height=2in,width=3in]{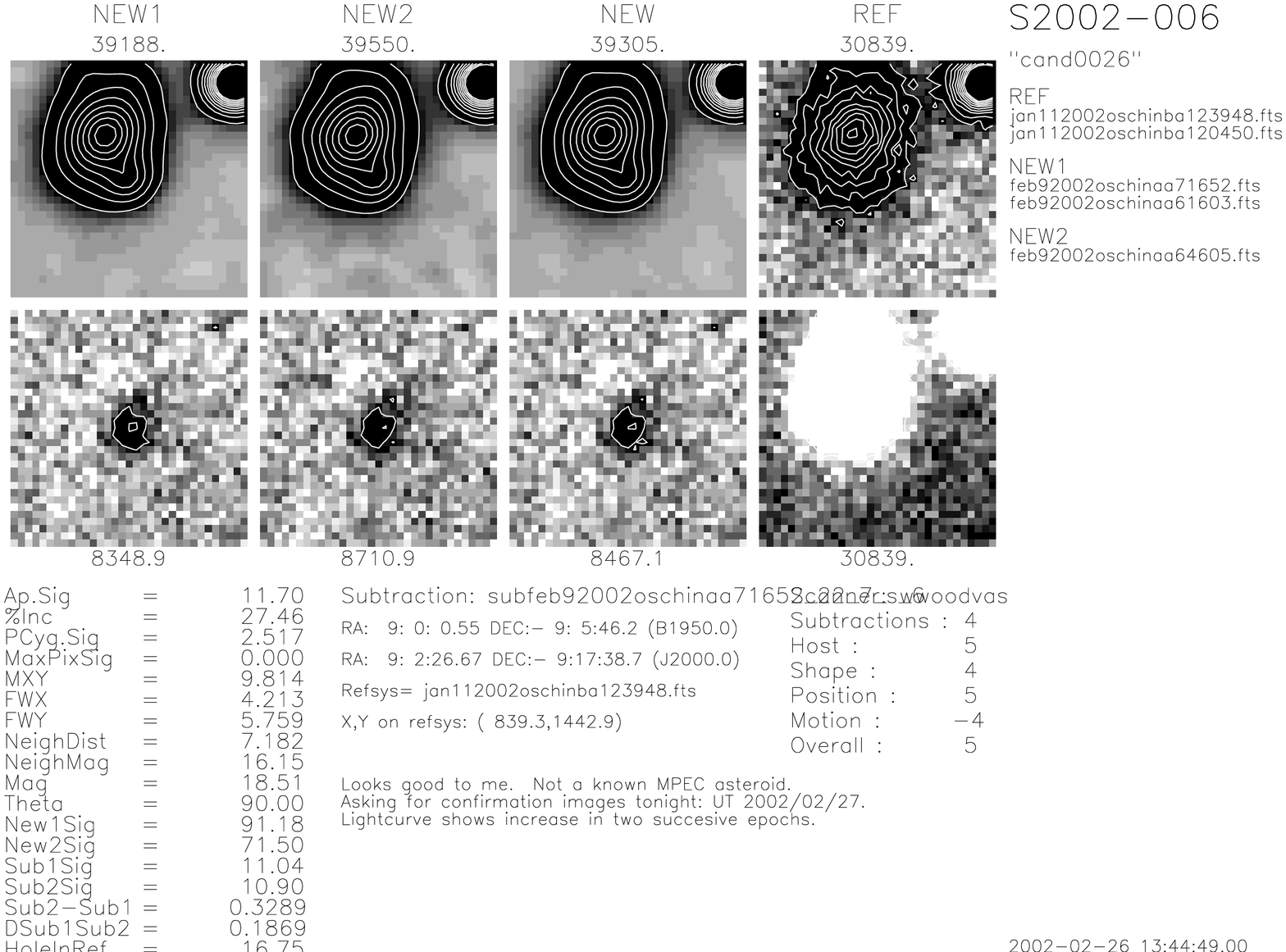}\label{fig:sn2002bk_discovery}}
\hspace{0.3in}
\subfigure[2002bk]{\includegraphics[height=2in]{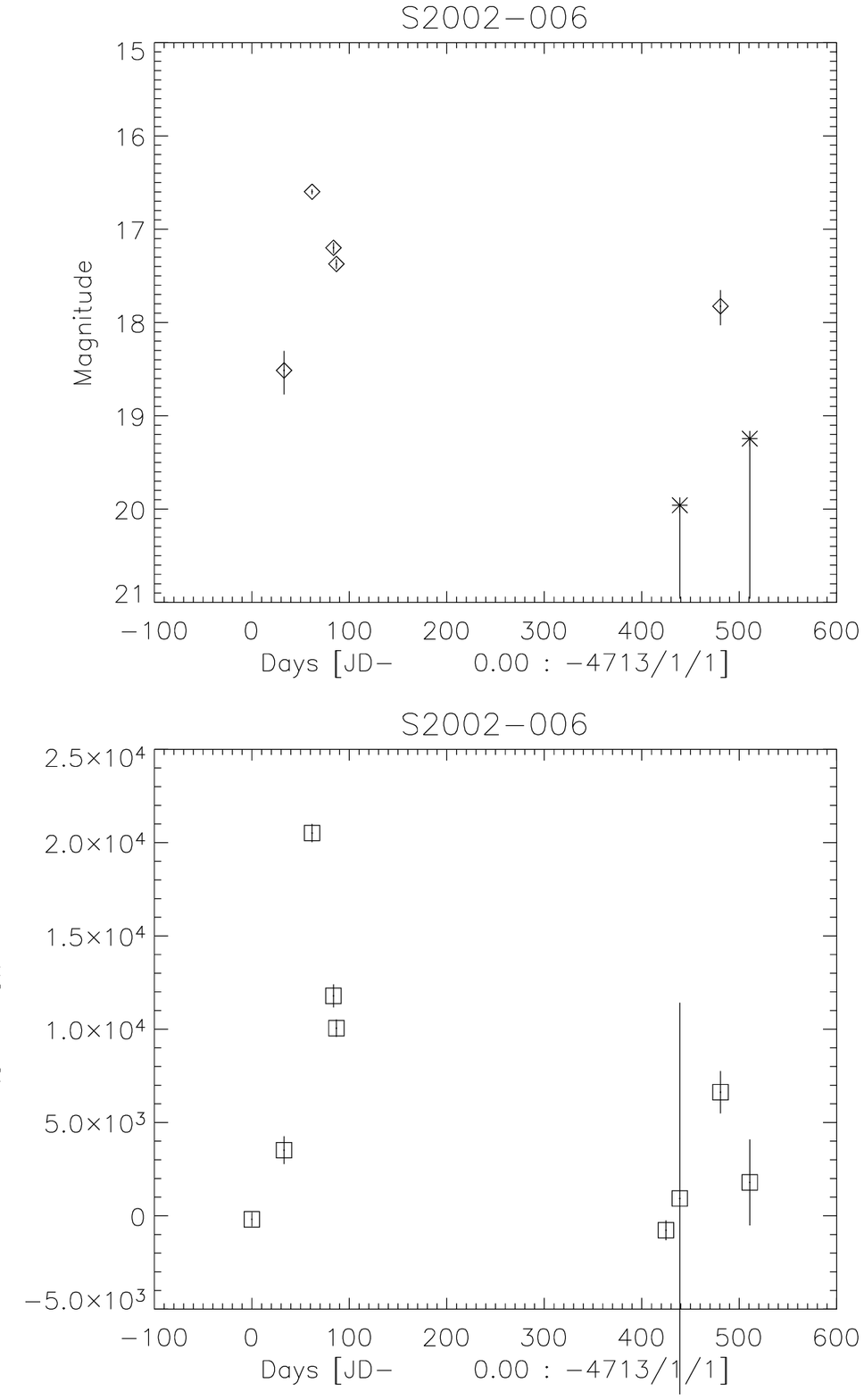}\label{fig:sn2002bk_lightcurve}}
\vspace{0.3in}
\subfigure[2002br]{\includegraphics[angle=90,height=2in,width=3in]{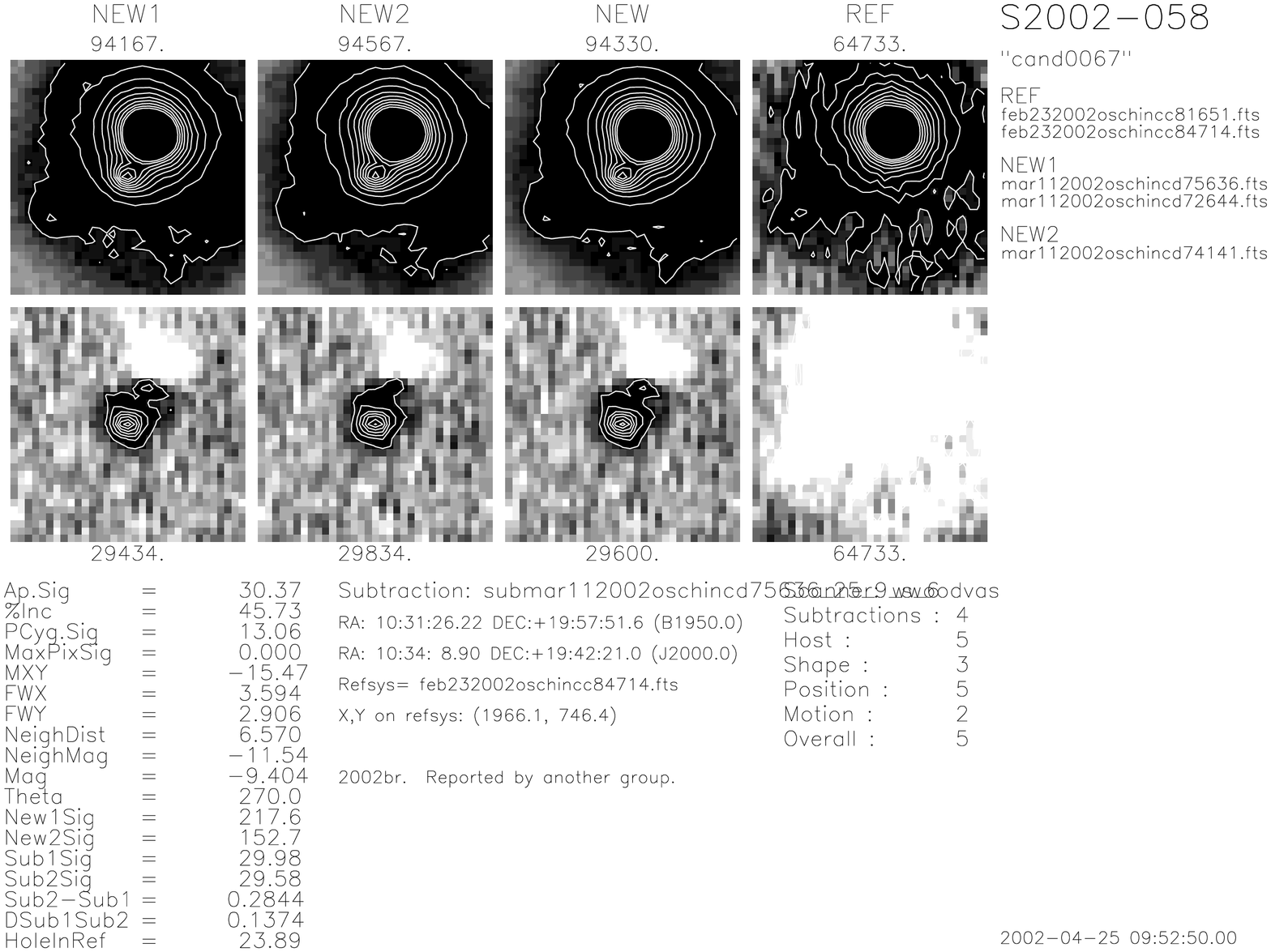}\label{fig:2002br_discovery}}
\hspace{0.3in}
\subfigure[2002br]{\includegraphics[height=2in]{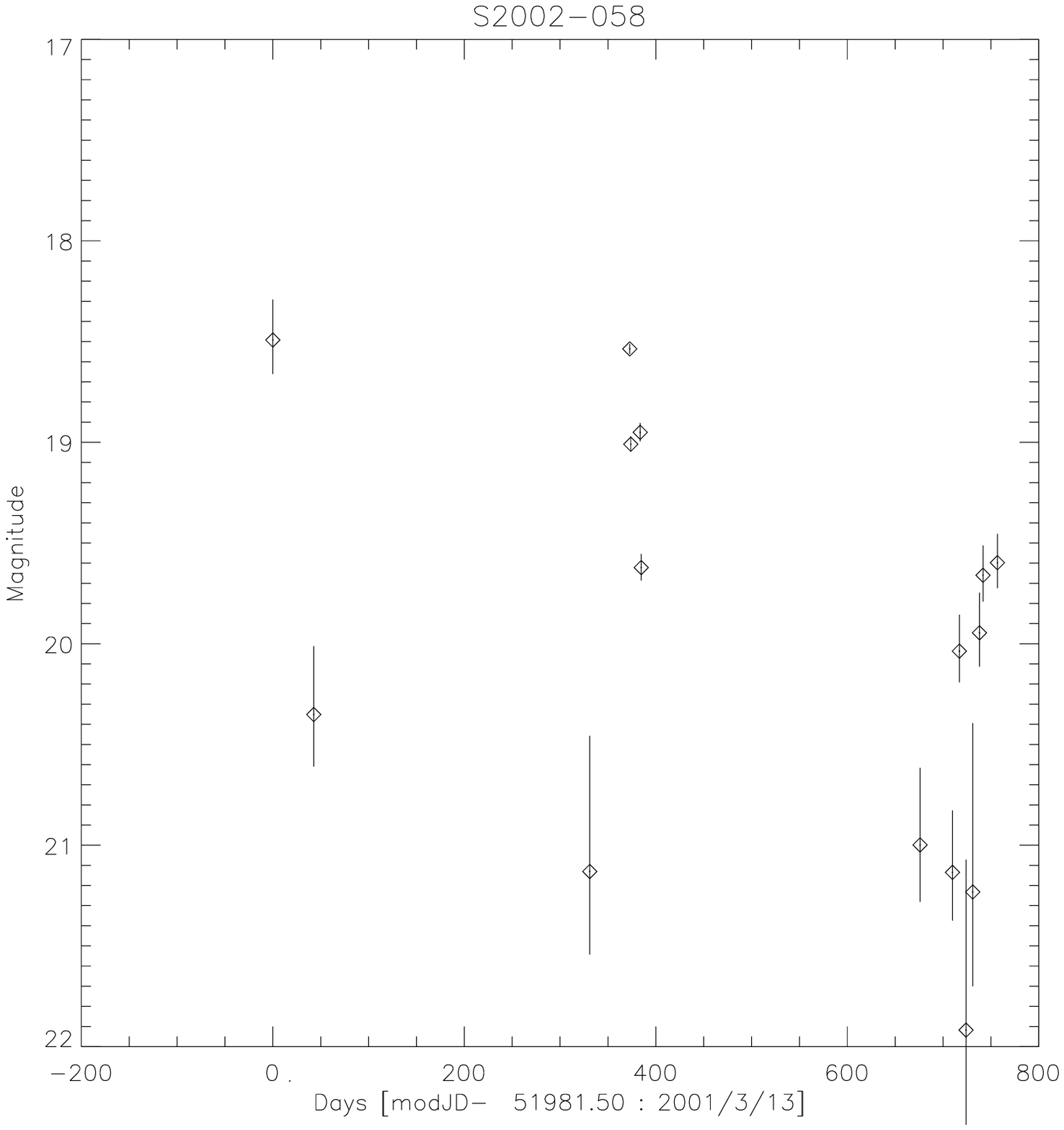}\label{fig:2002br_lightcurve}}
\vspace{0.3in}
\subfigure[2002cq]{\includegraphics[angle=90,height=2in,width=3in]{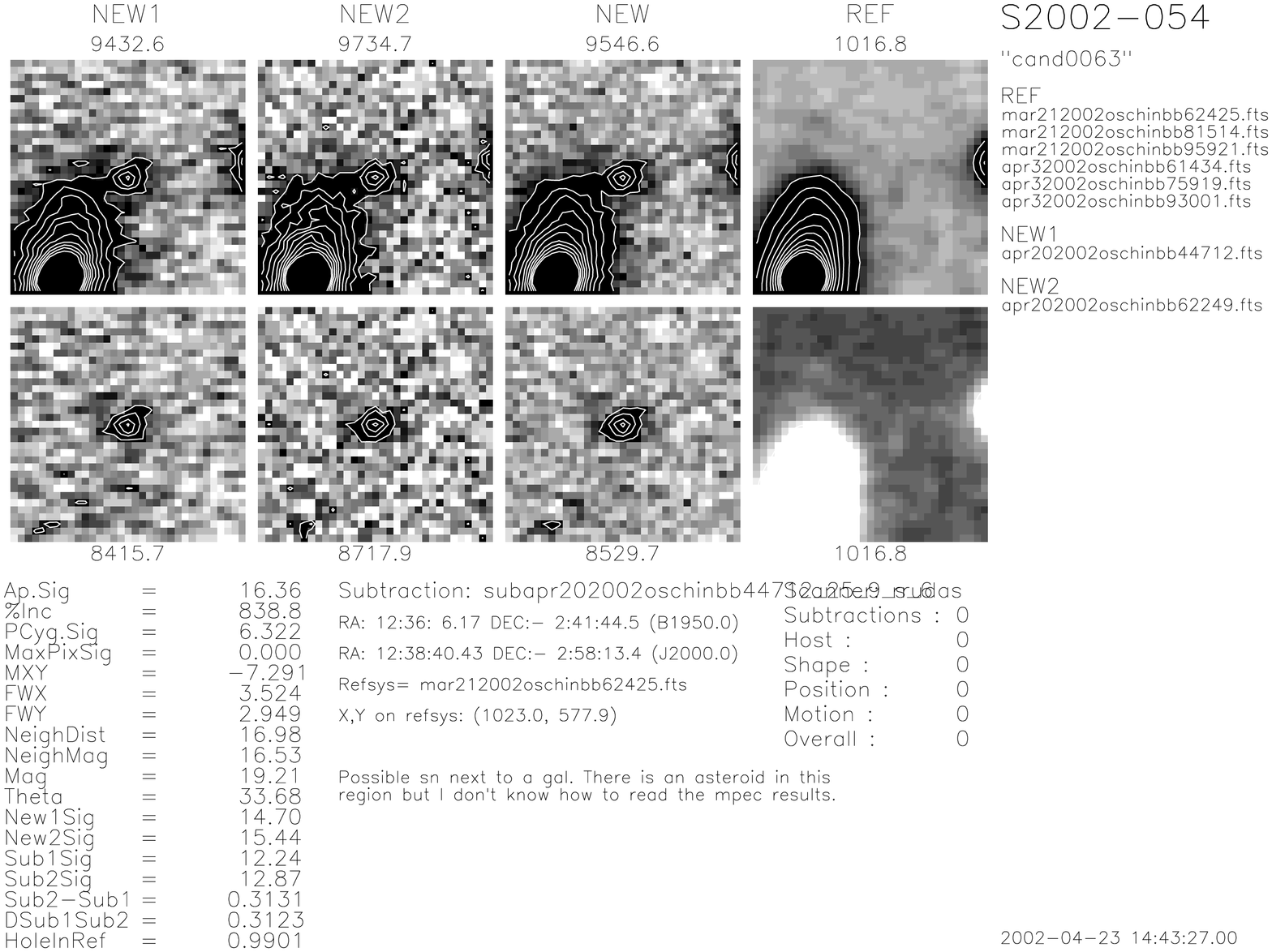}\label{fig:2002cq_discovery}}
\hspace{0.3in}
\subfigure[2002cq]{\includegraphics[height=2in]{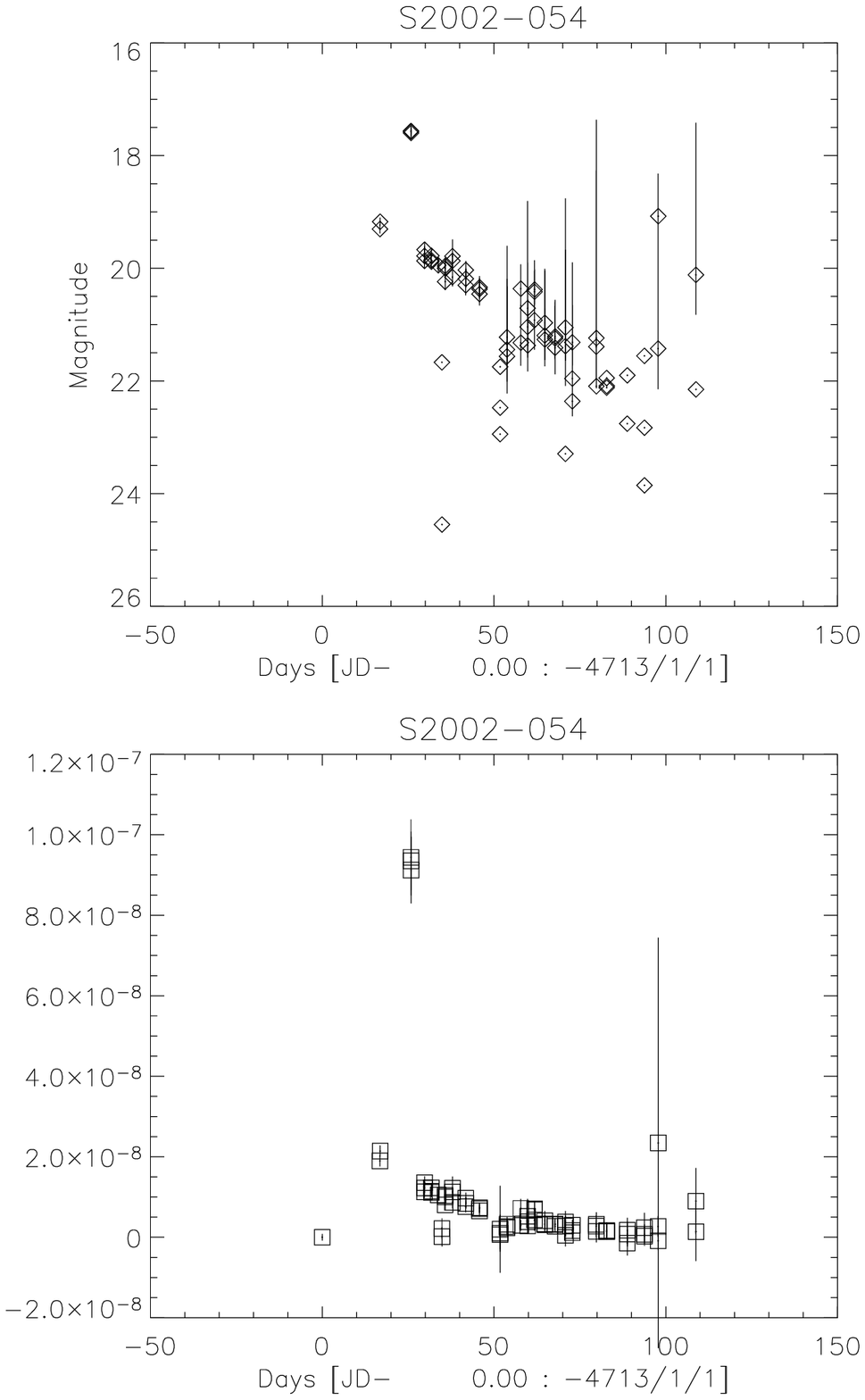}\label{fig:2002cq_lightcurve}}
\end{figure}

\clearpage\pagebreak
\begin{figure}
\subfigure[2002cx]{\includegraphics[angle=90,height=2in,width=3in]{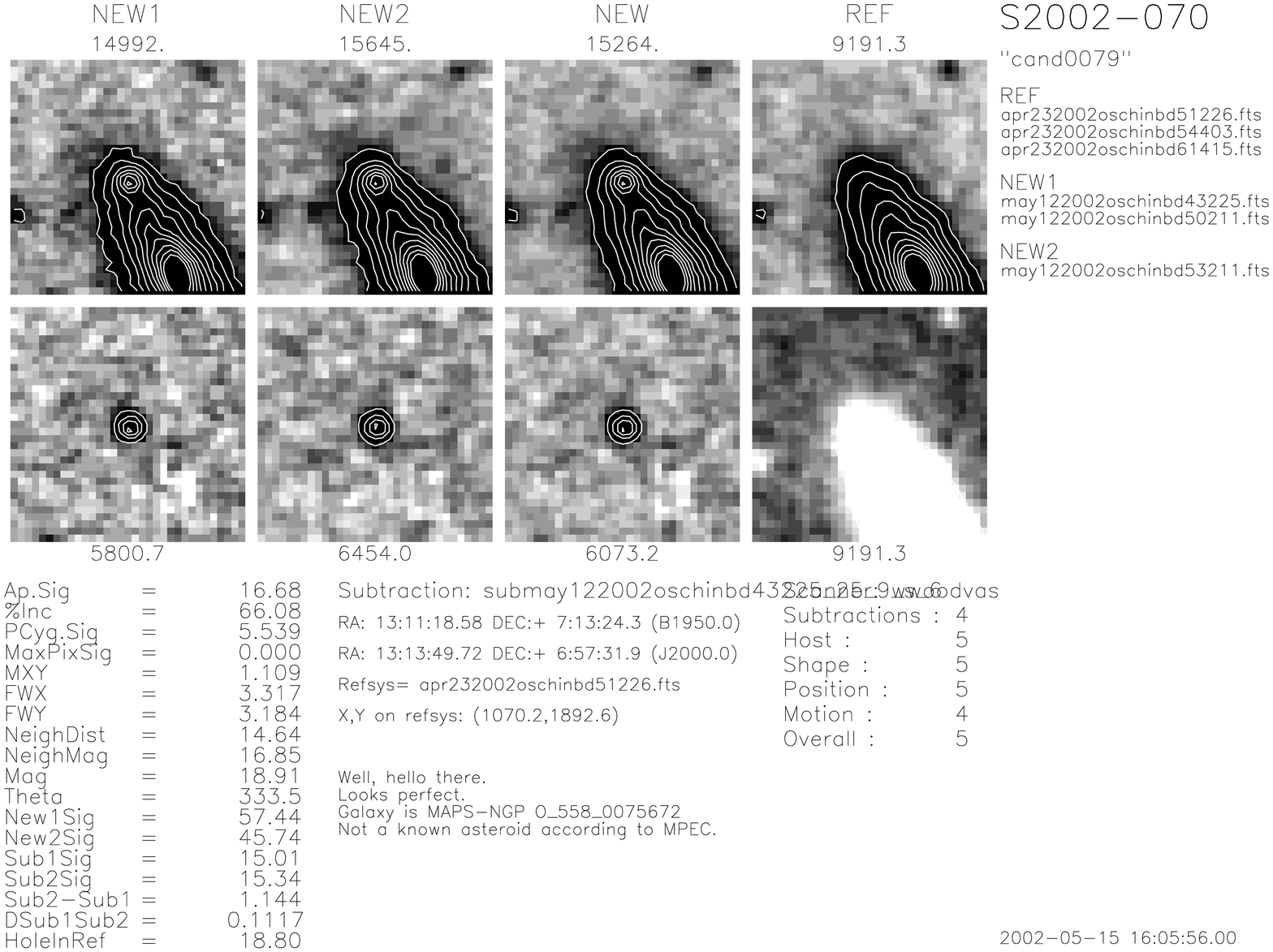}\label{fig:2002cx_discovery}}
\hspace{0.3in}
\subfigure[2002cx]{\includegraphics[height=2in]{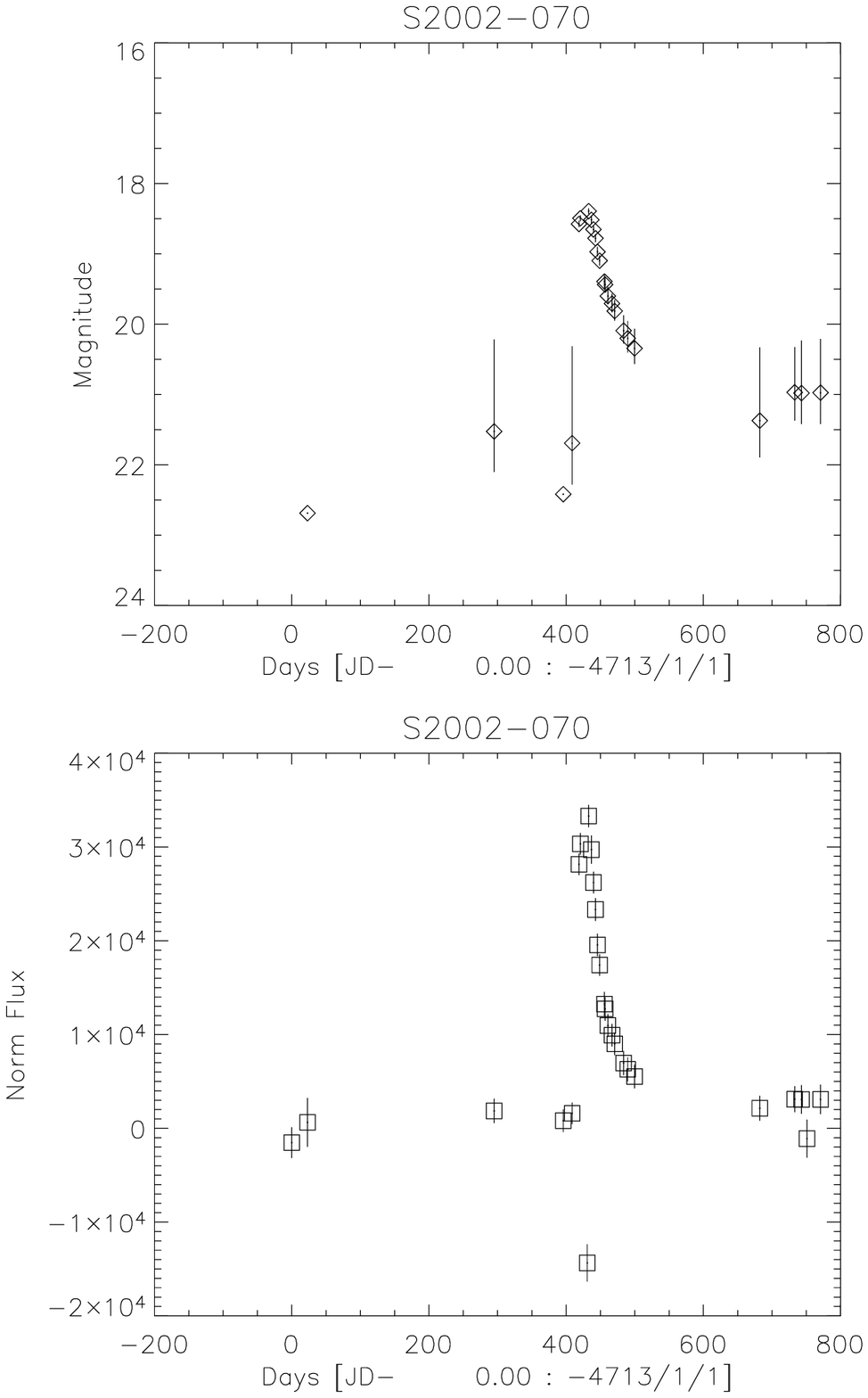}\label{fig:2002cx_lightcurve}}
\vspace{0.3in}
\subfigure[2002cz]{\includegraphics[angle=90,height=2in,width=3in]{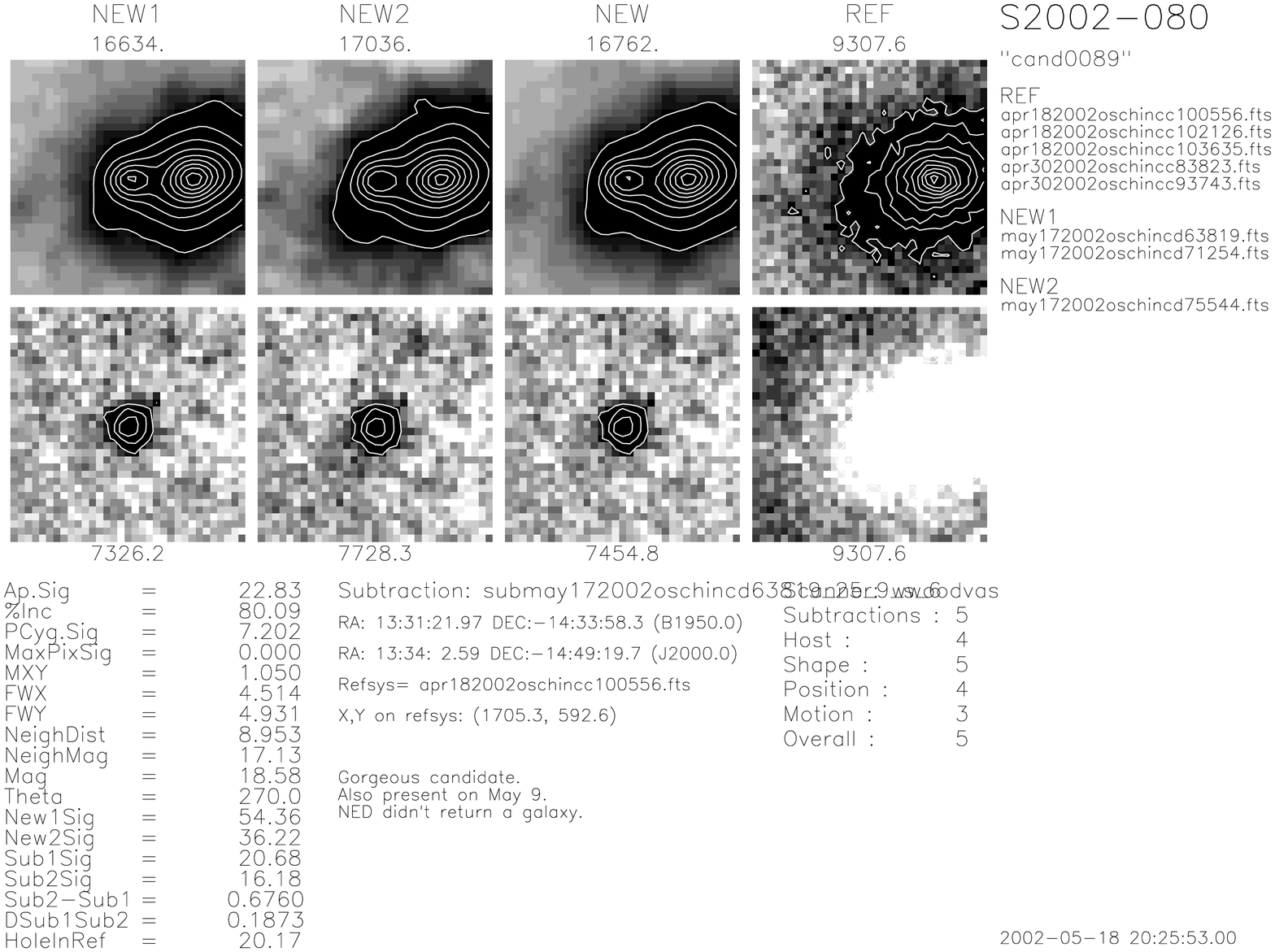}\label{fig:2002cz_discovery}}
\hspace{0.3in}
\subfigure[2002cz]{\includegraphics[height=2in]{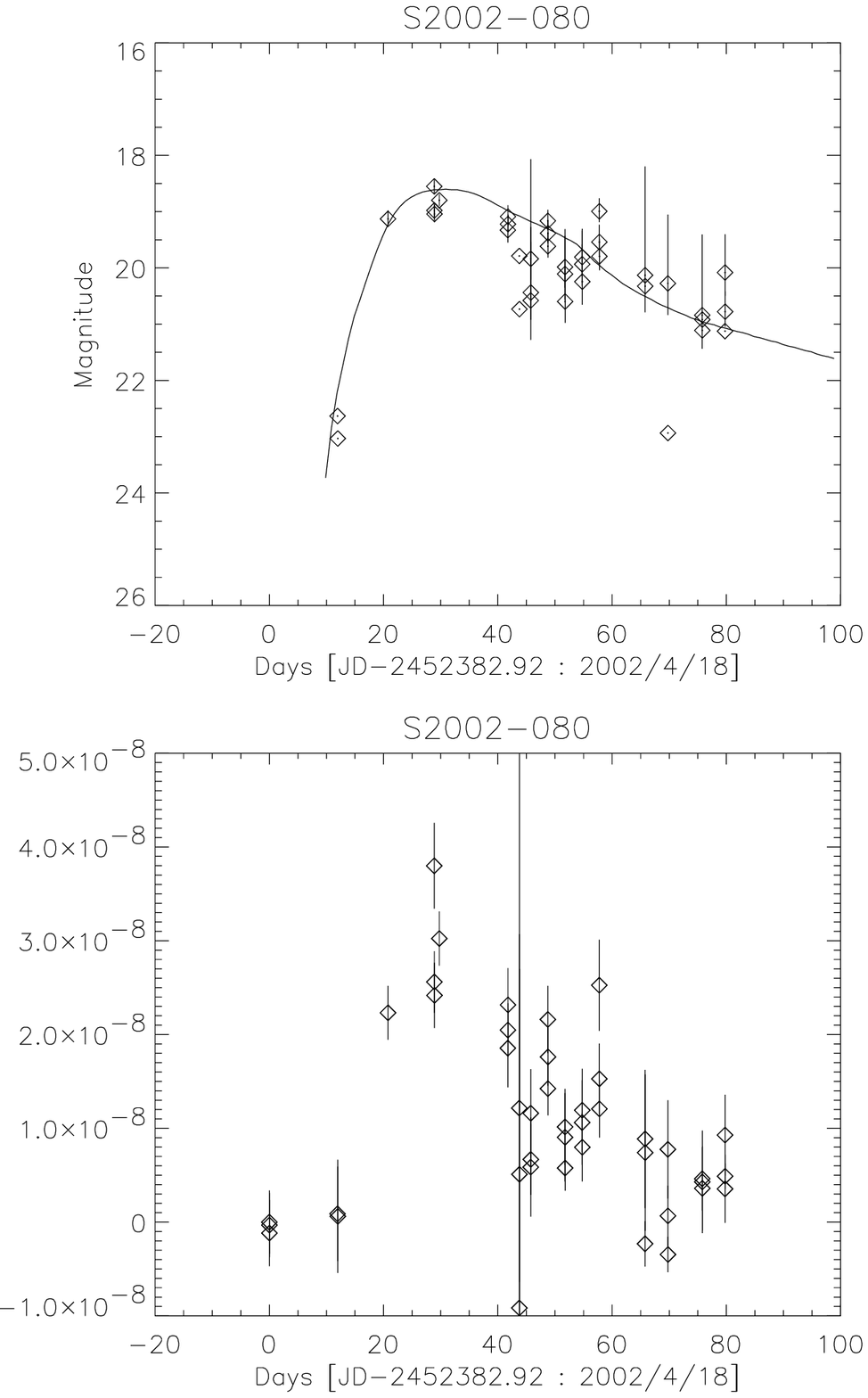}\label{fig:2002cz_lightcurve}}
\vspace{0.3in}
\subfigure[2002da]{\includegraphics[angle=90,height=2in,width=3in]{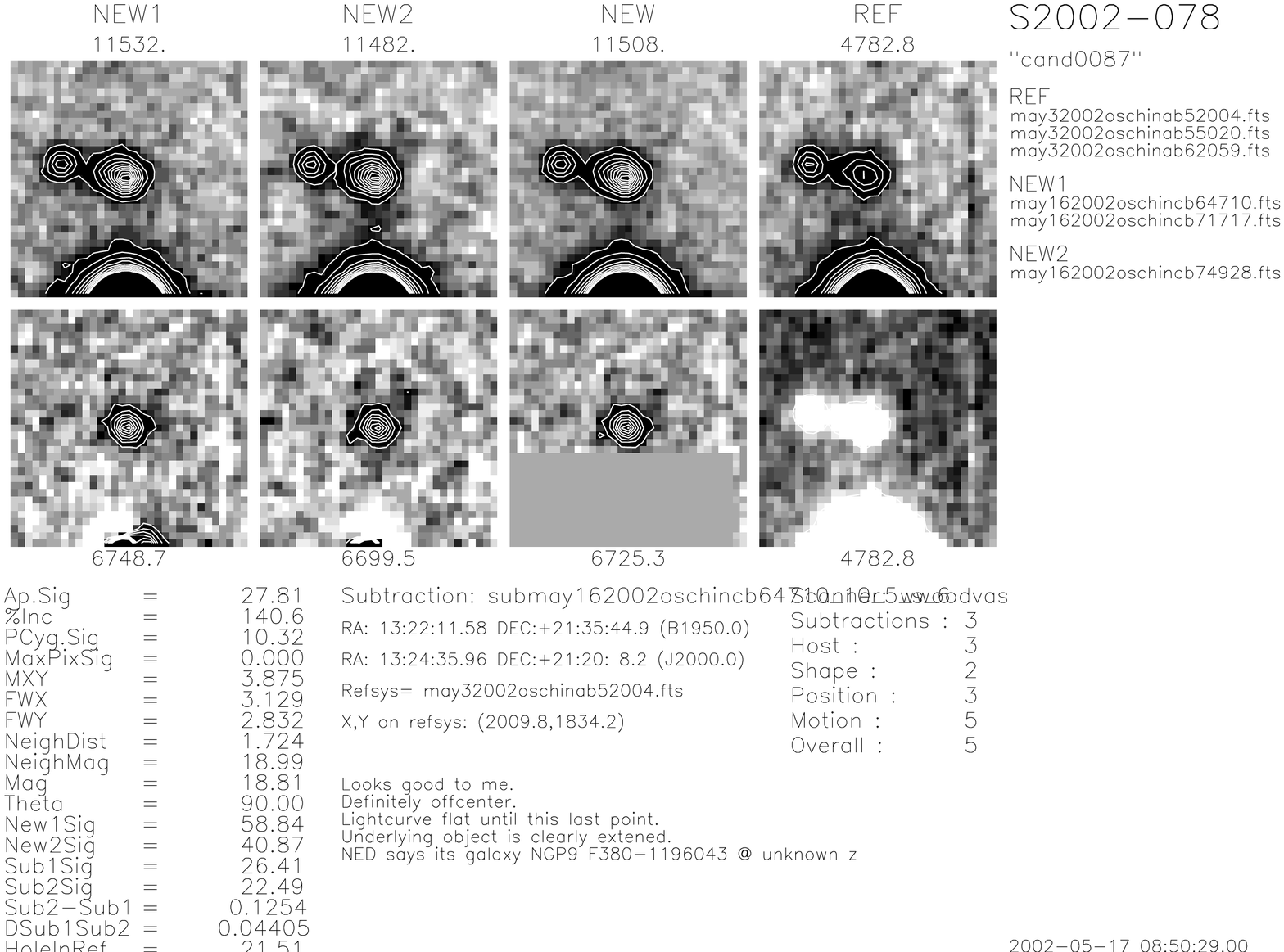}\label{fig:2002da_discovery}}
\hspace{0.3in}
\subfigure[2002da]{\includegraphics[height=2in]{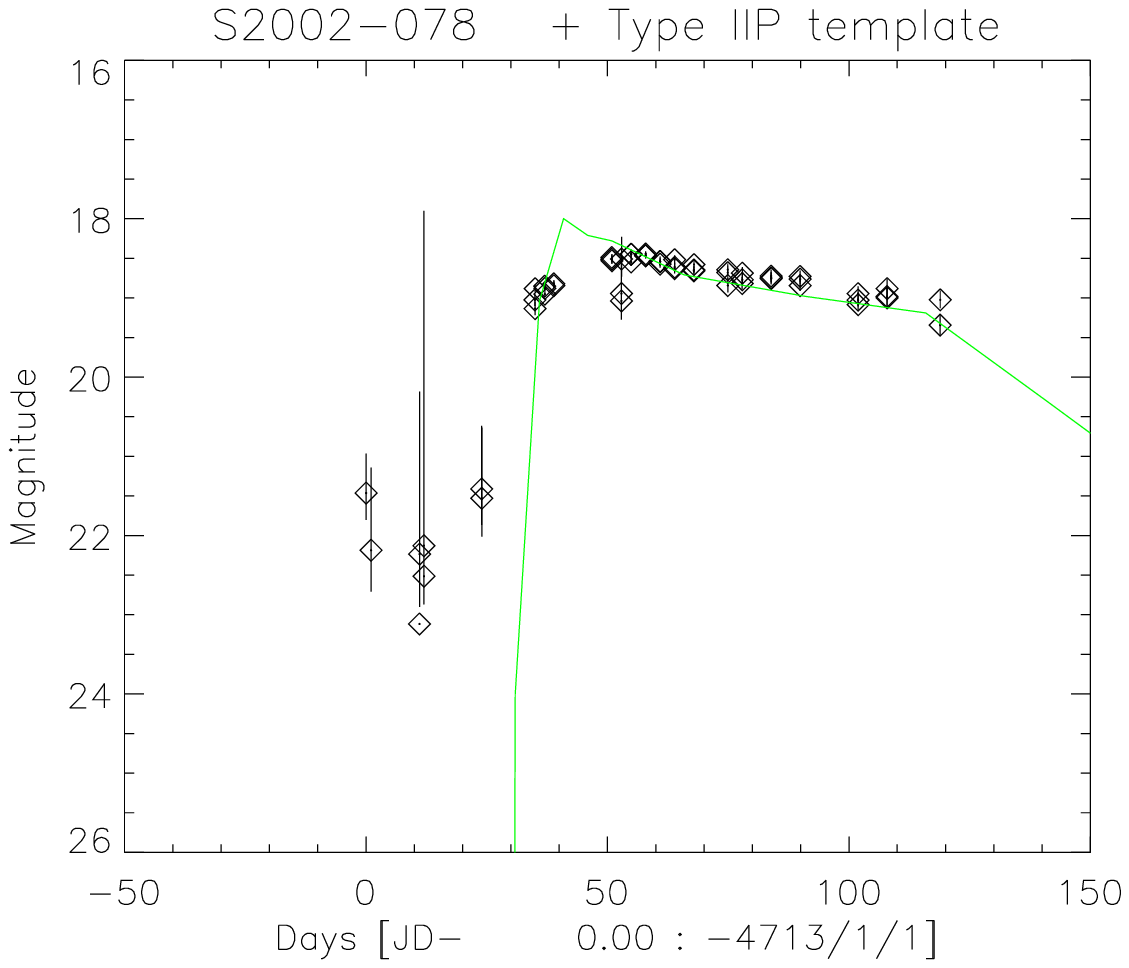}\label{fig:2002da_lightcurve}}
\end{figure}

\clearpage\pagebreak
\begin{figure}
\subfigure[2002dg]{\includegraphics[angle=90,height=2in,width=3in]{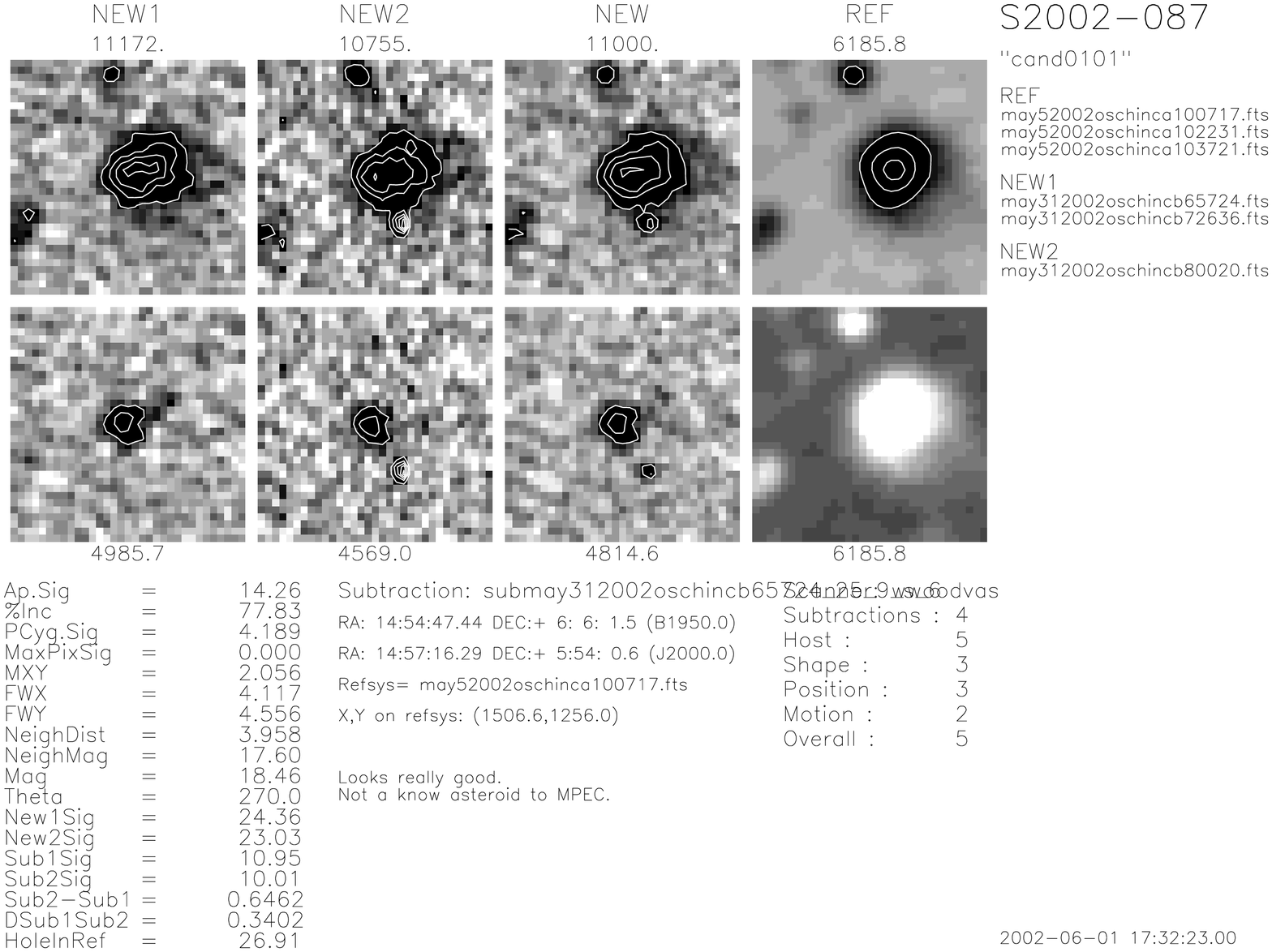}\label{fig:2002dg_discovery}}
\hspace{0.3in}
\subfigure[2002dg]{\includegraphics[height=2in]{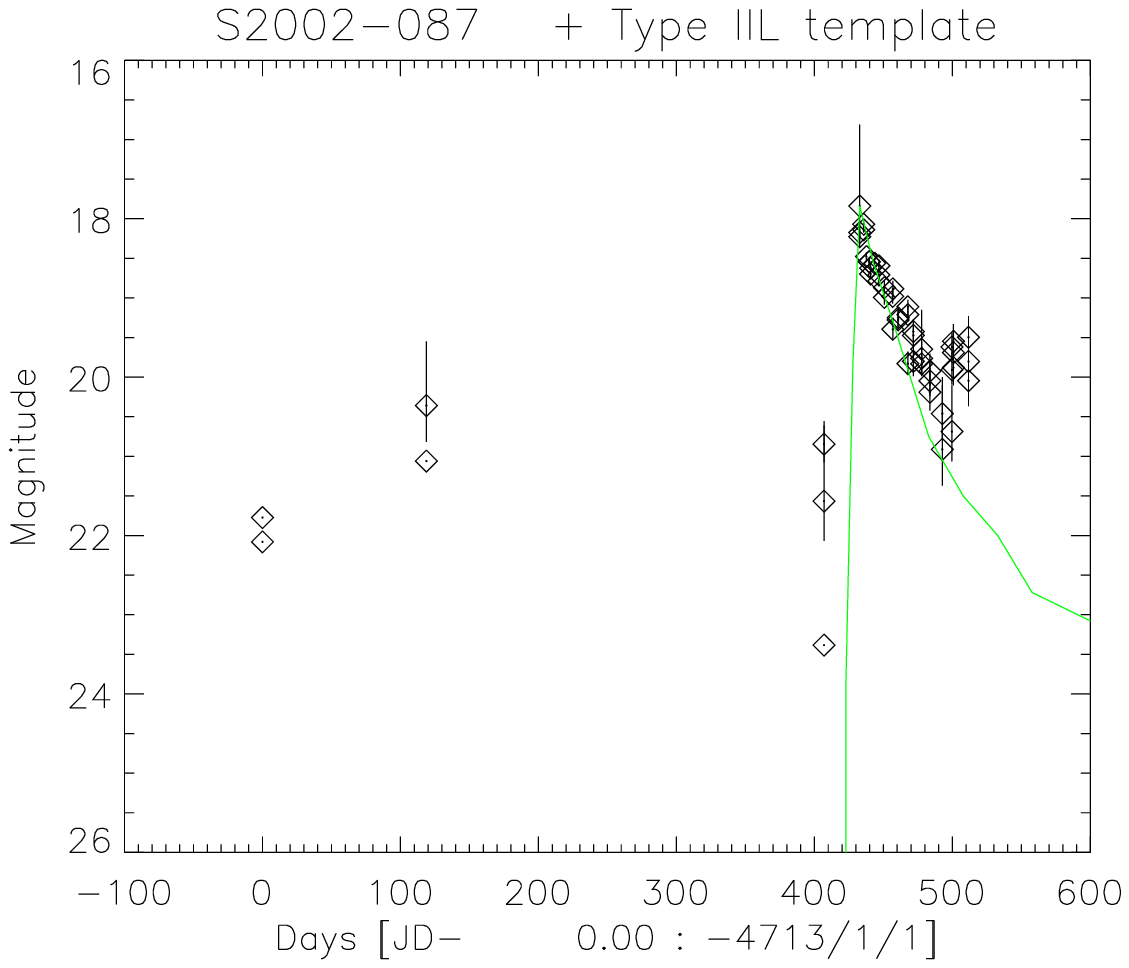}\label{fig:2002dg_lightcurve}}
\vspace{0.3in}
\subfigure[2002dh]{\includegraphics[angle=90,height=2in,width=3in]{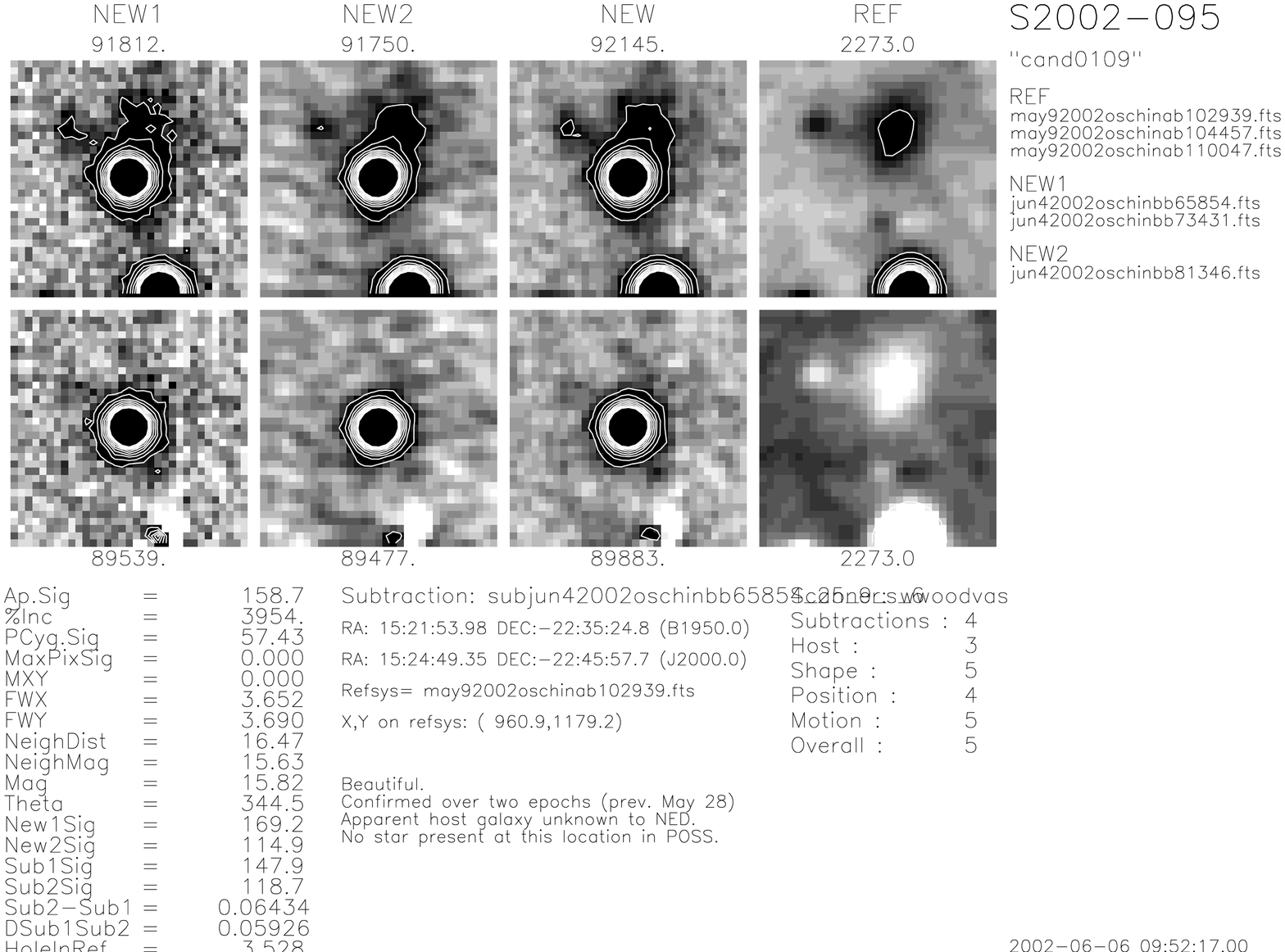}\label{fig:2002dh_discovery}}
\hspace{0.3in}
\subfigure[2002dh]{\includegraphics[height=2in]{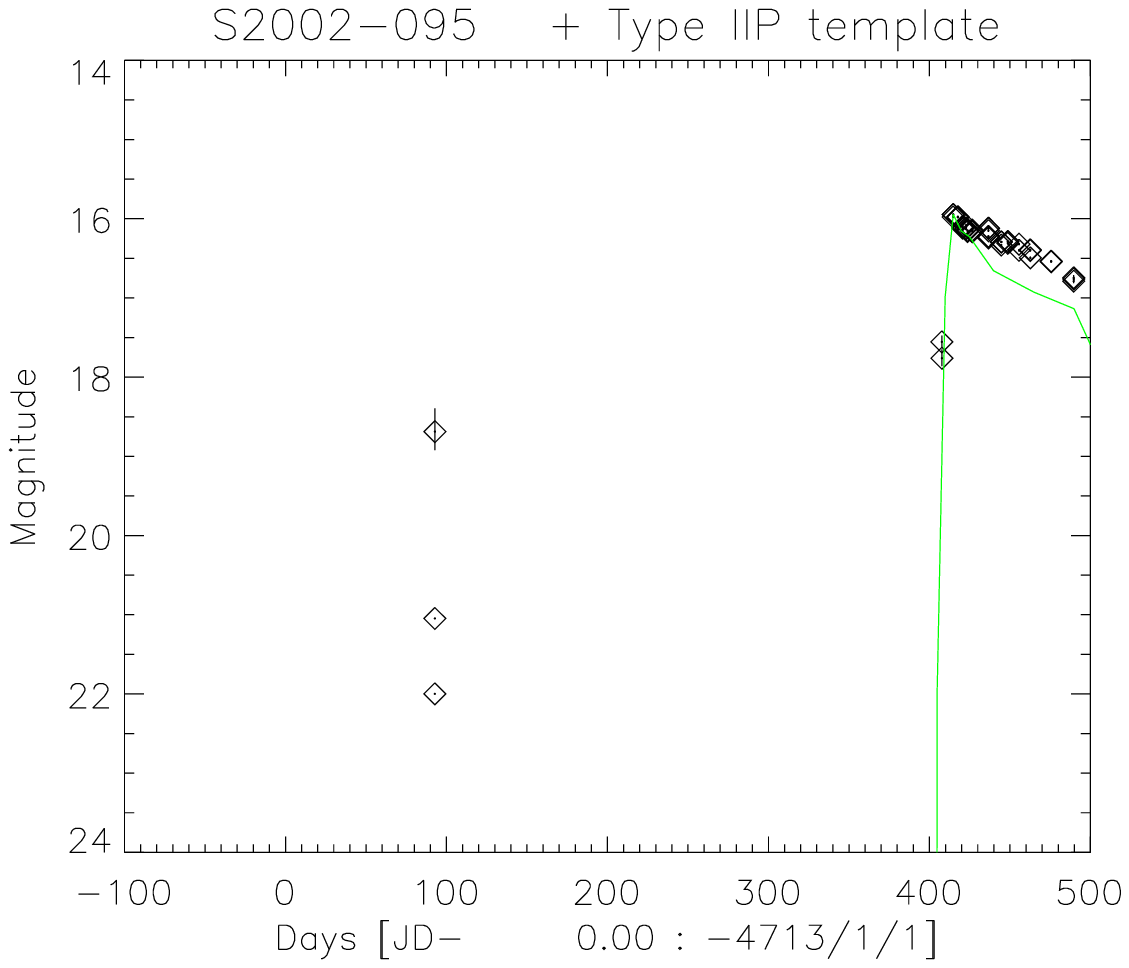}\label{fig:2002dh_lightcurve}}
\vspace{0.3in}
\subfigure[2002dy]{\includegraphics[angle=90,height=2in,width=3in]{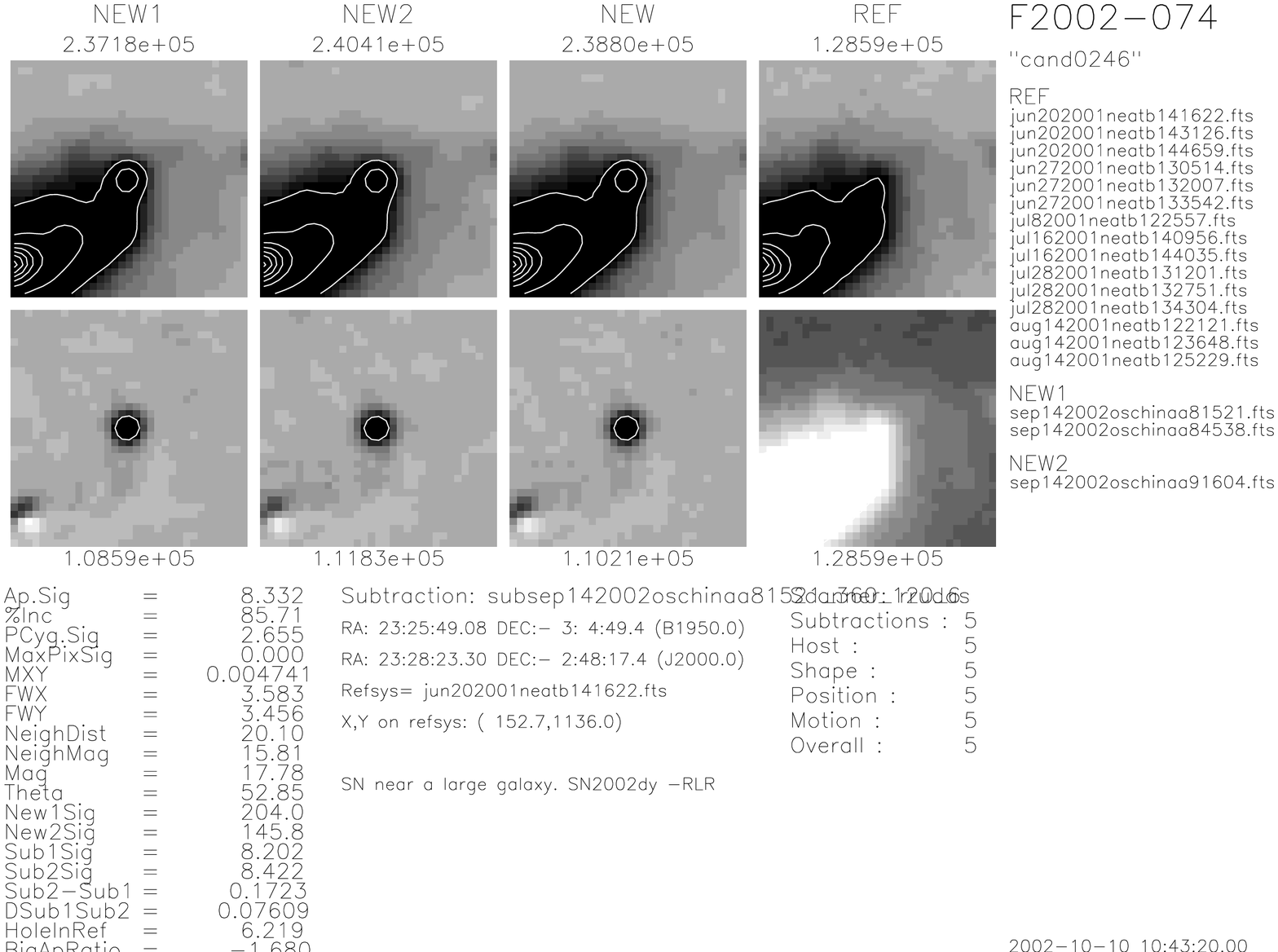}\label{fig:2002dy_discovery}}
\hspace{0.3in}
\subfigure[2002dy]{\includegraphics[height=2in]{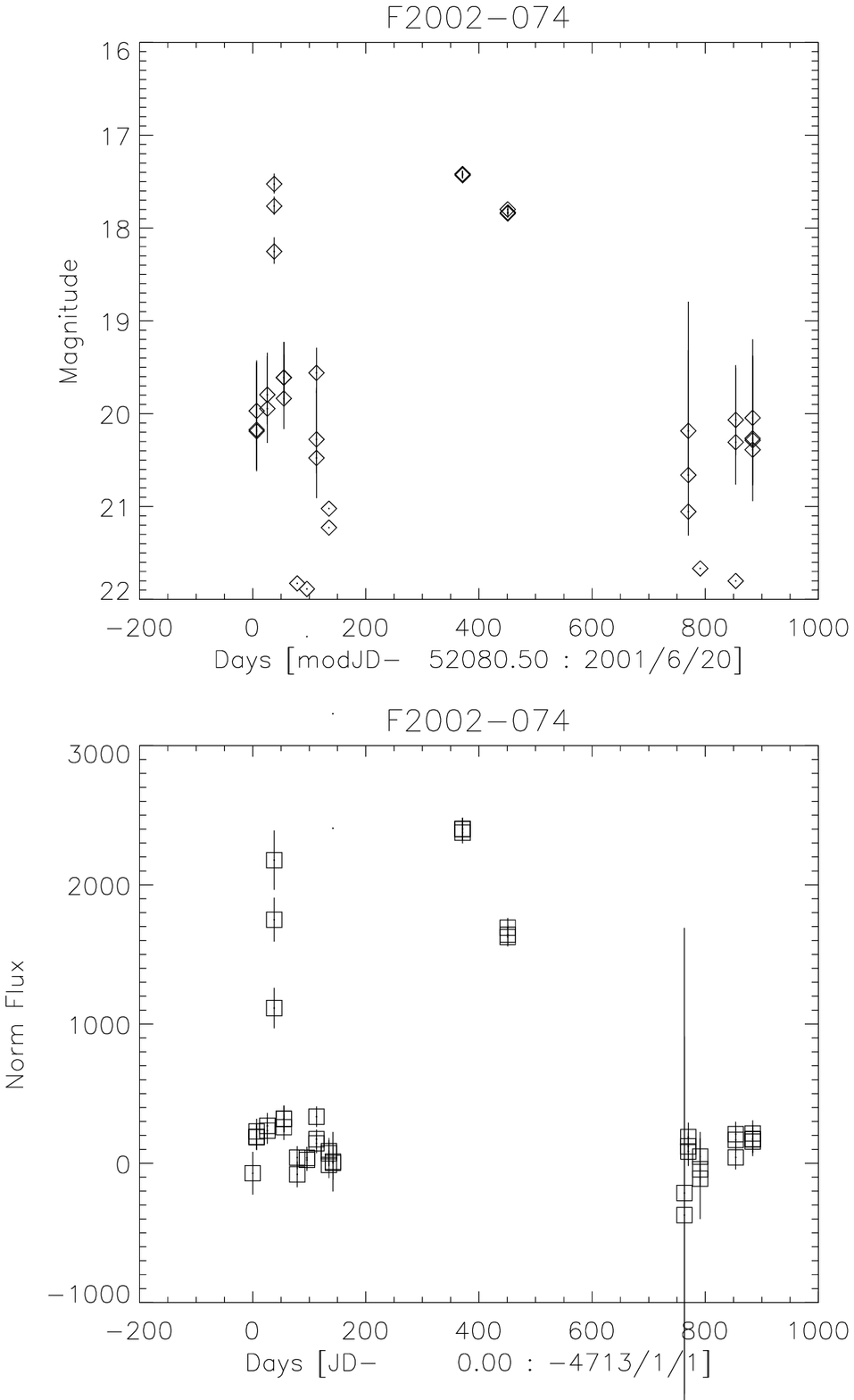}\label{fig:2002dy_lightcurve}}
\end{figure}

\clearpage\pagebreak
\begin{figure}
\subfigure[2002ek]{\includegraphics[angle=90,height=2in,width=3in]{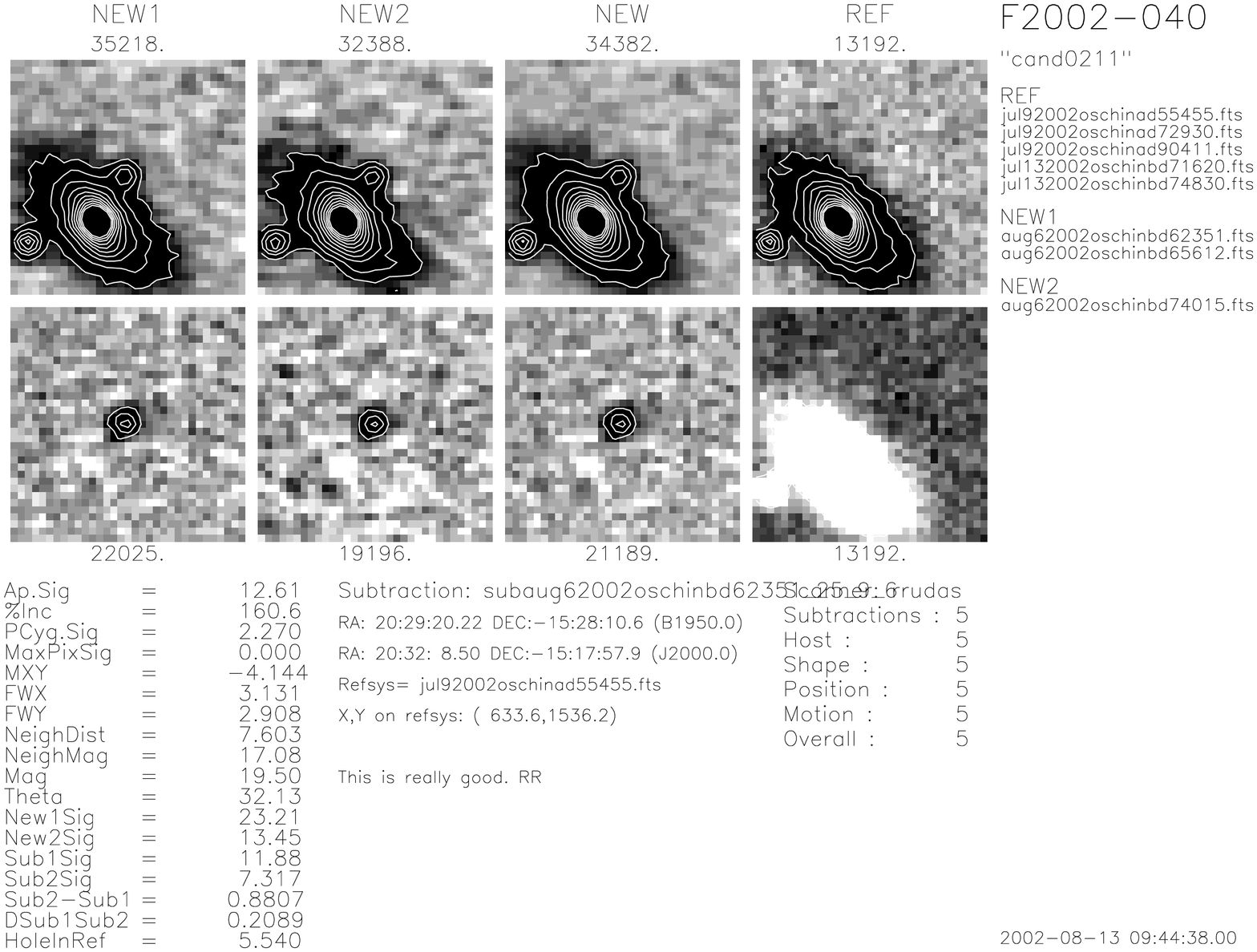}\label{fig:2002ek_discovery}} \quad
\subfigure[2002ek]{\includegraphics[height=2in]{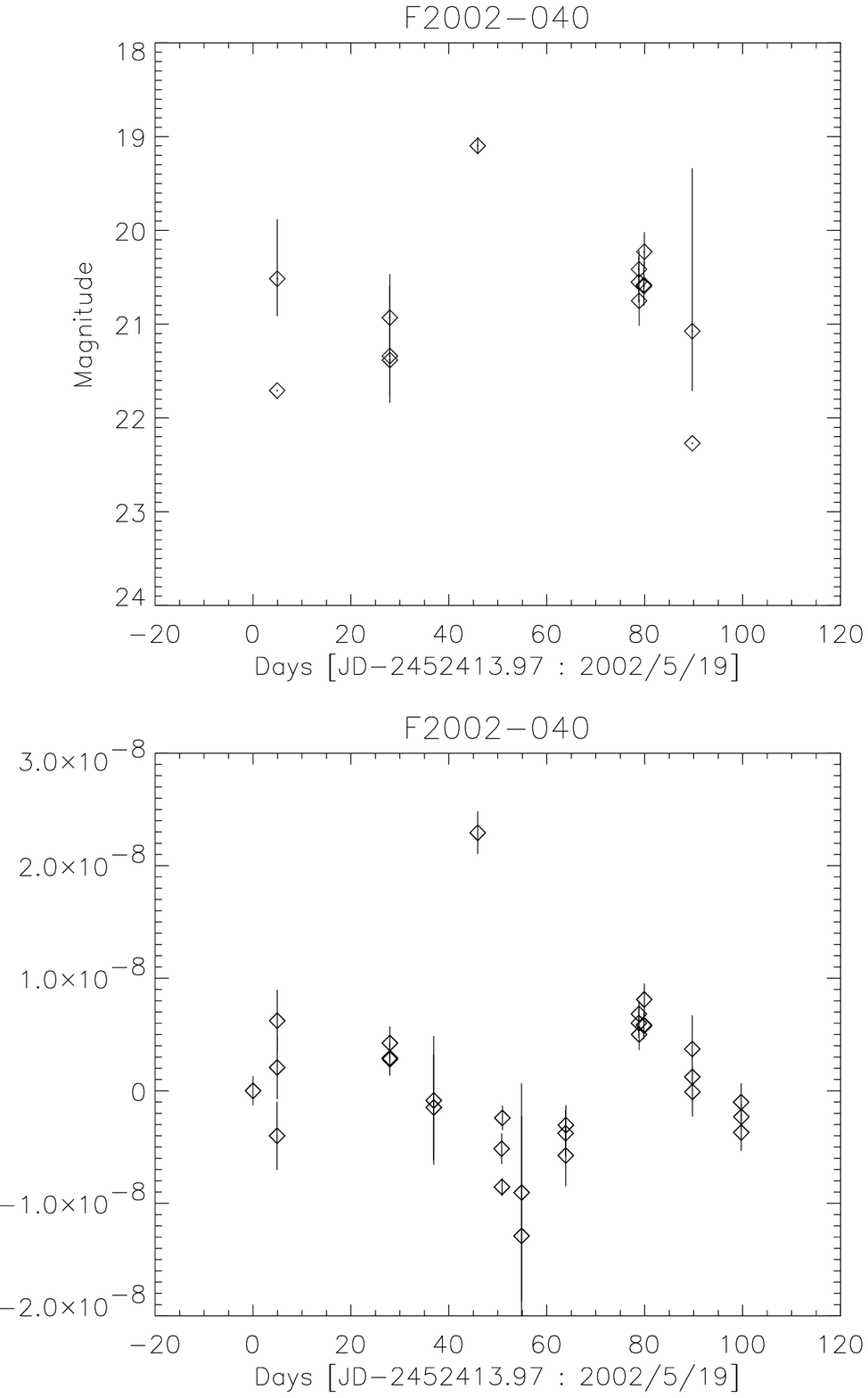}\label{fig:2002ek_lightcurve}}
\vspace{0.3in}
\subfigure[2002el]{\includegraphics[angle=90,height=2in,width=3in]{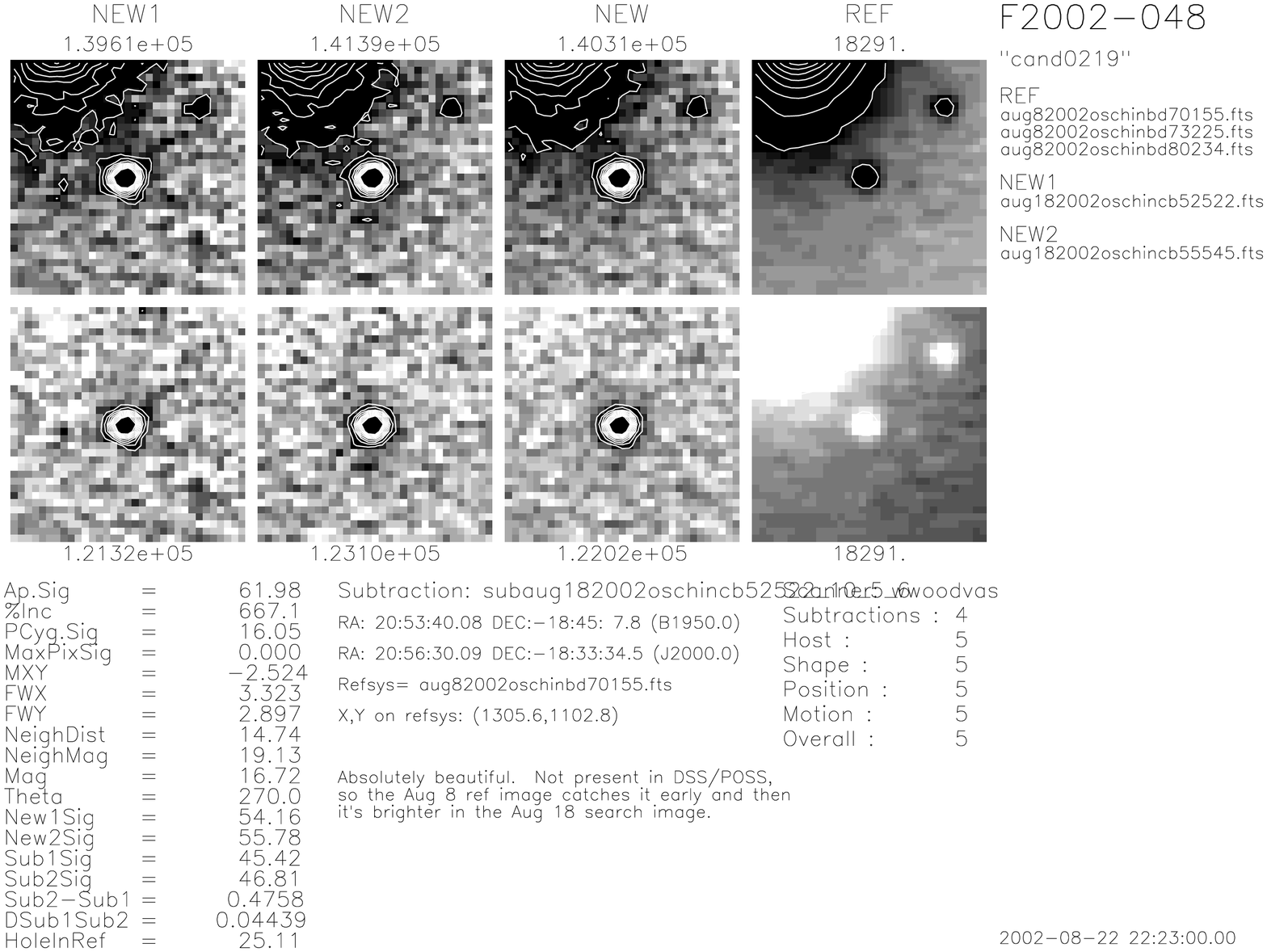}\label{fig:2002el_discovery}}
\hspace{0.3in}
\subfigure[2002el]{\includegraphics[height=2in]{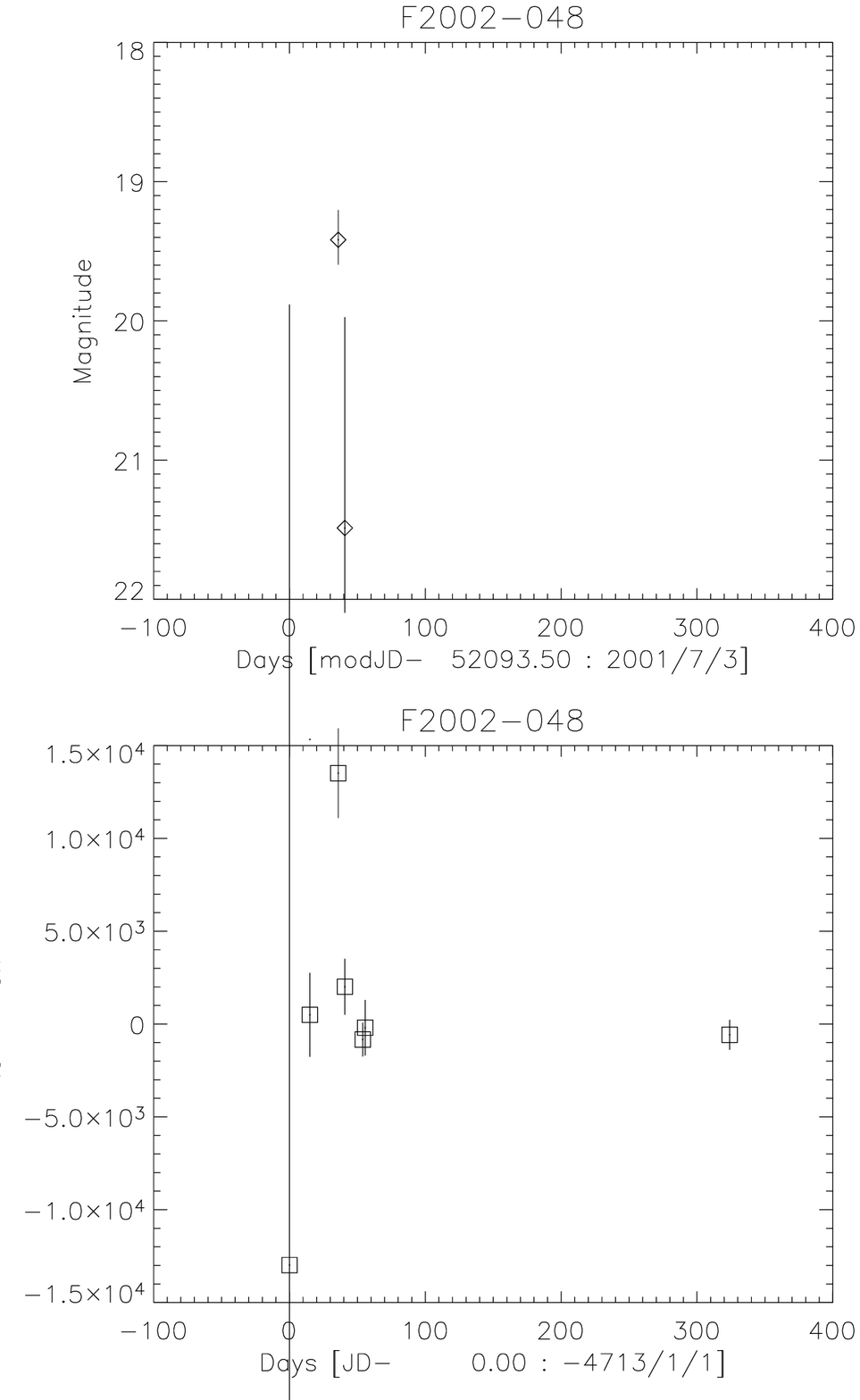}\label{fig:2002el_lightcurve}}
\vspace{0.3in}
\subfigure[2002ep]{\includegraphics[angle=90,height=2in,width=3in]{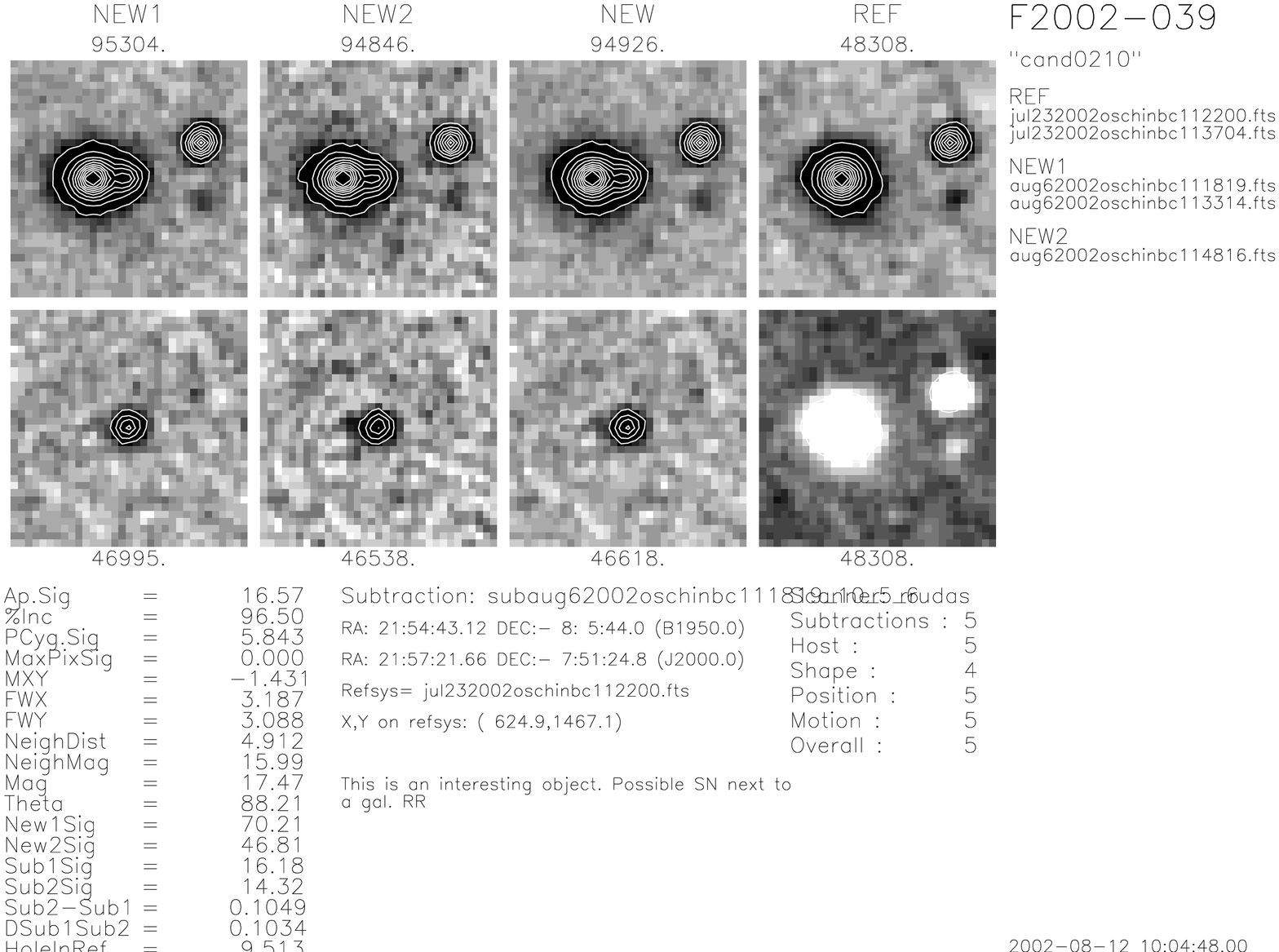}\label{fig:2002ep_discovery}}
\hspace{0.3in}
\subfigure[2002ep]{\includegraphics[height=2in]{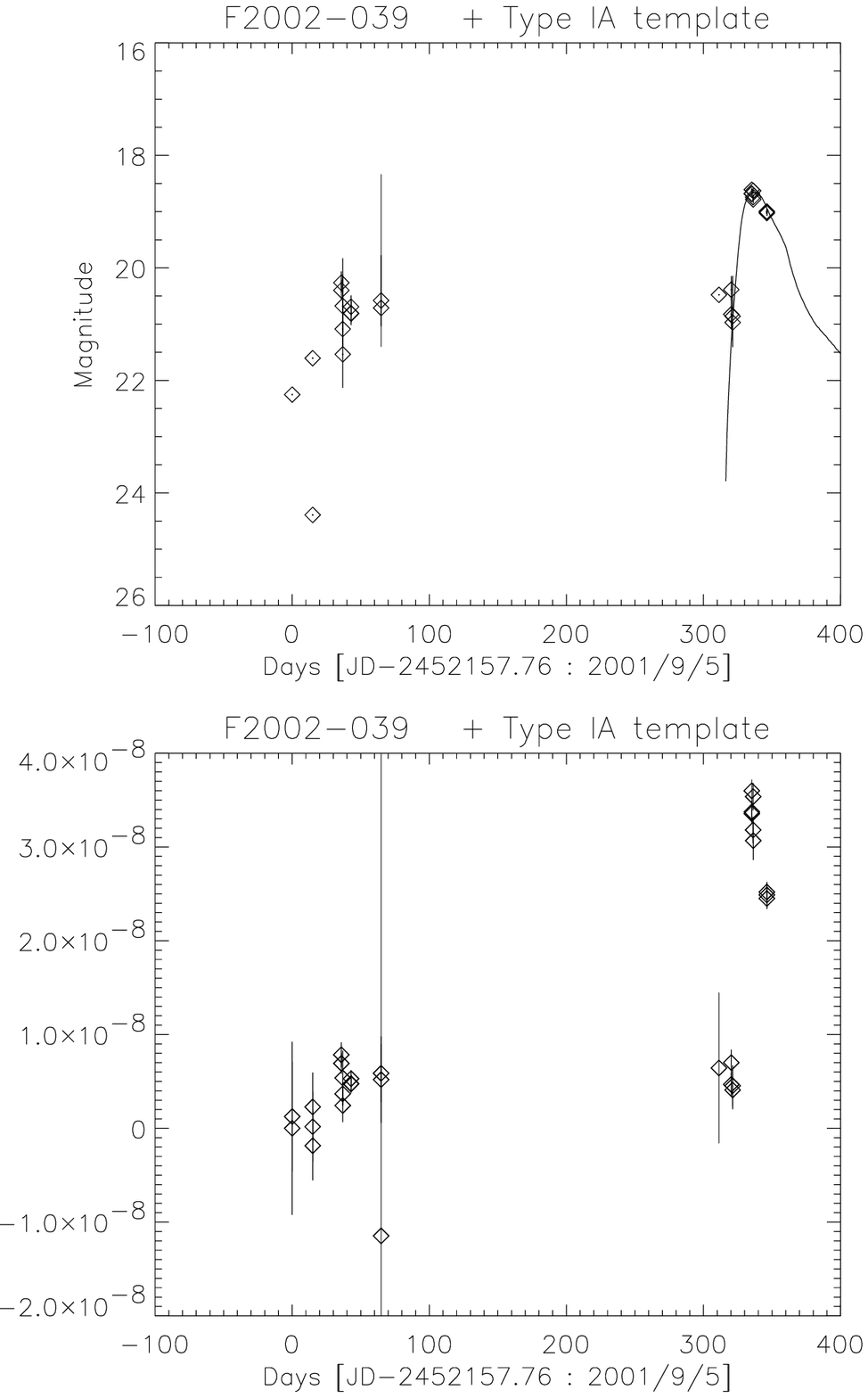}\label{fig:2002ep_lightcurve}}
\end{figure}

\clearpage\pagebreak
\begin{figure}
\subfigure[2002eq]{\includegraphics[angle=90,height=2in,width=3in]{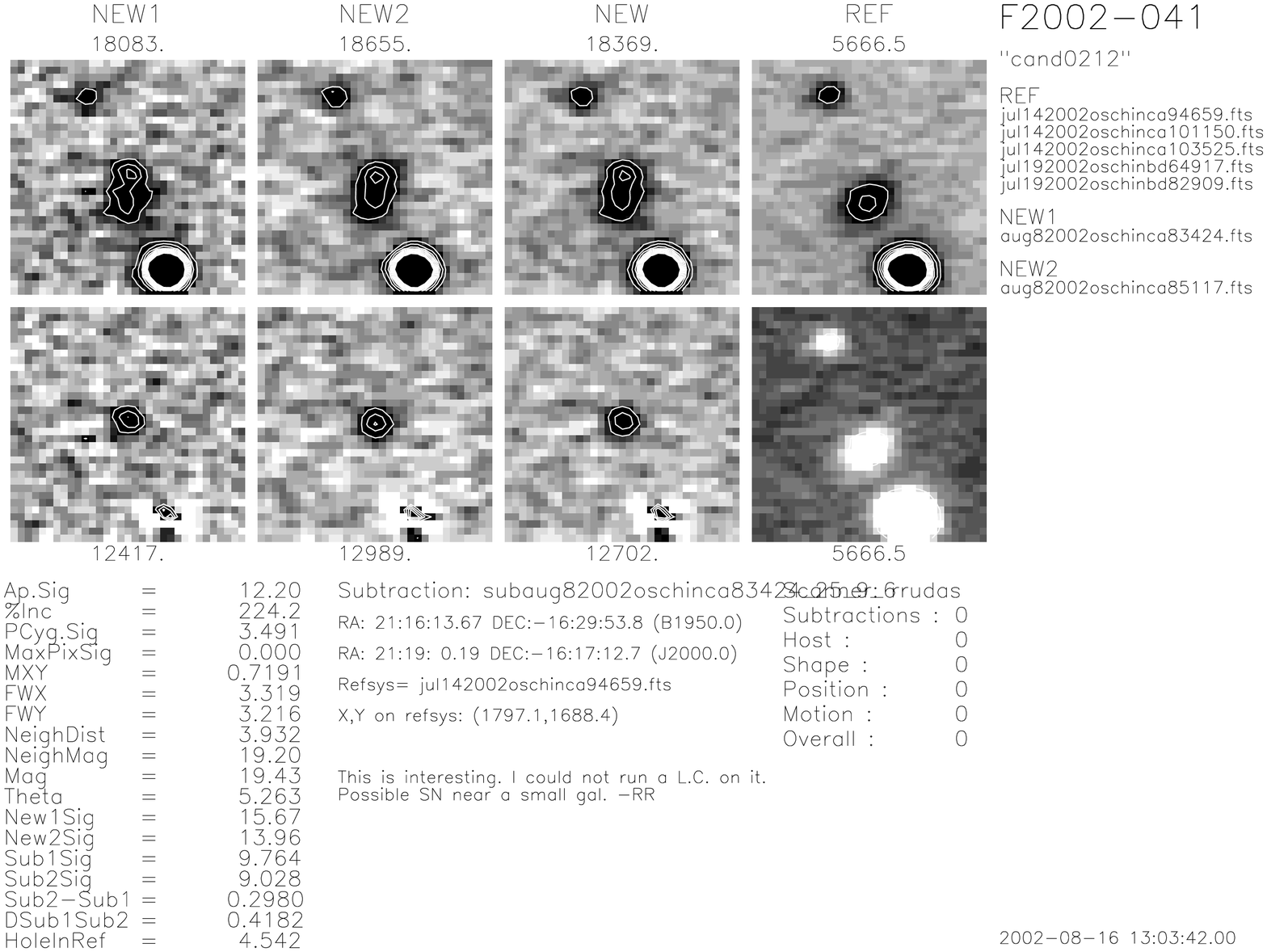}\label{fig:2002eq_discovery}}
\hspace{0.3in}
\subfigure[2002eq]{\includegraphics[height=2in]{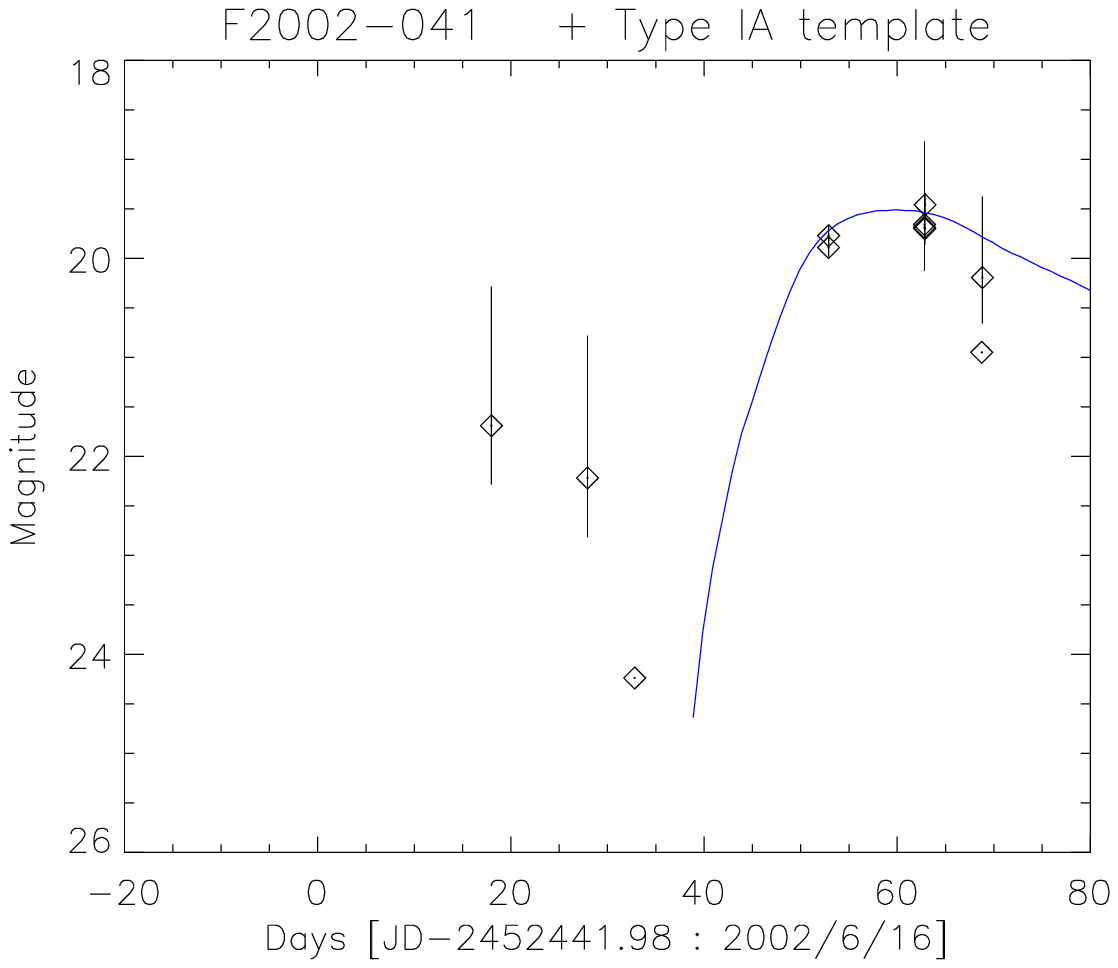}\label{fig:2002eq_lightcurve}}
\vspace{0.3in}
\subfigure[2002ev]{\includegraphics[angle=90,height=2in,width=3in]{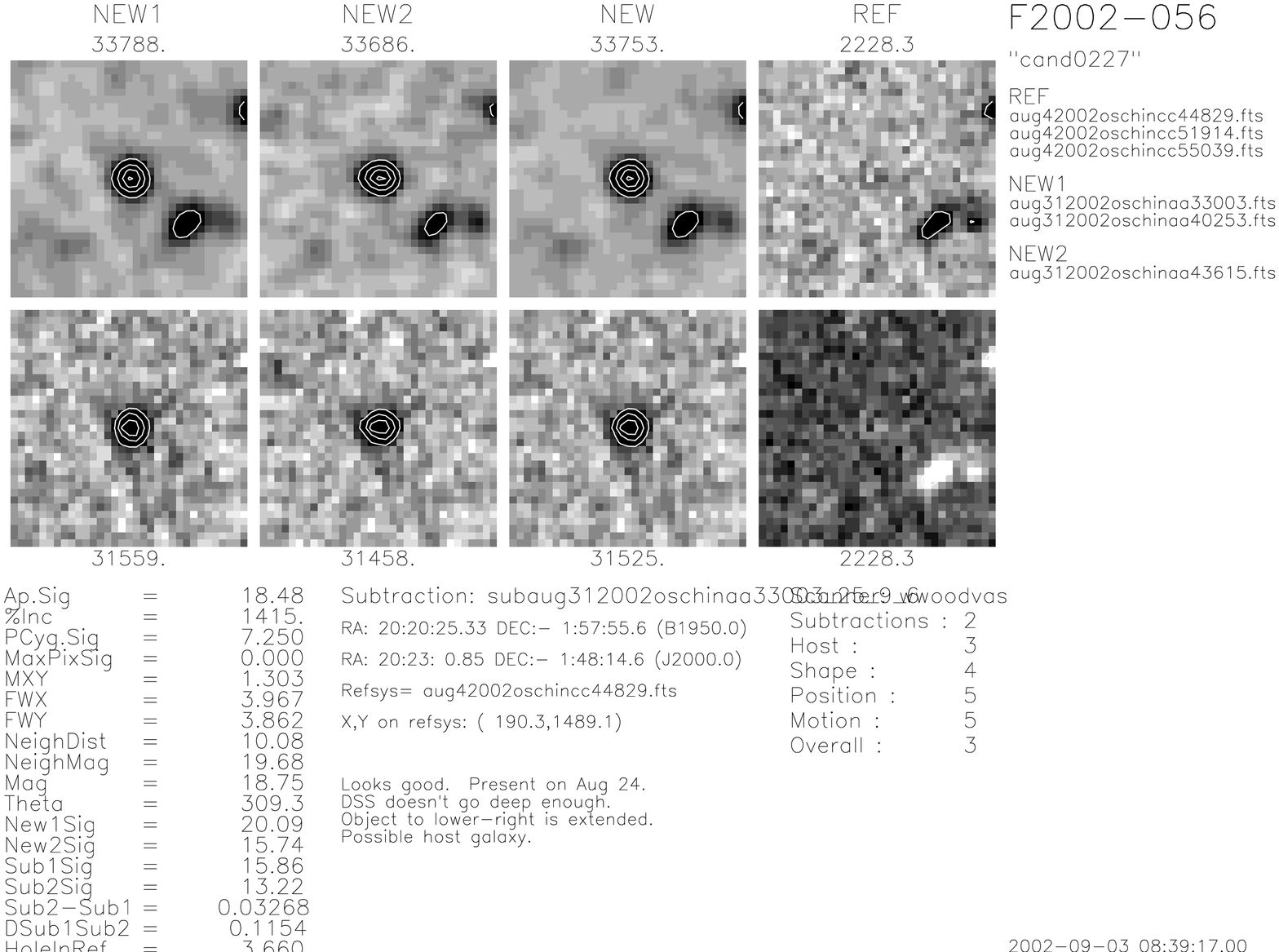}\label{fig:2002ev_discovery}}
\hspace{0.3in}
\subfigure[2002ev]{\includegraphics[height=2in]{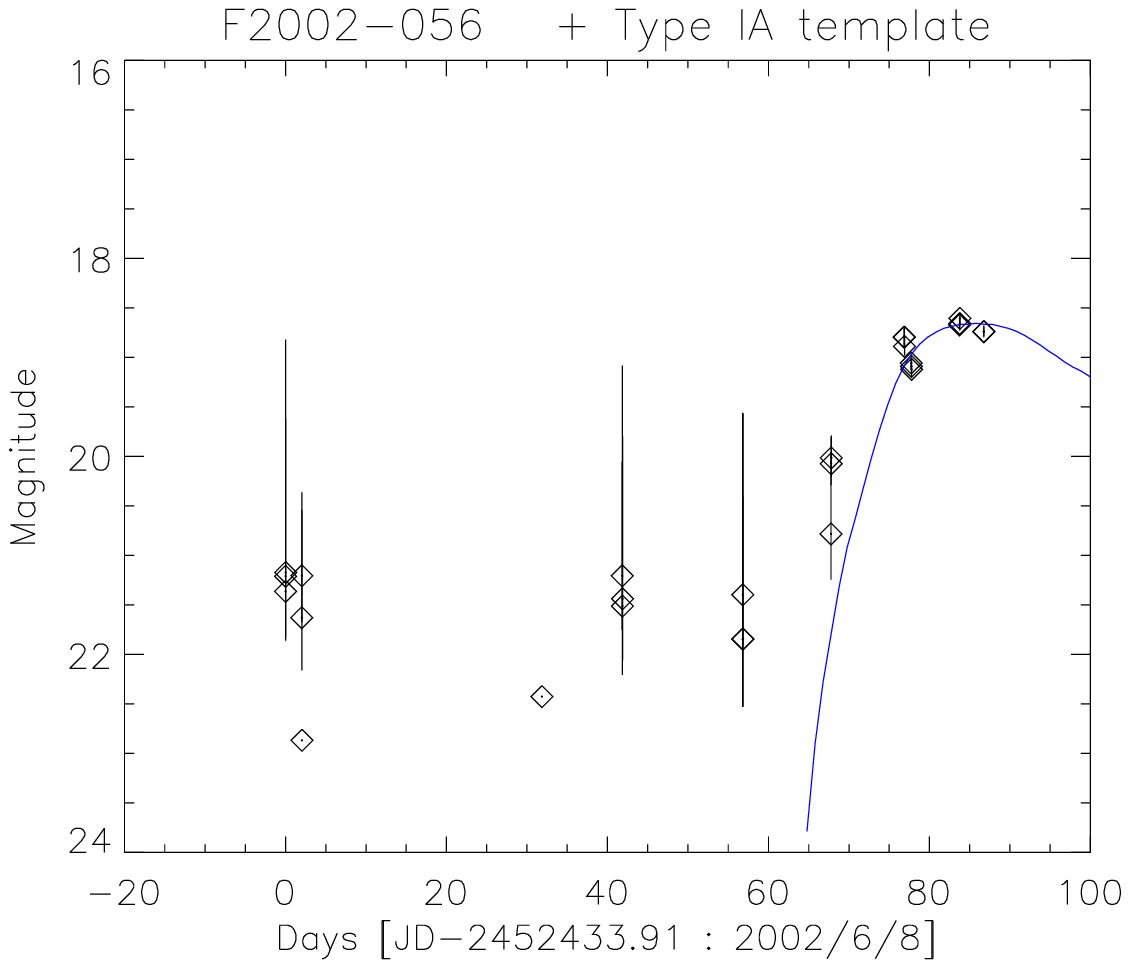}\label{fig:2002ev_lightcurve}}
\vspace{0.3in}
\subfigure[2002ew]{\includegraphics[angle=90,height=2in,width=3in]{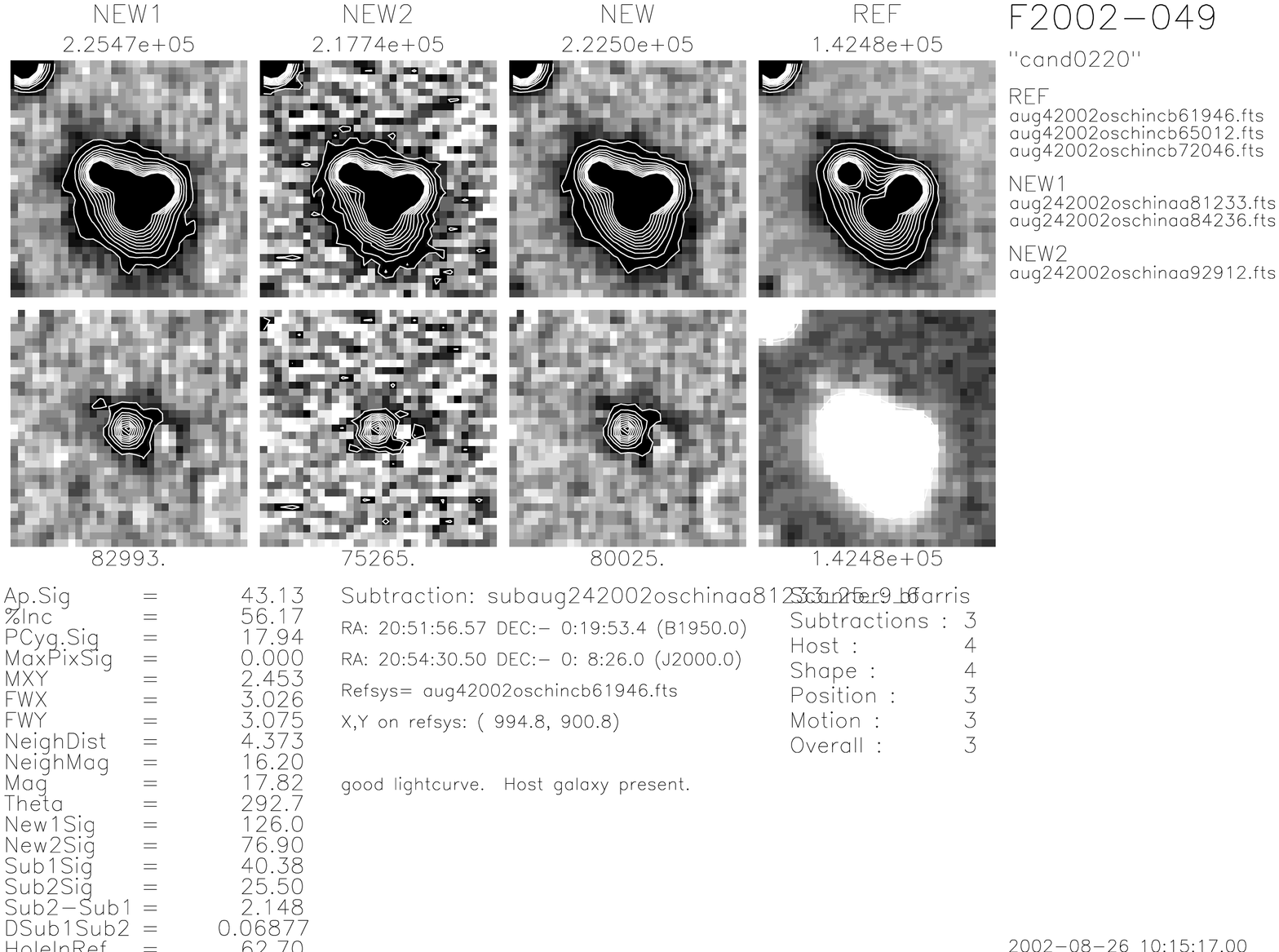}\label{fig:2002ew_discovery}}
\hspace{0.3in}
\subfigure[2002ew]{\includegraphics[height=2in]{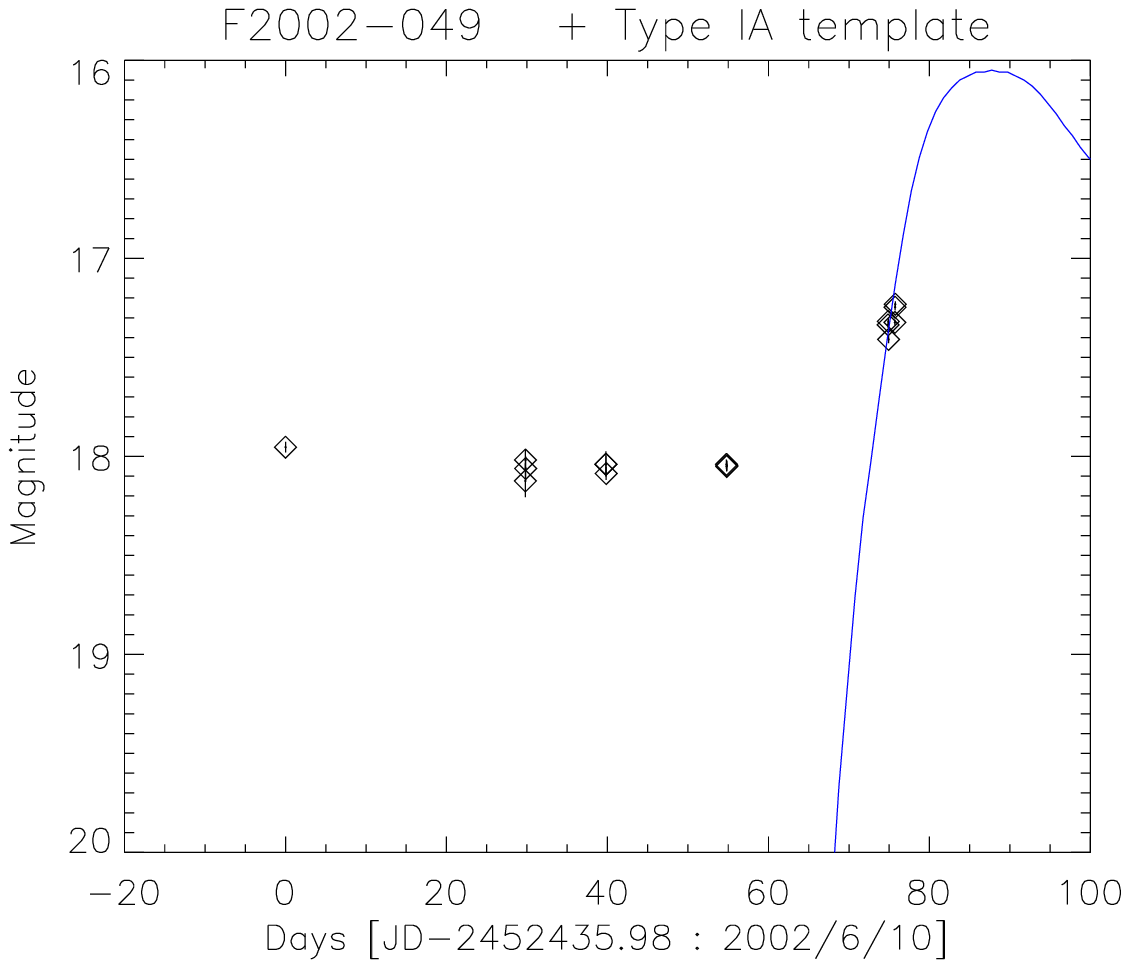}\label{fig:2002ew_lightcurve}}
\vspace{0.3in}
\end{figure}

\clearpage\pagebreak
\begin{figure}
\subfigure[2002ex]{\includegraphics[angle=90,height=2in,width=3in]{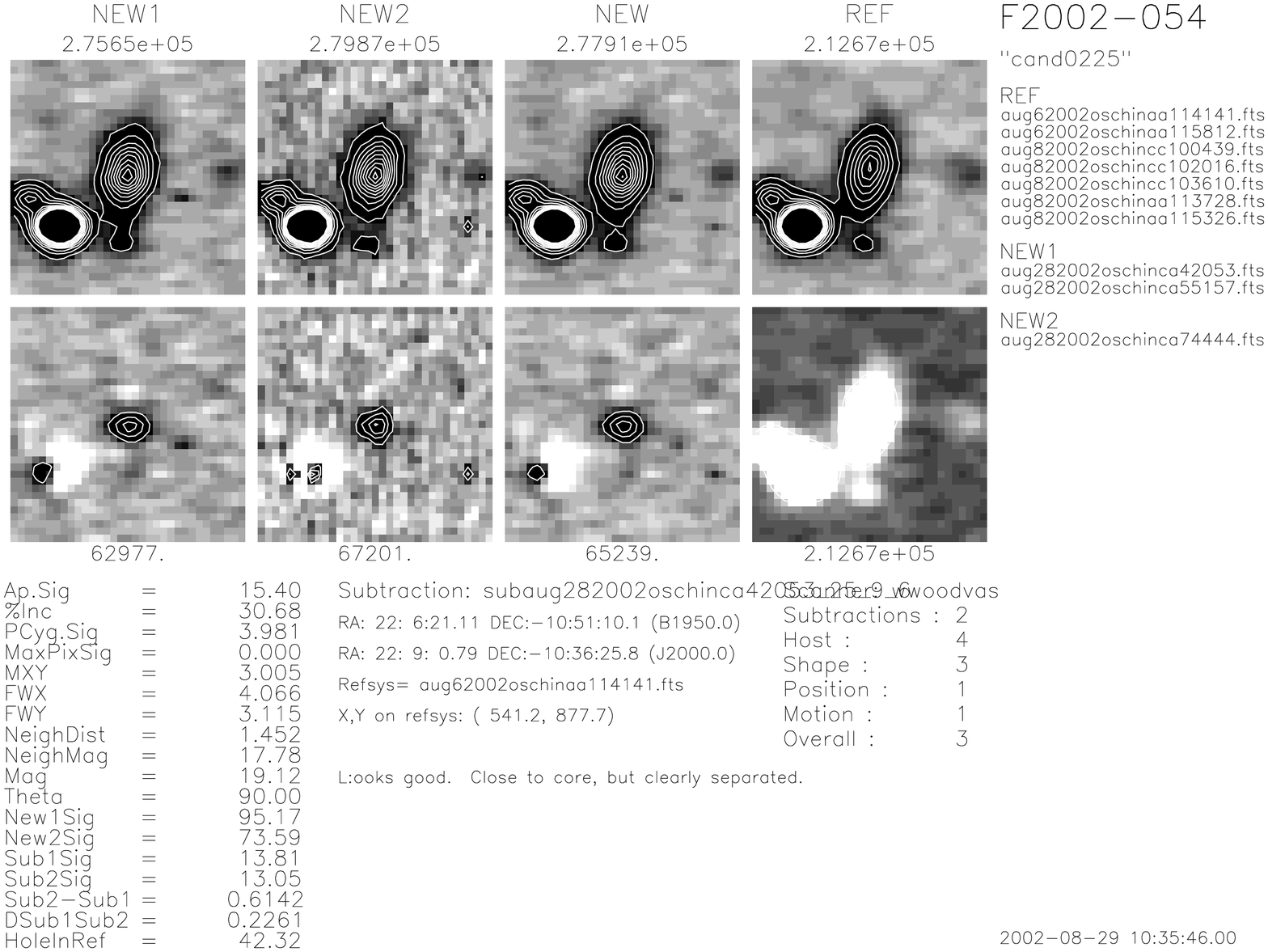}\label{fig:2002ex_discovery}}
\hspace{0.3in}
\subfigure[2002ex]{\includegraphics[height=2in]{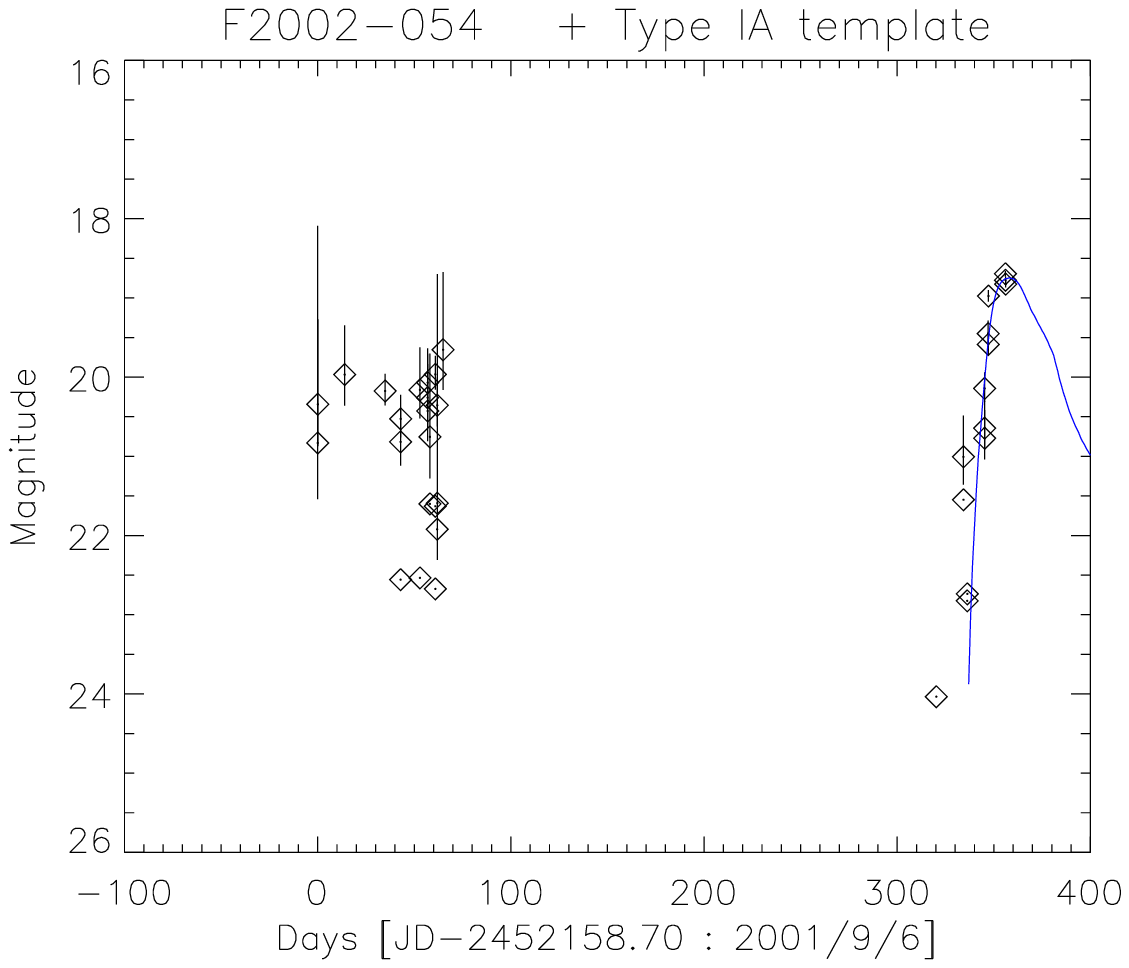}\label{fig:2002ex_lightcurve}}
\vspace{0.3in}
\subfigure[2002ez]{\includegraphics[angle=90,height=2in,width=3in]{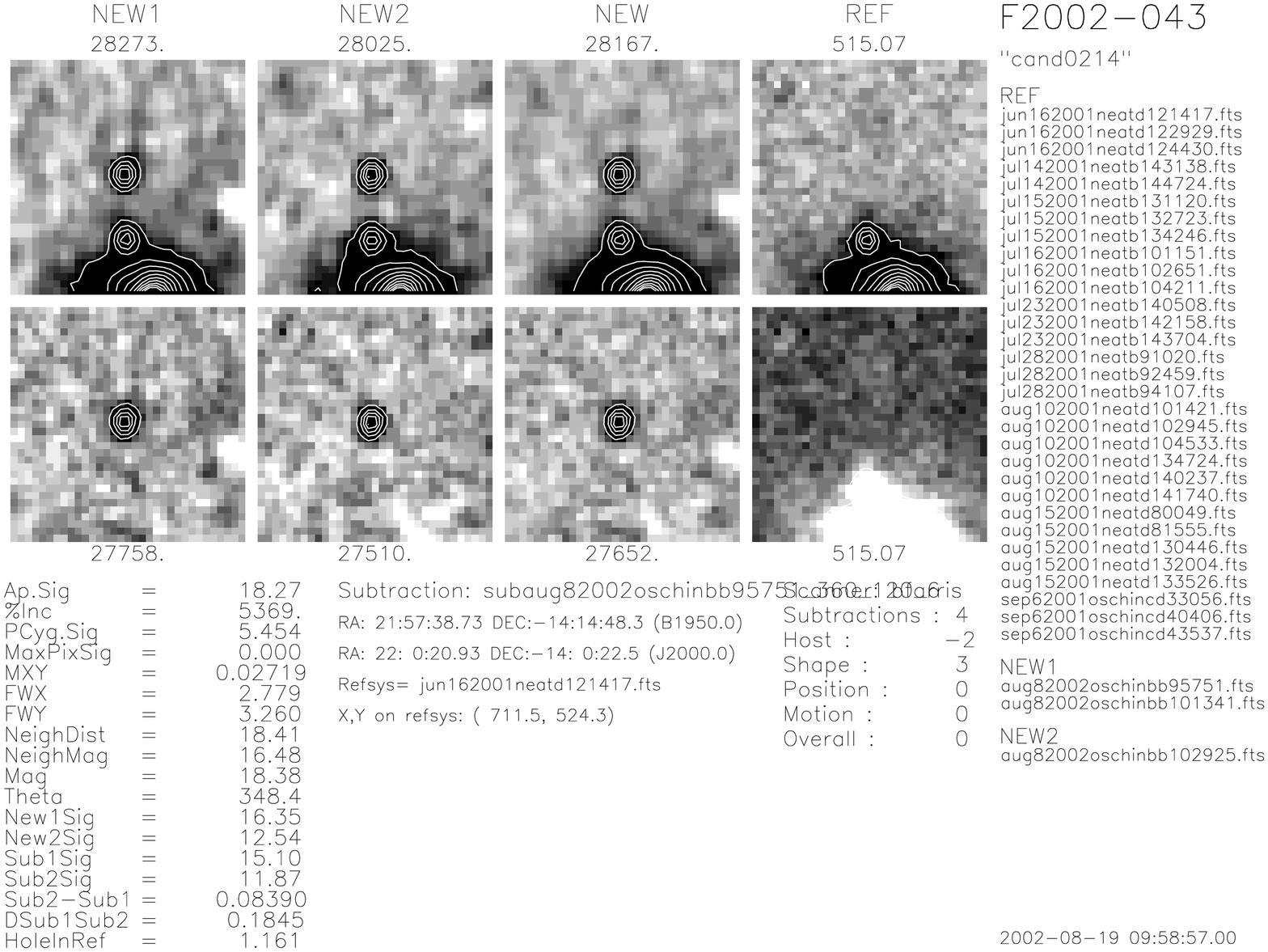}\label{fig:2002ez_discovery}}
\hspace{0.3in}
\subfigure[2002ez]{\includegraphics[height=2in]{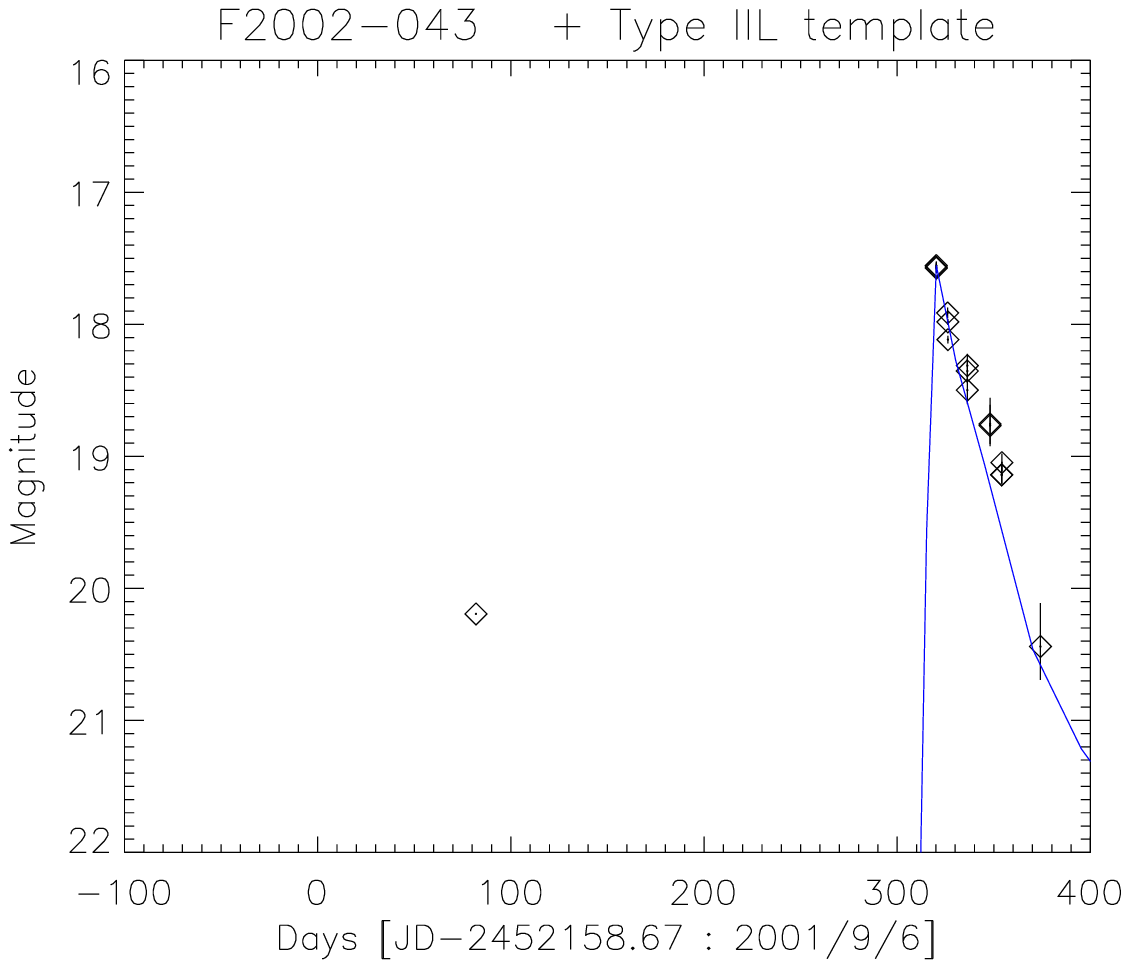}\label{fig:2002ez_lightcurve}}
\vspace{0.3in}
\subfigure[2002fa]{\includegraphics[angle=90,height=2in,width=3in]{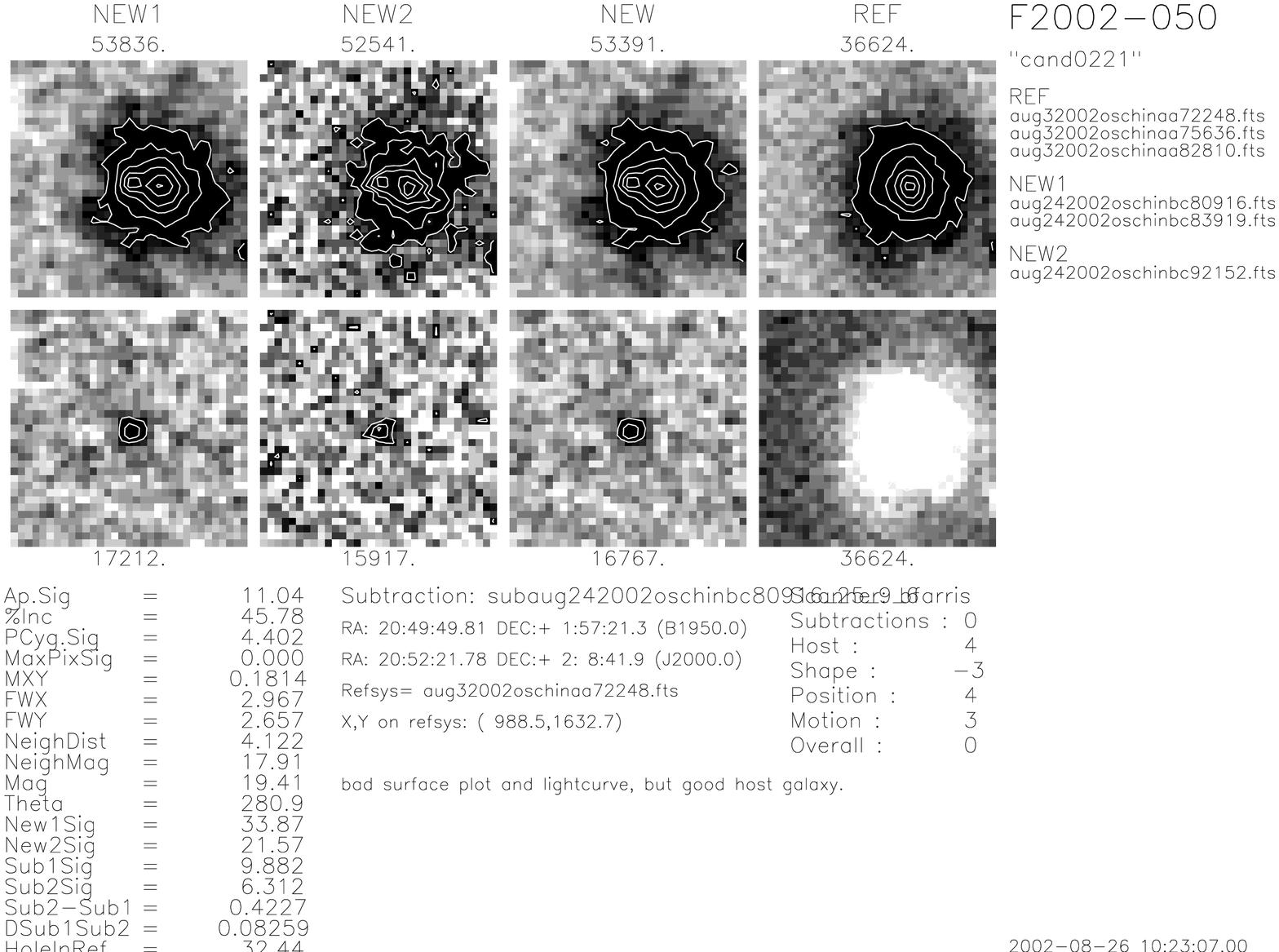}\label{fig:2002fa_discovery}}
\hspace{0.3in}
\subfigure[2002fa]{\includegraphics[height=2in]{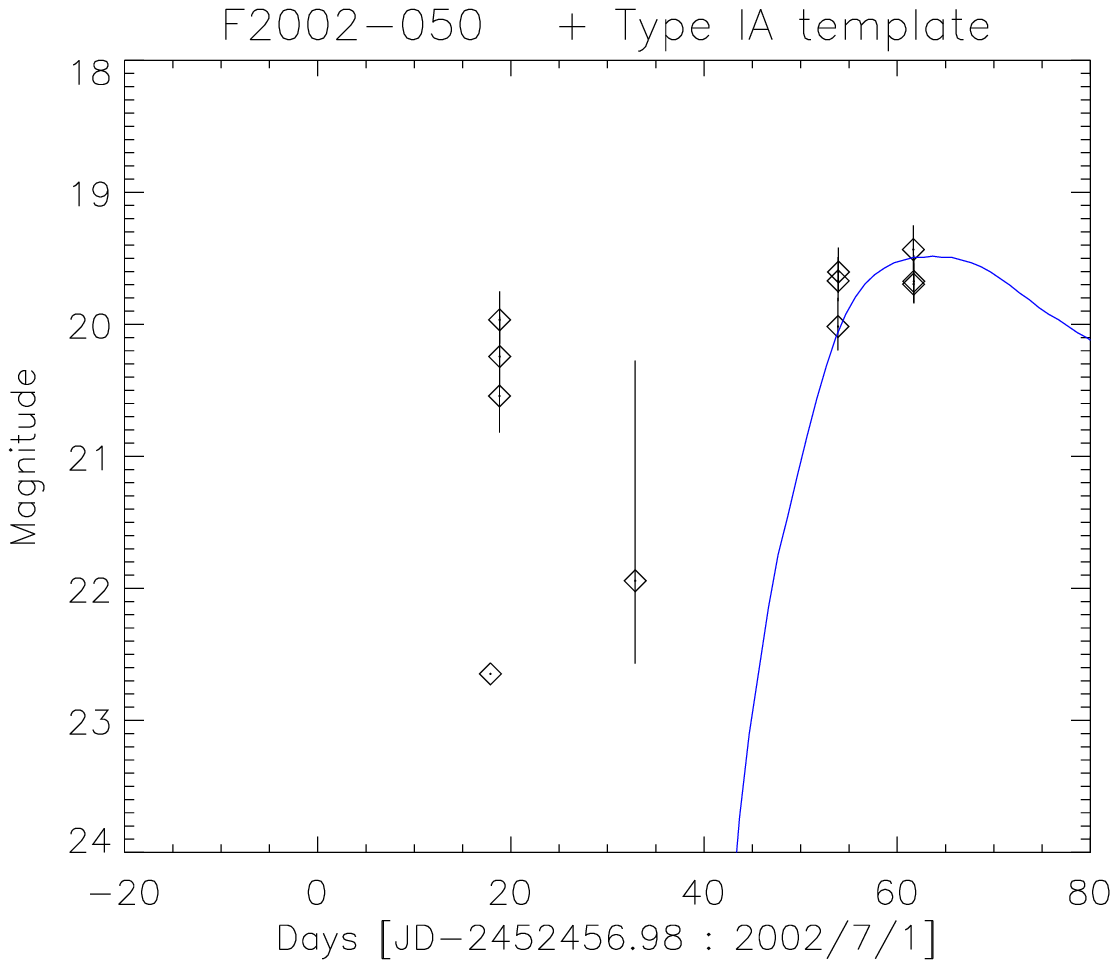}\label{fig:2002fa_lightcurve}}
\vspace{0.3in}
\end{figure}

\clearpage\pagebreak
\begin{figure}
\subfigure[2002fs]{\includegraphics[angle=90,height=2in,width=3in]{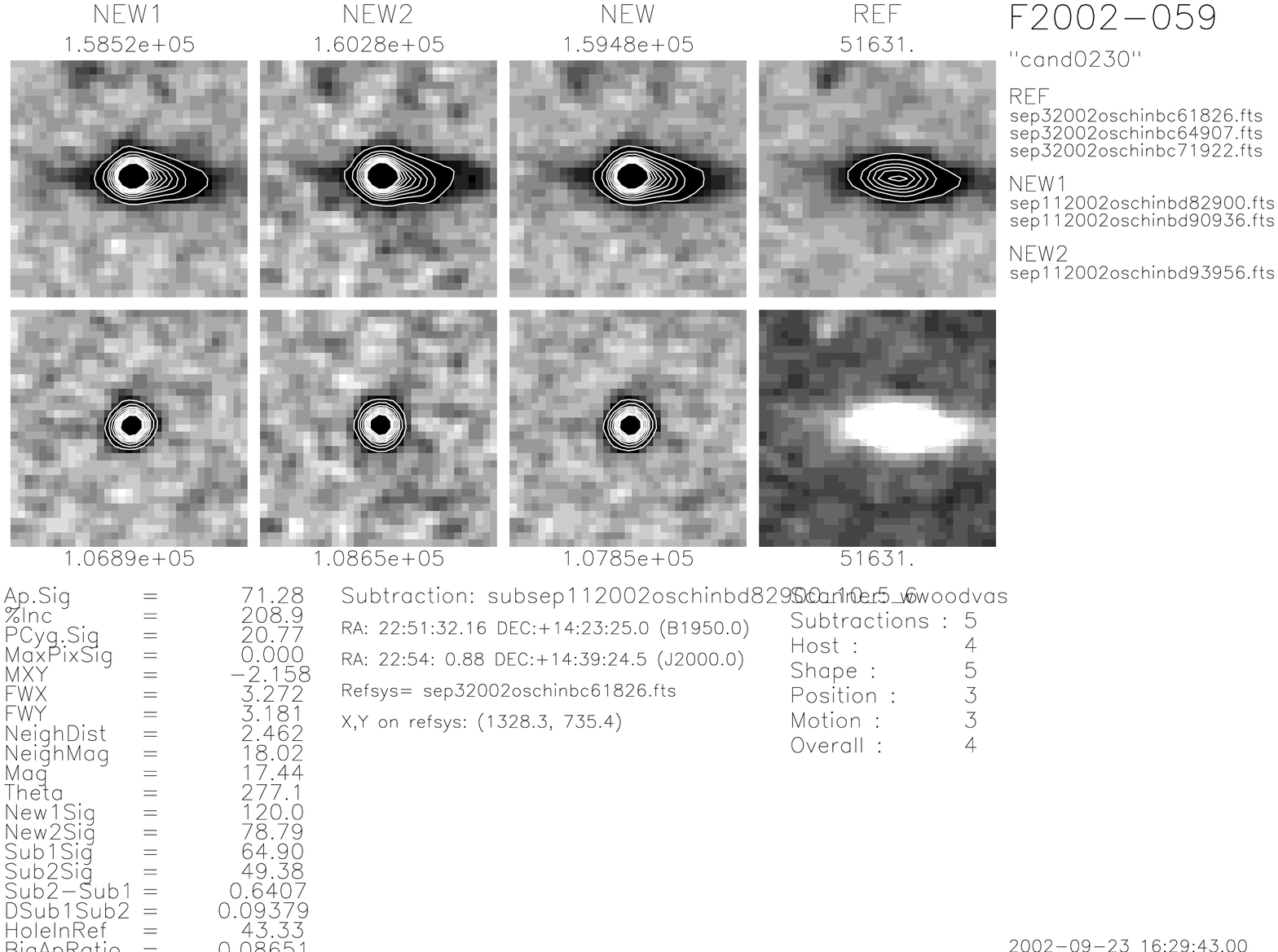}\label{fig:2002fs_discovery}}
\hspace{0.3in}
\subfigure[2002fs]{\includegraphics[height=2in]{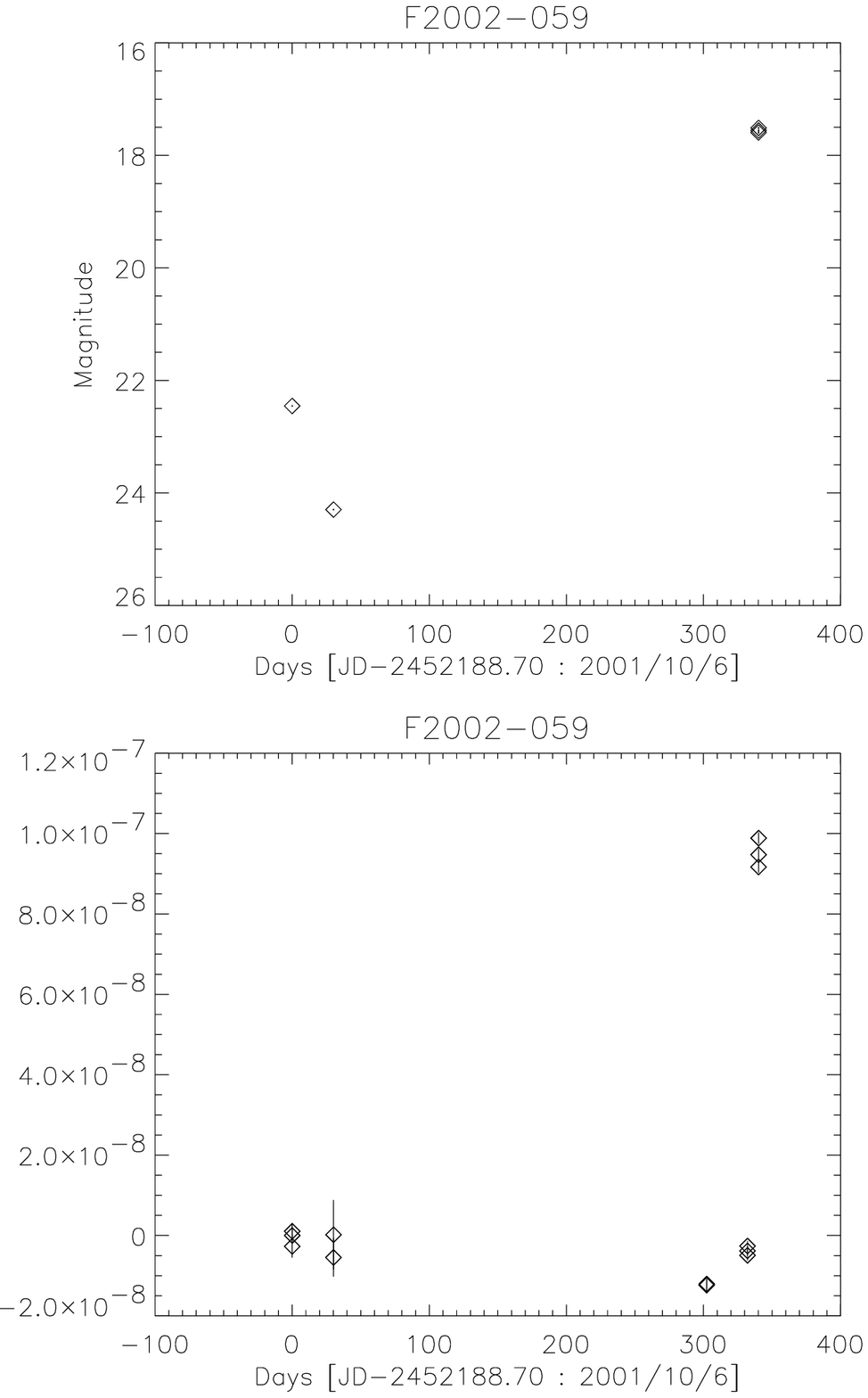}\label{fig:2002fs_lightcurve}}
\vspace{0.3in}
\subfigure[2002ft]{\includegraphics[angle=90,height=2in,width=3in]{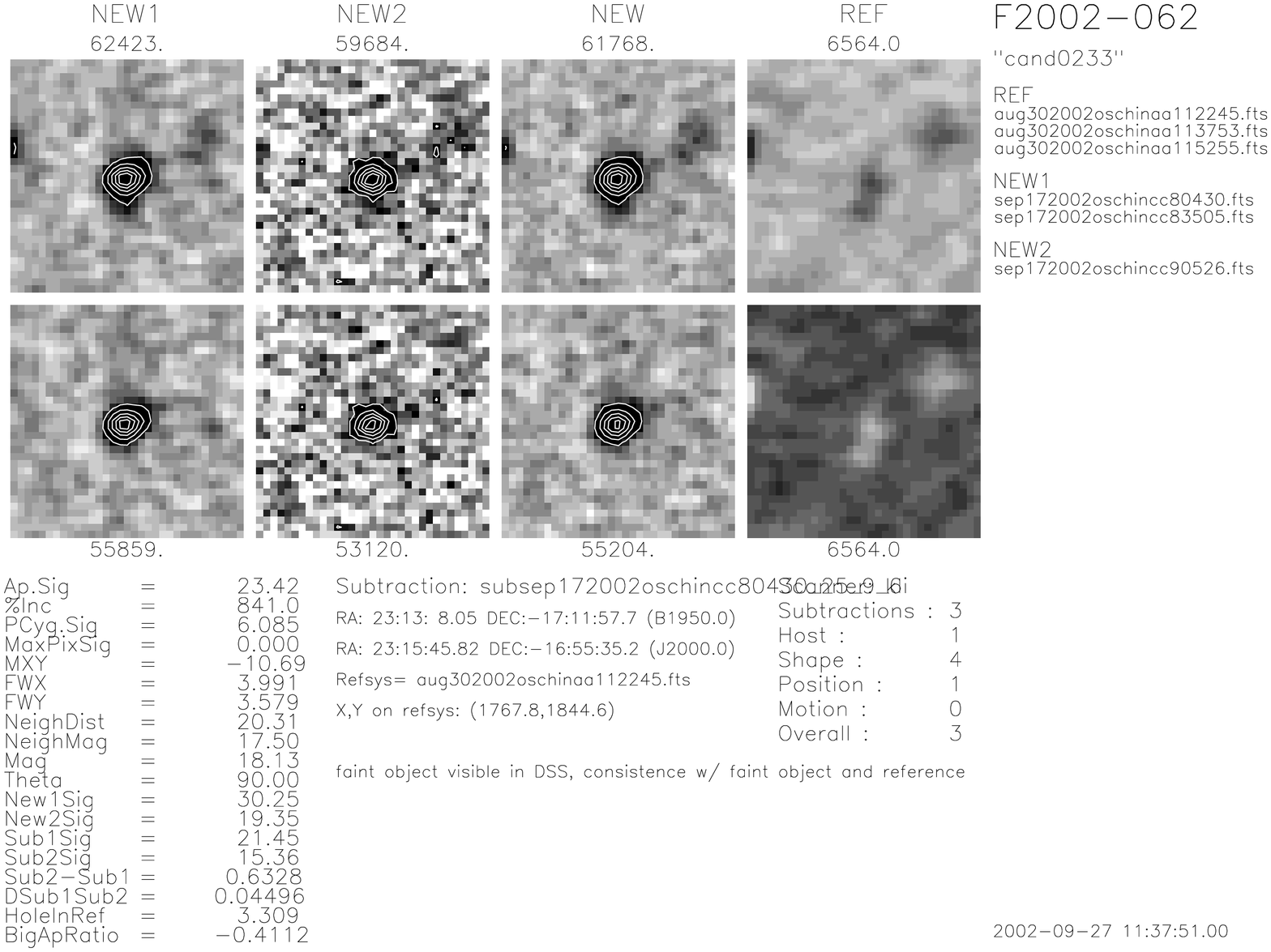}\label{fig:2002ft_discovery}}
\hspace{0.3in}
\subfigure[2002ft]{\includegraphics[height=2in]{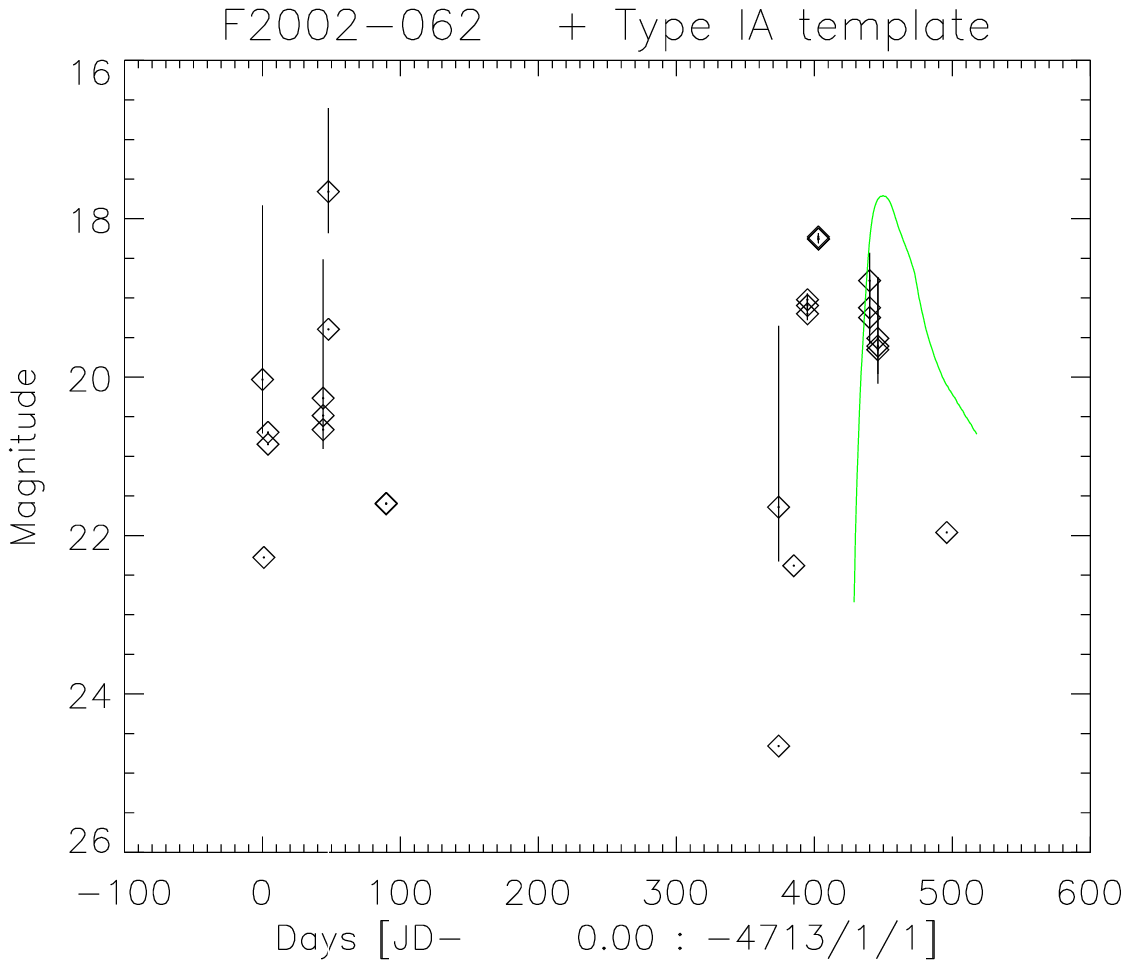}\label{fig:2002ft_lightcurve}}
\vspace{0.3in}
\subfigure[2002fu]{\includegraphics[angle=90,height=2in,width=3in]{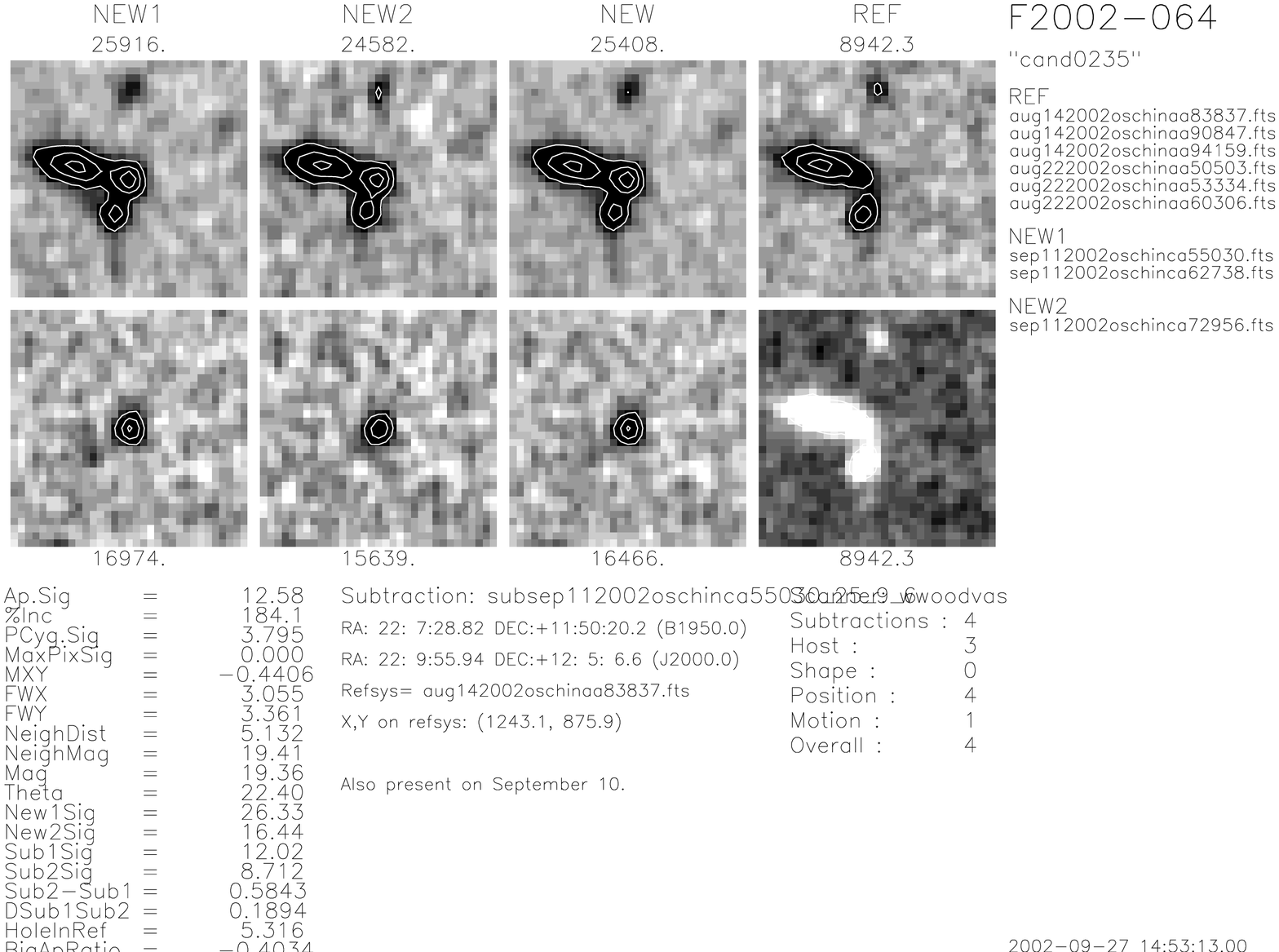}\label{fig:2002fu_discovery}}
\hspace{0.3in}
\subfigure[2002fu]{\includegraphics[height=2in]{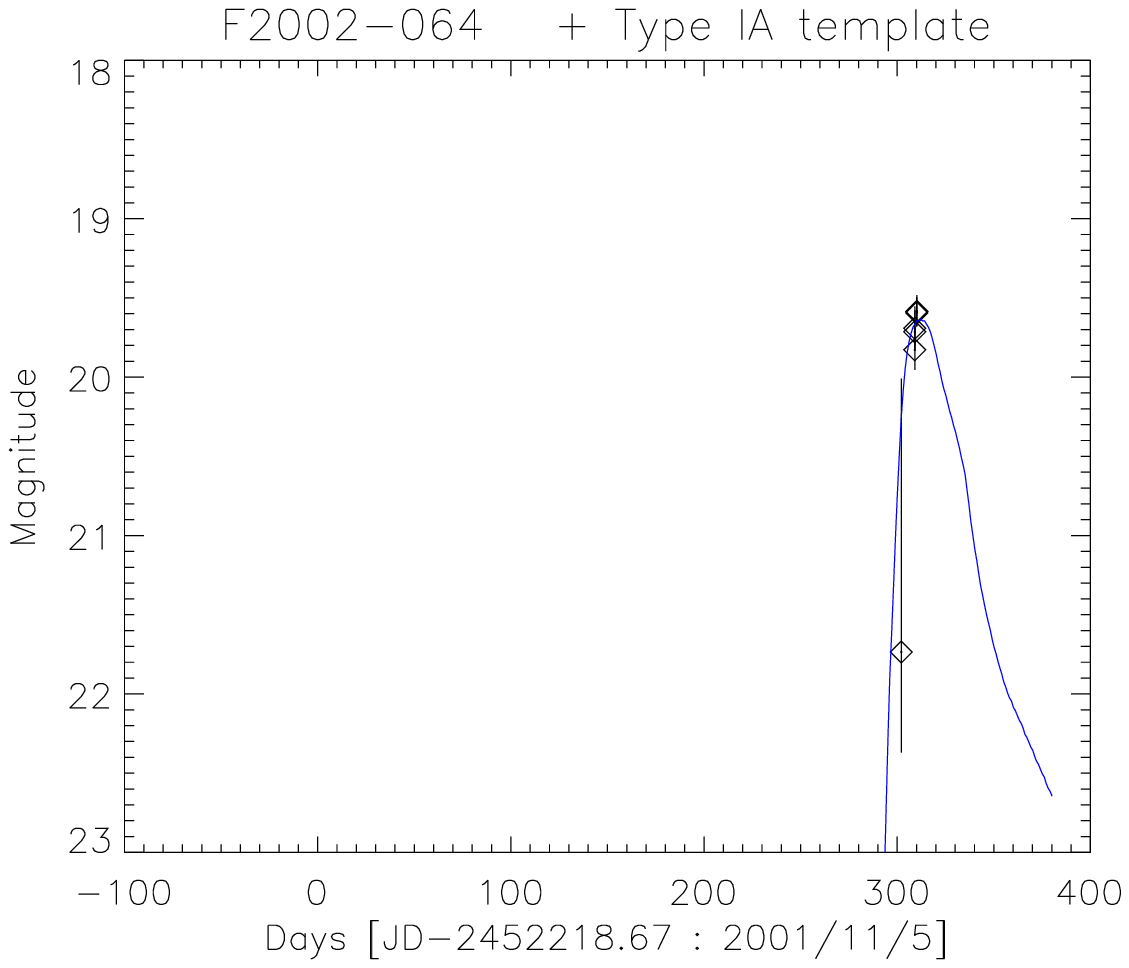}\label{fig:2002fu_lightcurve}}
\vspace{0.3in}
\end{figure}

\clearpage\pagebreak
\begin{figure}
\subfigure[2002gb]{\includegraphics[angle=90,height=2in,width=3in]{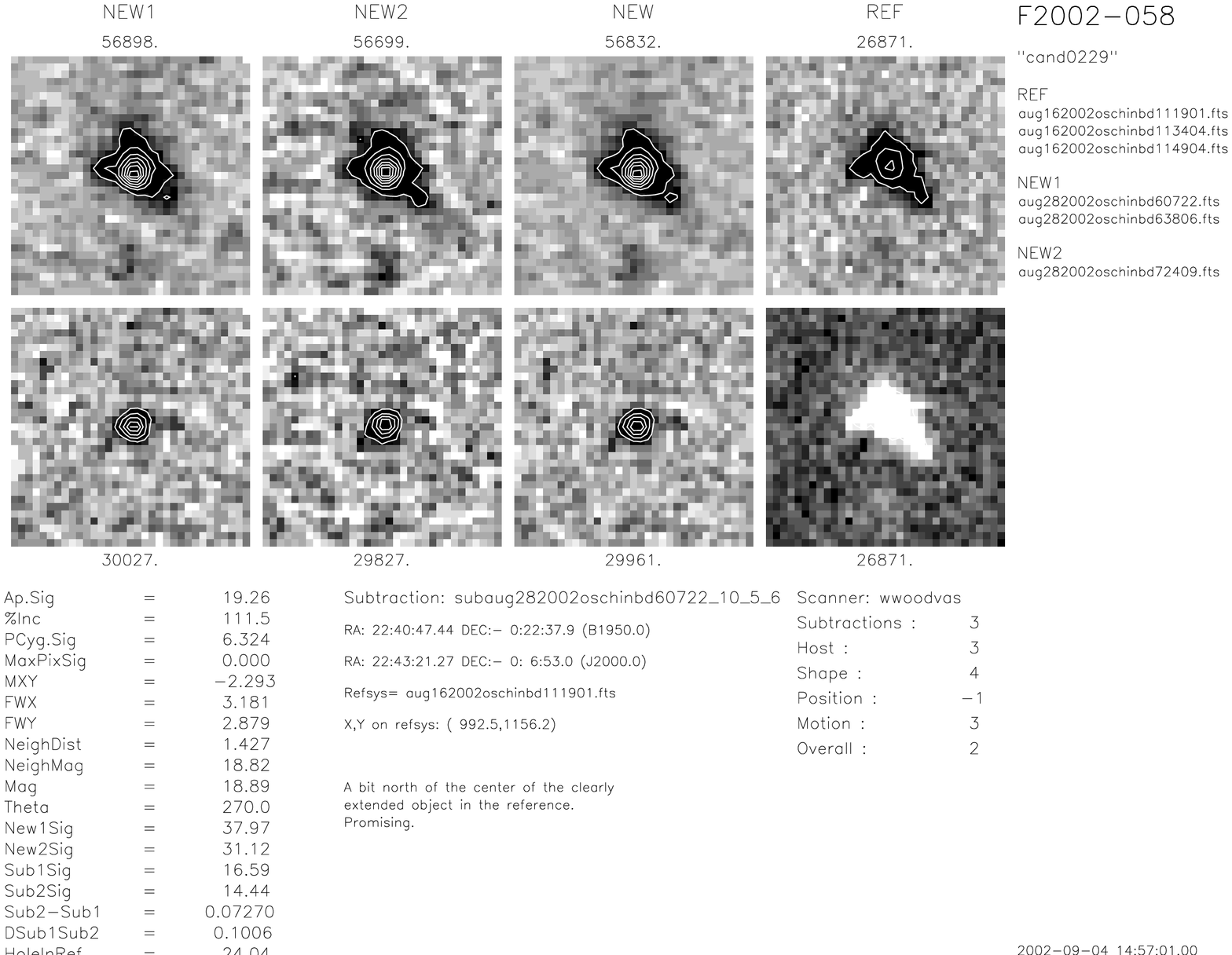}\label{fig:2002gb_discovery}}
\hspace{0.3in}
\subfigure[2002gb]{\includegraphics[height=2in]{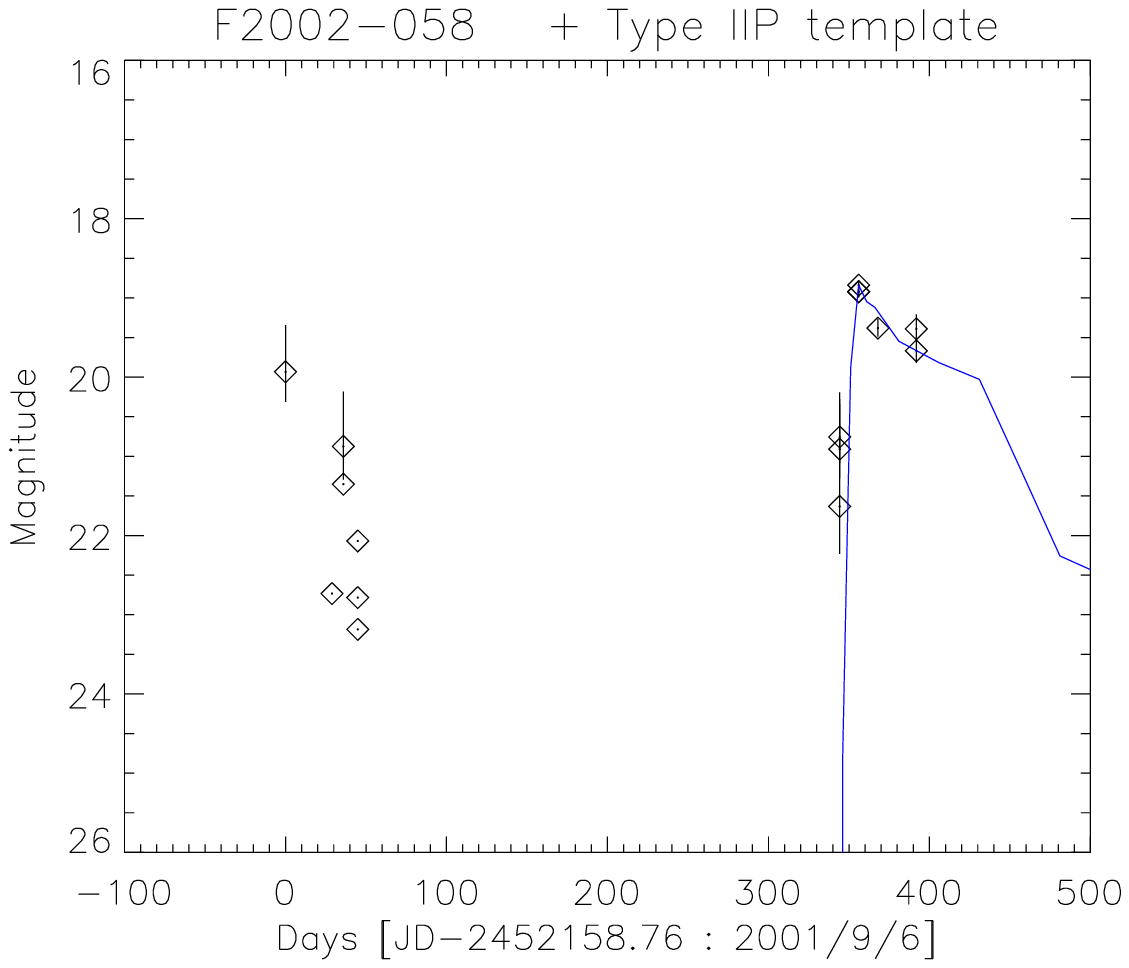}\label{fig:2002gb_lightcurve}}
\vspace{0.3in}
\subfigure[2002gd]{\includegraphics[angle=90,height=2in,width=3in]{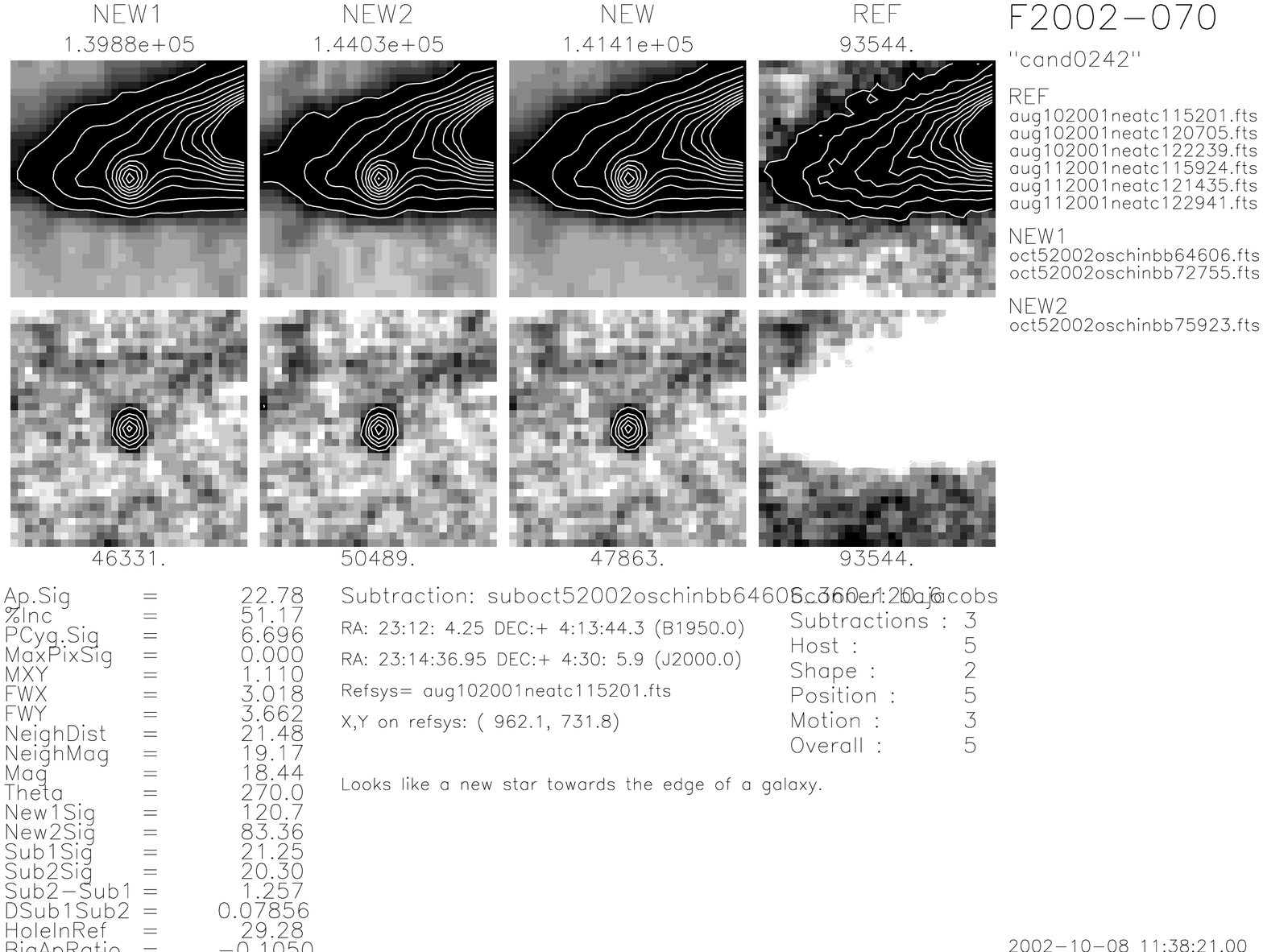}\label{fig:2002gd_discovery}}
\hspace{0.3in}
\subfigure[2002gd]{\includegraphics[height=2in]{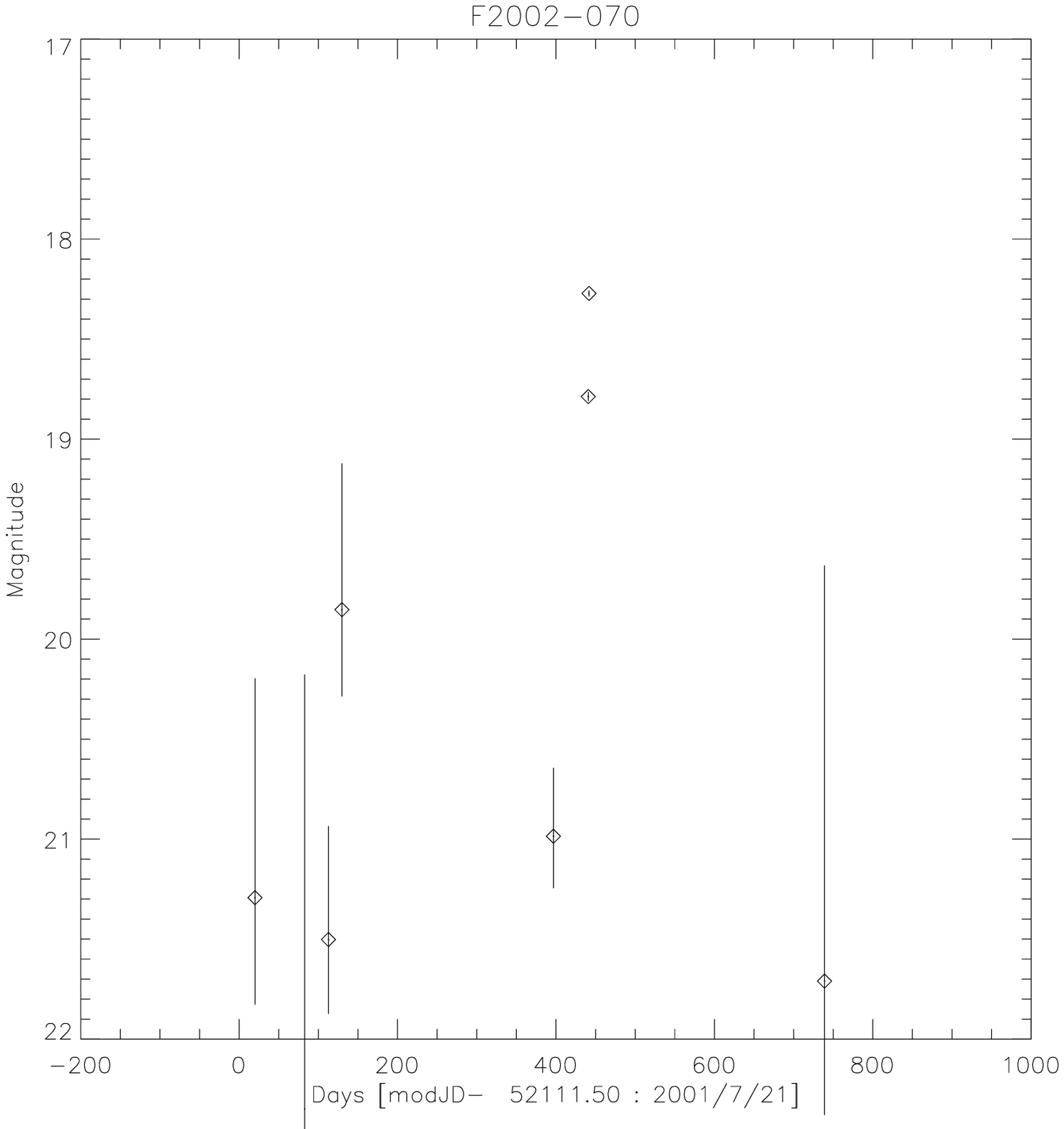}\label{fig:2002gd_lightcurve}}
\vspace{0.3in}
\subfigure[2002gf]{\includegraphics[angle=90,height=2in,width=3in]{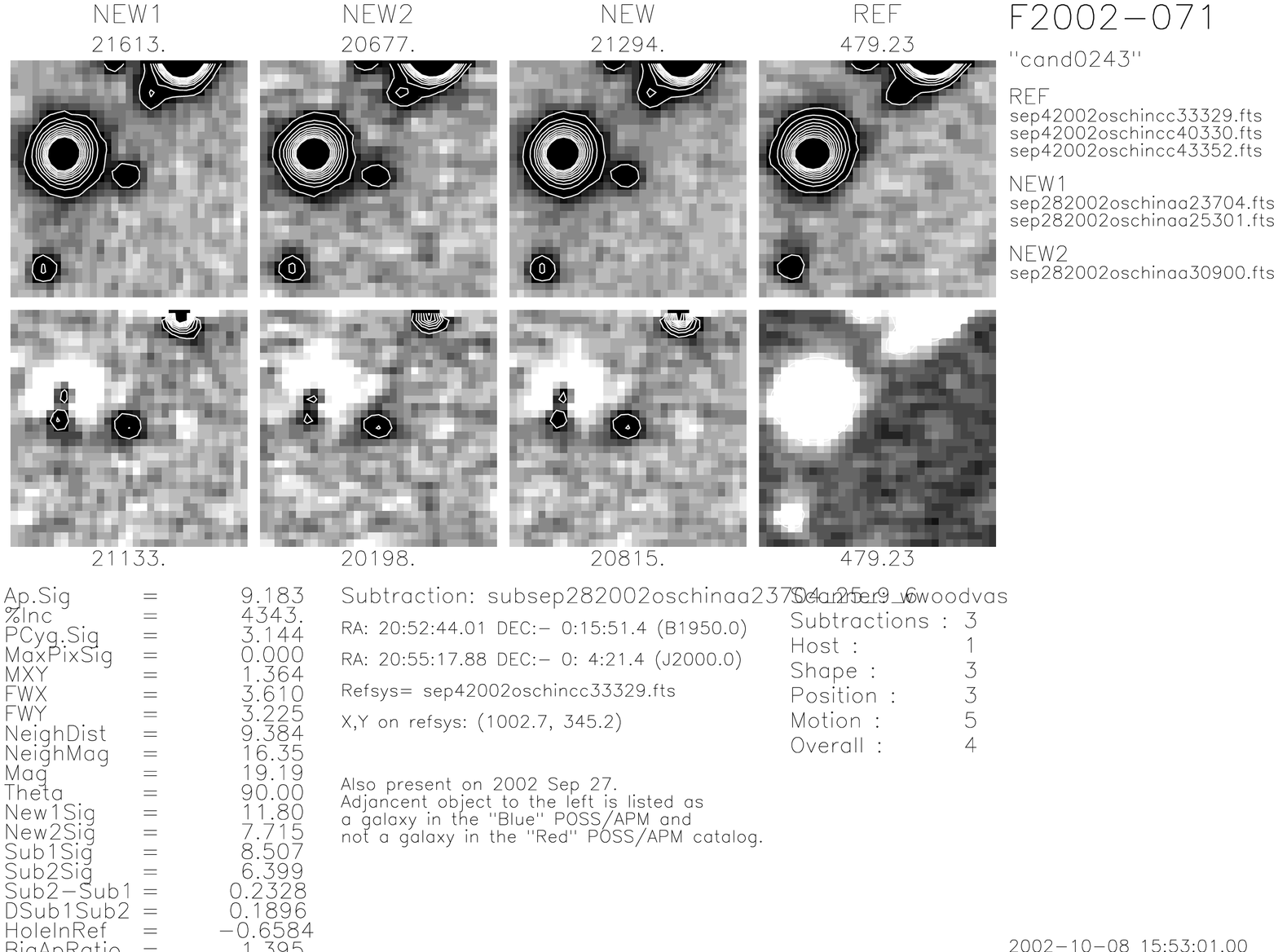}\label{fig:2002gf_discovery}}
\hspace{0.3in}
\subfigure[2002gf]{\includegraphics[height=2in]{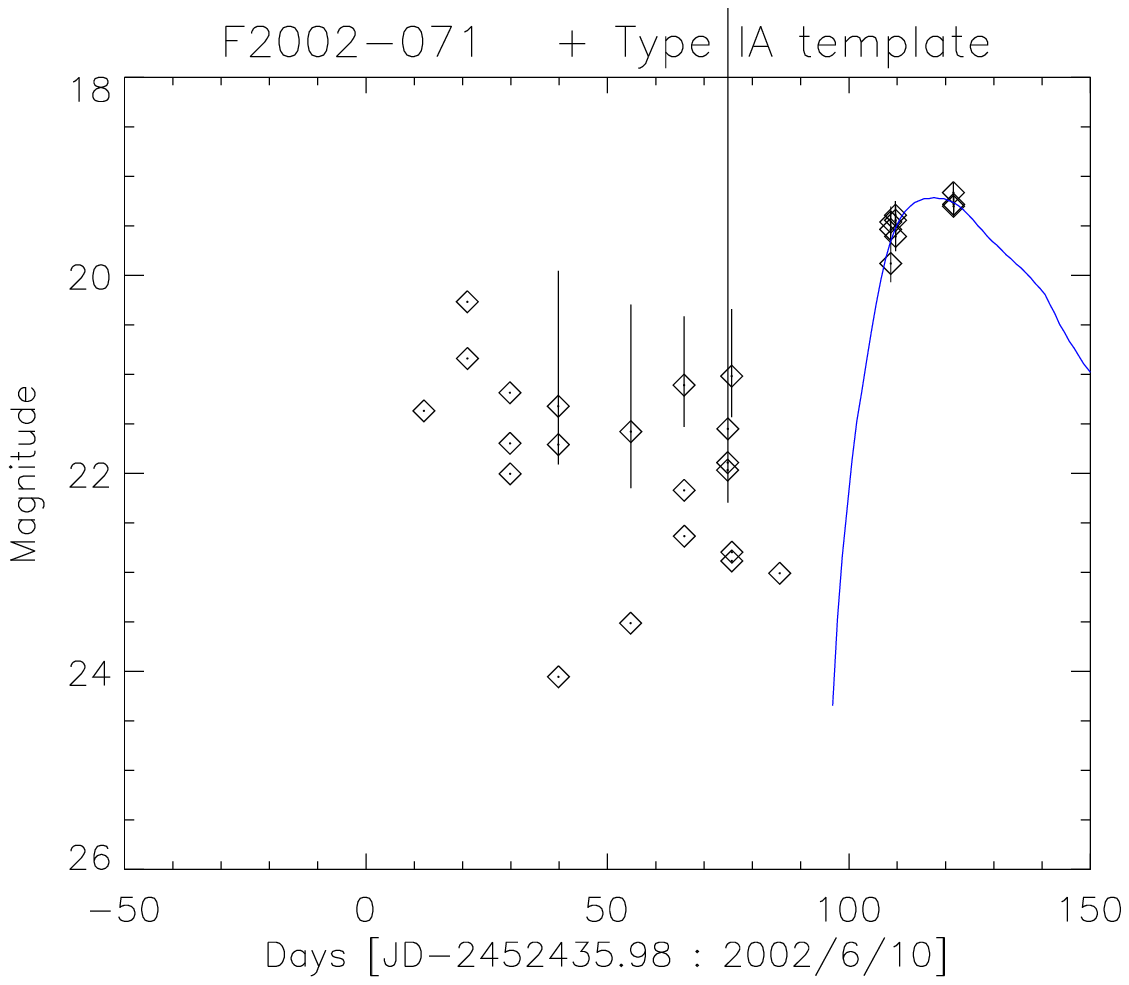}\label{fig:2002gf_lightcurve}}
\vspace{0.3in}
\end{figure}

\clearpage\pagebreak
\begin{figure}
\subfigure[2002gg]{\includegraphics[angle=90,height=2in,width=3in]{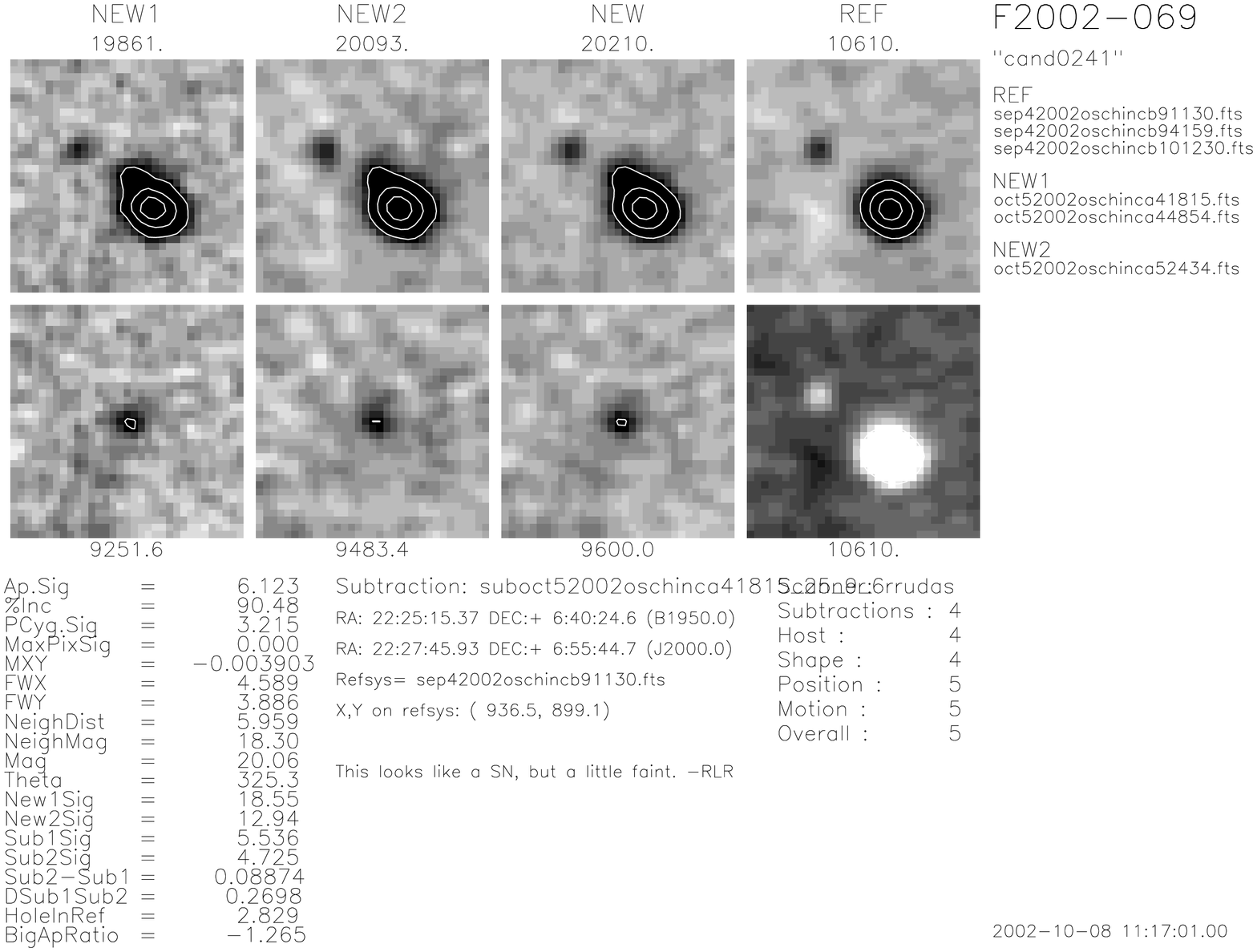}\label{fig:2002gg_discovery}}
\hspace{0.3in}
\subfigure[2002gg]{\includegraphics[height=2in]{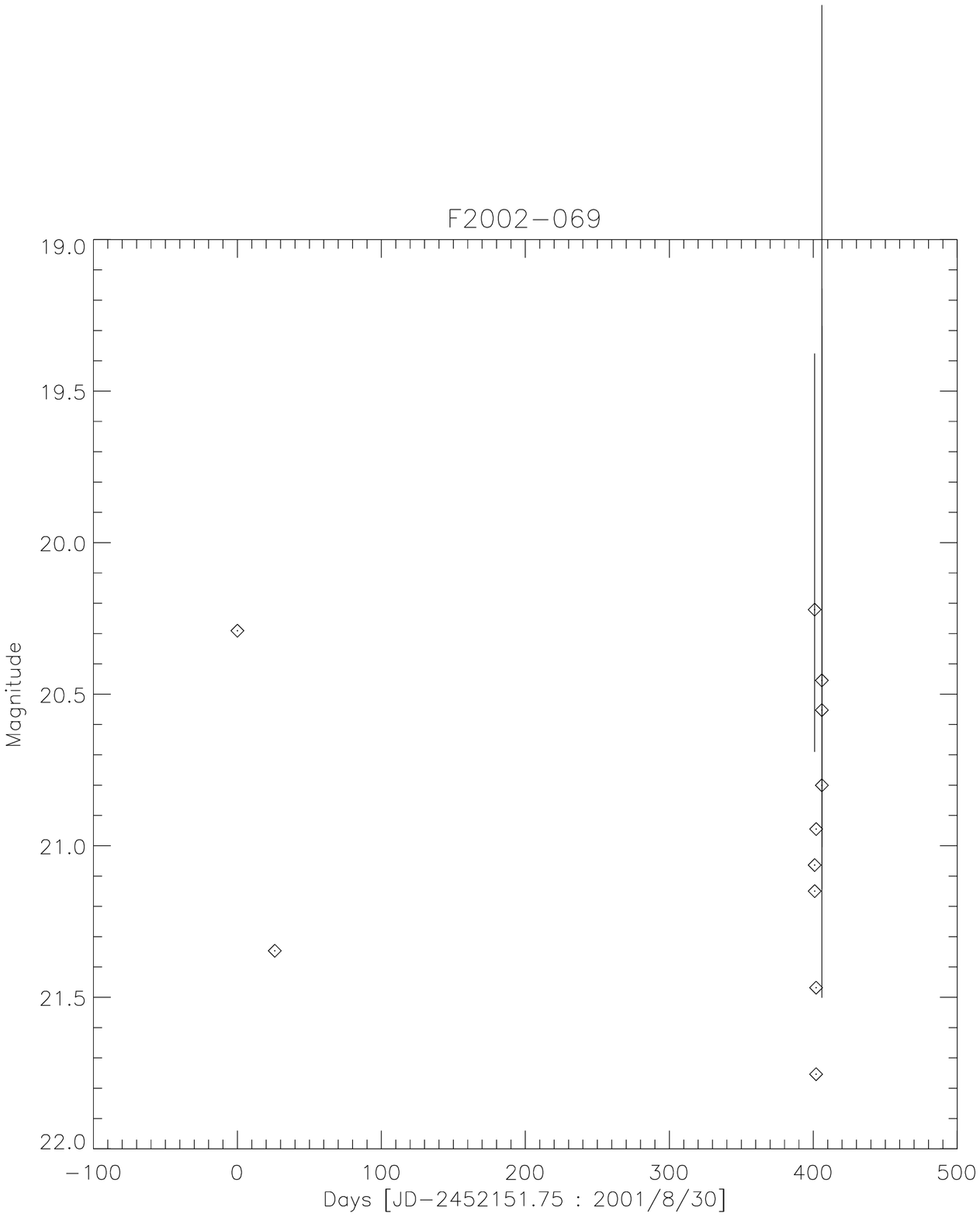}\label{fig:2002gg_lightcurve}}
\vspace{0.3in}
\subfigure[2002gh]{\includegraphics[angle=90,height=2in,width=3in]{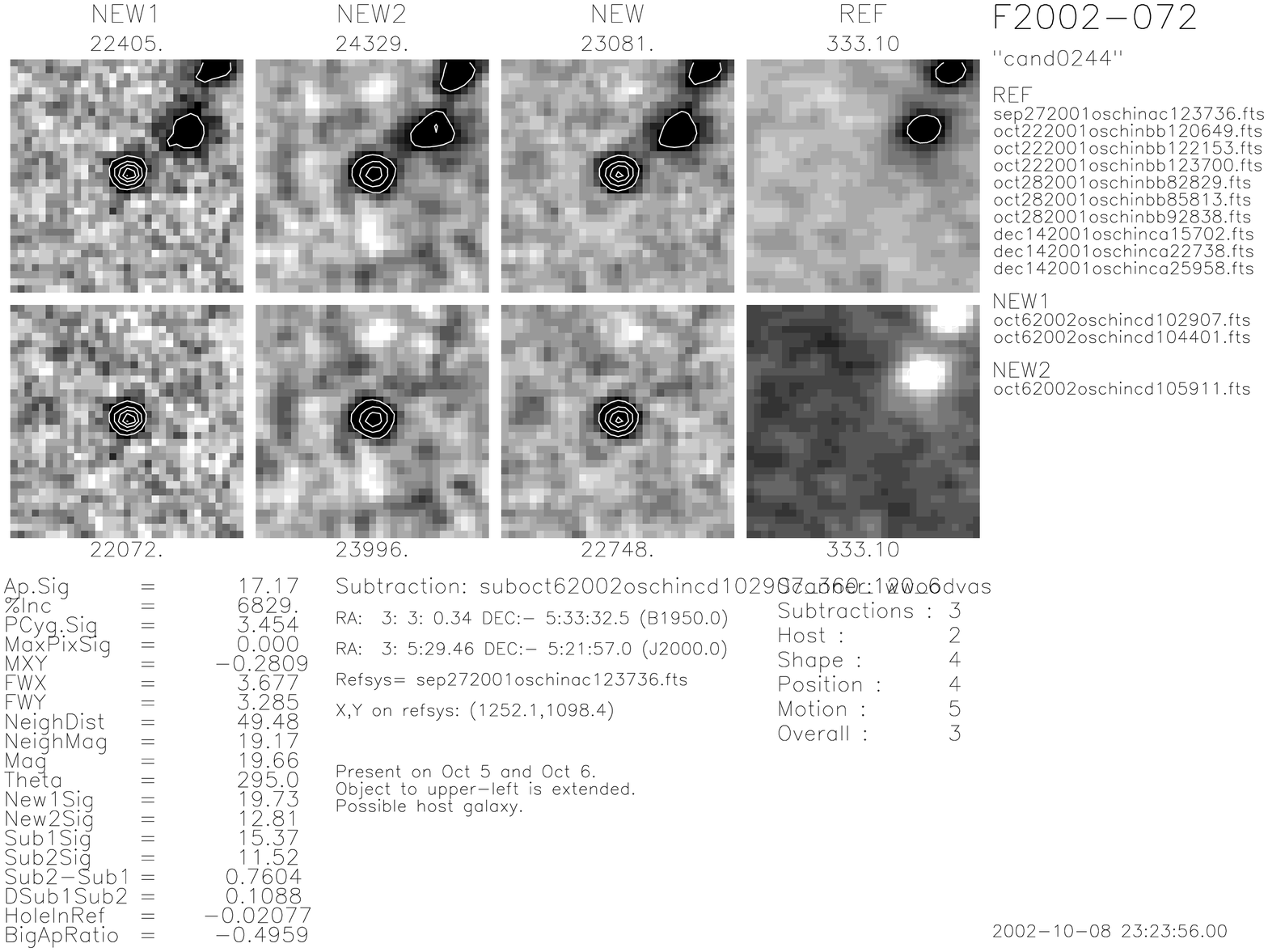}\label{fig:2002gh_discovery}}
\hspace{0.3in}
\subfigure[2002gh]{\includegraphics[height=2in]{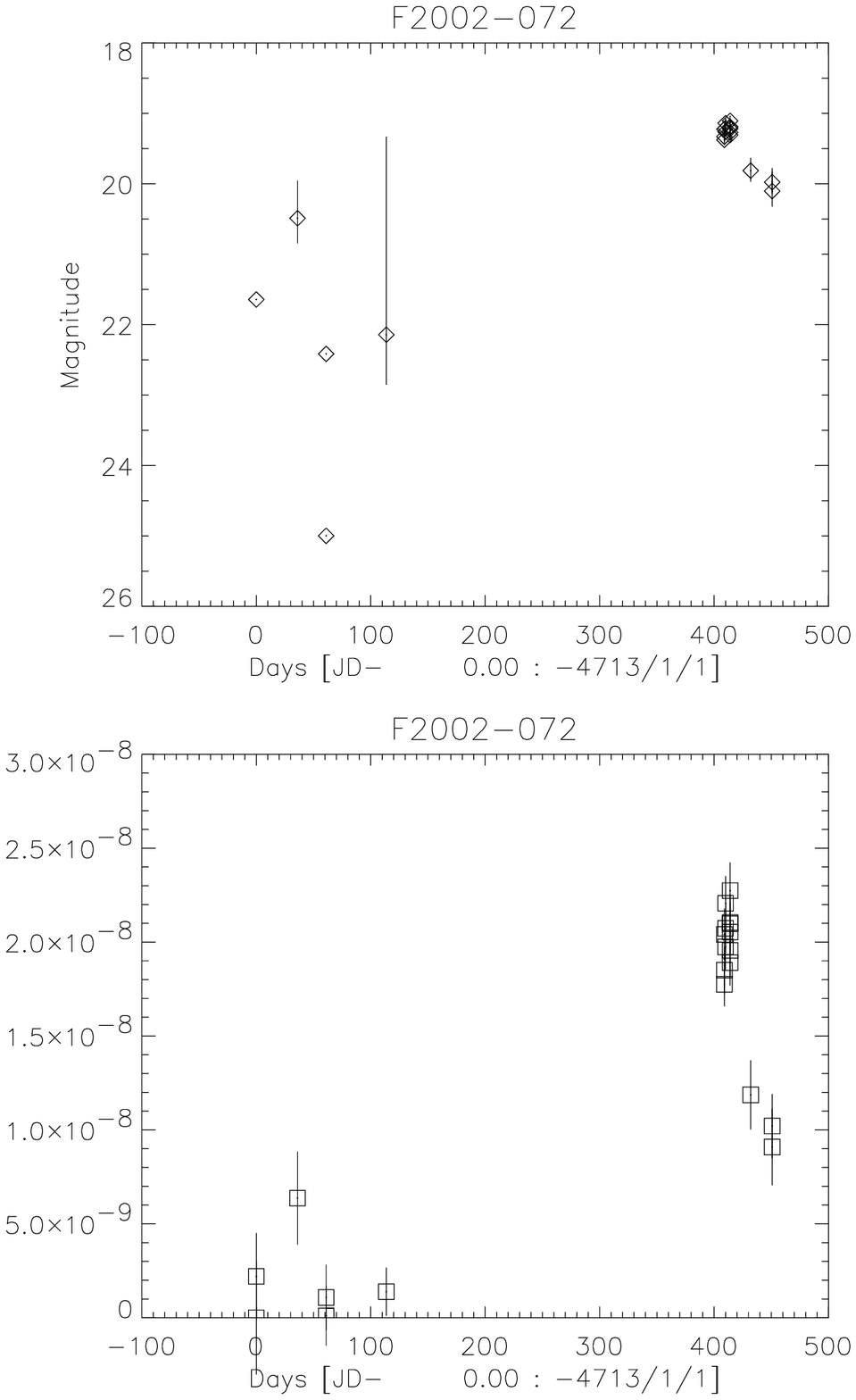}\label{fig:2002gh_lightcurve}}
\vspace{0.3in}
\subfigure[2002gx]{\includegraphics[angle=90,height=2in,width=3in]{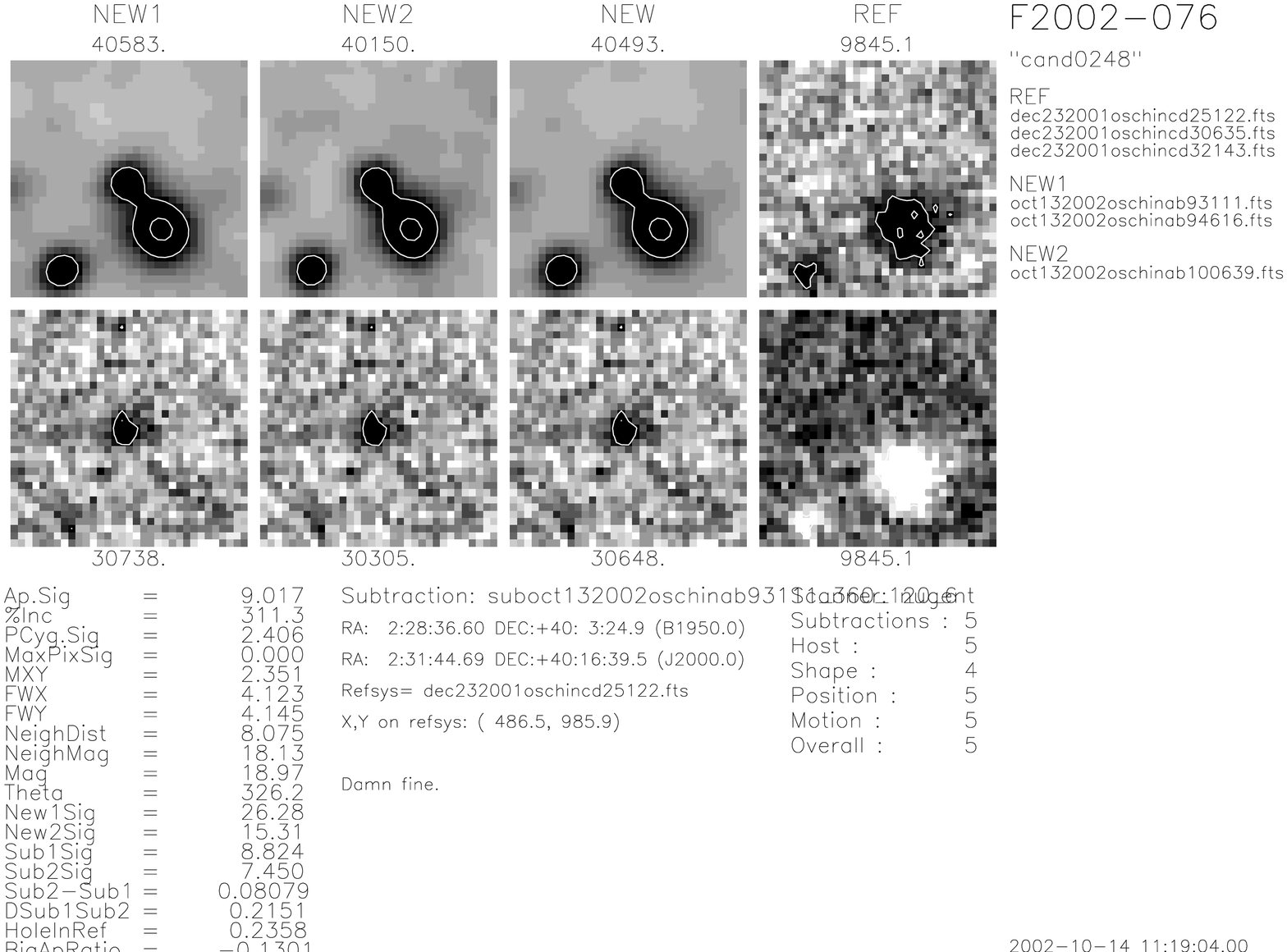}\label{fig:2002gx_discovery}}
\hspace{0.3in}
\subfigure[2002gx]{\includegraphics[height=2in]{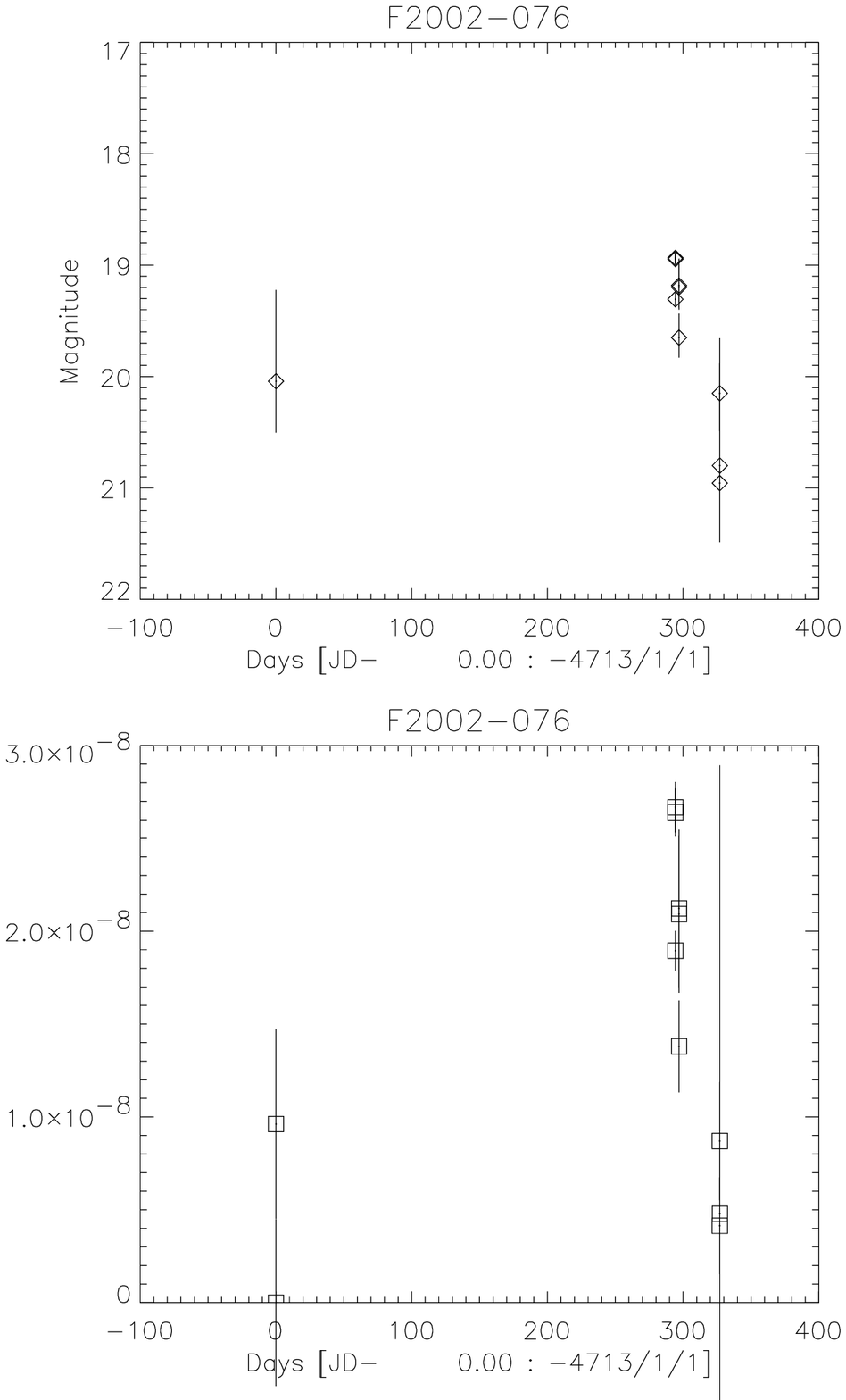}\label{fig:2002gx_lightcurve}}
\vspace{0.3in}
\end{figure}

\clearpage\pagebreak
\begin{figure}
\subfigure[2002gz]{\includegraphics[angle=90,height=2in,width=3in]{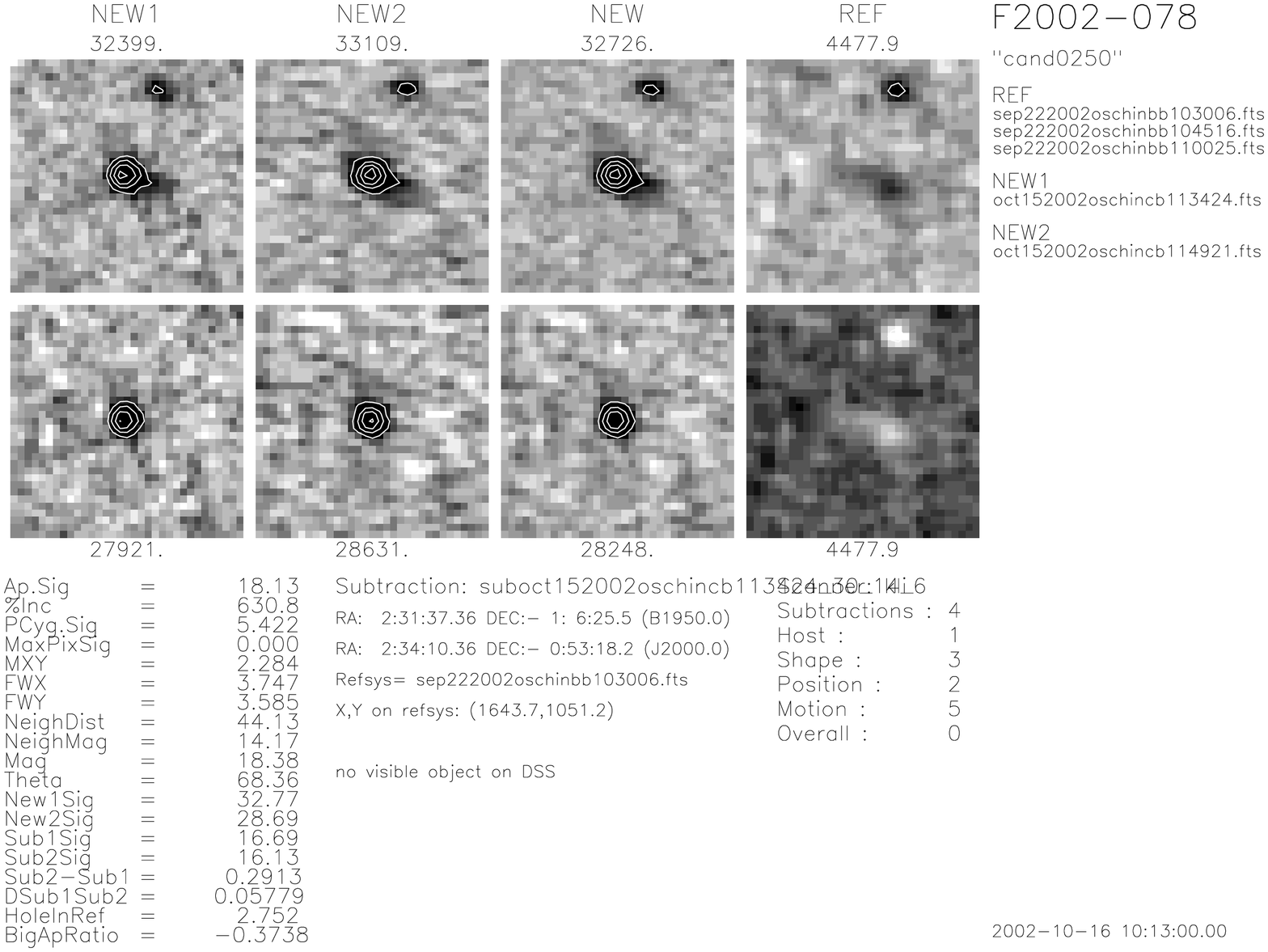}\label{fig:2002gz_discovery}}
\hspace{0.3in}
\subfigure[2002gz]{\includegraphics[height=2in]{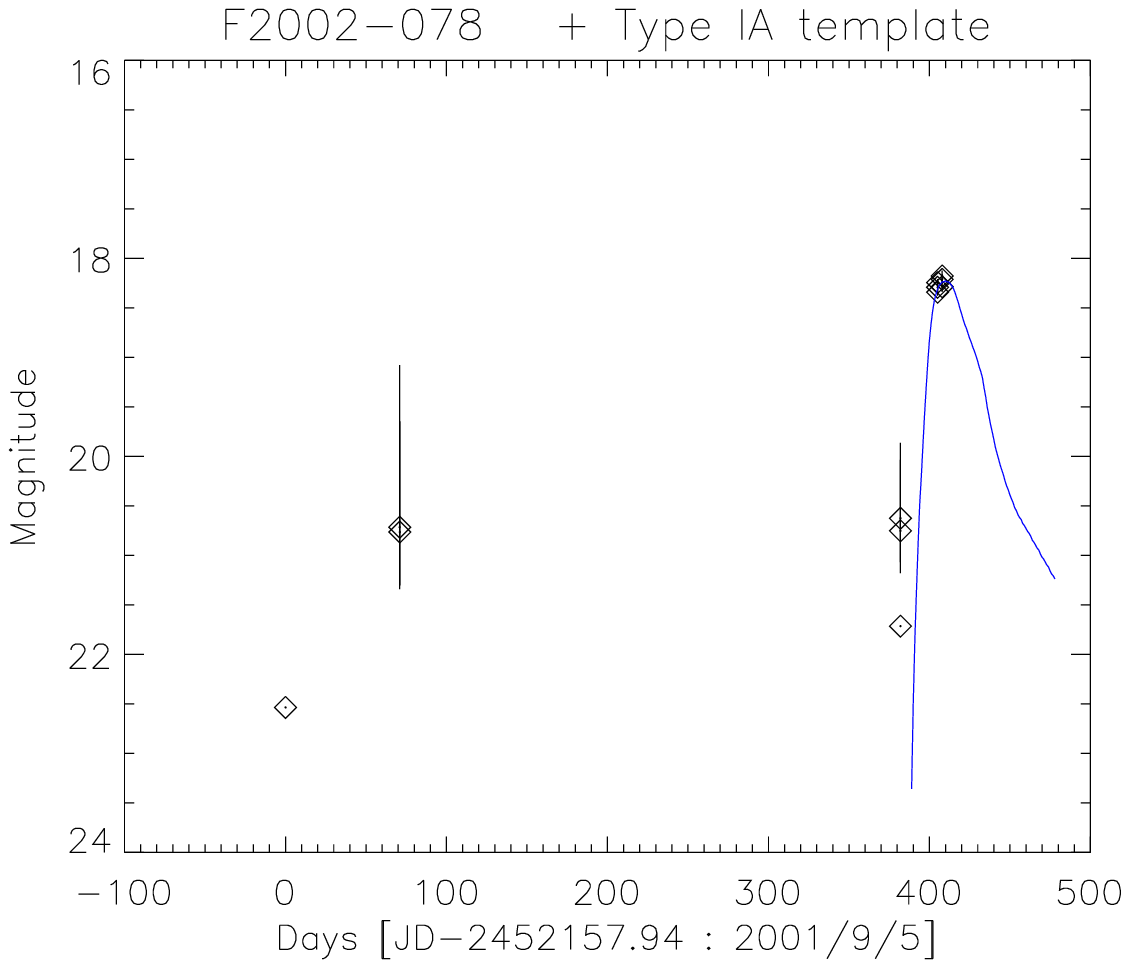}\label{fig:2002gz_lightcurve}}
\vspace{0.3in}
\subfigure[2002hb]{\includegraphics[angle=90,height=2in,width=3in]{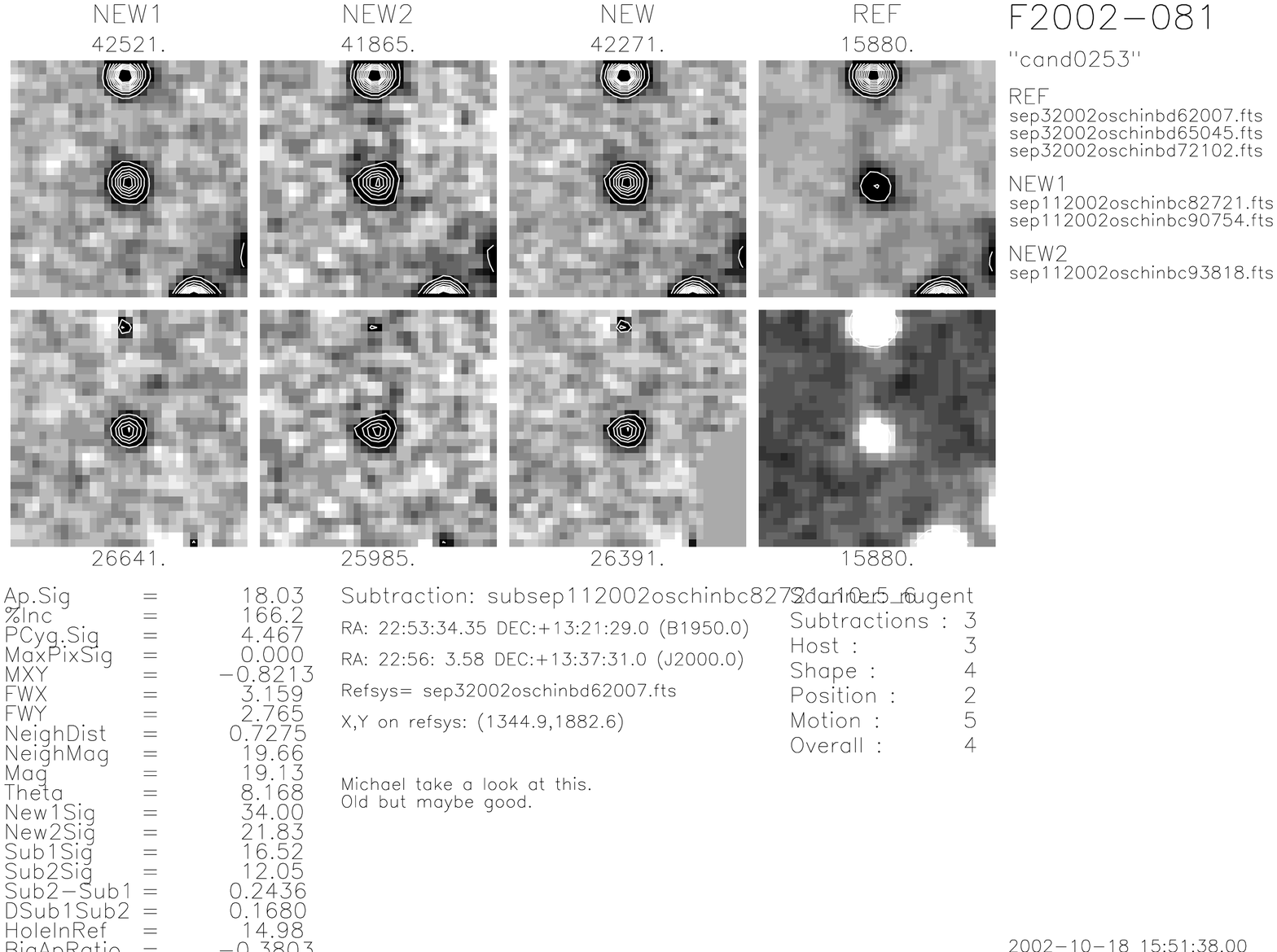}\label{fig:2002hb_discovery}}
\hspace{0.3in}
\subfigure[2002hb]{\includegraphics[height=2in]{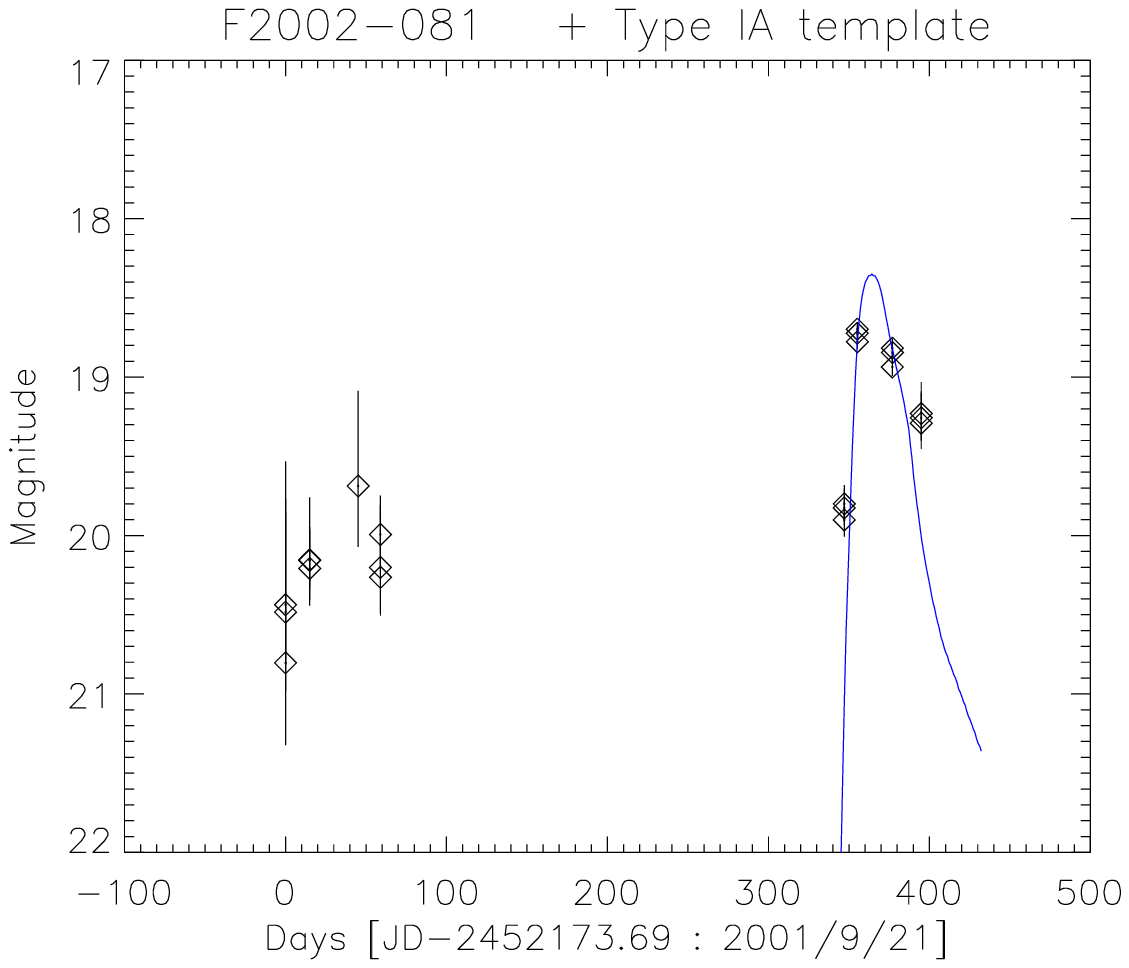}\label{fig:2002hb_lightcurve}}
\vspace{0.3in}
\subfigure[2002hf]{\includegraphics[angle=90,height=2in,width=3in]{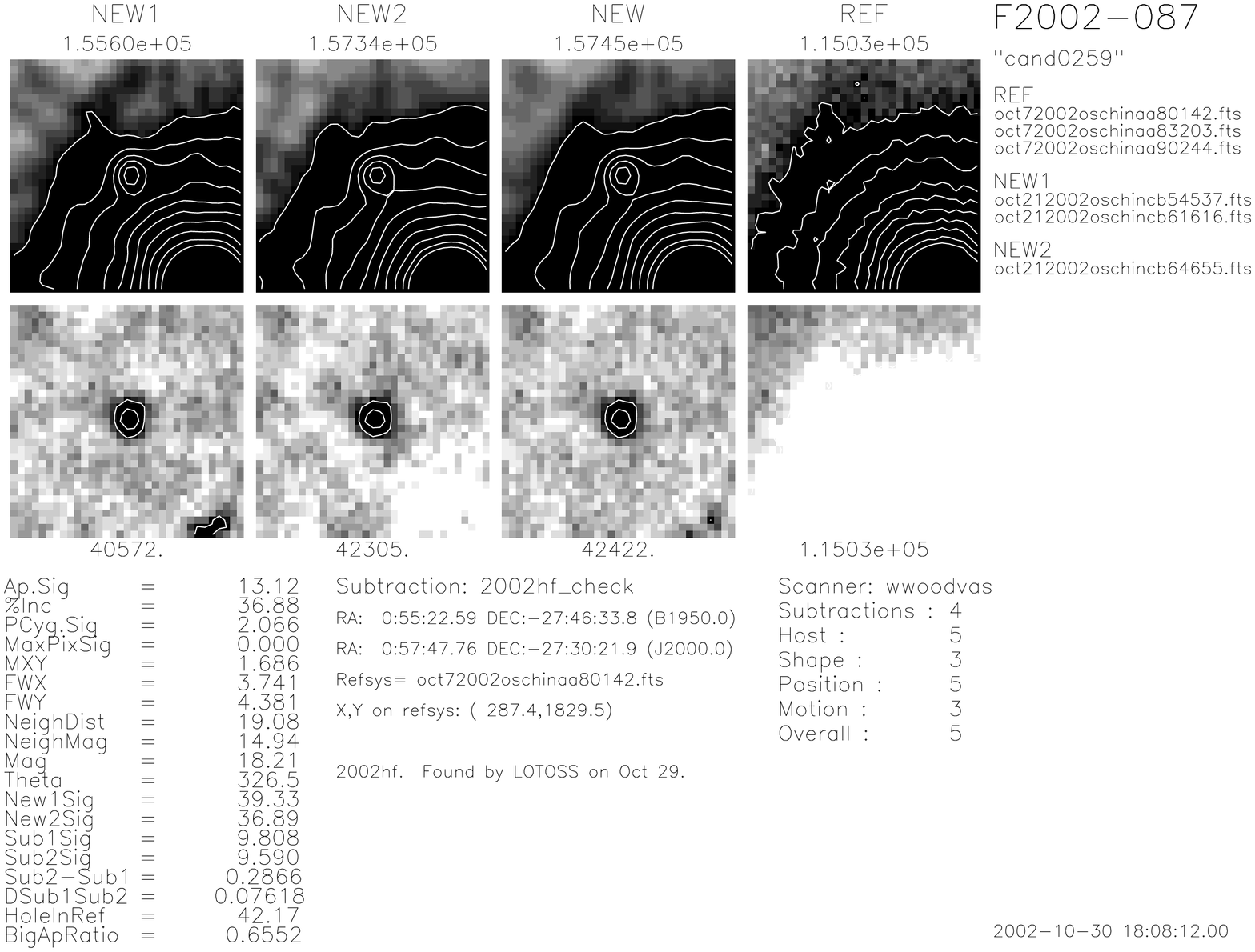}\label{fig:2002hf_discovery}}
\hspace{0.3in}
\subfigure[2002hf]{\includegraphics[height=2in]{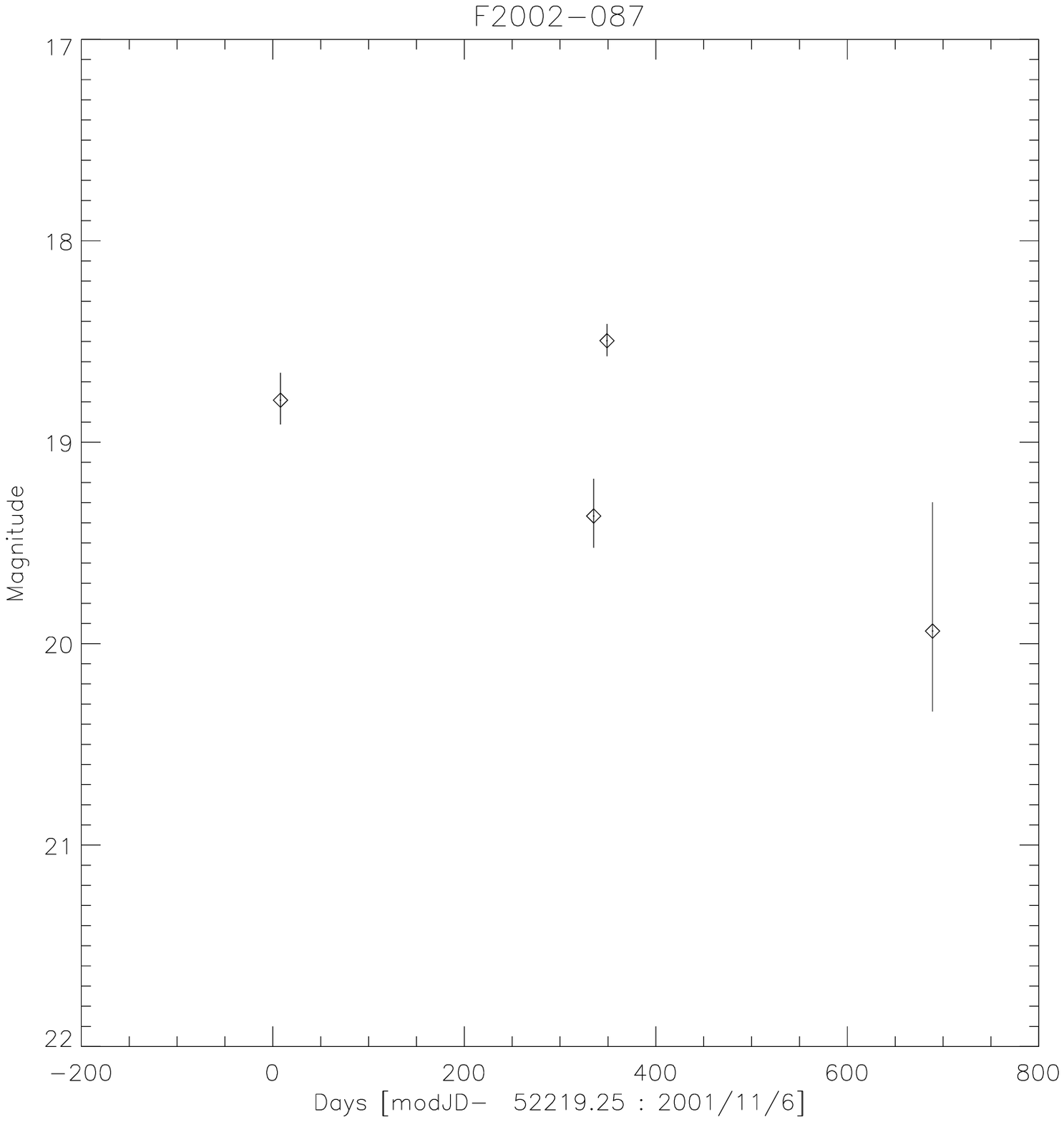}\label{fig:2002hf_lightcurve}}
\vspace{0.3in}
\end{figure}

\clearpage\pagebreak
\begin{figure}
\subfigure[2002hj]{\includegraphics[angle=90,height=2in,width=3in]{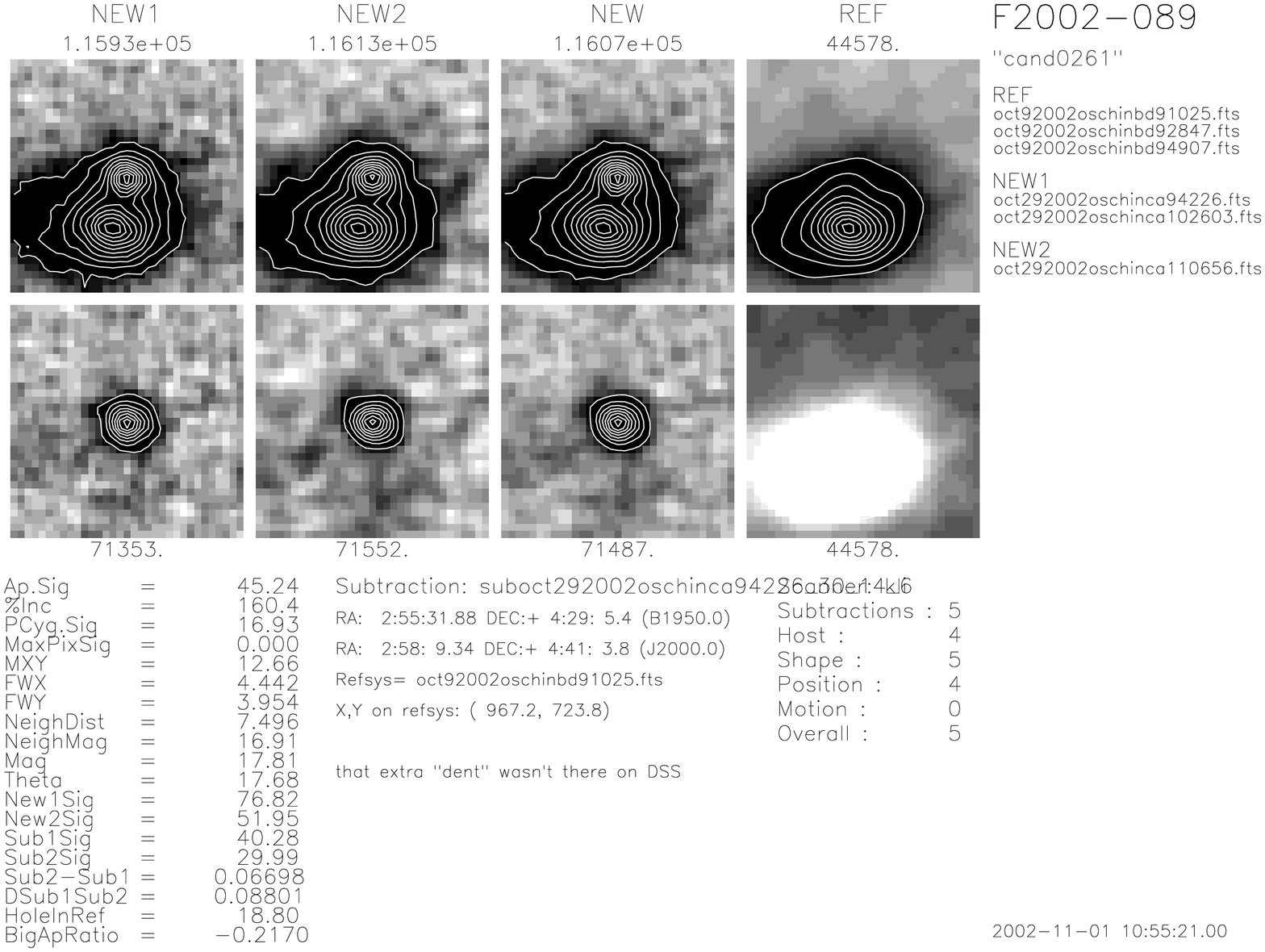}\label{fig:2002hj_discovery}}
\hspace{0.3in}
\subfigure[2002hj]{\includegraphics[height=2in]{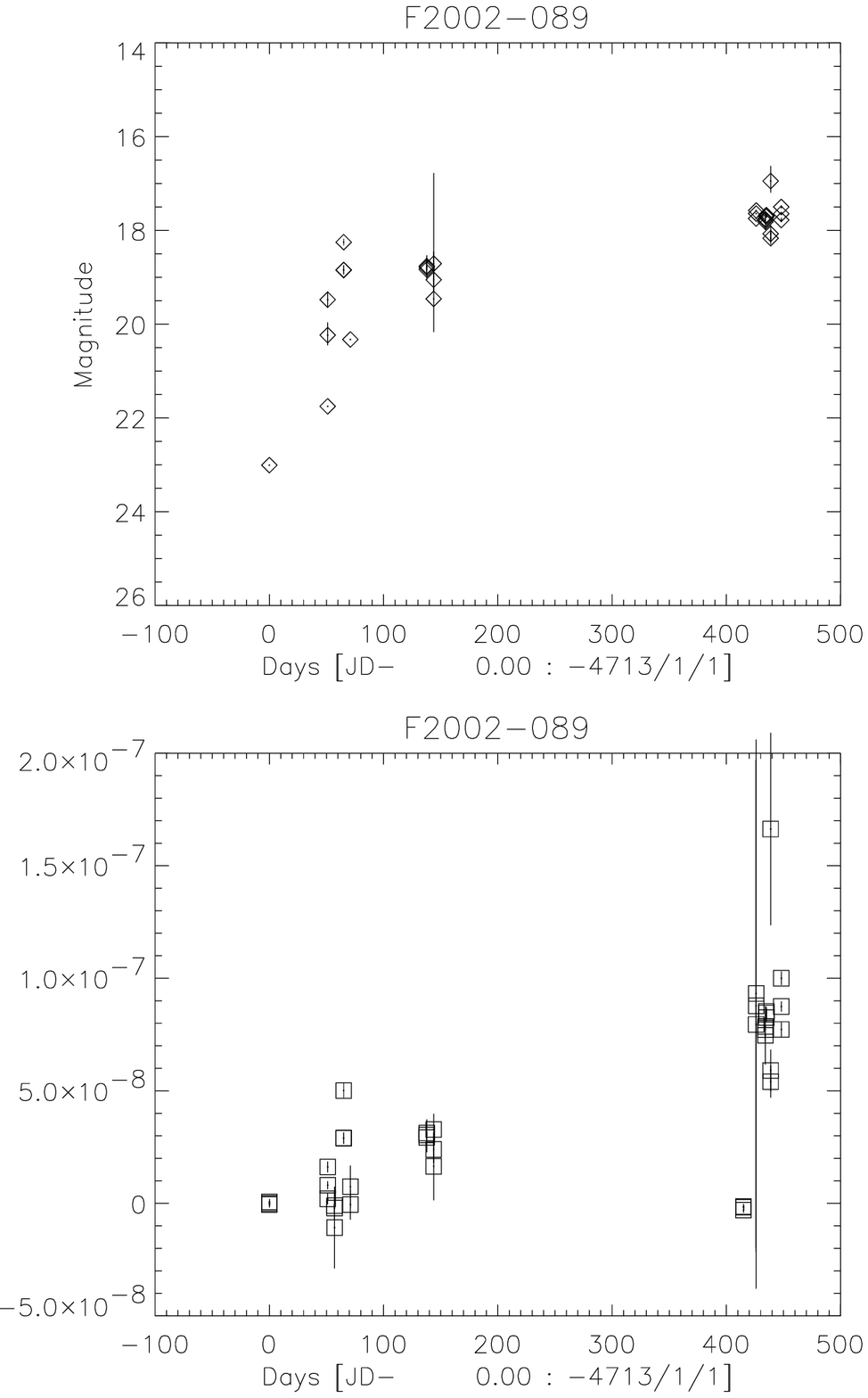}\label{fig:2002hj_lightcurve}}
\vspace{0.3in}
\subfigure[2002hw]{\includegraphics[angle=90,height=2in,width=3in]{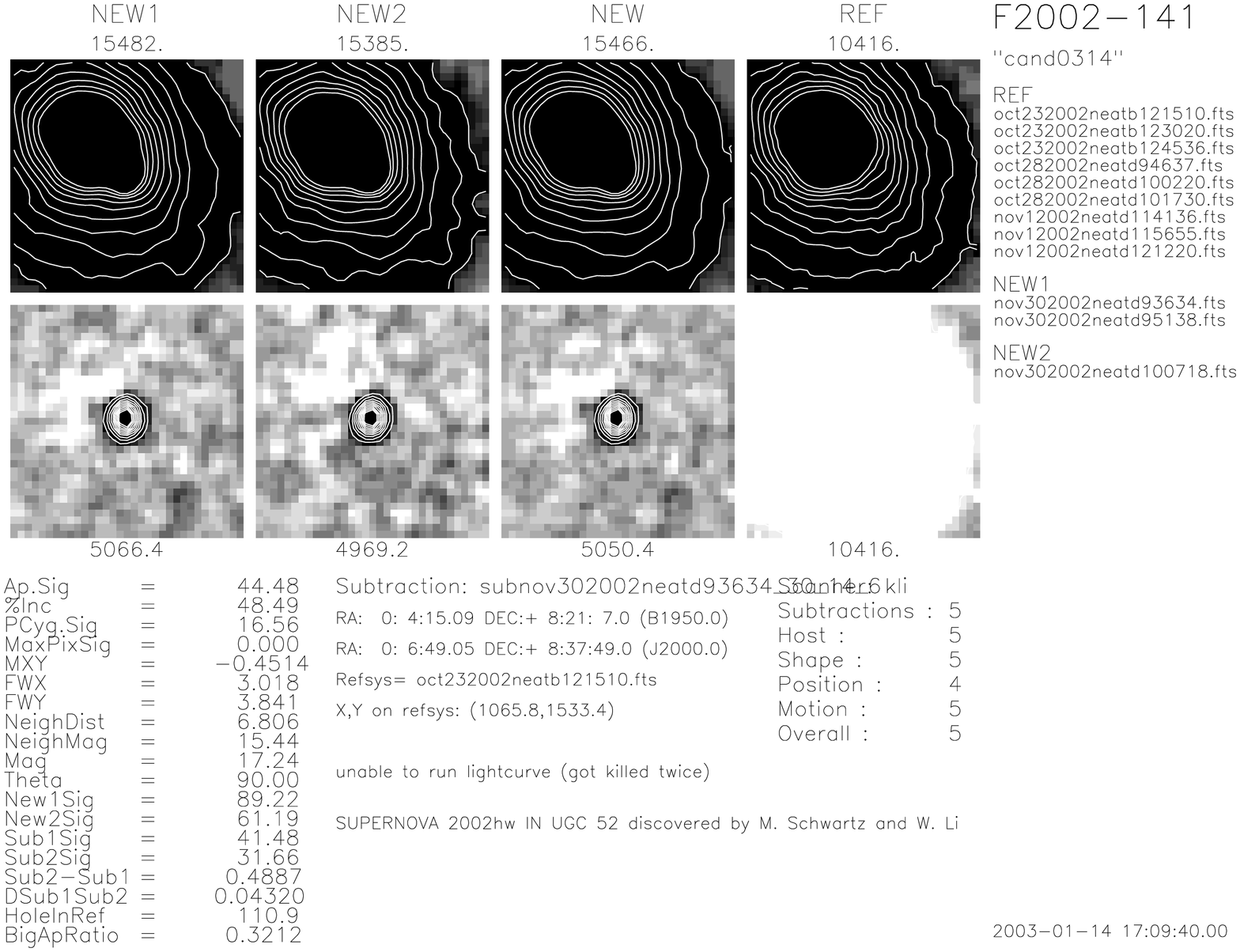}\label{fig:2002hw_discovery}}
\hspace{0.3in}
\subfigure[2002hw]{\includegraphics[height=2in]{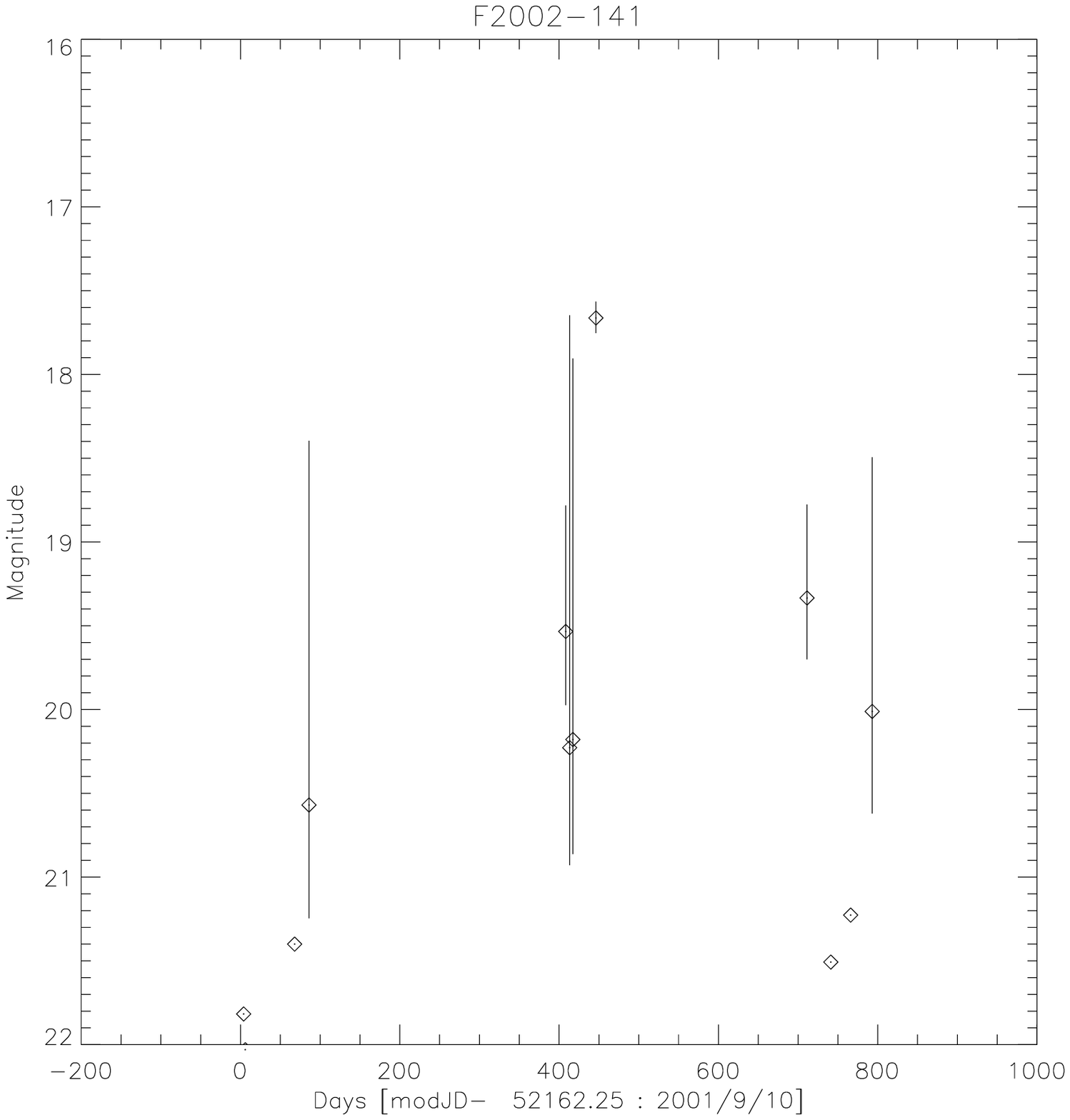}\label{fig:2002hw_lightcurve}}
\vspace{0.3in}
\subfigure[2002ia]{\includegraphics[angle=90,height=2in,width=3in]{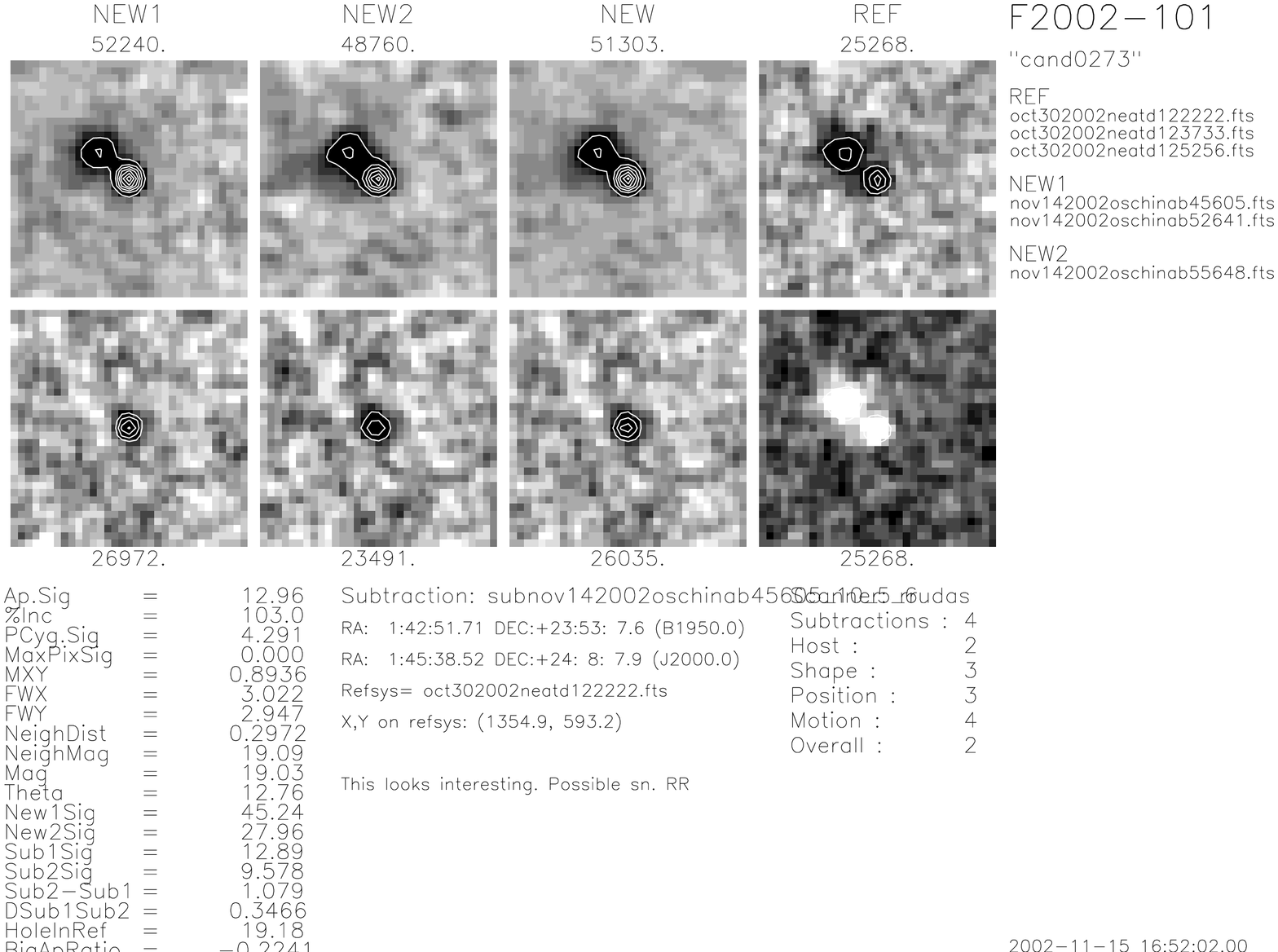}\label{fig:2002ia_discovery}}
\hspace{0.3in}
\subfigure[2002ia]{\includegraphics[height=2in]{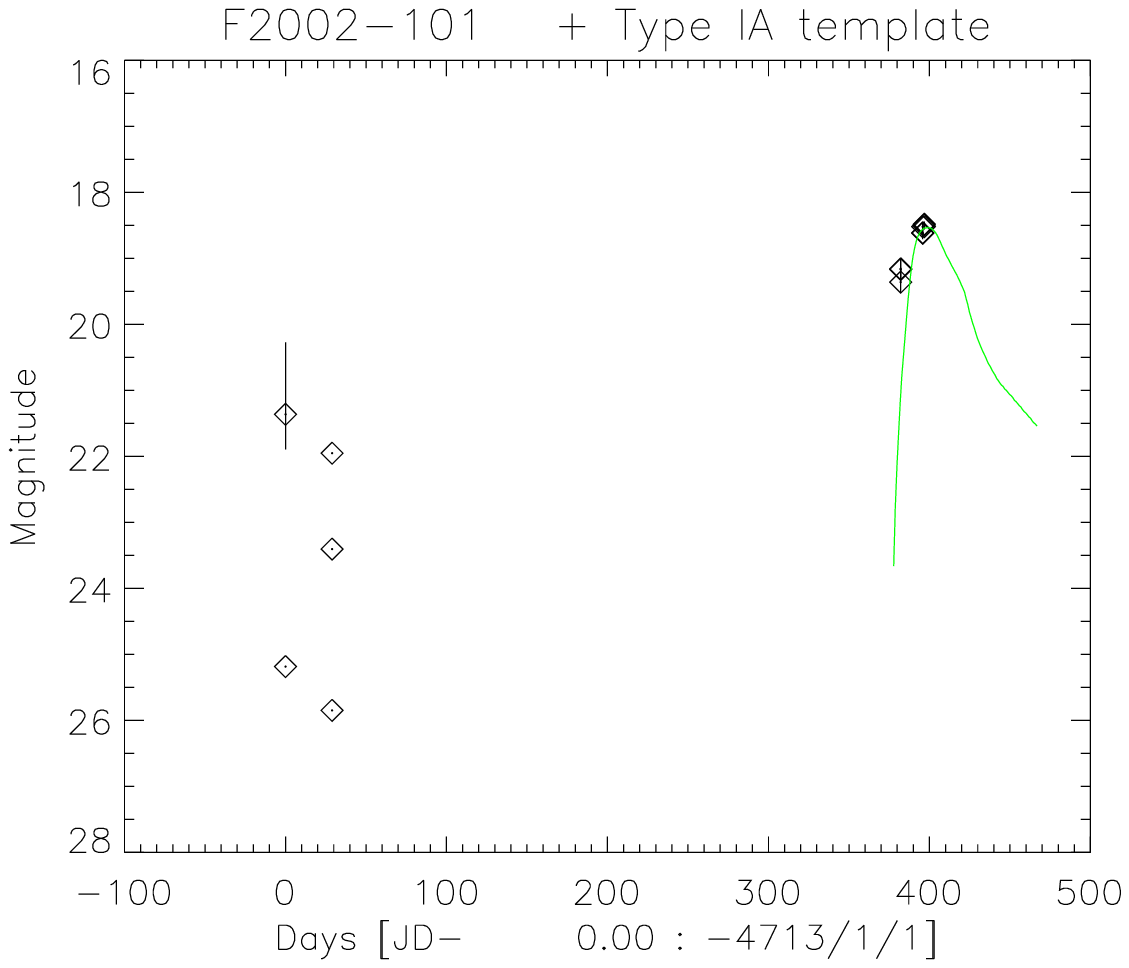}\label{fig:2002ia_lightcurve}}
\vspace{0.3in}
\end{figure}

\clearpage\pagebreak
\begin{figure}
\subfigure[2002ib]{\includegraphics[angle=90,height=2in,width=3in]{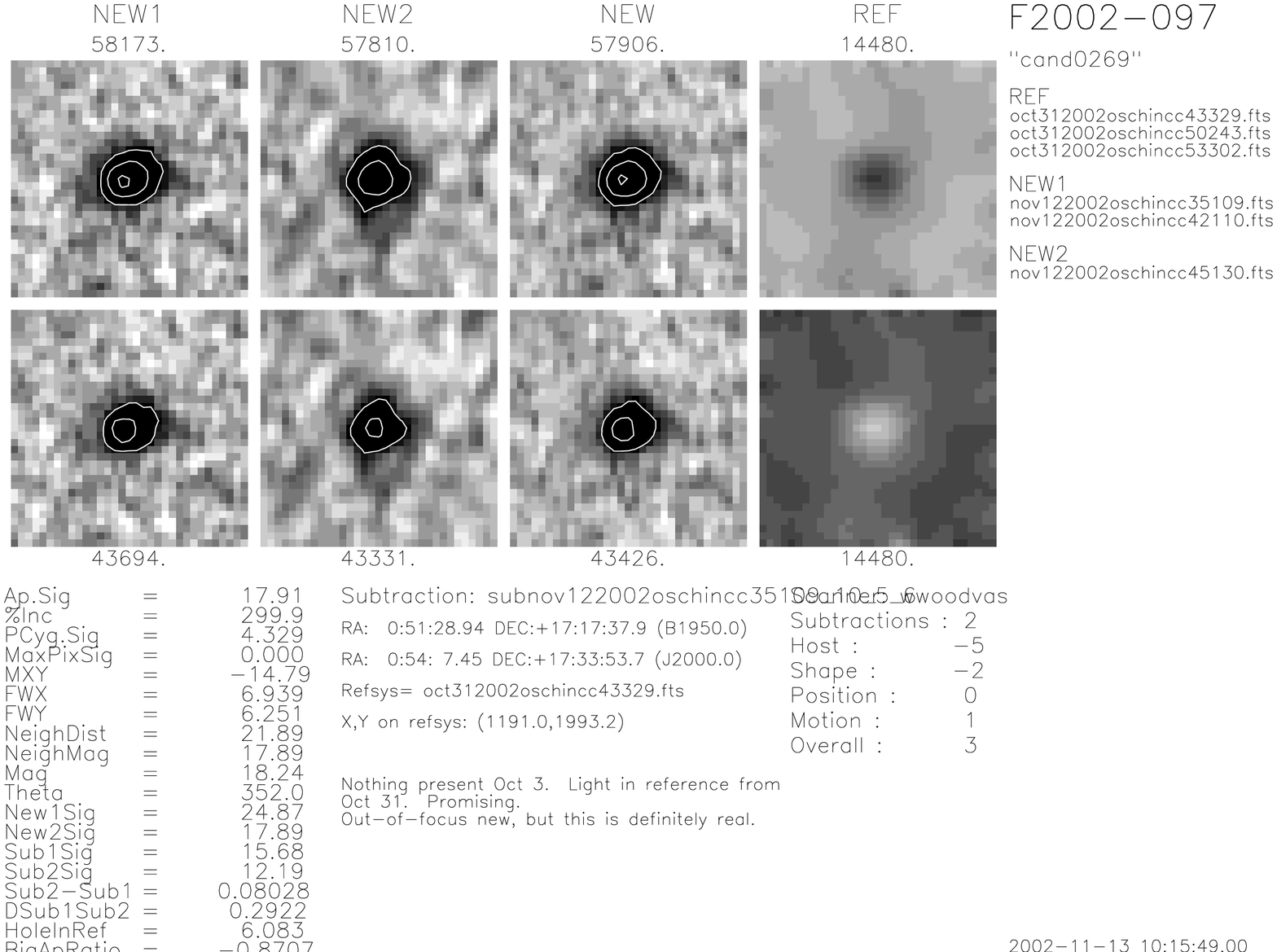}\label{fig:2002ib_discovery}}
\hspace{0.3in}
\subfigure[2002ib]{\includegraphics[height=2in]{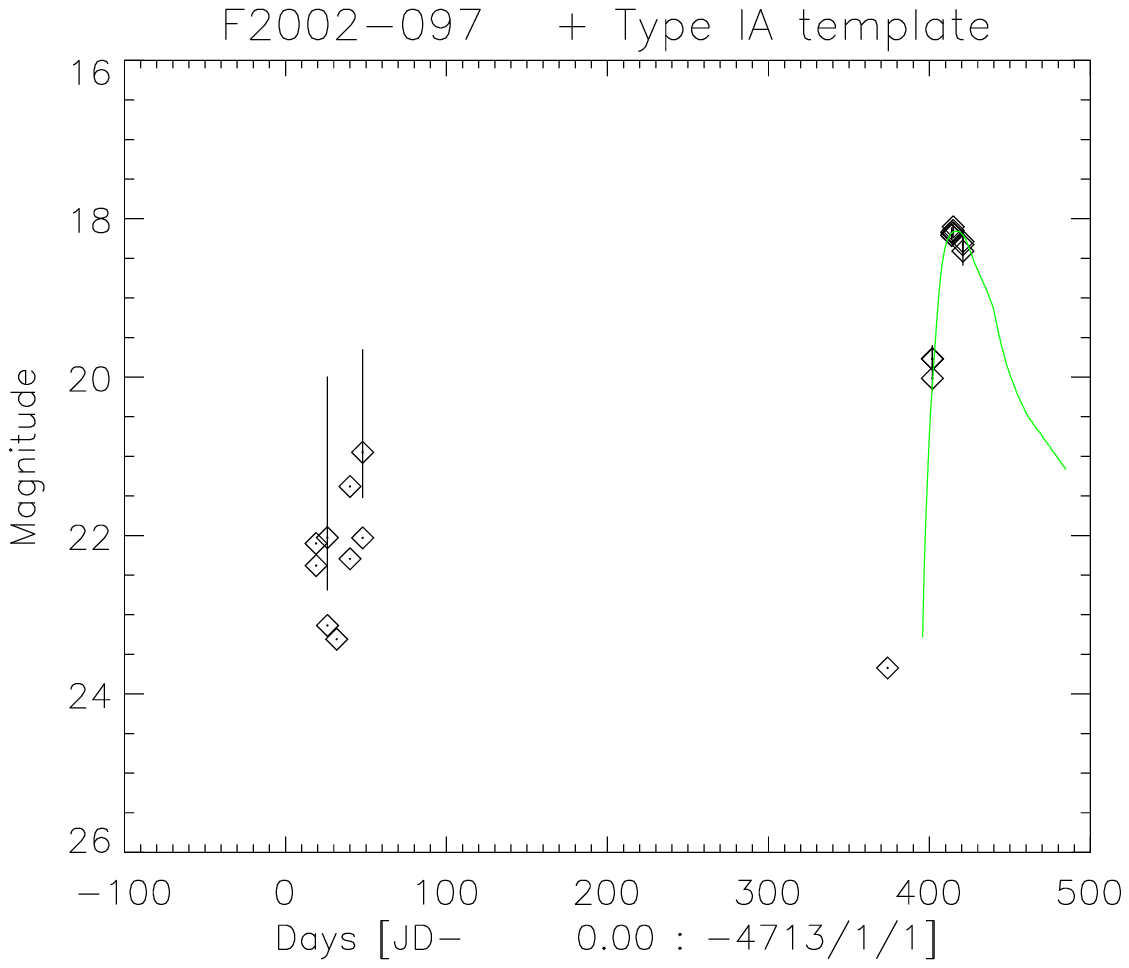}\label{fig:2002ib_lightcurve}}
\vspace{0.3in}
\subfigure[2002ic]{\includegraphics[angle=90,height=2in,width=3in]{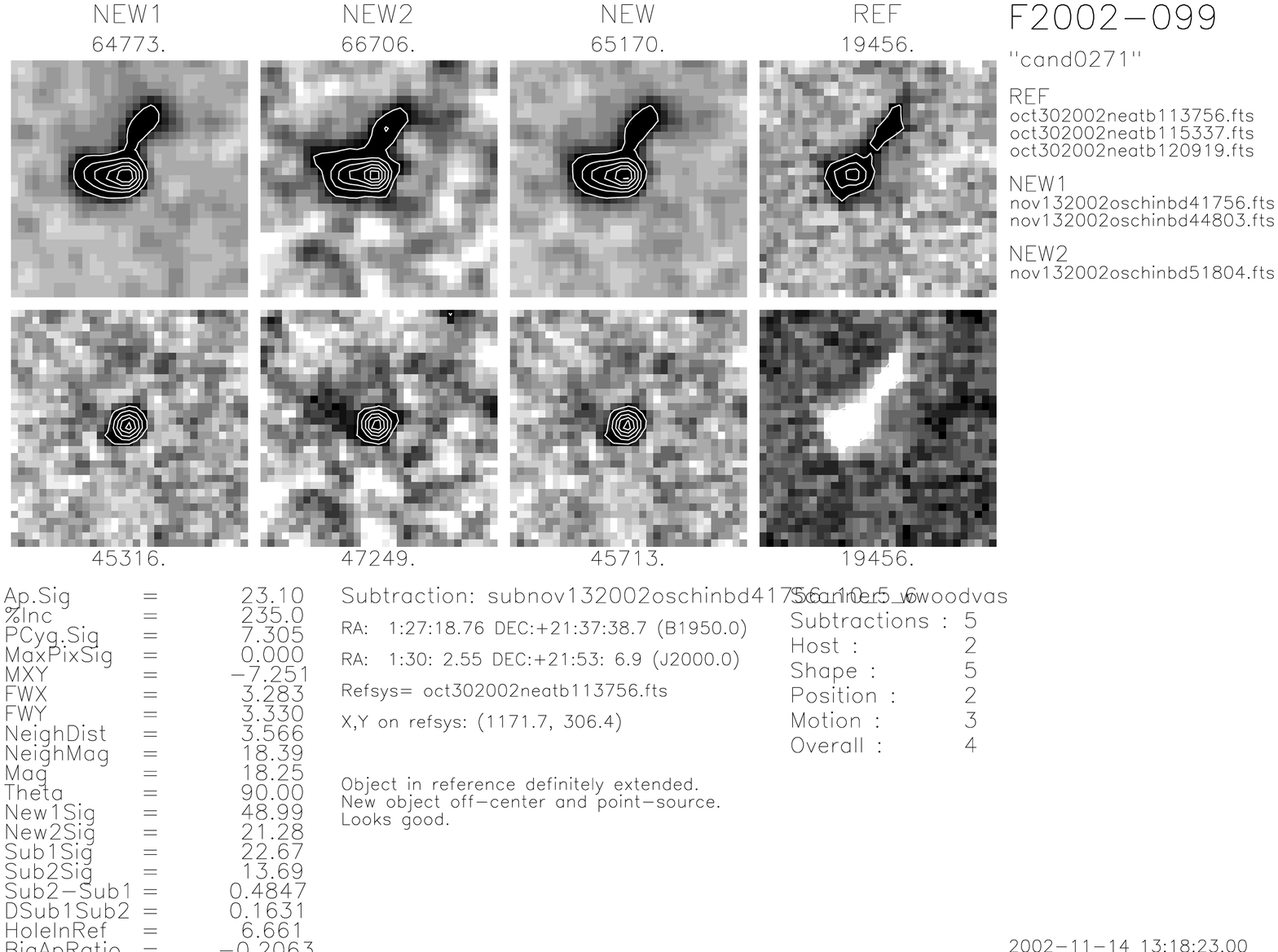}\label{fig:2002ic_discovery}}
\hspace{0.3in}
\subfigure[2002ic]{\includegraphics[height=2in]{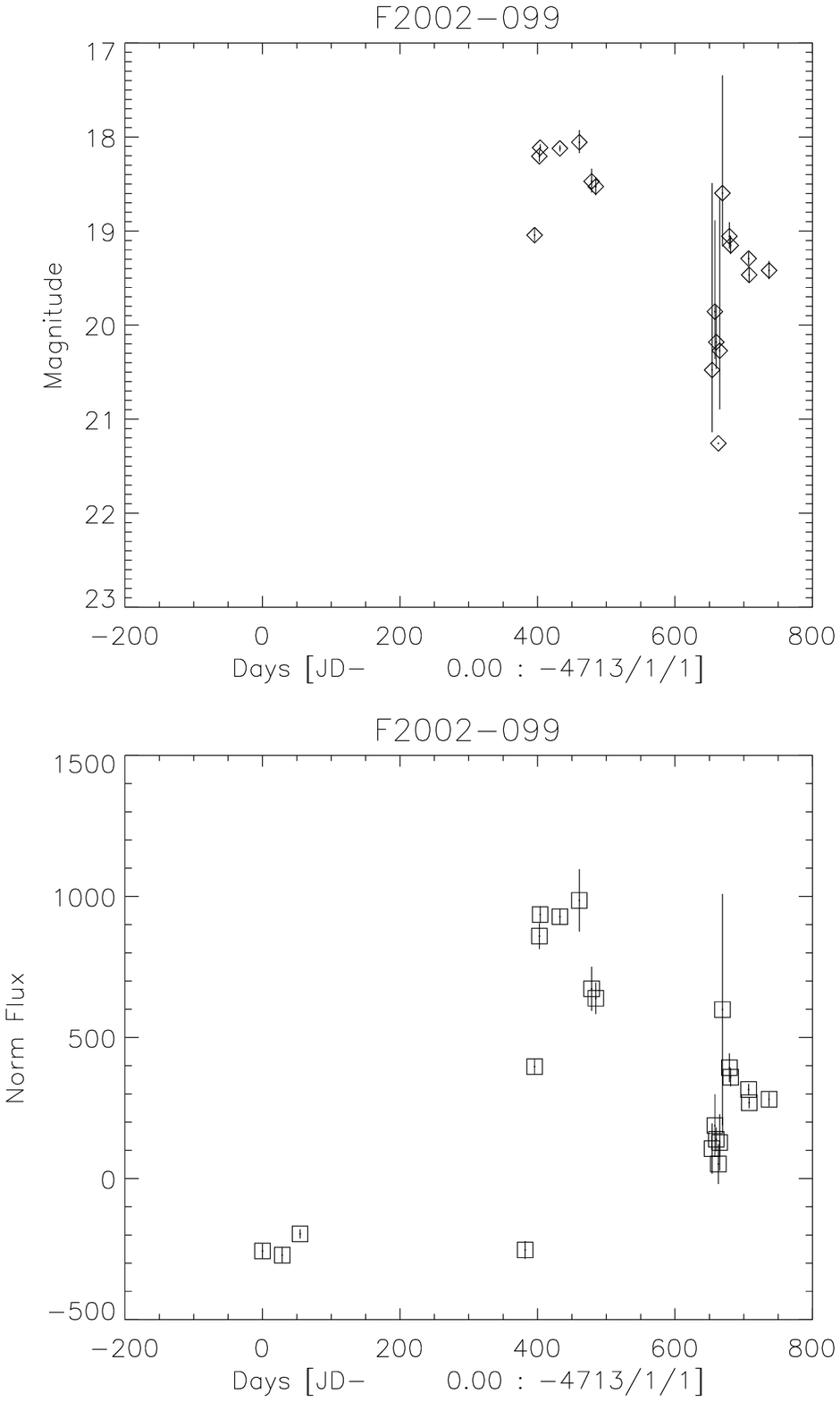}\label{fig:2002ic_lightcurve}}
\vspace{0.3in}
\subfigure[2002jf]{\includegraphics[angle=90,height=2in,width=3in]{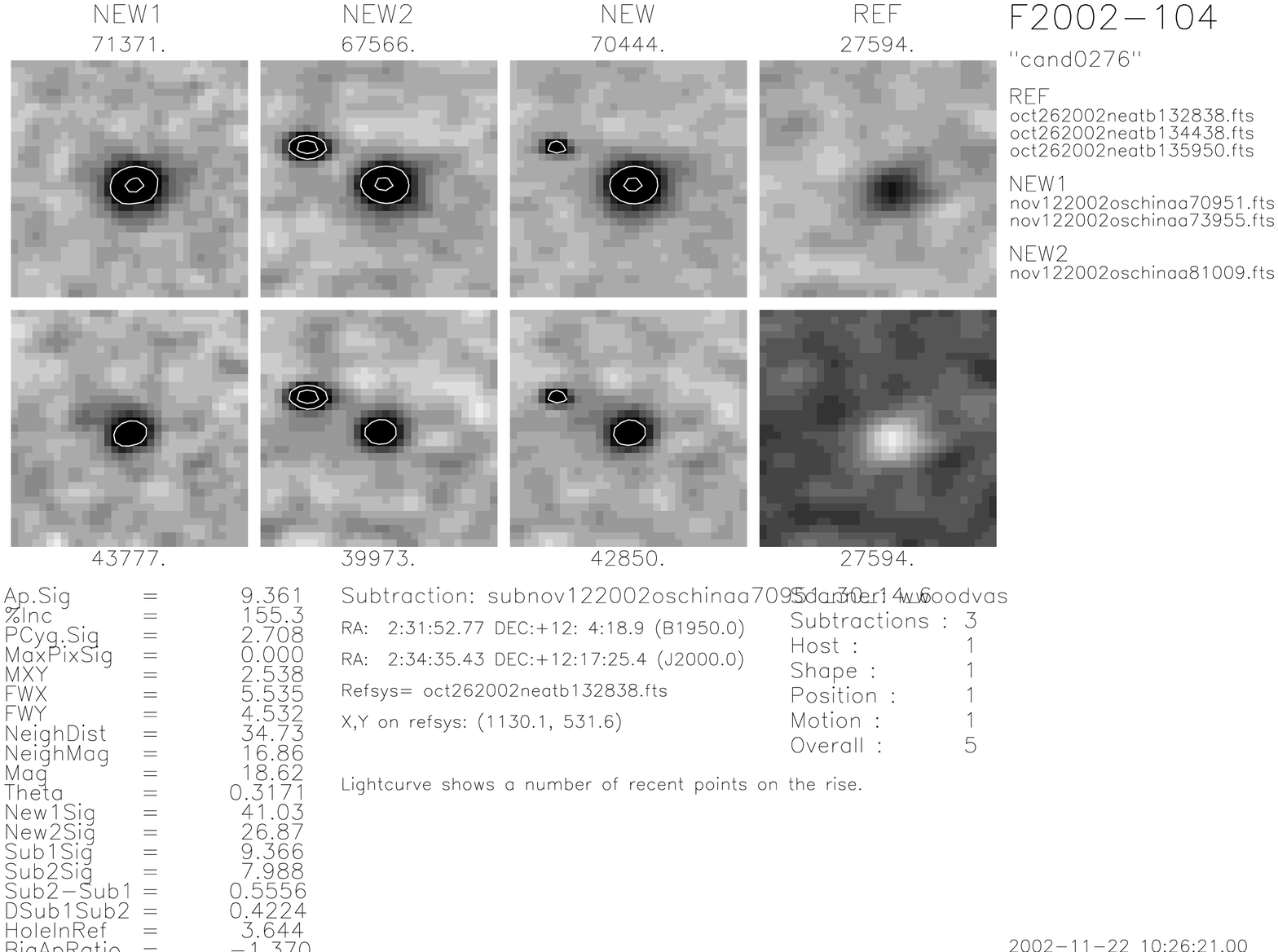}\label{fig:2002jf_discovery}}
\hspace{0.3in}
\subfigure[2002jf]{\includegraphics[height=2in]{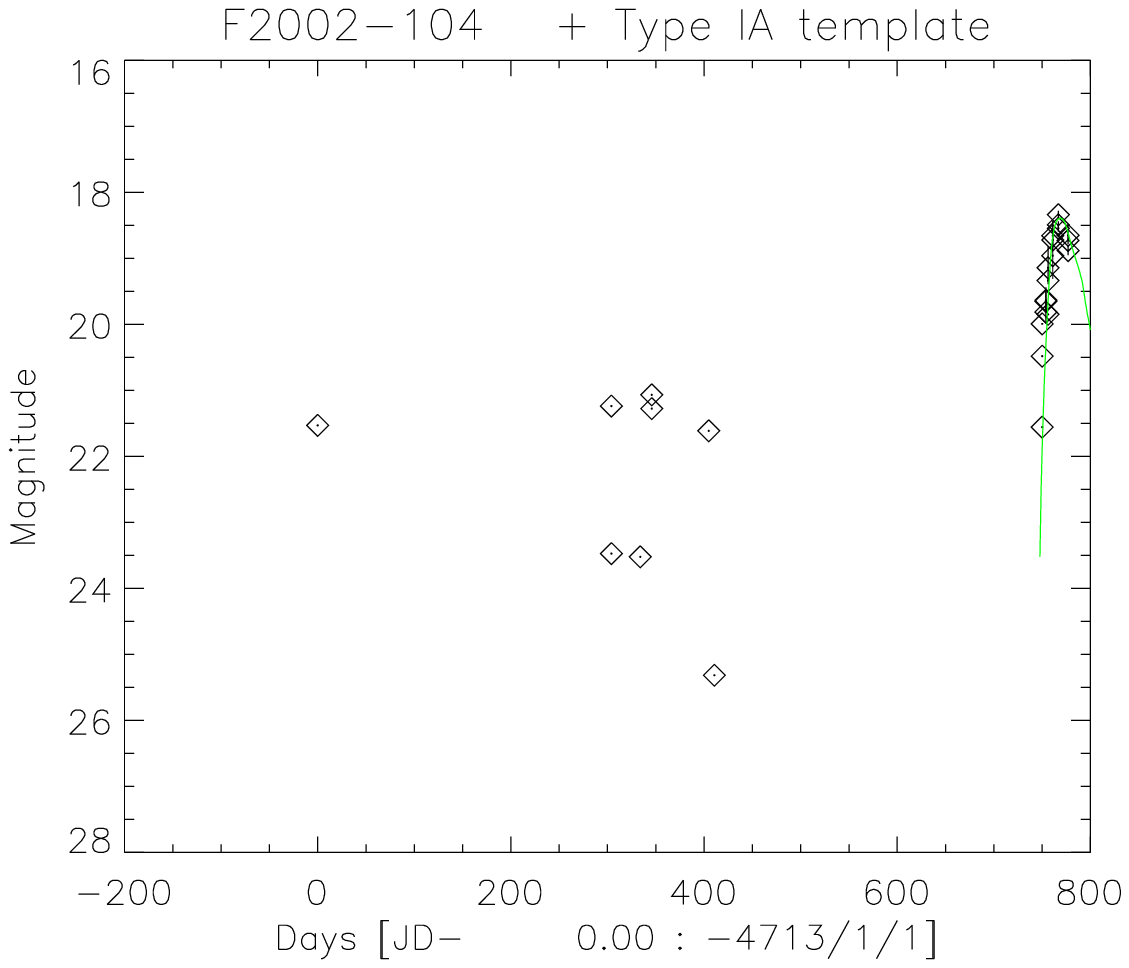}\label{fig:2002jf_lightcurve}}
\vspace{0.3in}
\end{figure}

\clearpage\pagebreak
\begin{figure}
\subfigure[2002jh]{\includegraphics[angle=90,height=2in,width=3in]{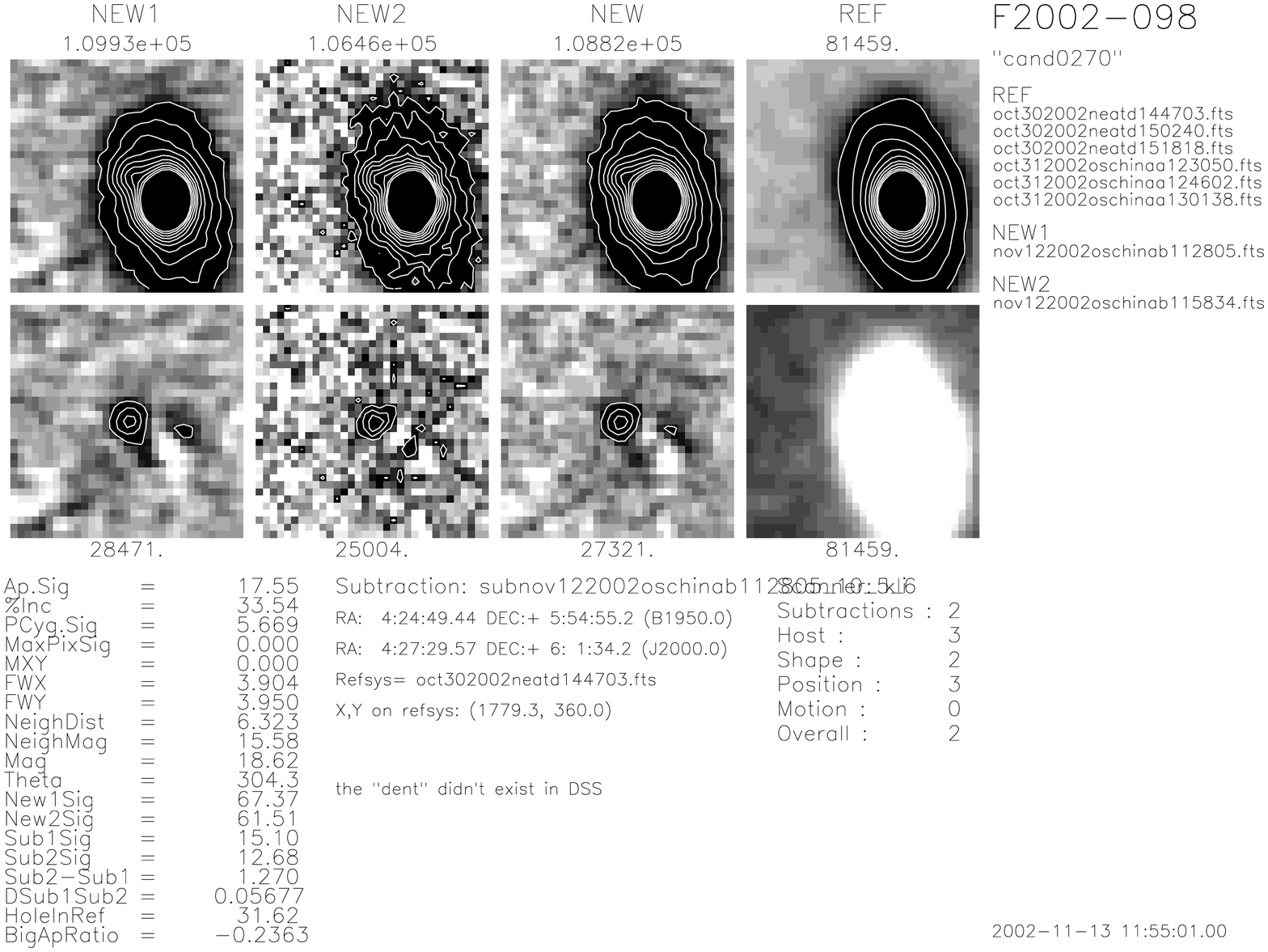}\label{fig:2002jh_discovery}}
\hspace{0.3in}
\subfigure[2002jh]{\includegraphics[height=2in]{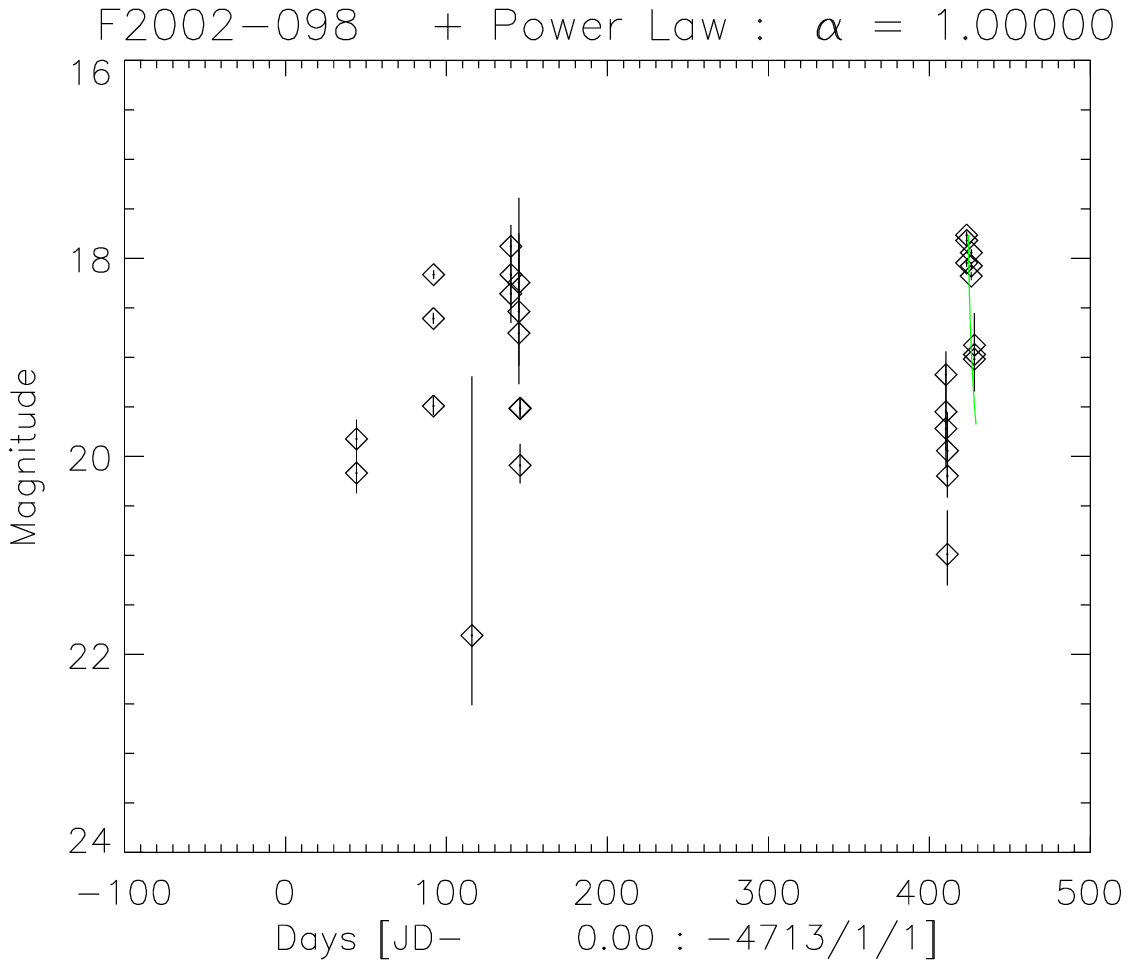}\label{fig:2002jh_lightcurve}}
\vspace{0.3in}
\subfigure[2002jk]{\includegraphics[angle=90,height=2in,width=3in]{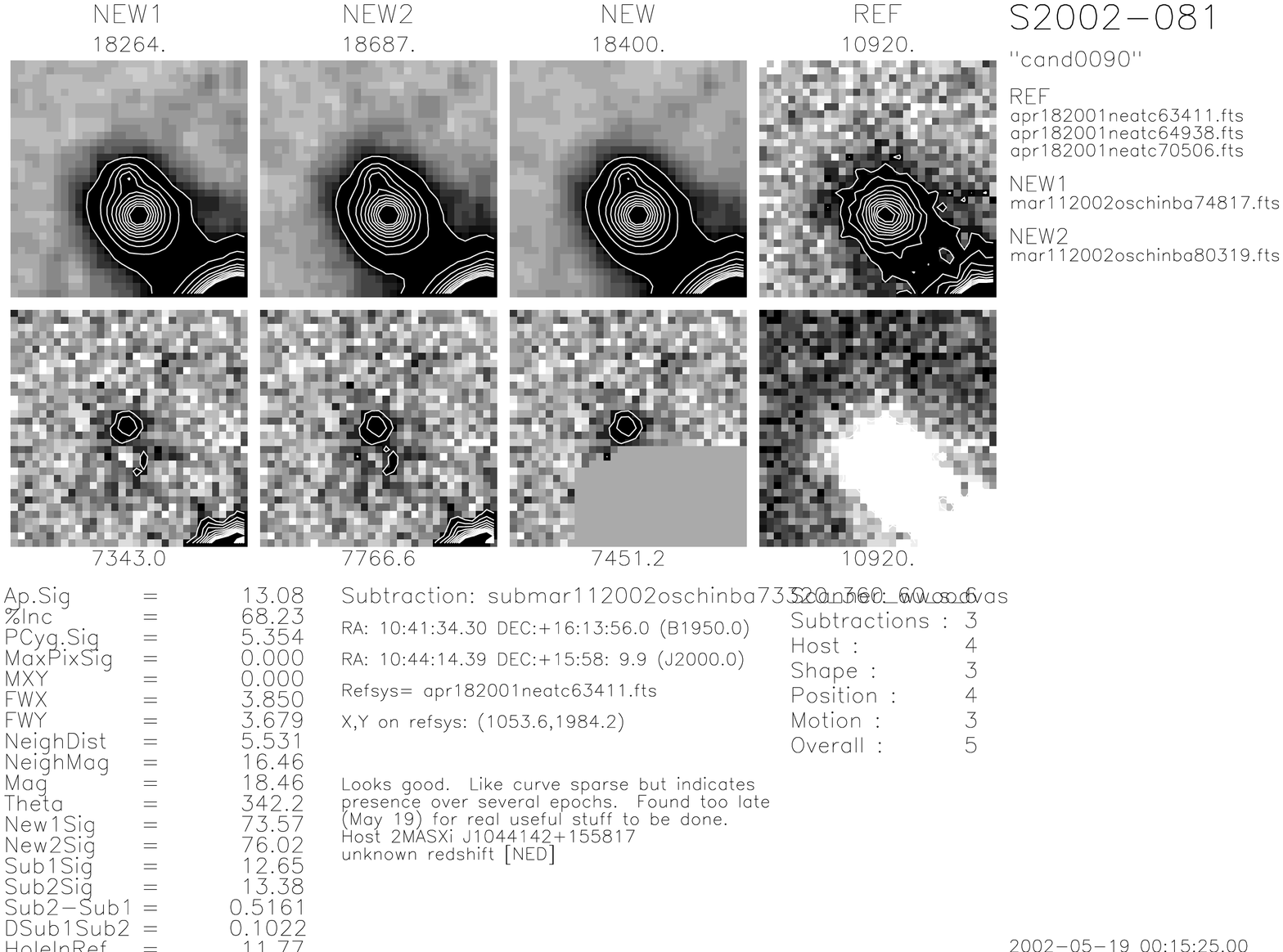}\label{fig:2002jk_discovery}}
\hspace{0.3in}
\subfigure[2002jk]{\includegraphics[height=2in]{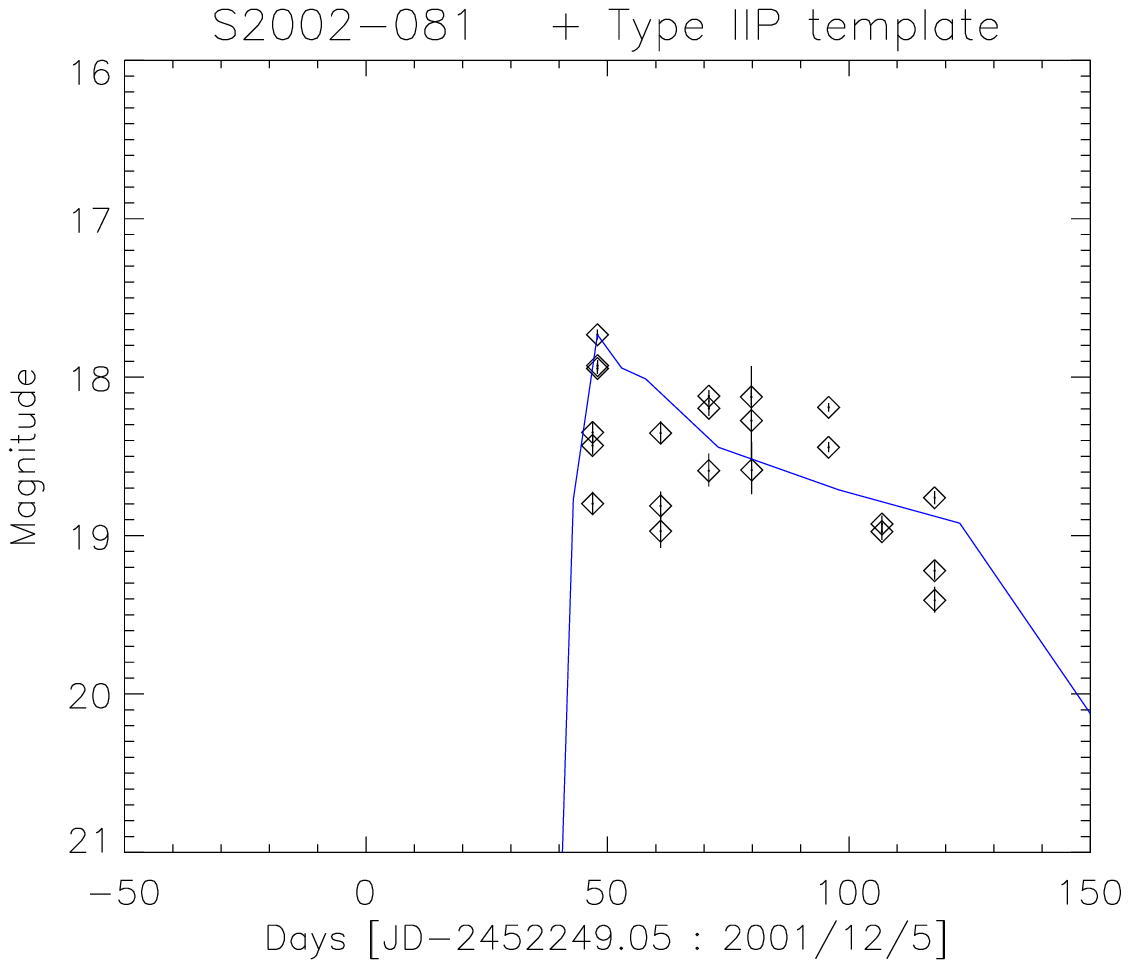}\label{fig:2002jk_lightcurve}}
\vspace{0.3in}
\subfigure[2002jl]{\includegraphics[angle=90,height=2in,width=3in]{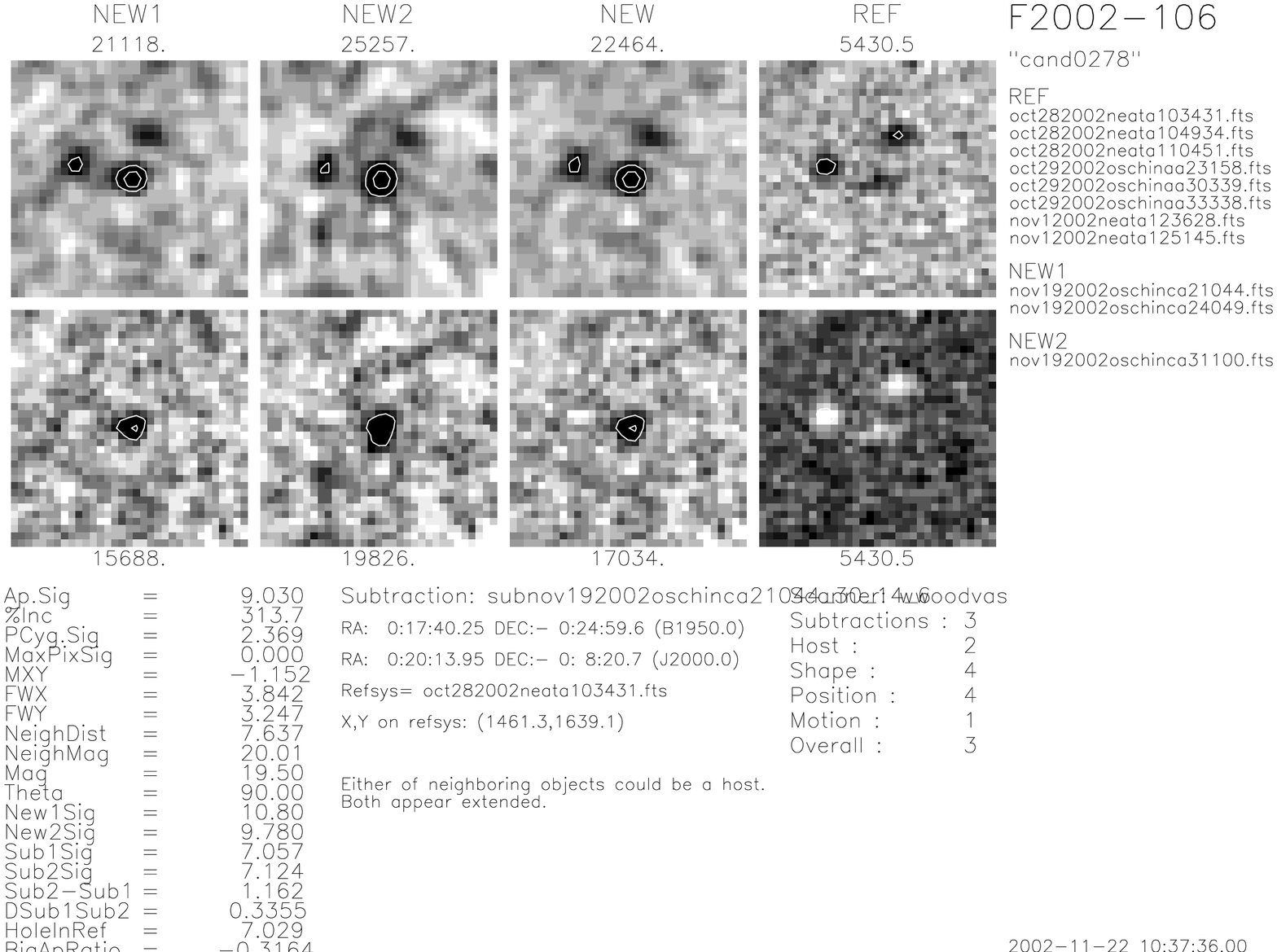}\label{fig:2002jl_discovery}}
\hspace{0.3in}
\subfigure[2002jl]{\includegraphics[height=2in]{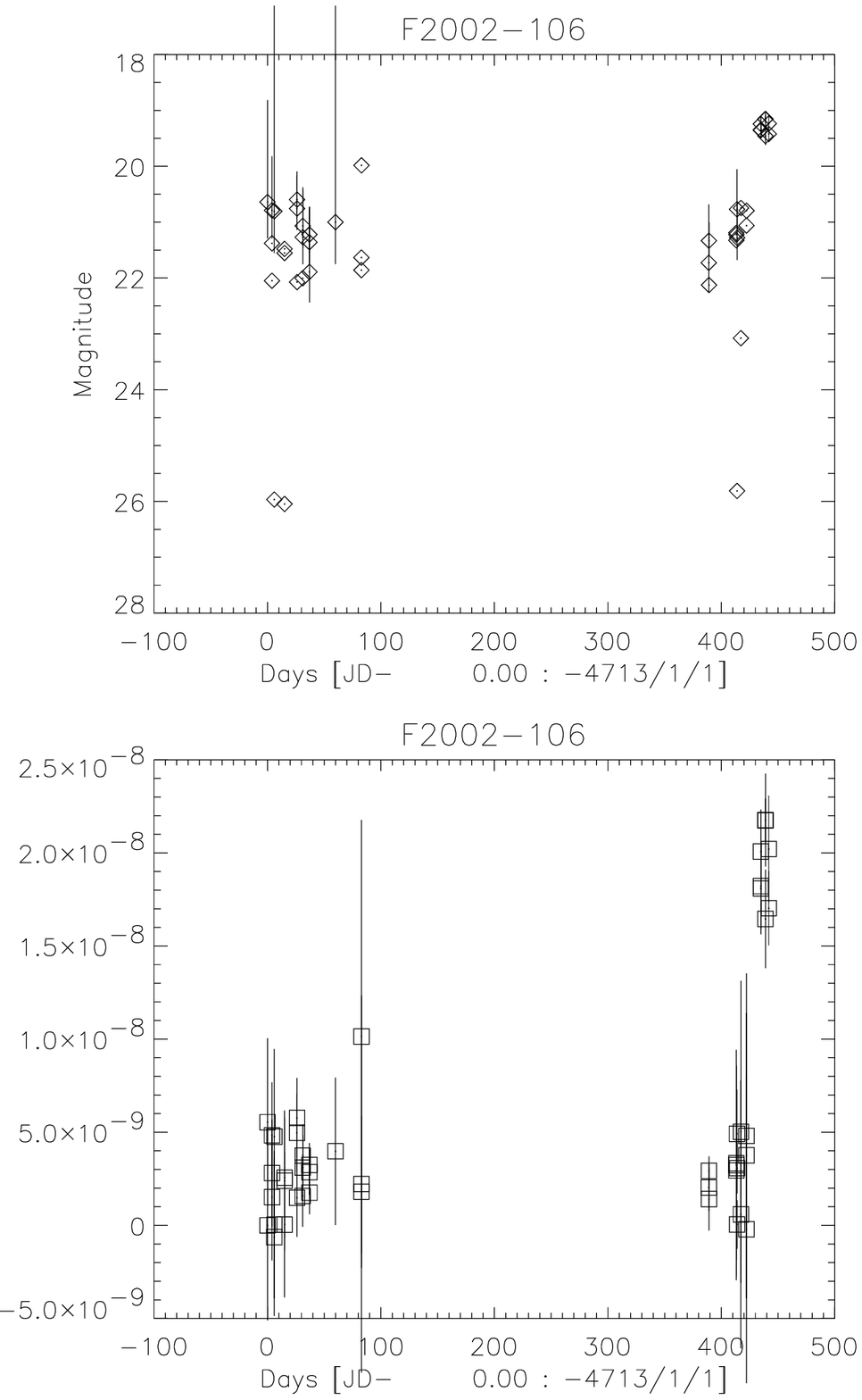}\label{fig:2002jl_lightcurve}}
\vspace{0.3in}
\end{figure}

\clearpage\pagebreak
\begin{figure}
\subfigure[2002jp]{\includegraphics[angle=90,height=2in,width=3in]{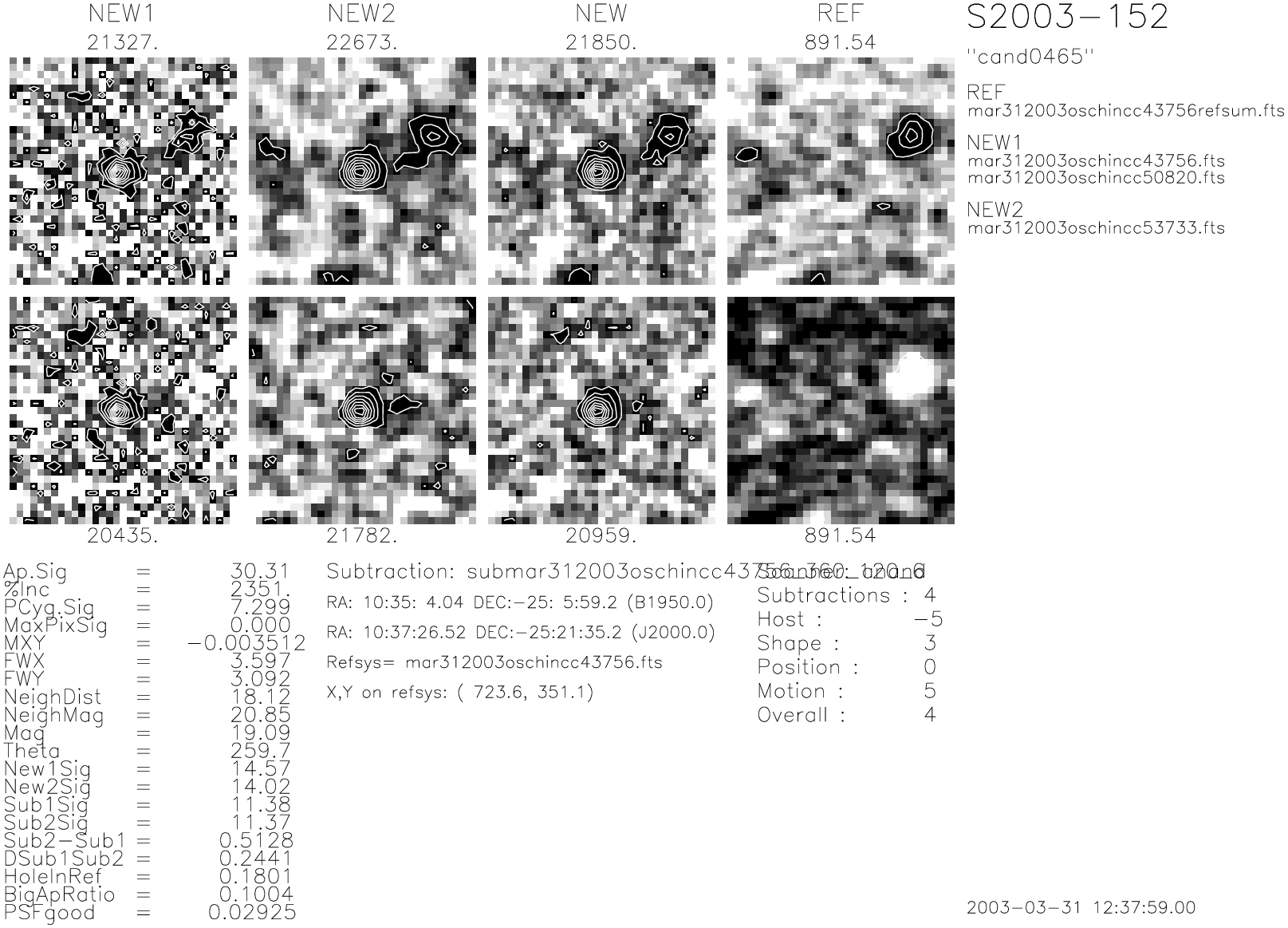}\label{fig:2002jp_discovery}}
\hspace{0.3in}
\subfigure[2002jp]{\includegraphics[height=2in]{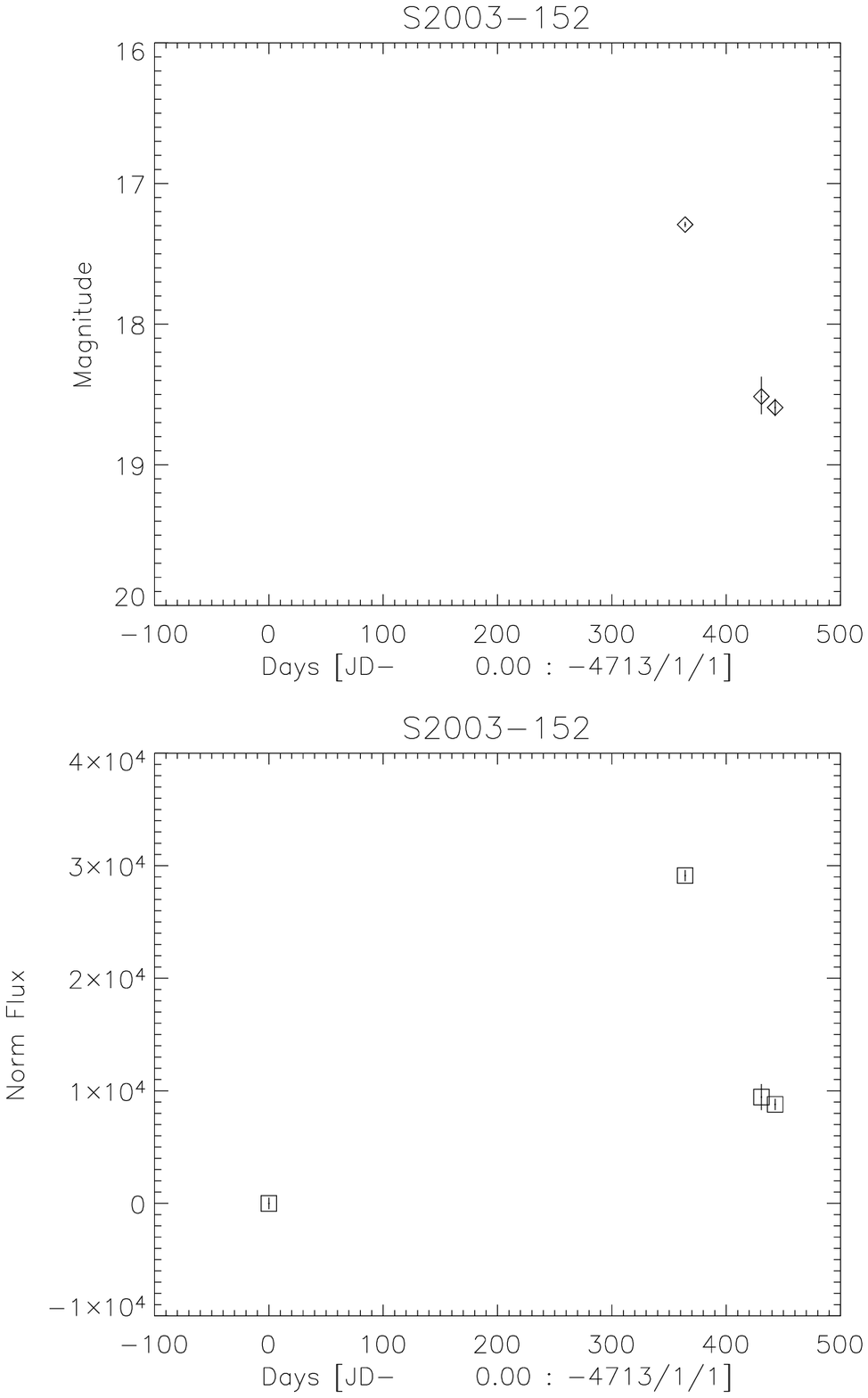}\label{fig:2002jp_lightcurve}}
\vspace{0.3in}
\subfigure[2002kj]{\includegraphics[angle=90,height=2in,width=3in]{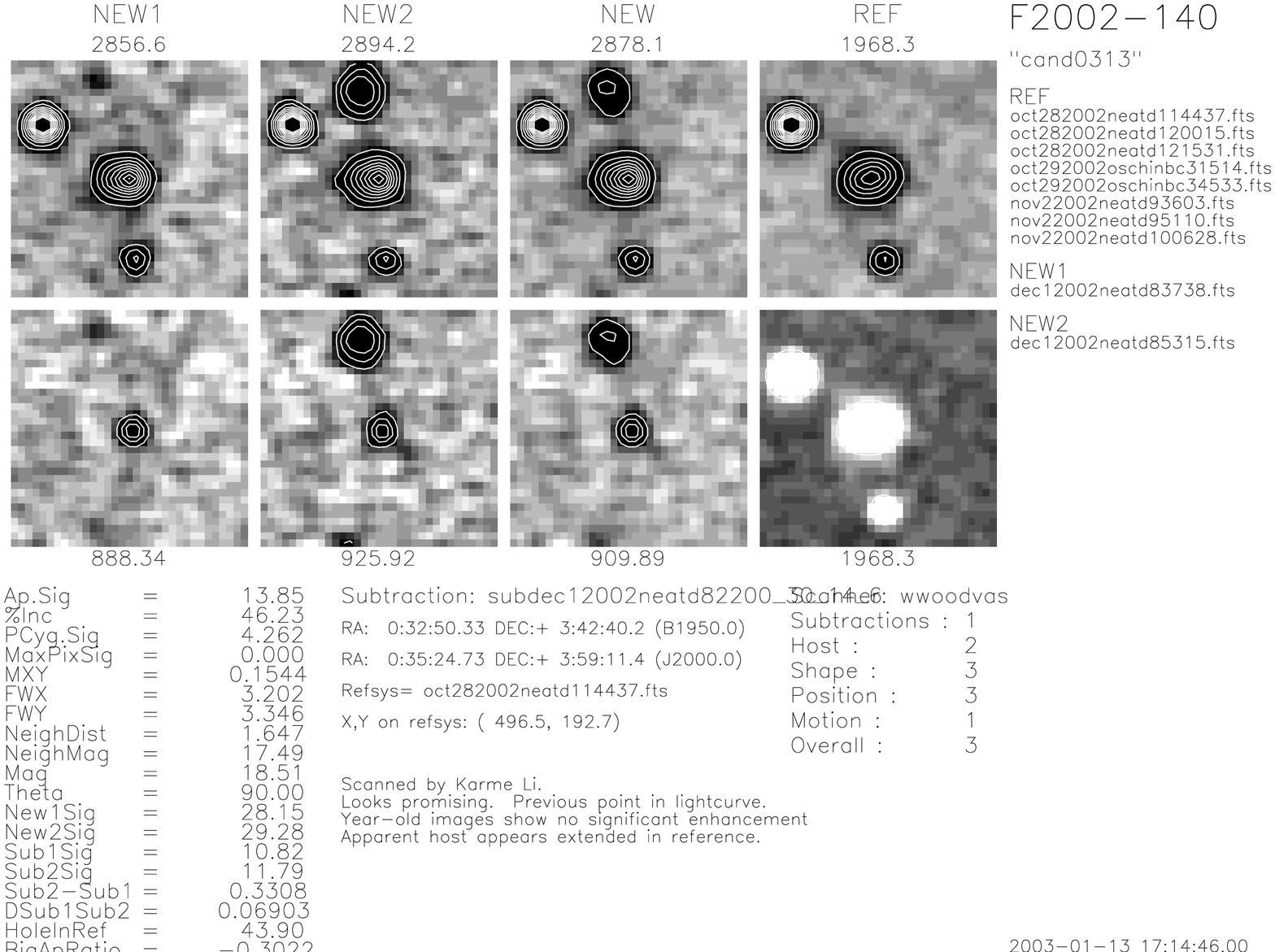}\label{fig:2002kj_discovery}}
\hspace{0.3in}
\subfigure[2002kj]{\includegraphics[height=2in]{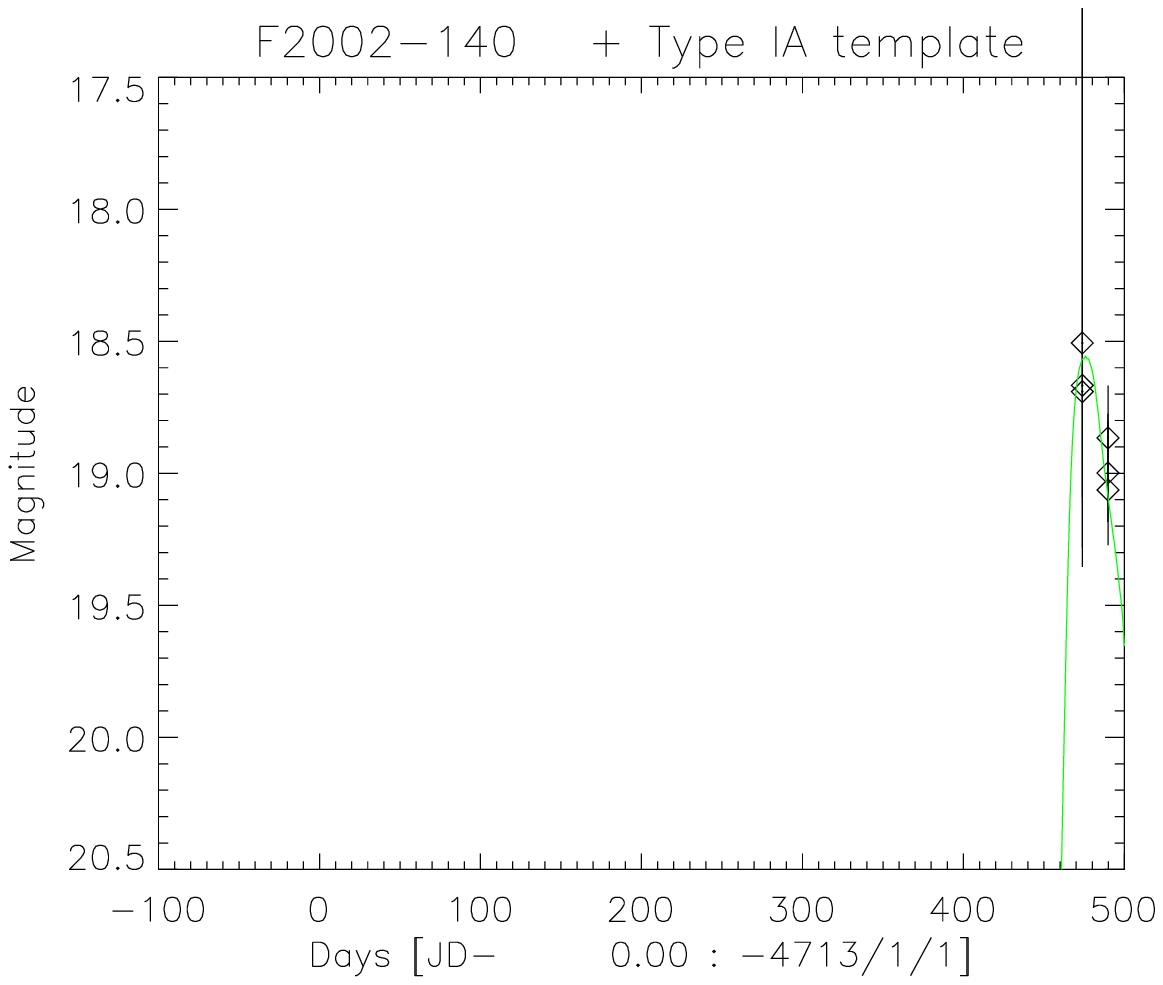}\label{fig:2002kj_lightcurve}}
\vspace{0.3in}
\subfigure[2002kk]{\includegraphics[angle=90,height=2in,width=3in]{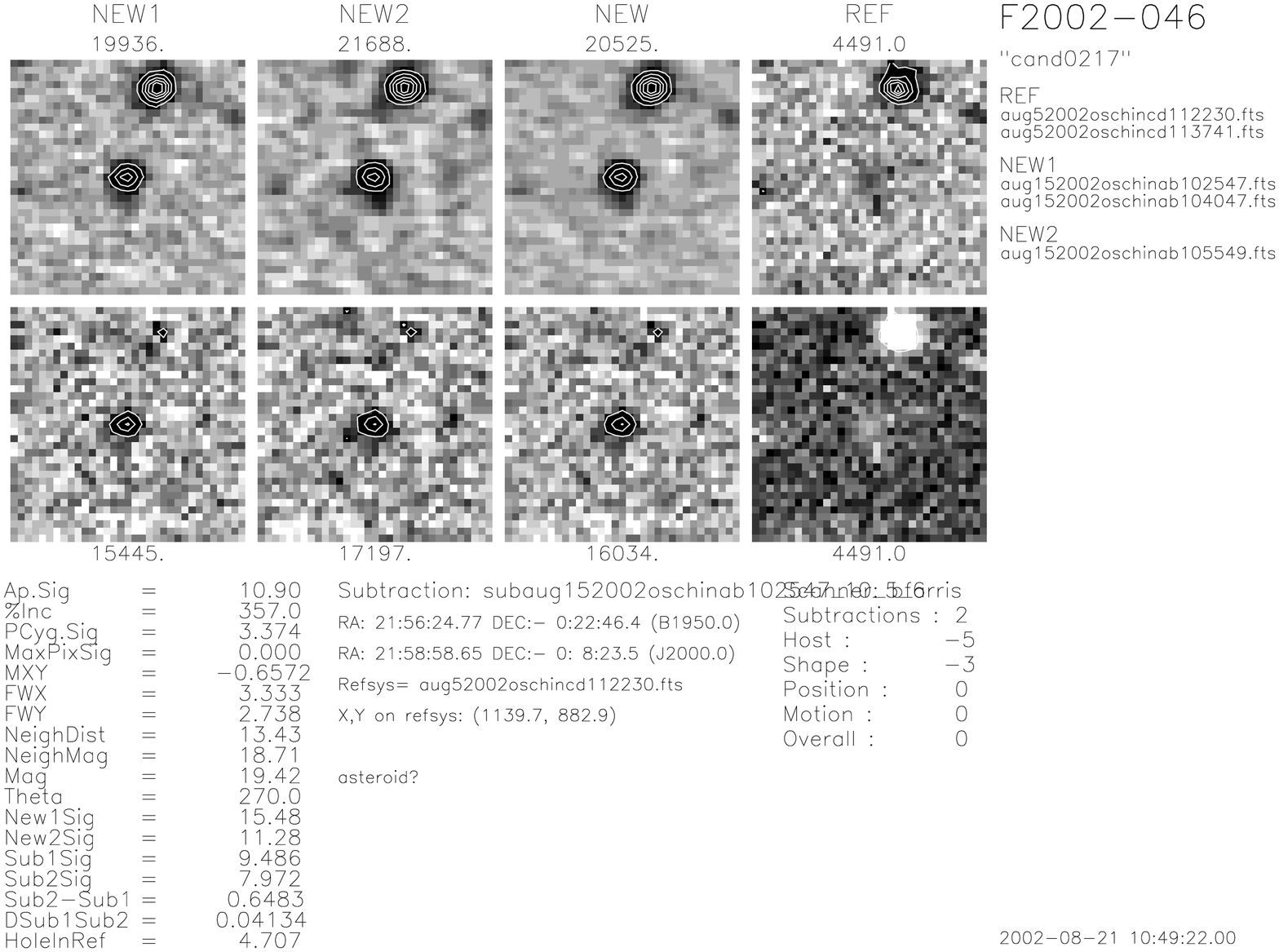}\label{fig:2002kk_discovery}}
\hspace{0.3in}
\subfigure[2002kk]{\includegraphics[height=2in]{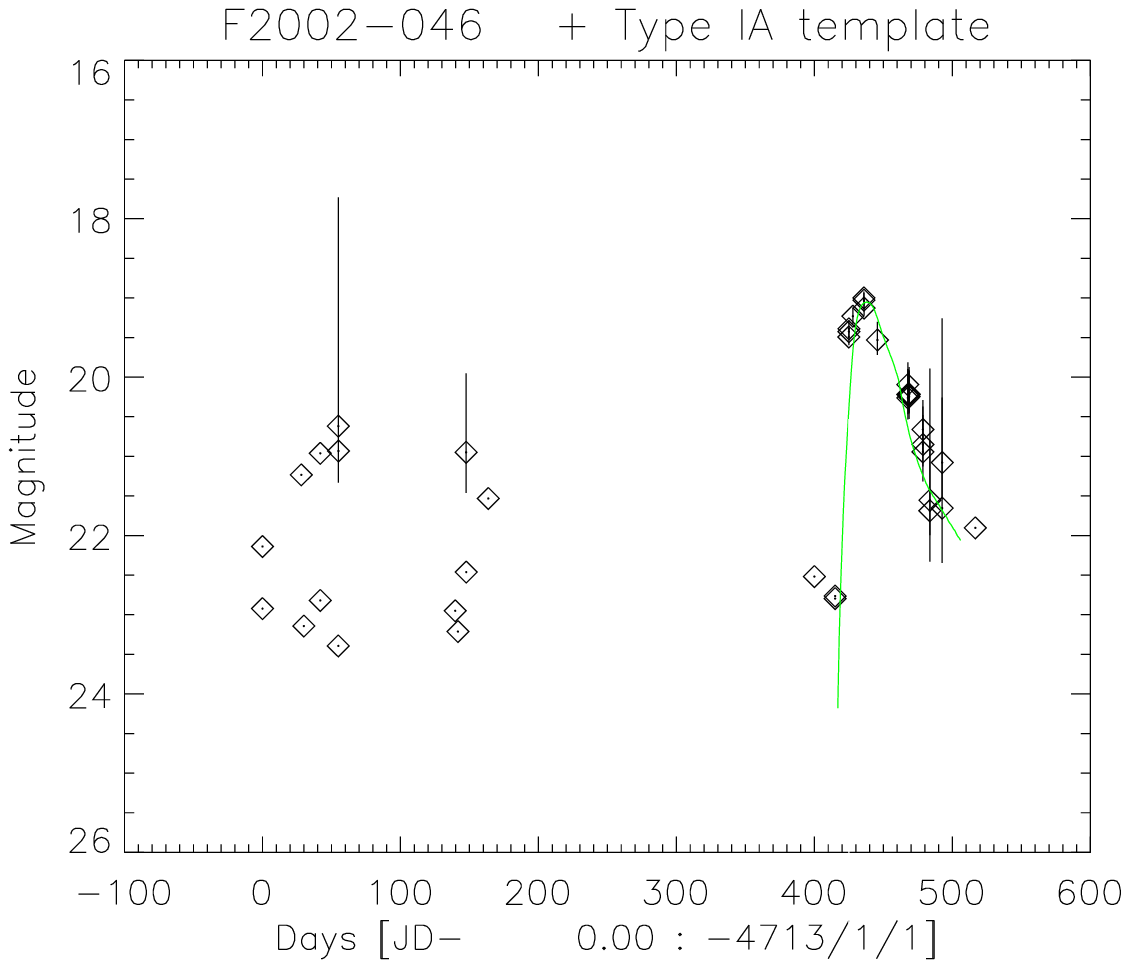}\label{fig:2002kk_lightcurve}}
\vspace{0.3in}
\end{figure}

\clearpage\pagebreak
\begin{figure}
\subfigure[2002le]{\includegraphics[angle=90,height=2in,width=3in]{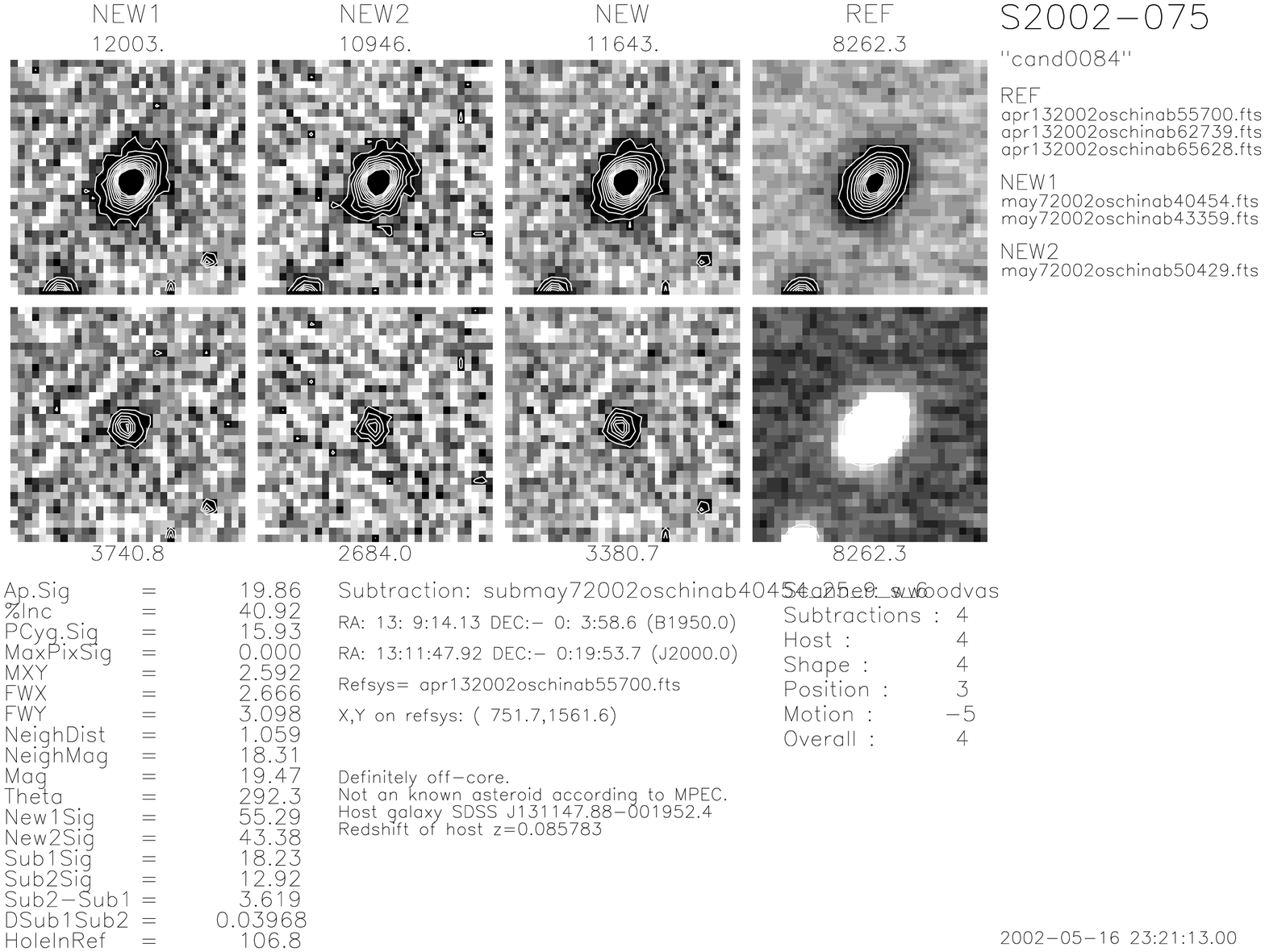}\label{fig:2002le_discovery}}
\hspace{0.3in}
\subfigure[2002le]{\includegraphics[height=2in]{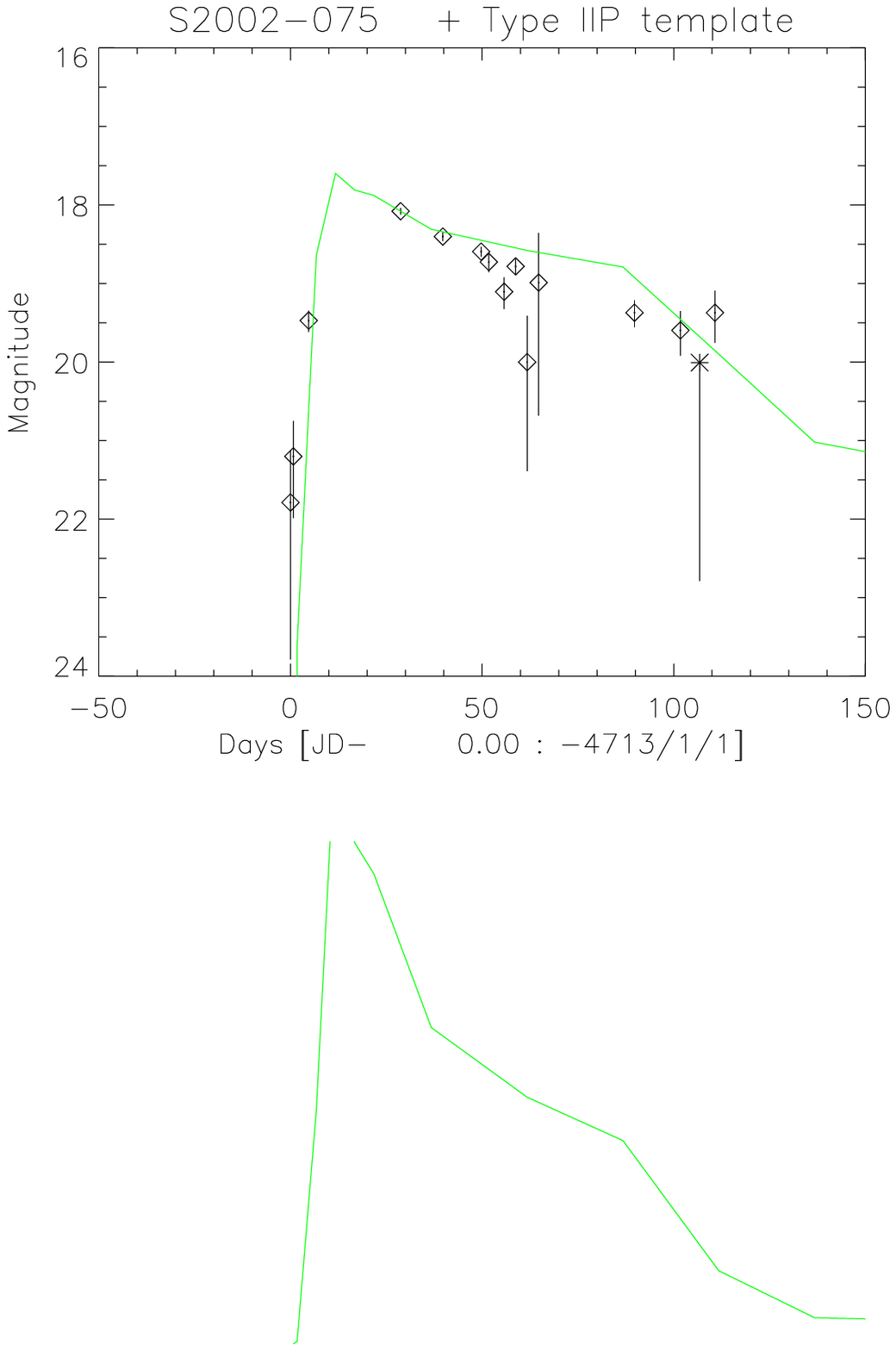}\label{fig:2002le_lightcurve}}
\vspace{0.3in}
\subfigure[2002lf]{\includegraphics[angle=90,height=2in,width=3in]{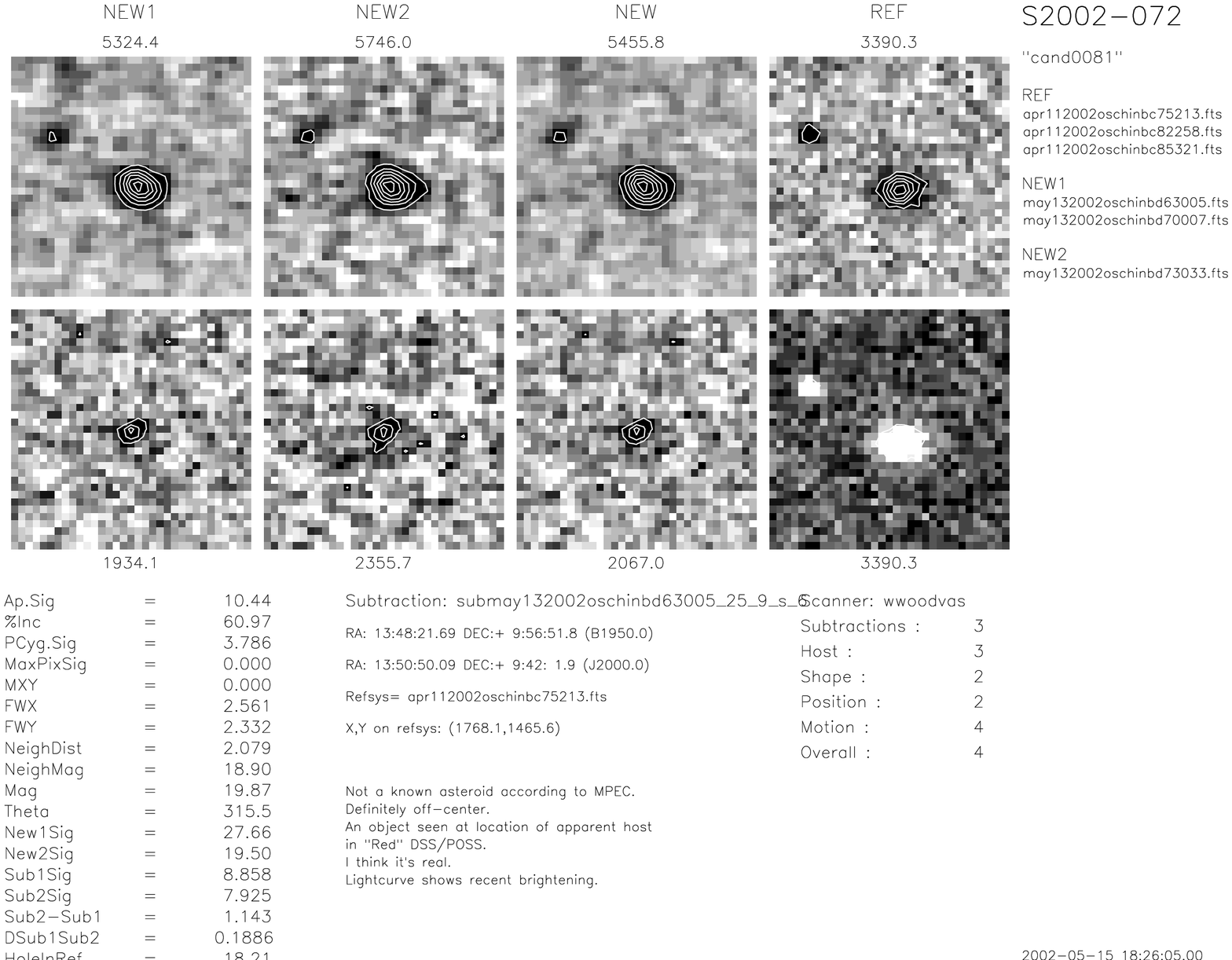}\label{fig:2002lf_discovery}}
\hspace{0.3in}
\subfigure[2002lf]{\includegraphics[height=2in]{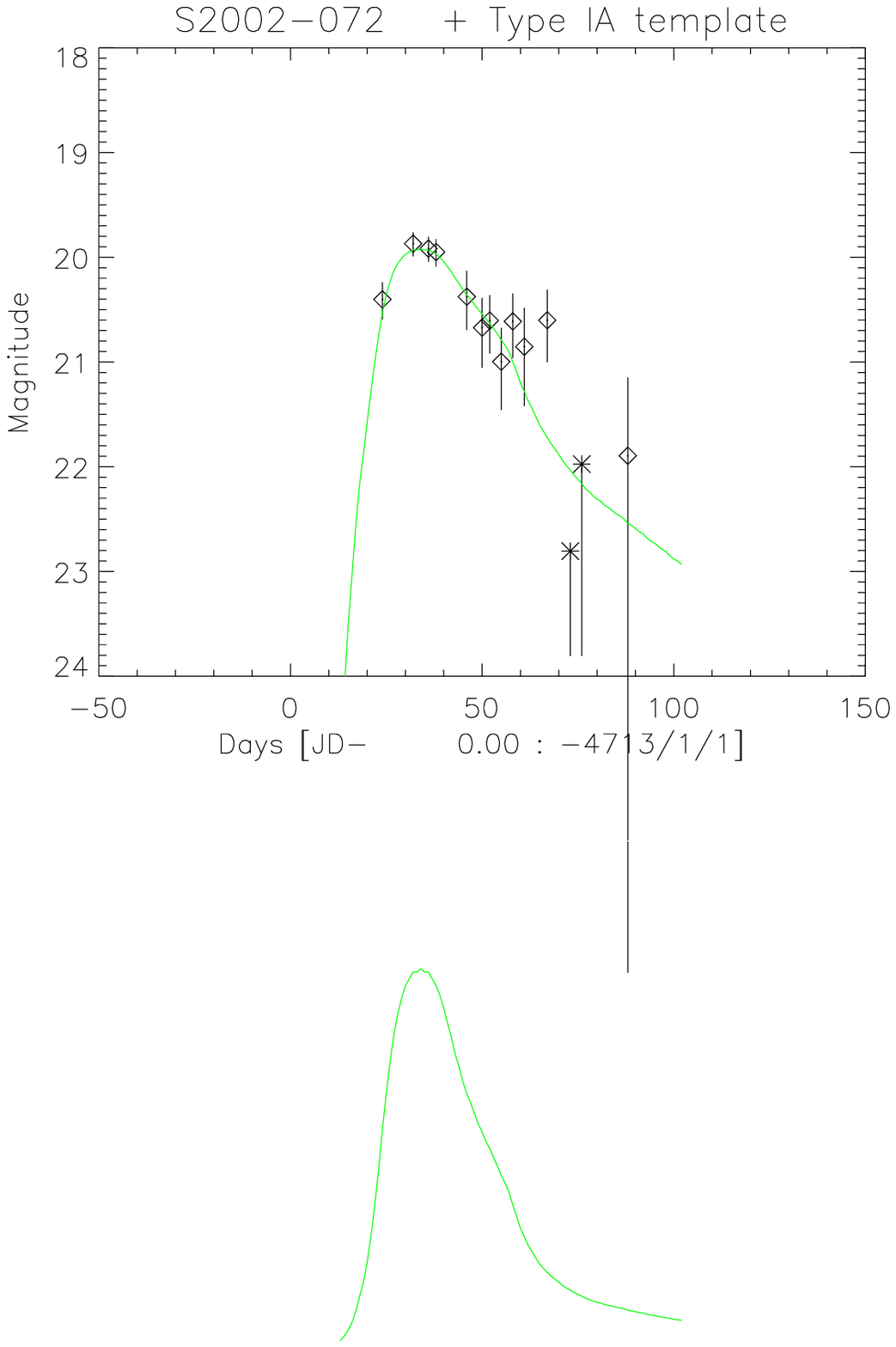}\label{fig:2002lf_lightcurve}}
\vspace{0.3in}
\subfigure[2003V]{\includegraphics[angle=90,height=2in,width=3in]{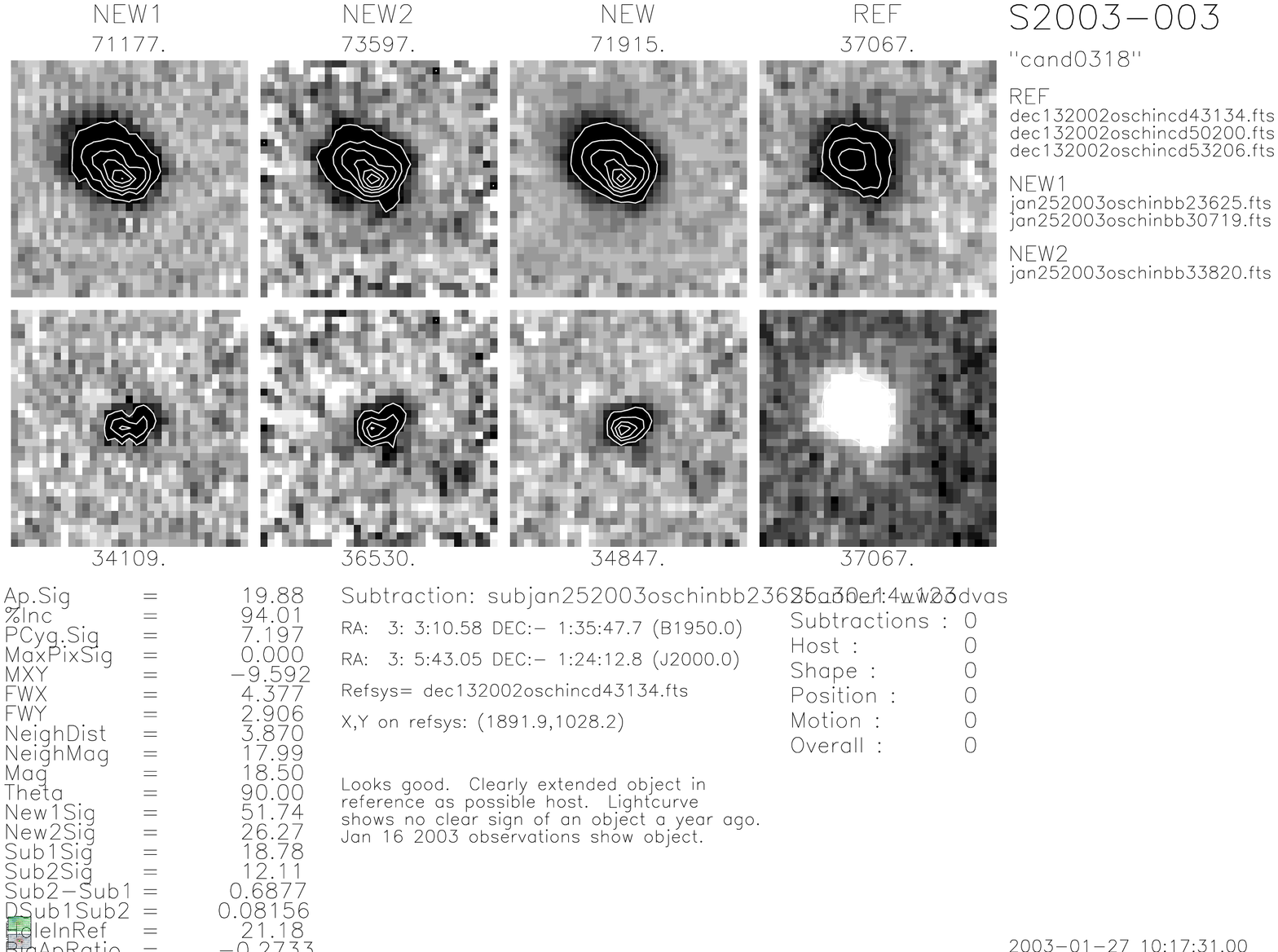}\label{fig:2003V_discovery}}
\hspace{0.3in}
\subfigure[2003V]{\includegraphics[height=2in]{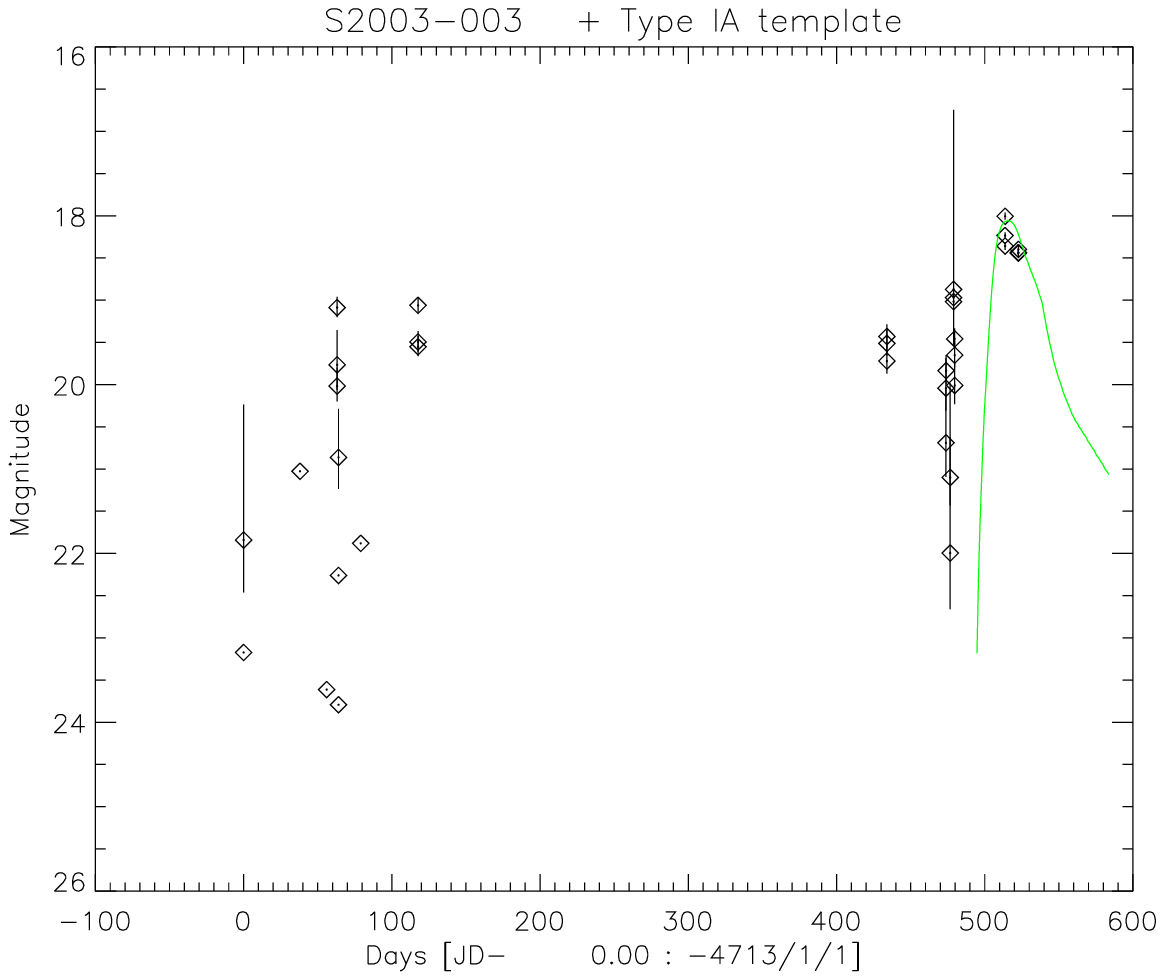}\label{fig:2003V_lightcurve}}
\vspace{0.3in}
\end{figure}

\clearpage\pagebreak
\begin{figure}
\subfigure[2003aa]{\includegraphics[angle=90,height=2in,width=3in]{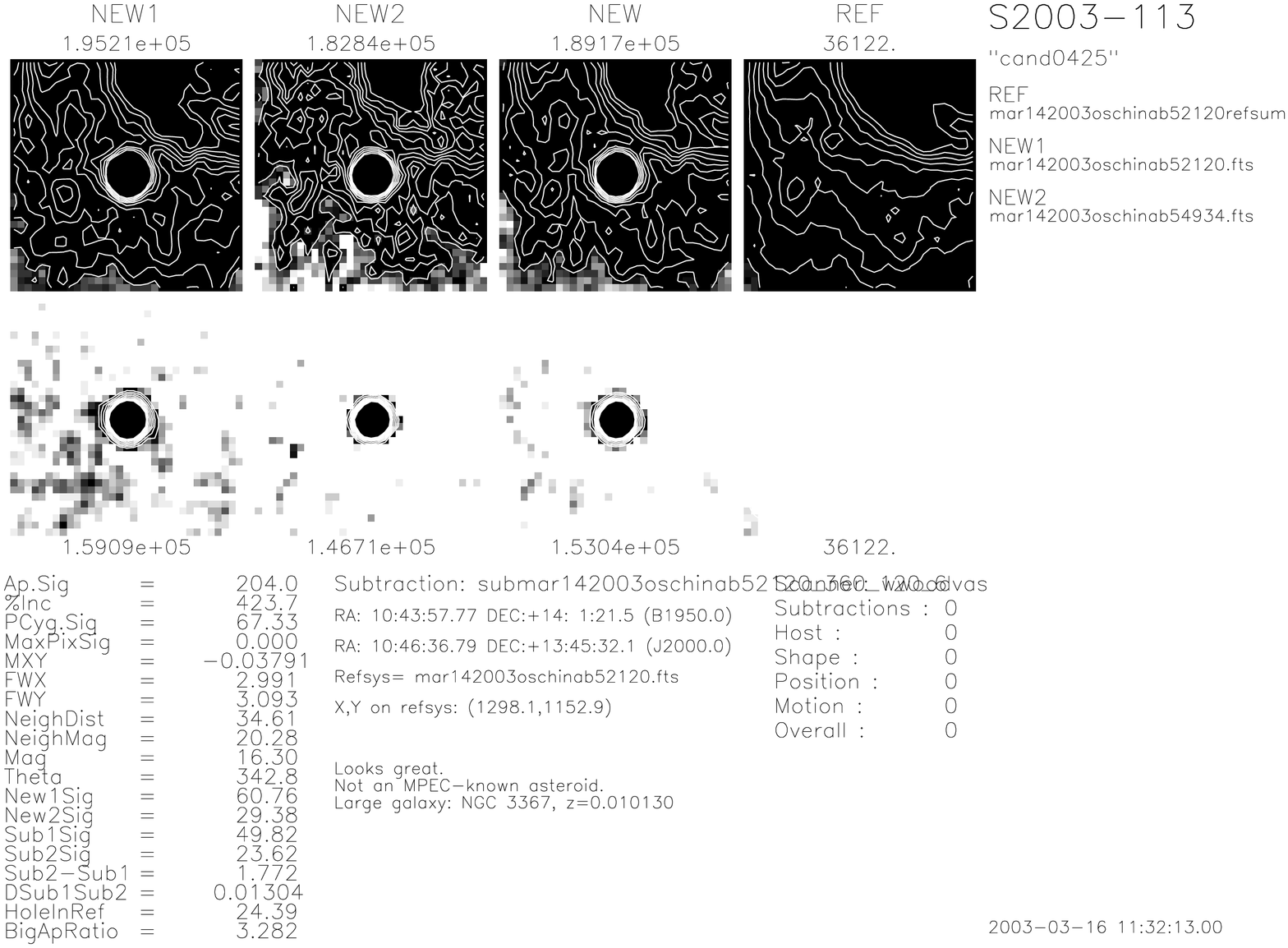}\label{fig:2003aa_discovery}}
\hspace{0.3in}
\subfigure[2003aa]{\includegraphics[height=2in]{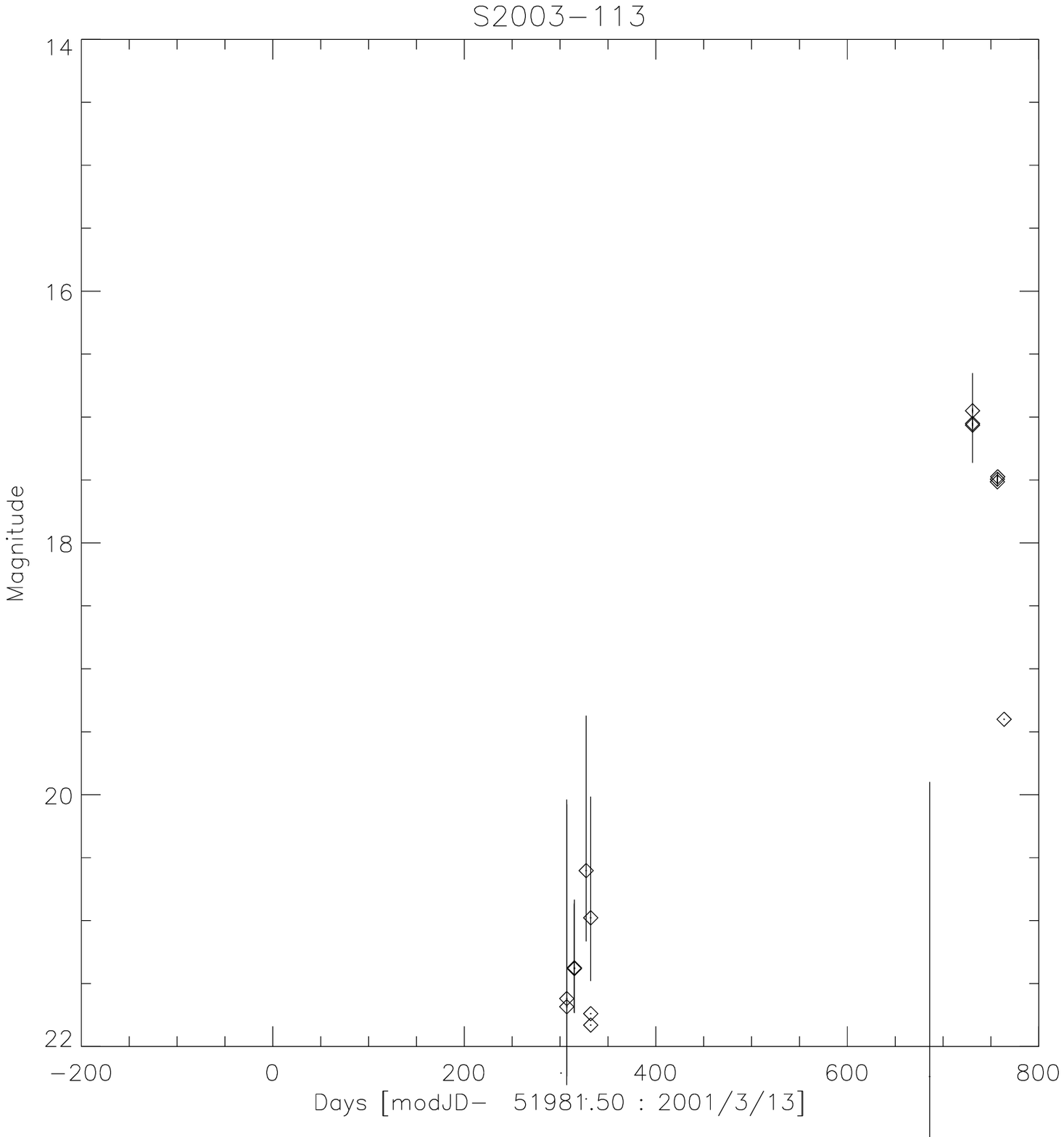}\label{fig:2003aa_lightcurve}}
\vspace{0.3in}
\subfigure[2003ab]{\includegraphics[angle=90,height=2in,width=3in]{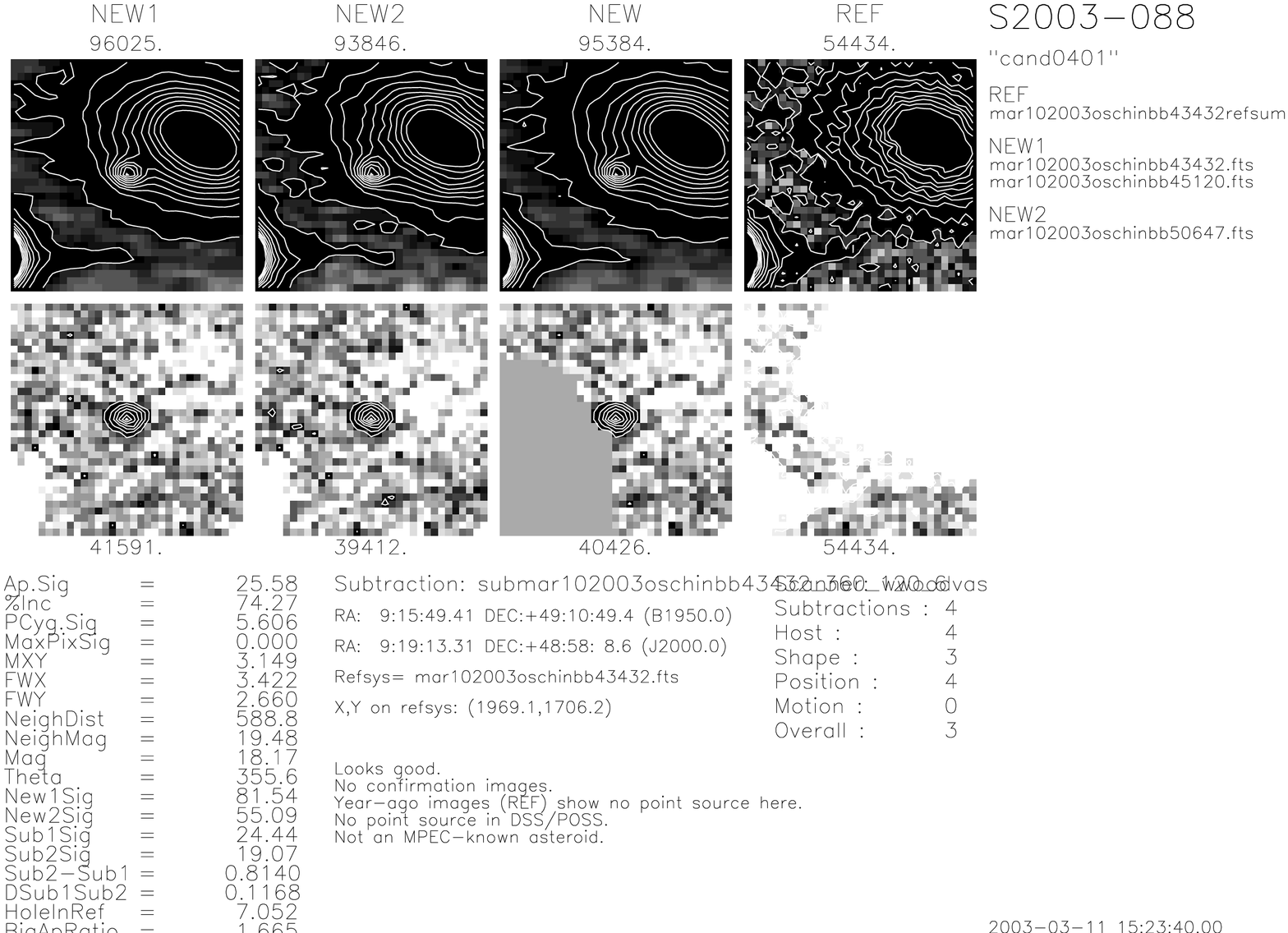}\label{fig:2003ab_discovery}}
\hspace{0.3in}
\subfigure[2003ab]{\includegraphics[height=2in]{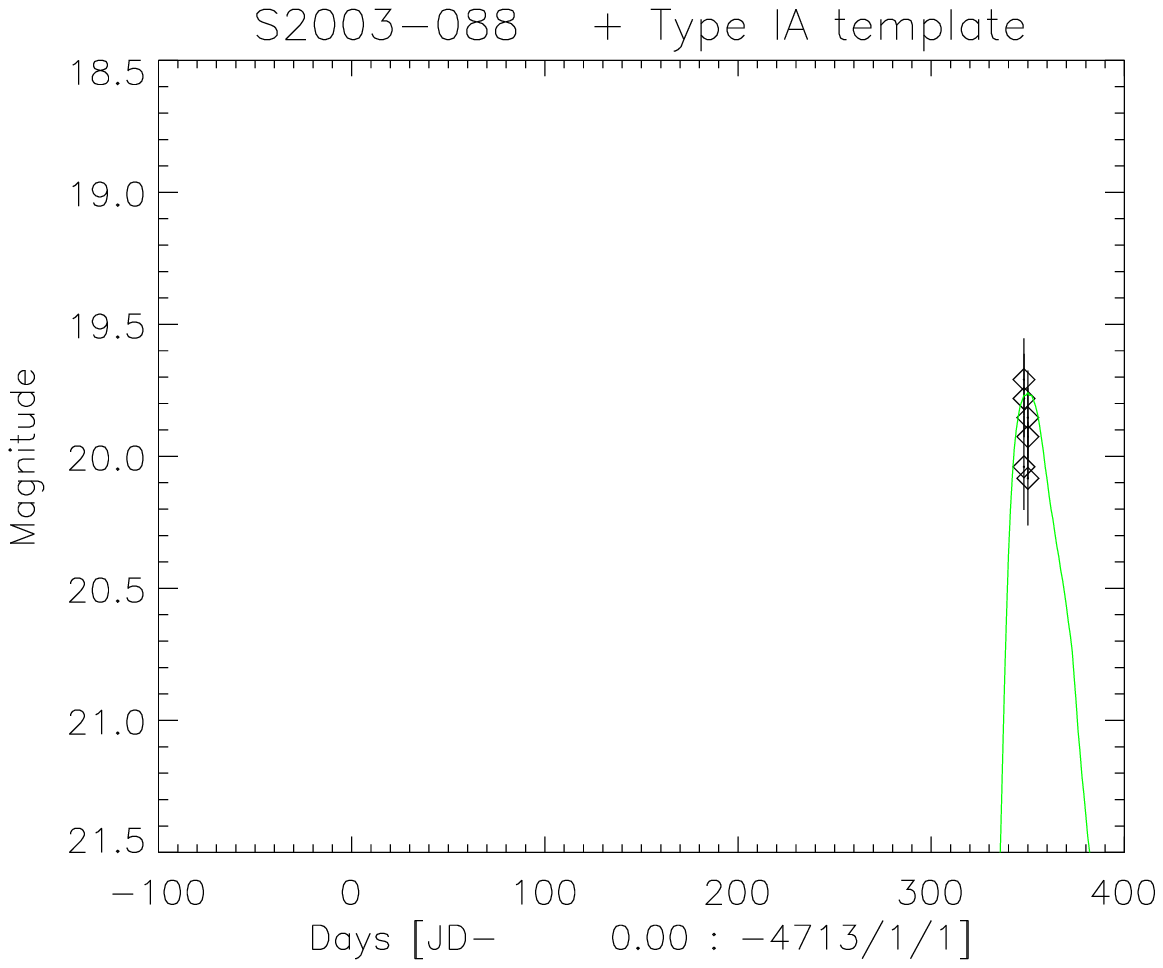}\label{fig:2003ab_lightcurve}}
\vspace{0.3in}
\subfigure[2003ae]{\includegraphics[angle=90,height=2in,width=3in]{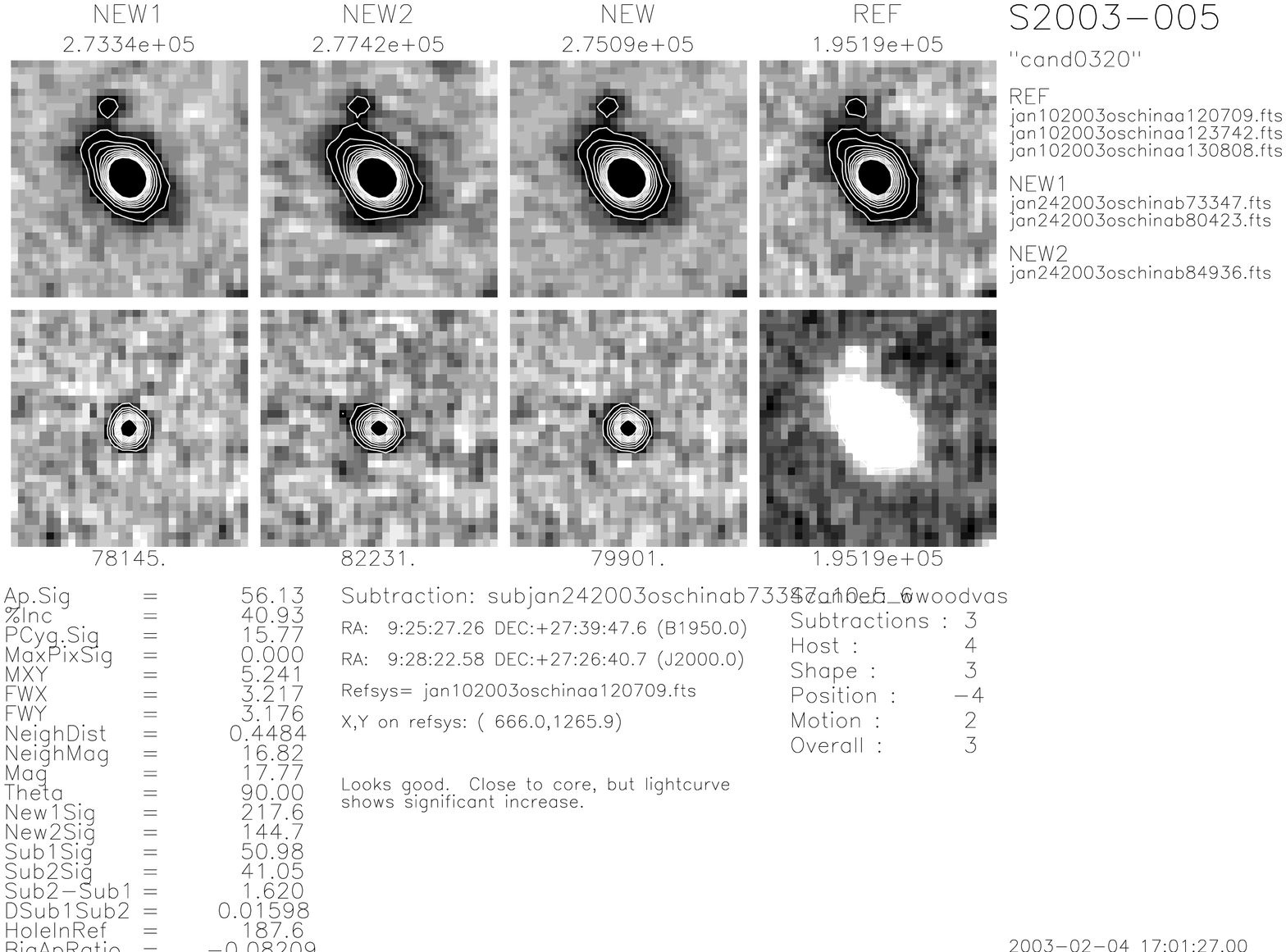}\label{fig:2003ae_discovery}}
\hspace{0.3in}
\subfigure[2003ae]{\includegraphics[height=2in]{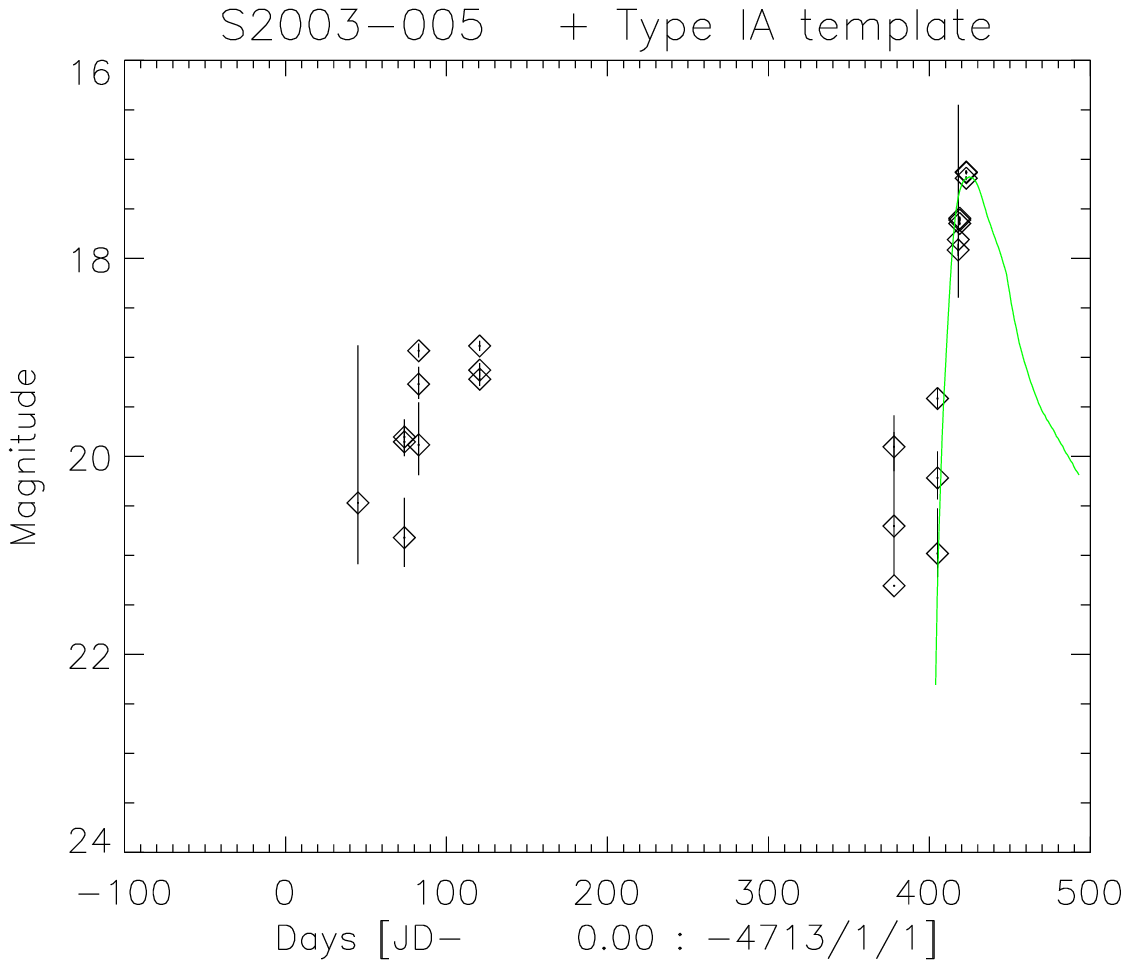}\label{fig:2003ae_lightcurve}}
\vspace{0.3in}
\end{figure}

\clearpage\pagebreak
\begin{figure}
\subfigure[2003af]{\includegraphics[angle=90,height=2in,width=3in]{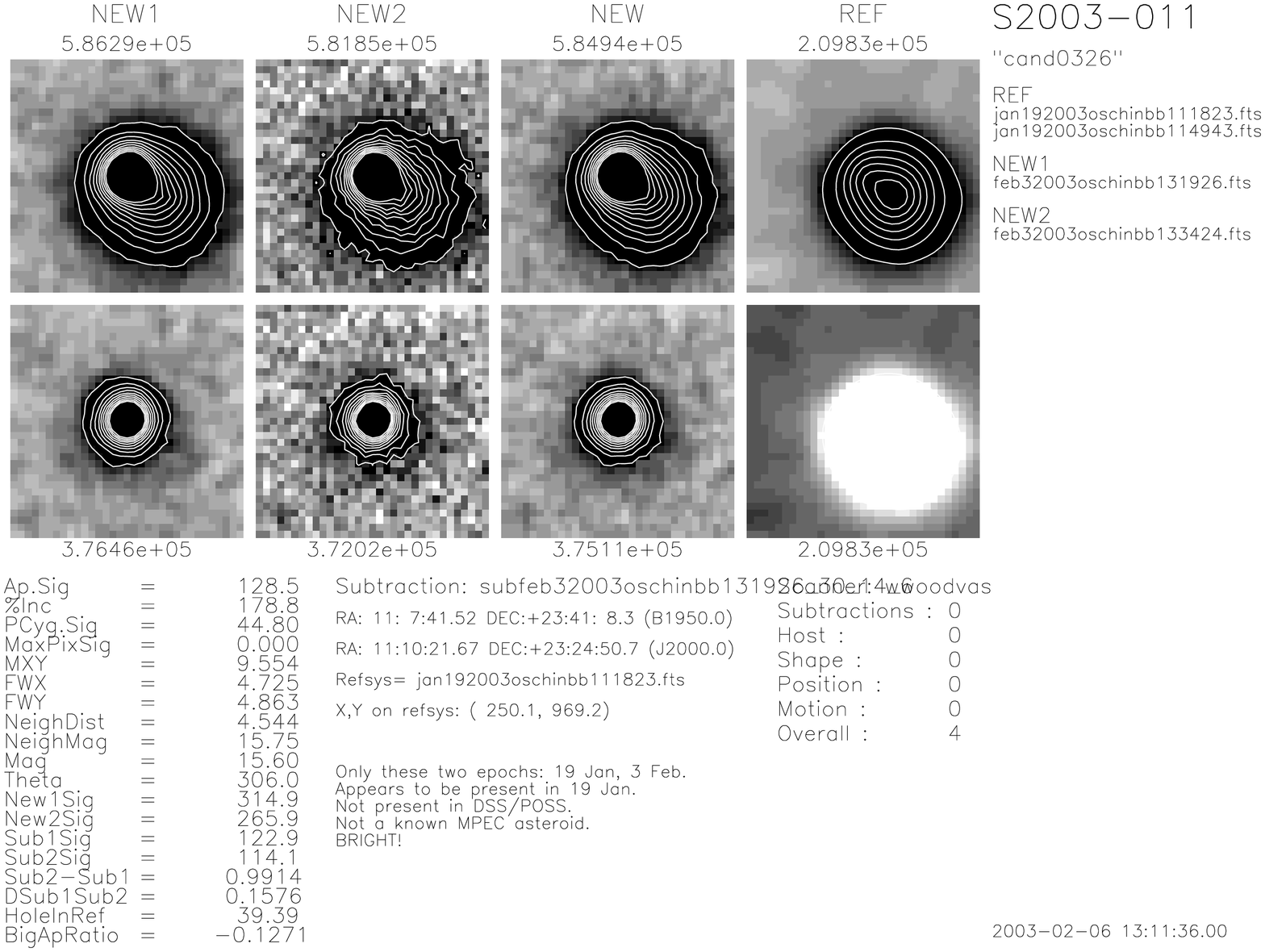}\label{fig:2003af_discovery}}
\hspace{0.3in}
\subfigure[2003af]{\includegraphics[height=2in]{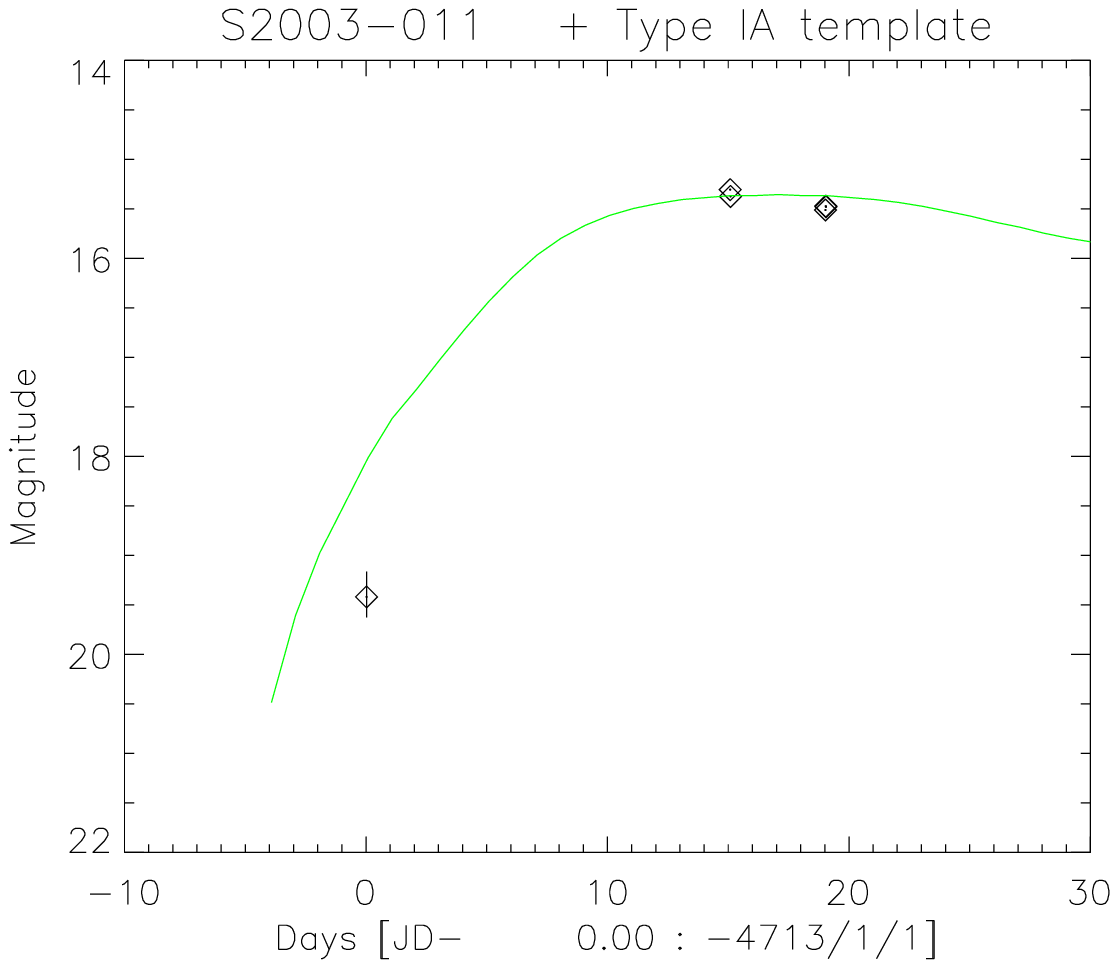}\label{fig:2003af_lightcurve}}
\vspace{0.3in}
\subfigure[2003ap]{\includegraphics[angle=90,height=2in,width=3in]{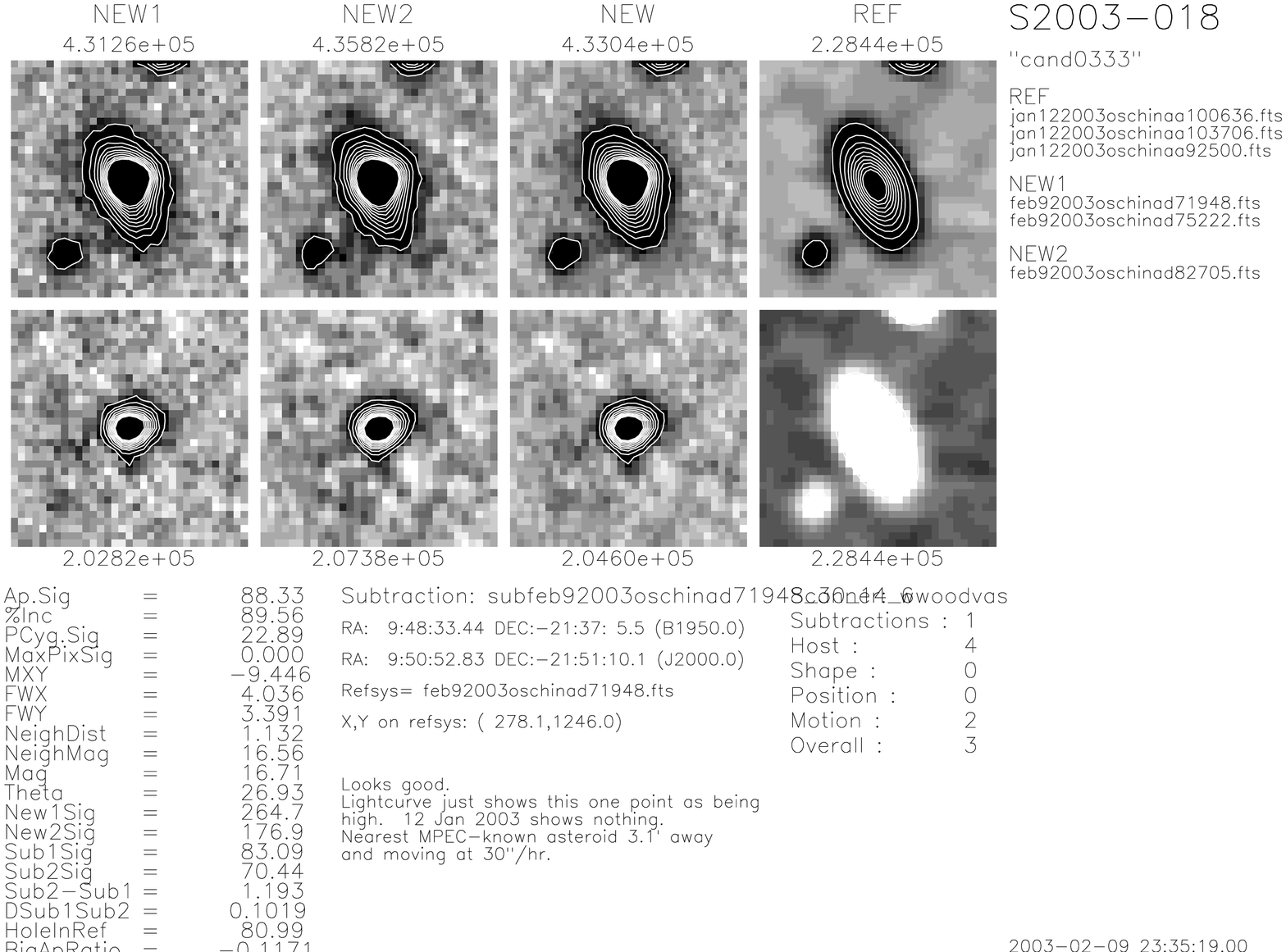}\label{fig:2003ap_discovery}}
\hspace{0.3in}
\subfigure[2003ap]{\includegraphics[height=2in]{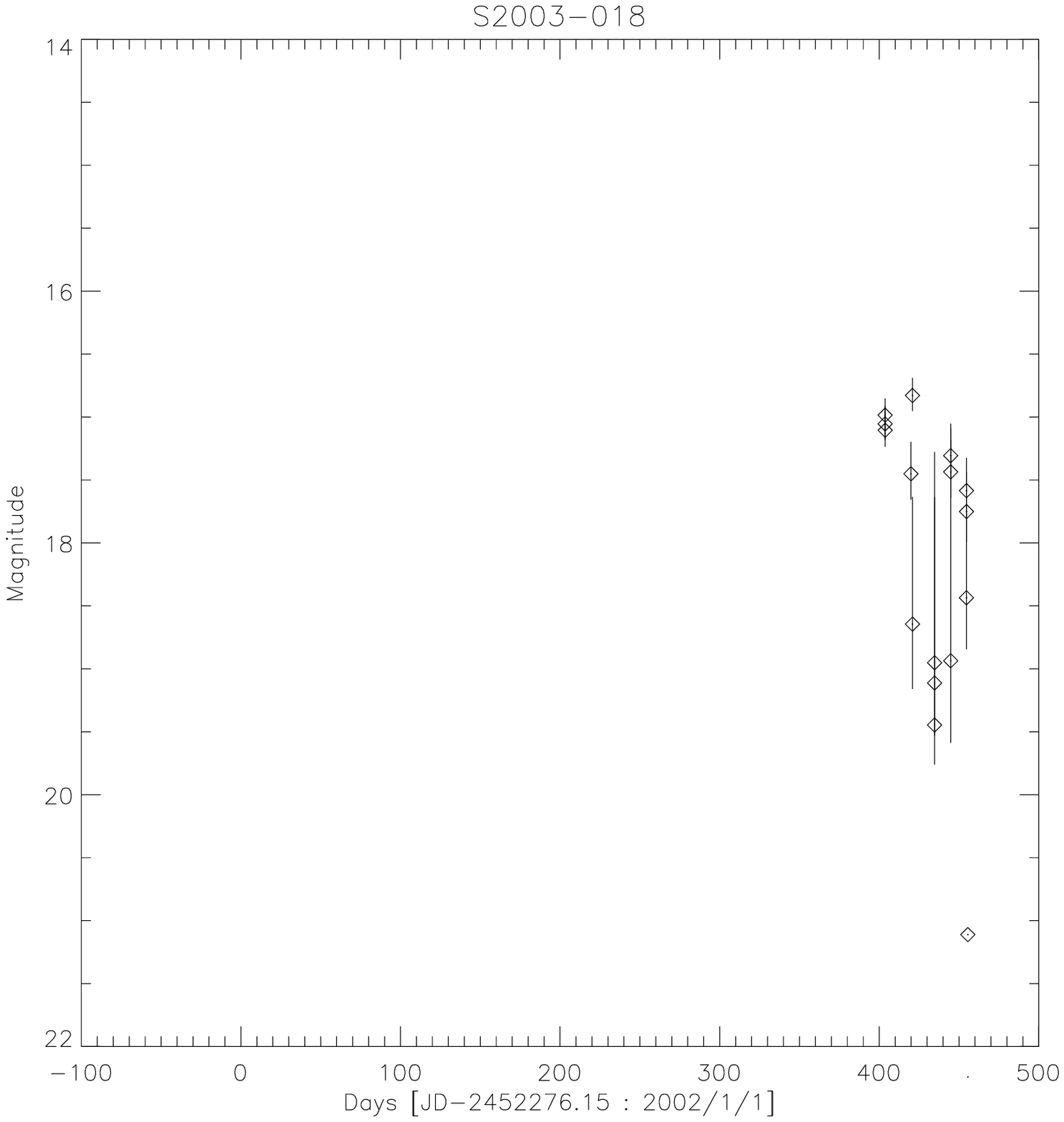}\label{fig:2003ap_lightcurve}}
\vspace{0.3in}
\subfigure[2003av]{\includegraphics[angle=90,height=2in,width=3in]{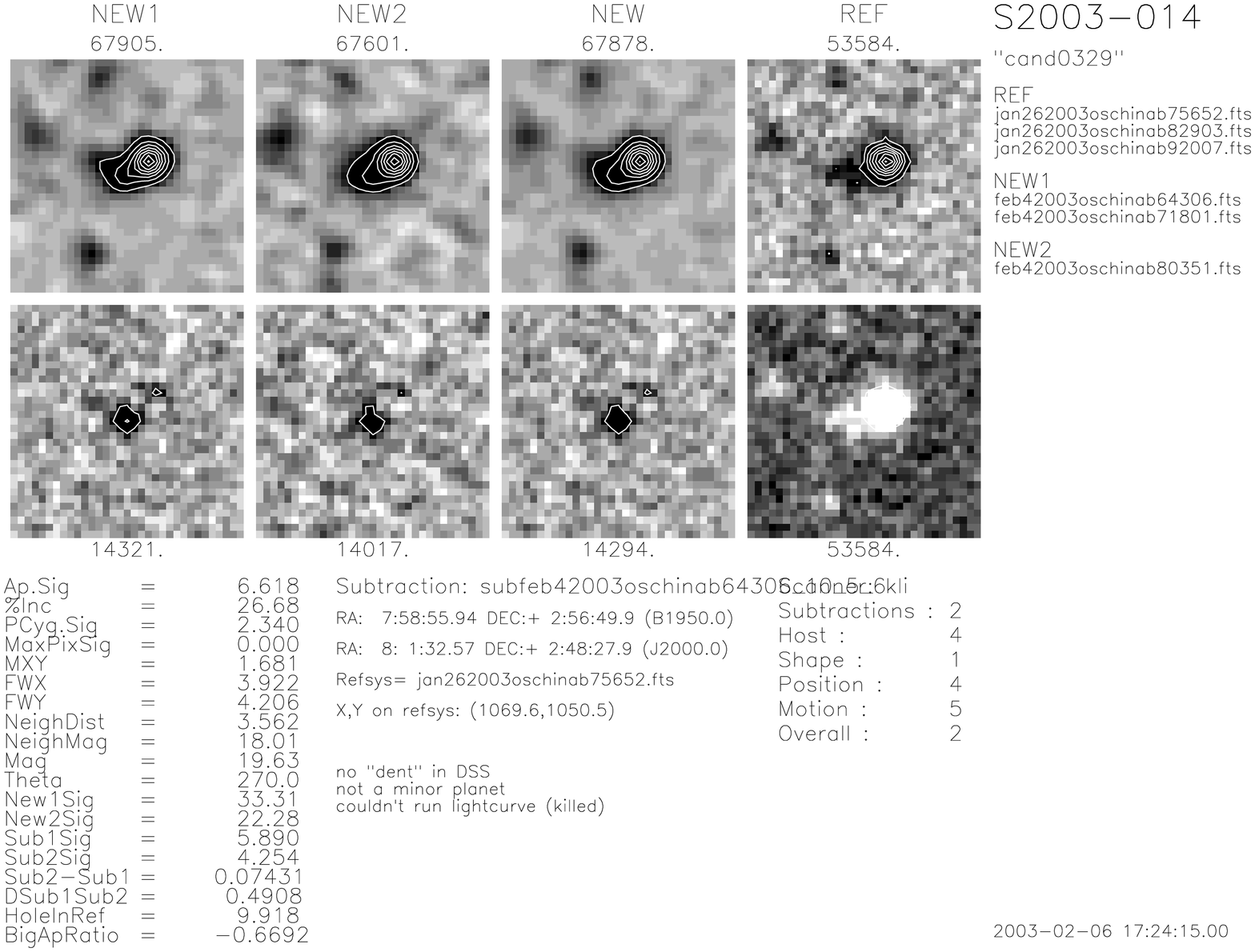}\label{fig:2003av_discovery}}
\hspace{0.3in}
\subfigure[2003av]{\includegraphics[height=2in]{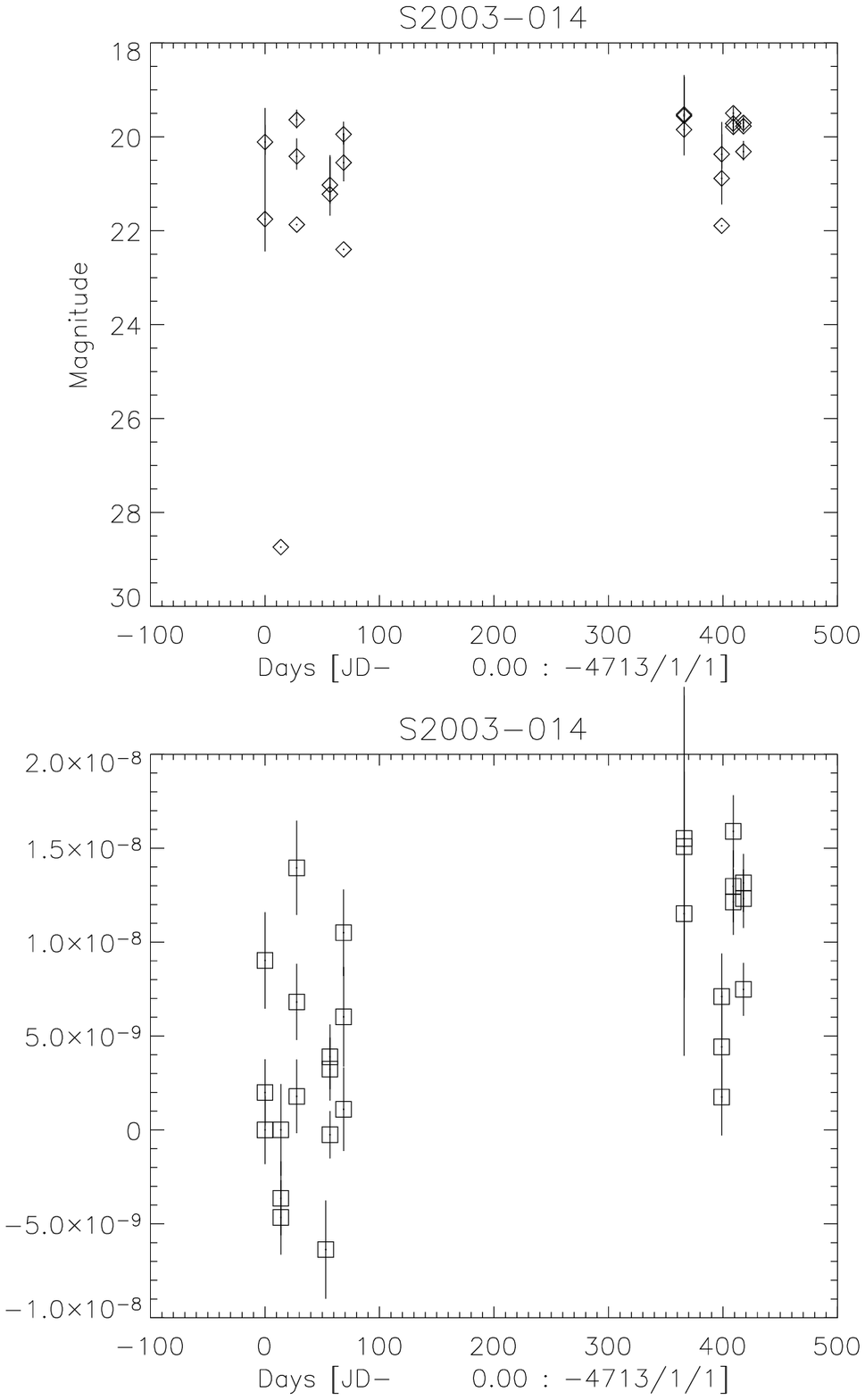}\label{fig:2003av_lightcurve}}
\vspace{0.3in}
\end{figure}

\clearpage\pagebreak
\begin{figure}
\subfigure[2003aw]{\includegraphics[angle=90,height=2in,width=3in]{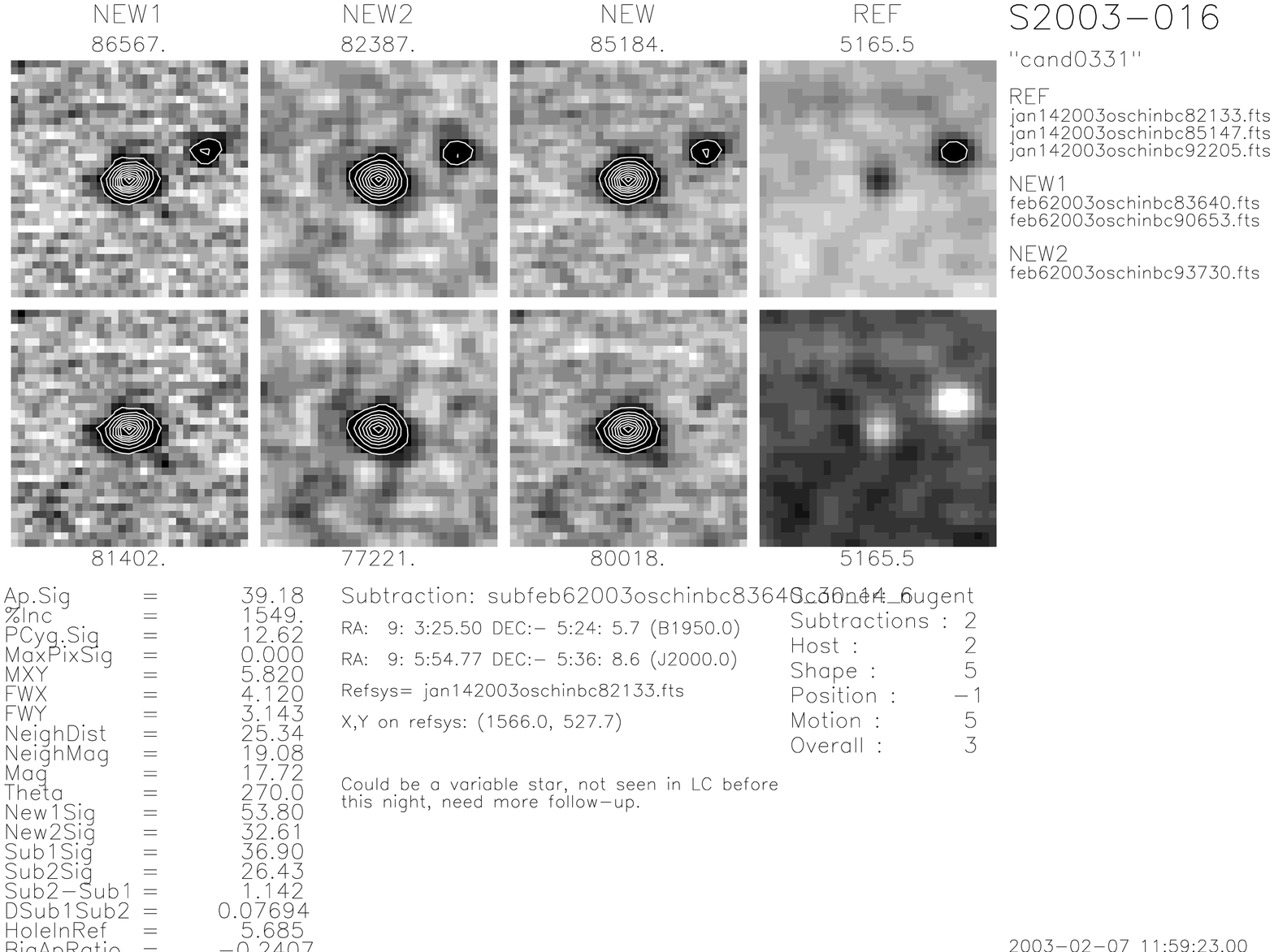}\label{fig:2003aw_discovery}}
\hspace{0.3in}
\subfigure[2003aw.  A dwarf nova.]{\includegraphics[height=2in]{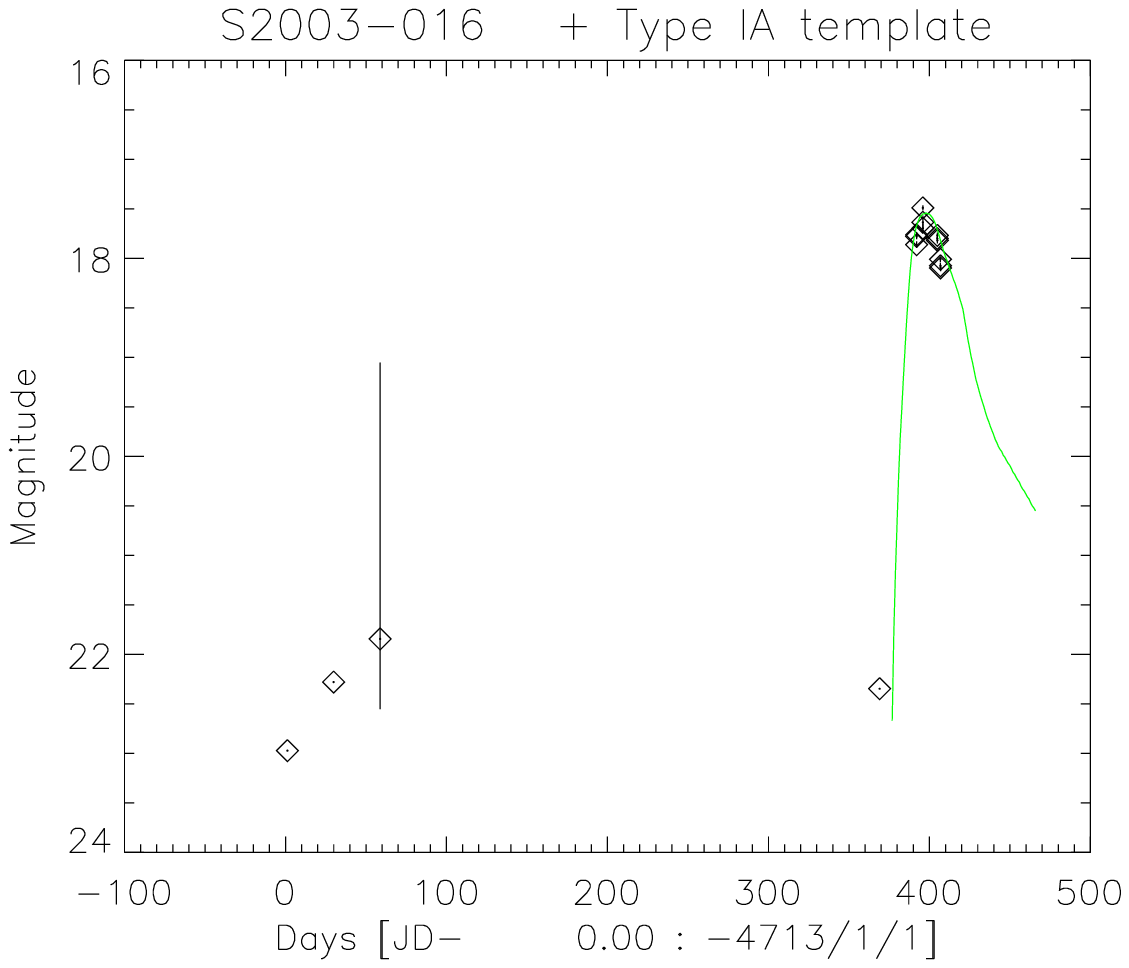}\label{fig:2003aw_lightcurve}}
\vspace{0.3in}
\subfigure[2003ax]{\includegraphics[angle=90,height=2in,width=3in]{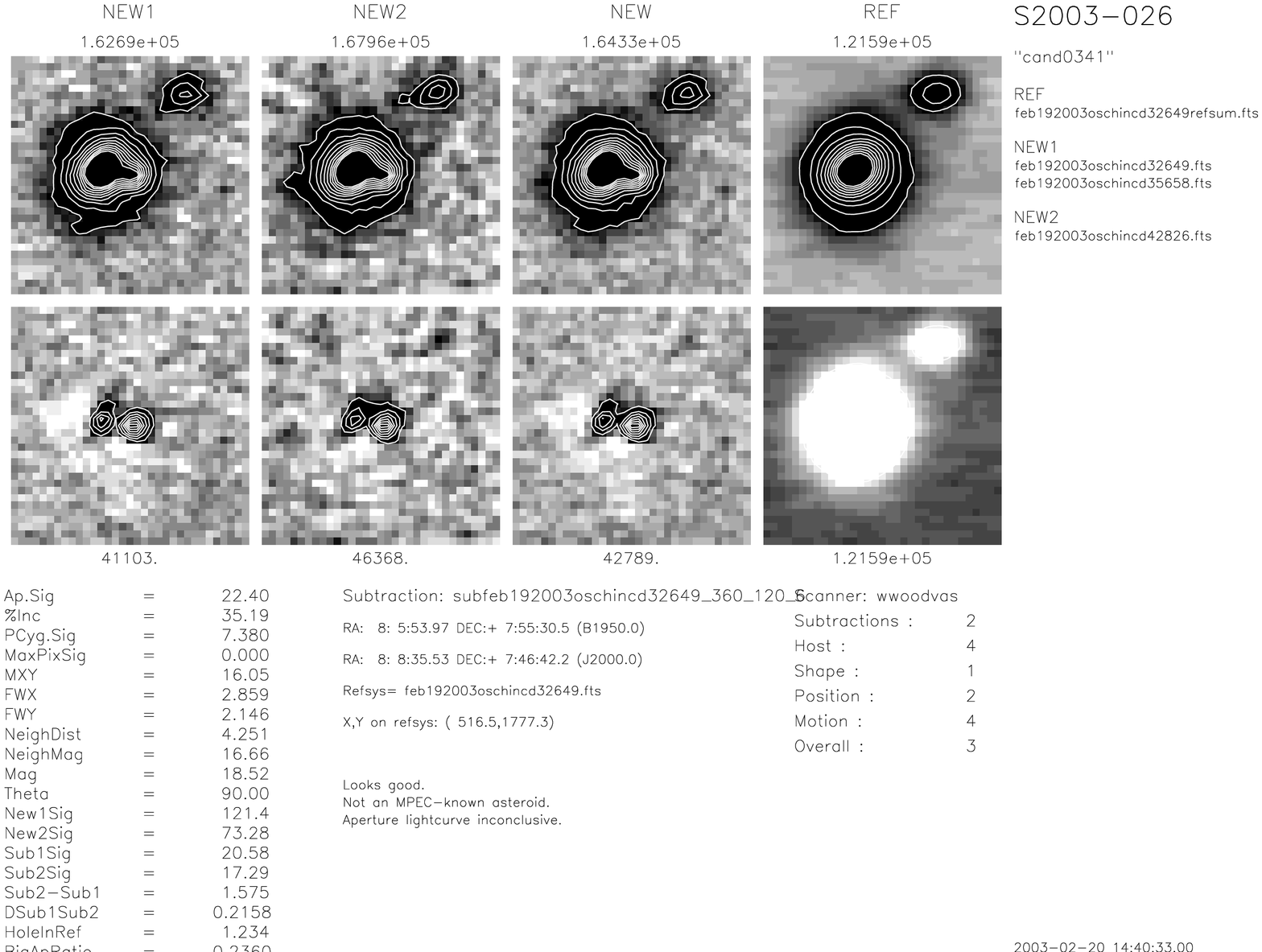}\label{fig:2003ax_discovery}}
\hspace{0.3in}
\subfigure[2003ax.  Bad refsys zeropoint.]{\includegraphics[height=2in]{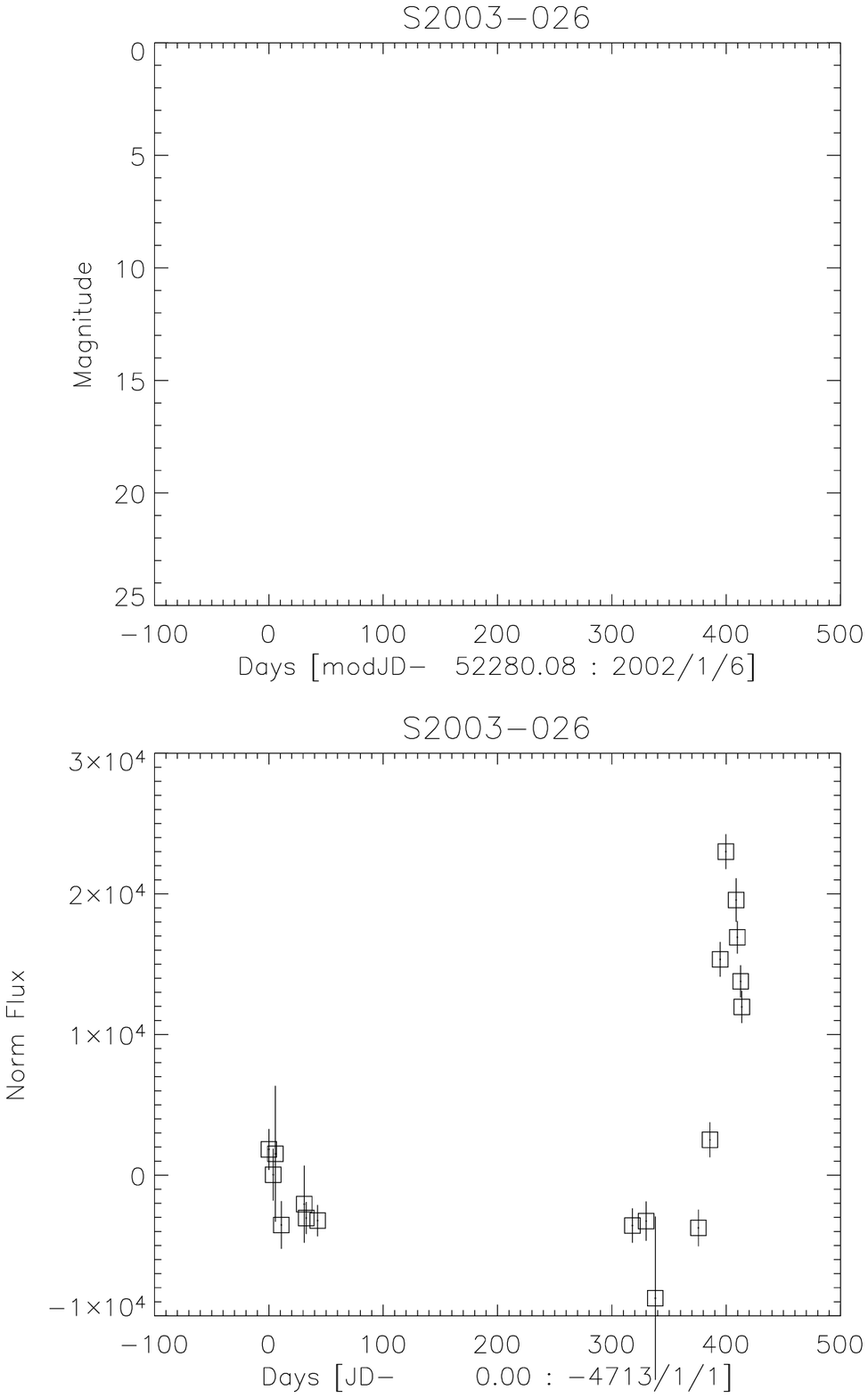}\label{fig:2003ax_lightcurve}}
\vspace{0.3in}
\subfigure[2003ay]{\includegraphics[angle=90,height=2in,width=3in]{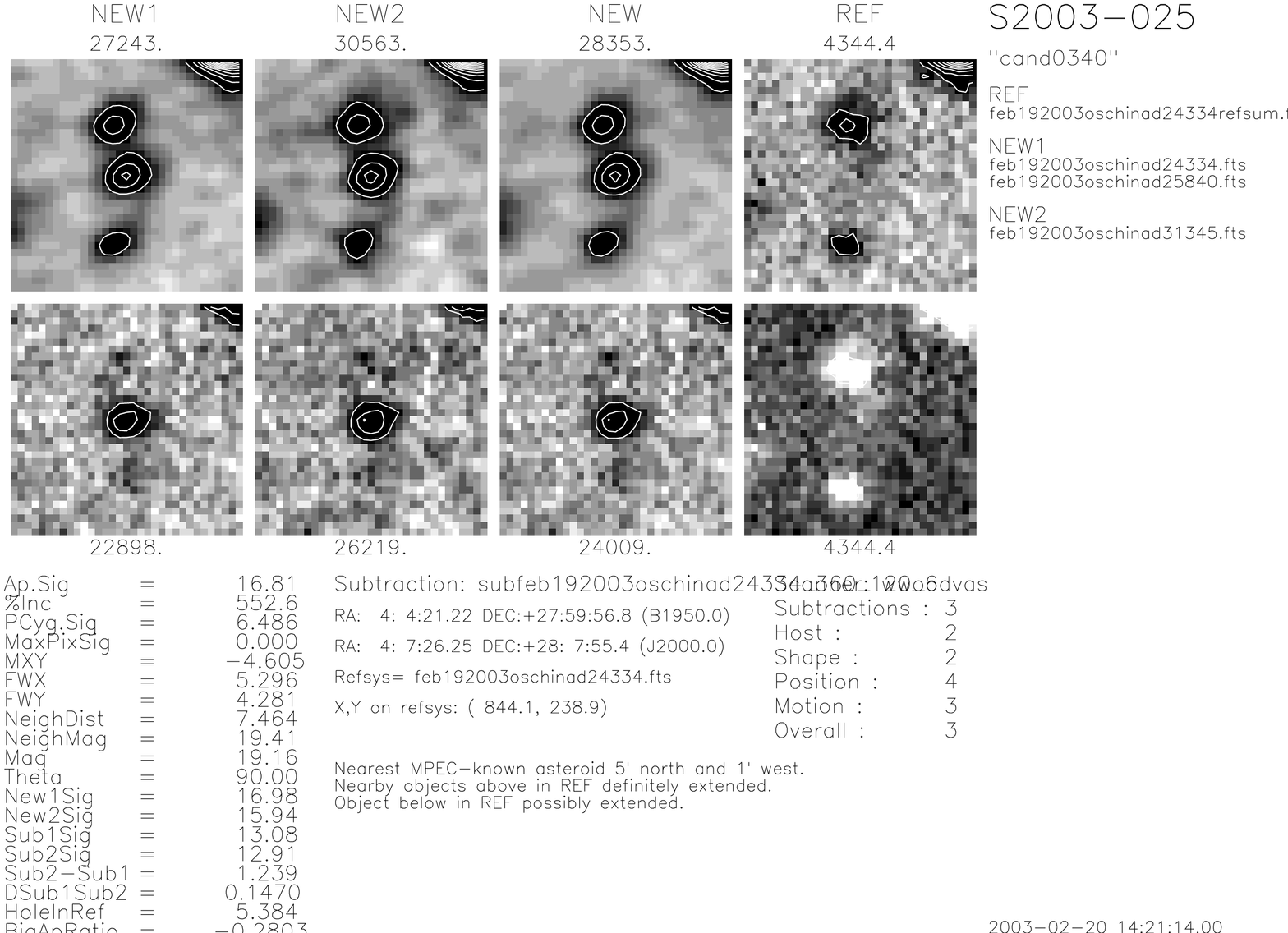}\label{fig:2003ay_discovery}}
\hspace{0.3in}
\subfigure[2003ay]{\includegraphics[height=2in]{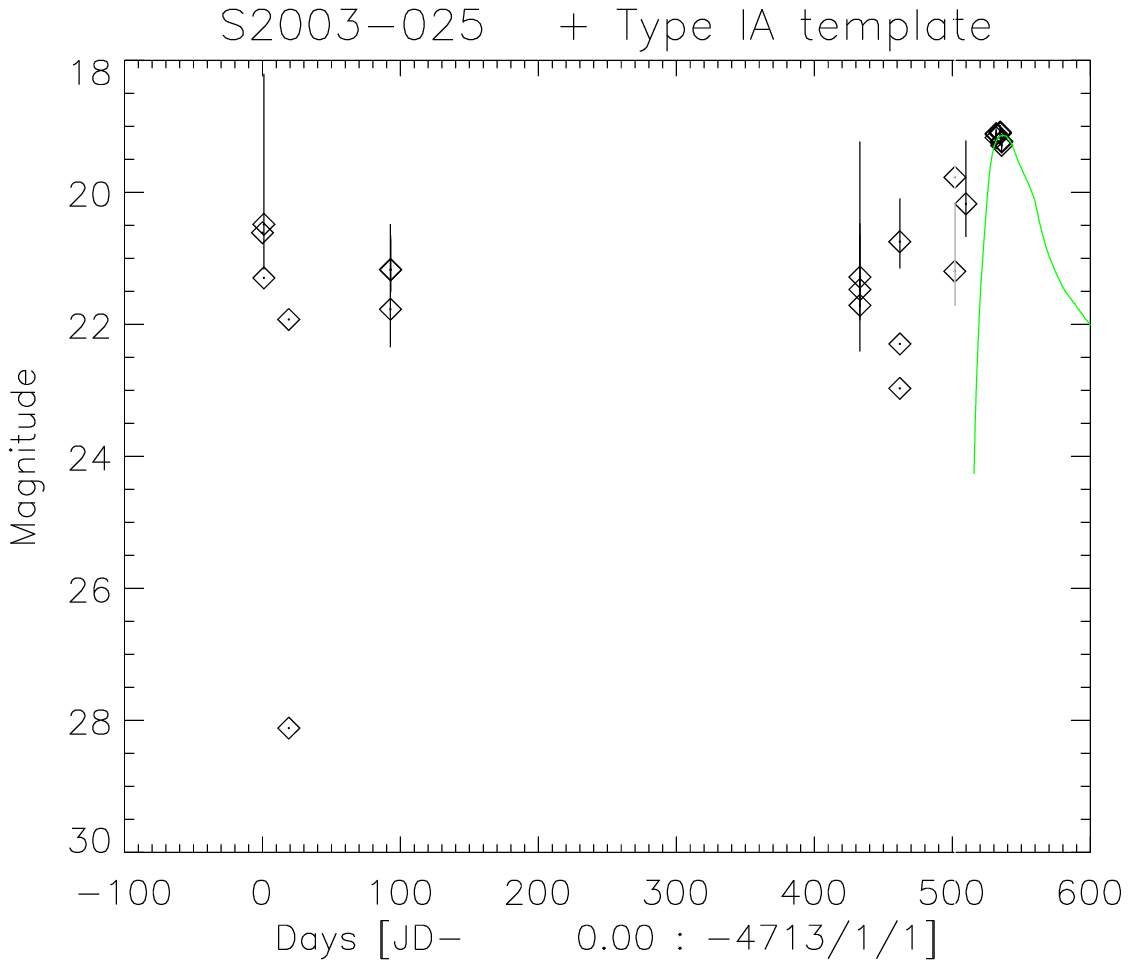}\label{fig:2003ay_lightcurve}}
\vspace{0.3in}
\end{figure}

\clearpage\pagebreak
\begin{figure}
\subfigure[2003bf]{\includegraphics[angle=90,height=2in,width=3in]{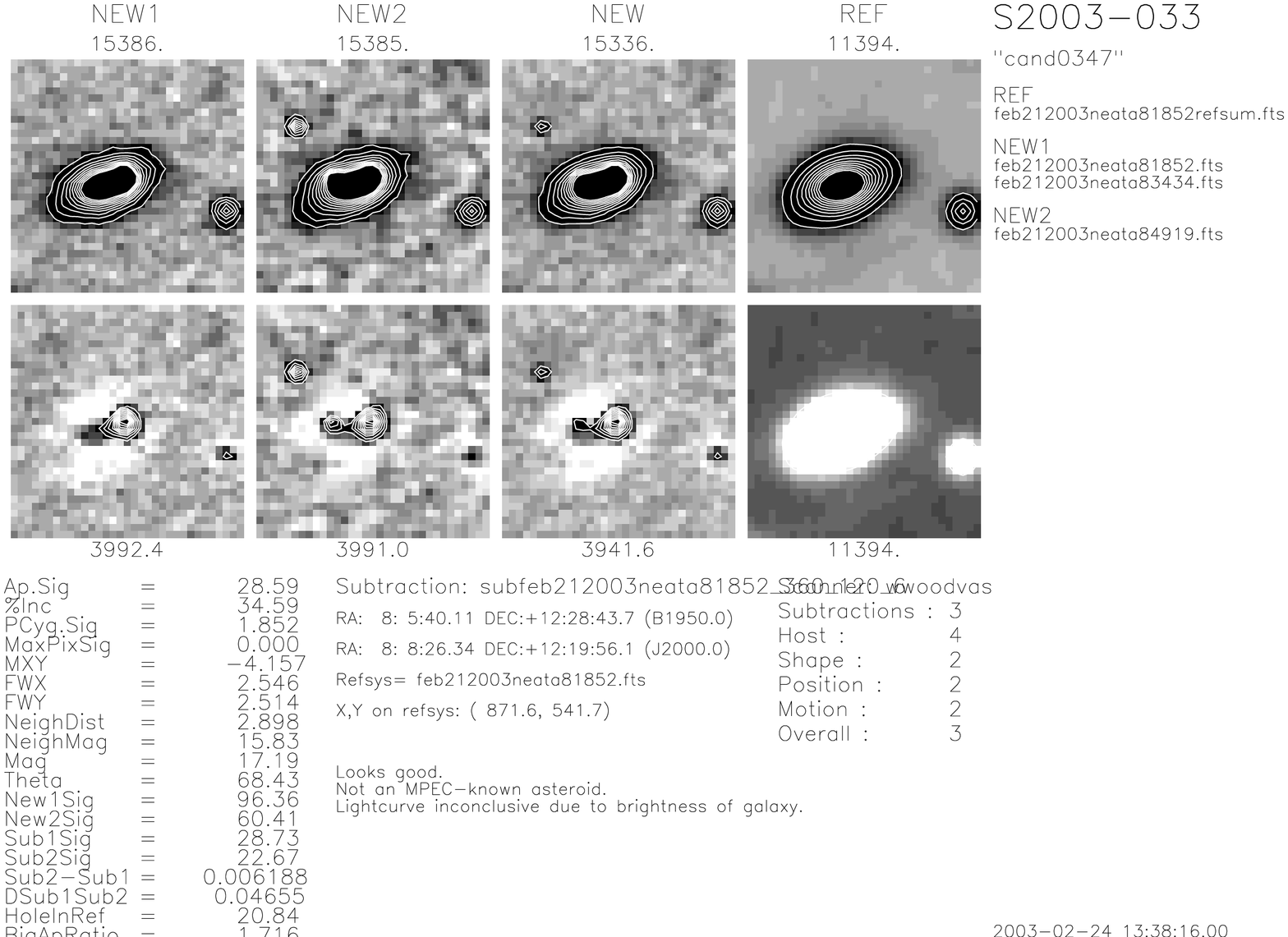}\label{fig:2003bf_discovery}}
\hspace{0.3in}
\subfigure[2003bf]{\includegraphics[height=2in]{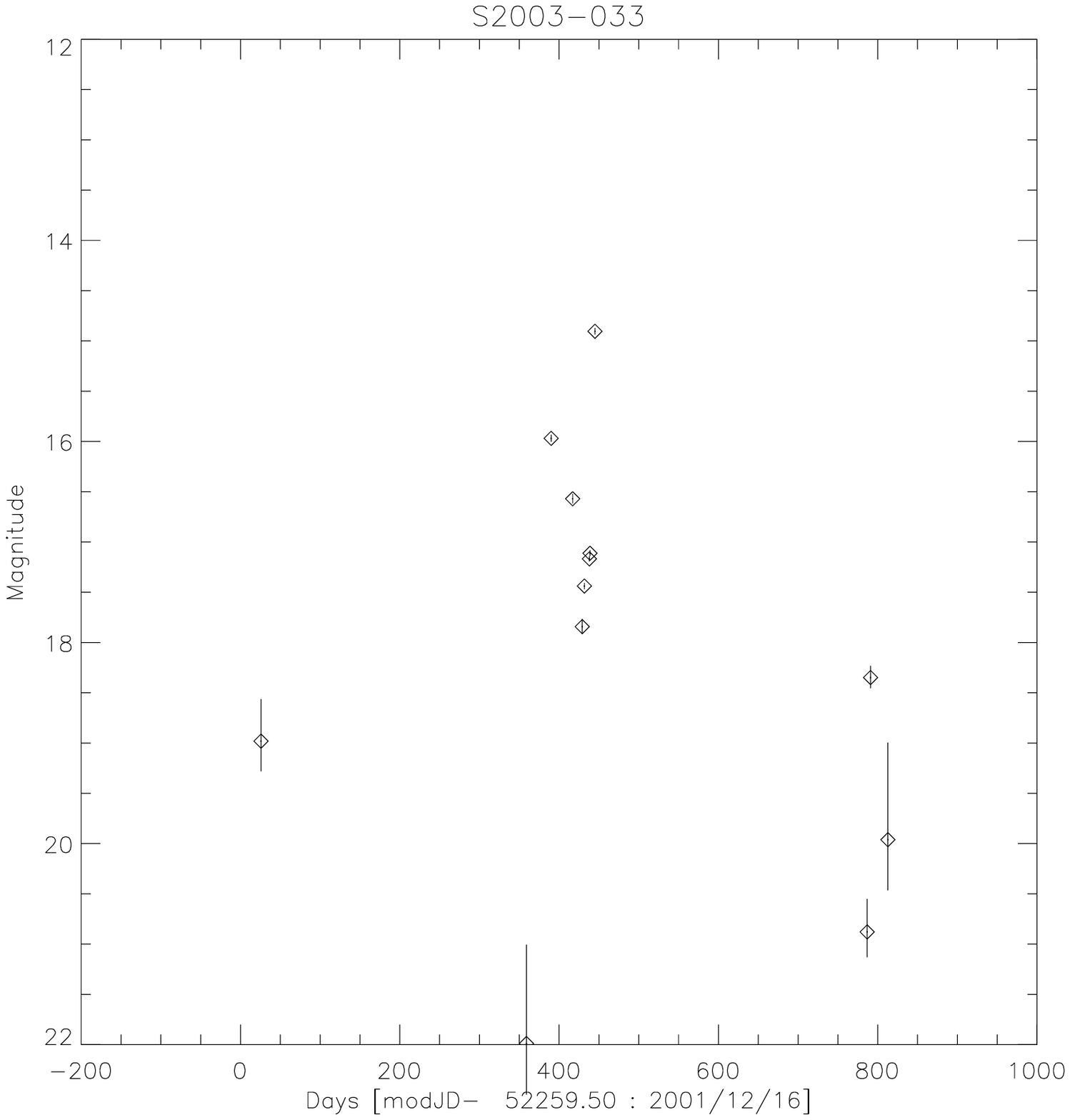}\label{fig:2003bf_lightcurve}}
\vspace{0.3in}
\subfigure[2003bh]{\includegraphics[angle=90,height=2in,width=3in]{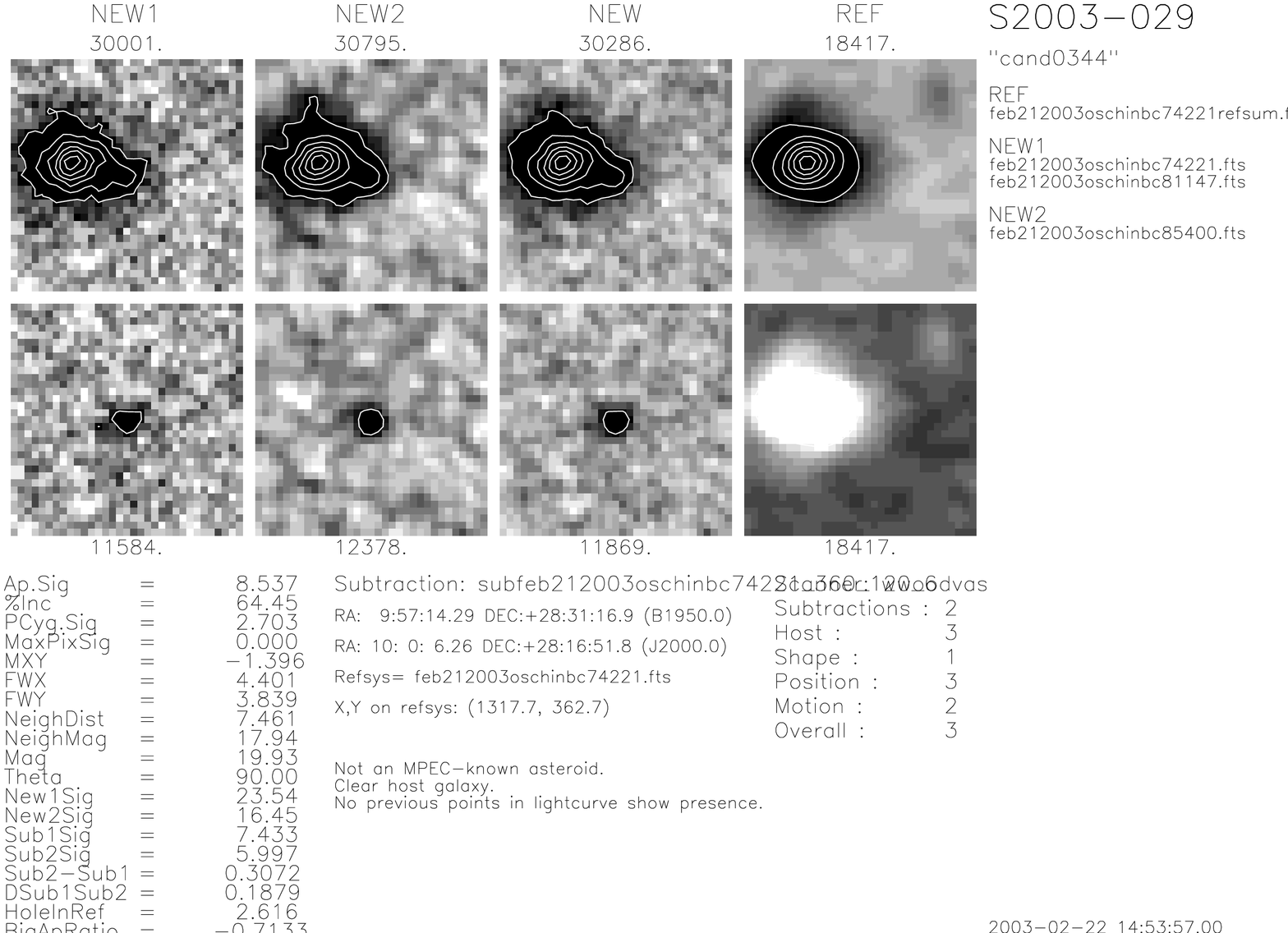}\label{fig:2003bh_discovery}}
\hspace{0.3in}
\subfigure[2003bh]{\includegraphics[height=2in]{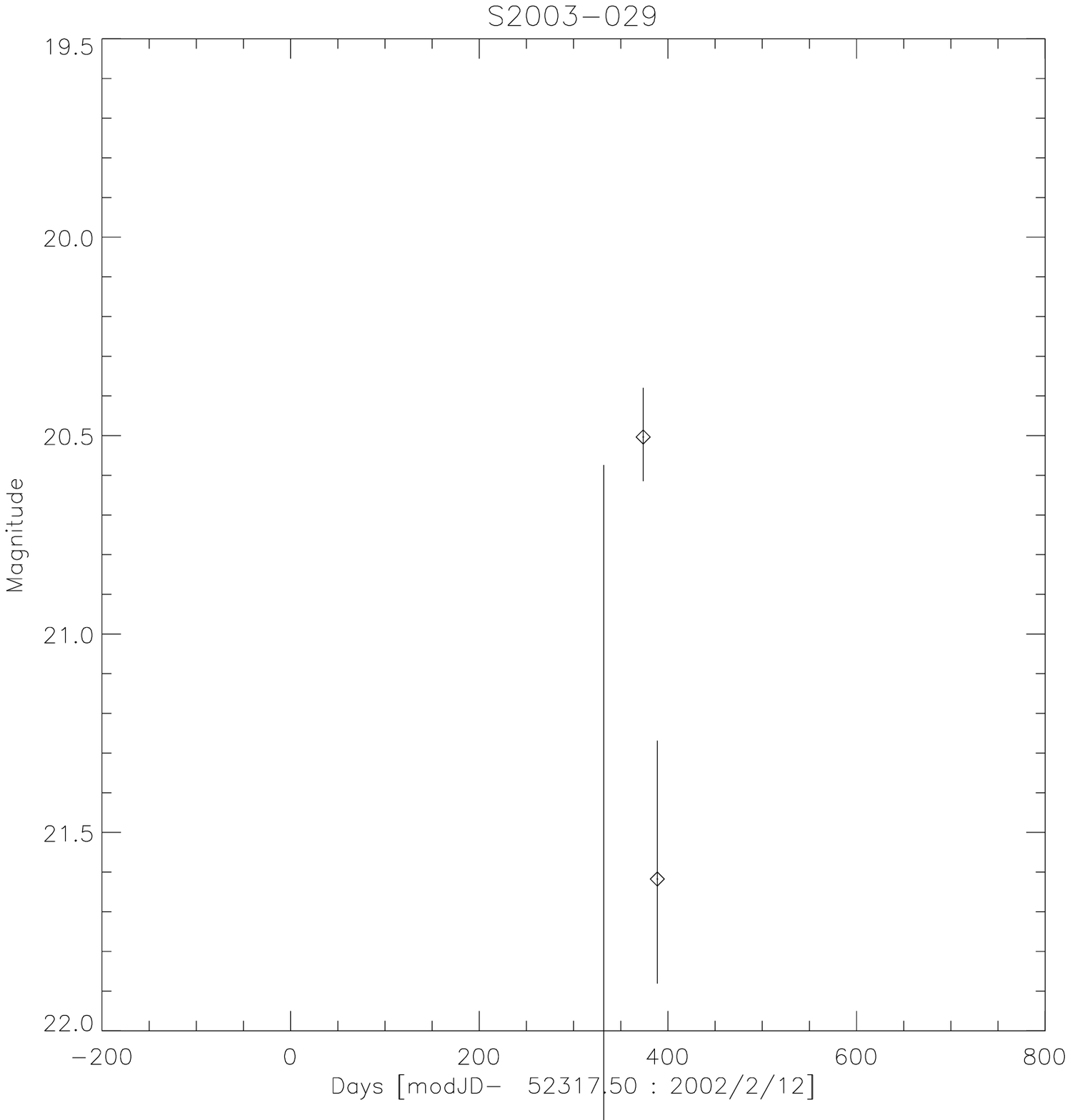}\label{fig:2003bh_lightcurve}}
\vspace{0.3in}
\subfigure[2003bi]{\includegraphics[angle=90,height=2in,width=3in]{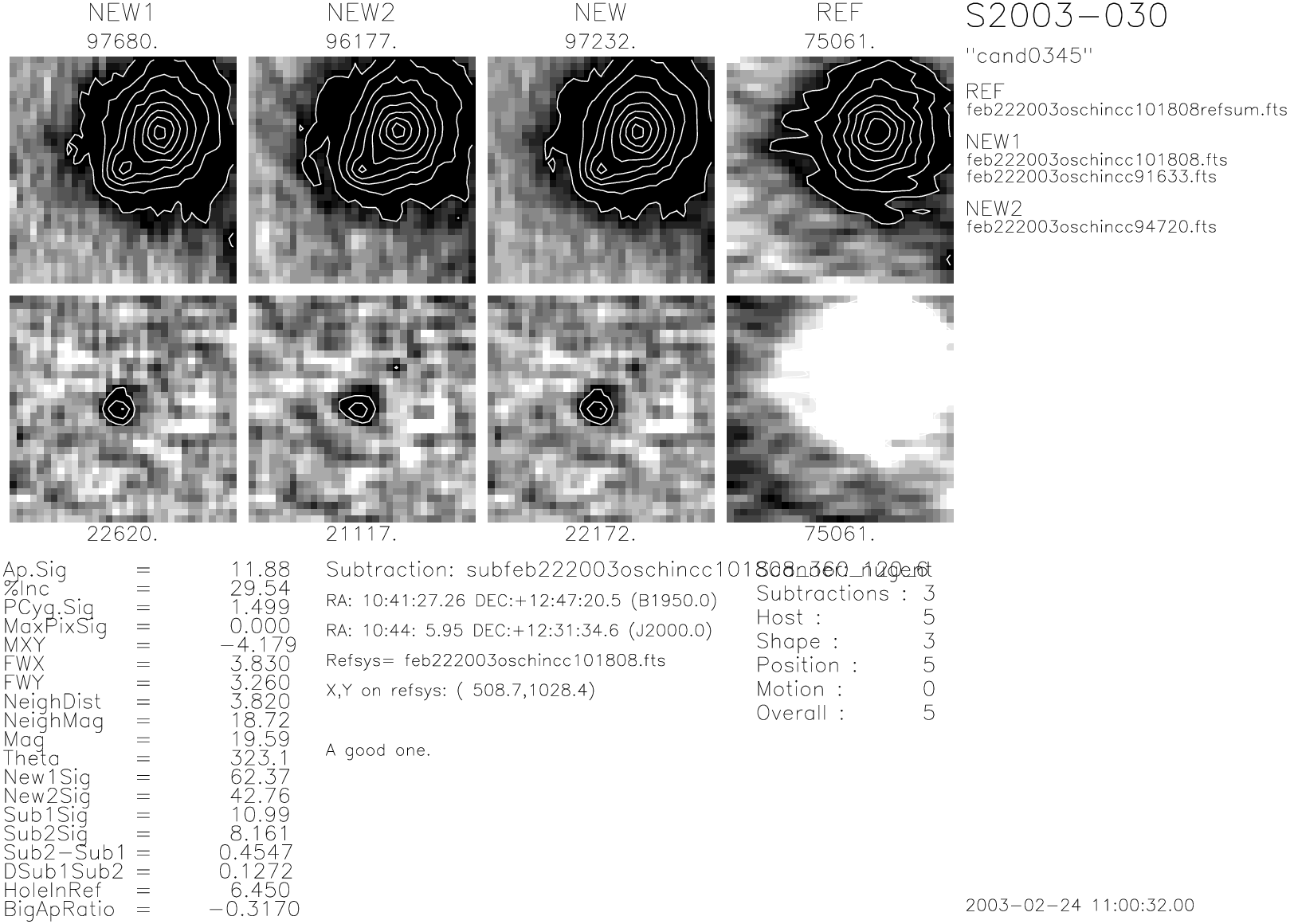}\label{fig:2003bi_discovery}}
\hspace{0.3in}
\subfigure[2003bi]{\includegraphics[height=2in]{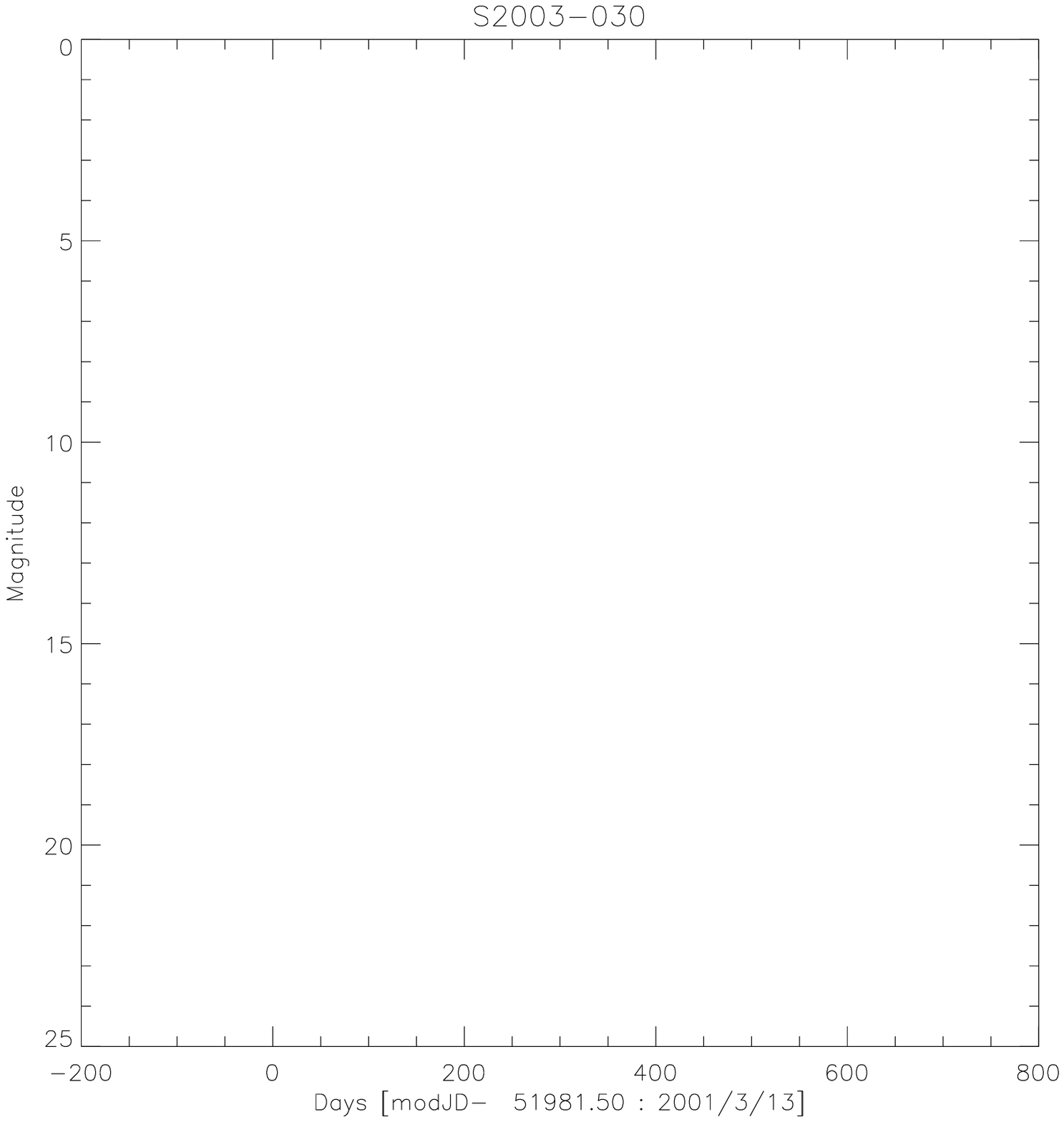}\label{fig:2003bi_lightcurve}}
\vspace{0.3in}
\end{figure}

\clearpage\pagebreak
\begin{figure}
\subfigure[2003bn]{\includegraphics[angle=90,height=2in,width=3in]{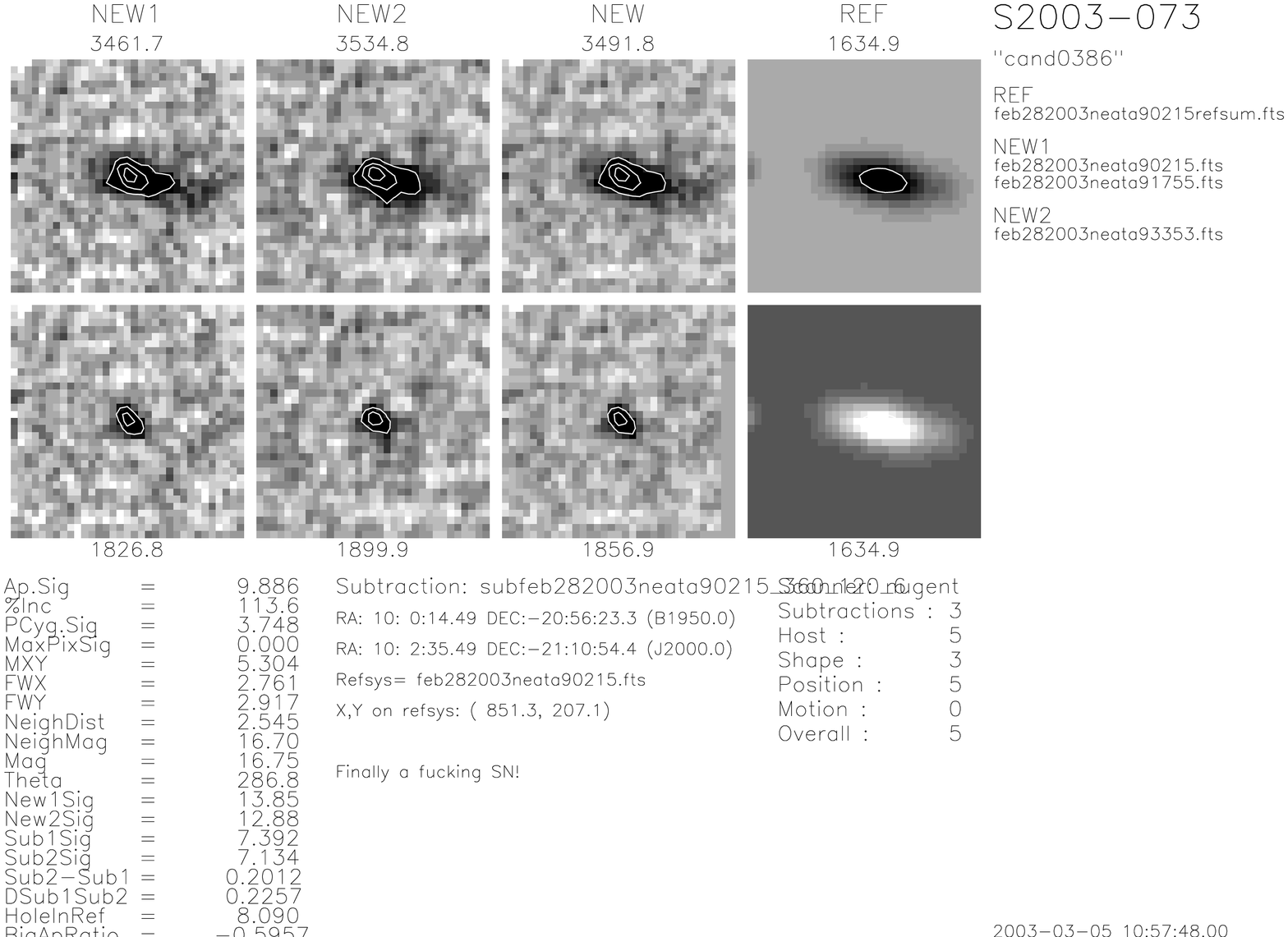}\label{fig:2003bn_discovery}}
\hspace{0.3in}
\subfigure[2003bn]{\includegraphics[height=2in]{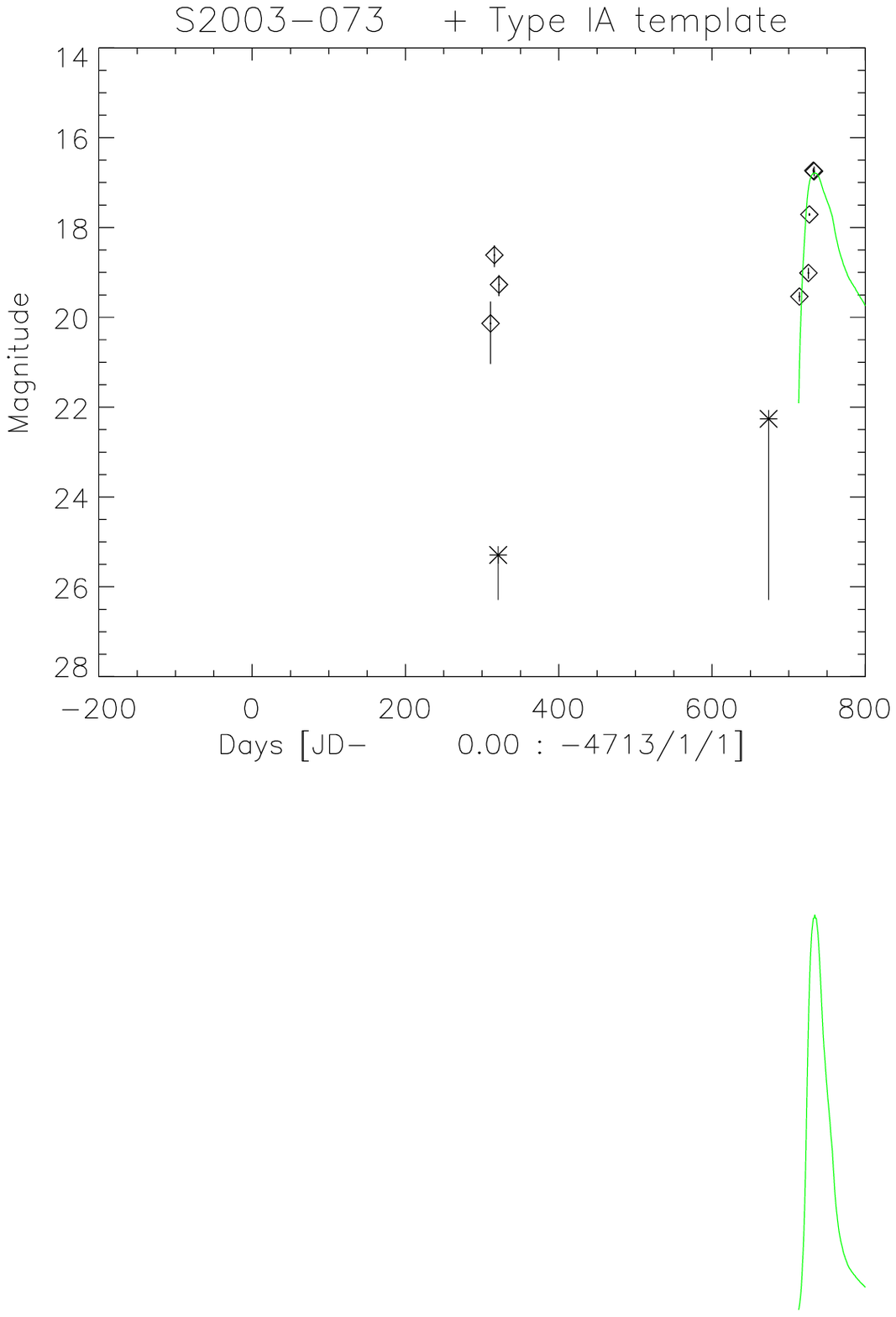}\label{fig:2003bn_lightcurve}}
\vspace{0.3in}
\subfigure[2003bo]{\includegraphics[angle=90,height=2in,width=3in]{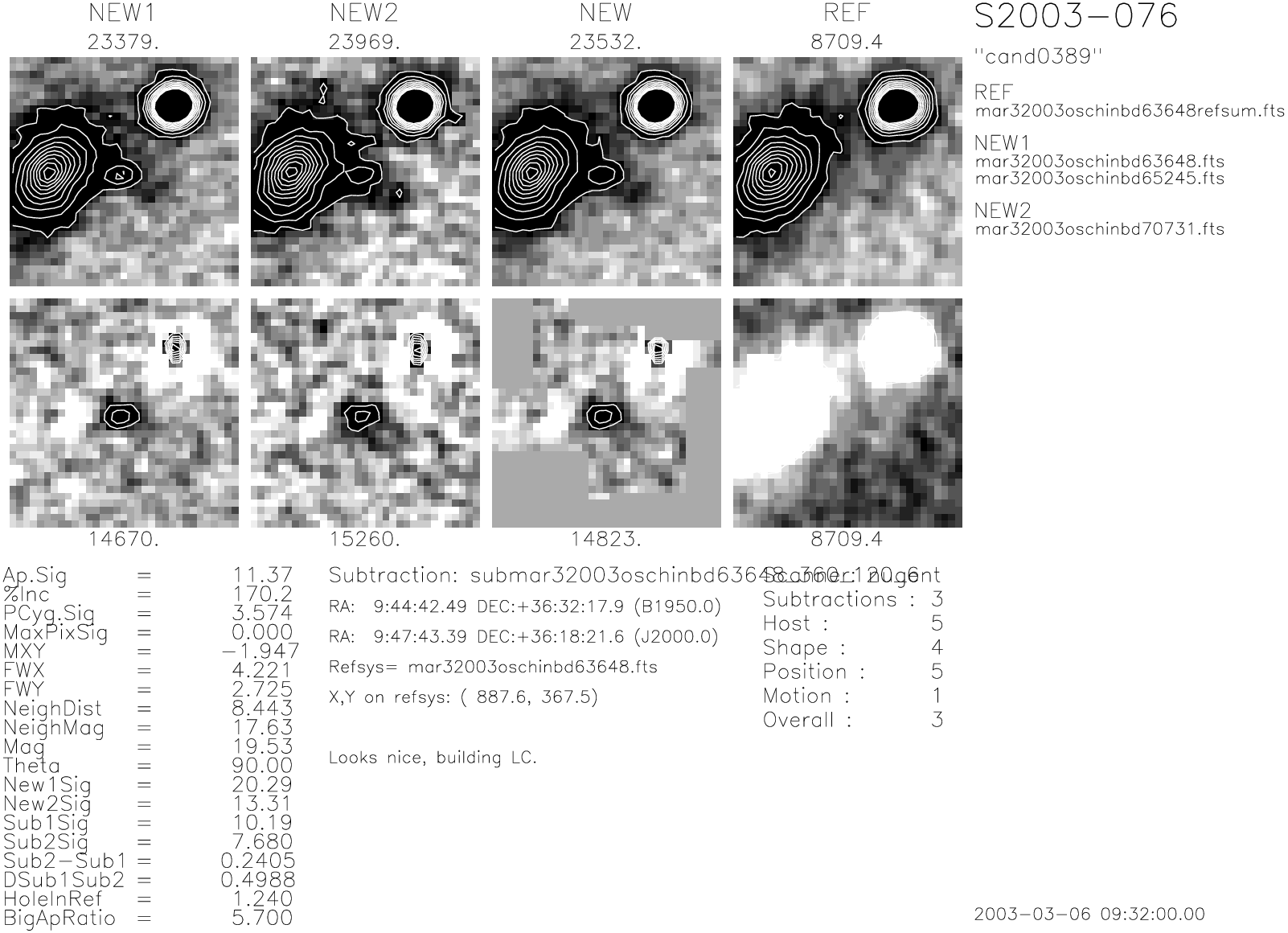}\label{fig:2003bo_discovery}}
\hspace{0.3in}
\subfigure[2003bo]{\includegraphics[height=2in]{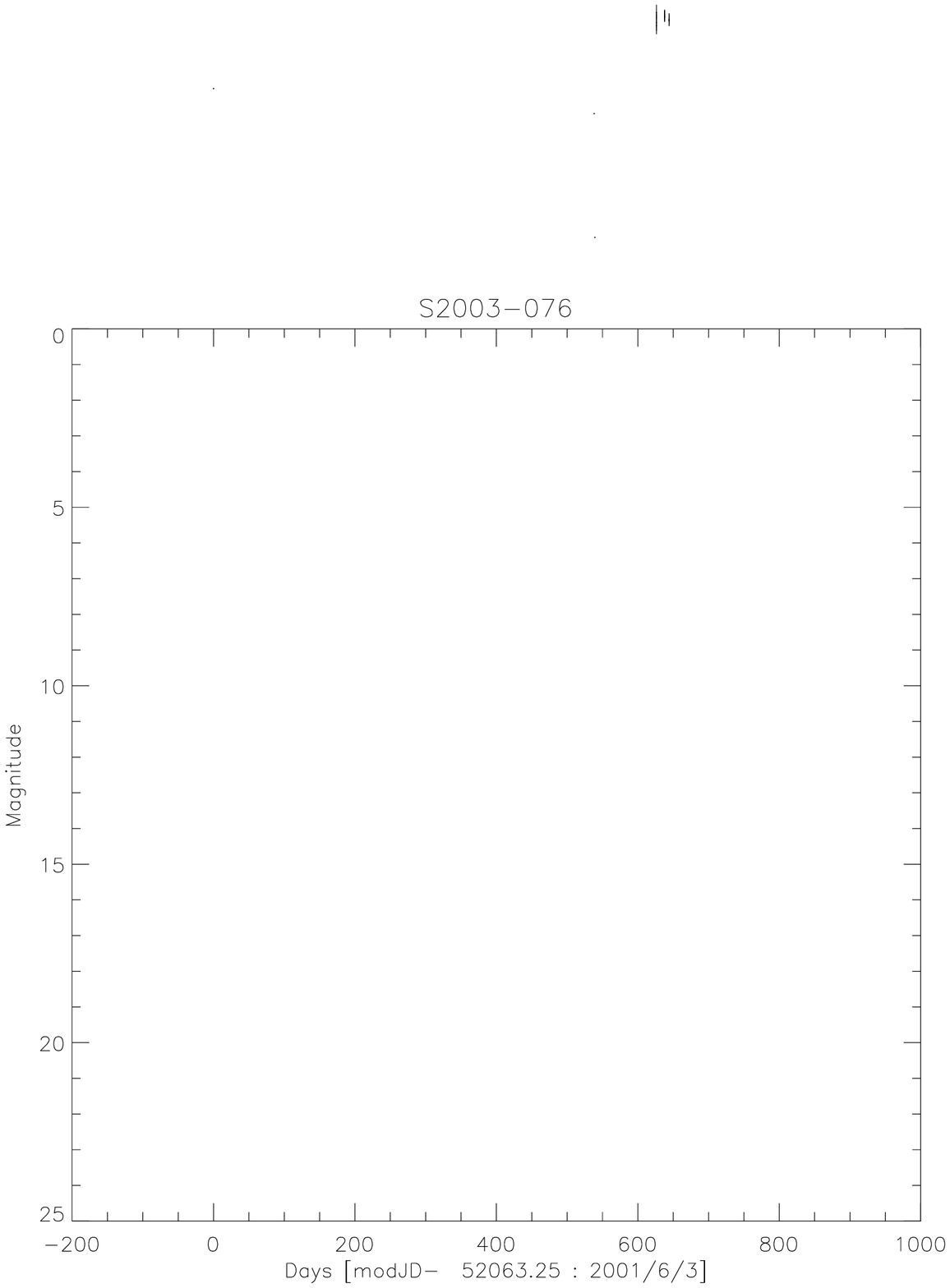}\label{fig:2003bo_lightcurve}}
\vspace{0.3in}
\subfigure[2003bp]{\includegraphics[angle=90,height=2in,width=3in]{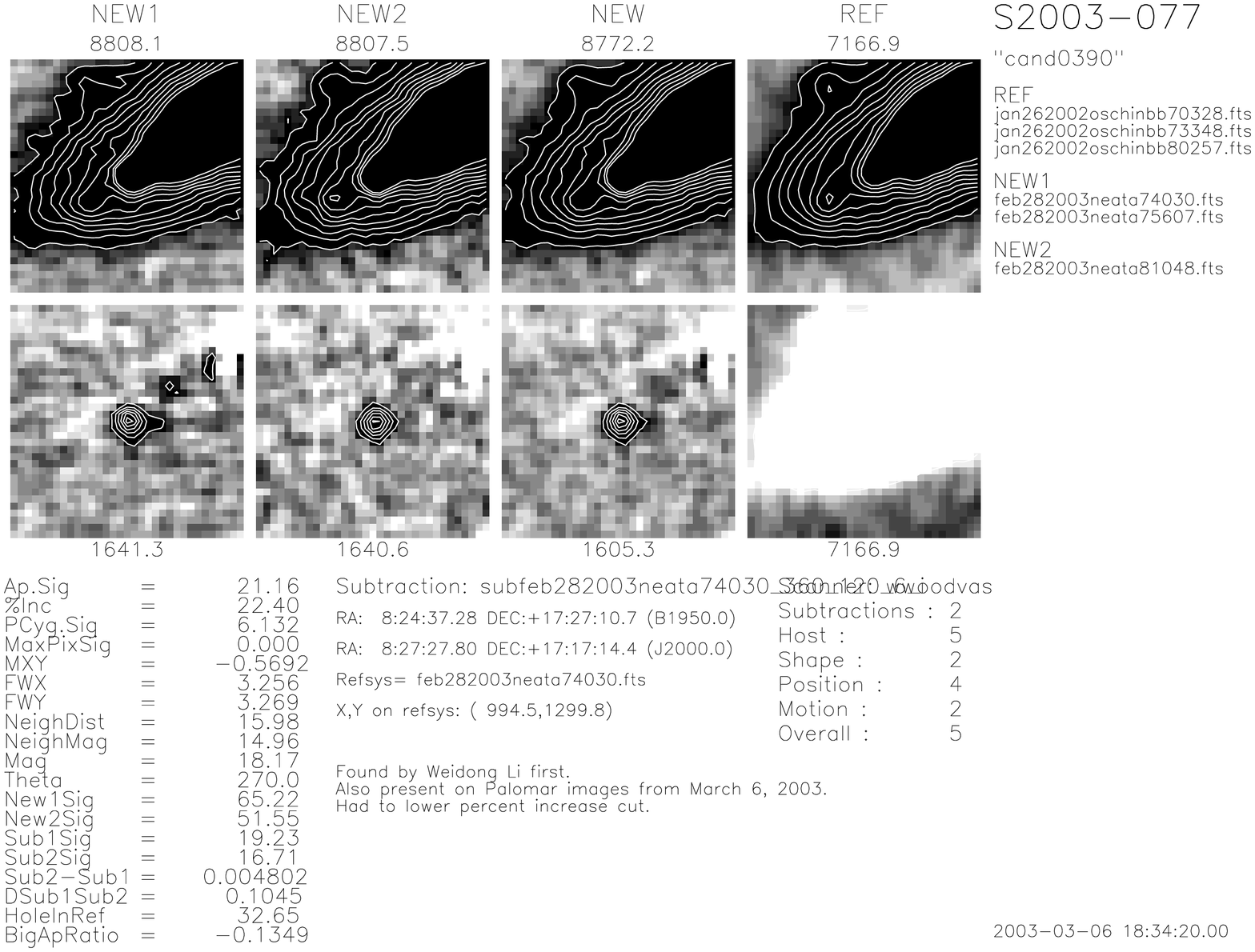}\label{fig:2003bp_discovery}}
\hspace{0.3in}
\subfigure[2003bp]{\includegraphics[height=2in]{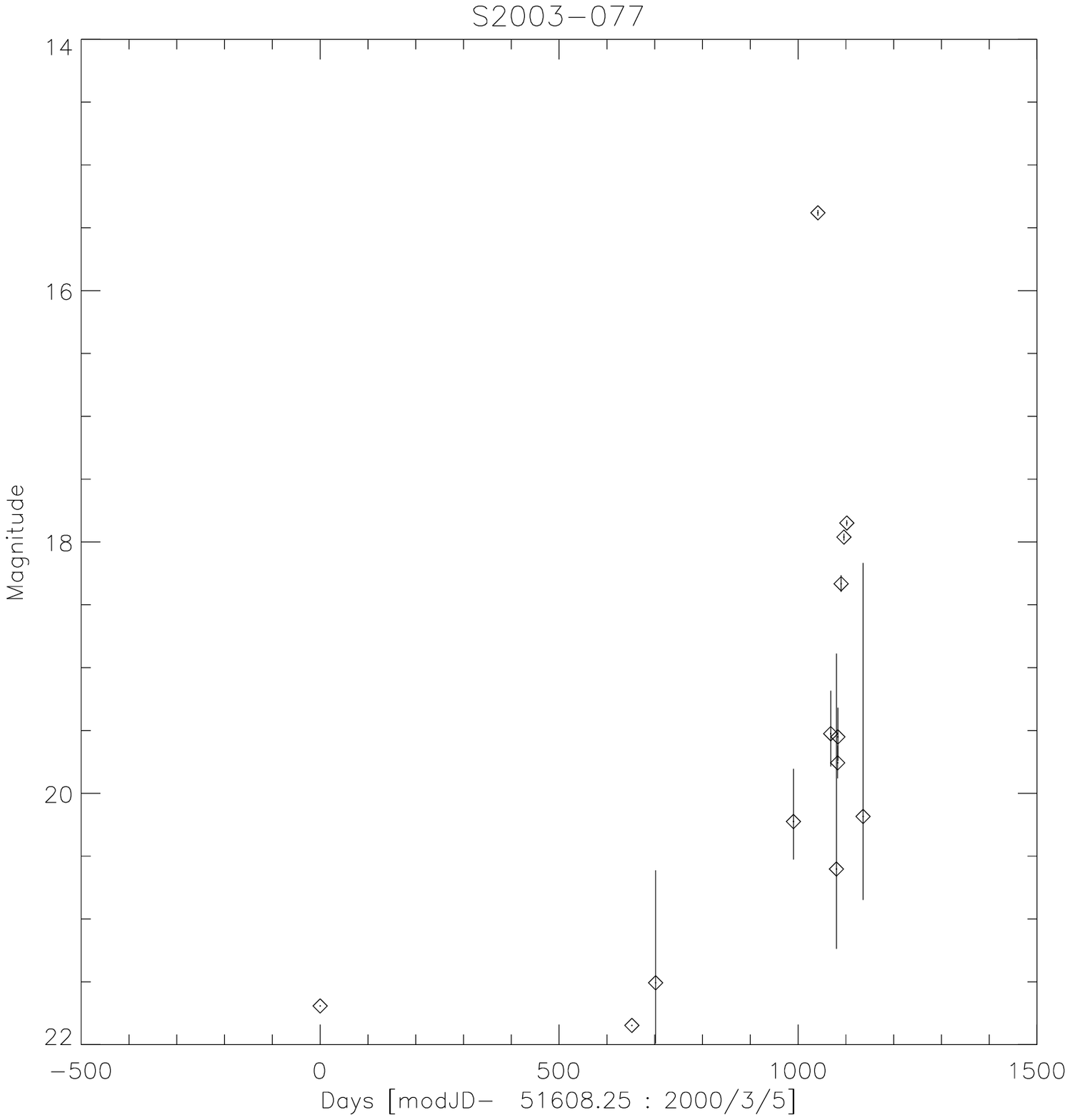}\label{fig:2003bp_lightcurve}}
\vspace{0.3in}
\end{figure}

\clearpage\pagebreak
\begin{figure}
\subfigure[2003bs]{\includegraphics[angle=90,height=2in,width=3in]{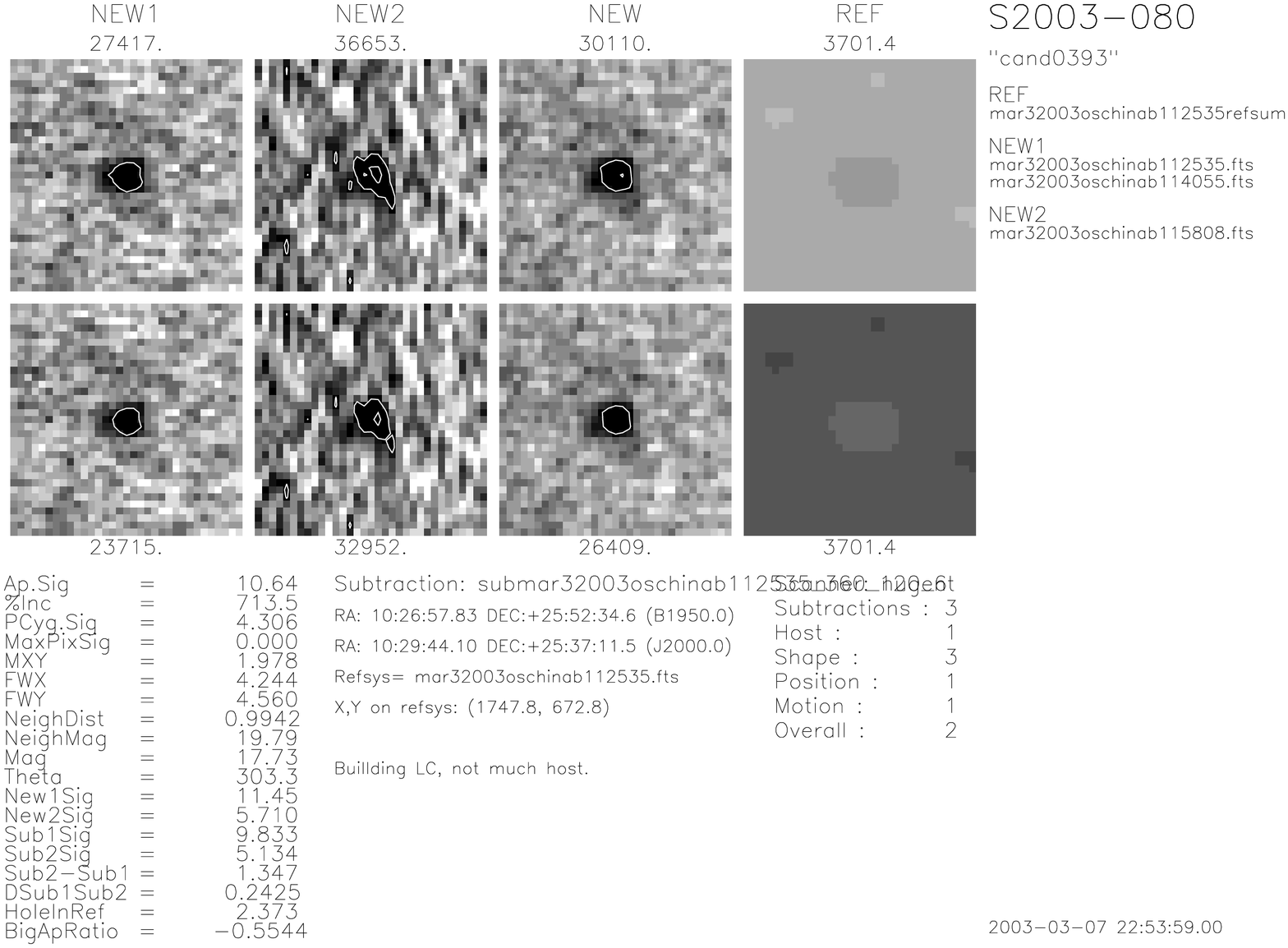}\label{fig:2003bs_discovery}}
\hspace{0.3in}
\subfigure[2003bs]{\includegraphics[height=2in]{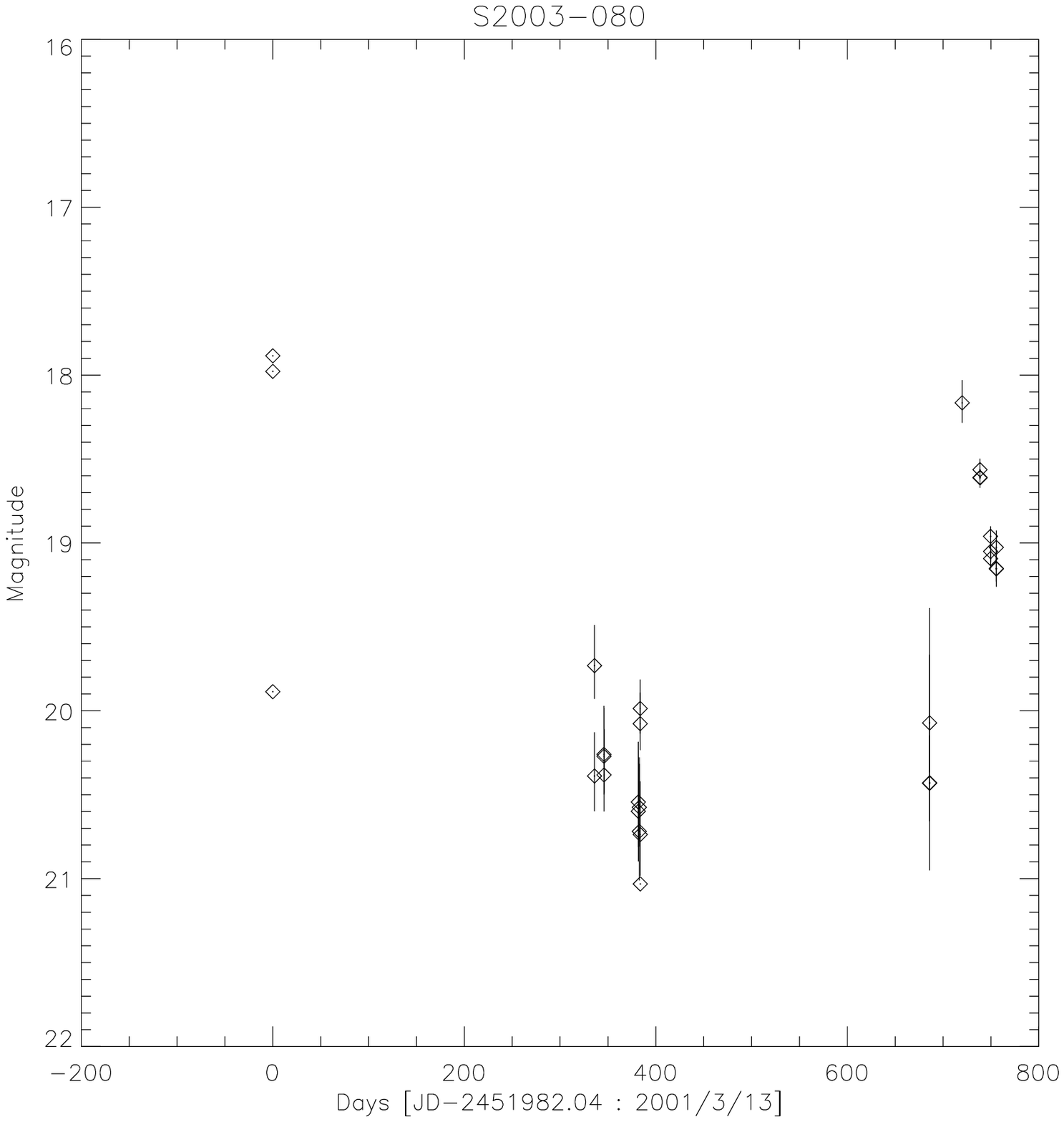}\label{fig:2003bs_lightcurve}}
\vspace{0.3in}
\subfigure[2003bt]{\includegraphics[angle=90,height=2in,width=3in]{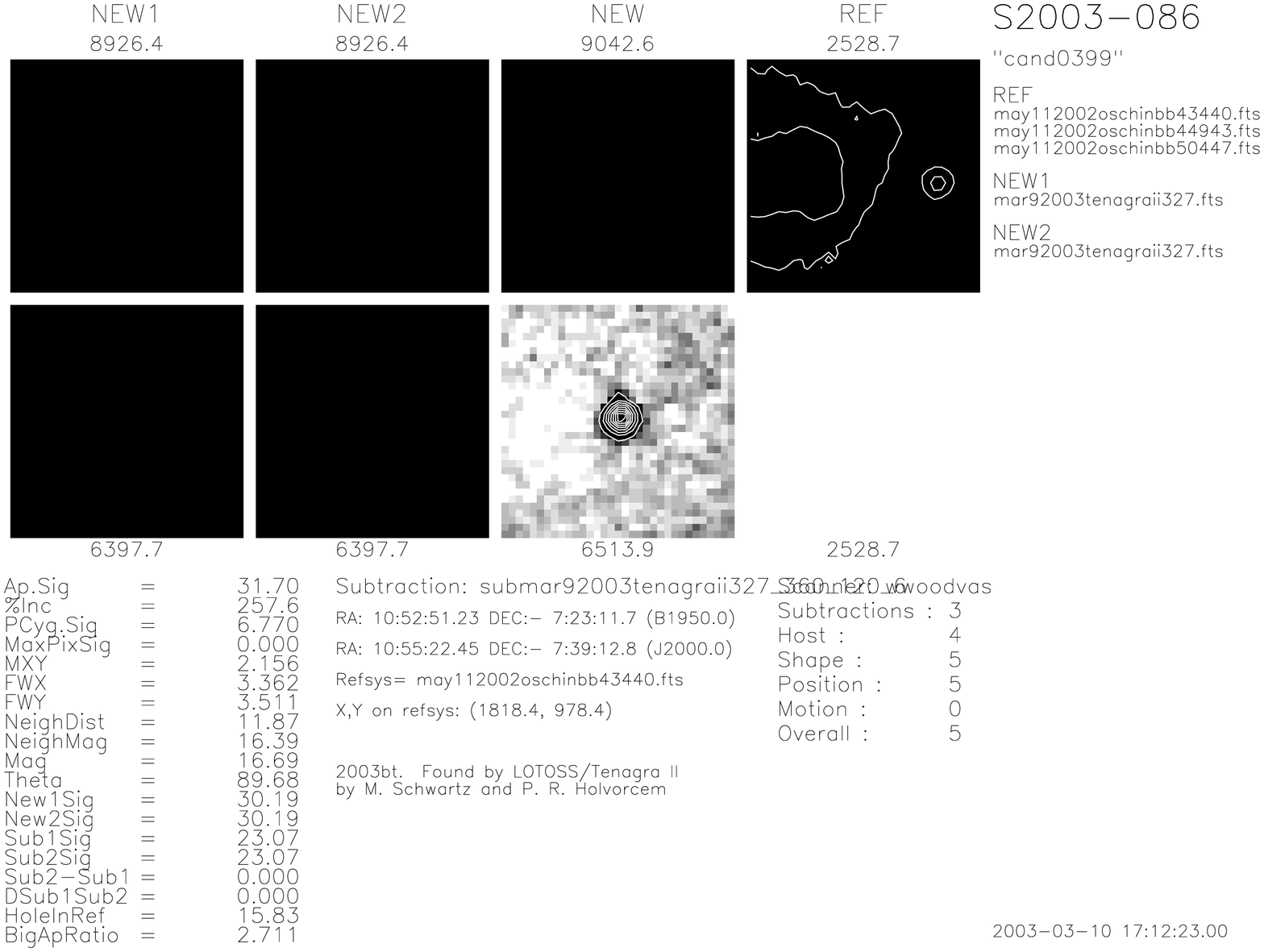}\label{fig:2003bt_discovery}}
\hspace{0.3in}
\subfigure[2003bt]{\includegraphics[height=2in]{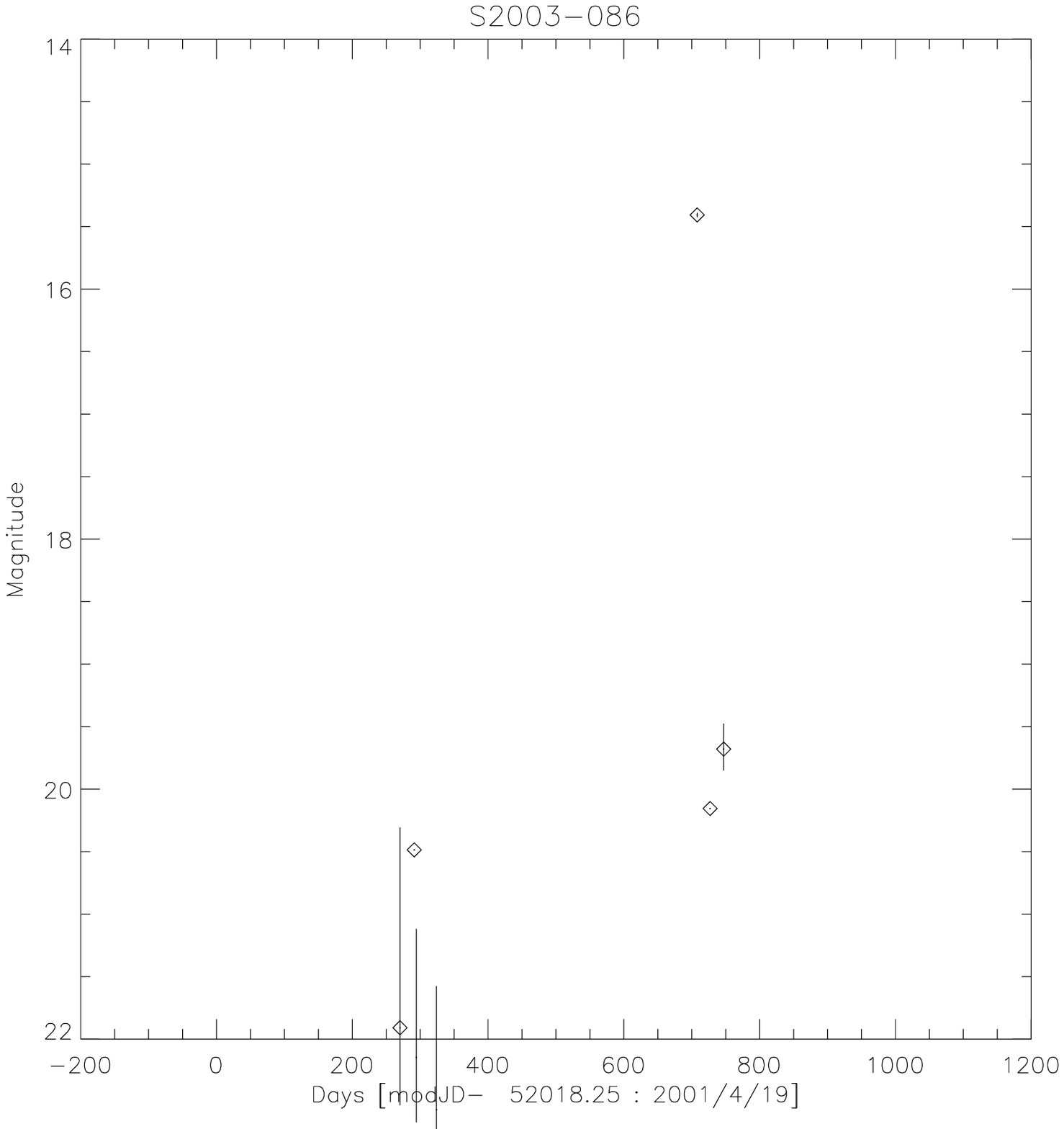}\label{fig:2003bt_lightcurve}}
\vspace{0.3in}
\subfigure[2003cc]{\includegraphics[angle=90,height=2in,width=3in]{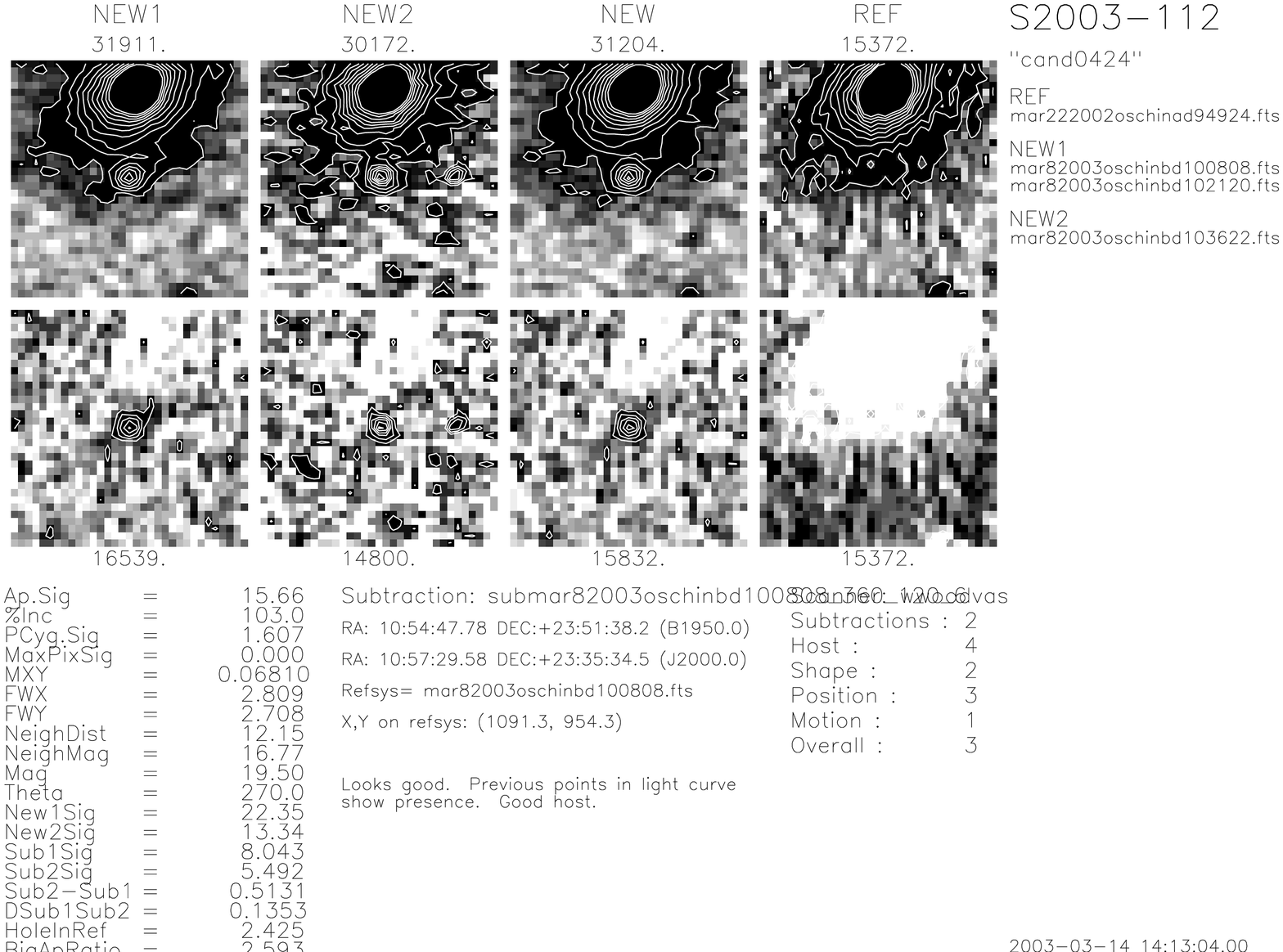}\label{fig:2003cc_discovery}}
\hspace{0.3in}
\subfigure[2003cc]{\includegraphics[height=2in]{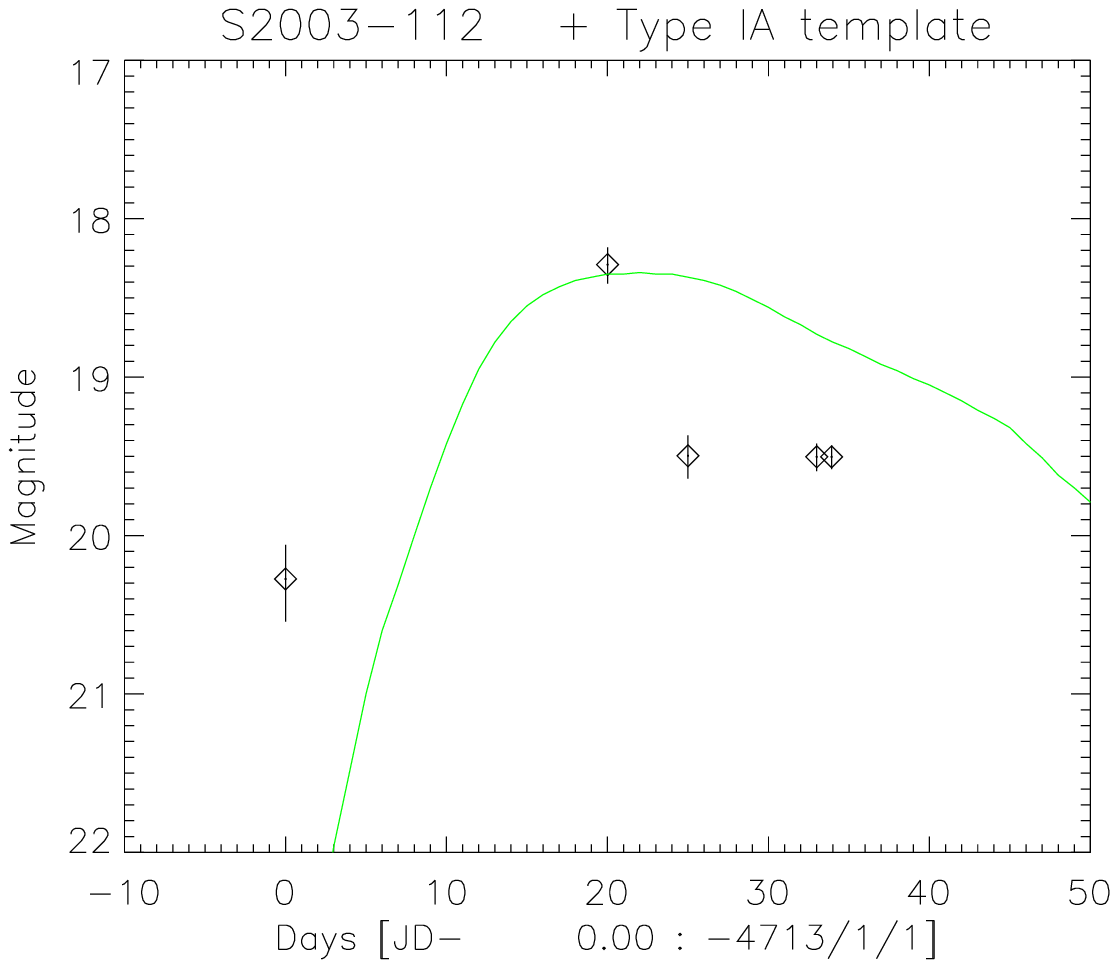}\label{fig:2003cc_lightcurve}}
\vspace{0.3in}
\end{figure}

\clearpage\pagebreak
\begin{figure}
\subfigure[2003cd]{\includegraphics[angle=90,height=2in,width=3in]{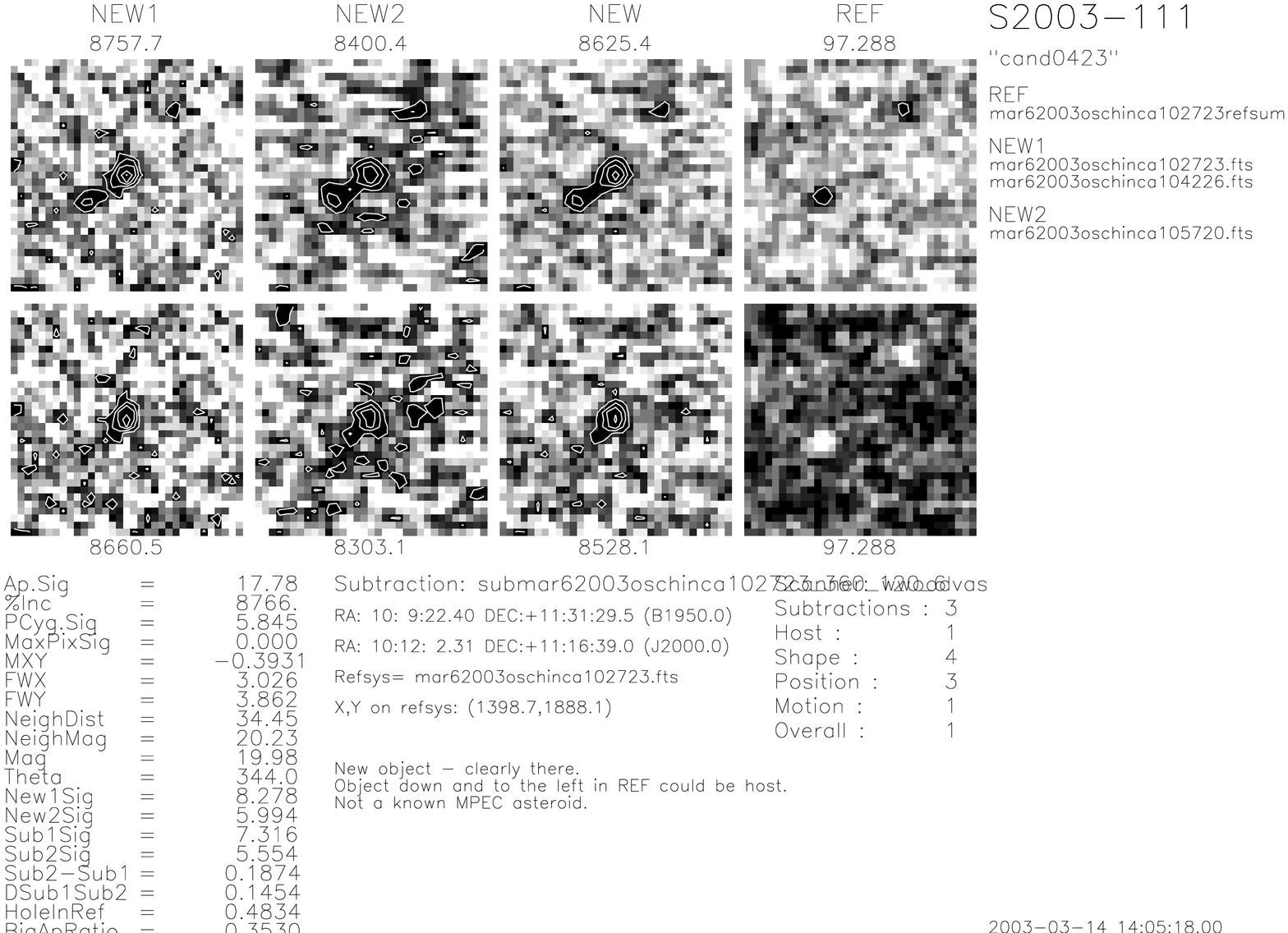}\label{fig:2003cd_discovery}}
\hspace{0.3in}
\subfigure[2003cd]{\includegraphics[height=2in]{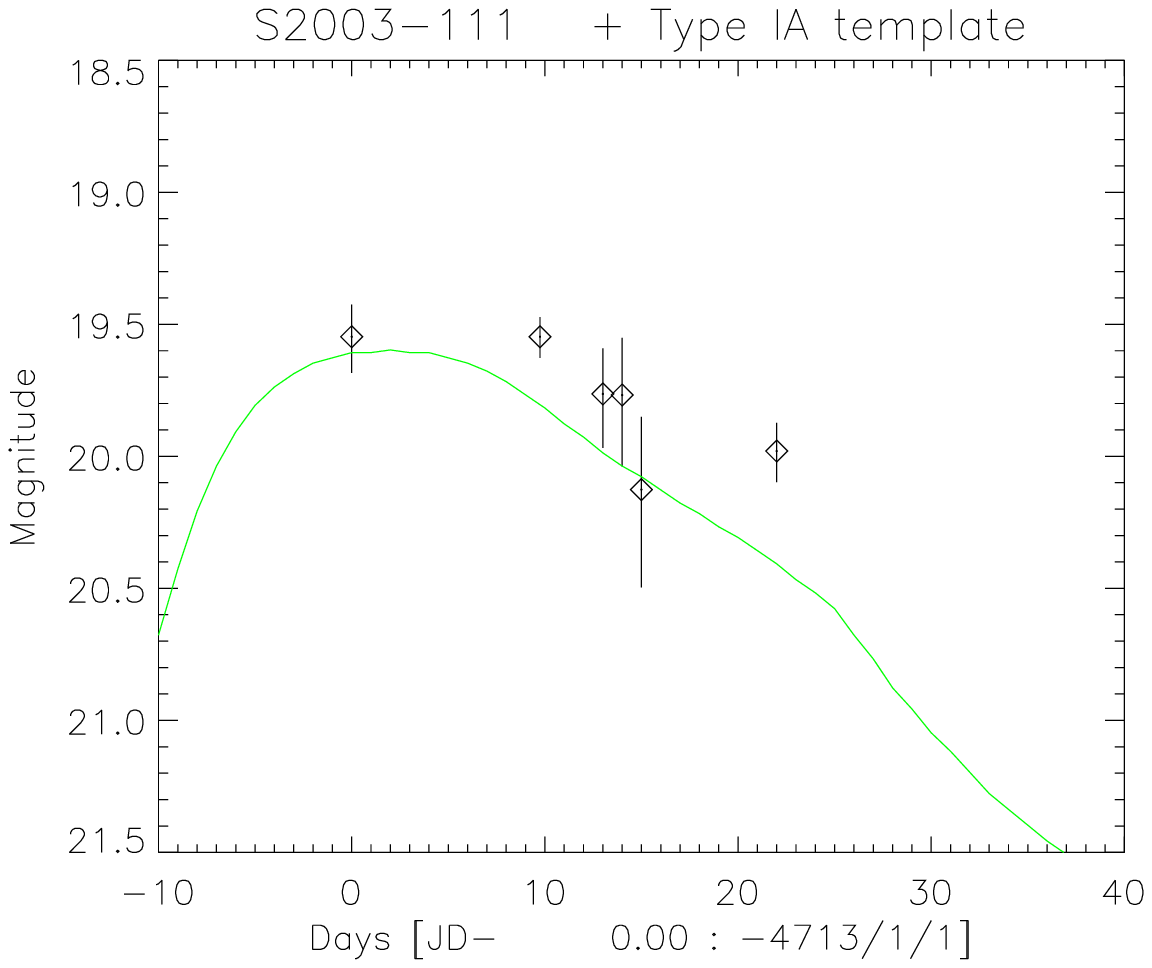}\label{fig:2003cd_lightcurve}}
\vspace{0.3in}
\subfigure[2003ce]{\includegraphics[angle=90,height=2in,width=3in]{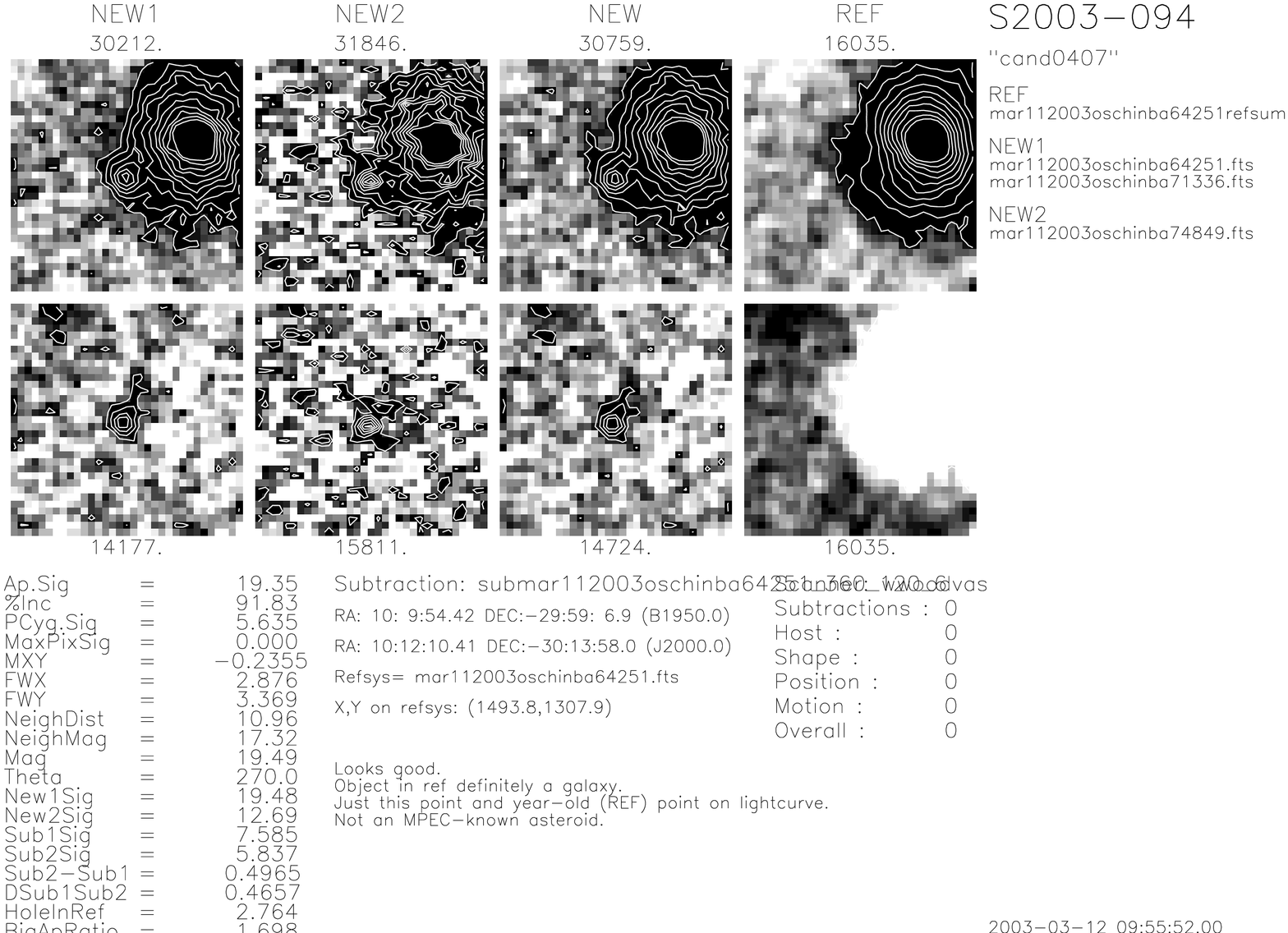}\label{fig:2003ce_discovery}}
\hspace{0.3in}
\subfigure[2003ce]{\includegraphics[height=2in]{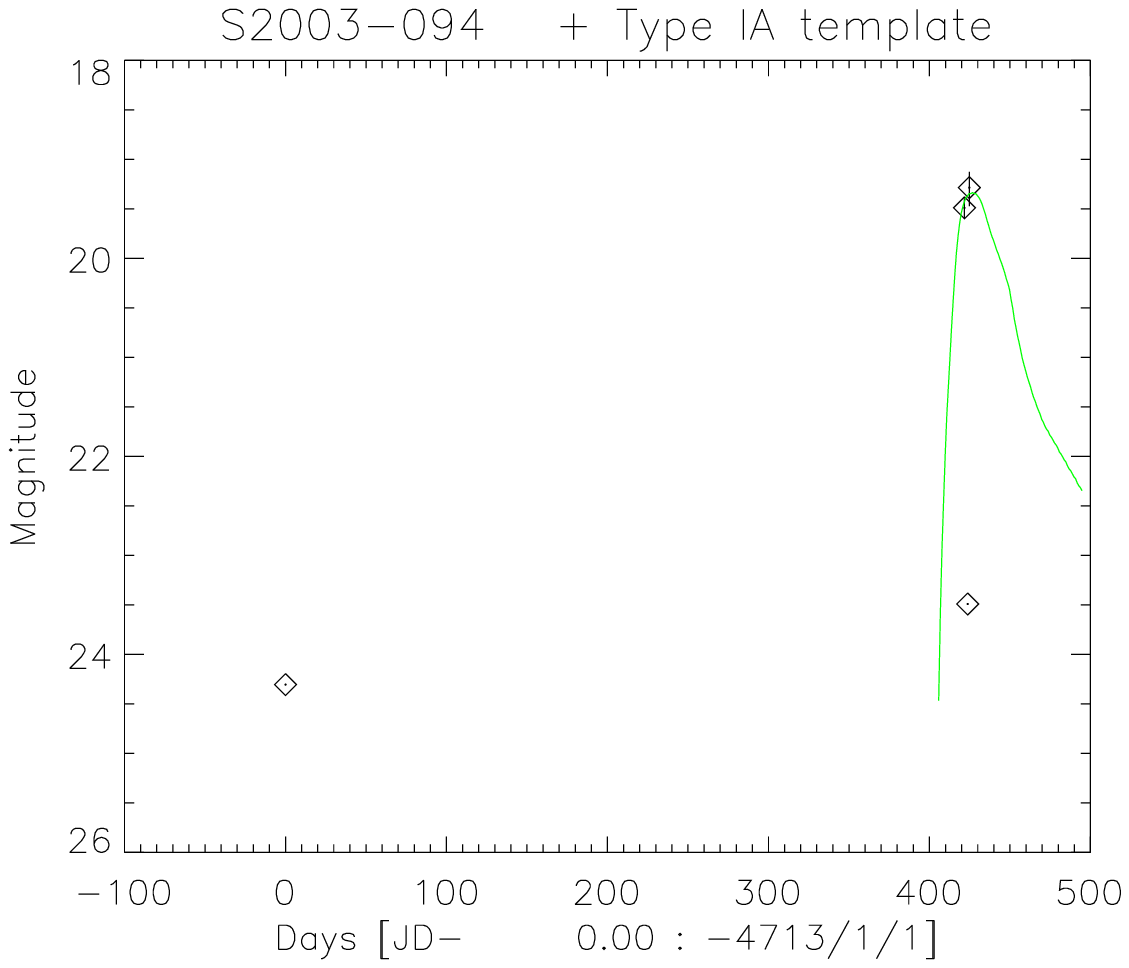}\label{fig:2003ce_lightcurve}}
\vspace{0.3in}
\subfigure[2003cf]{\includegraphics[angle=90,height=2in,width=3in]{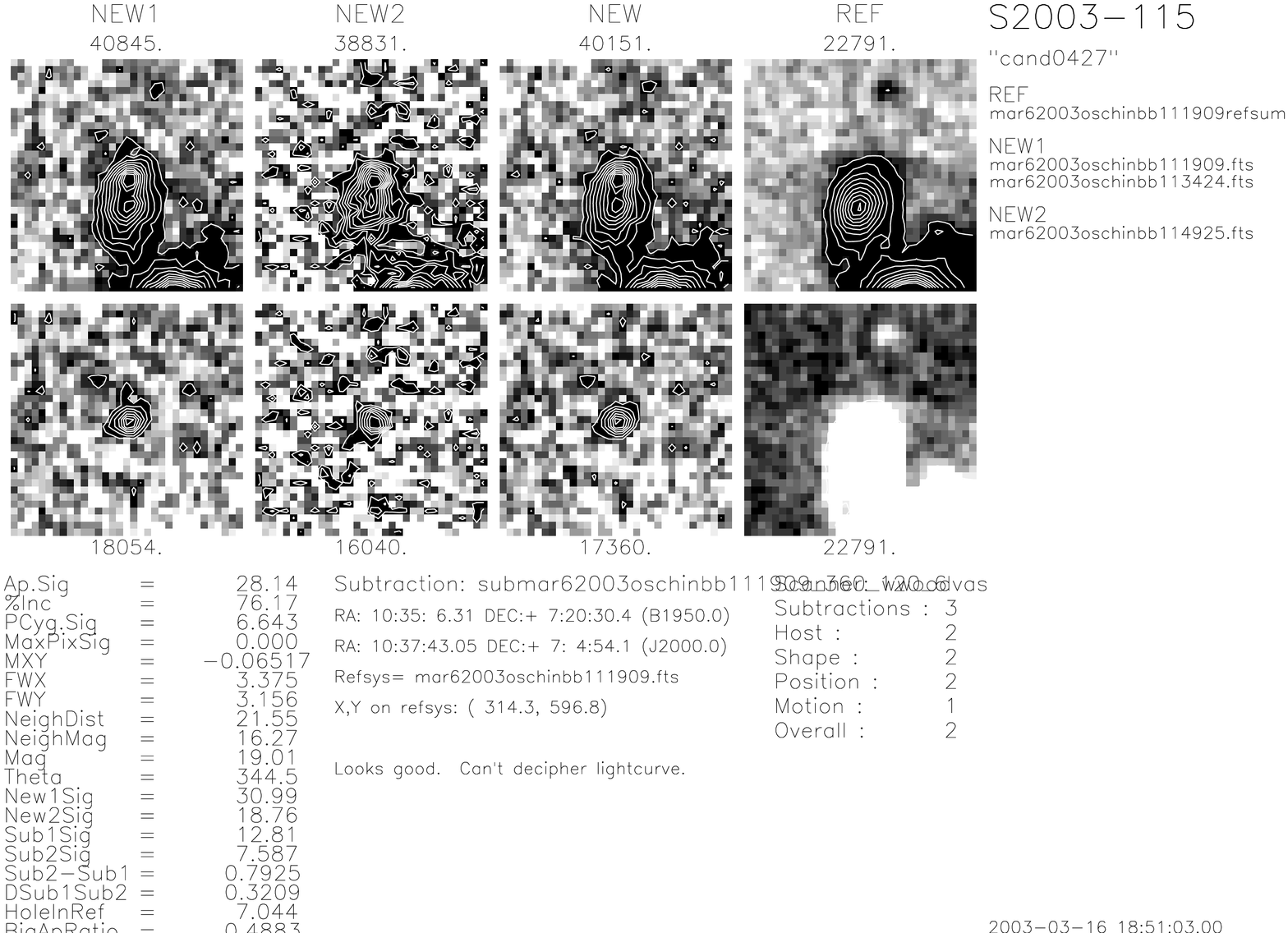}\label{fig:2003cf_discovery}}
\hspace{0.3in}
\subfigure[2003cf]{\includegraphics[height=2in]{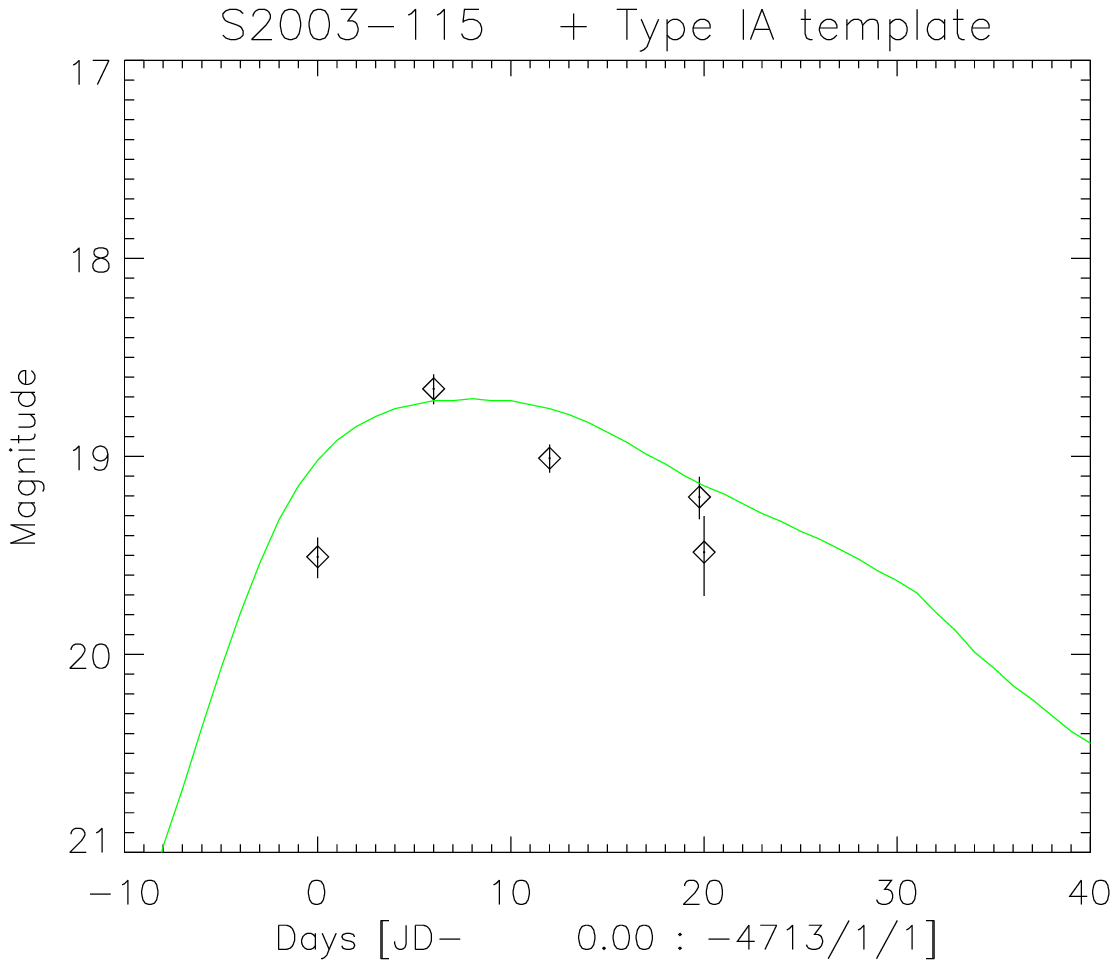}\label{fig:2003cf_lightcurve}}
\vspace{0.3in}
\end{figure}

\clearpage\pagebreak
\begin{figure}
\subfigure[2003cj]{\includegraphics[angle=90,height=2in,width=3in]{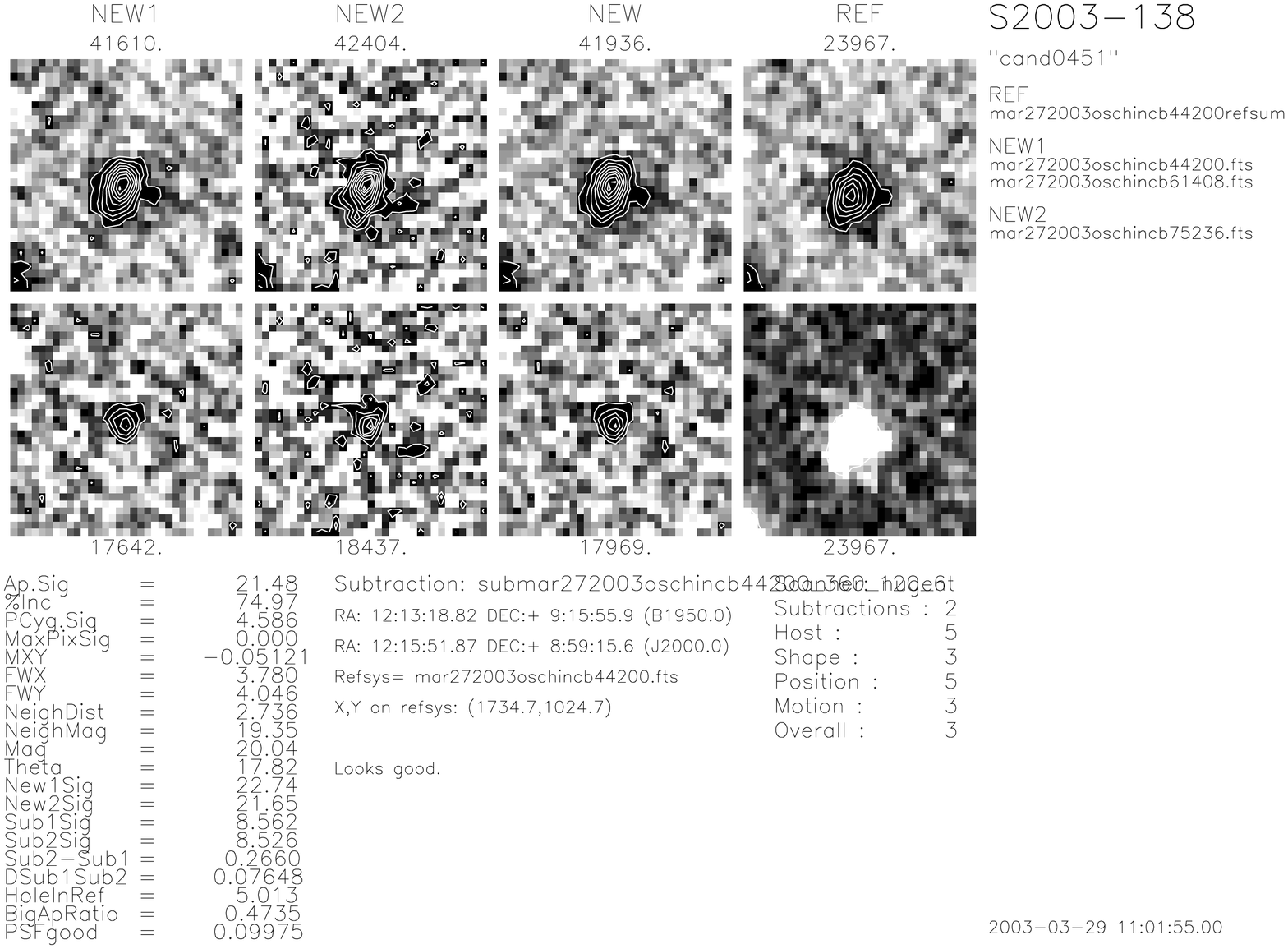}\label{fig:2003cj_discovery}}
\hspace{0.3in}
\subfigure[2003cj]{\includegraphics[height=2in]{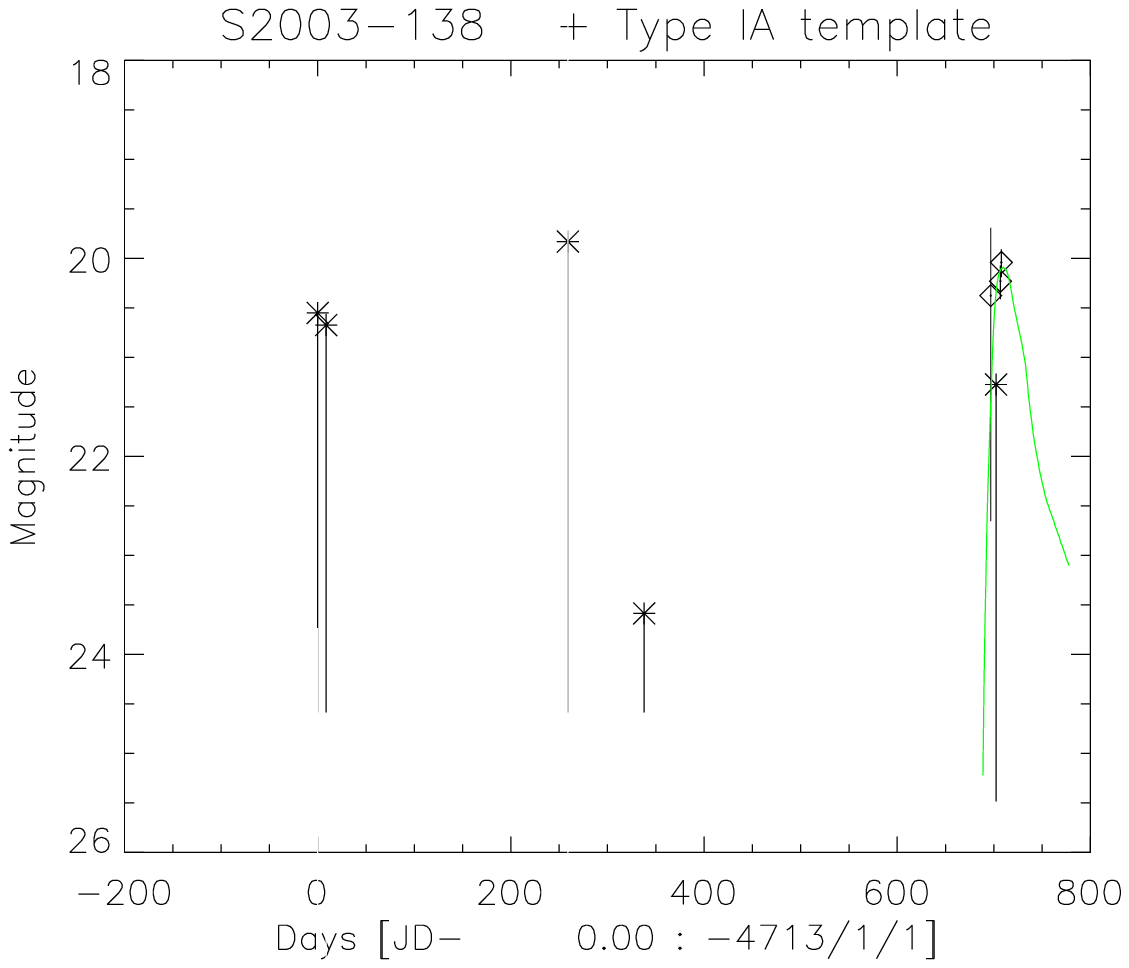}\label{fig:2003cj_lightcurve}}
\vspace{0.3in}
\subfigure[2003ck]{\includegraphics[angle=90,height=2in,width=3in]{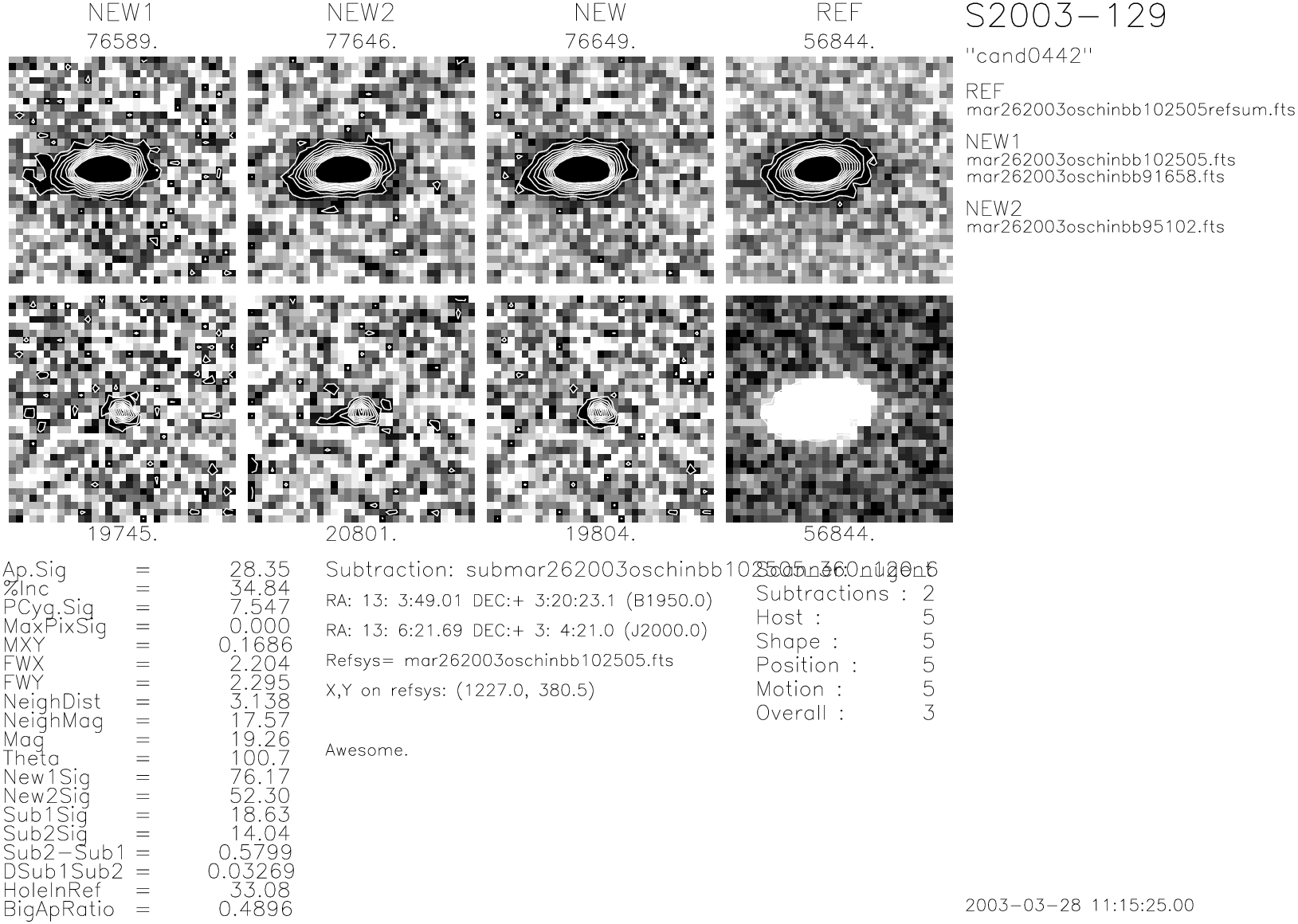}\label{fig:2003ck_discovery}}
\hspace{0.3in}
\subfigure[2003ck]{\includegraphics[height=2in]{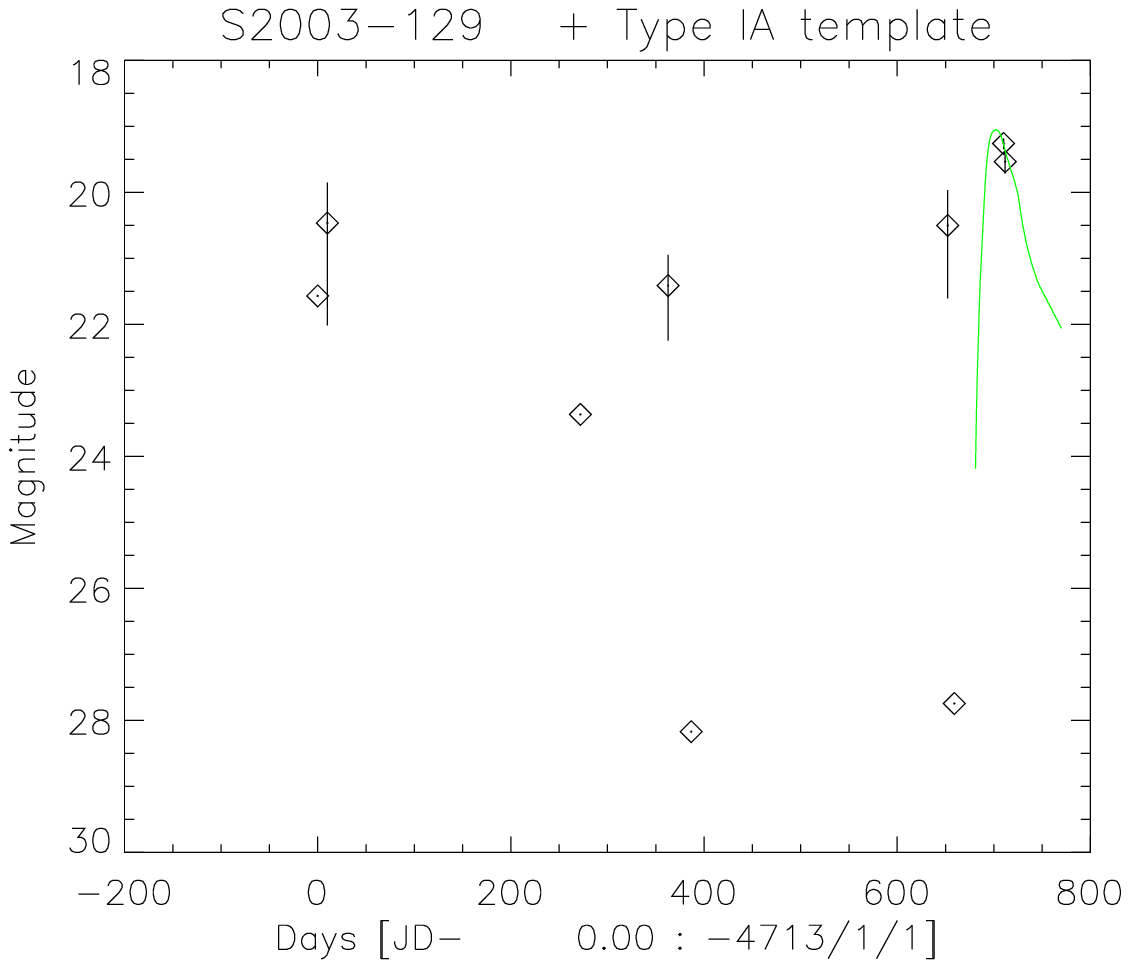}\label{fig:2003ck_lightcurve}}
\vspace{0.3in}
\subfigure[2003cl]{\includegraphics[angle=90,height=2in,width=3in]{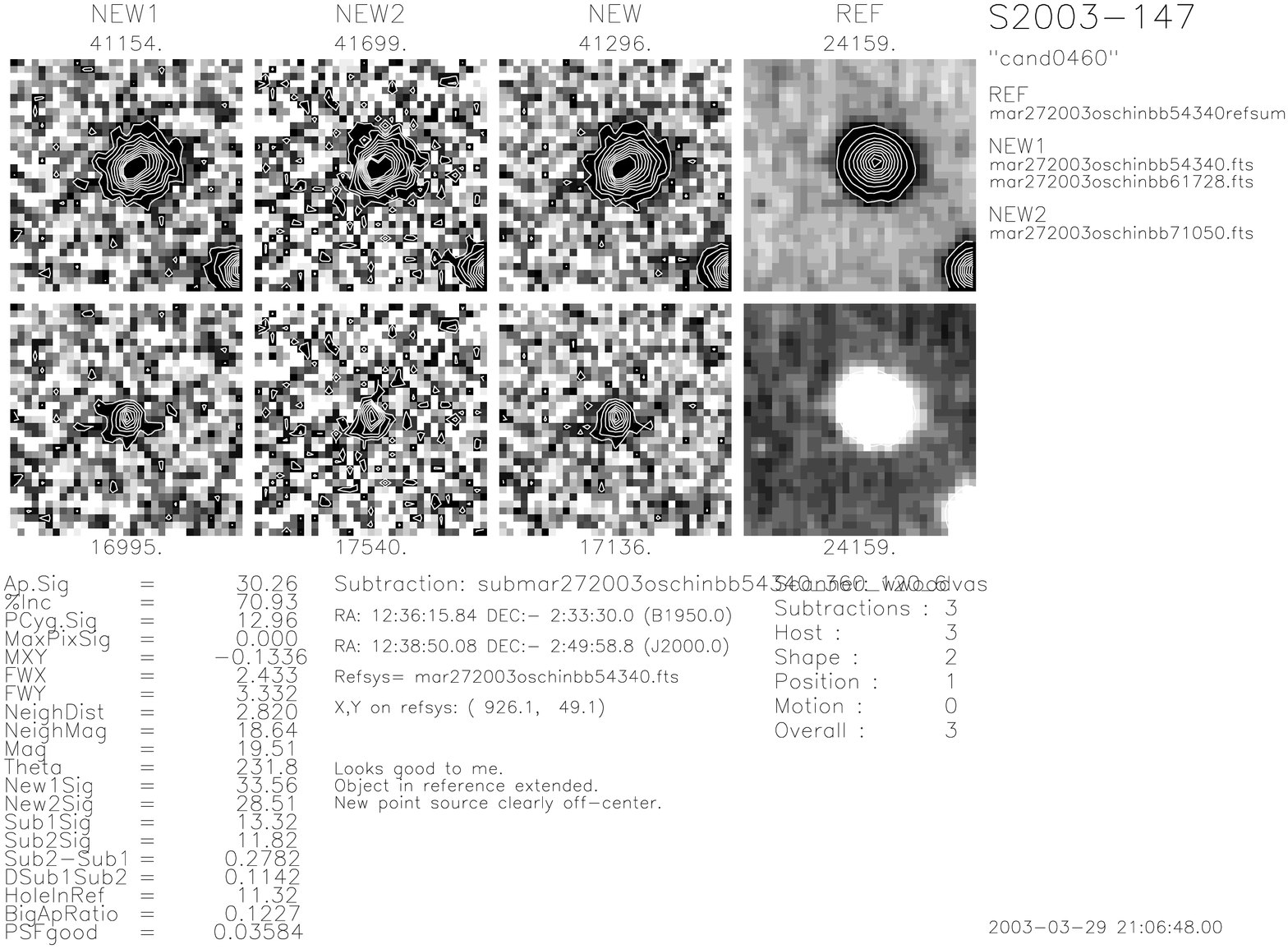}\label{fig:2003cl_discovery}}
\hspace{0.3in}
\subfigure[2003cl]{\includegraphics[height=2in]{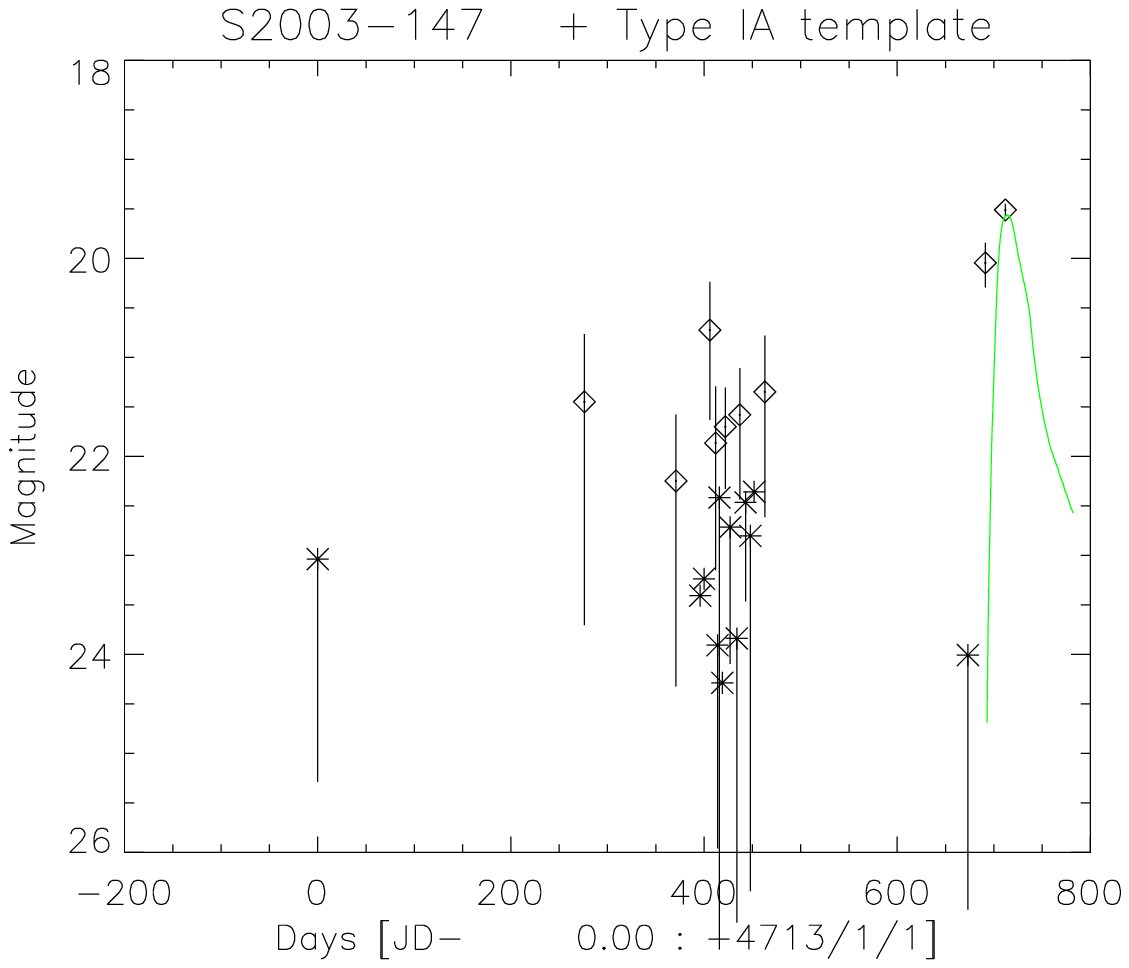}\label{fig:2003cl_lightcurve}}
\vspace{0.3in}
\end{figure}

\clearpage\pagebreak
\begin{figure}
\subfigure[2003cn]{\includegraphics[angle=90,height=2in,width=3in]{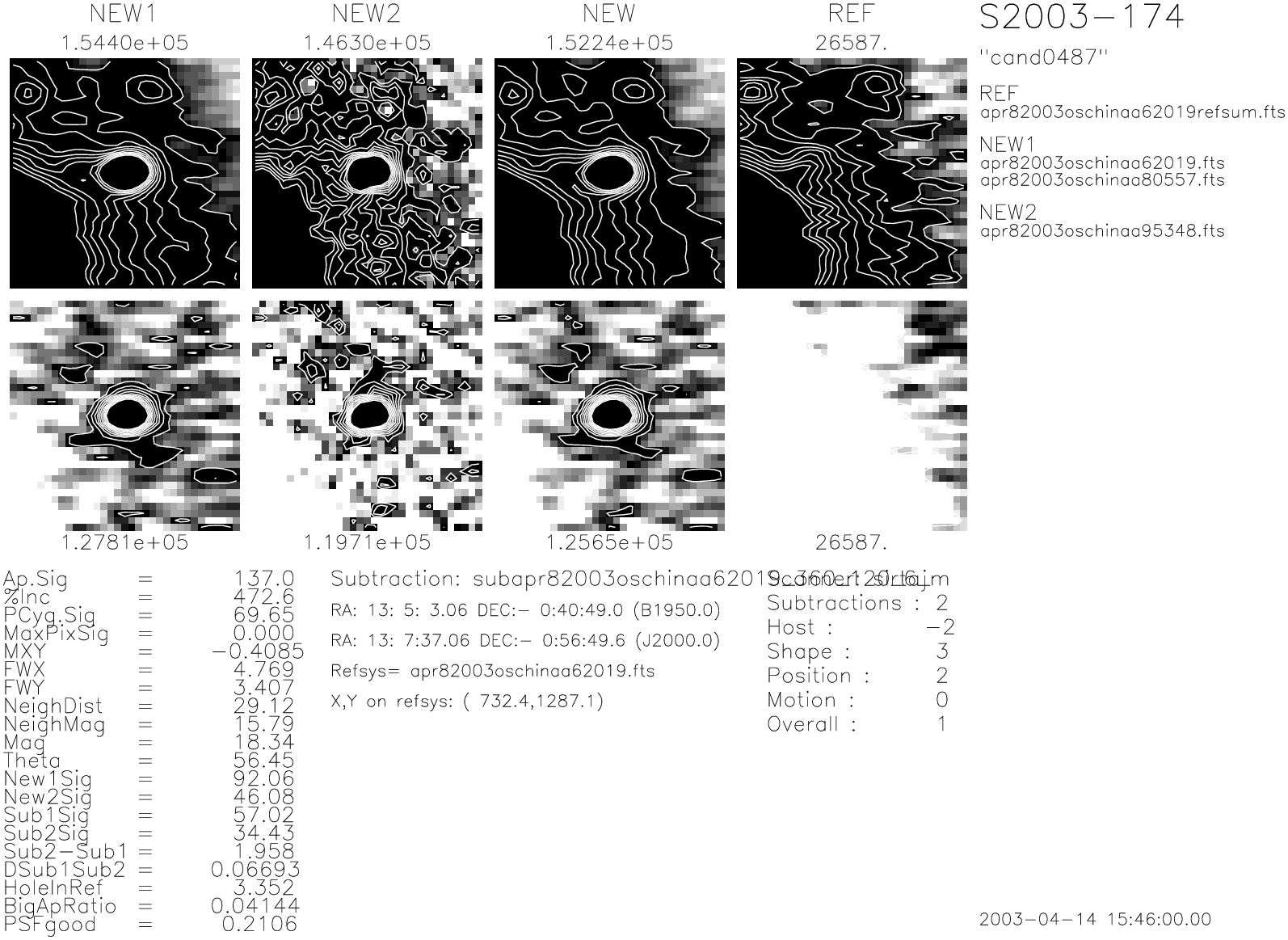}\label{fig:2003cn_discovery}}
\hspace{0.3in}
\subfigure[2003cn]{\includegraphics[height=2in]{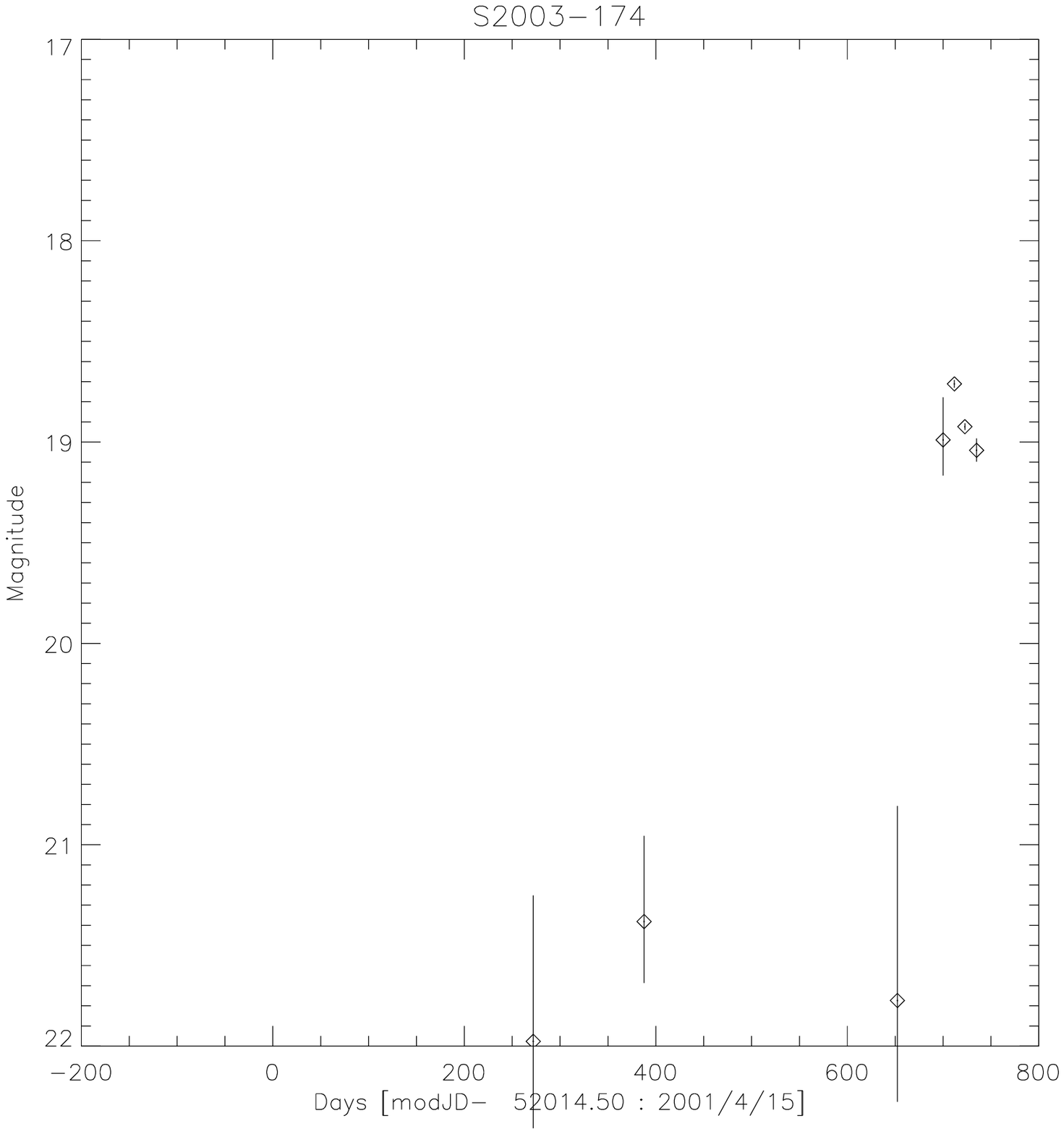}\label{fig:2003cn_lightcurve}}
\vspace{0.3in}
\subfigure[2003co]{\includegraphics[angle=90,height=2in,width=3in]{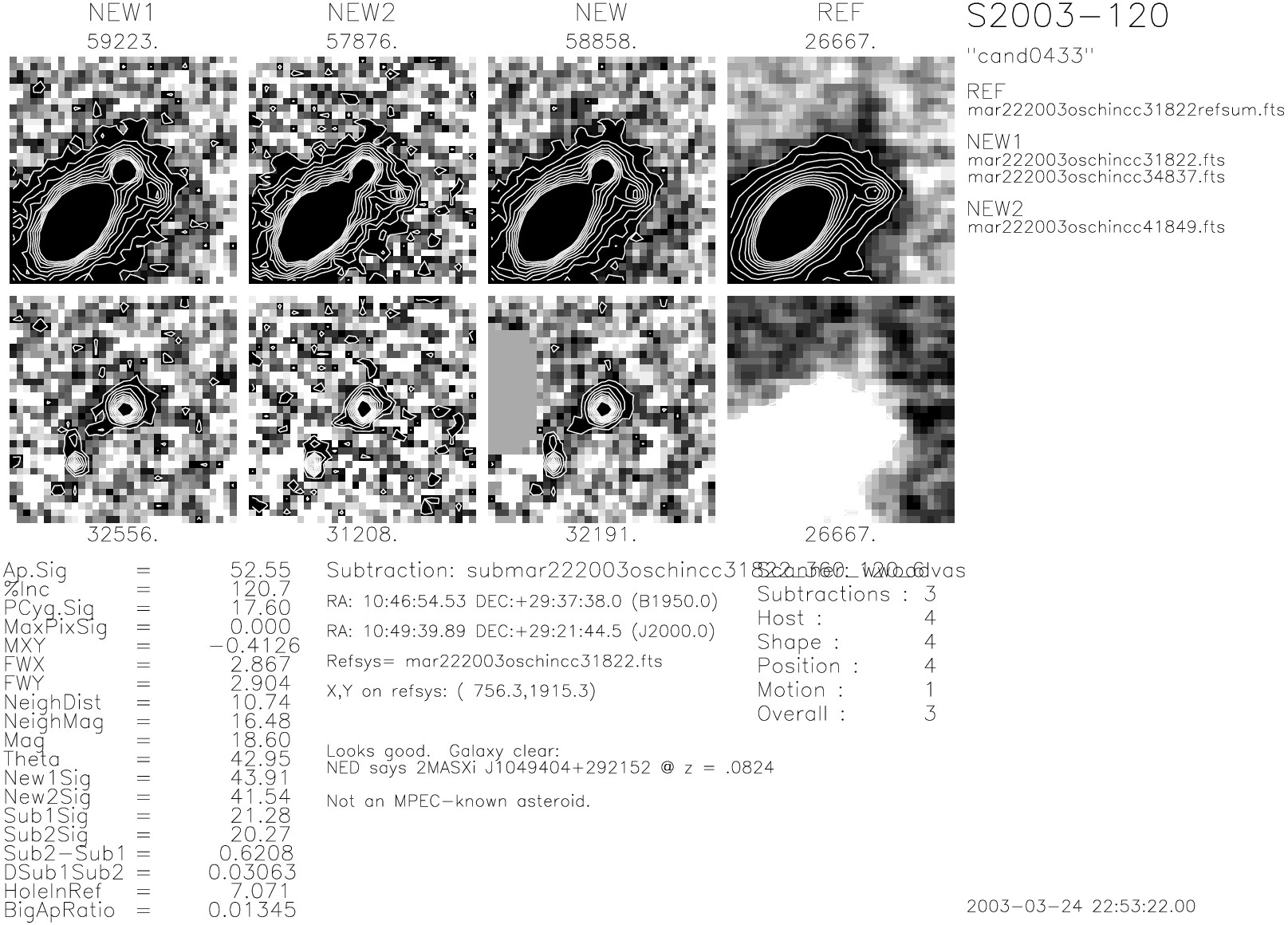}\label{fig:2003co_discovery}}
\hspace{0.3in}
\subfigure[2003co]{\includegraphics[height=2in]{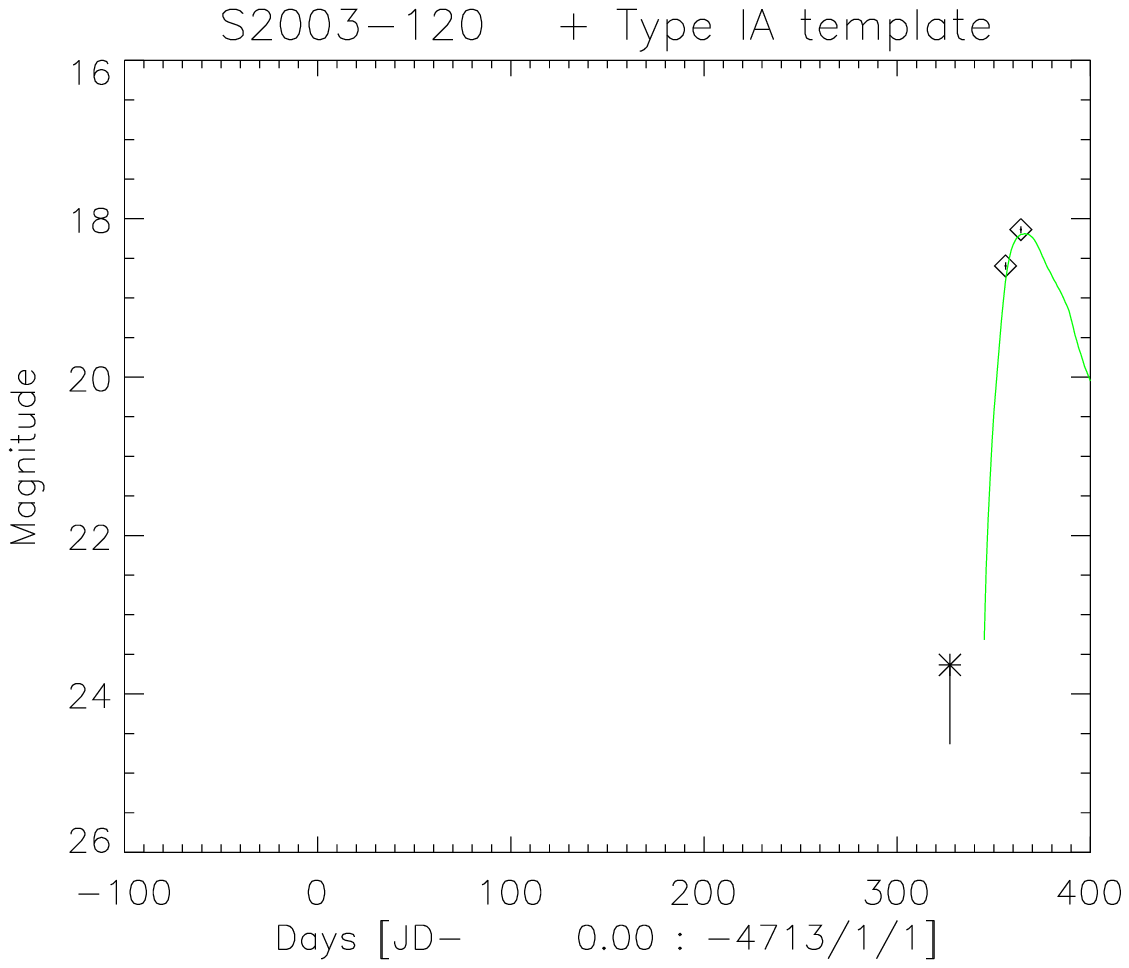}\label{fig:2003co_lightcurve}}
\vspace{0.3in}
\subfigure[2003cs]{\includegraphics[angle=90,height=2in,width=3in]{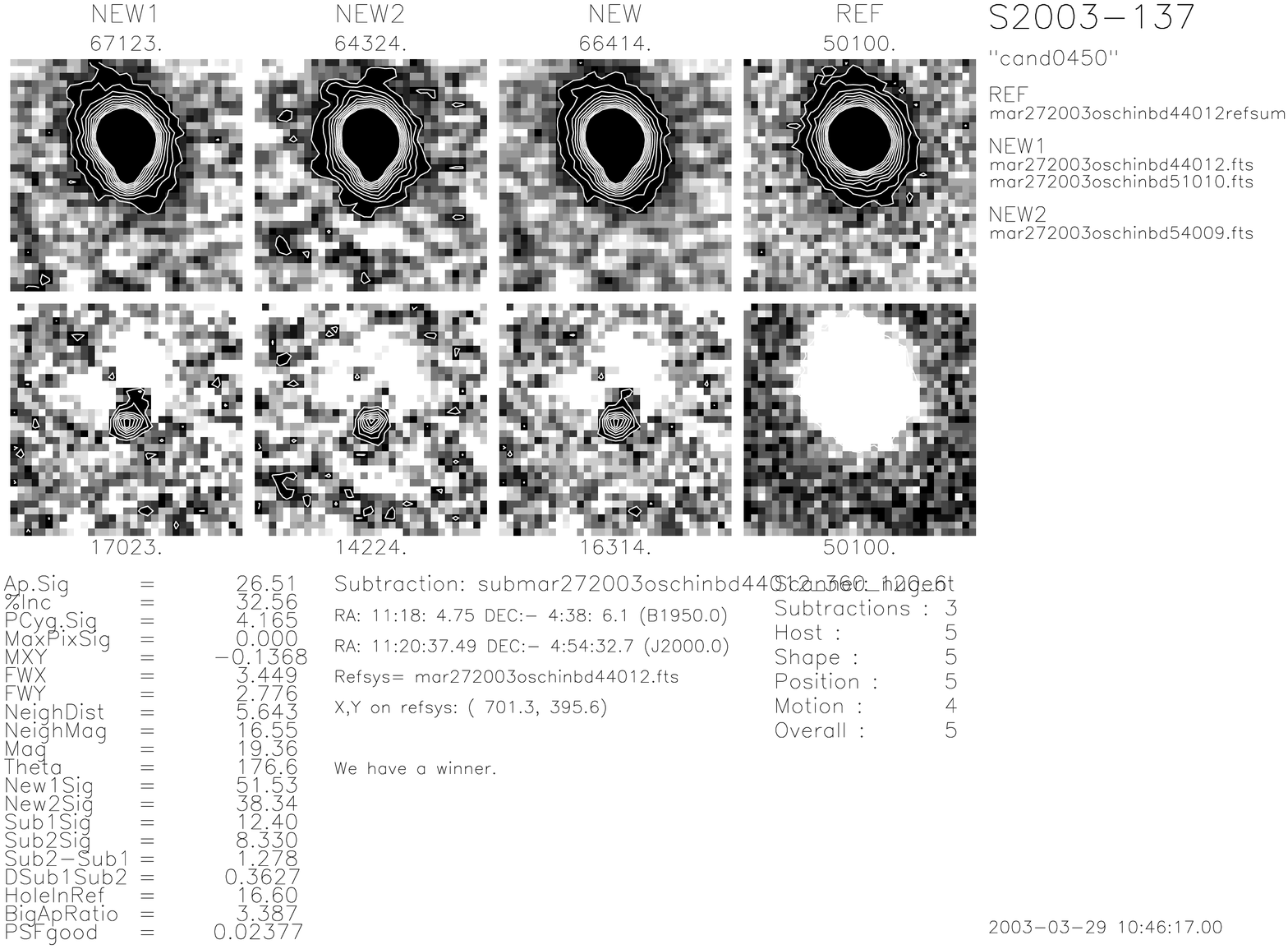}\label{fig:2003cs_discovery}}
\hspace{0.3in}
\subfigure[2003cs]{\includegraphics[height=2in]{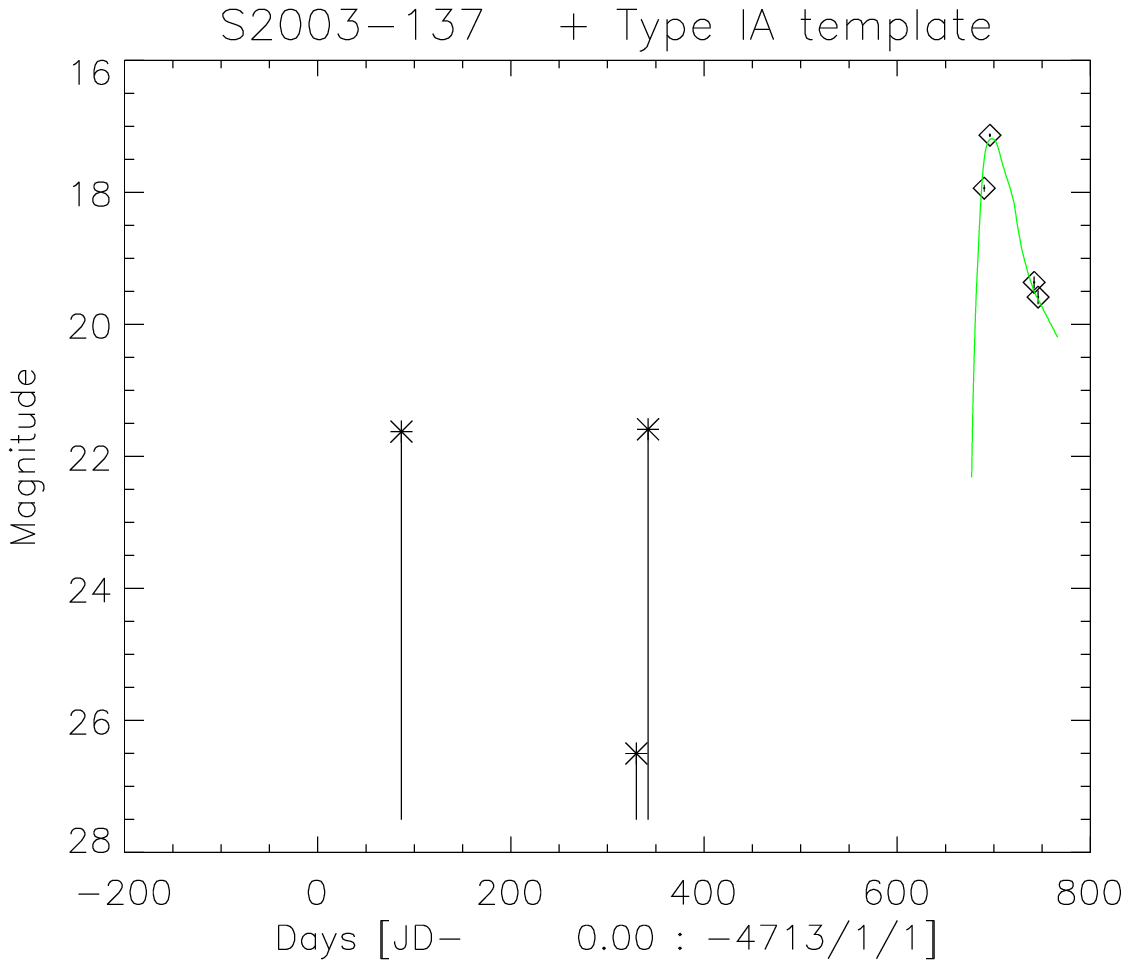}\label{fig:2003cs_lightcurve}}
\vspace{0.3in}
\end{figure}

\clearpage\pagebreak
\begin{figure}
\subfigure[2003ct]{\includegraphics[angle=90,height=2in,width=2in]{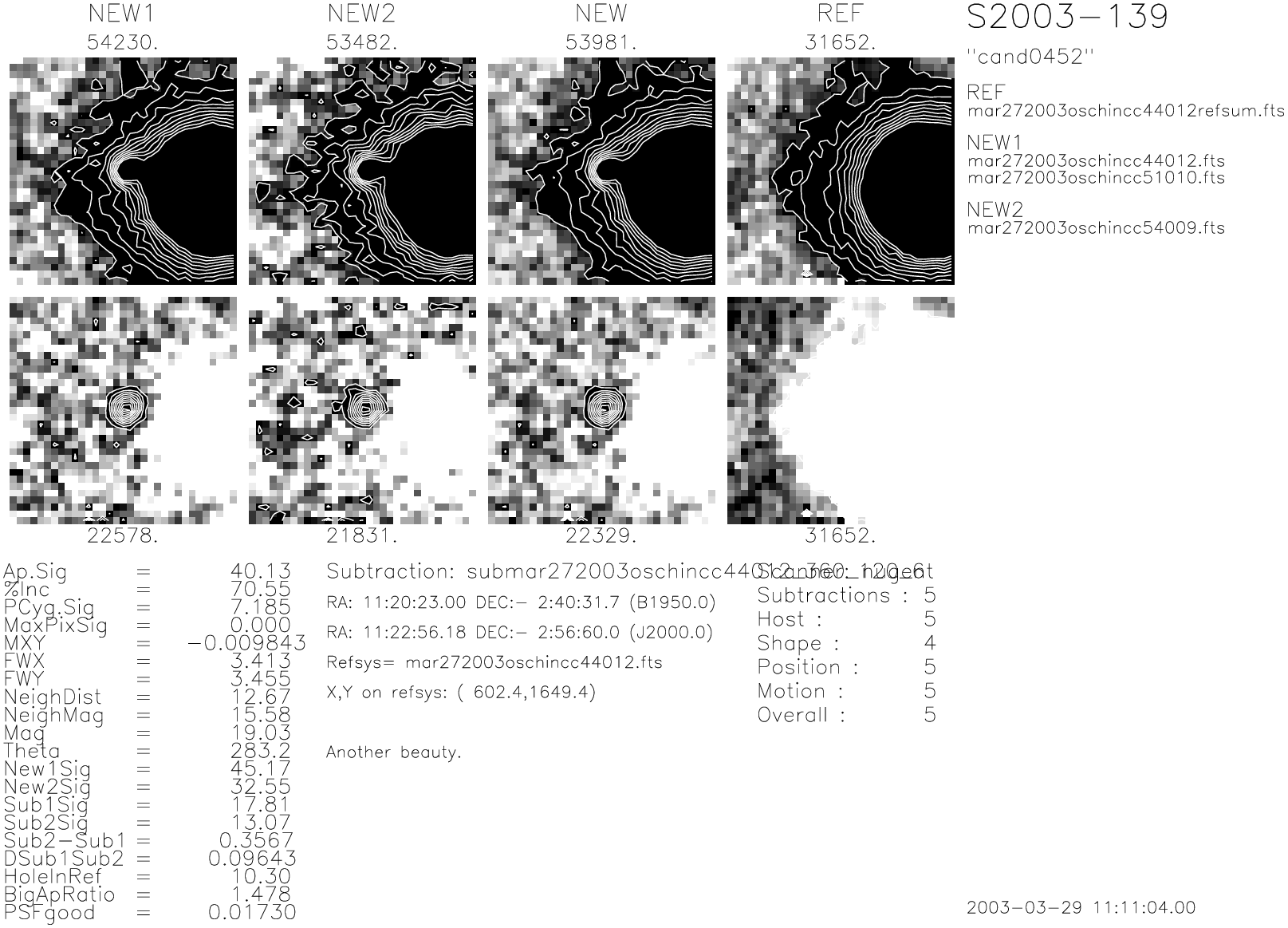}\label{fig:2003ct_discovery}}
\hspace{0.3in}
\subfigure[2003ct]{\includegraphics[width=2in]{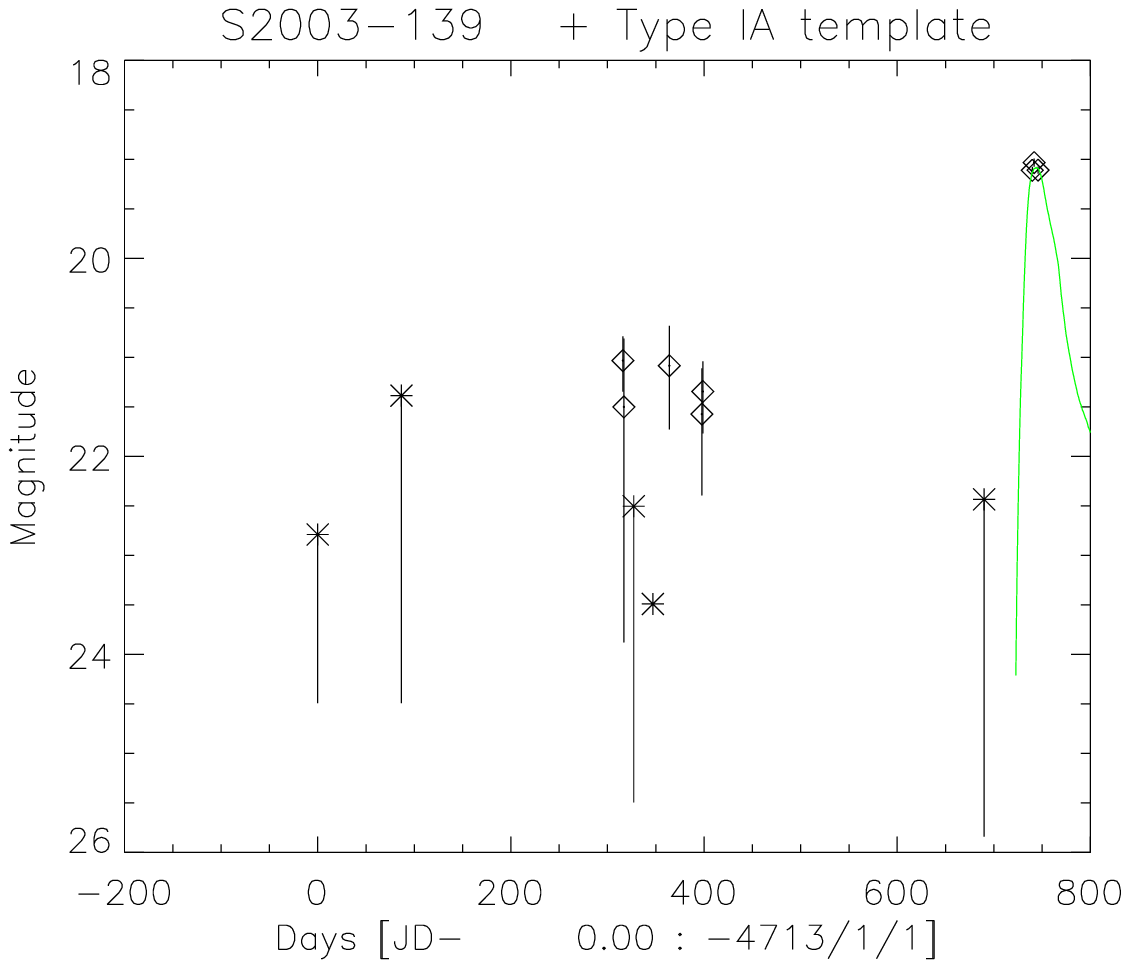}\label{fig:2003ct_lightcurve}}
\vspace{0.3in}
\subfigure[2003cu]{\includegraphics[angle=90,height=2in,width=2in]{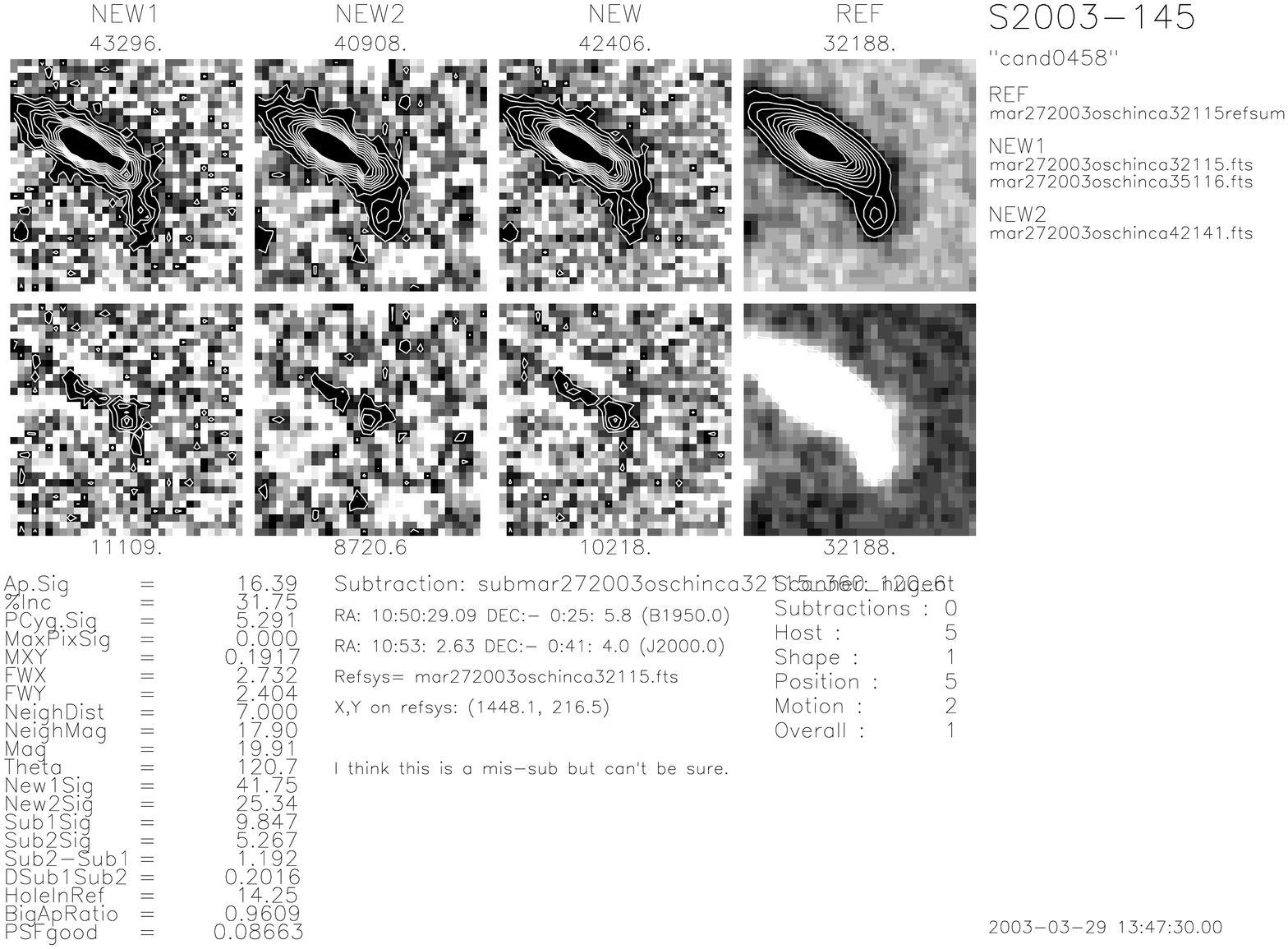}\label{fig:2003cu_discovery}}
\hspace{0.3in}
\subfigure[2003cu]{\includegraphics[width=2in]{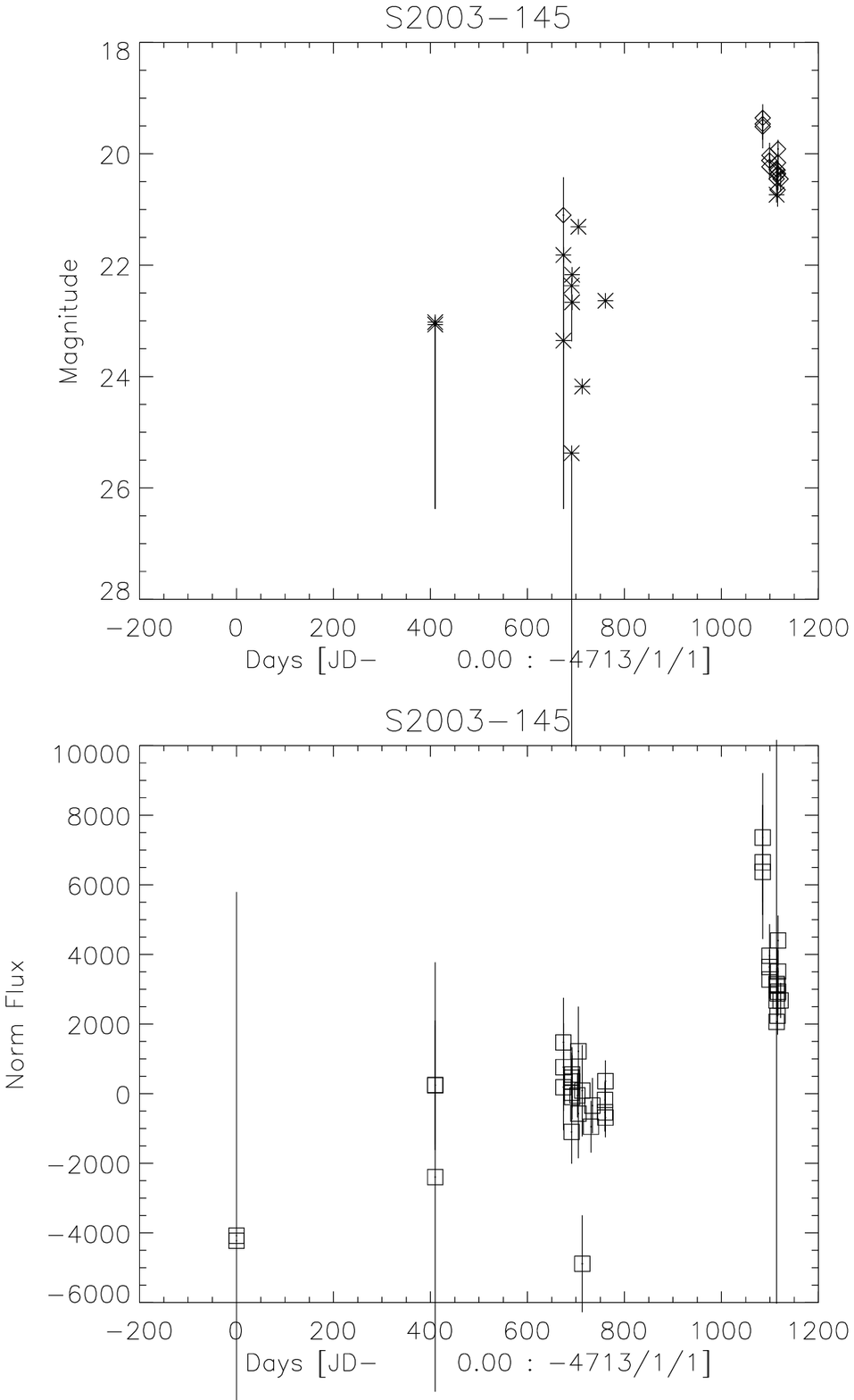}\label{fig:2003cu_lightcurve}}
\vspace{0.3in}
\subfigure[2003cv]{\includegraphics[angle=90,height=2in,width=2in]{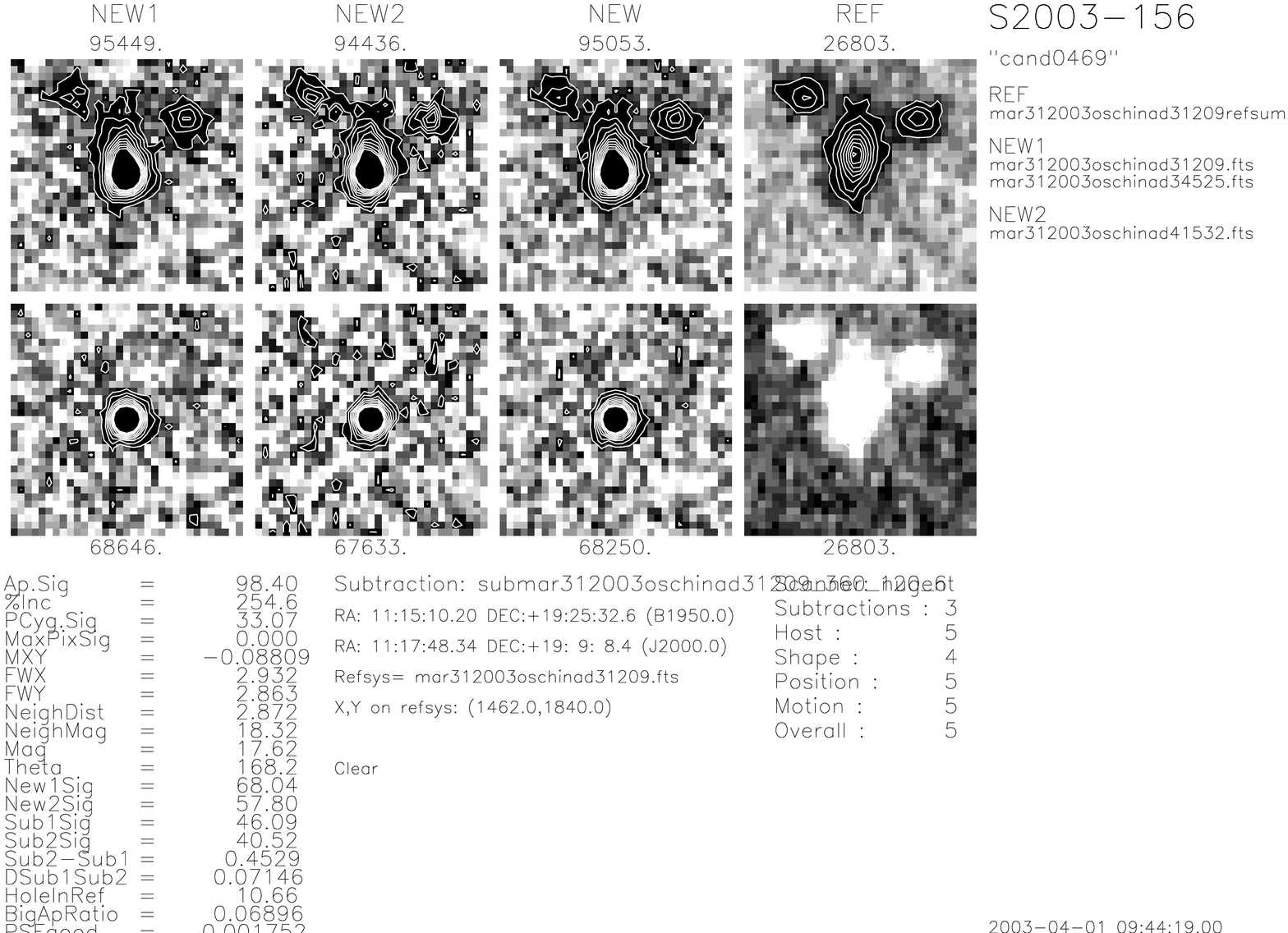}\label{fig:2003cv_discovery}}
\hspace{0.3in}
\subfigure[2003cv]{\includegraphics[width=2in]{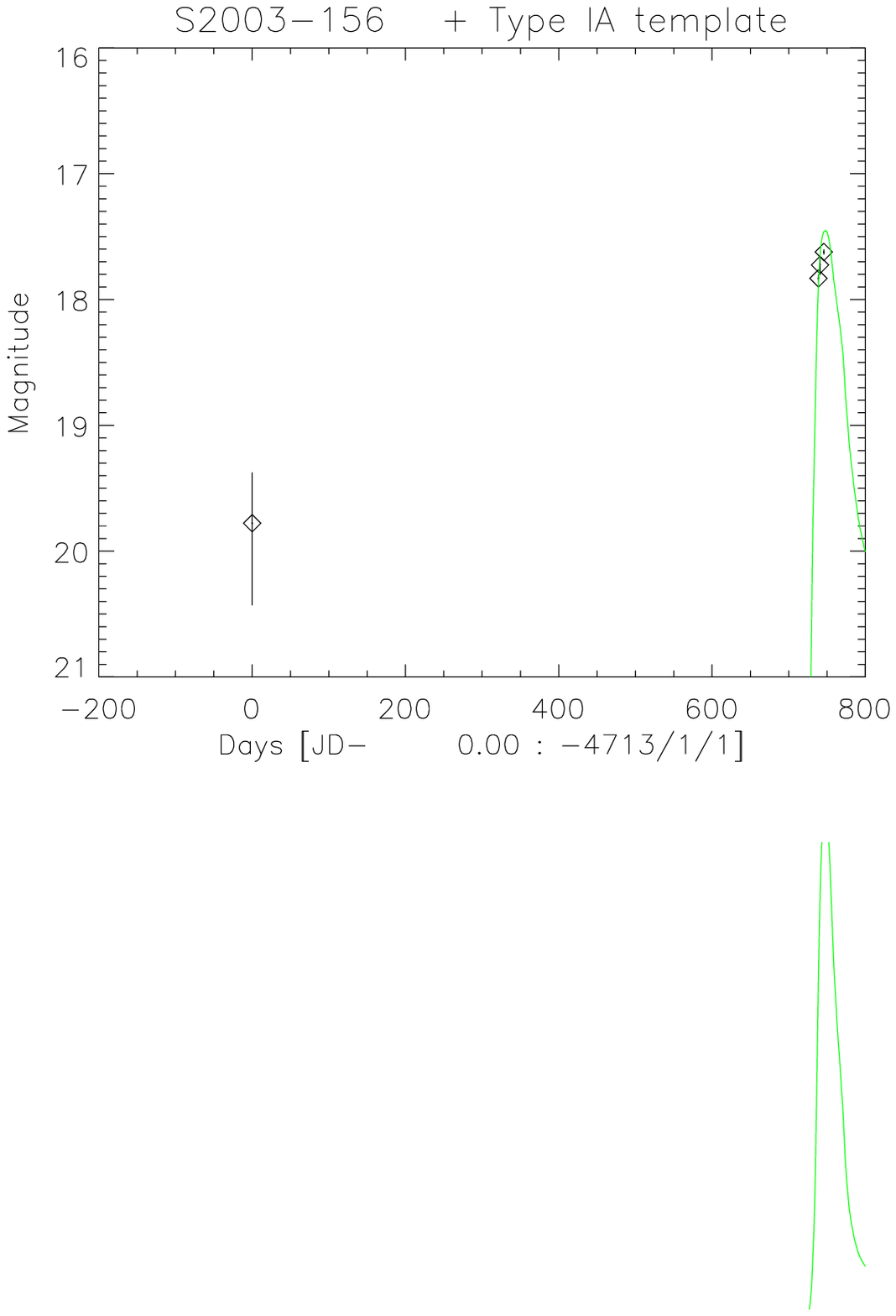}\label{fig:2003cv_lightcurve}}
\vspace{0.3in}
\end{figure}

\clearpage\pagebreak
\begin{figure}
\subfigure[2003cw]{\includegraphics[angle=90,height=2in,width=3in]{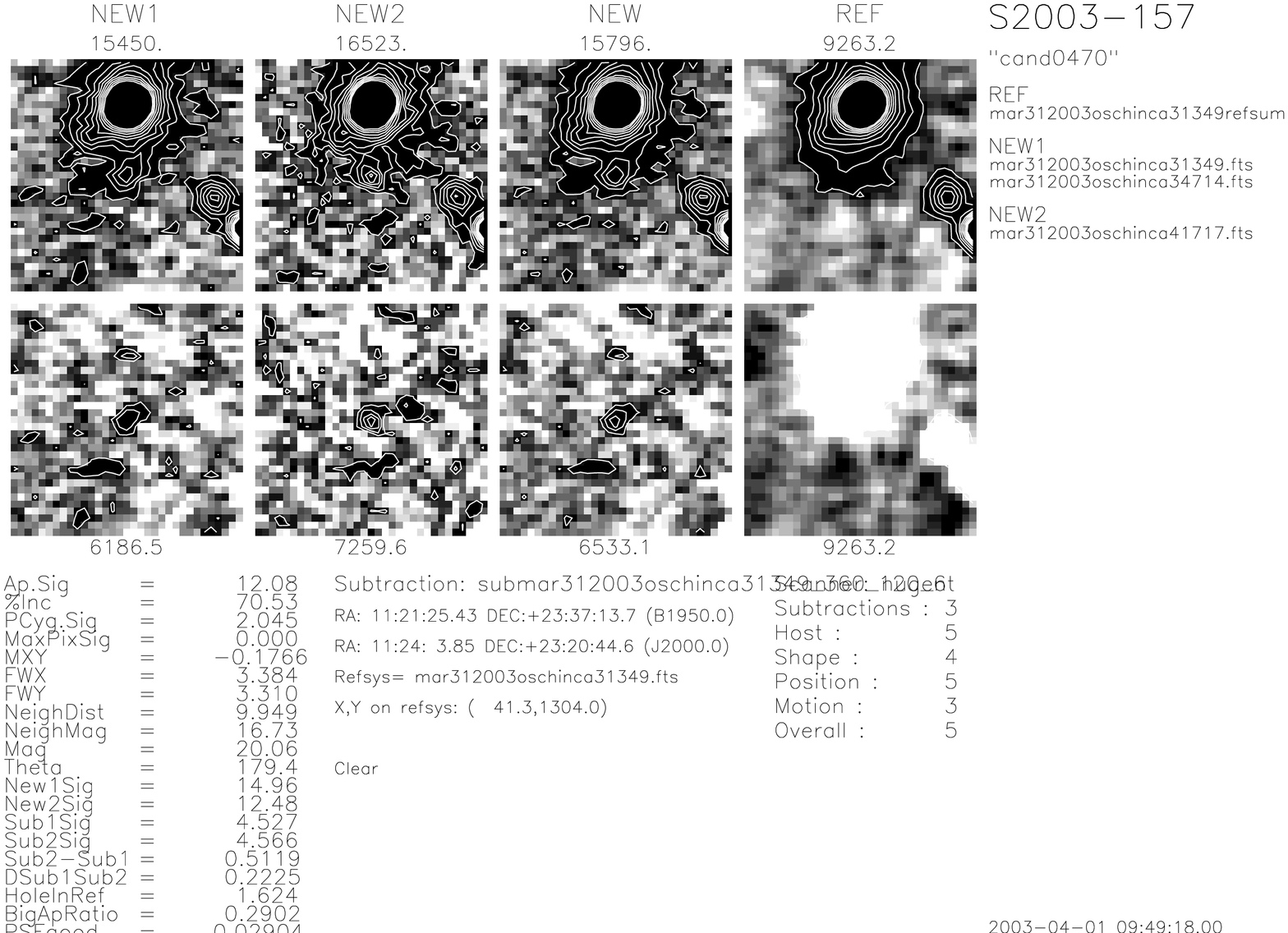}\label{fig:2003cw_discovery}}
\hspace{0.3in}
\subfigure[2003cw]{\includegraphics[height=2in]{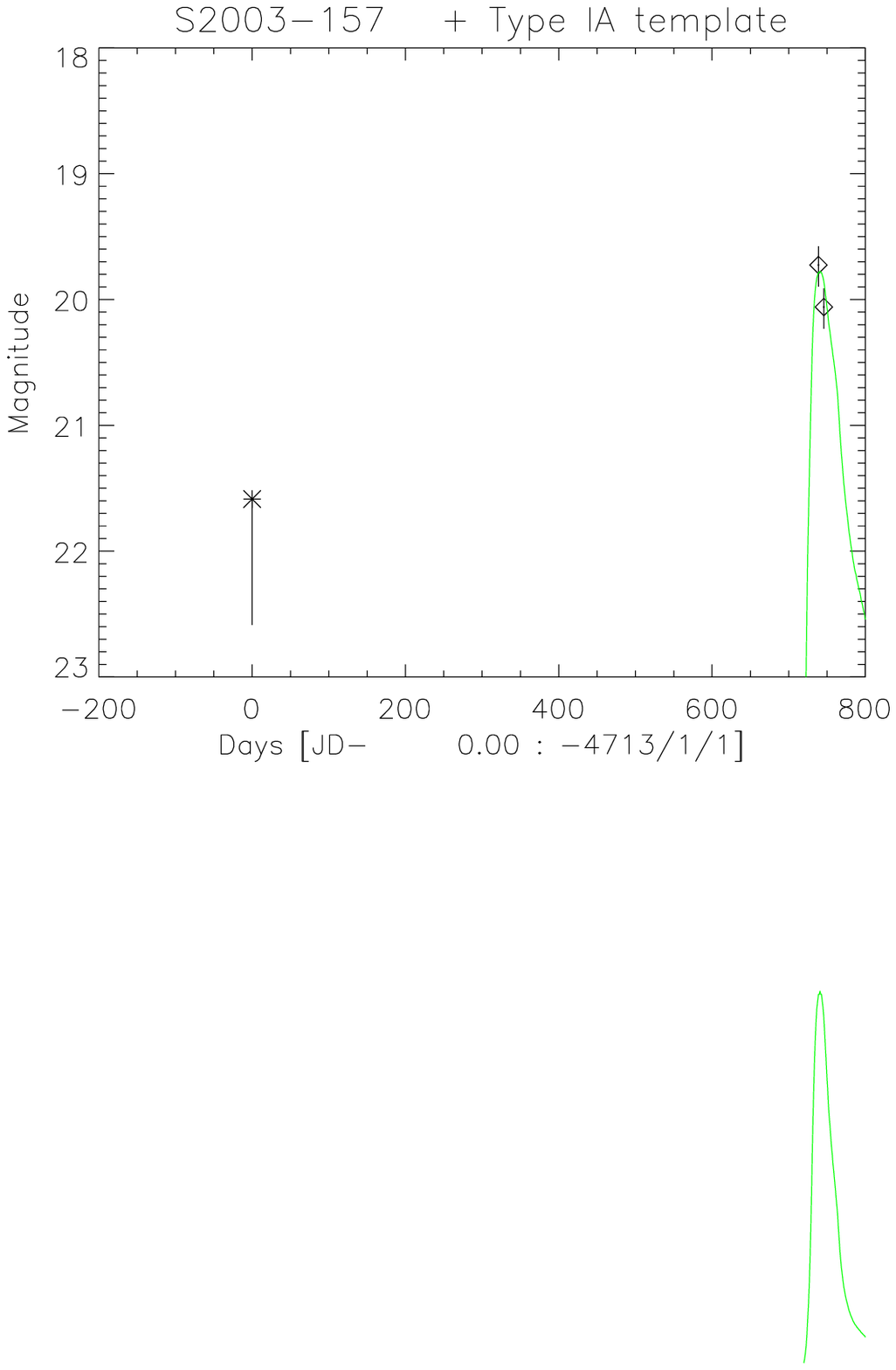}\label{fig:2003cw_lightcurve}}
\vspace{0.3in}
\subfigure[2003cx]{\includegraphics[angle=90,height=2in,width=3in]{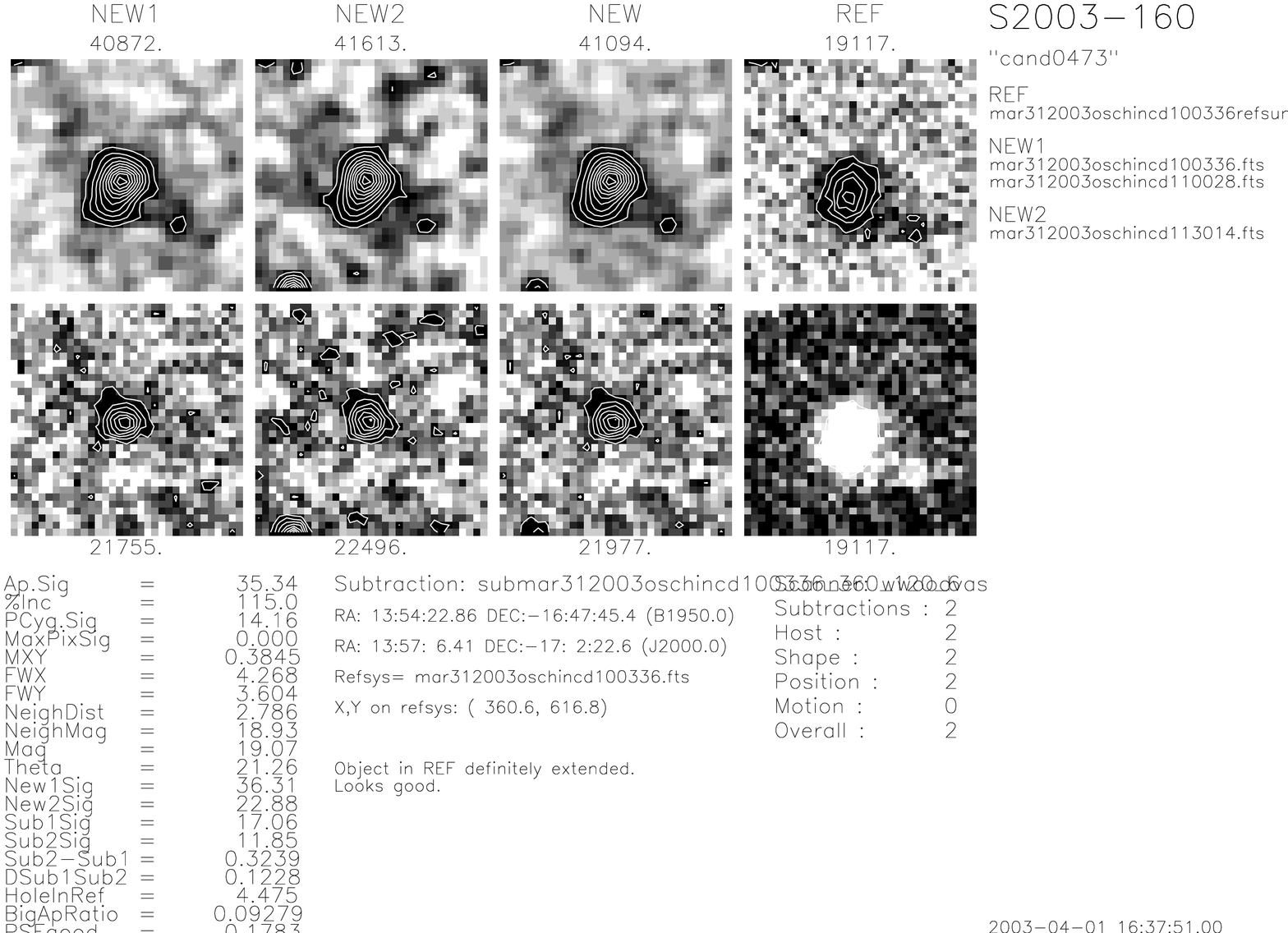}\label{fig:2003cx_discovery}}
\hspace{0.3in}
\subfigure[2003cx]{\includegraphics[height=2in]{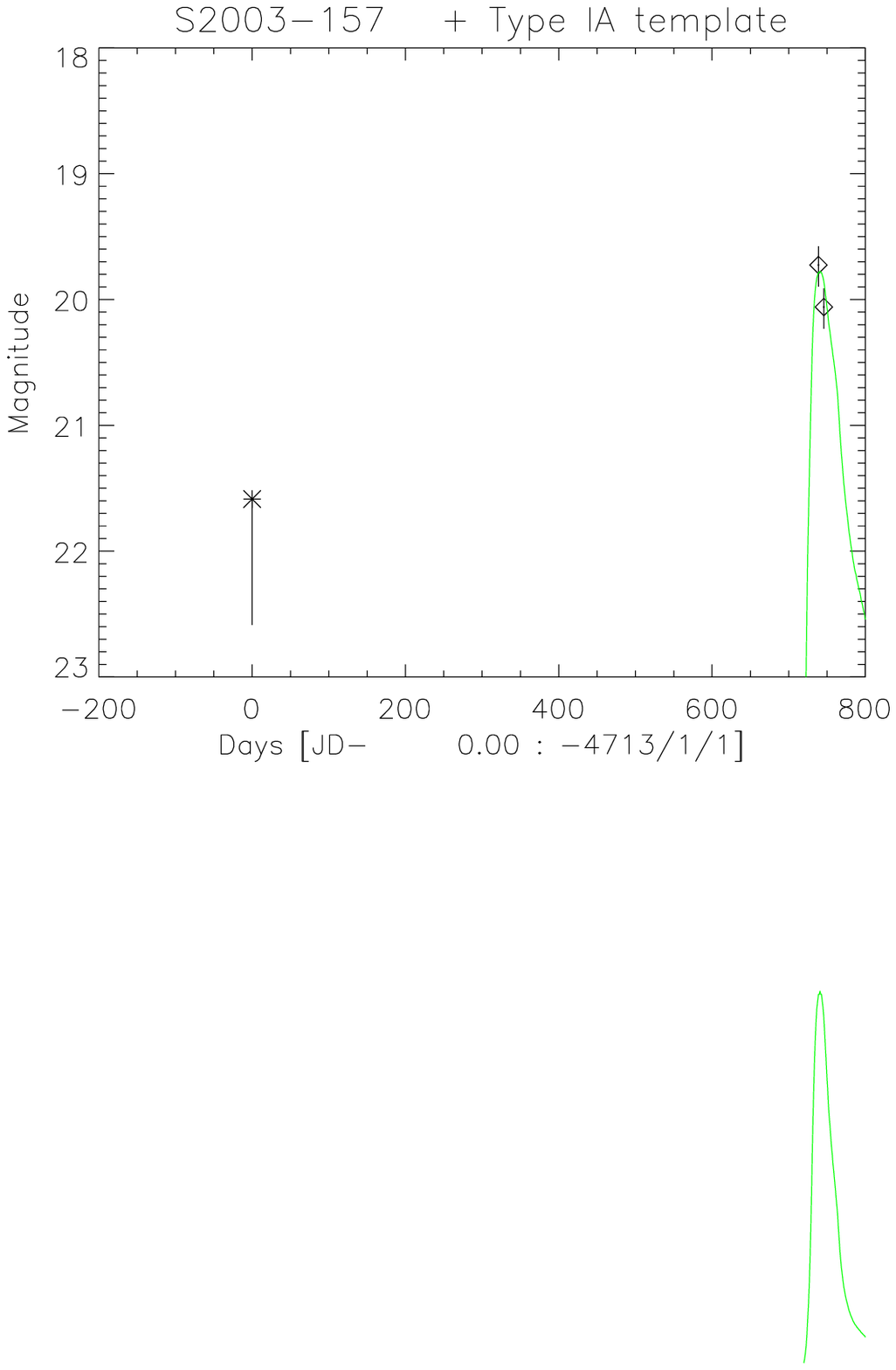}\label{fig:2003cx_lightcurve}}
\vspace{0.3in}
\subfigure[2003cy]{\includegraphics[angle=90,height=2in,width=3in]{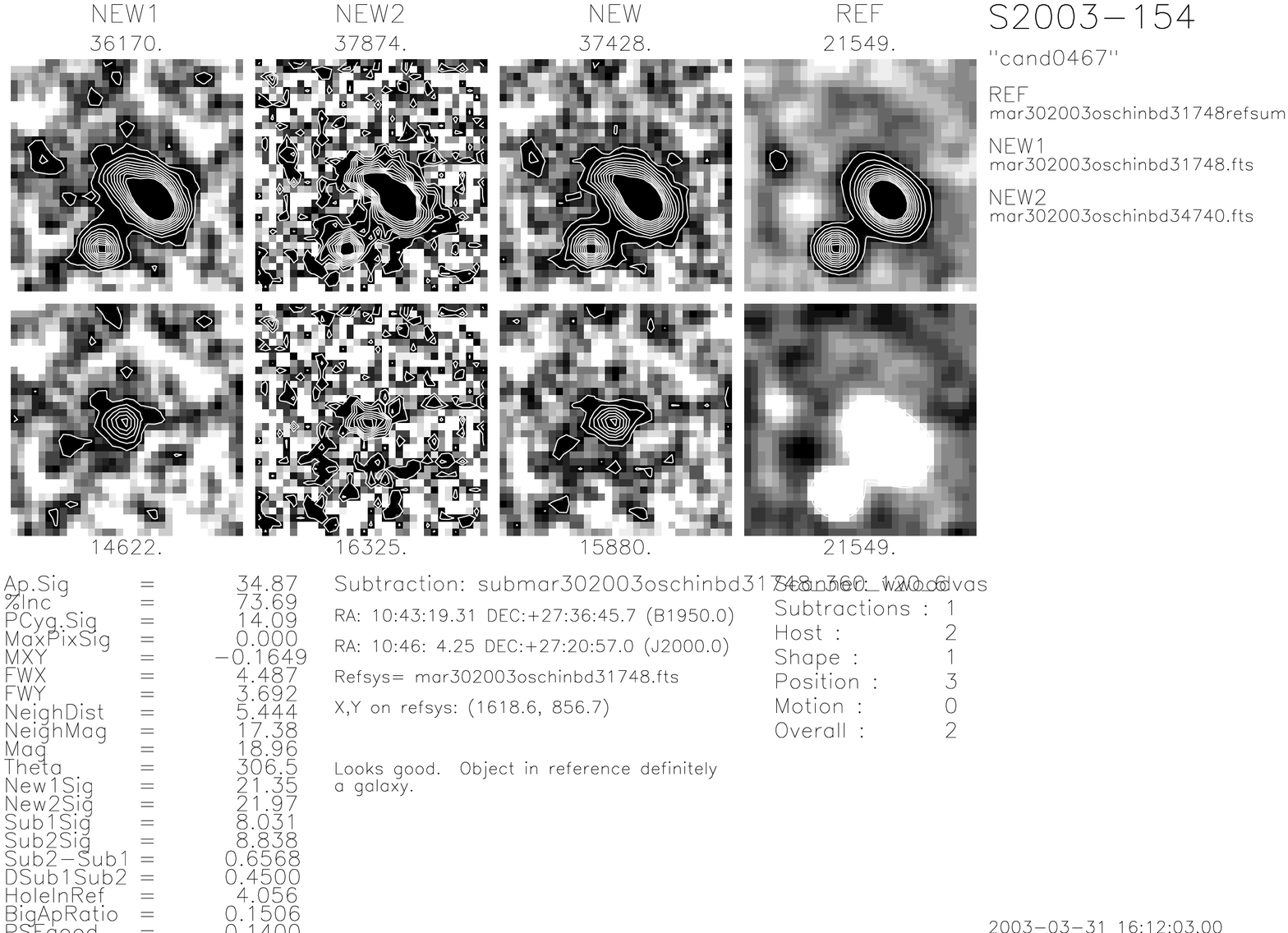}\label{fig:2003cy_discovery}}
\hspace{0.3in}
\subfigure[2003cy]{\includegraphics[height=2in]{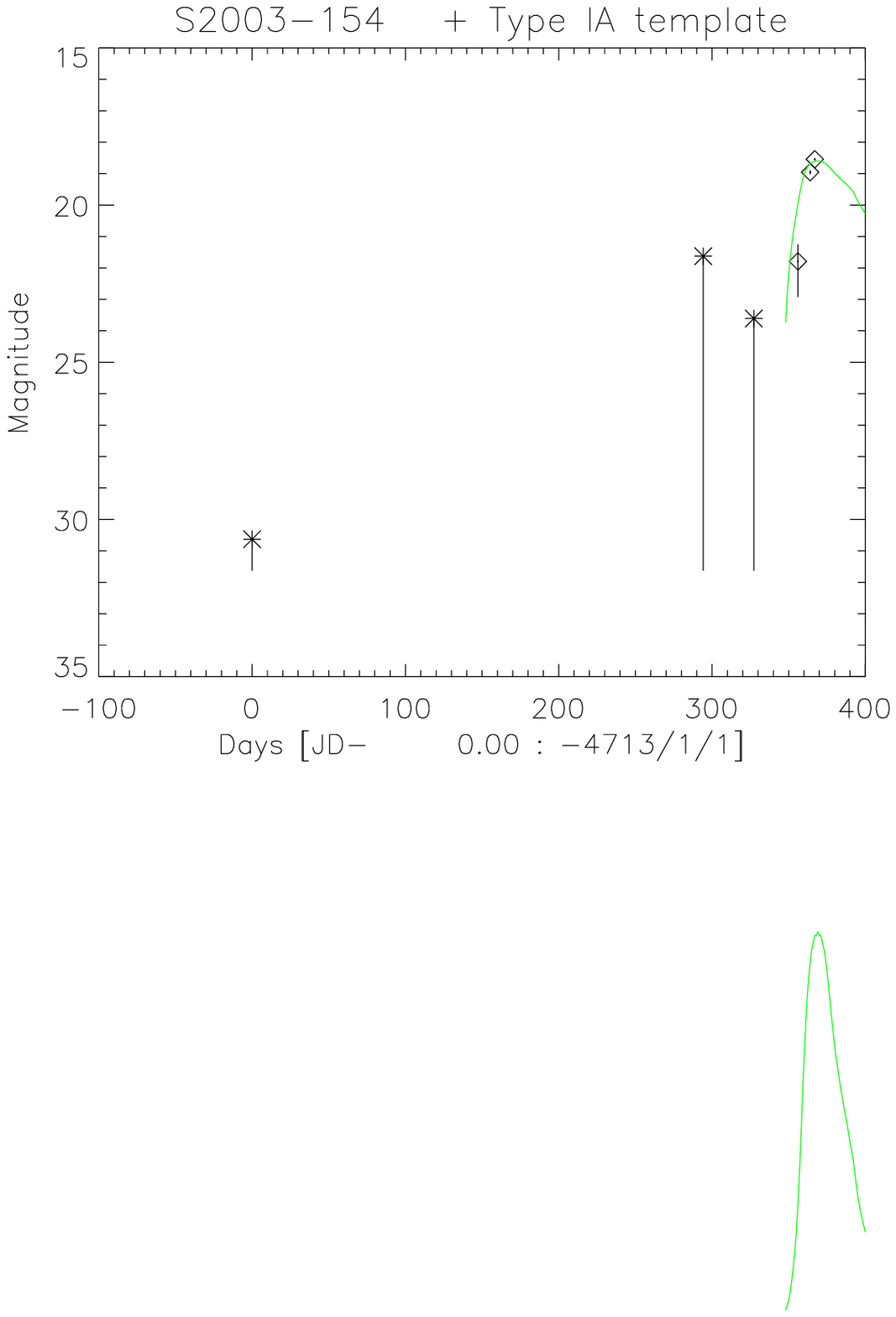}\label{fig:2003cy_lightcurve}}
\vspace{0.3in}
\end{figure}

\clearpage\pagebreak
\begin{figure}
\subfigure[2003cz]{\includegraphics[angle=90,height=2in,width=3in]{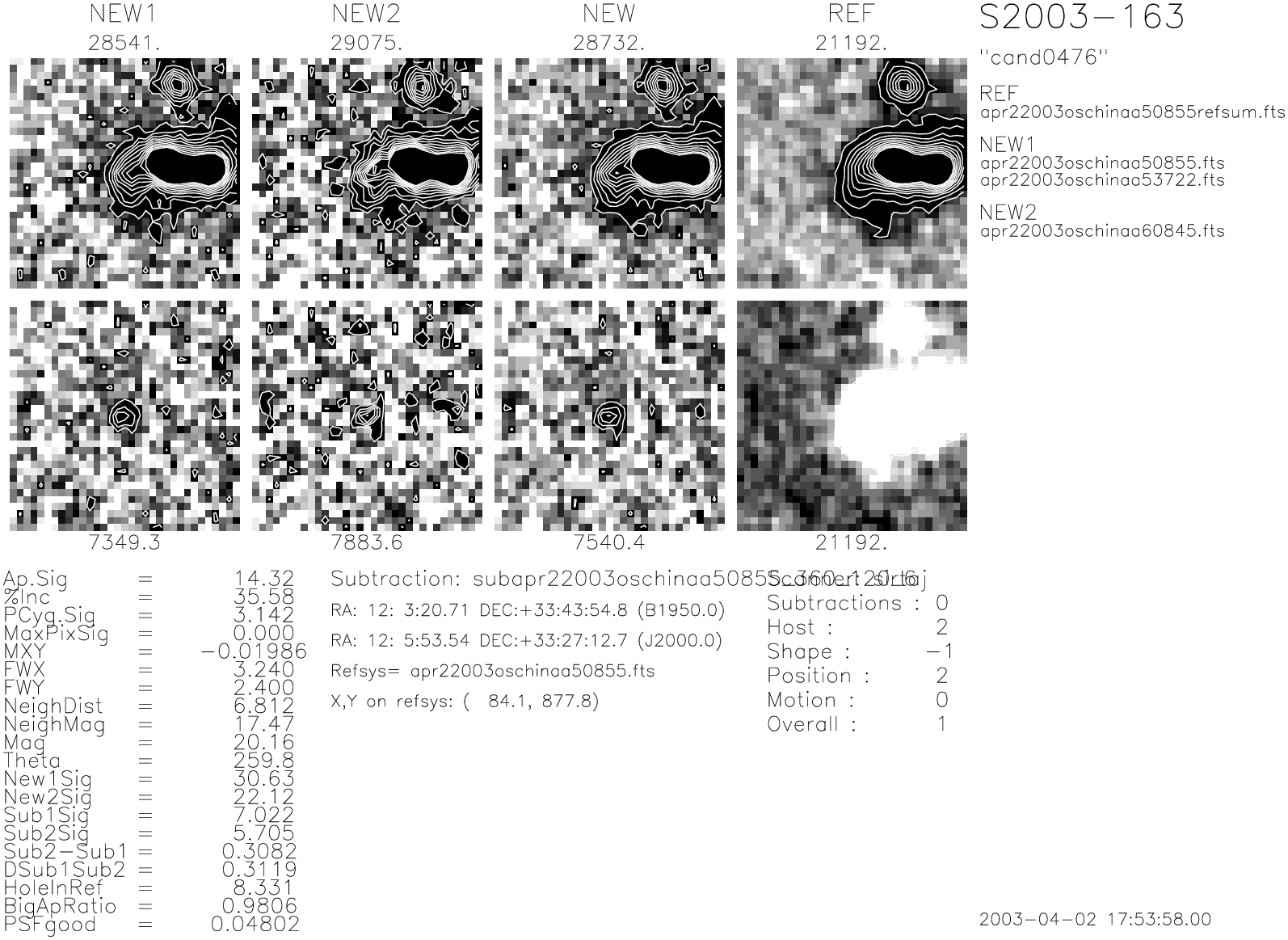}\label{fig:2003cz_discovery}}
\hspace{0.3in}
\subfigure[2003cz]{\includegraphics[height=2in]{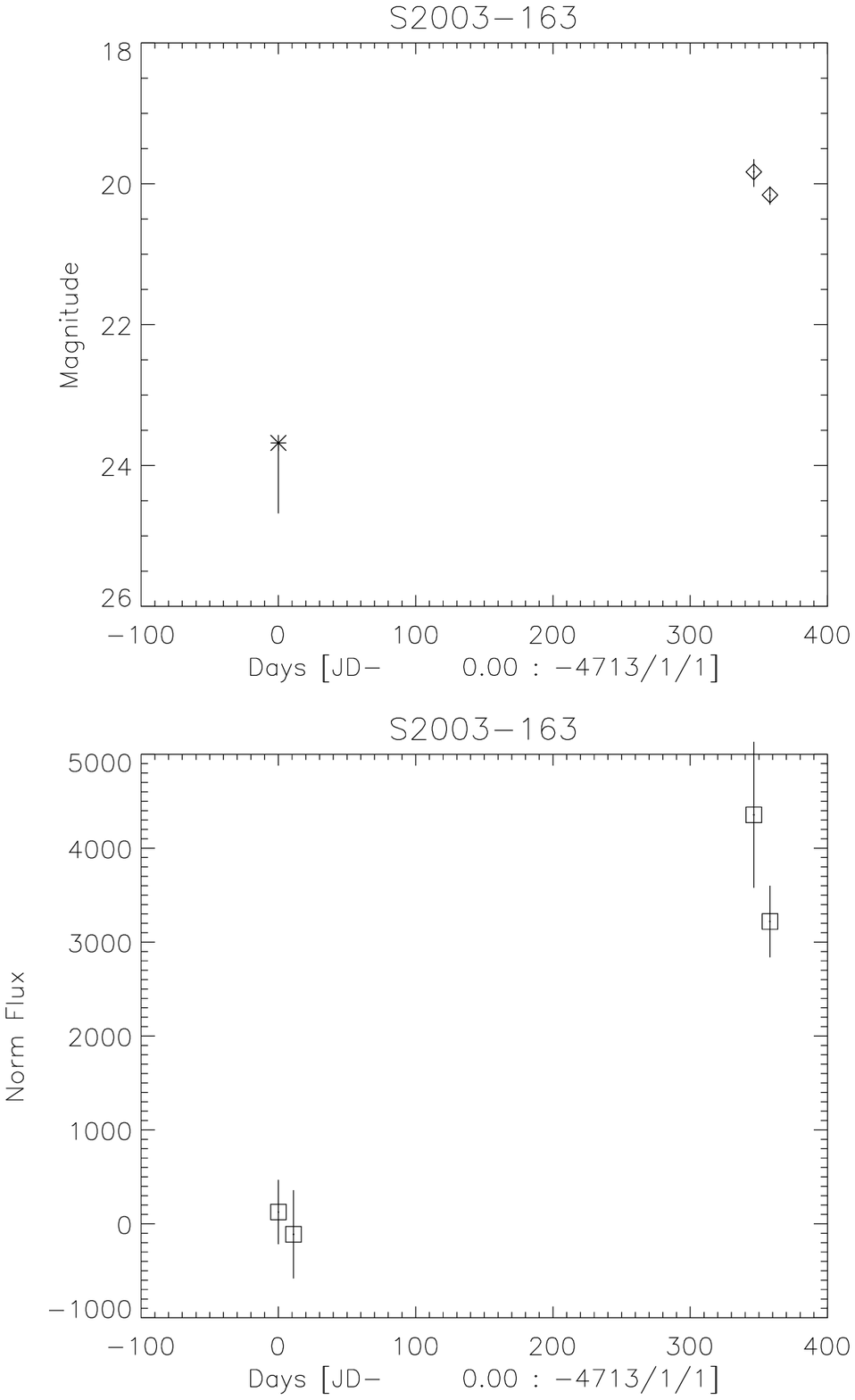}\label{fig:2003cz_lightcurve}}
\vspace{0.3in}
\subfigure[2003dc]{\includegraphics[angle=90,height=2in,width=3in]{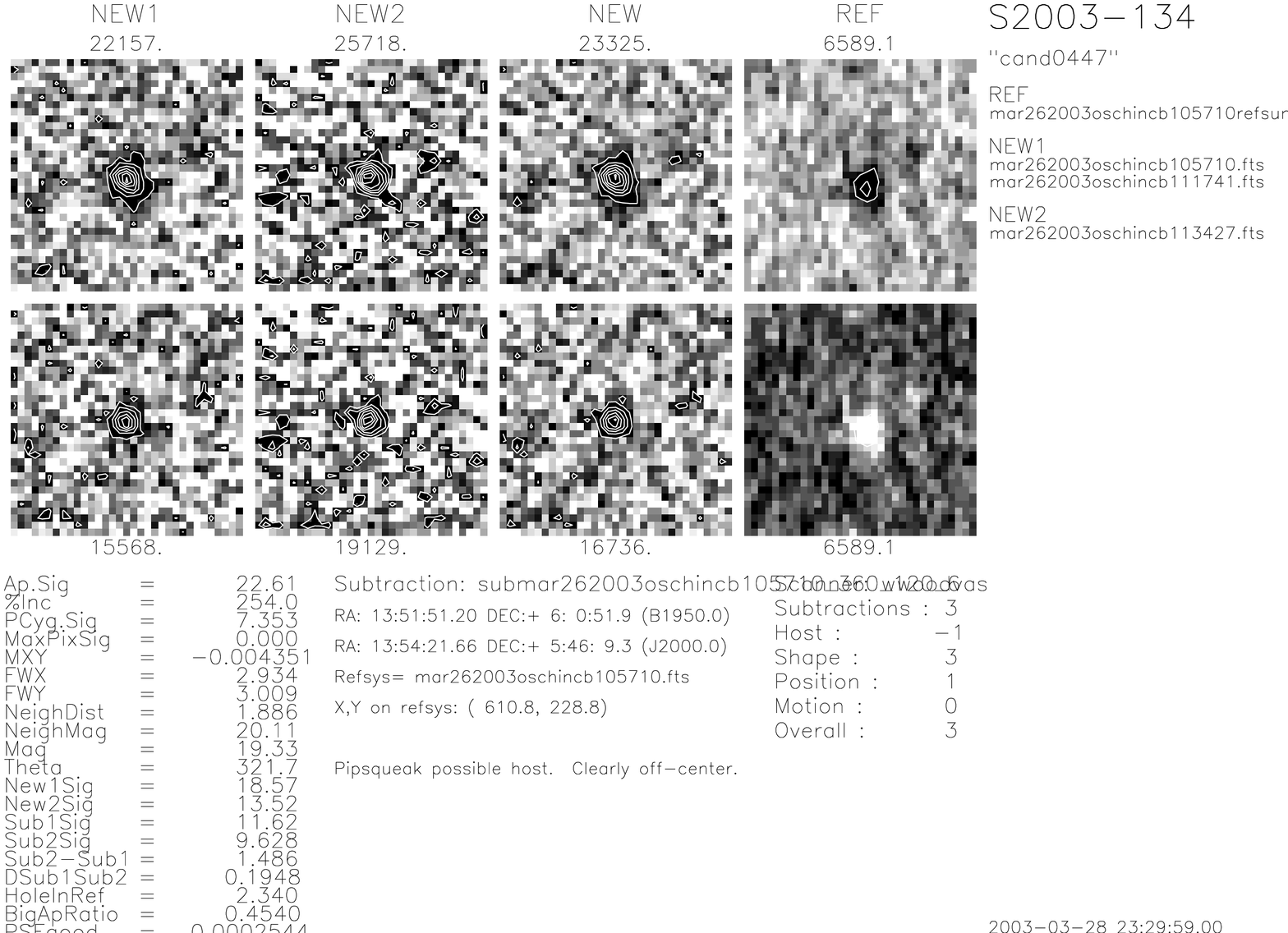}\label{fig:2003dc_discovery}}
\hspace{0.3in}
\subfigure[2003dc]{\includegraphics[height=2in]{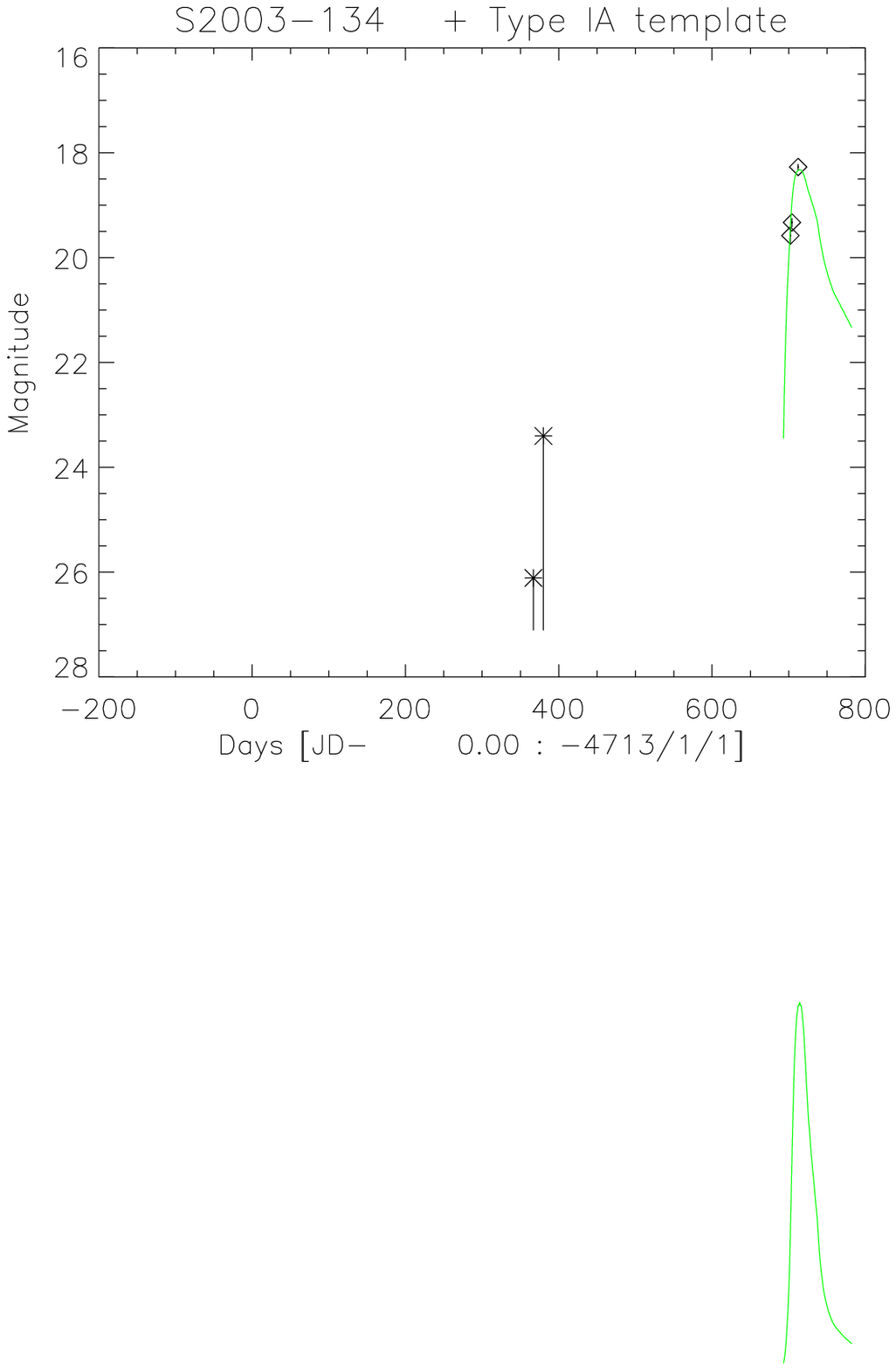}\label{fig:2003dc_lightcurve}}
\vspace{0.3in}
\subfigure[2003dd]{\includegraphics[angle=90,height=2in,width=3in]{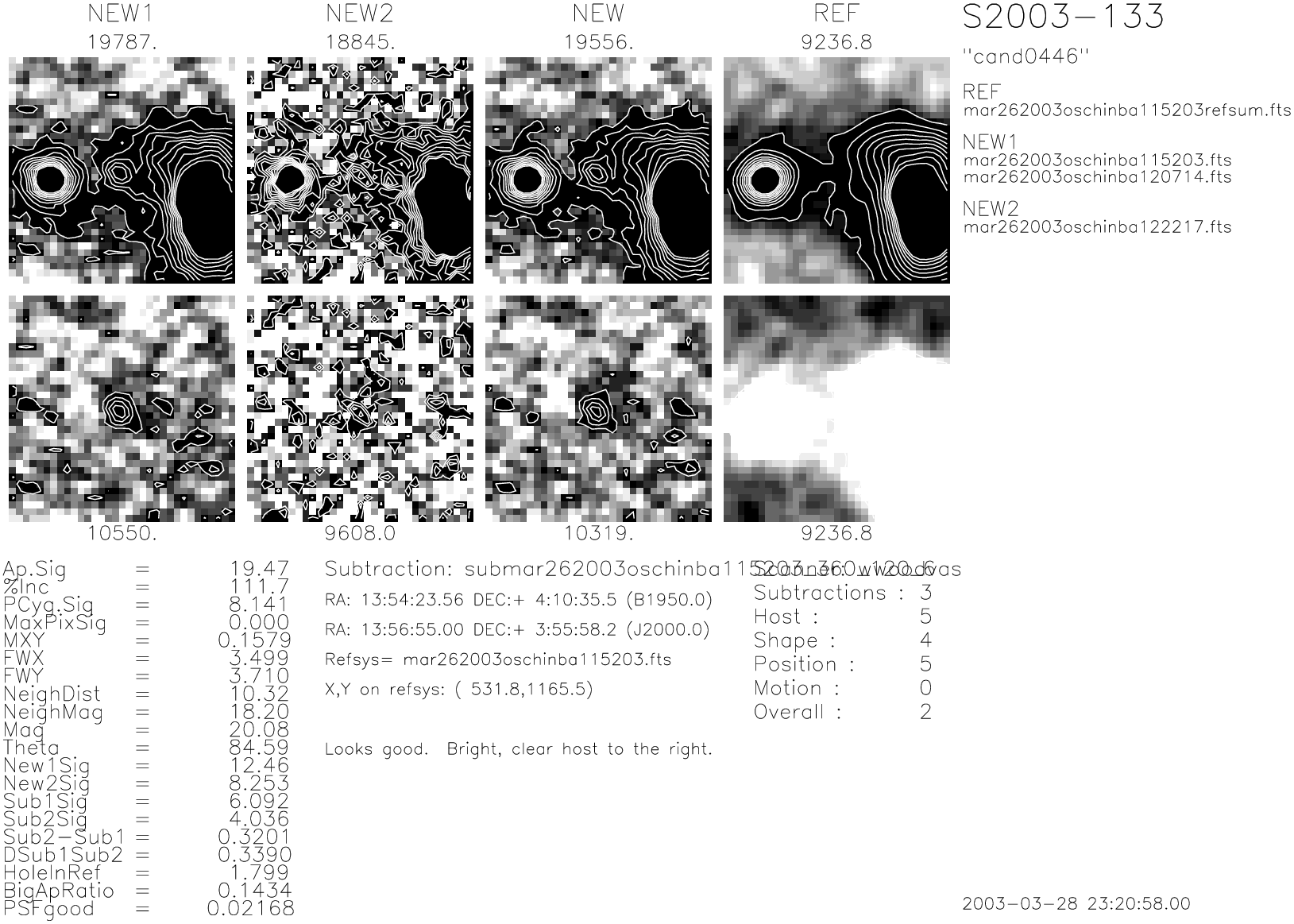}\label{fig:2003dd_discovery}}
\hspace{0.3in}
\subfigure[2003dd]{\includegraphics[height=2in]{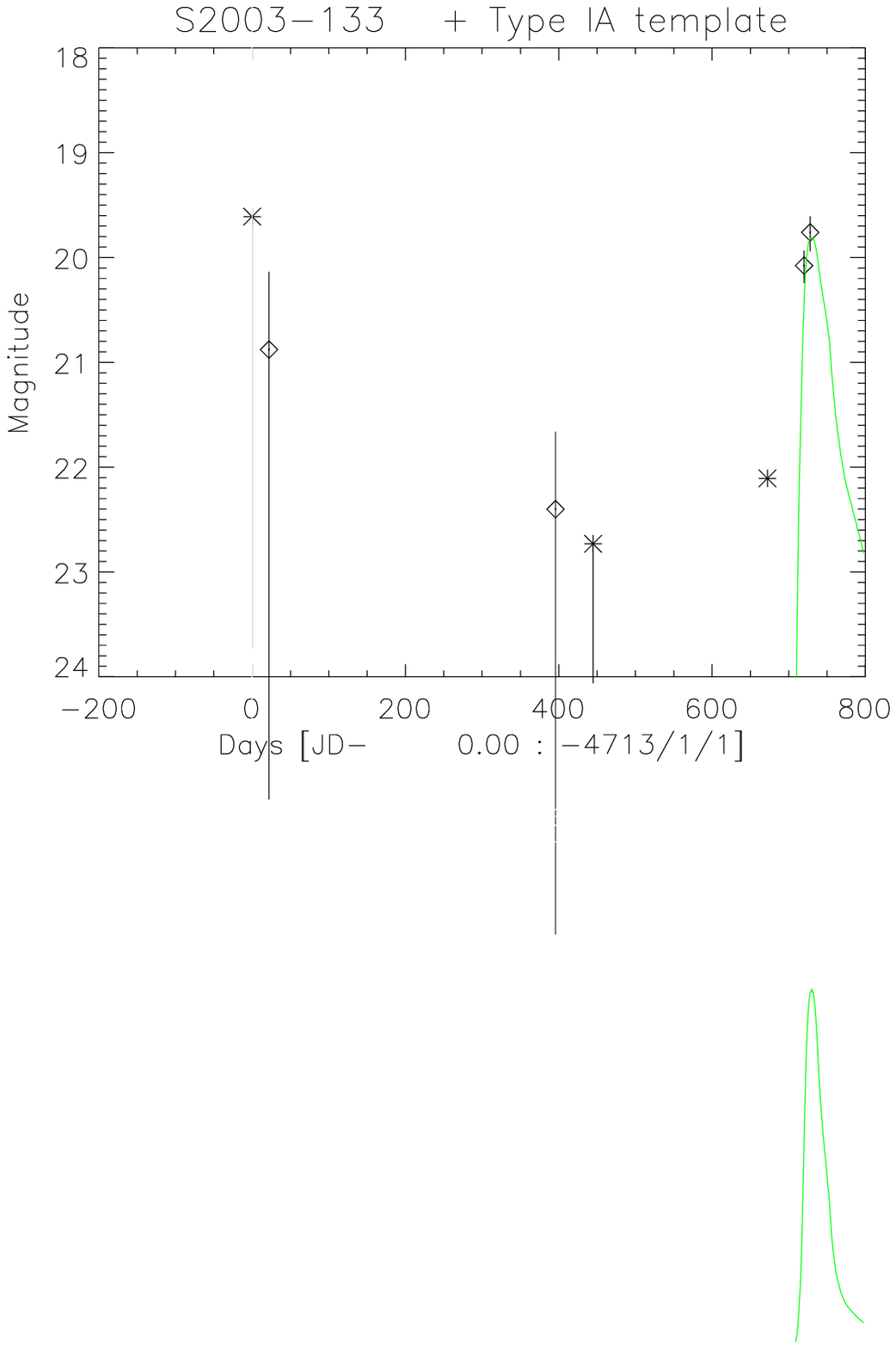}\label{fig:2003dd_lightcurve}}
\vspace{0.3in}
\end{figure}

\clearpage\pagebreak
\begin{figure}
\subfigure[2003de]{\includegraphics[angle=90,height=2in,width=3in]{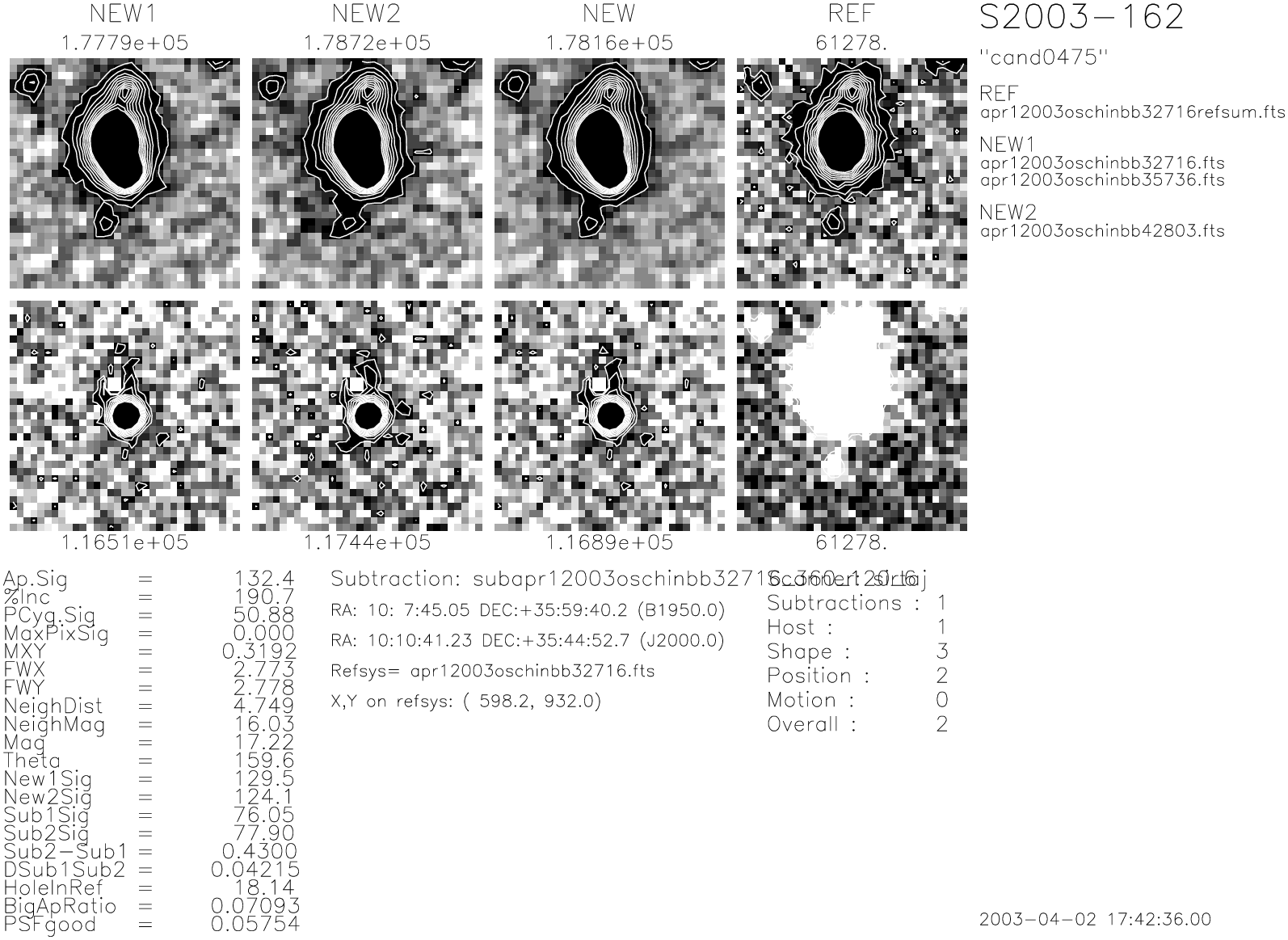}\label{fig:2003de_discovery}}
\hspace{0.3in}
\subfigure[2003de]{\includegraphics[height=2in]{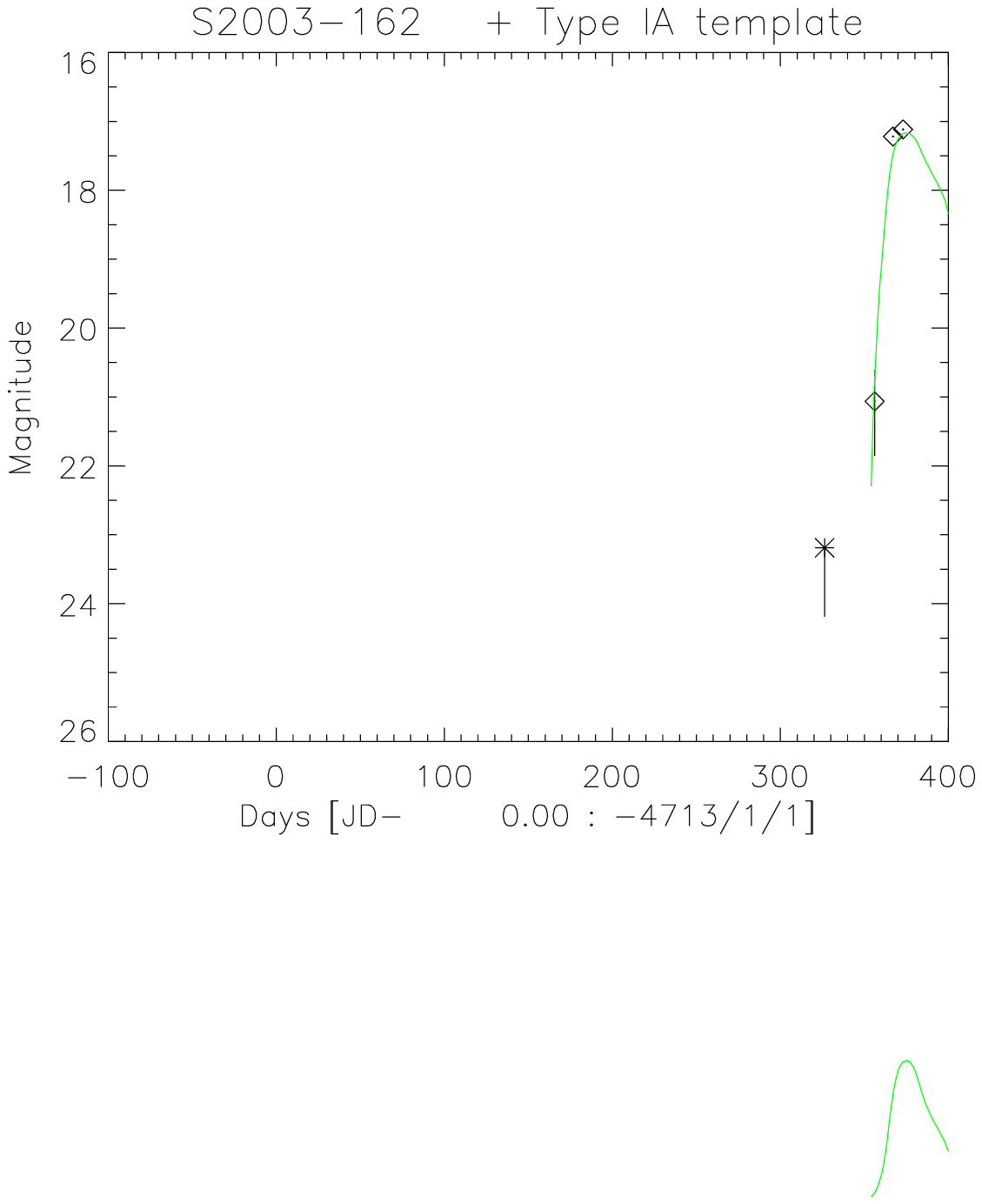}\label{fig:2003de_lightcurve}}
\vspace{0.3in}
\subfigure[2003df]{\includegraphics[angle=90,height=2in,width=3in]{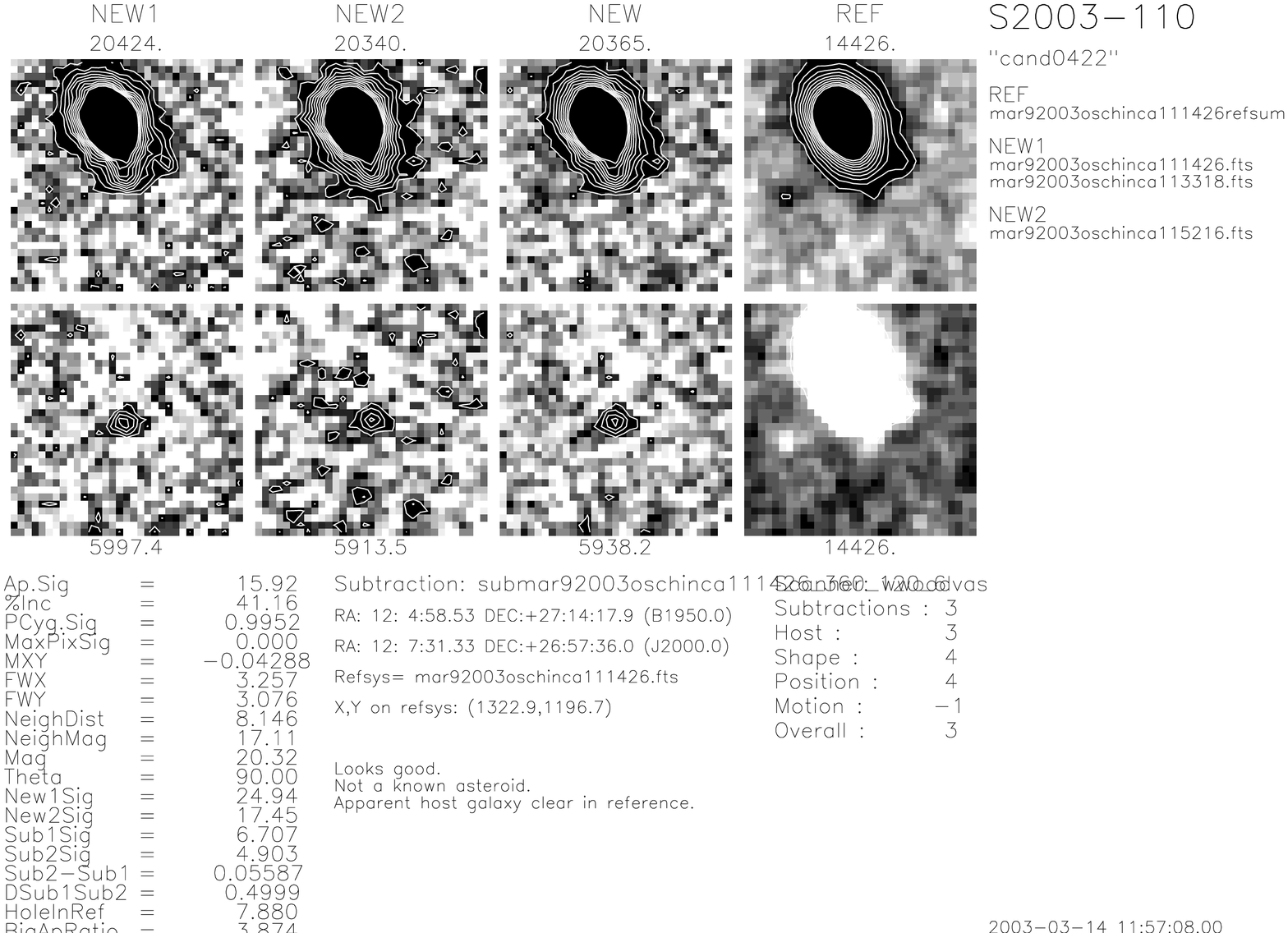}\label{fig:2003df_discovery}}
\hspace{0.3in}
\subfigure[2003df]{\includegraphics[height=2in]{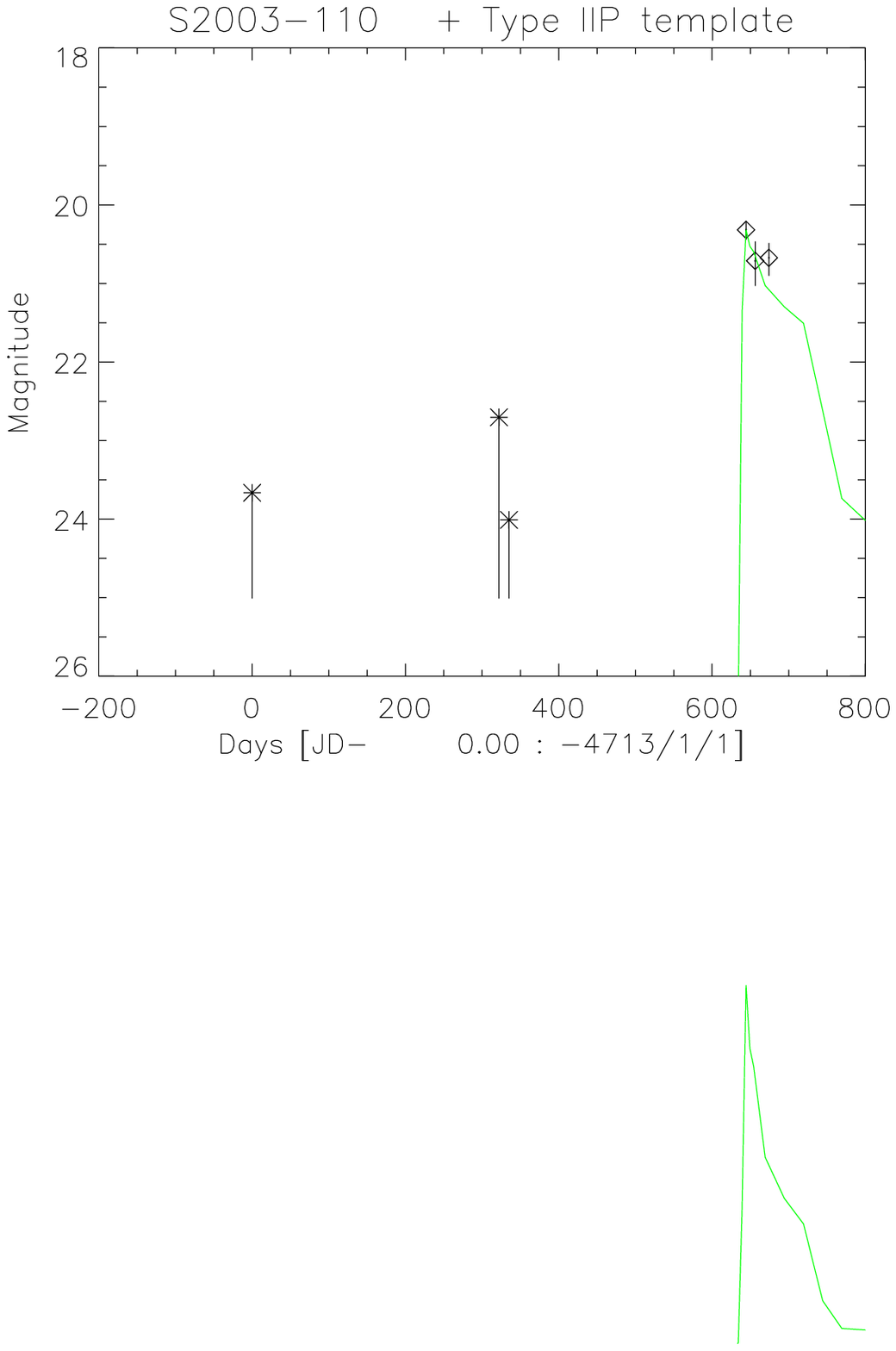}\label{fig:2003df_lightcurve}}
\vspace{0.3in}
\subfigure[2003di]{\includegraphics[angle=90,height=2in,width=3in]{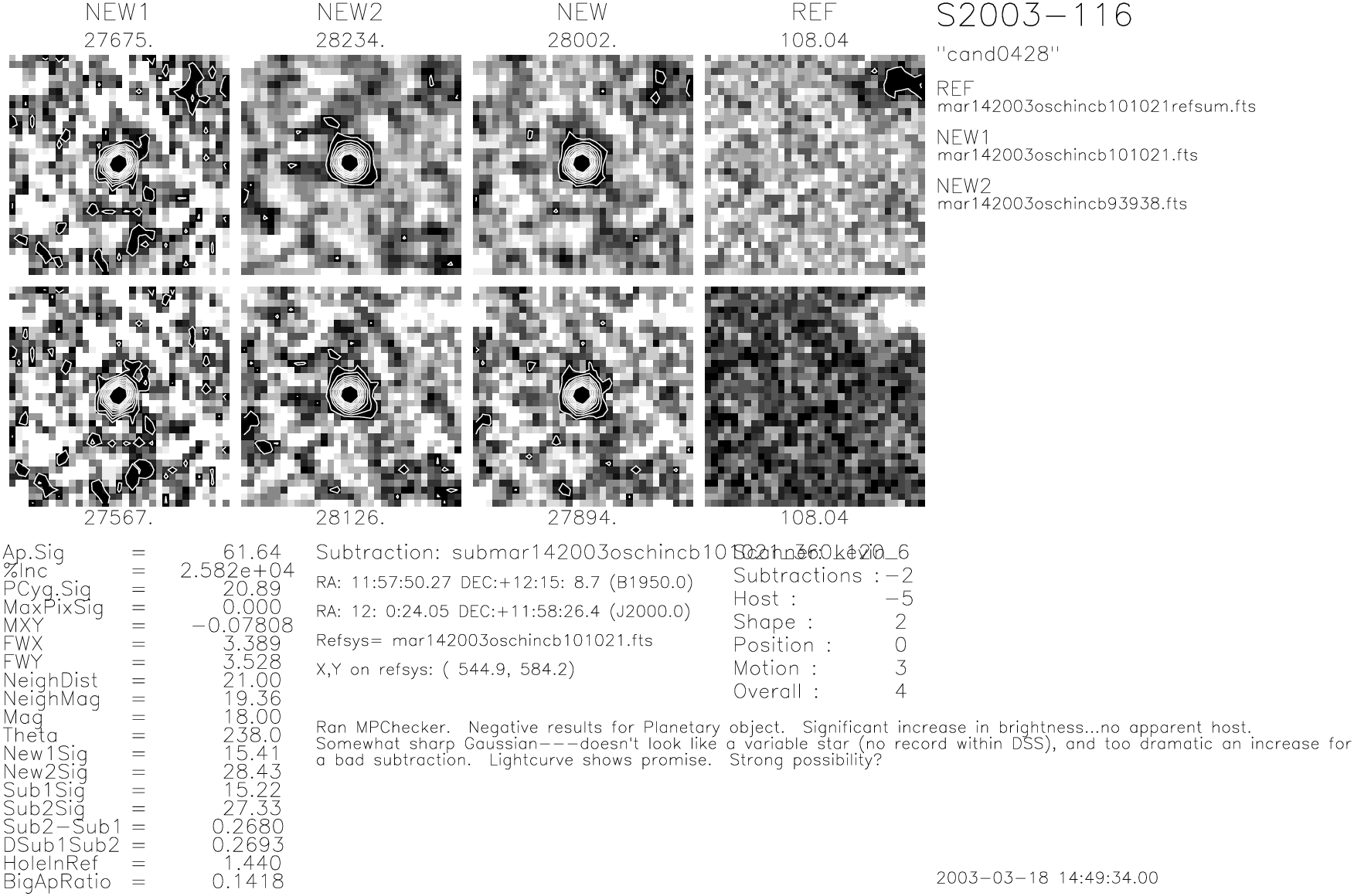}\label{fig:2003di_discovery}}
\hspace{0.3in}
\subfigure[2003di]{\includegraphics[height=2in]{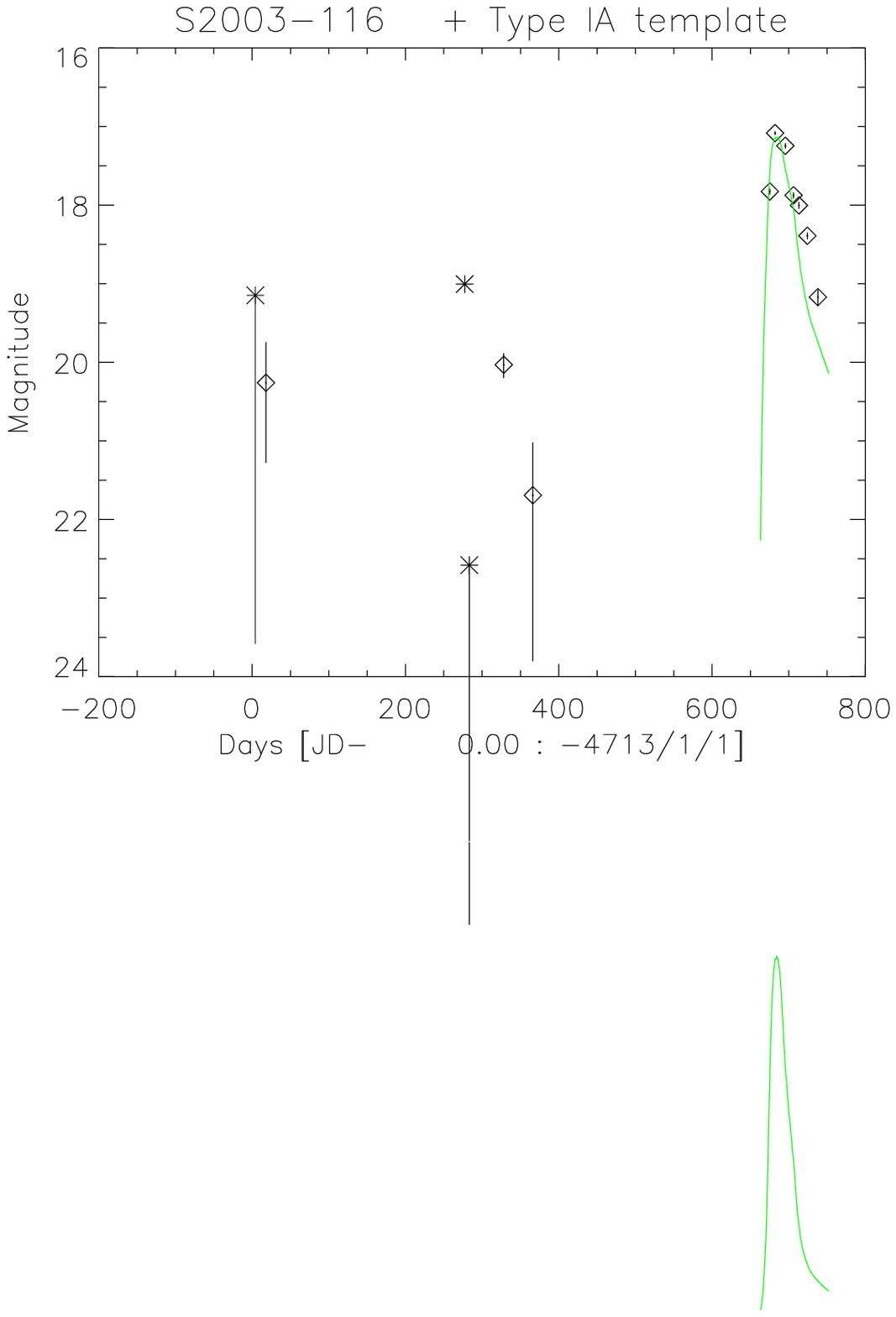}\label{fig:2003di_lightcurve}}
\vspace{0.3in}
\end{figure}

\clearpage\pagebreak
\begin{figure}
\subfigure[2003dj]{\includegraphics[angle=90,height=2in,width=3in]{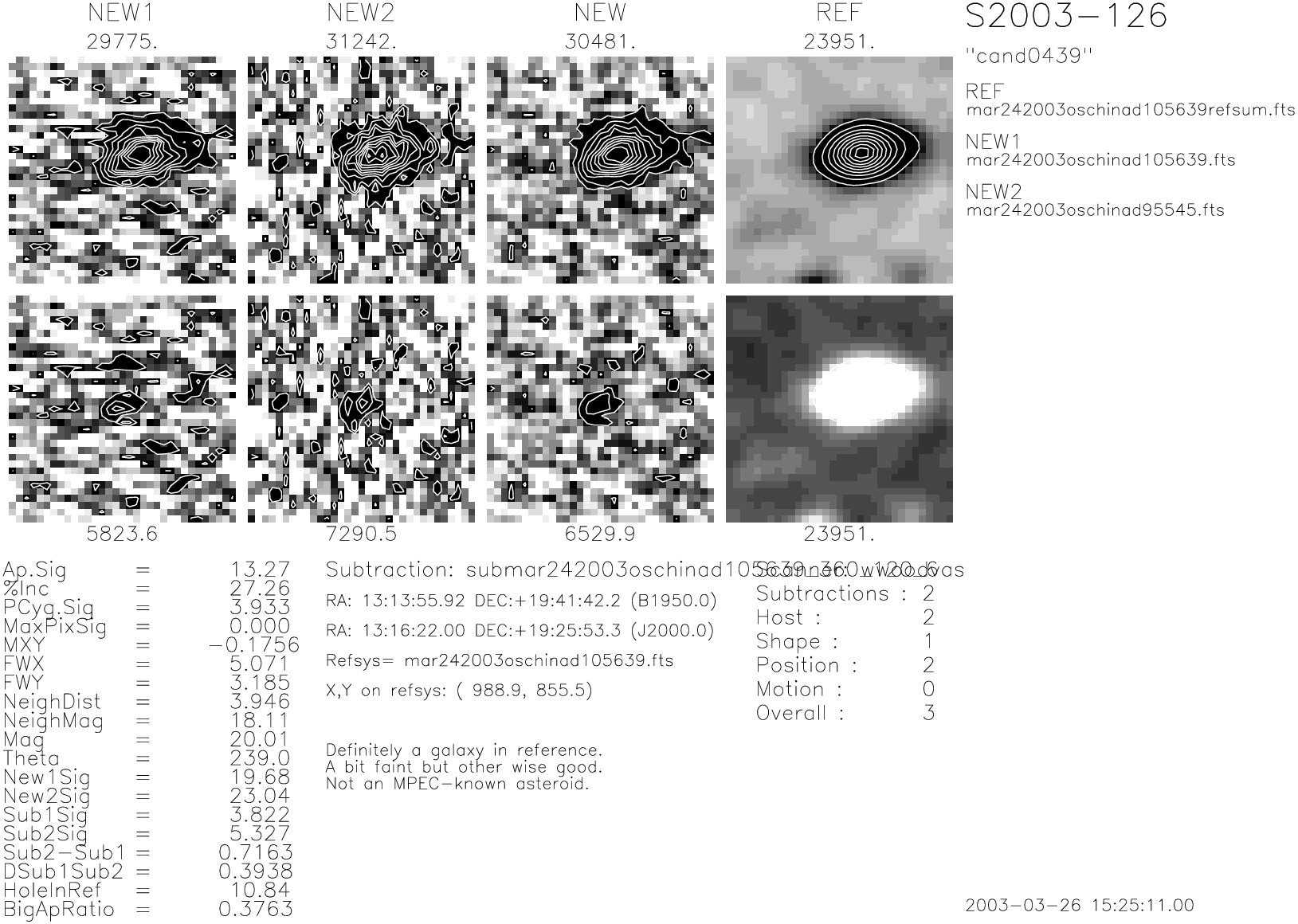}\label{fig:2003dj_discovery}}
\hspace{0.3in}
\subfigure[2003dj]{\includegraphics[height=2in]{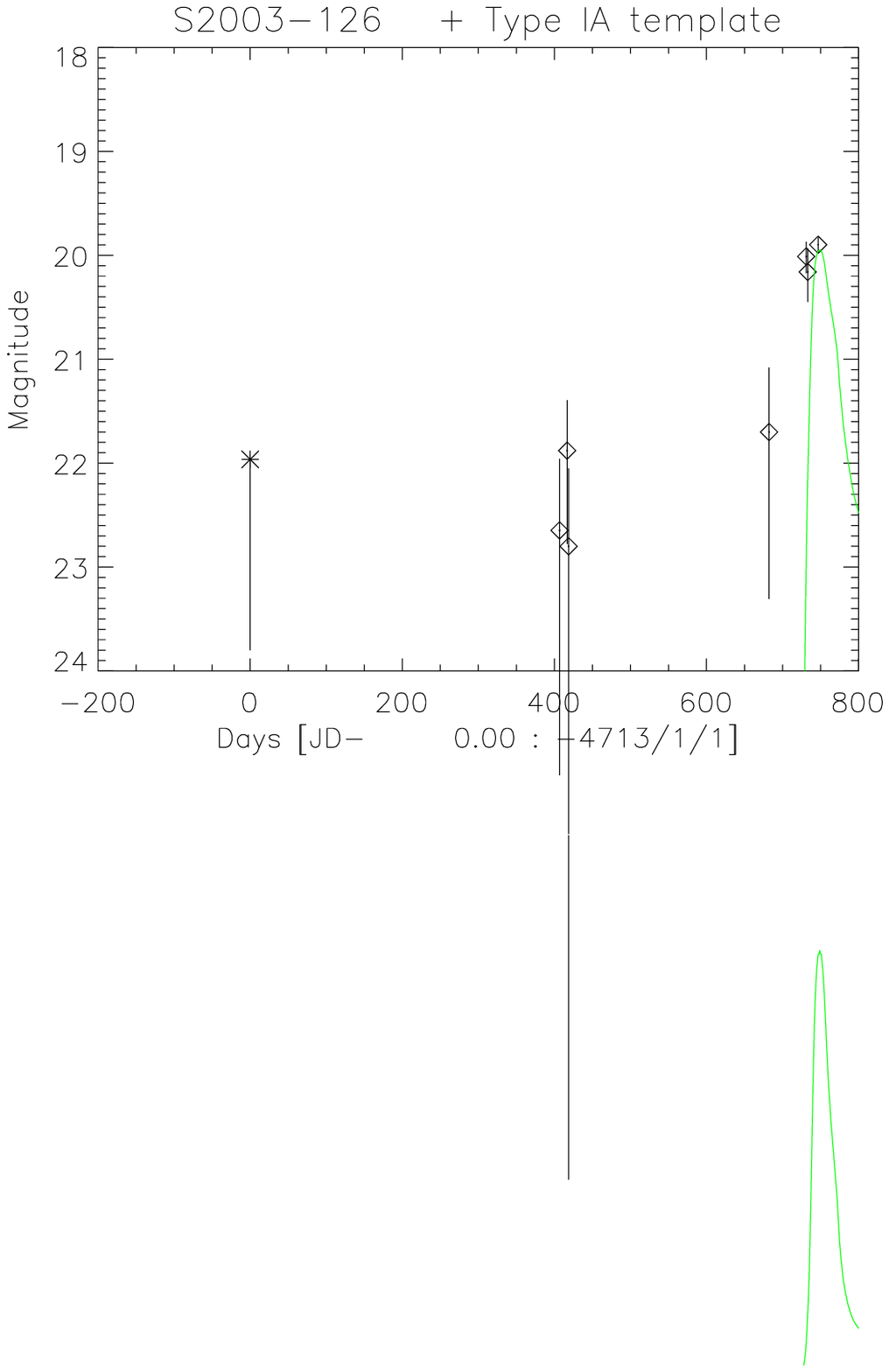}\label{fig:2003dj_lightcurve}}
\vspace{0.3in}
\subfigure[2003dk]{\includegraphics[angle=90,height=2in,width=3in]{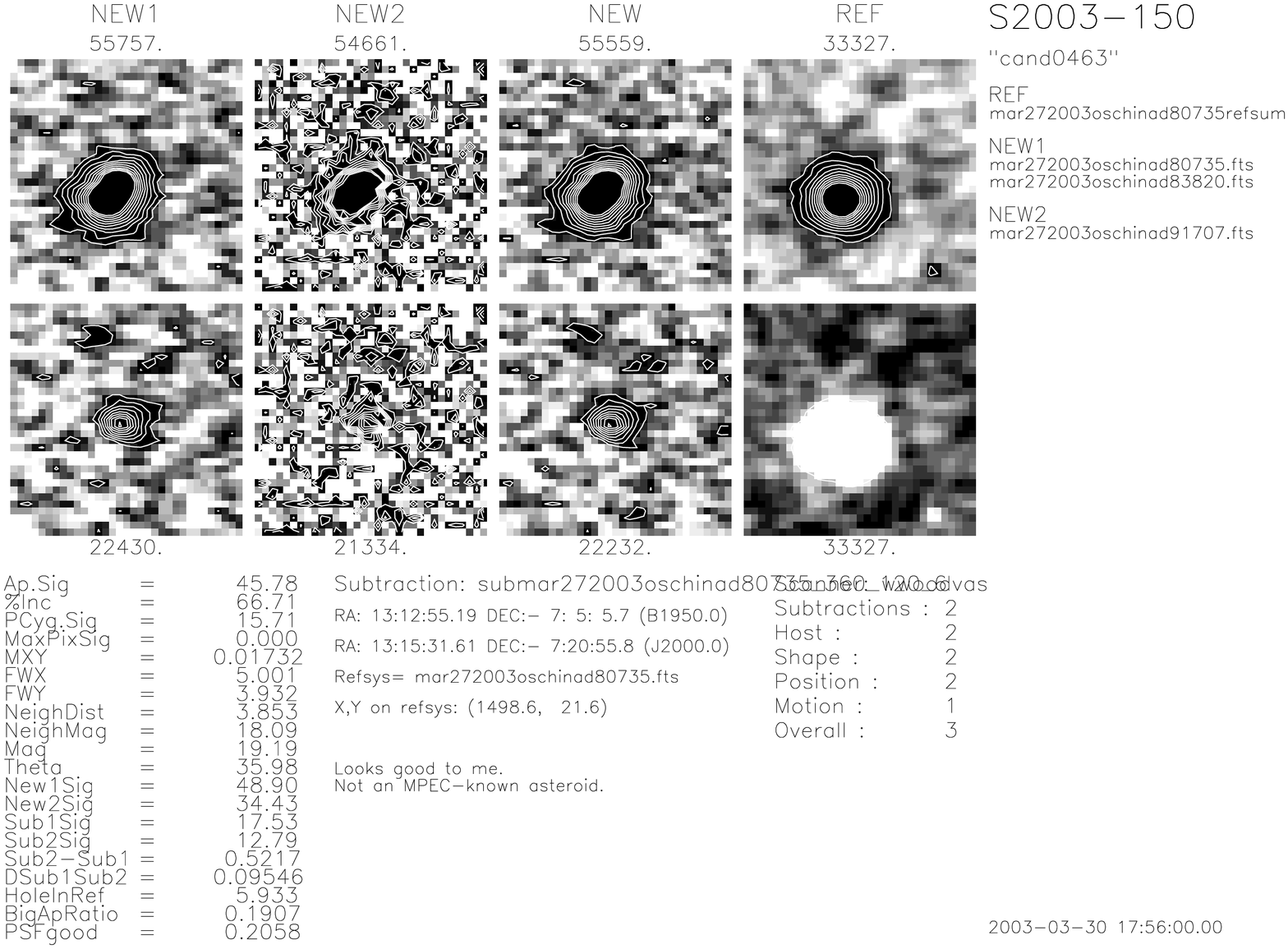}\label{fig:2003dk_discovery}}
\hspace{0.3in}
\subfigure[2003dk]{\includegraphics[height=2in]{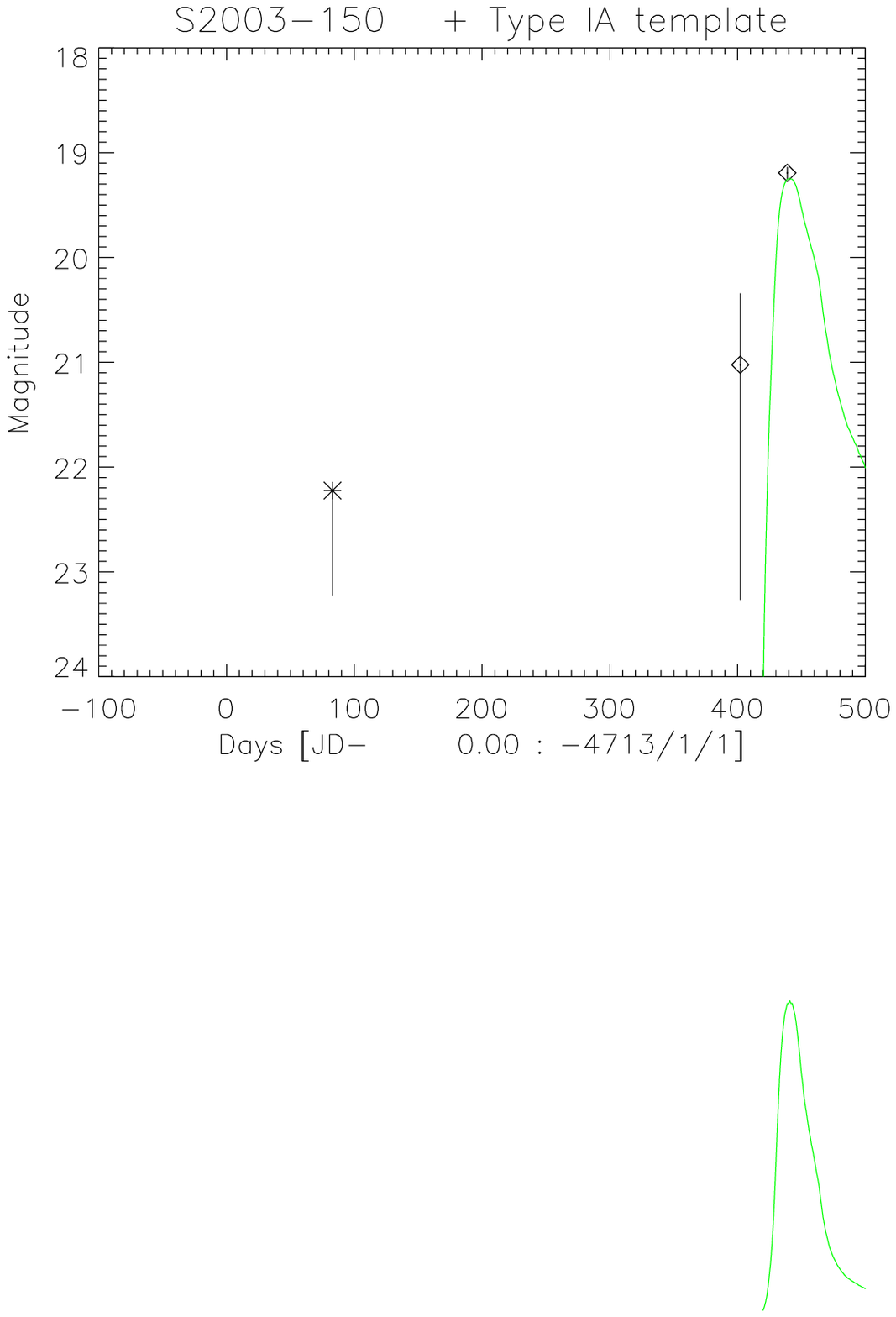}\label{fig:2003dk_lightcurve}}
\vspace{0.3in}
\subfigure[2003dm]{\includegraphics[angle=90,height=2in,width=3in]{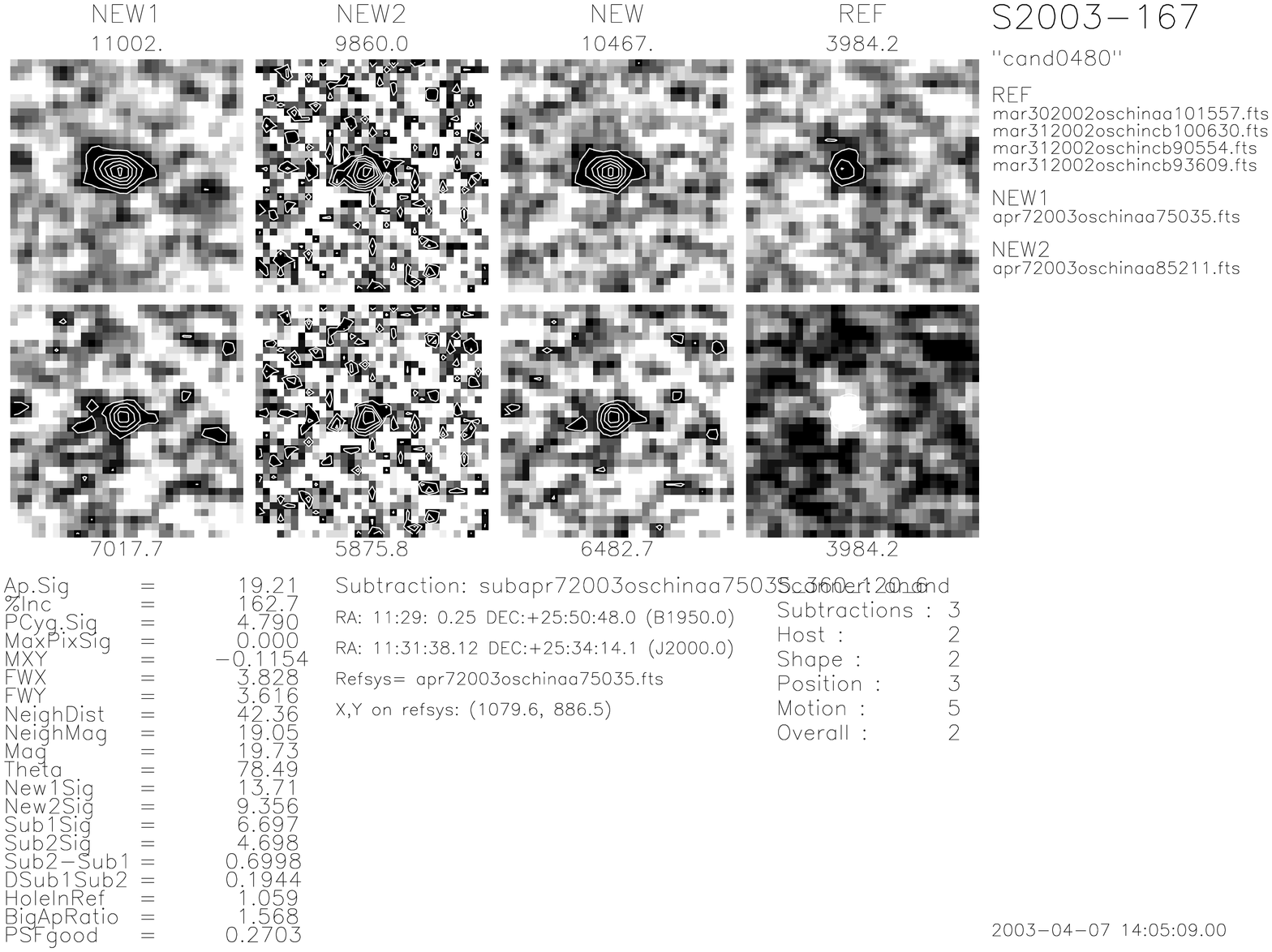}\label{fig:2003dm_discovery}}
\hspace{0.3in}
\subfigure[2003dm]{\includegraphics[height=2in]{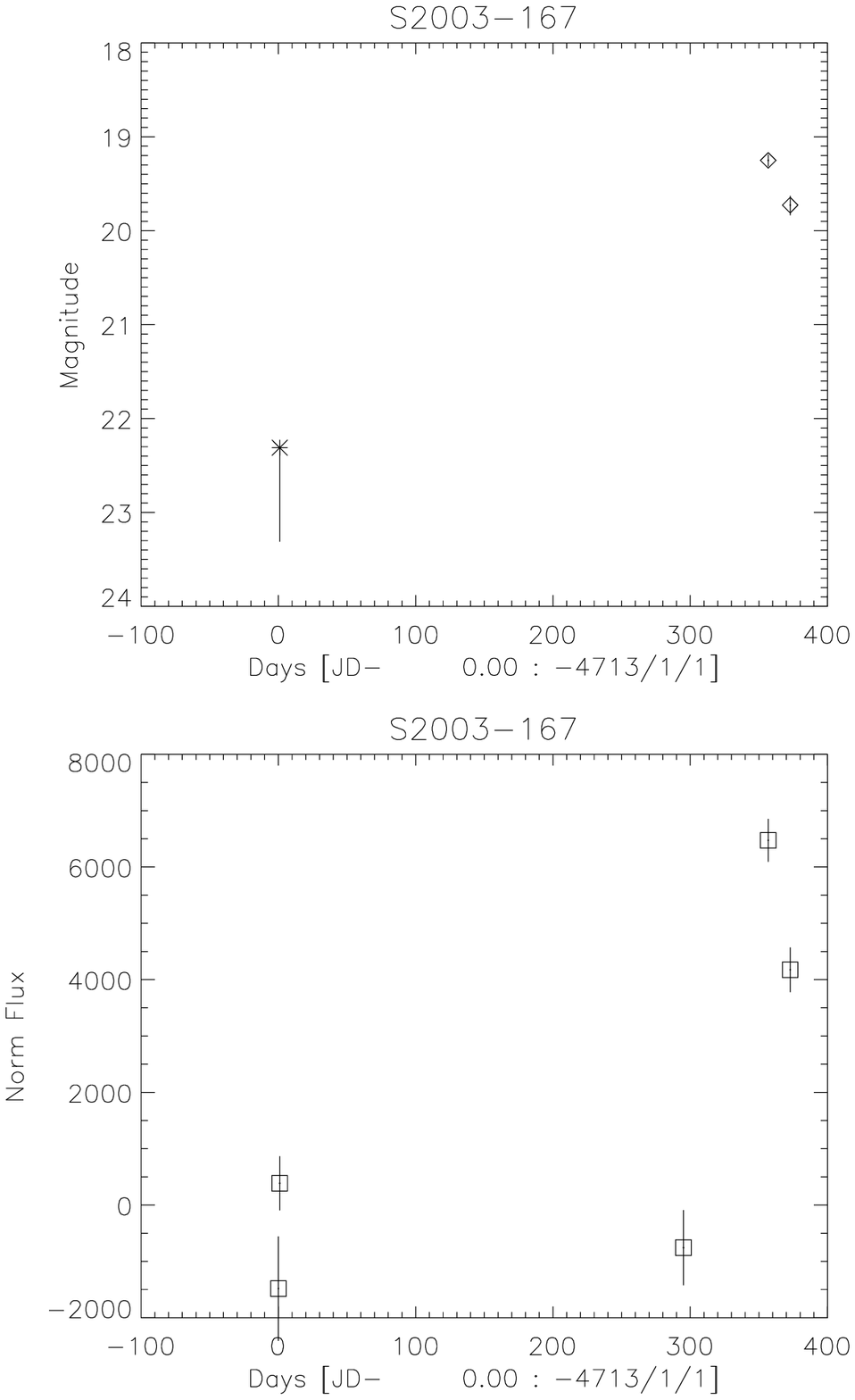}\label{fig:2003dm_lightcurve}}
\vspace{0.3in}
\end{figure}

\clearpage\pagebreak
\begin{figure}
\subfigure[2003dn]{\includegraphics[angle=90,height=2in,width=3in]{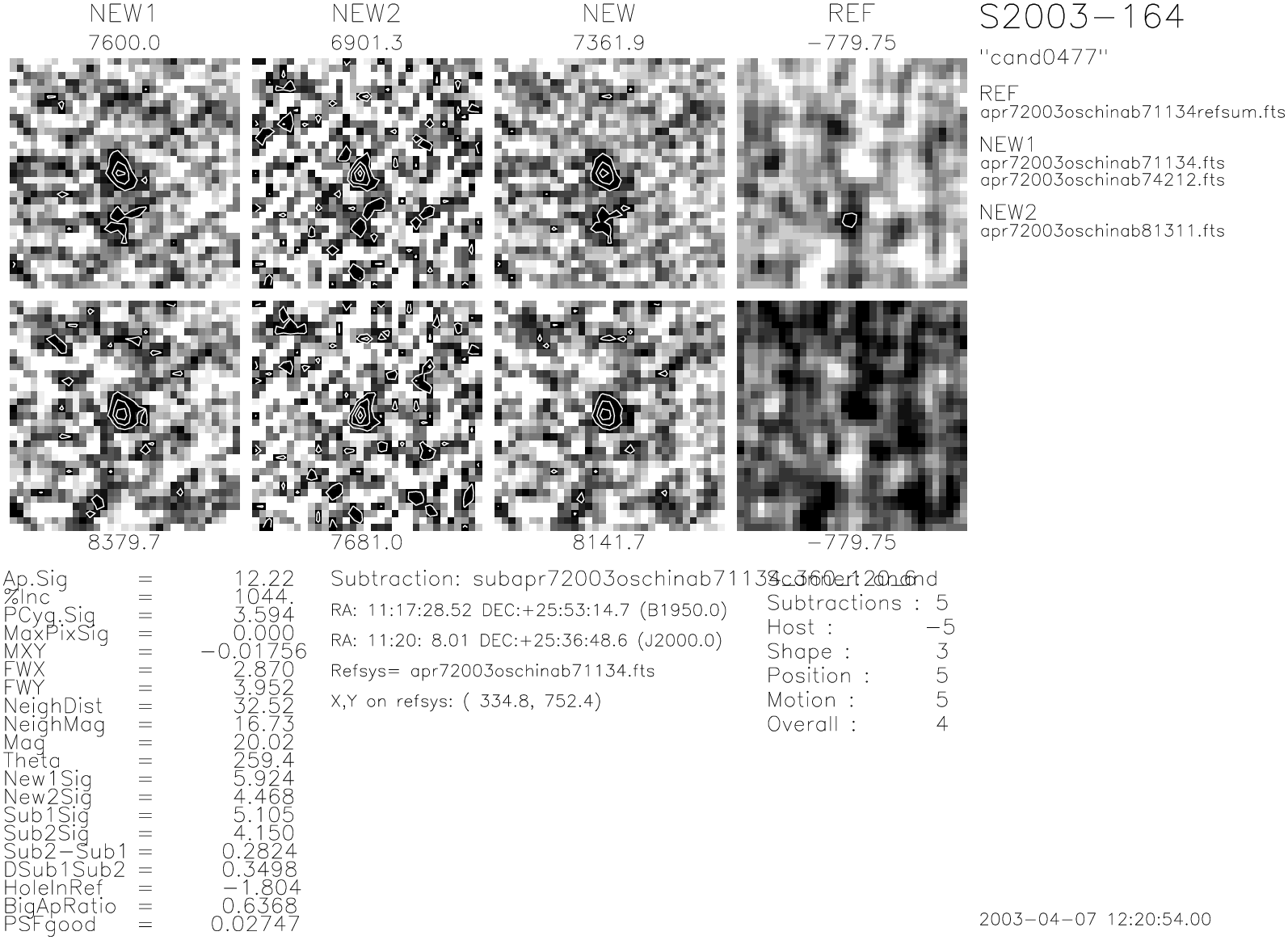}\label{fig:2003dn_discovery}}
\hspace{0.3in}
\subfigure[2003dn]{\includegraphics[height=2in]{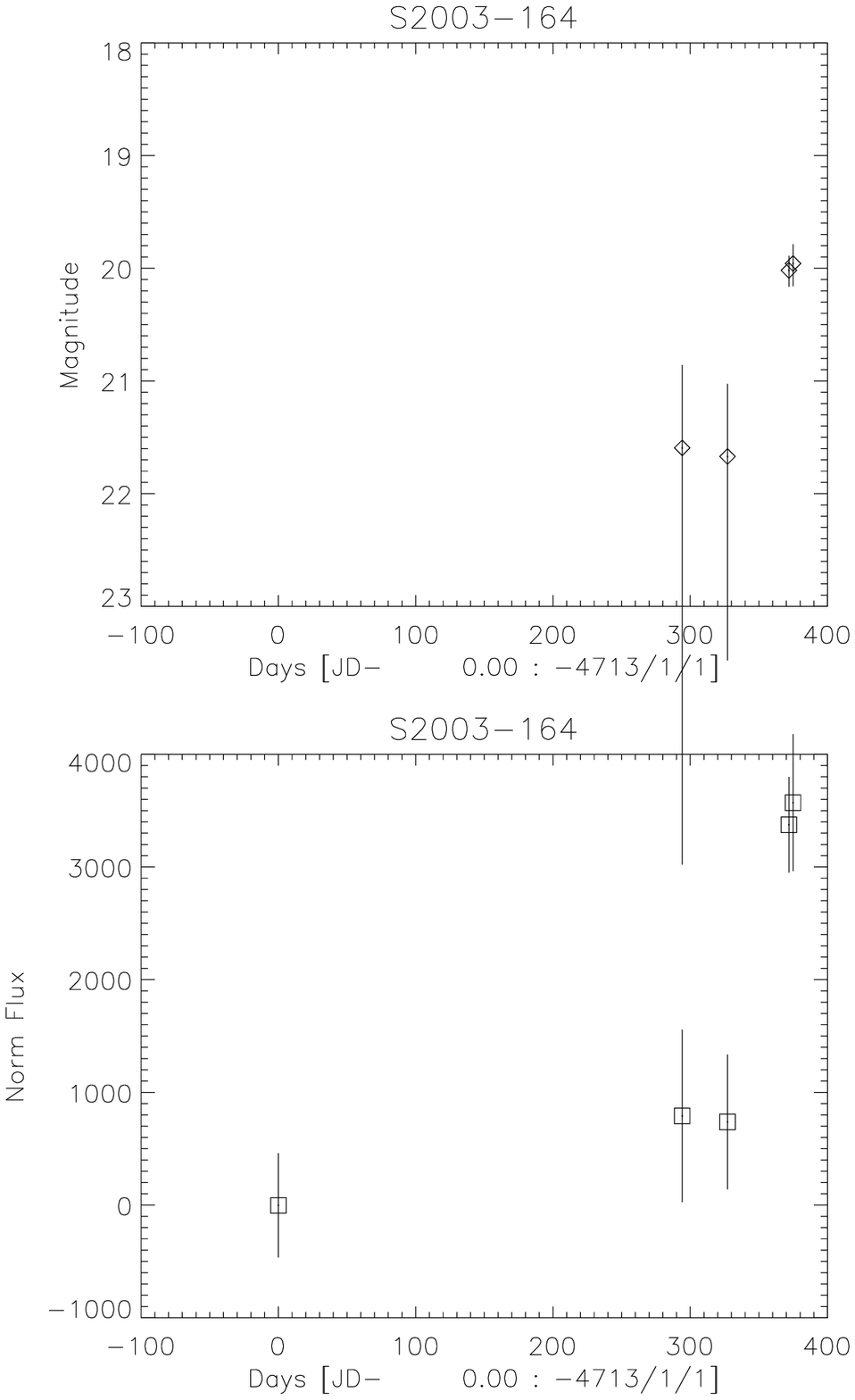}\label{fig:2003dn_lightcurve}}
\vspace{0.3in}
\subfigure[2003do]{\includegraphics[angle=90,height=2in,width=3in]{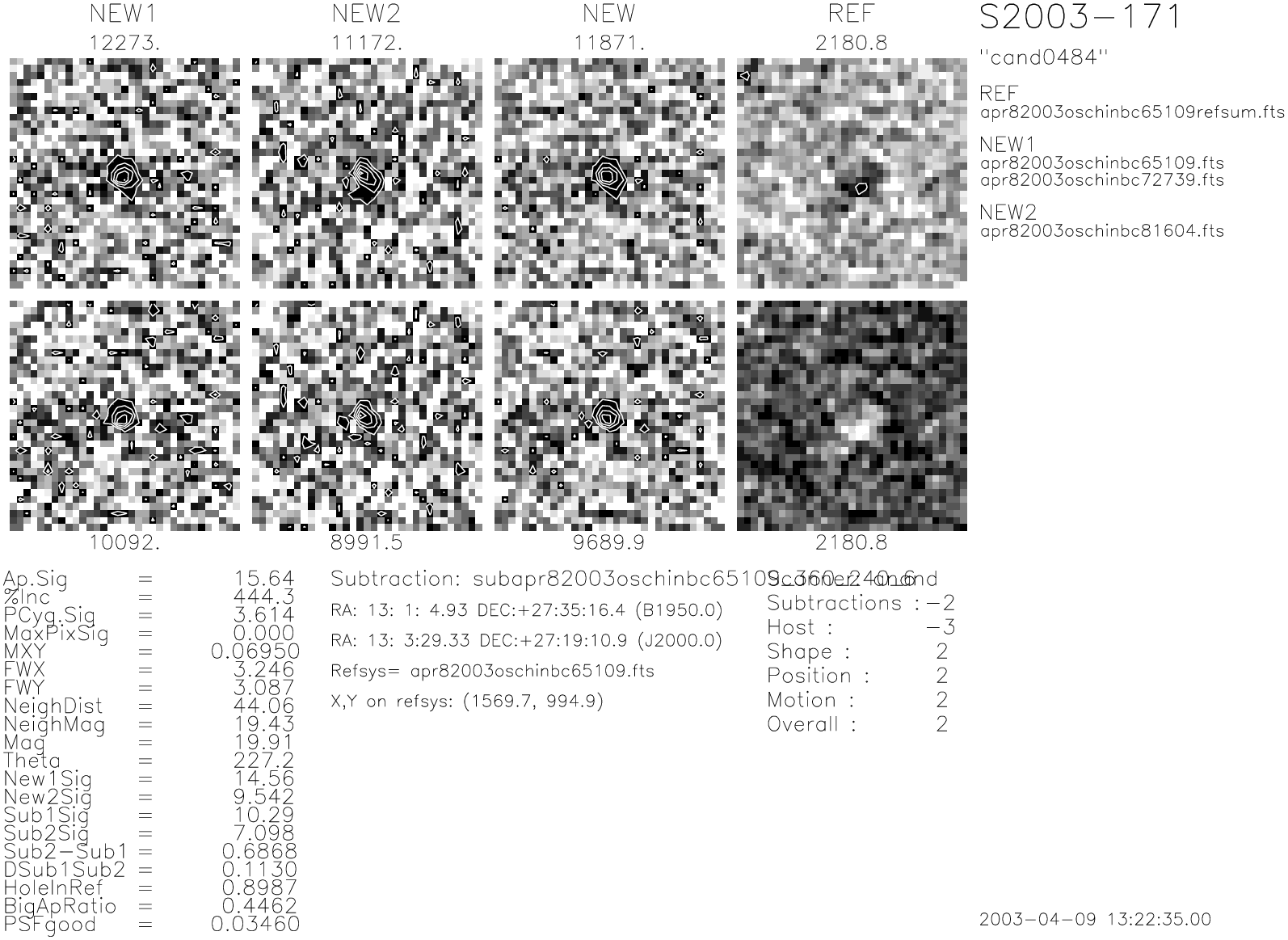}\label{fig:2003do_discovery}}
\hspace{0.3in}
\subfigure[2003do]{\includegraphics[height=2in]{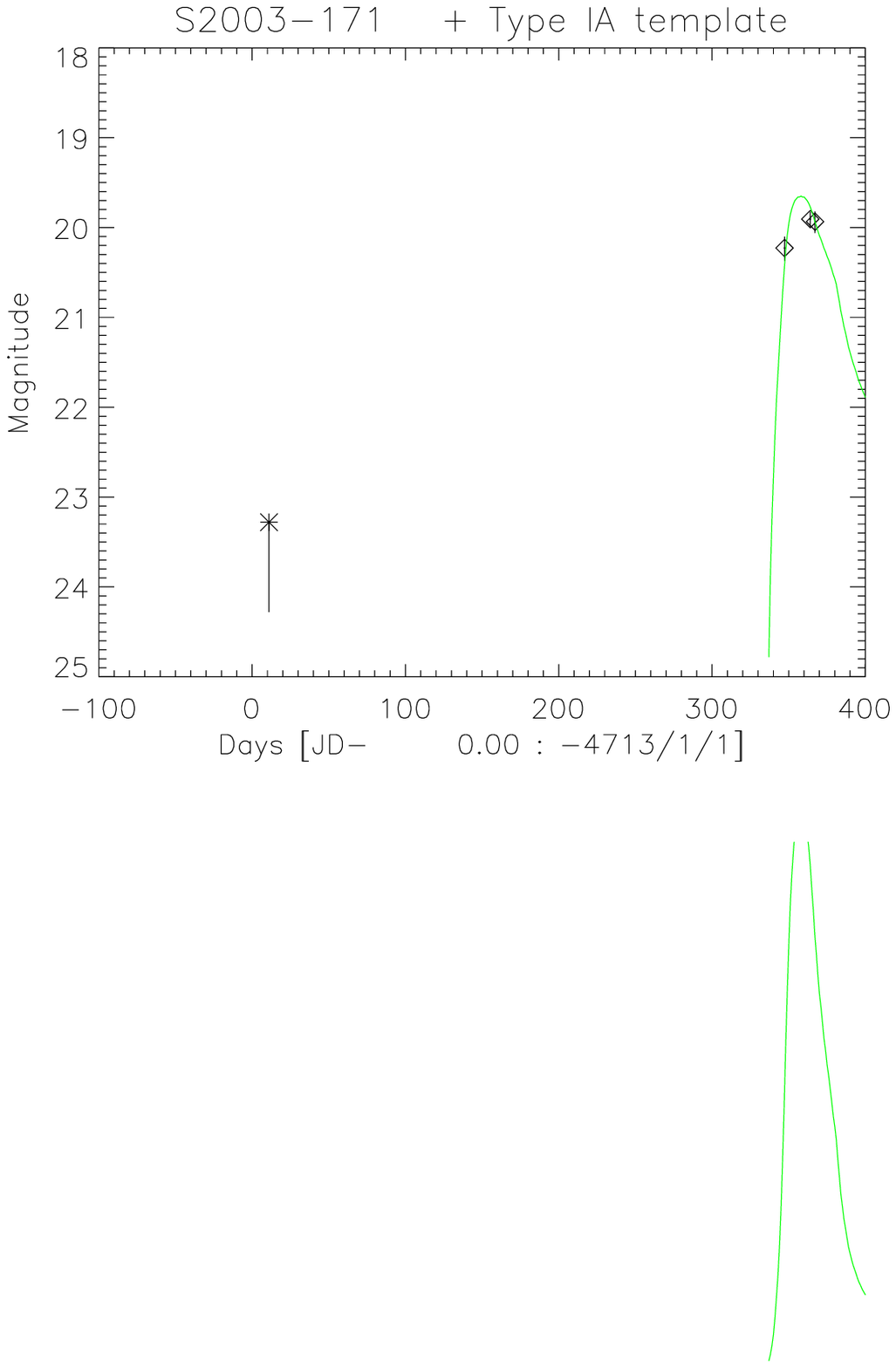}\label{fig:2003do_lightcurve}}
\vspace{0.3in}
\subfigure[2003dp]{\includegraphics[angle=90,height=2in,width=3in]{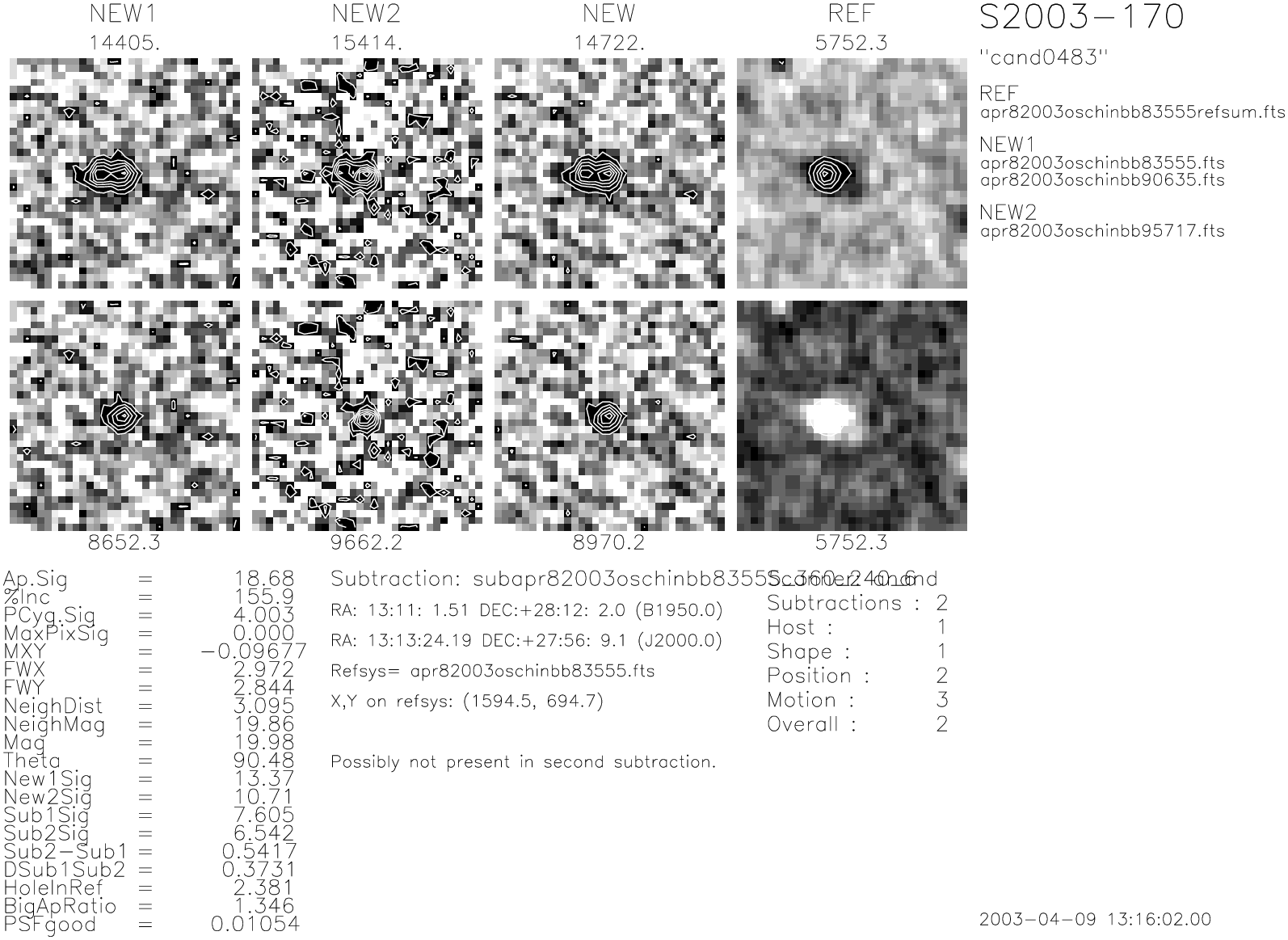}\label{fig:2003dp_discovery}}
\hspace{0.3in}
\subfigure[2003dp]{\includegraphics[height=2in]{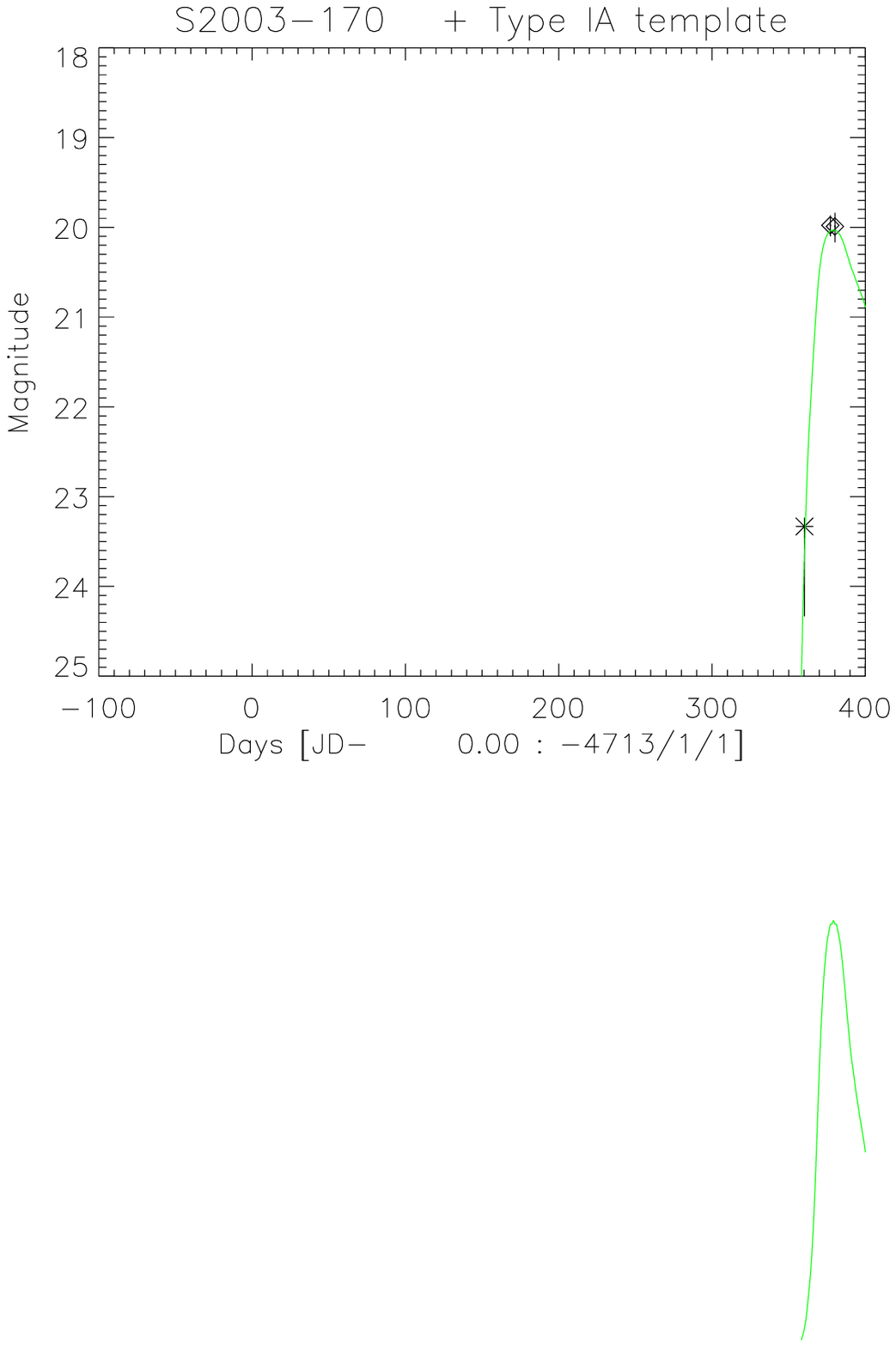}\label{fig:2003dp_lightcurve}}
\vspace{0.3in}
\end{figure}

\clearpage\pagebreak
\begin{figure}
\subfigure[2003dq]{\includegraphics[angle=90,height=2in,width=3in]{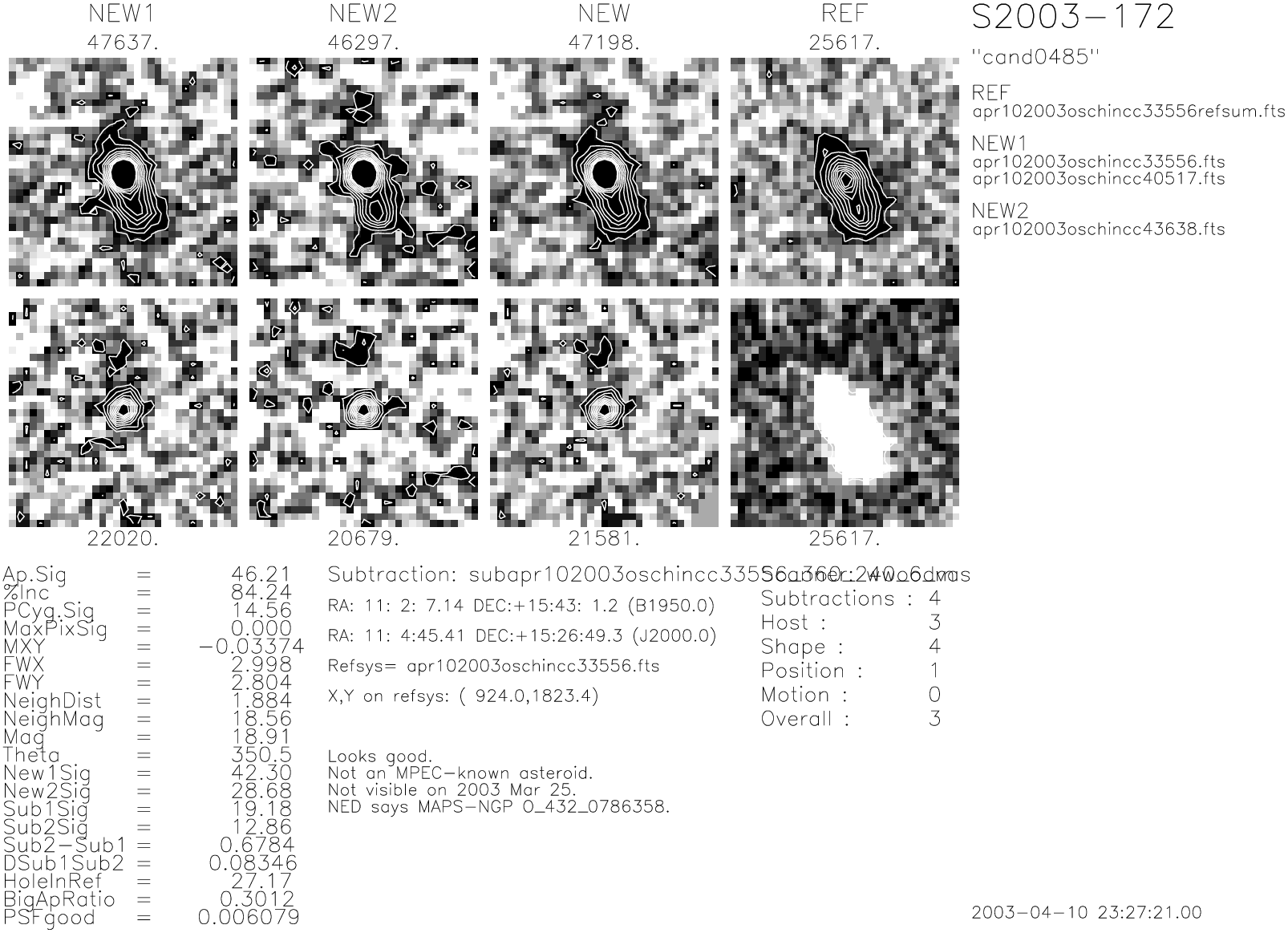}\label{fig:2003dq_discovery}}
\hspace{0.3in}
\subfigure[2003dq]{\includegraphics[height=2in]{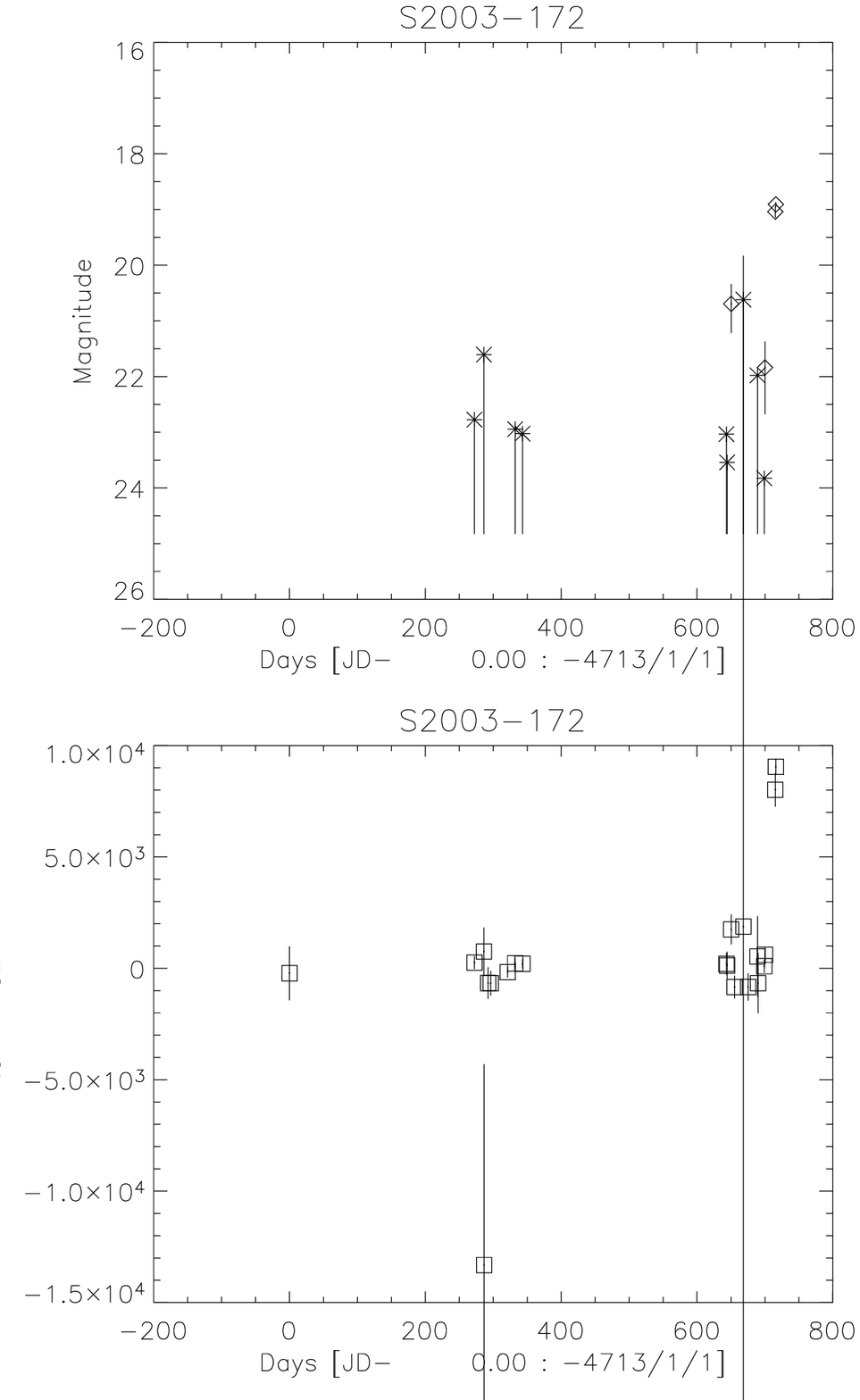}\label{fig:2003dq_lightcurve}}
\vspace{0.3in}
\subfigure[2003ee]{\includegraphics[angle=90,height=2in,width=3in]{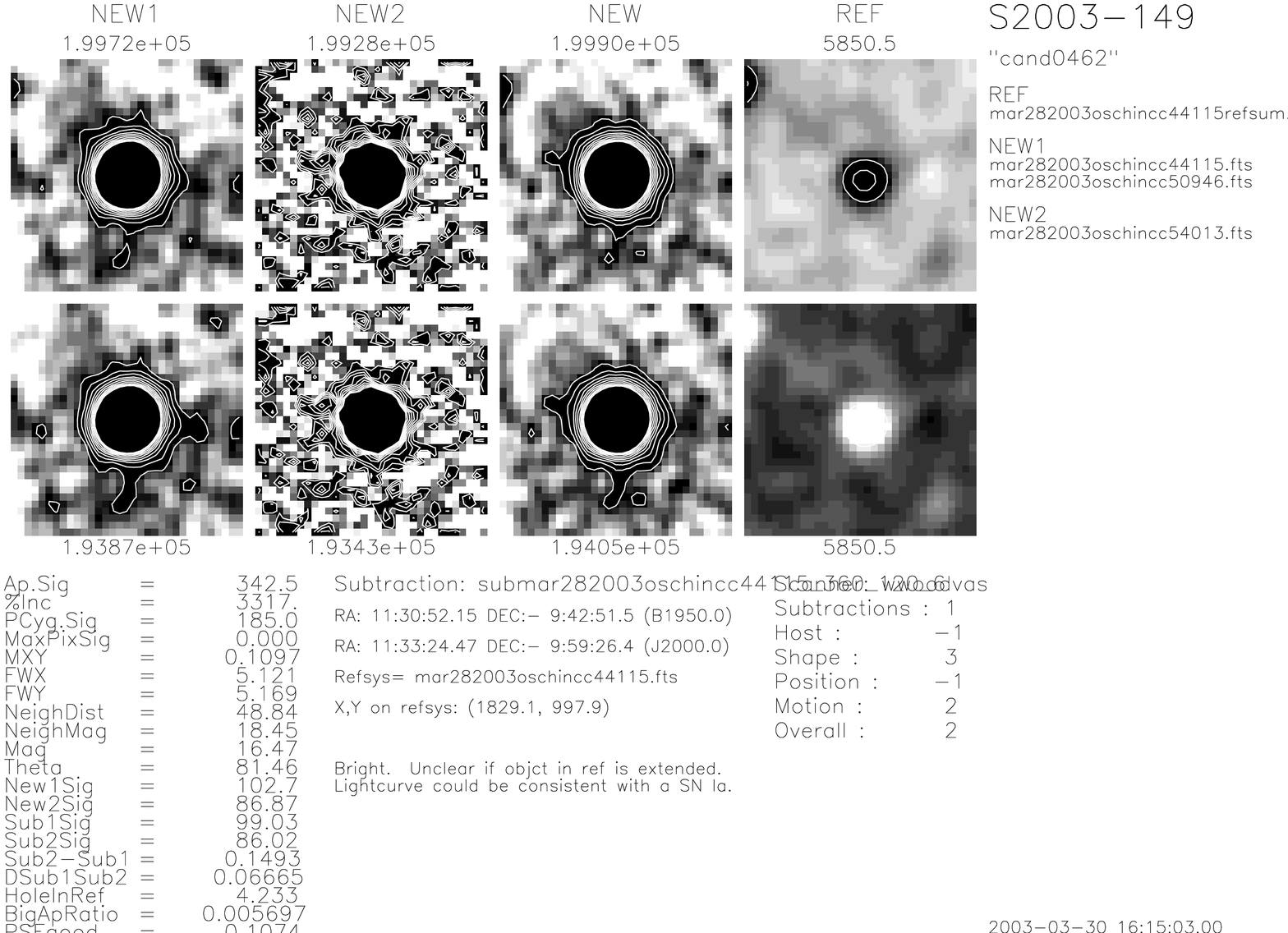}\label{fig:2003ee_discovery}}
\hspace{0.3in}
\subfigure[2003ee]{\includegraphics[height=2in]{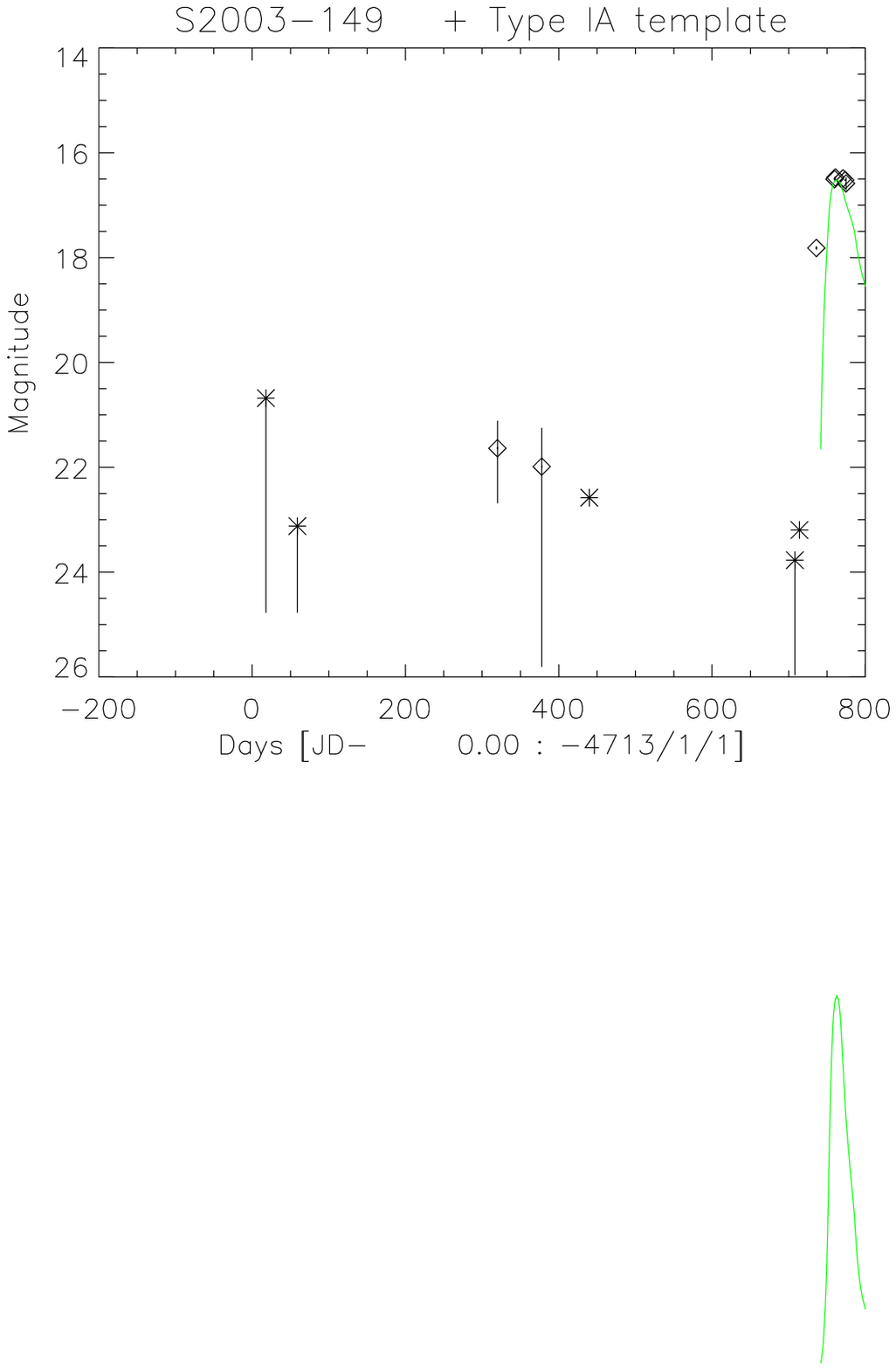}\label{fig:2003ee_lightcurve}}
\vspace{0.3in}
\subfigure[2003ef]{\includegraphics[angle=90,height=2in,width=3in]{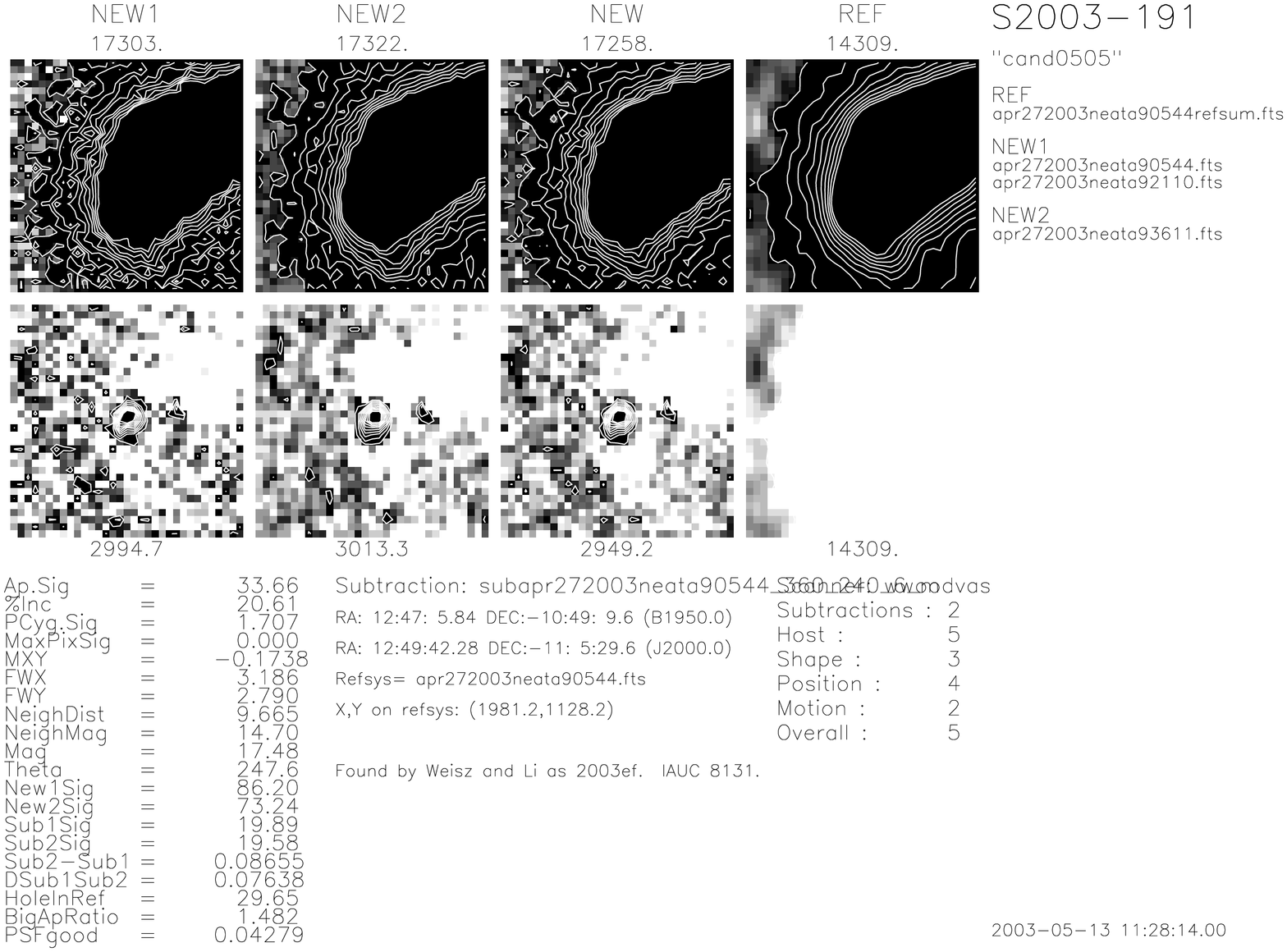}\label{fig:2003ef_discovery}}
\hspace{0.3in}
\subfigure[2003ef]{\includegraphics[height=2in]{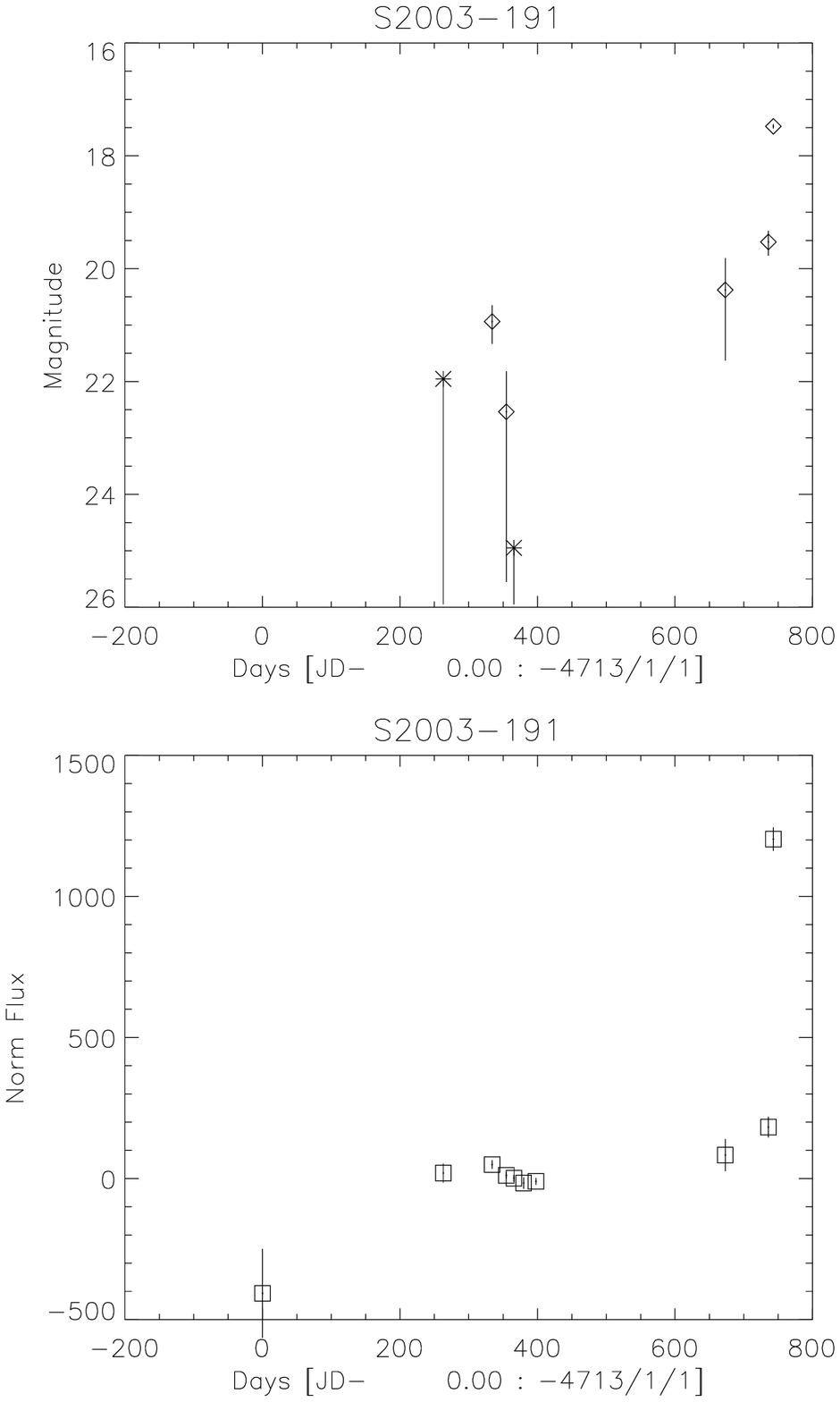}\label{fig:2003ef_lightcurve}}
\vspace{0.3in}
\end{figure}

\clearpage\pagebreak
\begin{figure}
\subfigure[2003eo]{\includegraphics[angle=90,height=2in,width=3in]{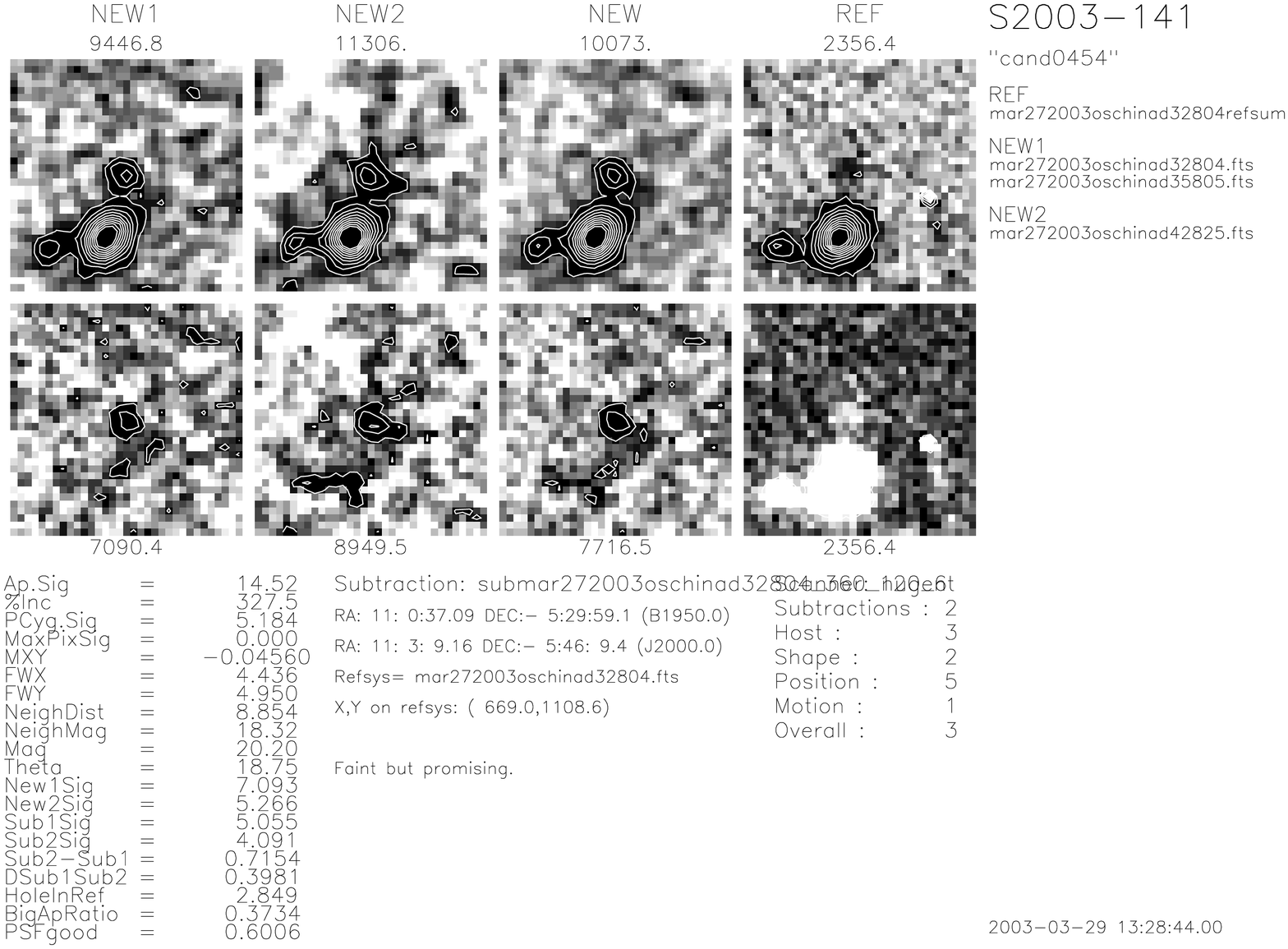}\label{fig:2003eo_discovery}}
\hspace{0.3in}
\subfigure[2003eo]{\includegraphics[height=2in]{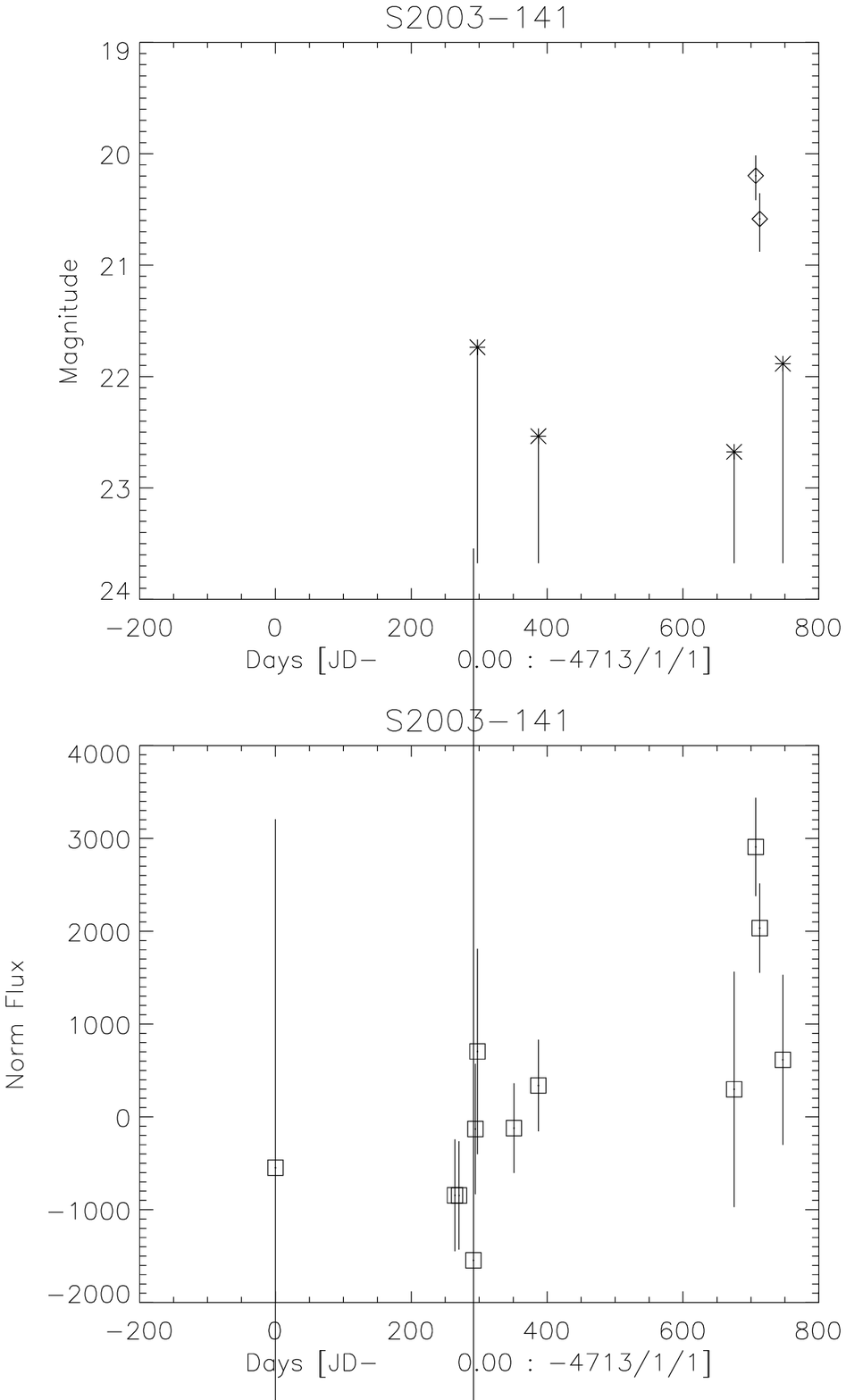}\label{fig:2003eo_lightcurve}}
\vspace{0.3in}
\subfigure[2003ex]{\includegraphics[angle=90,height=2in,width=3in]{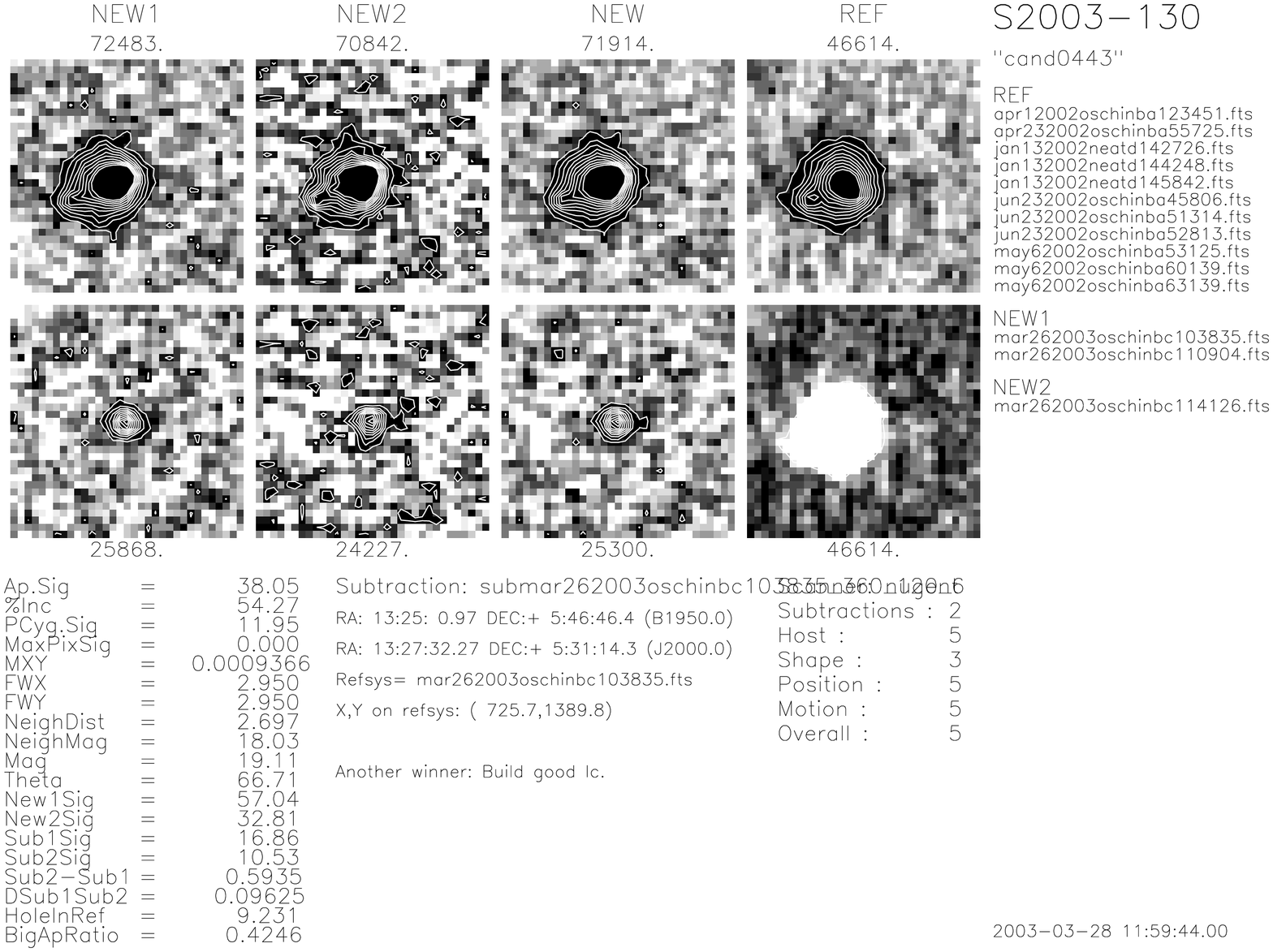}\label{fig:2003ex_discovery}}
\hspace{0.3in}
\subfigure[2003ex]{\includegraphics[height=2in]{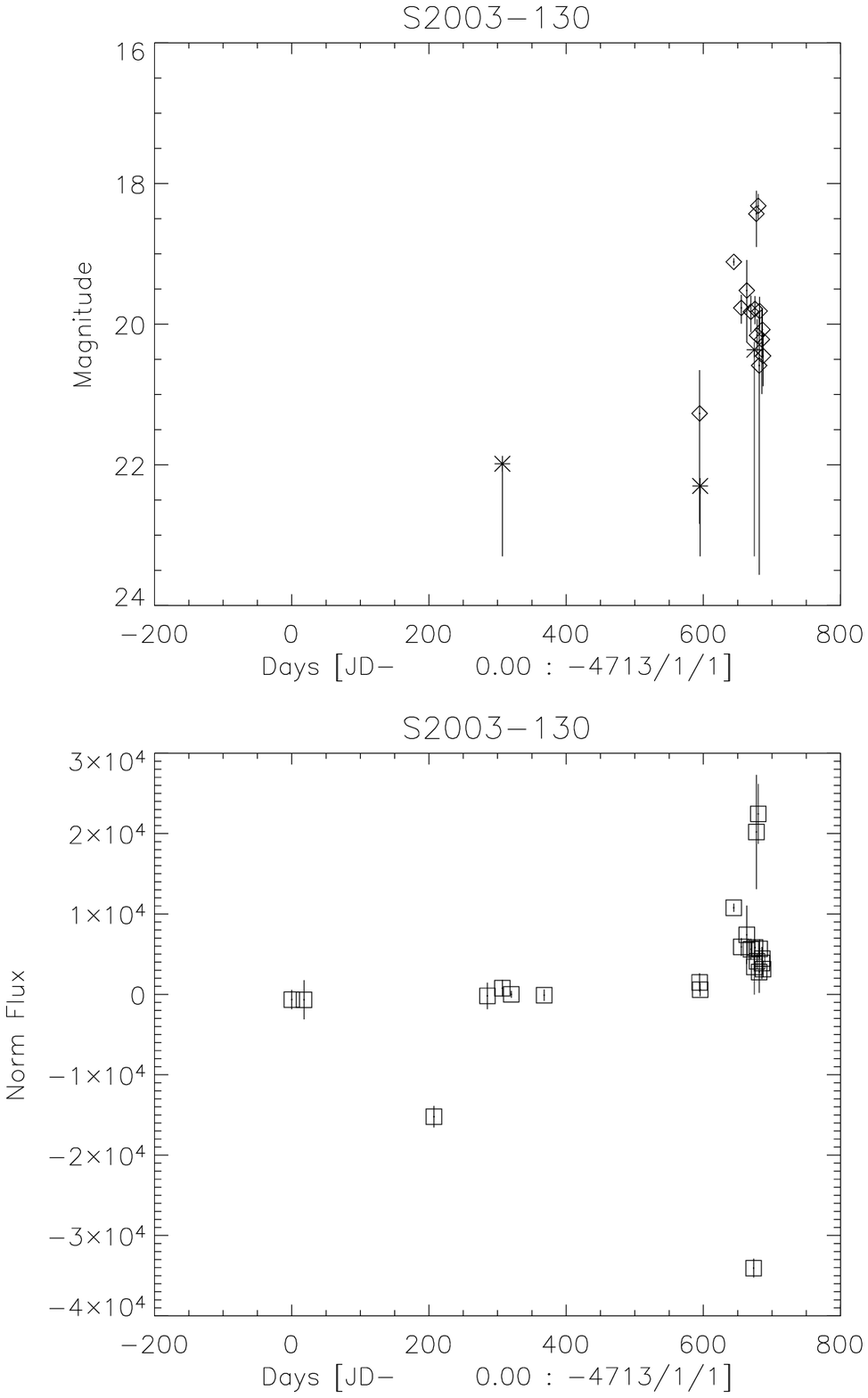}\label{fig:2003ex_lightcurve}}
\vspace{0.3in}
\subfigure[2003ey]{\includegraphics[angle=90,height=2in,width=3in]{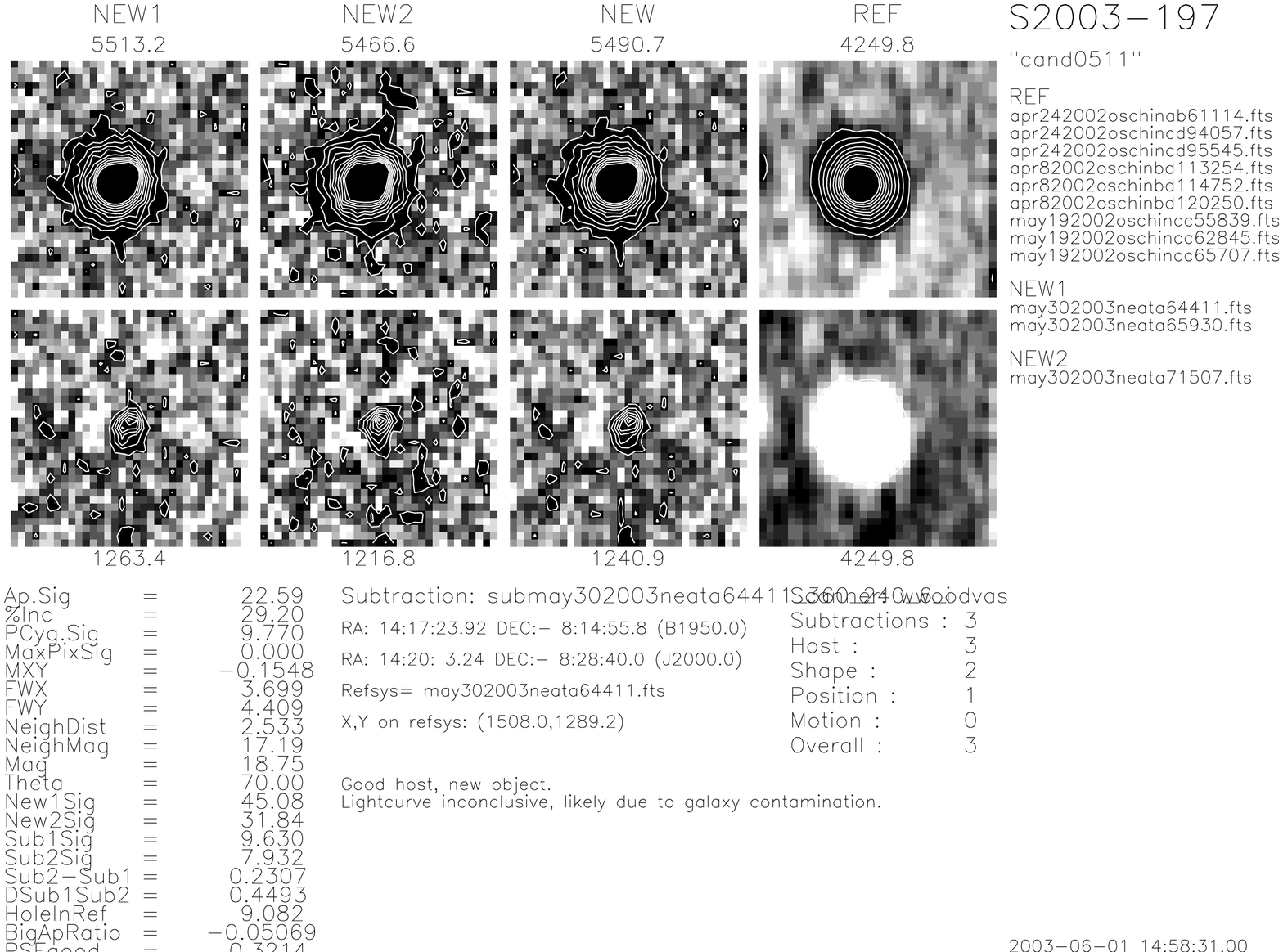}\label{fig:2003ey_discovery}}
\hspace{0.3in}
\subfigure[2003ey]{\includegraphics[height=2in]{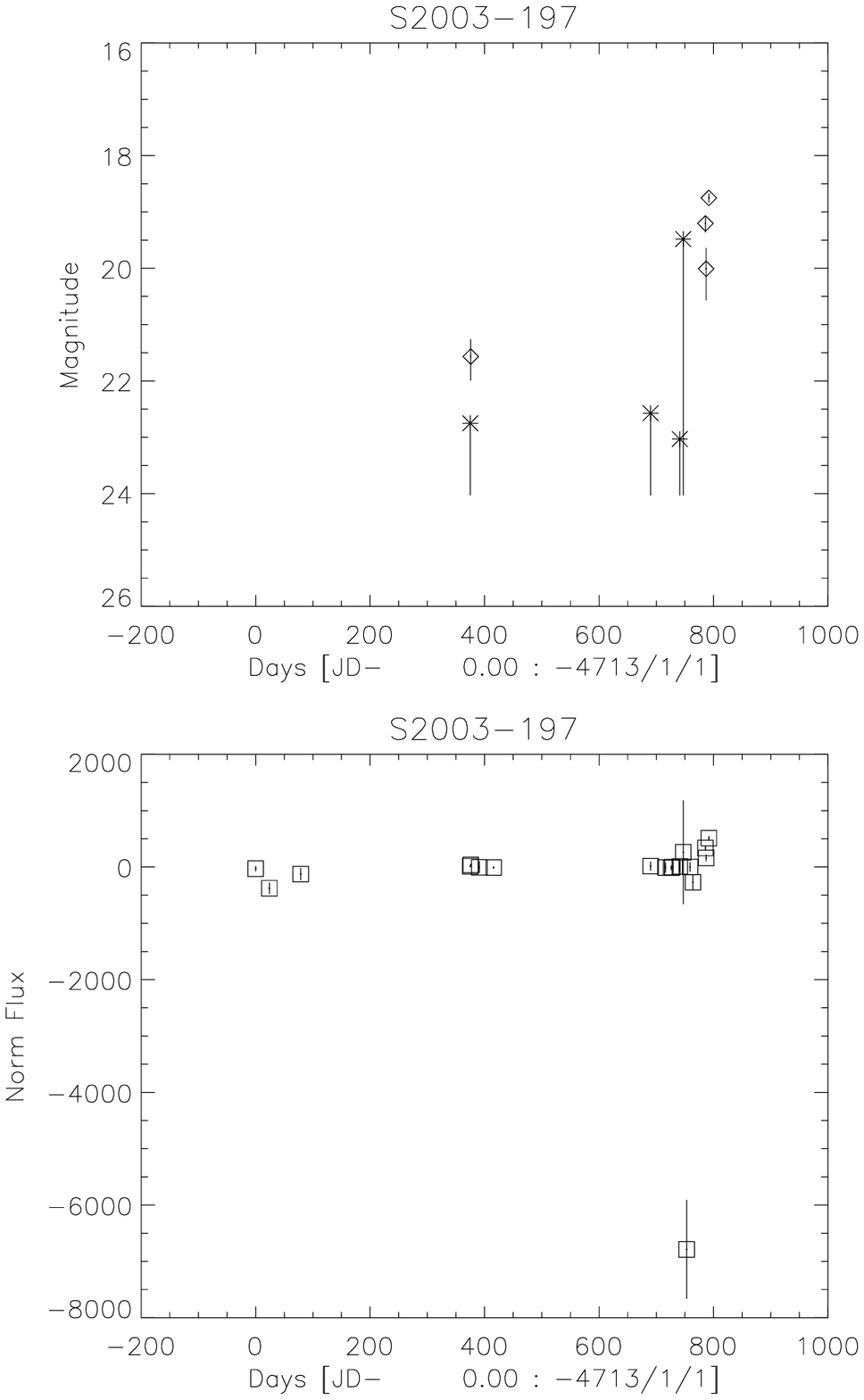}\label{fig:2003ey_lightcurve}}
\vspace{0.3in}
\end{figure}

\clearpage\pagebreak
\begin{figure}
\subfigure[2003gc]{\includegraphics[angle=90,height=2in,width=3in]{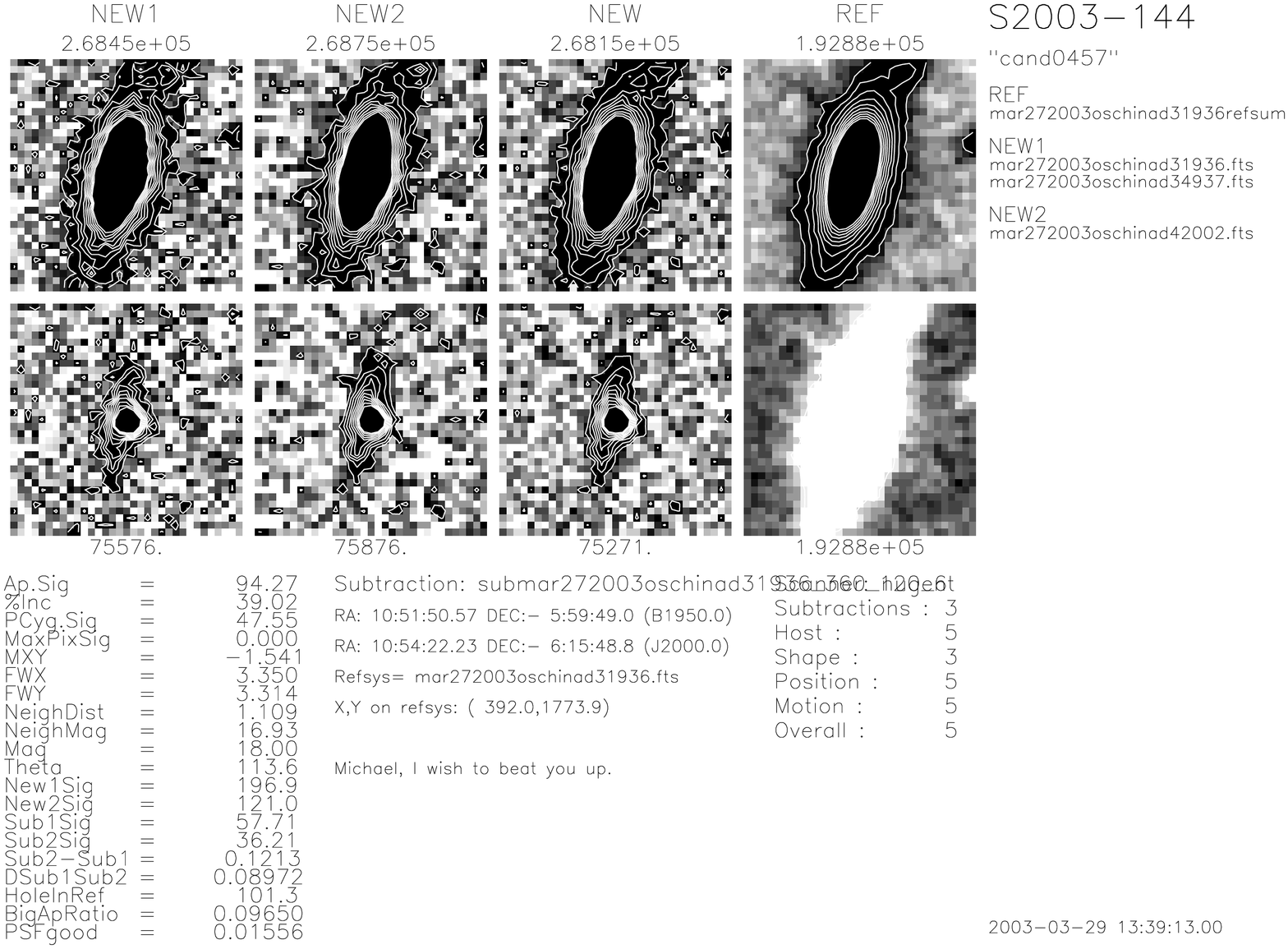}\label{fig:2003gc_discovery}}
\hspace{0.3in}
\subfigure[2003gc]{\includegraphics[height=2in]{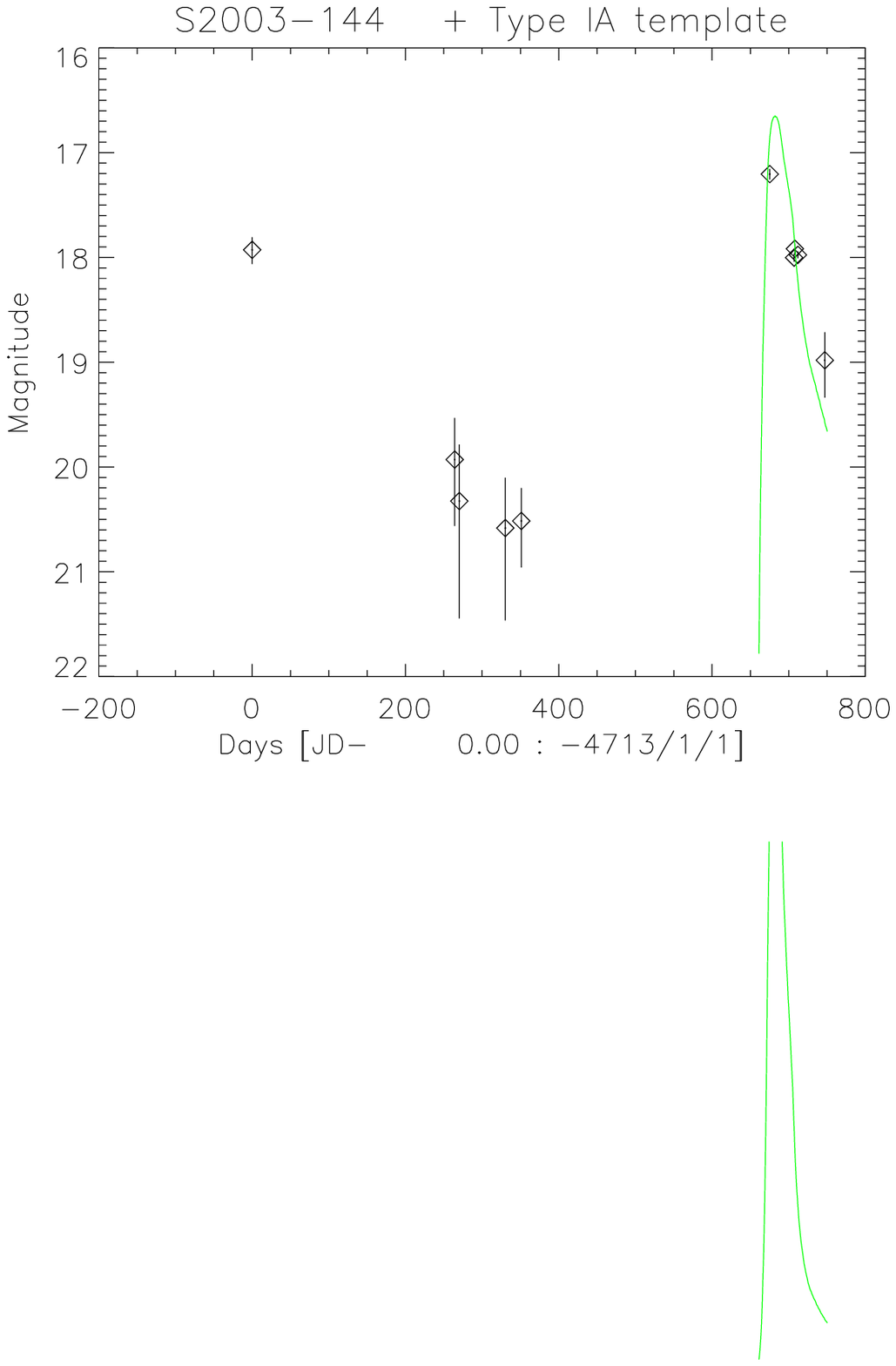}\label{fig:2003gc_lightcurve}}
\vspace{0.3in}
\subfigure[2003gi]{\includegraphics[angle=90,height=2in,width=3in]{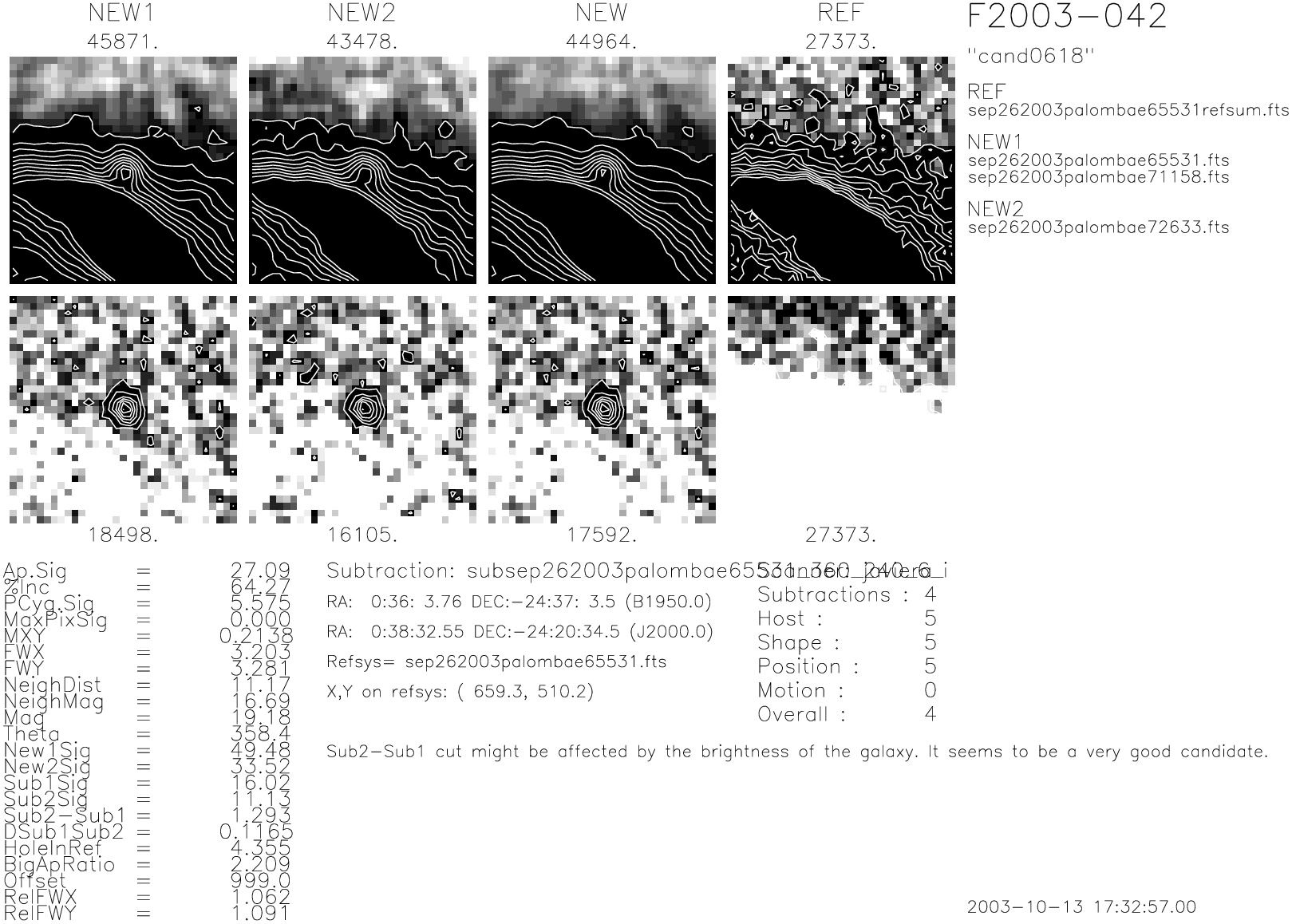}\label{fig:2003gi_discovery}}
\hspace{0.3in}
\subfigure[2003gi]{\includegraphics[height=2in]{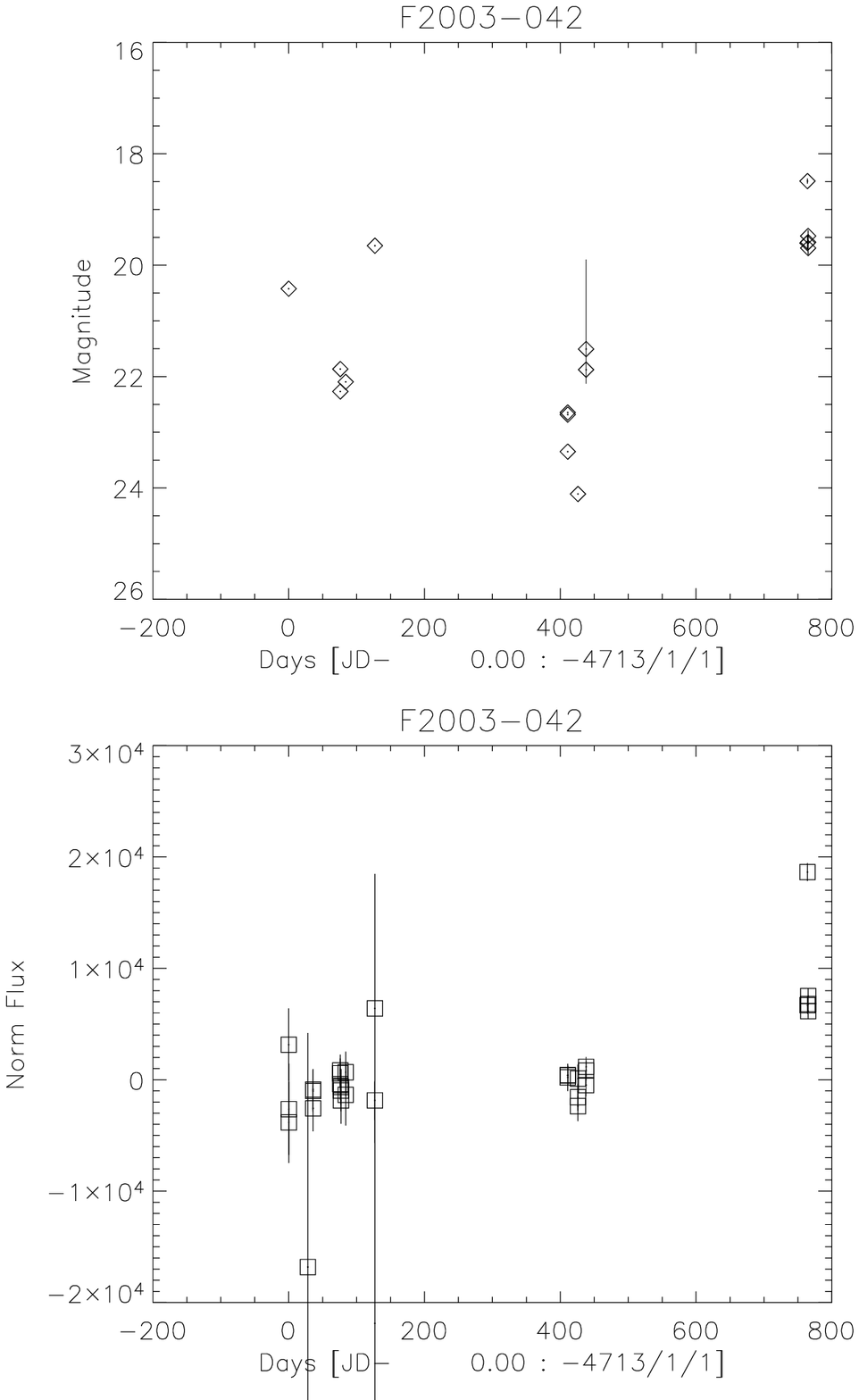}\label{fig:2003gi_lightcurve}}
\vspace{0.3in}
\subfigure[2003gt]{\includegraphics[angle=90,height=2in,width=3in]{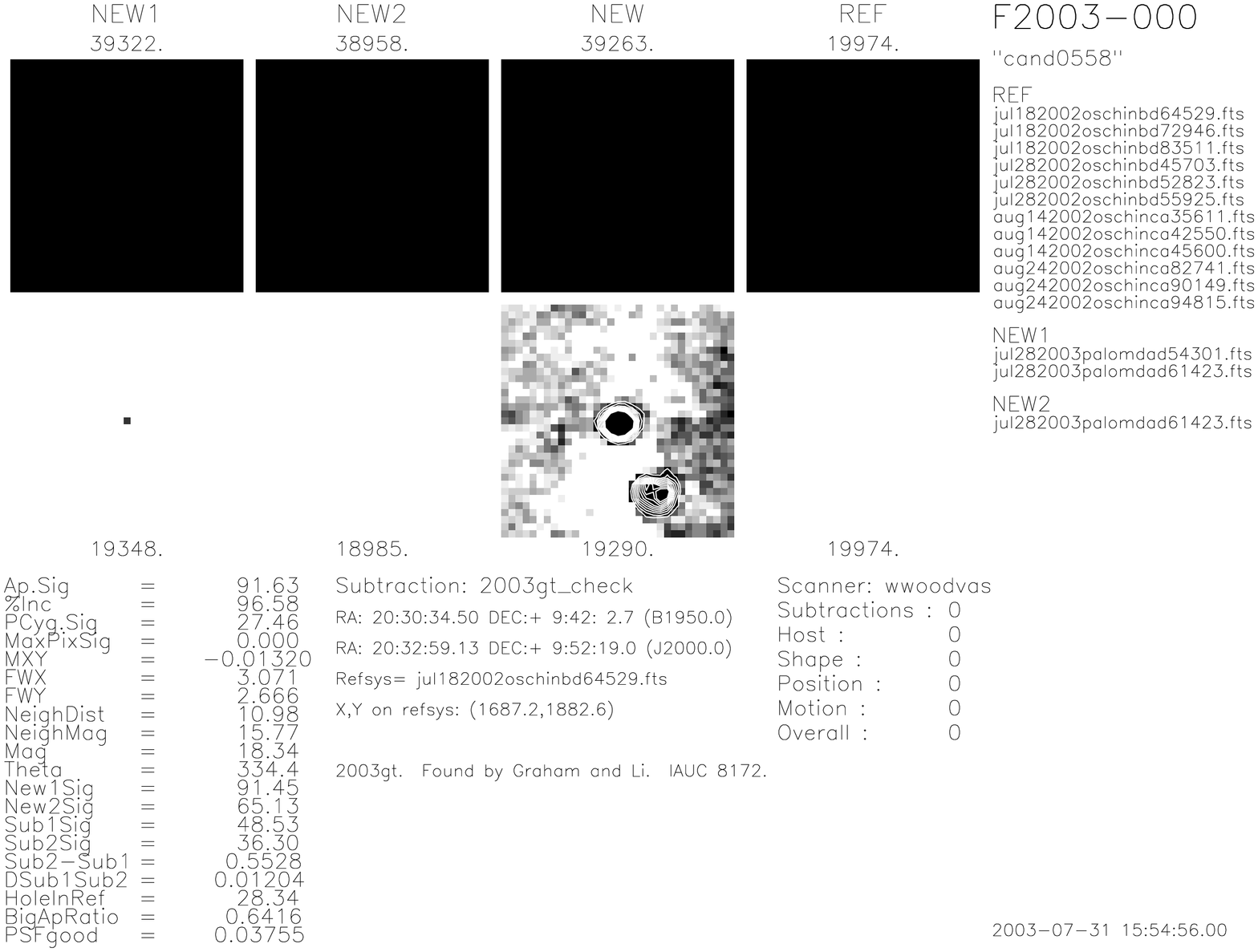}\label{fig:2003gt_discovery}}
\hspace{0.3in}
\subfigure[2003gt]{\includegraphics[height=2in]{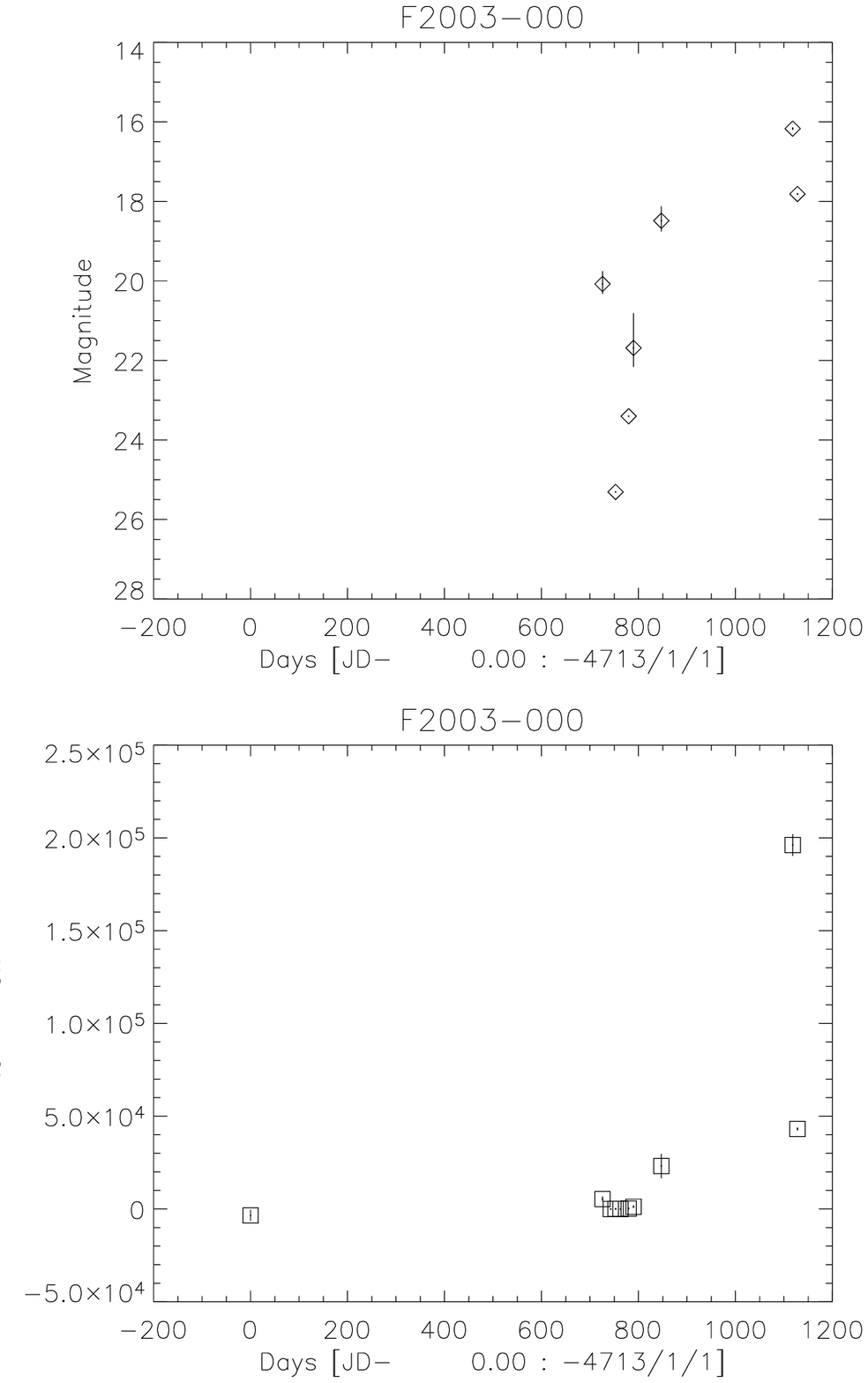}\label{fig:2003gt_lightcurve}}
\vspace{0.3in}

\end{figure}


\part{Supernova Rates}
\label{part:supernova_rates}
\chapter{Type Ia Supernova Rates from the SNfactory}
\label{chp:rates}

\section{Motivation}

Supernovae are a dramatic link to the star formation history (SFH) of
galaxies.
As discussed in Sec.~\ref{sec:intro_sn_rates}, the SN~Ia rate is
related to the SFH of the underlying galaxy population.
While core-collapse SNe trace the SFH within $\sim10^7$~years of the SN explosion, SNe~Ia are connected to the SFH through a delay time dependent on the formation mechanism for SNe~Ia 
(for example, see \citet{madau98b}, \citet{gal-yam04}, and \citet{strolger04}).
The delay time between the core-collapse SN rate and the
SN~Ia rate places constraints on SN~Ia progenitor models.
More generally, the relationship between the core-collapse and SN~Ia rates
provides insight into the initial mass fraction, chemical evolution,
and star formation rate and history of galaxies~\citep{madau98b,maoz04}.
The recent interest in SNe~Ia for cosmological studies has allowed for
and encouraged complementary studies of the rates of SNe~Ia to explore galaxy
evolution and star formation during the recent expansion history of
the Universe~\citep{pain02}.

\section{Introduction}

The history of supernova rates has been one of good effort and poor
statistics.  The central difficulty has lain in the lack of a sample of
a large number of supernovae from a well-understood and homogeneous
search of the sky.

Fritz Zwicky was a pioneer of supernova studies and contributed the
first papers on supernova rates~\citep{zwicky38,zwicky42}.  He used a
16$^{\prime\prime}$ Schmidt telescope on Mt. Palomar, California, to study a
controlled sample of 3000 galaxies brighter than 15th magnitude.  
\citet{zwicky38} introduced the control-time method of determining supernova
rates from a given search program.  In its most general form,
 the control-time method assigns a period of time
for which each exposure was sensitive to a supernova explosion.
This formalism compresses the variable sensitivities along 
the rise and fall of a typical supernova into one number.
Zwicky assumed a control-time of one
month for each observation in the studies mentioned.  While convenient
and appropriate for the state of knowledge of supernovae in 1938, the
difference in light curves between types of supernova can lead to
errors in estimation of the supernova rate, a possibility Zwicky 
considered and then dismissed in the context of his own study.
For example, assuming that every supernovae would be of Type I
results in an underestimate of the total supernova rate due to the fact
that Type II supernovae are intrinsically dimmer and thus fewer
are seen in a flux-limited survey.


To conduct an accurate rate estimation, one needs to model the light
curves of the supernova expected and then compare the fit of the model
to the data obtained.  This chapter will present an analysis of the
nearby ($z<0.1$) SN~Ia rate done in the manner of Zwicky's original
control-time method, a more generalized control-time analysis that
fully models the light curves of SNe~Ia, and, finally, the
preferred solution of using a simulated set of 200 million SNe~Ia
throughout the search coverage in time and area to directly measure
the actual sensitivities of the search images.

At present, the distant supernova rate is better understood than the
nearby supernova rate because of the way distant supernova searches
are conducted.  Searches for high-redshift SNe take deep exposures of
a couple of square degrees (or less) of the sky.  The volume covered
by this exposure samples all galaxies in the region, unlike the
nearby, directed supernova searches, which target known (and thus 
usually brighter) galaxies.  The small number of fields and easily
calculable control times characteristic of distant supernova searches
have allowed the determination of the supernova rate at $z=0.55$ to
$\sim15\%$~\citep{pain96,pain02}.  The very nearby supernova rate ($z
\sim 0.01$) is known to $\sim30\%$~\citep{muller92,cappellaro97,cappellaro99},
but the nearby supernova rate is only known to
$40$--$90\%$~\citep{hardin00,blanc04}.

\citet{cappellaro97} presented the results of five SN surveys in
an attempt to estimate the nearby supernova rate.  While the paper
did a reasonable job with data at hand, its conclusions were a bit
speculative because of the limitations of the searches.
\citet{cappellaro99} updated this attempt by focusing purely on the
search of \citet{evans97} as a consistent, homogeneous sample of
supernovae.  The greatest limitation of this second paper is that
Evans' search was only of known and generally bright galaxies.  
As a result,
the rate in the general field of space is poorly determined
from Evans' data.  This galaxy-selection effect is
a common limitation of nearby supernova searches.  In contrast, the
large-sky survey of the Nearby Supernova Factory (SNfactory; see Part.~\ref{part:snfactory}) represents a major improvement in
the ability to calculate general supernova rates.

\section{Supernova Rates from Supernova Searches}

Supernova rates are traditionally expressed in supernova units (SNu), where 
\linebreak
1 SNu~=~1~supernova~/~$10^{10}$~$L_B\Sun$~/~century (for example, see \citet{cappellaro88}), 
which is loosely speaking one supernova per galaxy per century.  
This unit is chosen because there is
evidence that the supernova rate is proportional to
luminosity~\citep{tammann70}.  In this way, the SNu allows one to compare
different galaxies in a standard manner.  Most supernova searches to
date have been directed-galaxy searches and so the SNu has been a way
to standardize the known-galaxy supernova rate to the supernova rate in the overall Universe.  This chapter presents a determination of the SN~Ia rate from a sample of the prototype SNfactory search.

The SNfactory search is an areal search that covers half of the sky
without regard to known galaxies (except our own---see Fig.~\ref{fig:neat_sky_coverage} for the typical NEAT coverage).  For this reason, this study will quote the SN~Ia
rate per unit volume.  
Using a galaxy luminosity density function from the Sloan Digital Sky Survey~\citep{blanton03a}, 
the SN~Ia rate will be translated to SNu for
comparison with other recent
studies~\citep{cappellaro97,cappellaro99,hardin00,pain02,dahlen04}.
The convention used in \citet{pain02} will be adopted here
with the SN~Ia rate expressed per volume denoted by $r_V$ and the
rate per luminosity denoted by $r_L$.

\begin{figure}
\includegraphics[angle=270]{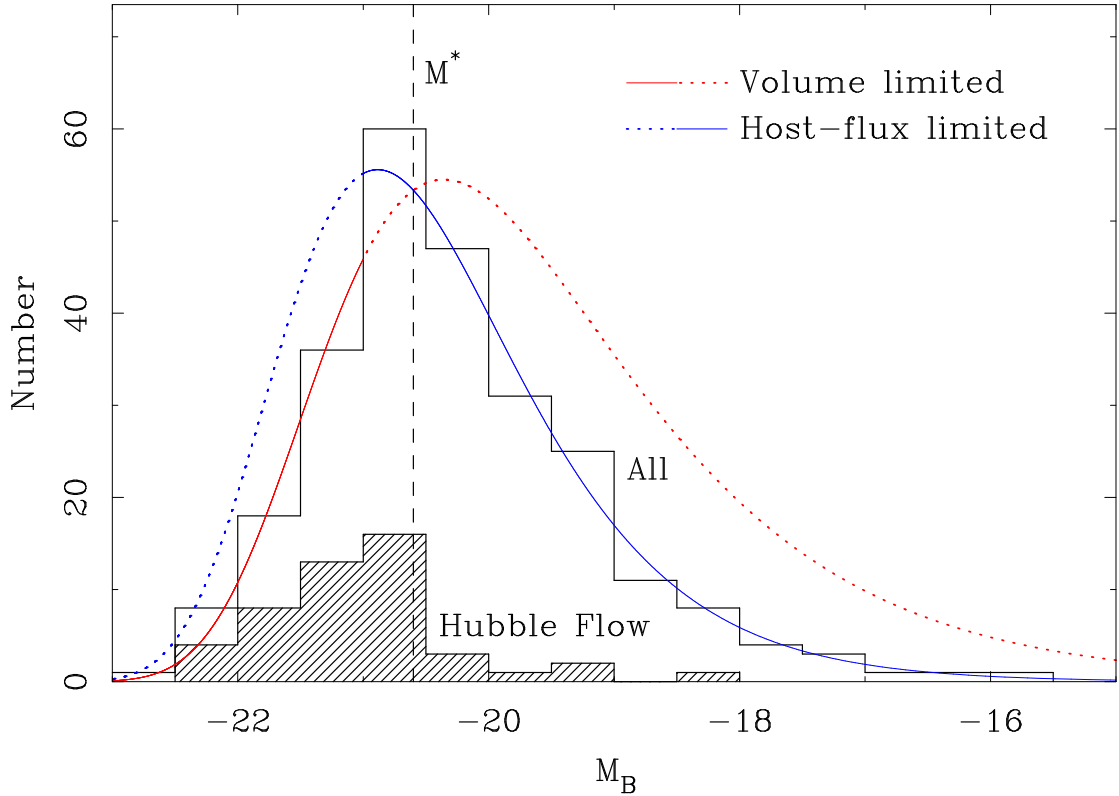}
\caption{The host galaxy luminosity distribution for local supernovae 
from the Asiago catalog.  The solid line shows the distribution
of all galaxies and is consistent with a volume-limited sample for the
brightest galaxies and a flux-limited sample for intrinsically fainter
galaxies.  The shaded histogram shows the luminosities of galaxies in
the Hubble flow.  Note the sharp cutoff for Hubble-flow galaxies with
$M_B < M^\star$.  This cutoff demonstrates the significant bias in the host
galaxy source population for Hubble-flow SN searches using known
galaxies.  The different results expected for volume-limited and galaxy-flux-limited searches are indicated by the ``Volume limited'' and ``Host-flux limited'' lines.  The ``All'' histogram indicates the SNe found prior to the SNfactory project and is consistent with a volume limited search for bright galaxies but host-flux limited for fainter galaxies.  The SNfactory will conduct an essentially supernova-flux
limited survey, 
which will result in galaxies selected following the ``Volume
limited'' curve (but see Sec.~\ref{sec:percent_increase}).
This survey will thus have better representation of galaxies at the faint end 
of the luminosity function than previous nearby supernova surveys,
particularly out in the smooth Hubble flow.
(Figure courtesy of Greg Aldering.)}
\label{fig:asiago_blum}
\end{figure}

In the case of known-galaxy searches, supernova
rates are generally quoted per blue luminosity as galaxy blue luminosity is a
measurable quantity believed to be correlated with the
supernova rate and available to normalize the rates.  It is
important to bear in mind, however, that the blue luminosity of a
galaxy as a proportion of total luminosity depends on its type
\citep{cappellaro99}.  In addition, known-galaxy searches are strongly
biased toward finding supernovae only in galaxies on the brighter end
of the galaxy luminosity function.  Fig.~\ref{fig:asiago_blum} shows
this effect in the host galaxy luminosity distribution for local supernovae.

In contrast, a ``blind'' search, scanning large swathes of sky,
provides an excellent opportunity to calculate supernova rates from a
much more complete sample of host galaxy luminosity types.  The galaxy
sample is determined by the area and flux limit for detection of supernovae,
rather than the luminosities of the galaxies, as it would be for a
known-galaxy search.  Because the spread of SN~Ia luminosities is much
smaller than that of galaxies, and their light curves are well
understood, the limiting magnitude at the faint limit of a ``blind''
search can be determined with relative confidence
(but see Sec.~\ref{sec:percent_increase}).

When calculating a supernova rate, the most salient factor is the
control time of the search, $T(z)$, or the amount of time the search
is sensitive to a supernova as a function of redshift.  The
control time includes all relevant search efficiencies and
observation cadences.  To compare with observations, the
number of supernovae found as a function of redshift, $N(z)$,
is compared to the predicted value given the rate as a function
of redshift, $r(z)$, and the volume and control time:
\begin{equation}
\Delta\,N(z) = \frac{r(z)}{1+z}\Delta\,V(z) T(z).
\label{eq:snrate_z}
\end{equation}
The analyses presented in this chapter assume a constant $r(z)$ for
the redshift range of interest, $z<0.1$, so the calculation of 
the SN rate, $r$, reduces to 
\begin{equation}
r = \frac{\Delta\,N(z)}{\Delta\,V(z) T(z)} (1+z).
\label{eq:snrate}
\end{equation}
$N(z)$ comes immediately from the result of the search.  $V(z)$ is
dependent on the exact choice of cosmological model---here taken to be
\OM$=0.3$, \OL$=0.7$--but out to a redshift of
$z=0.1$, this dependence is relatively unimportant (see
Sec.~\ref{eq:sys_cosmological_parameters}).  The control time, $T(z)$,
is the challenging quantity to calculate because it is intrinsically tied
to the light-curve behavior of the supernova population, the
sensitivity of the search images, the observation pattern of the
search, and the selection process for finding supernovae.  In this
analysis, a SN~Ia light curve similar to that of \citet{goldhaber01} is
used.  The candidate selection criteria are described in
Sec.~\ref{sec:scorecuts}.  Finally, the cadence and a quick measure of sensitivity of the
search is contained in the \code{subng} table of the SNfactory
database.  These last two factors represent the biggest challenge and
greatest strength in using the SNfactory search to calculate SN rates.

\section{The Nearby Supernova Factory Pilot Sample}

The Nearby Supernova Factory pilot search presents a well-studied and
well-understood large-area ($\sim20,000\sqdeg$) survey of the nearby
Universe ($z~\lesssim0.1$).  This data set provides a rich opportunity
for estimating the nearby rate of supernovae of all types.  With a
well-characterized sample and an automated detection strategy, the
efficiencies of the search can be determined with very good accuracy.

Analyzing the SN~Ia rate for the SNfactory using the control-time
methodology of \citet{zwicky38} is complicated by the varying
sensitivity of the SNfactory search images and the differing patterns
of repetition of the fields on the sky.  Zwicky observed fields with
gaps of at least one month and assumed that each observation was
sensitive to supernovae that went off in the galaxies under study in
the previous month.  Our knowledge of supernova light curves has
improved since 1938, and a more precise determination of the
sensitivity to a particular class of supernova can be made through the
use of light curve templates.  However, the varying sensitivity of the
SNfactory images makes the calculation of an effective control time
complicated.  Other supernova rate calculations have either assumed a
constant limiting
magnitude~\citep{zwicky38,zwicky42,richmond98,cappellaro99} or have
had only one search observation of a field~\citep{pain02}.

This analysis of rates from the SNfactory will begin with a statistical analysis of the SNe described in Chapter~\ref{chp:supernovae_found}.  The redshift, epoch, and type of the SNe found will be discussed in Sec.~\ref{sec:redshifts} and \ref{sec:epoch} (also see Sec.~\ref{sec:typing}).
After this discussion of the SNfactory sample, 
the first results presented in this chapter will make the assumption
of a constant limiting magnitude for a given field by only including
subtractions that are at least as deep as the chosen nominal limiting
magnitude and only including supernova discovered at a brighter
magnitude.  
Finally, the full analysis will be
presented, which includes a simulation that propagates a database
of 200 million SNe~Ia through the SNfactory search pipeline
to properly measure the true control time of the search.

\section{Percentage Increase Effect}
\label{sec:percent_increase}

Of the various scores described in Sec.~\ref{sec:scorecuts},
the one that has the largest effect on the calculation of supernova
rates is the score that measures the percentage increase of the 
measured candidate flux over the light in the reference image.
While light from the host galaxy also affects the signal-to-noise of the SNe, 
the percent increase cut used by the automatic candidate
classification has a more significant impact
on the effective sensitivity of the search.
The precise effect of this cut is a complicated function
dependent on the spatial distribution of host fluxes in the aperture of the SN,
which is dependent on the underlying host galaxy population.
While the light from
the host galaxy will also increase the noise in the aperture on the
subtraction and lead to additional suppression of faint supernova on
galaxies, the percentage increase threshold is the primary factor
in reducing the number of supernovae found, particularly at
very early and late epochs.

The percentage increase requirement can be seen to have an effect on
the limiting magnitude in the following simple calculation.  Assume
that a SN~Ia at maximum light is roughly as bright as its host galaxy
and that the radial distribution of supernova in a galaxy results in
the detection of supernova at half of the galaxy light on average.
Since, roughly speaking, a supernova is as bright as an average
galaxy, it will be fainter than 25\% of the total flux of its host
galaxy for all but 20 days or so of its light curve.  Thus, the
percent increase requirement of $>25\%$ used by the SNfactory 
(see Sec.~\ref{sec:scorecuts}) has a significant impact
on the number of supernovae found once galaxies become comparable to
the size of the aperture used in the calculation of the percent
increase score.  The impact of the percent-increase cut is an
even more significant effect for the intrinsically dimmer
core-collapse SNe.

Originally developed for use in the SCP high-redshift searches to help
discriminate against active galactic nuclei (AGNs), the percent
increase score was inherited by the SNfactory and became useful in
discriminating against variable and mis-subtracted stars.  However,
the impact on the search sensitivity of using this score with $25\%$
threshold was not realized until late 2003.  The largest effect of the
percent increase cut for the calculation of SN is in the 
after-maximum-light tail, 
but the SNfactory is only interested in SNe found before
maximum light, so the effect of this cut on the science goals is
significantly less than the effect on the rate calculation.  In modeling
the supernova rate from the SNfactory prototype search, a search that
included this cut, the effect of the percent increase cut has been
included into the calculations by using the galaxy population model 
discussed in Sec.~\ref{sec:galaxy_modeling}.  Ultimately a lower percentage increase cut will be
used in conjunction with additional scores (see Sec.~\ref{sec:scorecuts}) to help eliminate variable
stars while not eliminating so many supernovae at redshifts $> 0.05$
and/or at a very early phase (see Fig.~\ref{fig:efficiency_fke6_norm}).

Properly accounting for the effect of the percent increase cut in
measuring the efficiency of a supernova search requires an
understanding of the galaxy population in the volume searched for
supernovae.
One can try to simulate this effect from the search data by using all of
the galaxies on the images.  However, for the SNfactory NEAT images, it
is not possible to classify and detect the galaxies at the fainter and
more distants redshifts covered by the search, where the bulk of the
supernovae are found, so it is necessary to model the relevant galaxy populations for this search to understand the observe distribution of SNe found in the survey.

\section{Redshift Distribution of SNfactory Supernovae}
\label{sec:redshifts}

The expected redshift distribution for an large area survey,
such as the SNfactory search, is
primarily a function of the limiting magnitude, or depth to which one
is sensitive to those objects.  In the case of supernovae, the
brightness of the survey object varies with time.  Thus, near the
limits of the detection threshold, the search is sensitive only to supernovae
near their maximum brightness.  To convert from this magnitude limited
search to a volume limited search one can impose a cutoff magnitude.
This cutoff is the point at which one should see all supernova within a
nominal volume defined by the magnitude of the search object.

\begin{figure}
\begin{center}
\plotone{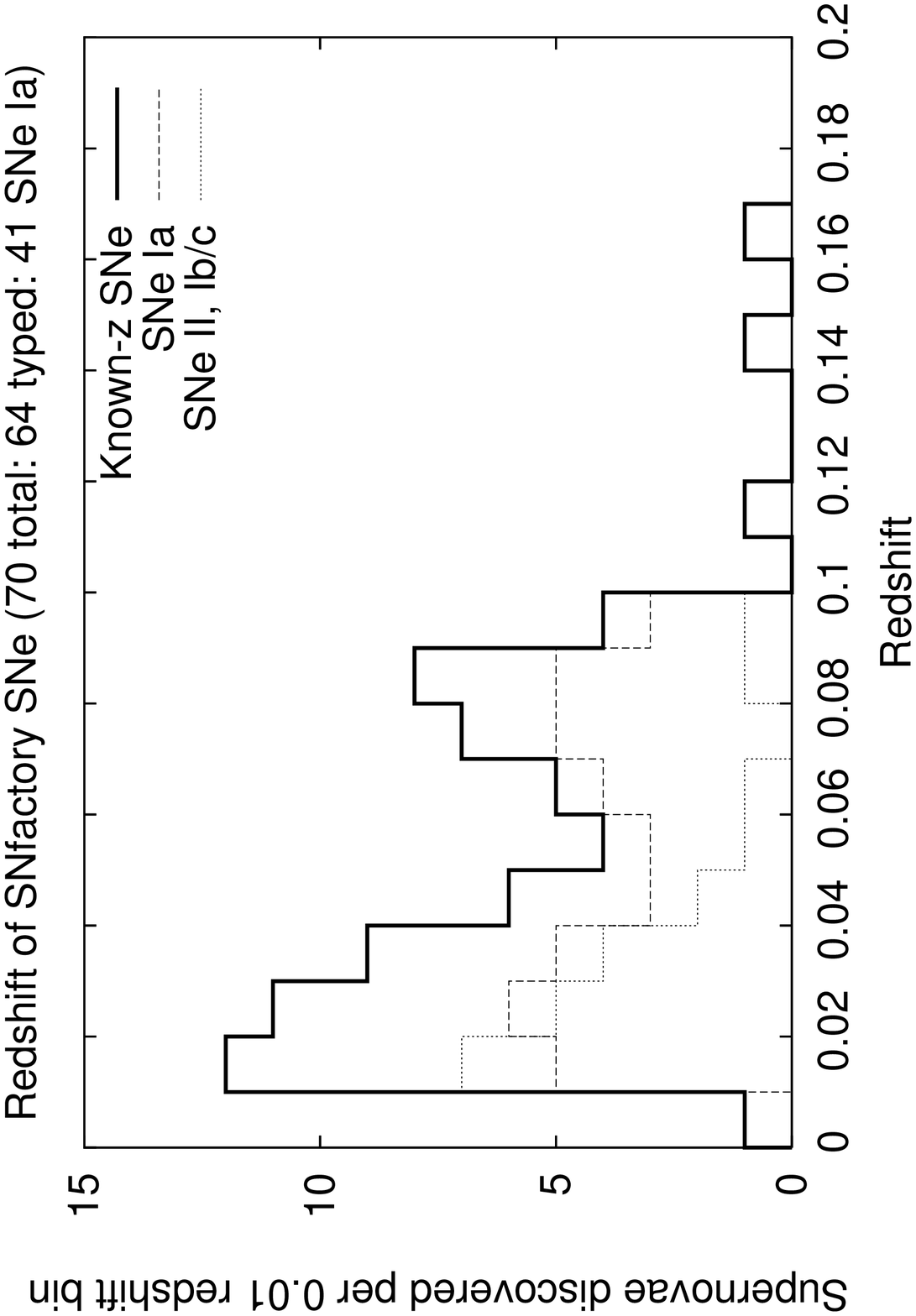}
\end{center}
\caption{The redshift distribution of 70~SNe with known redshifts found by the
SNfactory automated search pipeline from the beginning of the search
in 2002 through June 2003.  Note that as this search was dependent
on the community for most of these redshifts, there could be an
observer bias in this plot toward spectroscopically confirming
brighter and thus nearer supernovae.  See Fig.~\ref{fig:discmags_z}
for more detail on this effect.  The types of supernovae with known
types and redshift are shown by the dashed (SN~Ia) and dotted (other SN) lines.}
\label{fig:redshifts}
\end{figure}

The number of supernovae found at different redshifts (see
Fig.~\ref{fig:redshifts}) is a function of the increasing volume
with redshift and the decreasing time that one can observe a supernova
during its rise and fall.  At lower redshifts the increasing volume
dominates and the number of supernova discovered at a given redshift
increases with redshift.  But, as the observable time the supernova is
brighter than the limiting magnitude of the search falls quickly near
the limiting redshift, the curve rolls over and falls to zero at the
limiting redshift.

\begin{figure}
\plotone{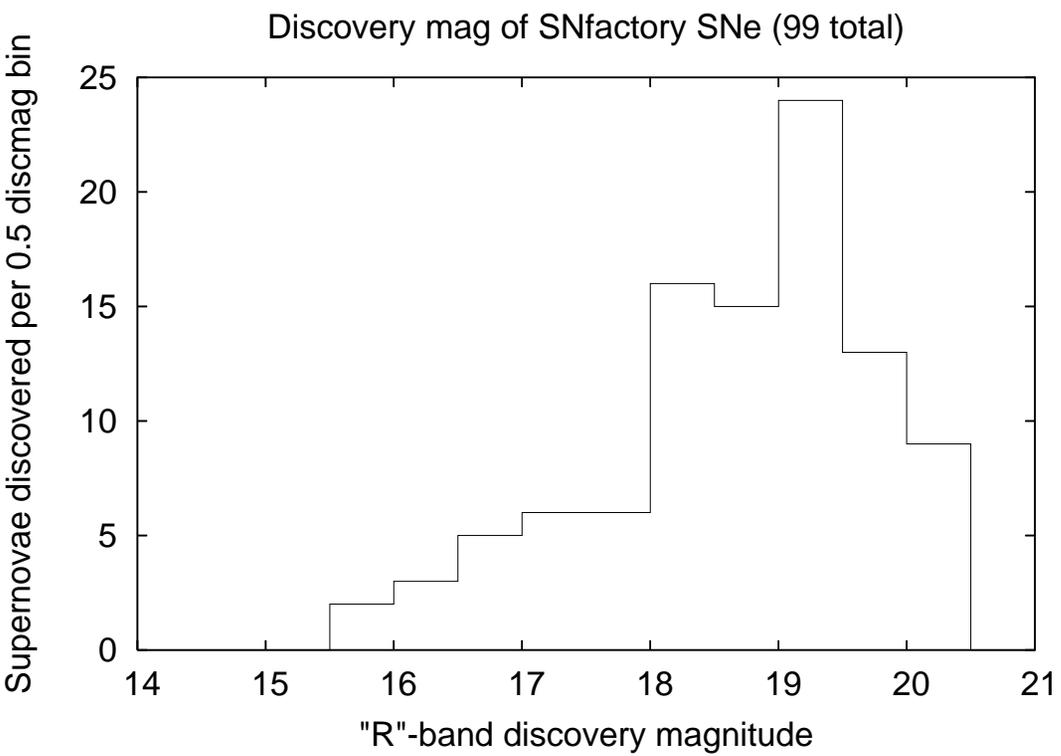}
\caption{The distribution of discovery magnitudes for the 99 SNe found
by the SNfactory automated search pipeline from the beginning of the
search in 2002 through June 2003.  }
\label{fig:discmags}
\end{figure}

Fig.~\ref{fig:discmags} shows the histogram of discovery magnitudes,
$m$, for all of the supernovae found in the SNfactory data set.
This plot gives a sense of the efficiency of the search as a function of magnitude, as
one would expect the number of SNe to go up as $r^3$.  As the distance
modulus, $\mu$ increases as $5\log_{10}{r^2}$, and the magnitude, $m$,
is equal to $M+\mu$, the number of supernovae as a function of
magnitude increases as $m^{3/2}$.
This increase with redshift due to increasing volume is then modified by the limiting magnitude
sensitivity of the search.  As images and conditions vary from night
to night, the limiting magnitude for the prototype SNfactory search varies from $19.5$ to $20.5$.  This
variation in sensitivity blurs out the faint edge of the redshift
distribution as seen in Fig.~\ref{fig:redshifts}.

\begin{figure}
\plotone{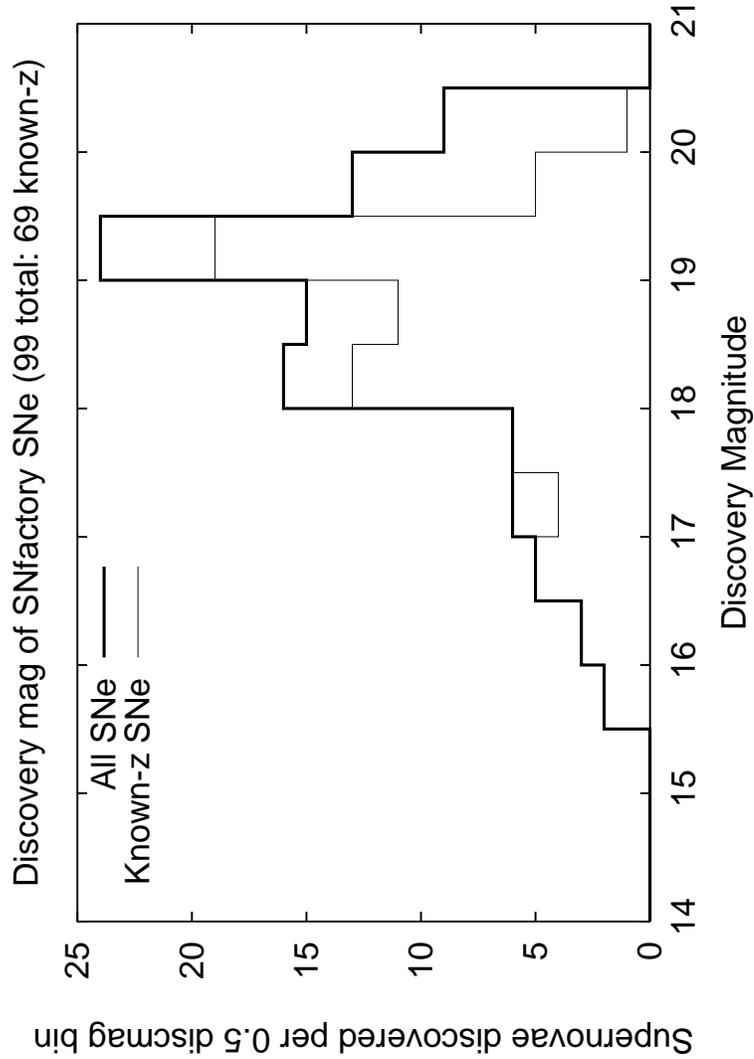}
\caption{The histogram of Fig.~\ref{fig:discmags} is shown here in the thick, dark line with the supernovae with known 
redshifts (which roughly corresponds to the supernovae
spectroscopically confirmed by others) in the thin, light line.  Note how the
coverage fraction is reasonably constant until a discovery magnitude of $19.5$.    SN~2002br, a SNe~Ia, was discovered on a subtraction with a bad zeropoint and had an unusable discover magnitude.  Thus, there are only 69 SNe with good redshifts and discovery magnitudes instead of the 70 SNe shown in Fig.~\ref{fig:redshifts}.}
\label{fig:discmags_z}
\end{figure}

Fig.~\ref{fig:discmags_z} shows the spectroscopic confirmation of
supernovae as a function of discovery magnitude.  There is no apparent
bias toward brighter supernovae for supernovae discovered at a
magnitude $<~19.5$.  Therefore, the rest of this analysis will assume
that the typing of supernovae in the SNfactory search is unbiased up
to magnitude $19.5$.


\begin{figure}
\plotone{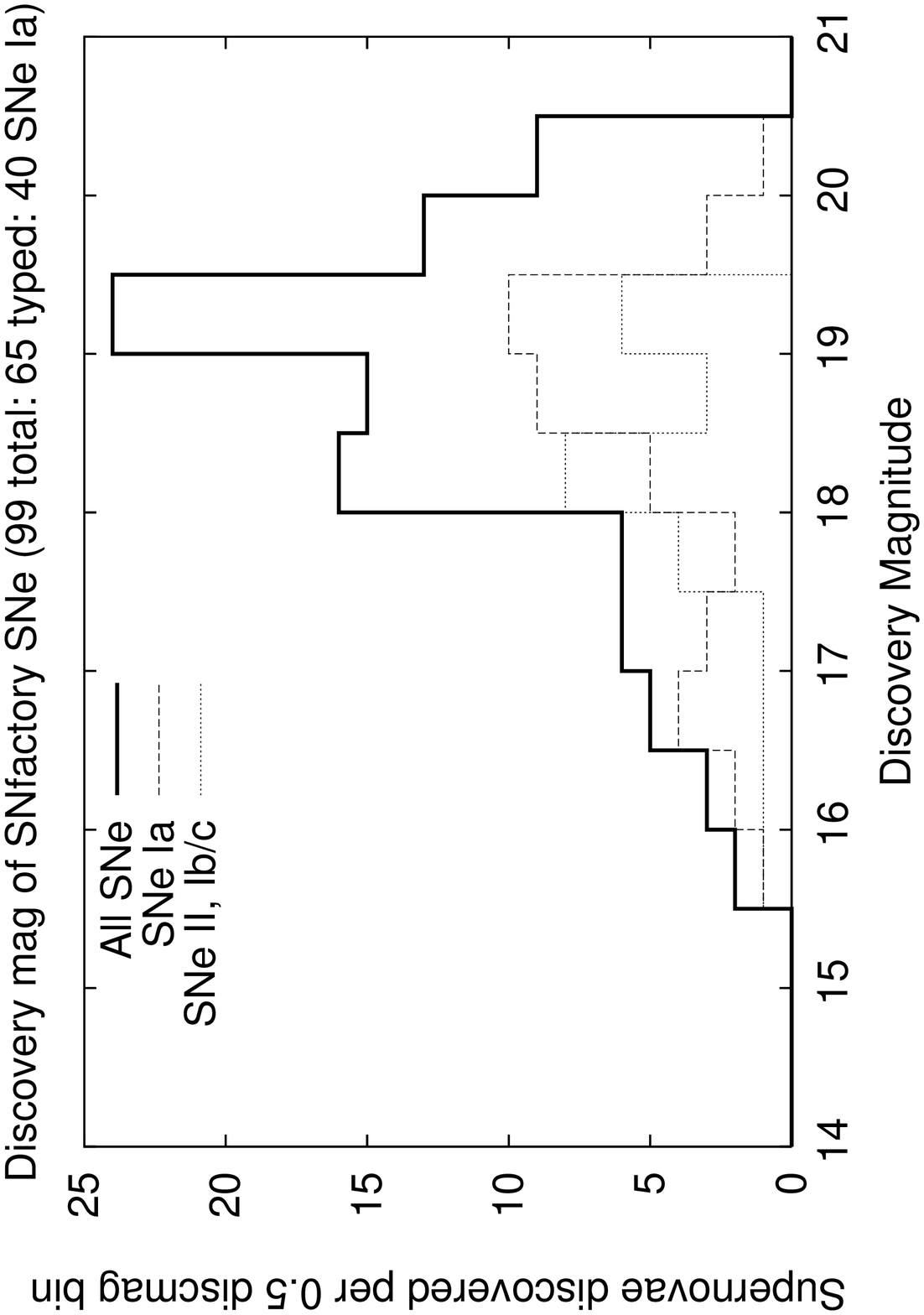}
\caption{The histogram of Fig.~\ref{fig:discmags} is shown here in the thick, dark line with the supernovae with known 
types (which generally corresponds to the supernovae
spectroscopically confirmed by others) in dashed (SNe~Ia) and dotted (core-collapse) lines.  The statistics are too low to interpret the apparent deficit of core-collapse SNe found $>19.5$~magnitude.  [See note in Fig.~\ref{fig:discmags_z} regarding the number of SNe~Ia shown here.]}
\label{fig:discmags_type}
\end{figure}

\section{Discovery Epoch of SNfactory SNe~Ia}
\label{sec:epoch}

\begin{figure}
\plotone{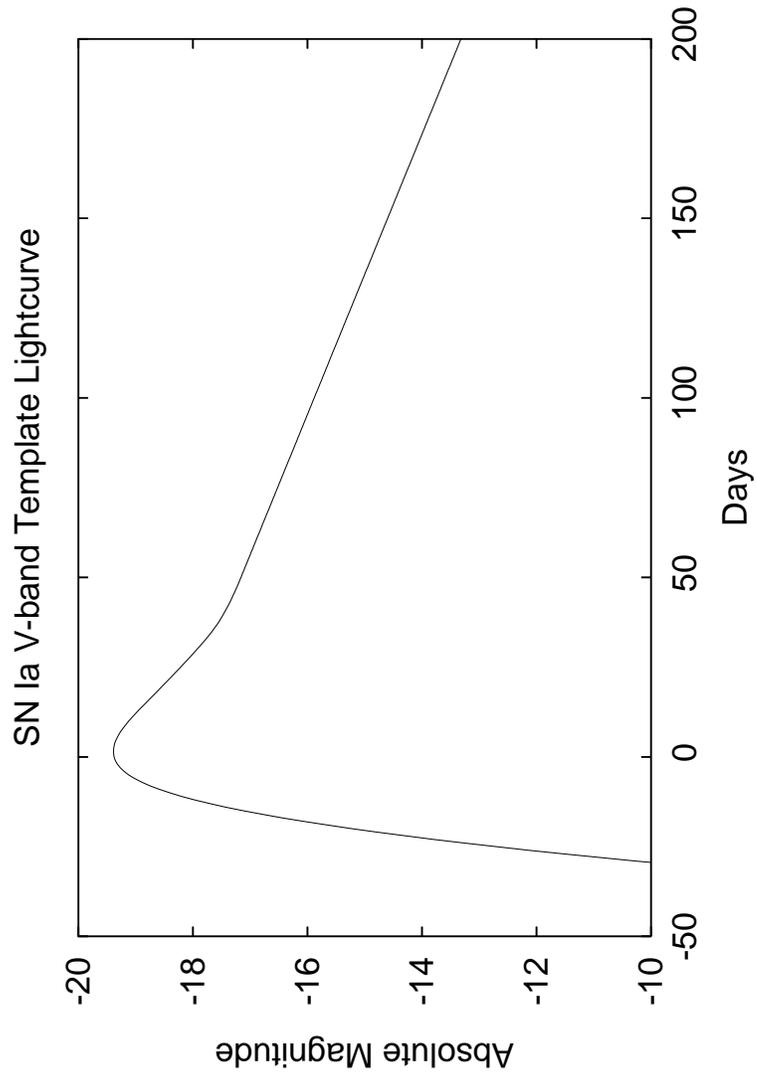}
\caption{The SNIa V-band template used in Sec.~\ref{sec:epoch} and in Fig.~\ref{fig:estimate_phase}~\&~\ref{fig:estimate_rates}.  A linear extrapolation of $0.026$mag/day has been used for dates greater than 70 days after maximum B-band light.  Light curve courtesy of Lifan Wang.}
\label{fig:snia_v_template_wang}
\end{figure}

\begin{figure}
\plotone{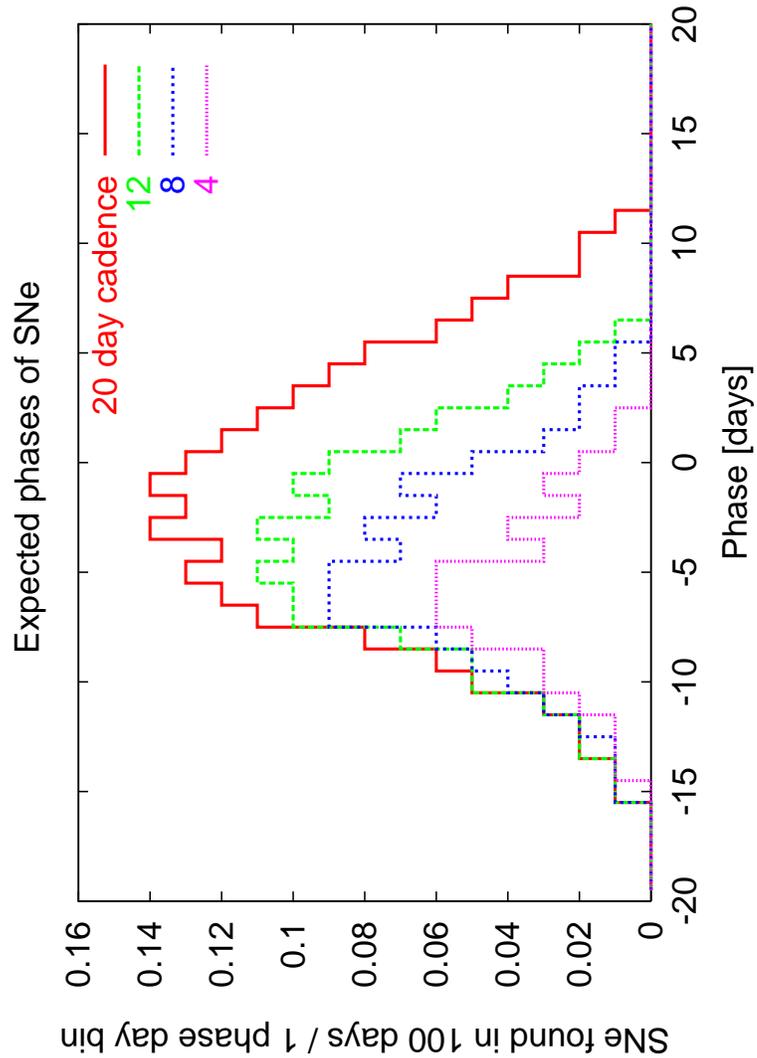}
\caption{A simulation of the expected number of supernovae found at a given phase for different
sky coverage cadences.  Shorter sky coverage cadences result in fewer overall supernovae and the same number of early supernovae.  See text (Sec.~\ref{sec:epoch}) for details of this simulation.}
\label{fig:estimate_phase}
\end{figure}

\begin{figure}
\plotone{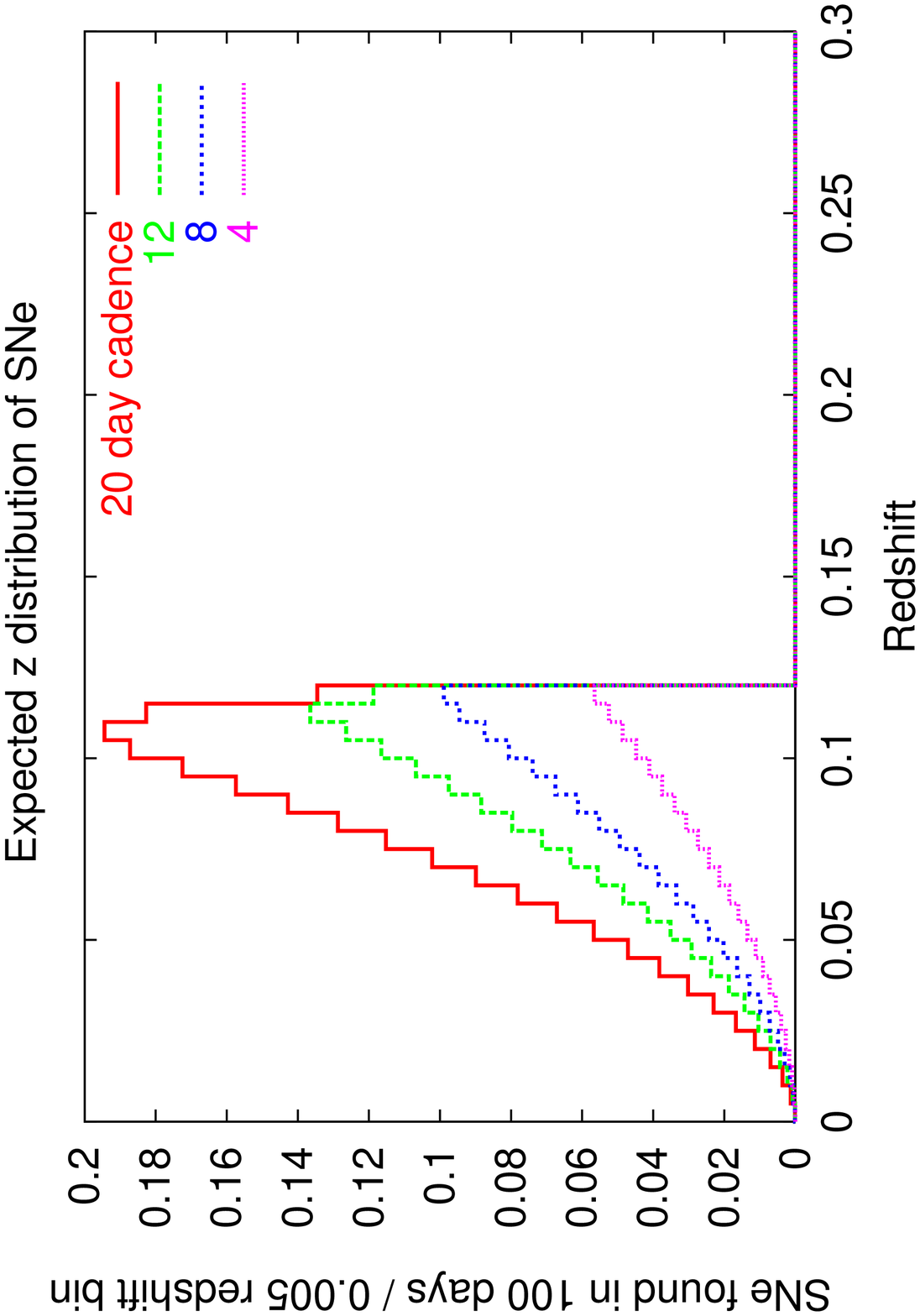}
\caption{A simulation of the expected number of supernovae found at different redshifts for
different sky coverage cadences.  In reality, the sharp cutoff is smoothed out by the dispersion in SN~Ia light curves.  See text (Sec.~\ref{sec:epoch}) for details of this simulation.}
\label{fig:estimate_rates}
\end{figure}

As the goal of the SNfactory project is to discover supernovae early
enough to study them in detail, it is important to find the supernovae
soon enough after their explosion to follow them through the rise
and fall of their optical light curves.  
Fig.~\ref{fig:estimate_phase} shows the results of a
simulation of the epoch of discovery for supernovae discovered by a
search with a limiting magnitude of $19.5$.  To compare the different
cadences, 
a SN~Ia rate of $r_V = 2\times10^{-5}$~SNe~Mpc$^{-3}$~yr$^{-1}$ 
and a total time of 100 days of observation were assumed.
This SN~Ia rate was multiplied by the effective volume
to get the number of supernovae.  The effective volume was determined
by calculating the control time that a SN~Ia would be
visible at a given redshift and by multiplying the control time by the volume 
element at that redshift.  A Euclidean 3-D universe was assumed with
a Hubble constant of $H_0 = 72$~km~s$^{-1}$~Mpc$^{-1}$.
A representative V-band light curve prepared by Lifan Wang was used to model
the rise and fall of the SNe~Ia (see Fig.~\ref{fig:snia_v_template_wang}).
The nightly coverage for the NEAT4GEN12 Palomar
camera, $A=500\sqdeg$, was used to constrain the
effect of the different sky coverage cadences with a constant
limiting magnitude of $19.5$.  For any detector that
can cover more than $\frac{1}{365}$ of the available sky, $\Sigma$, in a night,
there is a maximum effective cadence, $C_\mathrm{max}$, beyond which 
the telescope would be repeating observations faster than the 
specified cadence:
\begin{equation}
C_\mathrm{max} = \frac{\Sigma}{A-\Delta\,\Sigma},
\label{eq:max_cadence}
\end{equation}
where $\Delta\,\Sigma=\frac{\Sigma}{365.25}$ represents the new sky available
each night due to the motion of the Earth along its orbit.
Considering cadence values larger than the optimal value can be useful for simulating the effects
of weather and other interruptions, but it is important to consider
this idling limit when comparing cadences above and below this limit.
For a sky coverage of $A=500\sqdeg/\mathrm{day}$ a night and a total available sky of $\Sigma=\sim10000\sqdeg$
a night, $\Delta\,Sigma=\frac{10000\sqdeg}{365.25~\mathrm{days}}$, and Eq.~\ref{eq:max_cadence} gives the optimal cadence as approximately 20~days.  Thus,
this is the maximum cadence shown in
Figs.~\ref{fig:estimate_phase}~\&~\ref{fig:estimate_rates}.
For very short sky coverage cadences, one has to be content with
covering less sky, resulting in the discovery of fewer supernovae overall.

\begin{figure}
\plotone{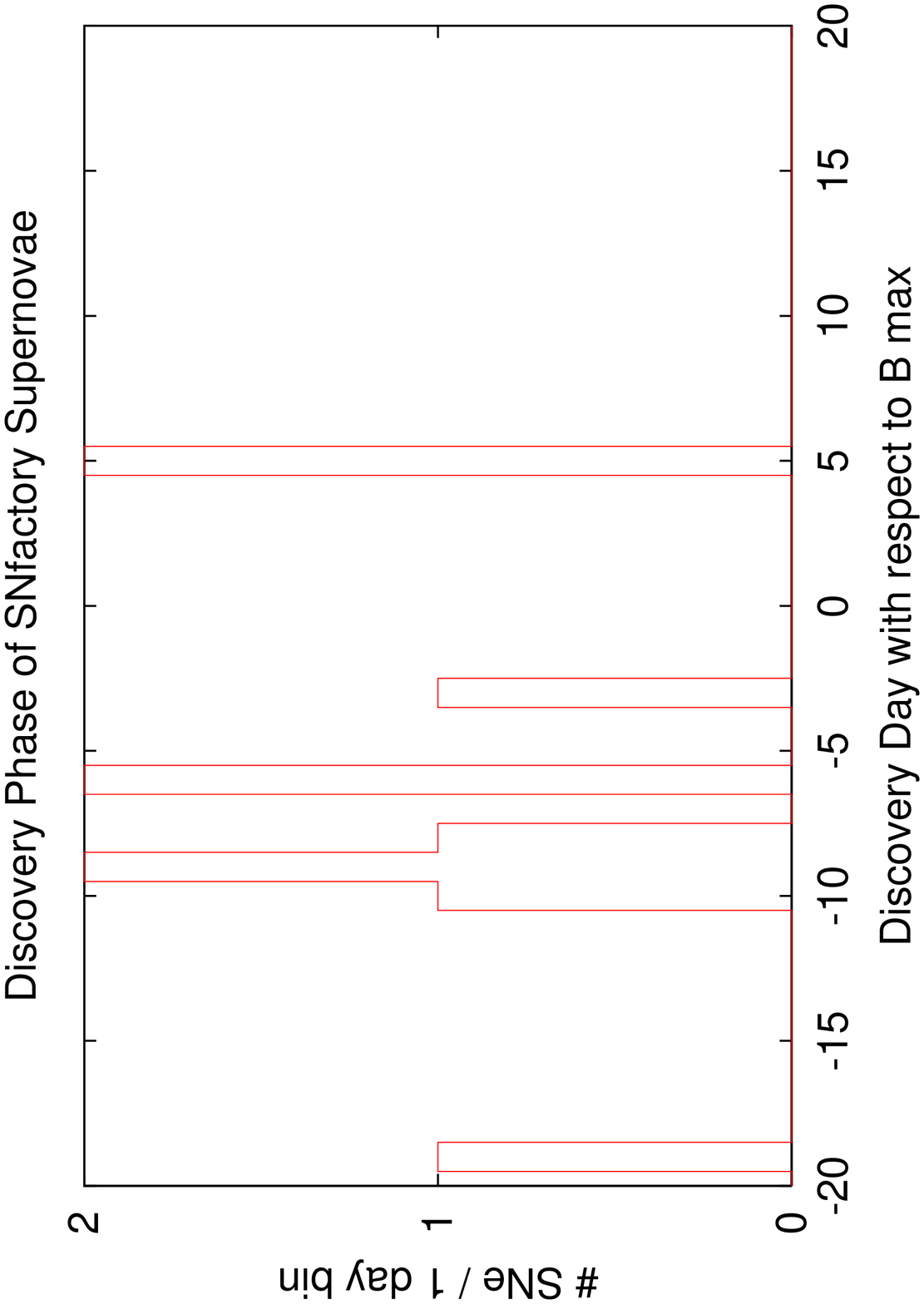}
\caption{The distribution of discovery epochs for the SNe~Ia found in
the SNfactory data set with well-sampled lightcurves sufficient to
determine the date of maximum light using the SCP light-curve fitting
code \code{snminuit}.}
\label{fig:phase}
\end{figure}

\begin{figure}
\plotone{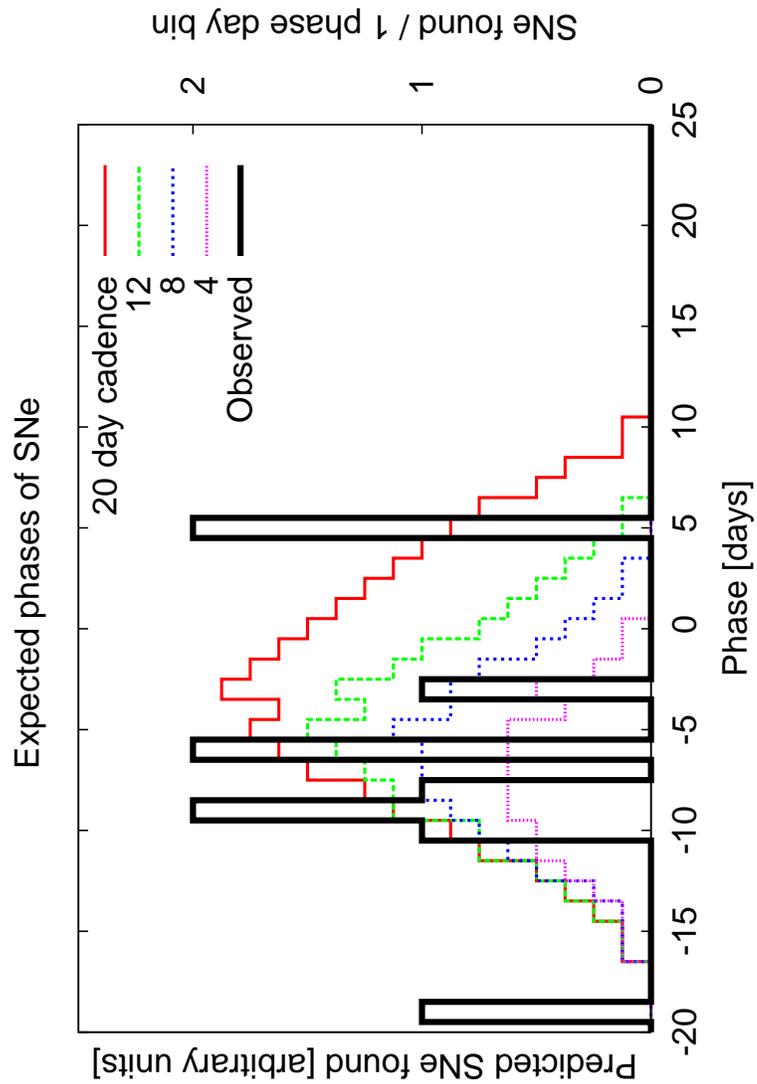}
\caption{Fig.~\ref{fig:estimate_phase} compared with Fig.~\ref{fig:phase}.  
The distribution of discovery epoch for the SNe~Ia found in
the SNfactory data set (heavy line) compared with the model described in
Section~\ref{sec:redshifts}.}
\label{fig:compare_model_phase}
\end{figure}

The actual discovery phase of SNfactory SNe~Ia with sufficient
coverage to determine a good date of maximum is given in
Fig.~\ref{fig:phase}.  These phases were determined from NEAT
follow-up observations of the SNfactory SNe~Ia by fitting the observed
light curves with the same
\code{snminuit} program used by the SCP to fit the time of maximum B-band
brightness, maximum magnitude, and stretch of SNe~Ia cosmological
studies.  
Fig.~\ref{fig:compare_model_phase} shows a comparison of Fig.~\ref{fig:estimate_phase}~\&~\ref{fig:phase}.  
Note that the Poisson noise prevents a meaningful comparison of the observed and model distributions.
Fig.~\ref{fig:compare_simsne_phase} shows a comparison of the discovery
phases expected from the simulation described in
Sec.~\ref{sec:montecarlo} with the known phases observed supernovae
as described in Fig.~\ref{fig:phase}.

\begin{figure}
\plotone{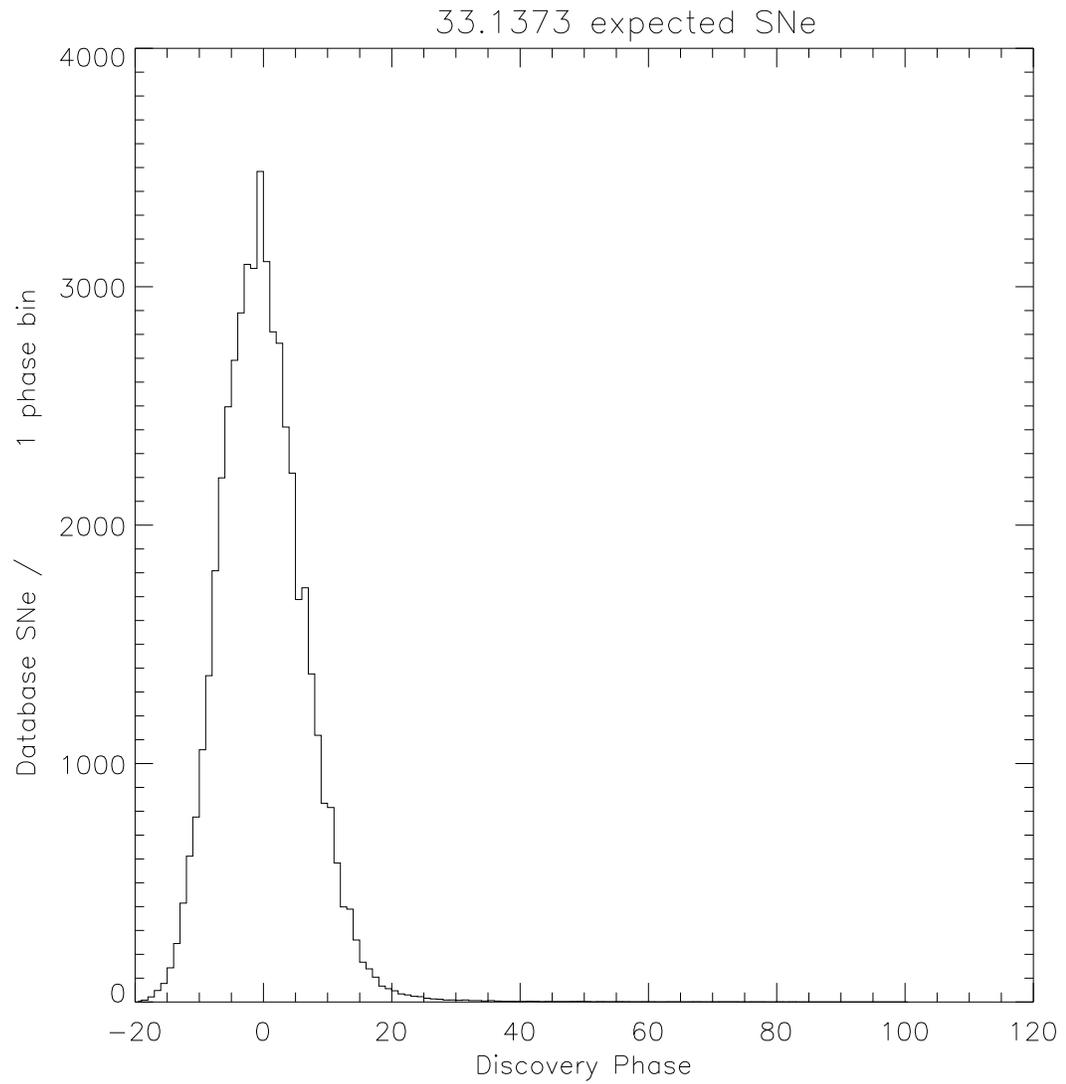}
\caption{The discovery epoch distribution for the Monte Carlo model
described in Sec.~\ref{sec:montecarlo}.}
\label{fig:simsne_phase}
\end{figure}

\begin{figure}
\plotone{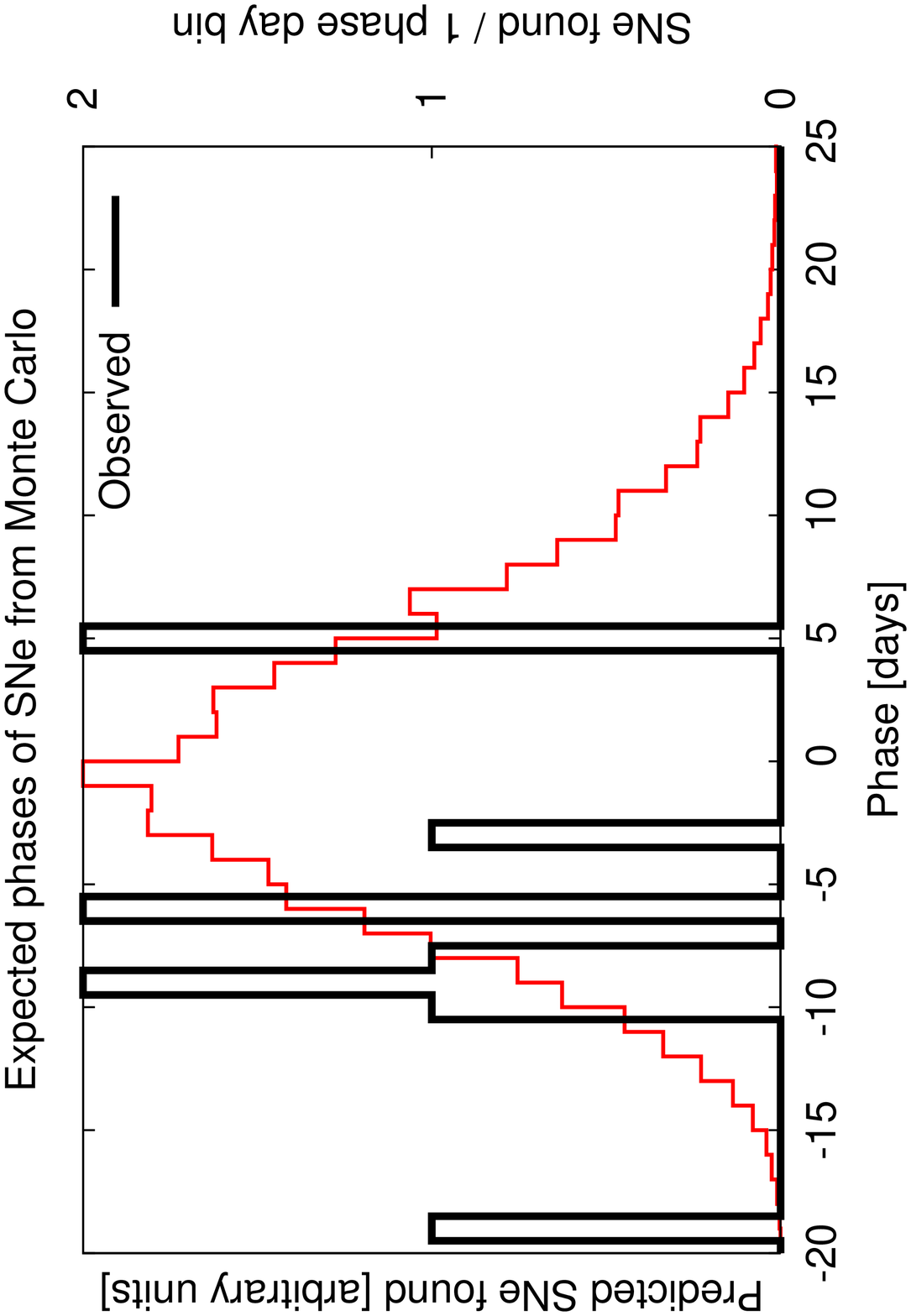}
\caption{The discovery epoch distribution for the Monte Carlo model
described in Sec.~\ref{sec:montecarlo} and shown in
Fig.~\ref{fig:simsne_phase} compared with the phases of the actual
supernovae found shown in Fig.~\ref{fig:phase}.}
\label{fig:compare_simsne_phase}
\end{figure}





\section{A Quick SN~Ia Rate Calculation}
\label{sec:rates_quick}

This section presents a quick calculation of the SN~Ia rate from a
sub-sample of the SNfactory prototype search to illustrate the basic
ingredients introduced in Eq.~\ref{eq:snrate}.  The following section,
Sec.~\ref{sec:montecarlo}, will present a more accurate and
significantly more involved calculation of the SN~Ia rate from a portion of 
the SNfactory prototype search.

In the month of March, 2003, there were $392.54\sqdeg$ worth of images subtracted to a
limiting magnitude $<19.5$.  There were 32~SNe found with a discovery
magnitude $<19.5$ during this time.  These subtractions were generally
done against year-old reference images.  Incorrectly taking a control
time of 1~month for every square degree searched, one gets 32~SNe /
$392.54\sqdeg$ / 1~month brighter than $19.5$ magnitude.  Converting
$19.5$ to a distance requires the use of the distance modulus,
\begin{equation}
\mu = m-M = 5 \log_{10} (r/[\mathrm{Mpc}]) + 25.
\end{equation}
The distance, $r$, can then be expressed as
\begin{equation}
r = 10^{(\mu -25)/5}~\mathrm{Mpc}.
\end{equation}
Assuming a maximum SN~Ia brightness of $-19.5$~magnitudes, or $\mu = 19.5 - (-19.5) = 39.0$ , gives
\begin{equation}
r = 10^{(39.0-25)/5}~\mathrm{Mpc} = 631~\mathrm{Mpc}.
\end{equation}
Integrating the solid angle of areal coverage gives a volume, $V$, of 
\begin{eqnarray}
     V & = & \frac{\Omega}{3} r^3,
\end{eqnarray}
where
\begin{eqnarray}
\Omega & = & 4\pi \frac{392.54\sqdeg}{41,253\sqdeg} = 0.1196,
\end{eqnarray}
and thus
\begin{eqnarray}
     V & = & \frac{0.1196}{3} (631~\mathrm{Mpc})^3.
\end{eqnarray}

Out of the $n_\mathrm{total}=65$ supernovae typed during
the SNfactory prototype search phase, $n_\mathrm{Ia}=42$ were SNe~Ia.
Integrating over Eq.~\ref{eq:snrate} by taking
$(\frac{n_\mathrm{Ia}}{n_\mathrm{total}})~32\approx20$ as the number of SNe~Ia discovered in March, 2003, assuming a constant control time out to the limiting magnitude of $-19.5$, and neglecting the time-dilation factor, $1+z$, results in a SN~Ia rate of 
\begin{equation}
r_V = \frac{20~\mathrm{SN}}{(\frac{0.1196}{3})(631~\mathrm{Mpc})^3 (1~\mathrm{month})} \frac{12~\mathrm{months}}{1~\mathrm{year}} = 2.4 \times 10^{-5}~\mathrm{SN}/\mathrm{Mpc}^3/\mathrm{year}
\end{equation}


There are clearly many oversimplications in this calculation.  
First,
there has been no accounting for repeated observations of the same
field.  Second, there has been no true calculation of control time.
These factors will decrease the volume searched and thus increase
the rate for the same number of supernova.

The next method will present a similar analysis to that above, but
will take into account the repeated observations of the same fields
during March 2003 as per the NEAT observing strategy.  A
one-month control time will still be assumed for each field.
This areal
accounting divides the sky up into equal area
sections and increments the count of each section for every subtractions
that overlaps that area by at least $50\%$.  As
part of a given subtraction may be masked, it will
be difficult to calculate the exact overlap.  To account for this masking,
the contribution of each subtraction is normalized
by ensuring that the amount of area added matches the
area of the subtraction.

A more detailed analysis with areal tracking reveals
there is little duplication of areal coverage in the previous
calculation.  Using bin sizes of $0.5$ degrees on a side, placing
each subtraction in the appropriate bin, and adding up the areas of the
subtractions in each bin to a maximum of the bin area ($0.25$~\sqdeg)
gives $347.51$~\sqdeg of unique area searched.
The updated supernovae rate from
this calculation is then
\begin{eqnarray}
r_V & = & \frac{20~\mathrm{SN}}{\mathrm{month}} \left(\frac{4}{3}\pi \frac{347.51\sqdeg}{41,253\sqdeg} (631~\mathrm{Mpc})^3\right)^{-1} \frac{12~\mathrm{months}}{1~\mathrm{year}} \\
r_V & = & 2.84\times 10^{-5}~\mathrm{SN} / \mathrm{yr} / \mathrm{Mpc}^3 \\ 
\end{eqnarray}

\section{The Right Way: Control-Time Monte Carlo Simulation}
\label{sec:montecarlo}

To integrate the sampling of the control times and efficiency
measurements, a database of simulated supernovae was created.
Two hundred million SNe~Ia were included in this database,
covering the sky from $+50$ to $-50$ degrees in declination, 
at all RA values, and in the redshift range from $z=0$ to $z=0.4$,
This sample
was created with a constructed SNe~Ia rate of 1~SN/Mpc$^3$/century.
Taking the literature SNe~Ia rate~\citep{cappellaro99,pain02} of
roughly $1\times10^{-4}$~SN/Mpc$^3$/century, roughly 10,000
supernovae were simulated for each supernova expected from the search.
%
The SNe~Ia were created with dates of maximum B-band light spread over
four years representing 2001--2004.
The approach introduced in this section is dubbed the Control-Time Monte
Carlo method because it departs from other methods in its proper
temporal tracking of simulated supernovae in the control-time
measurement.
The explicit dates given to simulated supernovae allowed the 
supernovae found in different subtractions of the same area of sky
to be tracked appropriately and not double-counted.
The simulated SNe~Ia were placed randomly
over the sky with a neighbor-avoidance check to ensure a spacing of at
least 10\arcsec between simulated SNe.  This strategy avoided object
confusion in the subtractions performed with the simulated SNe~Ia.
The number of objects
found scales as the number of objects in the sample while the
number of coincident object rates scales as the square of the number
of objects.
Thus, the probabilities of observing coincident supernovae in the Universe
are far smaller than in an unconstrained sample of
200~million objects across half of the sky.

The subtractions are rerun with the simulated supernovae added
from this sample.  Each supernova in the RA and Dec
range of the subtraction was placed on the constituent reference
and search images of the subtraction.
The SN~Ia flux is calculated based on the JD of each particular image
and the redshift and stretch of the simulated supernova.
For each image, a normalized PSF was
constructed from the good stars on the image.  This PSF was multiplied
by the flux appropriate to the magnitude of the simulated object and
added to the image.  These images were then used in a new
simulated subtraction based on the original subtraction to measure the
efficiency of the original subtraction.
The simulated subtraction used the original good star list from the
original subtraction in the calculation of flux ratios, PSF
widths, and other properties, just as was done in the original
subtraction.  These subtractions duplicated the steps and parameters
used in the original subtraction.


For each a subtraction, a list of the fake
candidates was saved to a file of the form
\file{$<$subtraction\_name$>$.genfakes.nc} in the
\code{DEEPIMAGEPATH}.
For this study, just SNe~Ia were considered, but a future analysis
should include different assumptions of core-collapse rates and
magnitude dispersions in addition to the SN~Ia rates.

To analyze the efficiency of each subtraction, a pair of IDL
routines,
\code{analyzefakes.pro} and \code{analyzeallfakes.pro}, were written
along with some 
helper scripts, to calculate the number of SNe~Ia expected for a
given supernova rate.  
Any desired model of the SN rate as a function of redshift can be
reconstructed from the simulated supernovae by appropriate
re-weighting of the redshift distribution.  The number of simulated
SNe~Ia recovered with a given model of the SN~Ia rate as a function of
redshift was compared with the number of SNe~Ia actually found.  The
distribution of recovered simulated SNe~Ia as a function of redshift
and as a function of discovery magnitude is also compared with
Fig.~\ref{fig:discmags} and Fig.~\ref{fig:redshifts} to check
consistency with the sample.  For the histograms presented in this chapter with numbers of supernovae
in a given magnitude or redshift bin, a relative error is calculated
using Poisson statistics (for $n\gtrsim5$, $n_\sigma\approx
\sqrt{n}$).  However, the actual comparisons of model and observation
is to the overall number of supernovae found (i.e., this is an
unbinned analysis).  This total is itself governed by Poisson
statistics as it is still a counting statistic of uncorrelated events.
This total and associated uncertainty is then used when comparing the
agreement of the simulated curves with the observed rates to determine
if a given model is consistent with the SNfactory results.

 



\subsection{Stretch Range}
\label{sec:stretch_range}

The stretch of a SN~Ia affects its light curve and thus control time
for a given search.  To estimate the effect of SN~Ia of
different stretches on the calculated rate, the time evolution of the
template supernovae was adjusted by $t' = s t$ and the peak
V-magnitude was similarly changed by $M'_V = M_V - \alpha (s-1)$,
where $\alpha=1.18$ was taken from the value given for all SCP SNe~Ia
in Table 8, row 6, of \citet{knop03}.


The impact of different stretch values can be seen in its effects on
the control time and the limiting magnitude of the search.  A dimmer,
faster-rising and falling supernova has a reduced effective control
time as the supernova is visible for less time.  This effect scales
with the stretch value and holds equally at any redshift.
The visible time of the supernova is further reduced for a
lower-stretch SN~Ia as it is dimmer and thus below the limiting
magnitude of a given search for more of its light curve.  For the
Control-Time Monte Carlo method described in Sec.~\ref{sec:montecarlo}, a
Gaussian stretch distribution centered around 1 with a $\sigma=0.1$
was used to model the observed diversity in SN~Ia light curves.
Sec.~\ref{sec:sys_snia_diversity} discusses the effect of assuming
different stretch distribution models.

\subsection{Modeling of the Host Galaxy Population}
\label{sec:galaxy_modeling}

The percentage increase cut discussed in
Sec.~\ref{sec:percent_increase} highlights the need for a proper
understanding of the host galaxy population observed in the SNfactory
search.  The Sloan Digital Sky Survey provides the most complete
galaxy sample to date for estimating the galaxy population observed in
the SNfactory search~\citep{blanton03a,blanton03b,shen03}.

Assuming no evolution in galaxy properties out to a
redshift of $z=0.2$, the SDSS DR1 galaxy sample can be scaled to model
the host galaxy population of the SNfactory search.  
In the modeling of the effect of the underlying host galaxy on the
detection of a supernova in the search, the immediately crucial
parameters are the luminosity of the galaxy, the spatial luminosity
distribution of the galaxy, the redshift of the galaxy and the effective
seeing of the subtraction.
The first two
parameters are dependent on the galaxy population, obtainable from
\citet{shen03}; the redshift can be simulated assuming a constant 
galaxy population out to $z=0.2$; and the seeing of the subtraction,
dependent on the observational conditions of the images that went into
the subtraction, is recorded in the \code{subng} database table 
as part of the original subtraction.

Based on \citet{strauss02}, \citet{shen03} estimate their sample to 
be complete for galaxies with apparent magnitudes between
$r=15.0$ and $r=17.5$.  Due to the nature of the SDSS survey, 
imaging and spectroscopic data are available for all galaxies that
make up the complete \citet{shen03} sample.  
The fluxes of galaxies in the full redshift range of interest
($0.0<z<0.2$) can be reconstructed by scaling the radius used 
to measure aperture fluxes to the desired redshift.
Ensuring full coverage in the search redshift range is complicated by the
distribution of absolute magnitudes of the galaxies.  For each
absolute magnitude, there is a different maximum redshift at which it
is observable above the $r>17.5$ cutoff.  Assuming the
galaxy luminosity function does not evolve out to $z=0.2$,
a full sample can be constructed by weighting each supernova by the inverse
of its covered volume as defined by the $z_\mathrm{max}$ at which it
would be observable.  This was the approach taken by \citet{shen03}.
A similar technique is used here to arrive at a complete sample of galaxies
out to a given redshift from which to draw the supernovae.  It is
possible that the very faintest galaxies are not well accounted for by
this analysis, but, assuming that supernovae are strongly correlated with
luminosity, the potential absence of these galaxies should not significantly
affect these results.

The galaxy stamps from the sample described by \citet{shen03} were
retrieved from the SDSS archive server in the form of \code{fpAtlas}
files.  Image stamps and photometric calibration information were 
retrieved as part of an SQL query to the SDSS database.
The galaxy stamp images were analyzed using the same \code{fisofind}
code that is used in the standard SNfactory data reduction.  The
\code{fisofind} analysis gave the isophotal extent of the galaxy of
pixels at least $3~\sigma$ above the sky noise background of the SDSS
images.






For each simulated supernova, an amount of host-galaxy light appropriate for the
effective seeing of the subtraction and the redshift of the supernova
was drawn from a galaxy luminosity distribution function 
calculated from the \cite{shen03} sample.
The redshift of the supernovae was included by calculating the
appropriate surface brightness dimming and radial size decrease
for a cosmological model with \OM$=0.3$, \OL$=0.7$.
The search scores of each simulated supernova were re-calculated to include
the effect of the host-galaxy light.
To incorporate the galaxy flux background into the rate calculation,
a grid of aperture bins was constructed to record the probability
distribution of host-galaxy background light as a function
of aperture and redshift.
For each aperture bin, the flux for that aperture
was calculated for each galaxy pixel found by \code{fisofind}.
The results for all of the galaxies in the sample
were combined to construct an aperture ``brightness'' distribution for all galaxy
pixels.  
%

Because of this approach, the SN rates can not be calculated as a
function of volume without taking into consideration an assumed galaxy luminosity distribution.  However, having to include a galaxy luminosity function also makes it clear that supernova
searches are sensitive to the host galaxy population 
and thus an understanding of the true coverage of a
given search is dependent on an understanding of the galaxy
population underlying the supernovae.

%
%

\subsection{Milky Way Extinction}
\label{sec:milky_way_extinction}

Dust in our own galaxy results in reduced sensitivity
along certain lines of sight.  Unlike high-redshift searches, which
specifically choose low-dust regions, the SNfactory search covers a
large fraction of the sky and encompasses a wide range of Milky Way
extinctions.  The dust maps of \citet{schlegel98} were used to include
this effect in the Control-Time Monte Carlo simulation.

\begin{figure}
\plottwo{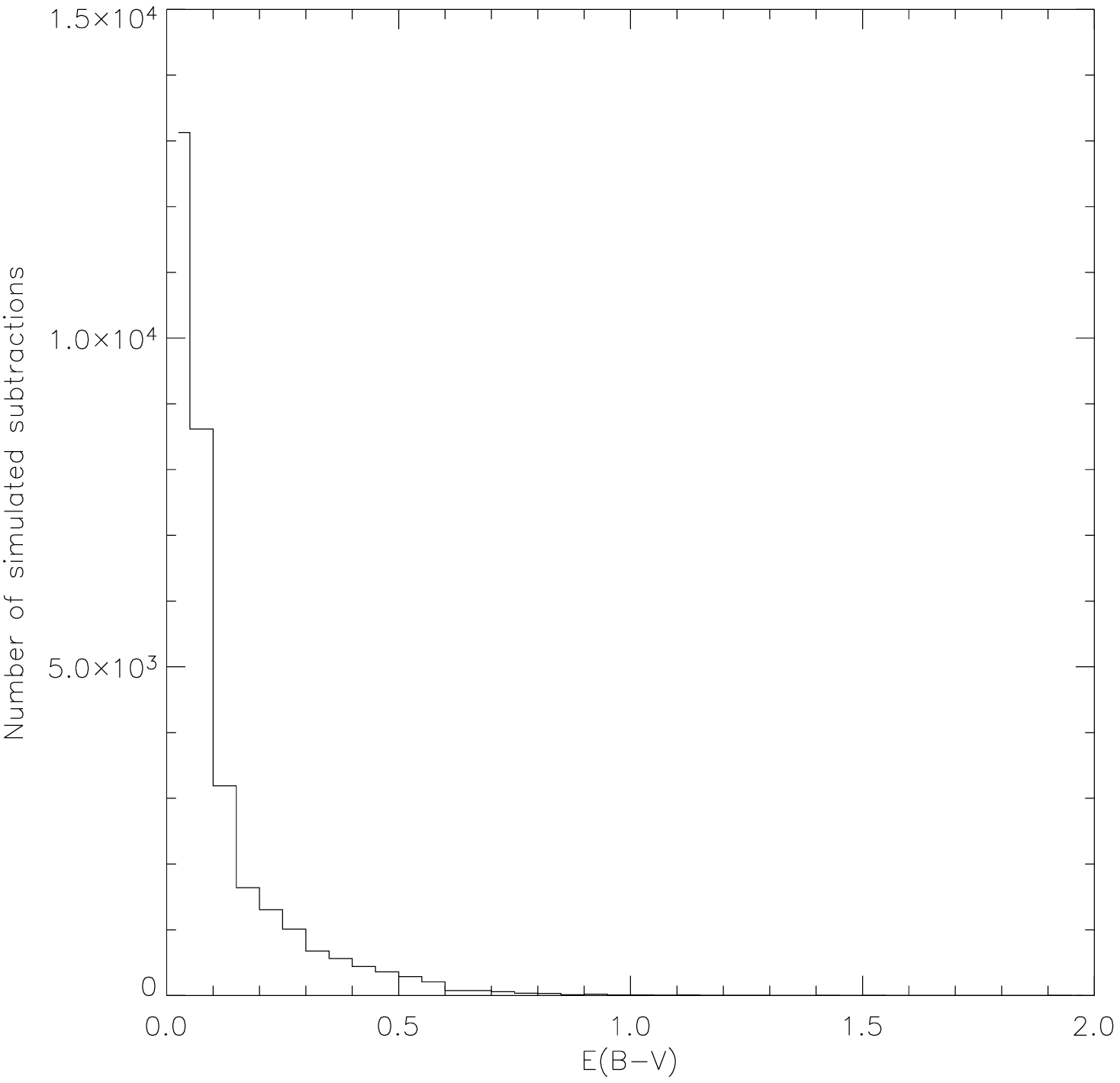}{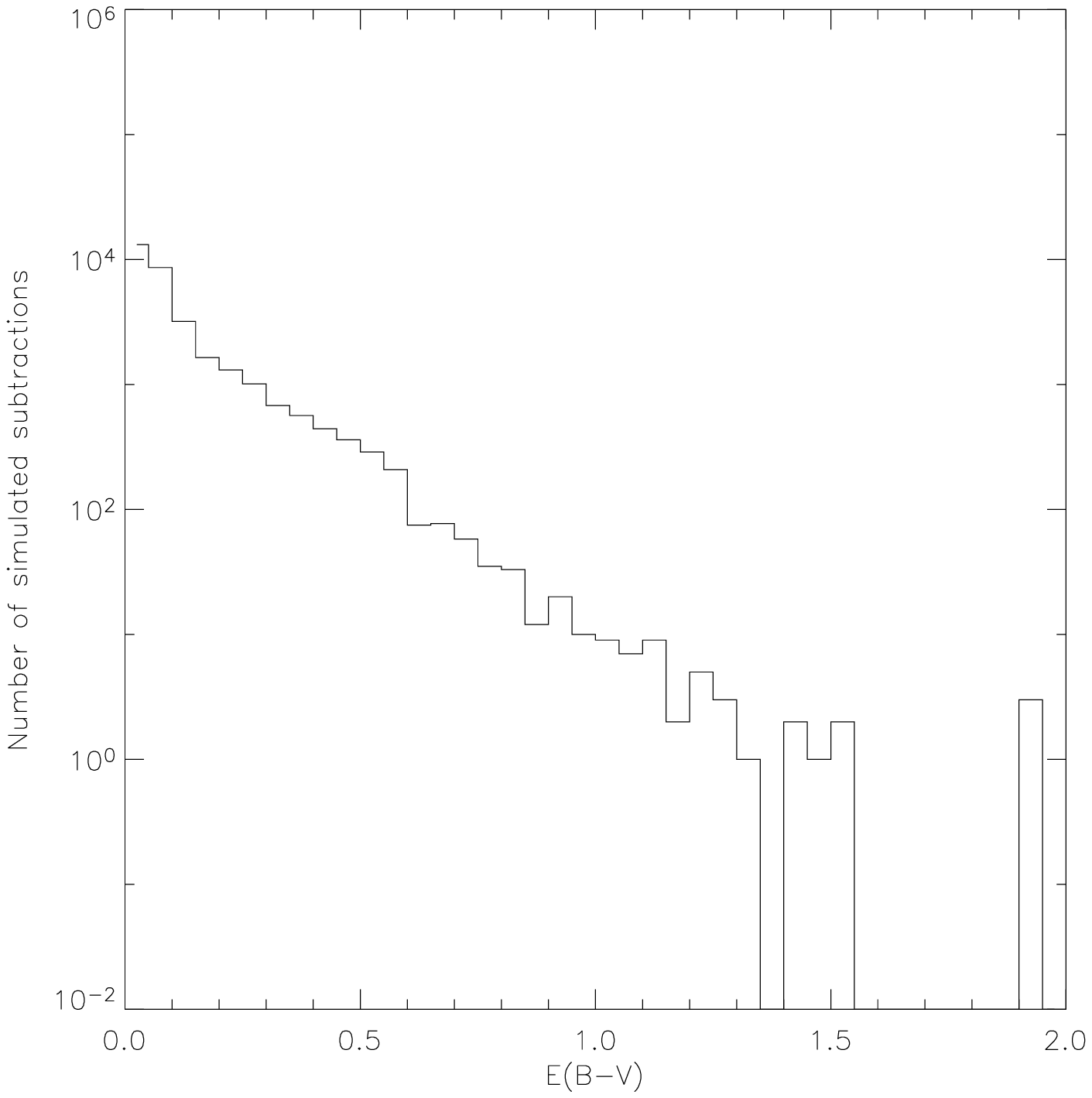}
\caption{The E(B-V) values for the RA and Dec values of the reference systems for the simulated subtractions used in the rates analysis presented in this chapter.  The linear-linear plot on the left illustrates that the majority of the fields have E(B-V) values of less than $0.1$. The linear-log plot on the right demonstrates the power-law nature of the dust distribution covered by this sample of the SNfactory survey.}
\label{fig:ext_E_BV}
\end{figure}

The dust maps\footnote{http://astro.berkeley.edu/davis/dust/index.html} 
were downloaded to PDSF\footnote{\file{/auto/snfactry6/DustMap\_SFD/}}
along with the companion IDL code to read the maps.\footnote{\file{/home/u/wwoodvas/local/deephome/snidlpro/CodeIDL/}}
A wrapper program\footnote{\code{/home/u/wwoodvas/local/deephome/snidlpro/SNIa/snia\_ext.pro}} 
was written to calculate the extinction for SNe~Ia observed with the NEAT system
response function as discussed in Appendix~\ref{apx:calibration} (see
Fig.~\ref{fig:nugent_response_curve_NEAT12GEN2}).  This program
measures the effect of an $R_V=3.1$ extinction law \citep{cardelli89} on
the magnitudes of the simulated supernova.  When placing simulated
supernova on the NEAT images\footnote{using the \code{subng} option \code{fke\_algorithm}$=6$}, a
spectrum of a SN~Ia at the appropriate phase (assuming stretch applies
to spectral phase as well as photometric phase) was convolved with the
NEAT system response.  The difference between applying an $R_V=3.1$ extinction
law,
\begin{equation}
A_\lambda = A_V / E(B-V) \times (0.574 - (0.527/R_V)) \times \lambda^{-1.61},
\label{eq:cardelli_A_lambda}
\end{equation}
and assuming no extinction was integrated over the wavelength
range of the NEAT detectors and converted to a $\Delta m$,
\begin{equation}
\Delta m = -2.5 \log{\frac{\int_{3000\AA}^{10000\AA}{10^{-0.5\times A_\lambda} r_{\mathrm{NEAT}} f_{\lambda} d\lambda}}{\int_{3000\AA}^{10000\AA} r_{\mathrm{NEAT}} f_{\lambda} d\lambda}},
\label{eq:cardelli_extinction}
\end{equation}
which was
then taken as the extinction for that line of sight and SN spectrum.
This extinction was applied to the simulated supernova magnitudes that 
were injected into the component images.
Fig.~\ref{fig:ext_E_BV} shows the
distribution of E(B-V) values for the subtractions in the sample of
the SNfactory survey considered here.

As can be seen in Fig.~\ref{fig:ext_E_BV}, the $E(B-V)$ values for the
NEAT fields searched ranged from a $0.005$--$0.6$.  It is
thus clearly important to include the modeling of the Milky Way
extinction in a detailed calculation of the SN rates from this
wide-field survey.

The contribution from host-galaxy dust extinction has been neglected
here. 
This leads to an underestimate of the true supernova rate.
The extinction due to host-galaxy dust is a general problem in
supernova rate studies and there is no clear solution.
However, as extinction is a line-of-sight effect, it
would apply equally to galaxies at different redshifts if the
evolution of host-galaxy dust with redshift were not significant.  
But, the dust content in galaxies does increase with redshift,
as it should follow the increasing star formation rate,
and a comparison of SN rates out to higher redshifts, $z>0.5$, 
will have to include modeling of host-galaxy dust extinction.
For more discussion of the likely amount of host-galaxy extinction for
different dust assumptions, see \citet{commins04}.

\subsection{Efficiency as Measured by Control-Time Monte Carlo}
\label{sec:efficiency}

\begin{figure}
\plotone{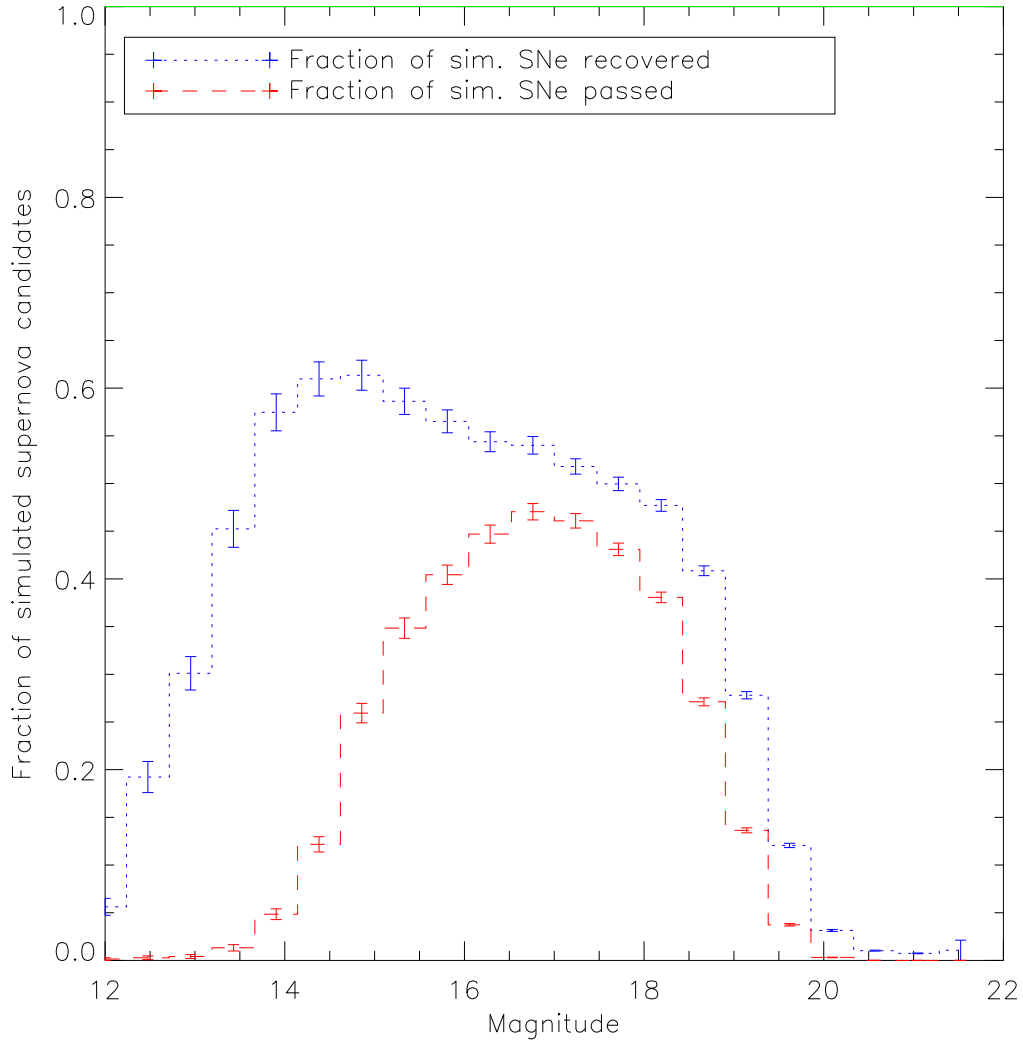}
\caption{The recovered and discovered fraction of database supernova as a 
function of magnitude as measured using the \code{analyzeallfakes.pro}
routine on the simulated subtractions run using the database of
200~million simulated SNe~Ia.  No correction for host galaxy light
and the effect of the percent-increase cut has been applied here (see Fig.~\ref{fig:efficiency_fke6_wgal}).  The low fraction in the recovered
discovered supernovae is mainly due to subtractions with short time
baselines where the majority of the simulated supernovae were dimmer
in the new images that in the reference images.  See text
(Sec.~\ref{sec:efficiency}) for more details.}
\label{fig:efficiency_fke6}
\end{figure}

The complete control-time measurement 
for the sample of the SNfactory prototype survey simulated to date
is shown in Fig.~\ref{fig:efficiency_fke6}~\&~\ref{fig:efficiency_fke6_wgal}.
These figures show the ``recovered'' fraction of simulated
supernovae---the simulated supernovae from the database that are
detected as objects in the subtraction---and the ``discovered''
fraction---the fraction of simulated supernovae from the database that
pass the default score cuts.
The dotted curves shows the fraction of supernovae from the simulated
supernova database that is recovered in the subtraction.  The cutoff
in the bright end is from detector saturation while the cutoff in the faint
end is from signal-to-noise loss.  The dashed curve underneath the
dotted curve
represents the number of recovered supernova that also pass the
standard SNfactory cuts (see Sec.~\ref{sec:scorecuts}).  
While these cuts have evolved over time, all but two of the
supernovae that have been found to date by the standard SNfactory
search pipeline pass the current default cuts.  The two exceptions are
SNe that were discovered with NEIGHDIST scores $<0.5\arcsec$;
as a result, these SNe were excluded from this calculation of SN rates.  The
structure observed in Fig.~\ref{fig:efficiency_fke6}~\&~\ref{fig:efficiency_fke6_wgal} is a combination
of the efficiencies of thousands of subtractions.  The sensitivity
variation between subtractions leads to the blurring of the edges of
both the detection and recovery limits.

\begin{figure}
\plotone{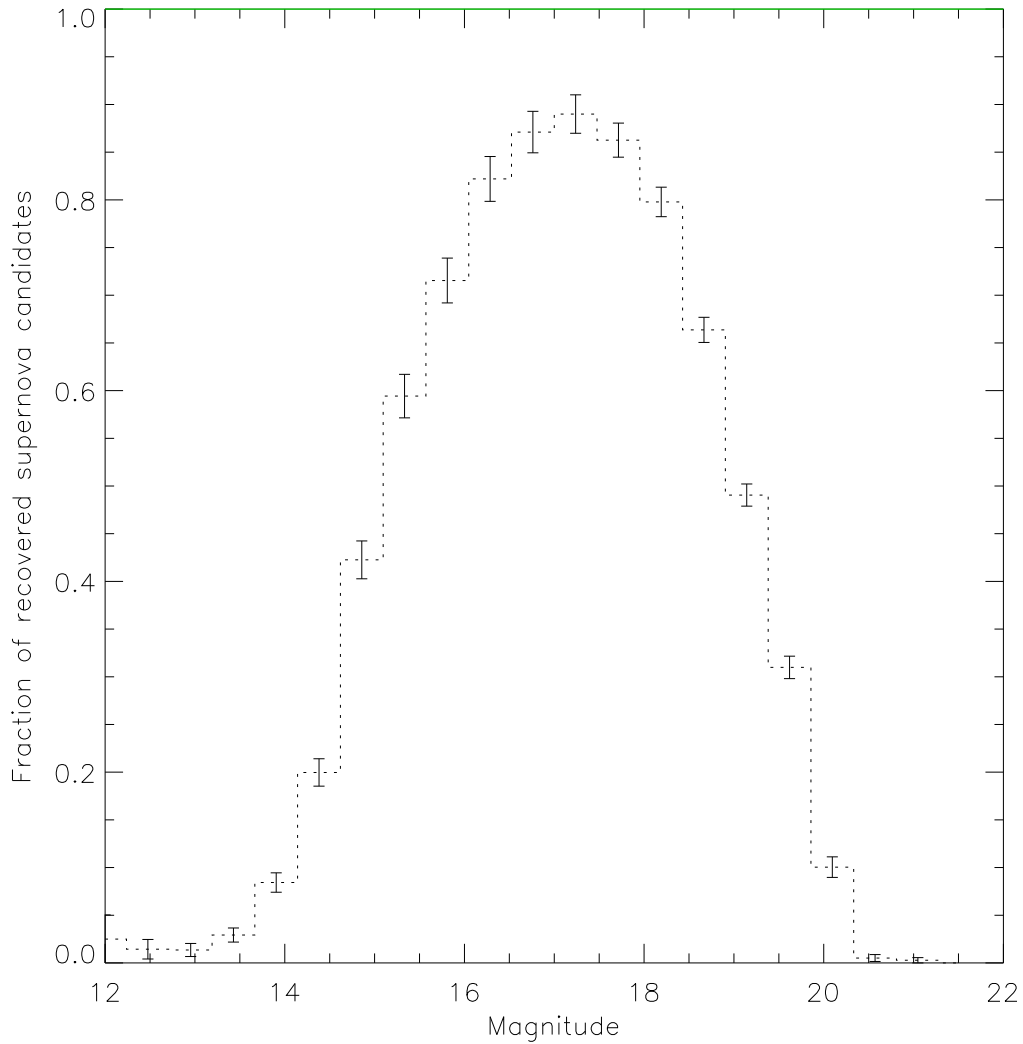}
\caption{The fraction of simulated SNe~Ia that pass the default score cuts as 
a function of magnitude as measured using the
\code{analyzeallfakes.pro} routine on the simulated subtractions run
using the database of 200~million simulated SNe~Ia. No correction for host galaxy light
and the effect of the percent increase score has been applied here (see Fig.~\ref{fig:efficiency_fke6_norm_wgal}).  This plot is analogous to what most other analyses consider the efficiency of a search.}
\label{fig:efficiency_fke6_norm}
\end{figure}

\begin{figure}
\plotone{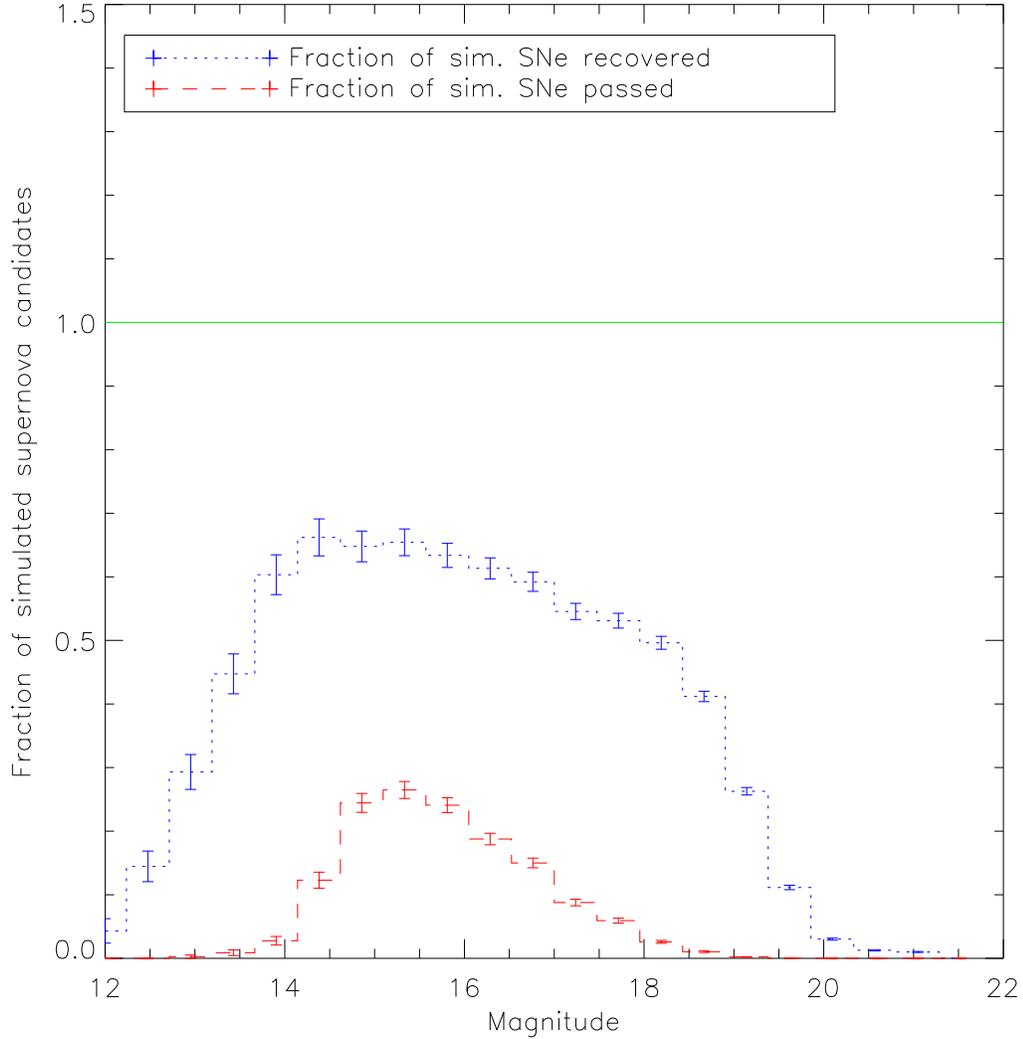}
\caption{The recovered and discovered fraction of database supernova as a 
function of magnitude as measured using the \code{analyzeallfakes.pro}
routine on the simulated subtractions run using the database of
200~million simulated SNe~Ia.  This plot includes the effect of host galaxy
light in the reference and its effect on the percent increase score.  The low fraction in the recovered
discovered supernovae is mainly due to subtractions with short time
baselines where the majority of the simulated supernovae were dimmer
in the new images that in the reference images.  See text
(Sec.~\ref{sec:efficiency}) for more details.}
\label{fig:efficiency_fke6_wgal}
\end{figure}

\begin{figure}
\plotone{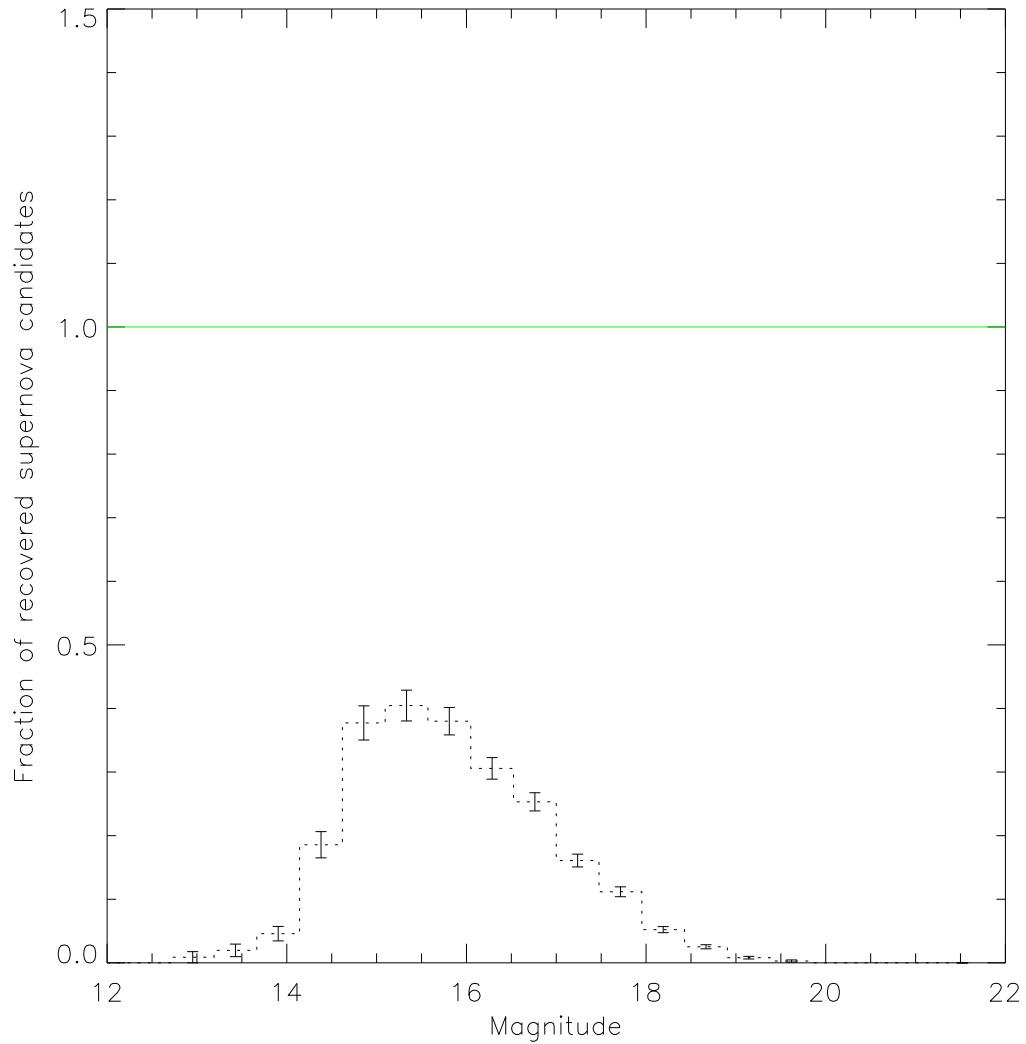}
\caption{The fraction of simulated SNe~Ia that pass the default score cuts as 
a function of magnitude as measured using the
\code{analyzeallfakes.pro} routine on the simulated subtractions run
using the database of 200~million simulated SNe~Ia.  The effect
of the galaxy light as discussed in Sec.~\ref{sec:percent_increase} is clear in comparison with Fig.~\ref{fig:efficiency_fke6_norm}.  This plot is analogous to what most other analyses consider the efficiency of a search.}
\label{fig:efficiency_fke6_norm_wgal}
\end{figure}

The efficiency of the SNfactory search as measured using the Control-Time
Monte Carlo method is shown in Fig.~\ref{fig:efficiency_fke6_norm}~\&~\ref{fig:efficiency_fke6_norm_wgal}.  This
figure shows the discovered fraction divided by the recovered fraction
and is analogous to the standard treatment of efficiencies in other
studies (e.g. \citet{pain02}), although those studies often include the
areal loss from masked regions in their efficiency numbers as well.

Fig.~\ref{fig:efficiency_fke6}~\&~\ref{fig:efficiency_fke6_norm} are
before including the effect of the host galaxy light through the
percent increase threshold.
Fig.~\ref{fig:efficiency_fke6_wgal}~\&~\ref{fig:efficiency_fke6_norm_wgal}
show the effect of the percent increase score in reducing the number
of supernovae found.

While the cutoff at the fainter end for the passed candidates
contrasted with the recovered candidates was not at all surprising,
the bright-end cutoff was.  This cutoff was due to an unexpected
effect related to verifying that potential candidates were the same
brightness in the constituent NEW images, properly weighted by the
different errors.  This score, SUB1MINSUB2, takes the difference in
properly normalized aperture flux in SUB1 and SUB2 and divides by the
quadrature sum of the errors.  However, this score does not include
the error in the calculation of the proper flux ratios to normalize
the aperture flux values.  The threshold that SUB1MINSUB2 must be less
than $2.5$ quickly begins to be comparable to the zeropoint
uncertainty at a magnitude of approximately $15$.  The simulated
supernovae exhibit this effect because of the uncertainty in the
zeropoint used to calculate the appropriate fluxes for placing the
simulated supernovae on the NEW images, while the actual sky objects
suffer from this effect when the images are adjusted for this measured
zeropoint difference.  The simulated and real supernova suffer from
this effect in the subtraction process and thus the effect on the
simulated supernovae does appropriately track the effect on sky
objects.

In addition to the obvious signal-to-noise threshold at the faint end
of the sensitivity, the dispersion in the score used to discriminate against moving
objects, DSUB1SUB2, becomes greater than the allowed thresholds due to
positional uncertainties.  Positional differences for the same object
on different images can come from the transformation between the
images, although normally images are matched to within a tenth of a
pixel RMS, or, at the faint end, from variations in the position found
as the center of the object due to noise fluctuations in the object.
As the noise is proportional to the square root of the signal, these
fluctuations are more pronounced for lower signal values.  The
simulated supernovae exhibit the same variation in DSUB1SUB2 because
noise is added to each simulated supernova as it is placed on the
individual images.

In general, this calculation of search efficiencies for the SNfactory has highlighted a
number of quantitative scores that had a more significant effect on
the search efficiencies than expected.  An in-depth study of both 
real and simulated supernovae should be undertaken to adjust the
scores and thresholds used to classify the objects found in
the subtractions.


\section{Systematic Uncertainties}

A thorough understanding of biases and systematic errors is a key part of a
proper calculation of SN rates.  
Possible sources of systematic uncertainty in this study include the
diversity of SN~Ia light curves, selection effects from the search
efficiency and supernova type identification, the luminosity
distribution of galaxies, host galaxy extinction, Milky Way
extinction, magnitude calibration of the search fields, 
cosmological parameters, Malmquist bias, and the Shaw effect.

\subsection{Diversity of SN~Ia Light Curves}
\label{sec:sys_snia_diversity}

Light curves of SN~Ia exhibit variation at the 30\% level in both peak
flux and rise and fall times~\citep{goldhaber01}.  This distribution has not been mapped
out with sufficient sampling to determine the nature of SN~Ia
variation.  For this analysis this variation was parameterized by the
stretch value~\citep{perlmutter97b,goldhaber01}.  The stretch of the
SNe~Ia was assumed to be normally distributed around a stretch value 1
with $\sigma=0.1$~\citep{perlmutter98a,pain02}.  For a further analysis,
the simulated supernovae used in the rate calculation could be
re-weighted to recreate any given model distribution of stretch values.
For example, the difference between assuming a uniform distribution
in stretch values from $0.9$ to $1.1$ and the normal distribution
with $\sigma=0.1$ resulted in a $0.1\%$ difference in the SN~Ia rate
measured in this study.
Taking twice this difference to represent the uncertainty in our current understanding
of SN~Ia diversity, the unknown variance of SN~Ia light curves contributes $\pm0.2\%$ to the
systematic uncertainty in the rates presented here.

\subsection{Luminosity Distribution of Galaxies}

The underlying galaxy population enters the SN~Ia rate per volume 
calculation through the modeling of the galaxy contribution
in the reference image.  Treatment of the galaxy luminosity distribution is discussed more thoroughly in
Secs.~\ref{sec:percent_increase}~\&~\ref{sec:galaxy_modeling}.
The dominant factor in the shifting of galaxies over the 
typical redshifts for this search, $z\sim0.05$, is $1+z$ from
the time dilation of the emission.  This effect was included 
in the aperture measurements used as the basis underlying
galaxy light modeling.

The conversion to SNu invokes the luminosity density of galaxies
to convert from Mpc$^3$ to $10^{10}$~L$_\Sun$.
Including an additional factor from the assumption that $g$-band\footnote{Technically, this $g$-band is the $^{0.1}g$-band discussed in \citet{blanton03a}.} 
galaxy magnitudes are equivalent $B$ galaxy magnitudes for the purposes of expressing galaxy brightness
in solar luminosities, the conversion to SNu introduces an additional
$\pm5\%$ systematic uncertainty on the SN~Ia rate in SNu, $r_L$.

\subsection{Search Efficiency and Type Identification}
\label{sec:typing}

The efficiency of the SNfactory search is well-understood and
controlled using the Monte Carlo approach described in
Sec.~\ref{sec:montecarlo} and thus the uncertainty from the search efficiency is taken
to be negligible.  However, the type identification of the SNfactory
prototype search requires a more detailed analysis to understand
selection effects in supernova confirmation and spectral
identification.  As the SNfactory SNIFS instrument was not running
during the prototype search and no spectroscopic follow-up was regularly
available for the search, the spectral identification of these
supernovae had to be provided by the community.  This might lead to
concerns that there was a selection bias in the supernovae that were
spectroscopically confirmed.  But, as discussed in
Sec.~\ref{sec:redshifts}, Fig.~\ref{fig:discmags_z} demonstrates that
there was no clear bias with respect to discovery magnitude for
supernovae discovered brighter than $19.5$~magnitudes.
In this analysis, only supernovae (both simulated and real) with
discovery magnitudes brighter than 19.5 magnitudes were considered.
As this calculation is for a constant supernova rate per physical
volume as a function of redshift, selection effects of the type of the observed supernovae
as a function of redshift are not important.

\subsection{Magnitude Calibration}

The magnitudes used for the simulated supernovae in this work were
calibrated to the search images through a calibration to the USNO~A1.0
catalog~\citep{usnoa1}.  This catalog has known systematic variations
between POSS-E fields of $0.2$ magnitudes.  However, the SNfactory
search covered a large number of POSS-E fields ($>100$).  Assuming
that these variations are uncorrelated and centered around $0$, then
this $\sim20\%$ flux uncertainty is reduced to 
$20\% / \sqrt{100} = 2\%$.  Thus, one can roughly include the uncertainties in
magnitude calibration as an additional source of systematic error on
the measured rate at the $\pm2\%$ level.

\subsection{Cosmological Parameters}
\label{eq:sys_cosmological_parameters}

The value of the cosmological parameters $H_0$, \OM and
\OL affect the calculation of distances and volume
elements for this survey.  While \citet{freedman01} measured
$H_0=72\pm8$~km/s/Mpc, the uncertainty
in $H_0$ is factored out of the SN~Ia rate in the $h^3$ term, for the SN~Ia rate per Mpc$^3$. 
In the expression of the SN~Ia rate in
terms of SNu there is one less apparent factor of $h$, but 
any interpretation of the result in terms of physical size
effectively reinstates the factor of $h$.
This leaves uncertainty in \OM and \OL as the
dominant source of cosmological uncertainty in the numeric $r_V$ value
found by this study.
Recent studies of SNe~Ia and
the cosmic microwave background have provided constraints on these
parameters to roughly $\pm0.05$~\citep{knop03,spergel03,riess04b} and
strongly indicate a flat Universe, \OM$+$\OL$=1$
(ignoring curvature).  While the variation in these parameters does
not translate easily to the rate per either volume or luminosity, its
effect can be estimated by considering the difference between 
the volume element at $z=0.1$ in an 
\OM$=0.25$, \OL$=0.75$ and an \OM$=0.35$, \OL$=0.65$ Universe.  
For this nearby search,
the uncertainty in \OM and \OL contributes
$\pm2.4\%$ to the systematic uncertainty in either $r_V$ or $r_L$.

\subsection{Extinction}

As described in Sec.~\ref{sec:milky_way_extinction}, the extinction
from our own galaxy is modeled as part of the simulations using the
dust maps of \citet{schlegel98}.  The extinction from the host galaxy
was not modeled in the simulations.  However, \citet{commins04} 
used a combination of theoretical models and observational data to show
that this effect does not evolve strongly with redshift, and so, in a
comparison with supernova rates at different redshift, the effect of
host galaxy extinction should not be corrected for~\citep{pain02}.
The absolute rate of supernovae as a function of underlying stellar
population is of interest for other purposes, such as studies of
chemical evolution of galaxies and initial mass and binary fractions
of stellar populations.  For these purposes careful attention 
should be paid to modeling the effect of host galaxy dust extinction
to arrive at the ``true'' supernova rate.

\subsection{Malmquist Bias}

Malmquist bias~\citep{malmquist24,malmquist36} is the simple
recognition that in a magnitude-limited survey, there is a bias
against finding dimmer objects at the limits of the survey.  Put in
the specific context of supernova, one will be biased toward finding
fewer fainter supernovae at the limits of one's redshift range.  This
effect is estimated here by doing a Monte Carlo study on the detection
efficiency as a function of magnitude to determine the limiting
redshift to which one has, in effect, conducted a volume-limited
survey~\citep{pain96,pain02}.  For the SNfactory sample, this
efficiency measurement is done on an image-by-image basis through the
Control-Time Monte Carlo method, and only supernova that fall within the
estimated volume sample are included.  Thus, this technique automatically
corrects for any Malmquist bias effect.

However, it should be noted that the SNe~Ia generally are not
sensitive to a strong Malmquist bias because their observed dispersion
is low.  Malmquist bias would be more of an issue for core-collapse
supernovae.


\subsection{The Shaw Effect}

In 1979, Shaw wrote a short, succinct article that presented evidence
for a significant bias against finding supernova near the centers of
galaxies in photographic surveys~\citep{shaw79}.  The basic issue was
that the nucleus of many galaxies is saturated on a photographic
plate.  As a result, in this process, supernovae can not be
detected within that region.
However, the first-ever CCD search for supernovae \citep{muller92} 
found no evidence for a Shaw effect.

\citet{wang97} conducted a study of ``Supernovae and their Host Galaxies''
and found that there were differences in the observed properties and
types of supernova at varying projected radial distances from the
center of the host galaxy.  This distance is referred to as the
projected galactocentric distance (PGD).  \citet{wang97} confirmed the effect
noted in \citet{shaw79} and also found that there was a deficit of
supernova at large PGD in nearby galaxies, possibly due to searches
with fields of view smaller than some of the galaxies surveyed.

\begin{figure}
\plotone{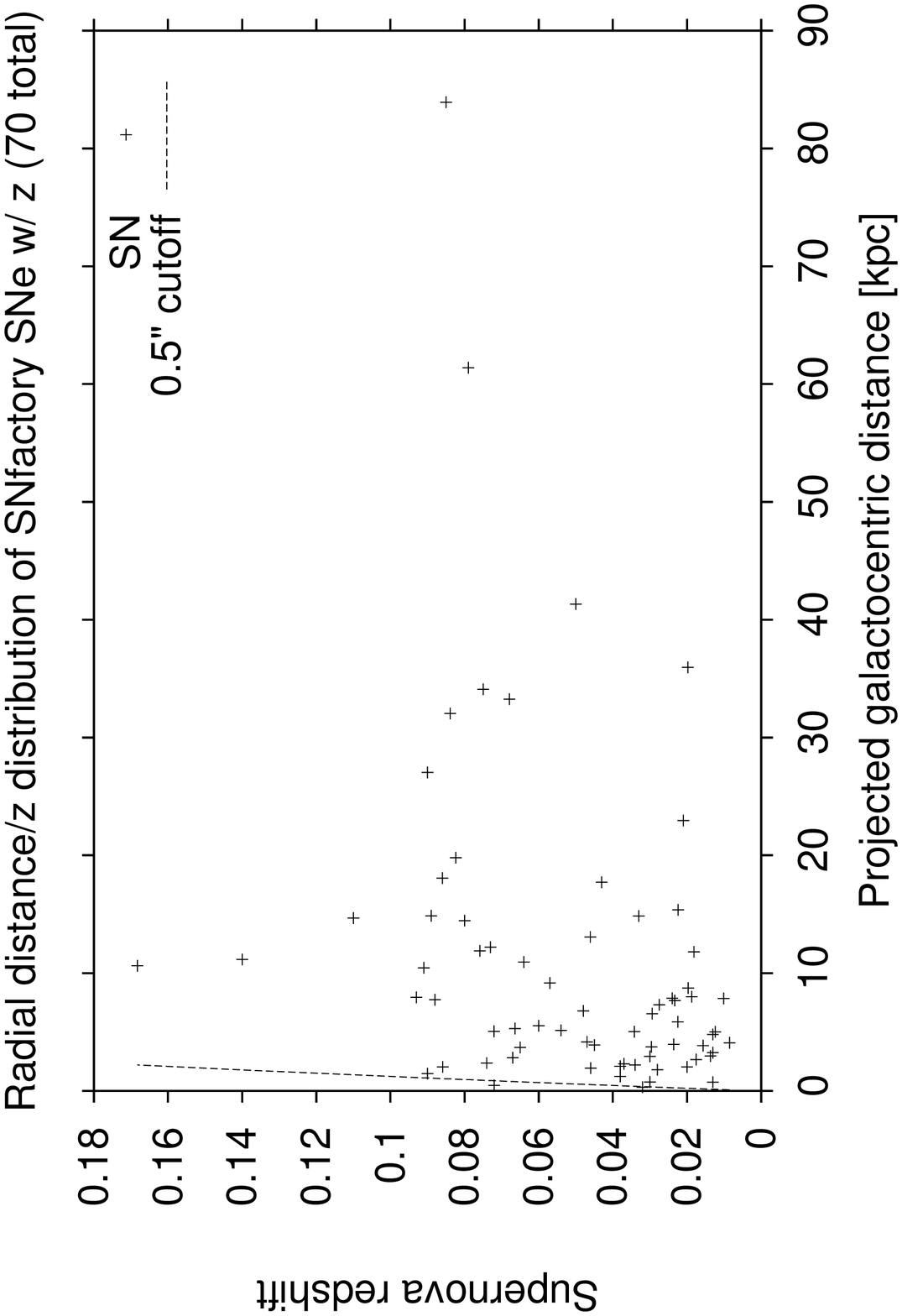}
\caption{The redshift versus projected galactocentric distance for a
sample of 58~SNe with known redshifts found by the SNfactory search.
No evidence for the Shaw effect is observed.  The $0.5$'' cutoff used by
the SNfactory detection software is shown by the dashed line at the far left.  }
\label{fig:pdgs_kpc_z}
\end{figure}

Fig.~\ref{fig:pdgs_kpc_z} shows a scatter plot of redshift versus PGD
for the 70 SNfactory supernova with known redshifts (as of July 1,
2003).  This plot shows no evidence of the Shaw effect.  Compare with
Fig. 1 of \citet{shaw79} and note the absence of any region of
avoidance in Fig.~\ref{fig:pdgs_kpc_z}.  The higher redshift regions
are sparsely filled due to the limits of the search.

There is, however, a known bias in the PGD found as a function of
redshift in the SNfactory search due to the finite angular resolution
ability of the search telescopes.  A $0.5$ pixel cutoff is used in the
candidate screening on distance to the nearest object in the
reference.  This cutoff results in roughly a $0.5$'' threshold (for the Palomar
48$^{\prime\prime}$ 3-CCD detector; for the MSSS 1.2-m, $0.67$''; and for the Palomar
48$^{\prime\prime}$ QUESTII, $0.44$'') that increases in physical distance with
increasing redshift, as shown in Fig.~\ref{fig:pdgs_kpc_z}.

In line with \citet{wang97}, one could interpret
Fig.~\ref{fig:pdgs_kpc_z} as also demonstrating a bias against finding
supernova at large physical separation from the center of the apparent
host in the nearby sample.  However, the SNfactory/NEAT survey uses images on the
order of $0.25\deg \times 0.25\deg$, so this deficit should not be a result of
a limited field of view, as proposed by \citet{wang97} as an
explanation for the effect in their study.
We note that the SNfactory software masks out the entirety of any object with a
sufficient number of saturated pixels to mask out bright stars that
may contaminate large regions of the CCD.  This same procedure will
mask the entirety of a galaxy with a saturated region such as the core
or a superimposed bright star.  However, the number of galaxies
at these low redshifts is very sparse and so there are fewer
supernovae at those redshifts for a given solid angle covered.

\begin{figure}
\plotone{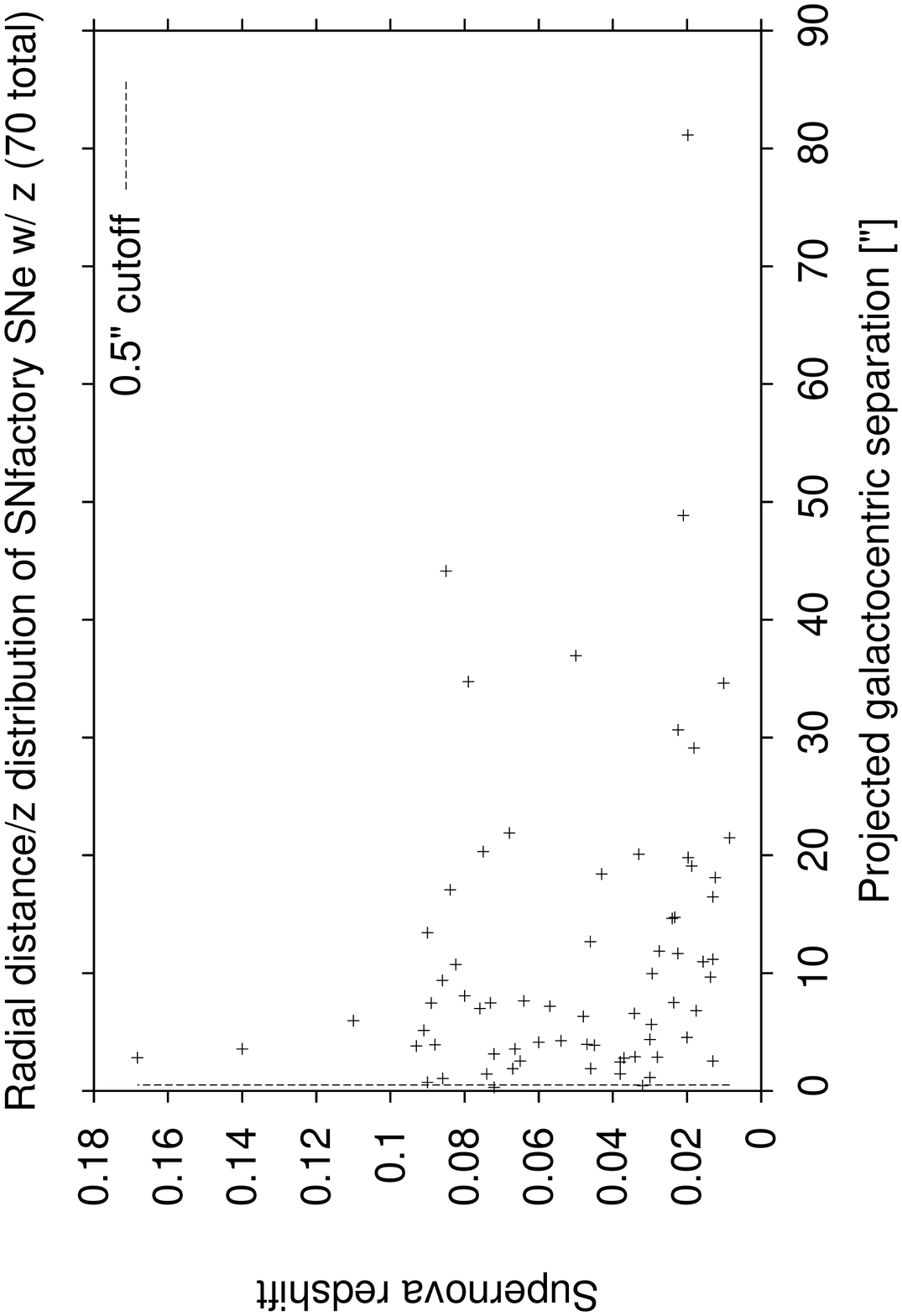}
\caption{The redshift versus projected galactocentric angular
separation for a sample of 58~SNe with known redshifts found by the
SNfactory search.  The $0.5$'' cutoff used by the detection software due
to seeing considerations is shown by the dashed line at the far left.  }
\label{fig:pdgs_sec_z}
\end{figure}

Fig.~\ref{fig:pdgs_sec_z} presents the same results as in
Fig.~\ref{fig:pdgs_kpc_z} but in terms of projected arcseconds from the
apparent host galaxy instead of projected physical distance.  This
representation
makes it clear that the selection in terms of angular distance shows
no clear trend as a function of $z$.  With some allowance for
fluctuation and statistical noise at any particular redshift, there
is no clear deficit at any redshift.  



\section{SN~Ia Rate}
\label{sec:snrate}

A subsample of the subtractions from the SNfactory prototype search from September 2002
through May 2003 was used to generate a preliminary rate measurement using the technique described above.
To avoid bias in the count of 
found supernovae, the simulated subtractions were
run and debugged without reference to which nights or subtractions had
real supernovae.

The simulated number of SNe~Ia from the search was compared with the
total effective number of SNe~Ia found in the sample considered here.
An analysis of the simulated subtractions found a predicted number of SNe~Ia of
\begin{equation}
n_\mathrm{Ia} = \snesimexpected \left[\frac{r_V}{1\times 10^{-4}~\mathrm{SNe~Ia}/\mathrm{yr}/\mathrm{Mpc}^3}\right],
\end{equation}
where the uncertainty comes from the finite number of simulated supernovae
used in the Control-Time Monte Carlo approach.
As $n_\mathrm{Ia}=\sneiaused$~SNe~Ia were actually found in this sample, this preliminary study obtains a 
SN~Ia rate of
\begin{equation}
r_V=\snrate~\mathrm{SNe~Ia/yr/Mpc}^3,
\label{eq:sniarate}
\end{equation}
where the listed uncertainties are statistical and systematic.  While
the statistical error bars remain significant in this analysis of a
sample of the SNfactory prototype search, the strength of the
Control-Time Monte Carlo approach is apparent in the small systematic
uncertainties.

\subsubsection{Conversion to SNu}

To convert Mpc$^3$ to SNu requires a measurement of the luminosity
density of galaxies in the sample.  The SDSS galaxy catalog provides
the most thorough measurement of this density to date.  \citet{blanton03a}
found a luminosity density of $(1.78\pm0.05)\times 10^8~h~L_{g \Sun}/$Mpc$^3$.
Taking $L_{g \Sun}\sim L_{B\Sun}$, and inverting this
luminosity density, one can convert from Mpc$^3$ to $10^{10}~L_{B\Sun}$,
\begin{equation}
\mathrm{SN~Ia~rate}~[\mathrm{SNu}] =
 \frac{56.2\pm2.8}{h} \frac{\mathrm{Mpc}^3}{10^{10} L_{B\Sun}}
 \mathrm{SN~Ia~rate}~[10^{-4} /\mathrm{yr}/\mathrm{Mpc}^3]
\end{equation}
Thus the SN~Ia rate can be expressed in SNu as 
\begin{equation}
r_L=\snrateSNu~\mathrm{SNu}.
\label{eq:sniarate_snu}
\end{equation}

\begin{figure}
\plotone{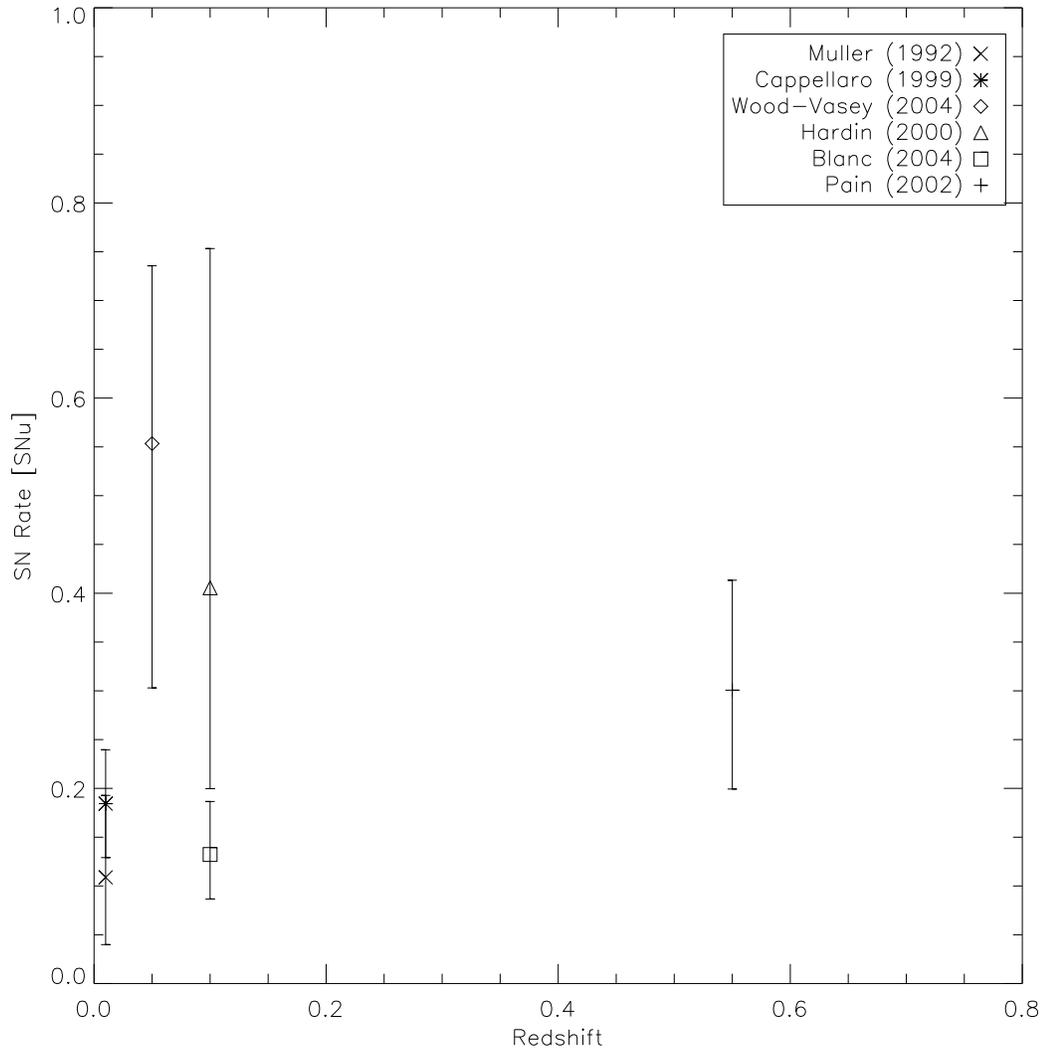}
\caption{A comparison of the results of this preliminary study with previous
SN~Ia rate measurements: \citet{muller92}, \citet{cappellaro99}, \citet{hardin00},
\citet{blanc02}, and \citet{pain02}.  
The statistical limits of this study will be improved with further
work to be published in 2005.
The results of each study
have been standardized to $H_0=72$~km~s$^{-1}$~Mpc$^{-1}$.
The rate of \citet{muller92} has been assumed to be for $z=0.01$.
}
\label{fig:snrate}
\end{figure}

As shown Fig.~\ref{fig:snrate}, the SN~Ia rate found here is
consistent with the results of 
\cite{muller92}, \cite{cappellaro99}, \citet{hardin00}, \citet{blanc02}, and
\citet{pain02}.  The statistical uncertainties on the current 
SNfactory rate measurement will be reduced with further analysis of
the SNfactory prototype search.
The current results from this and other nearby studies 
do not greatly constrain the SN~Ia rate as a function of redshift.
The improved constraints from a further analysis of the SNfactory
search together with the intermediate-redshift ($z\sim0.5$) SN~Ia rate 
of \citet{pain02}
will allow for an investigation of galaxy evolution and star
formation rates over the past 5~billion years.


\part{The Hybrid Supernova 2002ic}
\label{part:2002ic}
%





\hypersetup{
            pdftitle={Photometry of {SN~2002ic} and Implications for the Progenitor Mass-loss History},
            pdfauthor={W. Michael Wood-Vasey, UC Berkeley/LBNL and Lifan Wang, LBNL}
            bookmarks=true,
            bookmarksnumbered=true,
            colorlinks=true,
            linkcolor=blue
           }

\chapter{Photometry of SN~2002ic and Progenitor Mass-Loss}
\label{chp:2002ic}
{\em W.~M. Wood-Vasey and L. Wang and G. Aldering}

{\bf [The following was accepted for publication in ApJ, 2004.]}

\date{2004 May 6}

\section{Abstract}We present new pre-maximum and late-time optical photometry of the
Type~Ia/IIn supernova 2002ic. These observations are combined with the
published V-band magnitudes of~\citet{hamuy03b} and the
VLT spectrophotometry of~\citet{wang04} to construct the most extensive
light curve to date of this unusual supernova.  The observed flux at
late time is significantly higher relative to the flux at maximum than
that of any other observed Type~Ia supernova and continues to fade
very slowly a year after explosion.  Our analysis of the light curve
suggests that a non-Type Ia supernova component becomes prominent
$\sim20$~days after explosion.  Modeling of the non-Type Ia supernova
component as heating from the shock interaction of the supernova
ejecta with pre-existing circumstellar material suggests the presence
of a $\sim1.7\times10^{15}$~cm gap or trough between the progenitor system 
and the surrounding circumstellar material.
This gap could be due to significantly lower mass-loss
$\sim15~(\frac{v_w}{10~\mathrm{km/s}})^{-1}$~years 
prior to explosion or evacuation of the circumstellar material by a
low-density fast wind.  The latter is consistent with observed
properties of proto-planetary nebulae and with models of white-dwarf +
asymptotic giant branch star progenitor systems with the asymptotic
giant branch star in the proto-planetary nebula phase.

\section{Introduction}
\label{sec:introduction}

Historically, the fundamental division of supernova (SN) types was
defined by the absence (Type~I) or presence (Type~II) of hydrogen in
the observed spectrum.  Later refinements distinguished Type~Ia
supernovae from other types of supernovae by the presence of strong
silicon absorption features in their spectra
\citep{wheeler90,filippenko97}. Type~Ia supernovae (SNe~Ia) are
generally accepted to result from 
the thermonuclear burning of a white dwarf in a binary
system, whereas all the other types of supernovae are believed to be
produced by the collapse of the stellar core, an event which leads to
the formation of a neutron star or black hole.

While interaction with circumstellar material (CSM) has been observed
for many core-collapse supernovae, the search for evidence of CSM
around Type~Ia~SNe has so far been unsuccessful. \citet{cumming96}
reported high resolution spectra of SN~1994D and found an upper limit
for the pre-explosion mass-loss rate of
$\dot{M}\sim1.5\times10^{-5}~M_\Sun~\mathrm{yr}^{-1}$ for an assumed
wind speed of $v_w = 10$~km~s$^{-1}$.
However, they also note that this limit allows most of the expected
range of mass-loss rates from symbiotic systems
($\frac{\dot{M}}{v_{10}} \lesssim 2\times10^{-5}~M_\Sun$~yr$^{-1}$).
On the other
hand, the surprisingly strong high-velocity Ca~II absorption and
associated high degree of linear polarization observed in SN~2001el 
by \citet{wang03a} and the high velocity features in SN~2003du by
\citet{gerardy04} have led these authors 
to suggest that the high velocity Ca feature could be
the result of the interaction between the supernova ejecta and a CSM
disk.  About 0.01~M$_\Sun$ of material is required in the disk, and
the spatial extent of the disk must be small to be consistent with the
absence of narrow emission lines at around optical maximum
\citep{cumming96}.  Due to the strength of the Ca~II feature in SN~2001el,
\citet{wang03a} speculated that the disk of SN~2001el may have been over-abundant in Ca~II. In contrast, \citet{gerardy04} found that a standard solar
abundance of Ca~II is sufficient to explain the observed feature in
SN~2003du~\citep{gerardy04}, for which the high-velocity Ca~II feature
is significantly weaker than in SN~2001el.

Supernova 2002ic~\citep{iauc8019} is a very interesting event that shows both silicon
absorption~\citep{iauc8028} and hydrogen
emission~\citep{hamuy03b}. This SN is the first case for which there is
unambiguous evidence of the existence of circumstellar matter around a
SN~Ia and is therefore of great importance to the understanding of the
progenitor systems and explosion mechanisms of SNe~Ia~\citep{livio03}.  By studying
the spectral polarimetry and the light curve of the H$\alpha$ line,
\citet{wang04} found the spatial extent of the hydrogen-rich 
material to be as large as 10$^{17}$~cm and distributed in a quite
asymmetric configuration, most likely in the form of a flattened
disk. The implied total mass of the hydrogen-rich CSM is a few solar
masses.  Similar conclusions were reached by \citet{deng04}.

In this paper, we present new photometry of SN~2002ic and discuss the
implications for the interaction of the ejecta and the CSM.
Sec.~\ref{sec:processing} presents our data processing procedure and
calibration for our photometry of SN~2002ic.  In
Sec.~\ref{sec:lightcurve}, we discuss the light curve of SN~2002ic and
the immediate implications from our data.  A more in-depth
investigation and qualitative modeling of the light curve of SN~2002ic
as an interaction of a SN~Ia with surrounding CSM is presented in
Sec.~\ref{sec:modeling}.  Our discussion in Sec.~\ref{sec:discussion}
presents our interpretations of the structure of the CSM surrounding
SN~2002ic.  Finally, in Sec.~\ref{sec:conclusions} we present some
intriguing possibilities for the progenitor system of SN~2002ic and
speculate on other possible SN~2002ic-like events.

\section{Data Processing for SN~2002ic}
\label{sec:processing}

\subsection{Data processing and Discovery}
 

We discovered SN~2002ic on images from the NEAT team~\citep{pravdo99}
taken on the Samuel Oschin 1.2-m telescope on Mt.~Palomar, California.
In preparation for searching, the images were transmitted from the
telescope to the High-Performance Storage System (HPSS) at the
National Energy Research and Scientific Computer Center (NERSC) in
Oakland, California via the
HPWREN~\citep{hpwren}\footnote{http://hpwren.ucsd.edu} and
ESnet~\citep{esnet}\footnote{http://www.es.net} networks.  These data were then
automatically processed and reduced on the NERSC Parallel Distributed
System Facility (PDSF) using software written at Lawrence Berkeley
National Laboratory by WMWV and the Supernova Cosmology Project.

The first-level processing of the NEAT images involved decompression
and conversion from the NEAT internal format used for transfer to the
standard astronomical FITS format, subtraction of the dark current for
these thermoelectrically cooled CCDs, and flat-fielding with sky flats
constructed from a sample of the images from the same night.  These
processed images were then loaded into an image database, and archival
copies were stored on HPSS.  The images were further 
processed to remove the sky background.  An object-finding algorithm
was used to locate and classify the stars and galaxies in the fields.
The stars were then matched and calibrated against the USNO~A1.0
{POSS-E} catalog~\citep{usnoa1} to derive a magnitude zeropoint for each
image.  There were typically a few hundred USNO~A1.0 stars in each
0.25~\sq\degr\ image.

The supernova was discovered by subtracting PSF-matched historical
NEAT images from new images, then automatically detecting residual
sources for subsequent human inspection (see \citet{wood-vasey04a}).

\subsection{Photometry}

For analysis, we assembled all NEAT images, including later images
kindly taken at our request by the NEAT team.

Light curves were generated using aperture photometry scaled to the
effective seeing of each image.  A set of the 4 best-seeing
($<3$\arcsec) reference images was selected from among all NEAT
Palomar pre-explosion images from 2001 of SN~2002ic.  Multiple
reference images were chosen to better constrain any underlying galaxy
flux.  The differential flux in an aperture around SN~2002ic was
measured between each reference image and every other image of
SN~2002ic.  Aperture correction was performed to account for the
different seeing and pixel scales of the images.  The overall flux
ratio between each reference image and light-curve image was tracked
and normalized with respect to a primary reference image.  This
primary reference image was chosen from the reference images used for
the image subtraction on which SN~2002ic was originally discovered.
The flux differences calculated relative to each reference were
combined in a noise-weighted average for each image to yield an
average flux for the image.  As the observations were taken within a
span of less than one hour on each night, the results from the images
of a given night were averaged to produce a single light curve point
for that night.

The reference zeropoint calculated for the primary reference image
from the above USNO~A1.0 {POSS-E} calibration was used to set the
magnitudes for the rest of the measured fluxes.
Table~\ref{tab:2002ic_lightcurve} reports these magnitudes and
associated measurement uncertainties.  An overall systematic
uncertainty in the zeropoint calibration is not included in the listed
errors.  The USNO~A1.0 {POSS-E} catalog suffers from systematic
field-to-field errors of $\sim0.25$~magnitudes in the northern
hemisphere~\citep{usnoa1}.  The conversion of {POSS-E} magnitudes to
V-band magnitudes for a SN~Ia is relatively robust, as a SN~Ia near
maximum resembles a $\sim10,000$~K blackbody quite similar to Vega in
the wavelength range from $4,500$--$10,000$~\AA.  At late times,
the observations of~\citet{wang04} show that the smoothed spectrum of
SN~2002ic tracks that of Vega red-ward of 5,000~\AA.  We estimate
that, taken together, the calibration of our unfiltered observations
with these {POSS-E} magnitudes and the subsequent comparison with V-band
magnitudes
are susceptible to a $0.4$~magnitude systematic uncertainty.  Any such
systematic effect is constant for all data points and stems directly
from the magnitude calibration of the primary reference.

However, the observed NEAT {POSS-E} magnitudes show agreement with the
V-band magnitudes of \citet{hamuy03b} and the V-band magnitudes
obtained from integrating the spectrophotometry of \cite{wang04} to
significantly better than any $0.4$~magnitude systematic uncertainty
estimate.  This synthesized VLT photometry is presented in
Table~\ref{tab:2002ic_VLT}.  Comparing the photometry of SN~2002ic and
nearby reference star with a similar color ($B-R=0.3$), we find agreement
between the VLT V-band acquisition camera images and the NEAT images
to within $\pm 0.05$~magnitudes.  Given this good agreement, it appears
that our POSS-E-calibrated magnitudes for SN~2002ic can be used effectively as V-band
photometry points.

Mario Hamuy was kind enough to share his BVI secondary standard stars
from the field of SN~2002ic.  We attempted to use these stars to
calculate the color correction from our {POSS-E} magnitudes to V-band,
but our analysis predicted an adjustment of up to $+0.4$~magnitudes.
This was inconsistent with the
good agreement with the VLT magnitudes (calculated correction =
$+0.1$~mag) at late times and with the Hamuy V-band points after
maximum ($+0.4$~mag).  This disagreement is not fully understood.  We
note, however, that the colors of the secondary standard stars did not
extend far enough to the blue to cover the majority of the color range
of the supernova during our observations (a common problem when
observing hot objects such as supernovae).  In addition, as there is no 
color information from before maximum light, it is possible that
SN~2002ic does not follow the color evolution of a typical SN~Ia.

\begin{figure}
\plotone{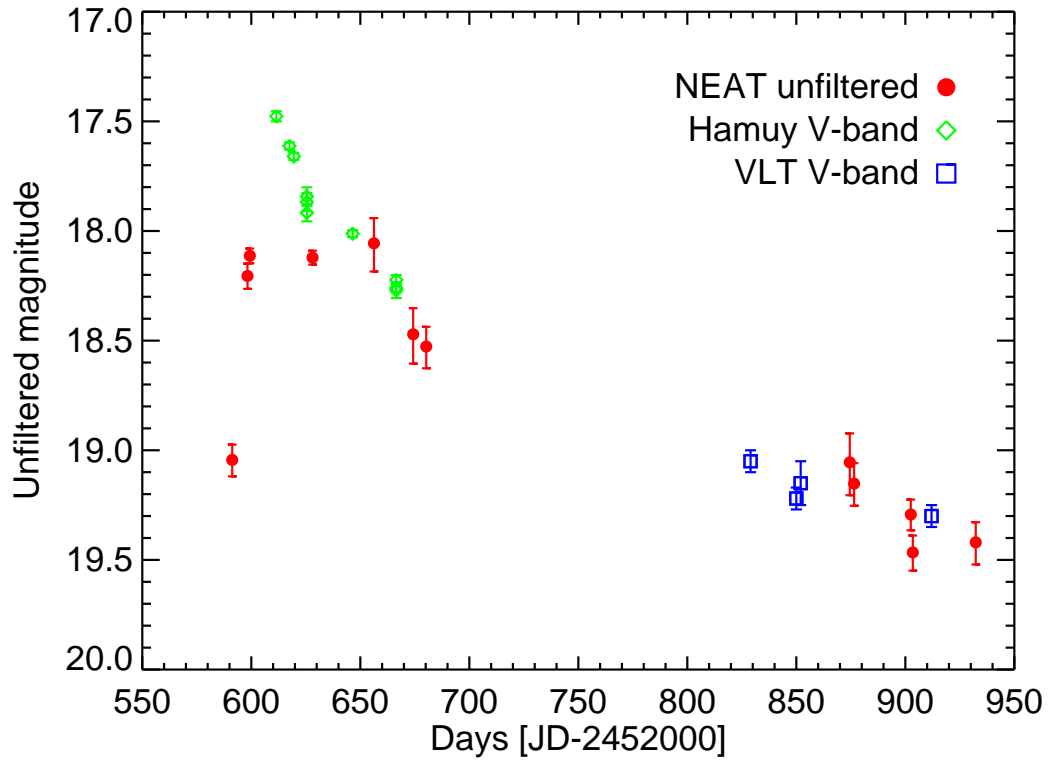}
\caption{The unfiltered optical light curve of SN~2002ic as observed by NEAT with
the Palomar 1.2-m and Haleakala 1.2-m telescopes (see
Table~\ref{tab:2002ic_lightcurve}).  The magnitudes have been
calibrated against the USNO-A1.0 {POSS-E} stars in the surrounding
field.  No color correction has been applied.  Also shown 
are the observed V-band magnitudes from~\citet{hamuy03b} and
V-band magnitudes from the spectrophotometry of~\citet{wang04}.}
\label{fig:2002ic_VLT}
\end{figure}

Our newly reported pre-maximum photometry points (see
Table~\ref{tab:2002ic_lightcurve} and Fig.~\ref{fig:2002ic_VLT}) are
invaluable for disentangling the SN and CSM components, which we now
proceed to do.

\section{Light Curve of SN~2002ic}
\label{sec:lightcurve}

The light curve of SN~2002ic is noticeably different from that of a
normal SN~Ia, as can be seen in Fig.~\ref{fig:2002ic_template}, and as
was first noted by \citet{iauc8151}.  The detection of hydrogen
emission lines in the spectra of SN~2002ic in combination with the
slow decay of the light curve is seen as evidence for interaction of
the SN ejecta and radiation with a hydrogen-rich
CSM~\citep{hamuy03b,wang04}.  The profile of the hydrogen emission
line and the flat light curves can be understood in the context of
Type~IIn supernovae as discussed in \citet{chugai02}, \citet{chugai04}, and references therein. 

The data presented here show that the
slow decay has continued $\sim320$~days after maximum at a rate of
$\sim0.004$~mag/day, a rate that is significantly slower than the
$0.01$~mag/day decay rate expected from Co$^{56}$ decay (also see
\citet{deng04}).  In addition, our early-time points show that the
light curve of SN~2002ic was consistent with a pure SN~Ia early in its
evolution.  This implies that there was 
a significant time delay between the explosion and development
of substantial radiation from the CSM interaction, possibly due to a 
a physical gap between the
progenitor explosion and the beginning of the CSM.  After maximum, we
note the existence of a second bump in the light curve, which is put in
clear relief by our photometry data on JD~$2452628.6$.  We interpret
this second bump as evidence for further structure in the CSM.

\section{Decomposition of SN Ia and CSM components}
\label{sec:decomposition}

\begin{figure}
\plotone{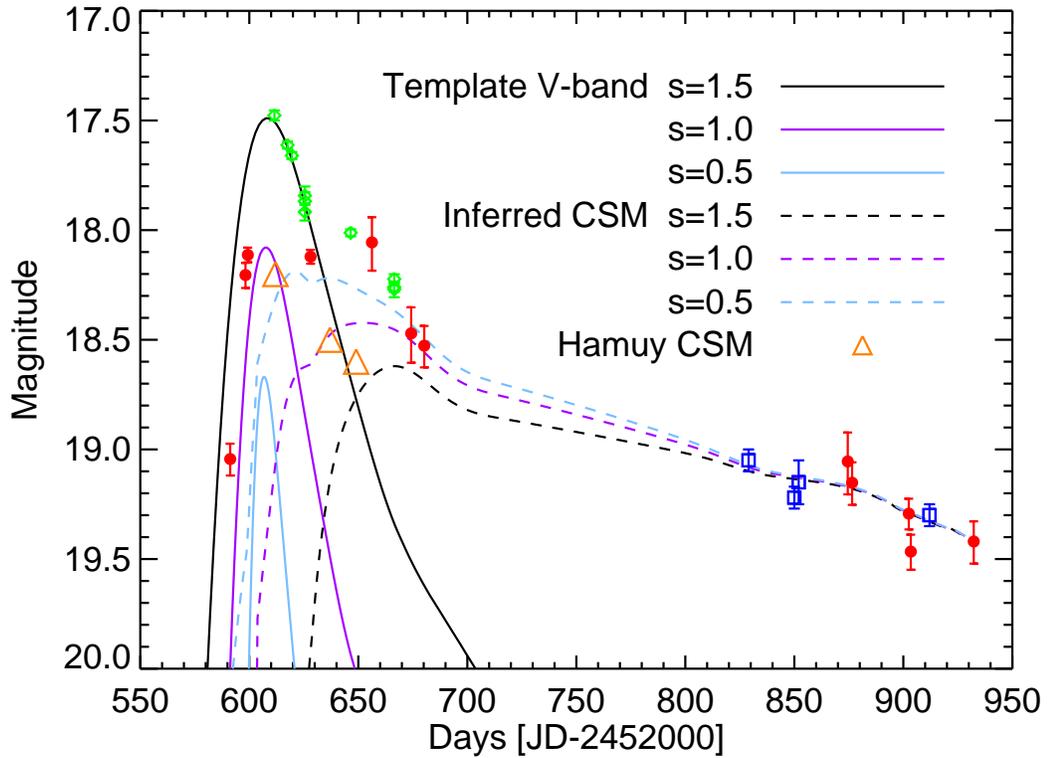}
\caption{
A template SN~Ia V-band light curve (solid lines -- stretch decreases from top to bottom line) shown for comparison 
with the photometric observations at several
stretch values, $s$, where the magnitude-stretch relation 
$\Delta m = 1.18~(1-s)$
has been applied.  The difference between the observed
photometry points and the template fit has been smoothed over a $50$-day window (dashed lines).  Note that an assumption of no CSM
contribution in the first $15$ days after maximum light 
(i.e. $s=1.5$)
is in conflict
with the spectroscopic measurements of \citet{hamuy03b} (open triangles--no error bars available).  }
\label{fig:2002ic_template}
\end{figure}

\citet{hamuy03b} performed a spectroscopic decomposition of the 
underlying supernova and ejecta-CSM interaction components.  We
perform here an analogous photometric decomposition.  To decompose the
observed light curve into the contributions from the SN
material and the shock-heated CSM, we first consider a range of
light curve stretch values~\citep{perlmutter97b}, using
the magnitude-stretch relation, $\Delta m =1.18~(1-s)$~\citep{knop03}, 
applied to the normal SN~Ia template light
curve of \citet{goldhaber01}; we consider the remaining flux as being
due to the SN~eject-CSM interaction (see
Fig.~\ref{fig:2002ic_template}).
At early times, the inferred contribution of the CSM is dependent on
the stretch of the template chosen, but at later times the CSM
component completely dominates for any value of the stretch parameter.  It is not possible to disentangle the
contribution of the CSM from that of the SN at maximum light, although
a normal SN~Ia at the redshift of SN~2002ic,
$z=0.0666$~\citep{hamuy03b}, corresponding to a distance modulus of
$37.44$ for an $H_0=72$~km/s/Mpc~\citep{freedman01}, would only generate about half of
the flux observed for SN~2002ic at maximum.
\citet{hamuy03b} find that SN~2002ic resembles SN~1991T
spectroscopically and note that SN~1991T/SN~1999aa-like events are
brighter a month after maximum light than explainable by the standard
stretch relation.  A SN~1991T-like event (stretch$=1.126$, $\Delta
m=0.15$, based on the template used in \citet{knop03} (A.J. Conley
2004, private communication)), would lie near the stretch$=1$ line of
Fig.~\ref{fig:2002ic_template}.  The light curve of SN~2002ic for the
first 50~days is thus much too luminous to be due entirely to a
91T-like supernova.  In addition, the spectroscopically-inferred
CSM-interaction contribution of \citet{hamuy03b} 
(open triangles in Fig.~\ref{fig:2002ic_template}) limits the SN
contribution at maximum to that expected from a normal SN~Ia.  After
50~days, SN~2002ic exhibits even more significant non-SN~Ia-like
behavior.

We next use the formalism of \citet{chevalier94} to fit a simple
interacting SN ejecta-CSM model to the observed data.  While 
\citet{chevalier94} focus on SNe~II, their formalism is generally
applicable to SNe ejecta interacting with a surrounding CSM.
We simultaneously fit the SN~Ia flux and the luminosity from the SN
ejecta-CSM interaction.  Our analysis allows us to infer the integrated radial
density distribution of the CSM surrounding SN~2002ic.  

\section{Simple Scaling of the SN Ejecta-CSM Interaction}
\label{sec:modeling}

Following the hydrodynamic models of \citet{chevalier94}, we assume a
power-law supernova ejecta density of
\begin{equation}
\rho_\mathrm{SN} \propto t^{n-3} r^{-n}
\label{eq:ejecta_density}
\end{equation}
where $t$ is the time since explosion, $r$ is the radius of the ejecta,
and $n$ is the power-law index of a radial fall-off in the ejecta density.
\cite{chevalier01} note that for SNe~Ia an exponential ejecta profile 
is perhaps preferred.  However, this profile does not yield an analytical
solution and so, for the moment, we proceed assuming a power-law profile.
In Sec.~\ref{sec:discussion} we explore the ramifications of an 
exponential ejecta profile.

\cite{chevalier94} give the time evolution of the shock-front radius, $R_s$, as
\begin{equation}
R_s = \left [ \frac{2}{(n-3)(n-4)} \frac{4\pi v_w}{\dot{M}} A \right]^{1/(n-2)} t^{(n-3)/(n-2)},
\label{eq:shock_radius}
\end{equation}
where $v_w$ is the velocity of the pre-explosion stellar wind,
$\dot{M}$ is the pre-explosion mass-loss rate, and
$A$ is a constant in the appropriate units for the given
power-law index $n$.

Taking the parameters in the square brackets as fixed constants, we can
calculate the shock velocity, $v_s$, as
\begin{equation}
v_s = \left [ \frac{2}{(n-3)(n-4)} \frac{4\pi v_w}{\dot{M}} A \right]^{1/(n-2)} \left(\frac{n-3}{n-2}\right) t^{-1/(n-2)}.
\label{eq:shock_velocity}
\end{equation}
Thus the shock velocity goes as 
\begin{equation}
v_s \propto t^{-\alpha},
\label{eq:vs_time}
\end{equation}
where
\begin{equation}
\alpha = \frac{1}{n-2}.
\label{eq:vs_time_alpha}
\end{equation}

We assume that the luminosity of the ejecta-CSM interaction is fed by
the energy imparted at the shock front and view the unshocked wind as
crossing the shock front with a velocity of $v_s + v_w \approx v_s$.
As the wind particles cross the shock front, they are thermalized and
their crossing kinetic energy, $\mathrm{K.E.}=\onehalf \rho_w v_s^2 dV$, is
converted to thermal energy.  Putting this in terms of the mass-loss
rate, $\dot{M}$, we can express the CSM density as
\begin{equation}
\rho_w = \frac{\dot{M}}{4\pi R_s^2 v_w},
\end{equation}
and we can calculate the energy available to be converted to luminosity, $L$, as
\begin{equation}
L = \alpha(\lambda,t) \frac{d}{dt}\mathrm{K.E.} = \alpha(\lambda,t) \frac{1}{2} \frac{\dot{M}}{4\pi R_s^2 v_w} v_s^2 dV 
= \alpha(\lambda,t) \frac{1}{2} \frac{\dot{M}}{4\pi R_s^2 v_w} v_s^2 v_s 4\pi R_s^2.
\end{equation}
The luminosity dependence on $R_s$ drops out and we have
\begin{equation}
L = \alpha(\lambda,t) \frac{d}{dt} \mathrm{K.E.} = \alpha(\lambda,t) \frac{1}{2} \frac{\dot{M}}{v_w} v_s^3.
\end{equation}
%
A key missing ingredient is a more detailed modeling of the kinetic
energy to optical luminosity conversion term, $\alpha(\lambda, t)$.
We note that the available kinetic energy is on the order of $1.6\times
10^{44}$~erg~s$^{-1}$ for $\dot{M} = 10^{-5}~M_\sun$~yr$^{-1}$, $v_s =
10^4$~km~s$^{-1}$, and $v_w = 10$~km~s$^{-1}$.  This implies a
conversion efficiency from shock interaction K.E. to luminosity of
50\%, given the luminosity, $1.6\times10^{44}$~erg~s$^{-1}$, of
SN~2002ic and the typical luminosity of a SN~Ia near maximum of
$0.8\times10^{44}$~erg~s$^{-1}$.  Assuming this constant conversion
produces reasonable agreement with the data, so we proceed with this
simple assumption.  Using Eq.~\ref{eq:vs_time} to give the time
dependence of $v_s$, we obtain the time dependence of the luminosity,
\begin{equation}
L \propto v_s^3 \propto t^{-3\alpha},
\end{equation}
which can be expressed in magnitude units as
\begin{equation}
m_\mathrm{ejecta-CSM} = C - \frac{5}{2} \log_{10}{t^{-3\alpha}} = C + \frac{15}{2}\alpha \log_{10}{t},
\end{equation}
where $C$ is a constant that incorporates $\dot{M}$,
$\rho_{\mathrm{SN}}$, $v_w$, $n$, and the appropriate units for those
parameters.
The difference in magnitude between two times, $t_1$ and $t_2$, then becomes
\begin{equation}
m_{t_2} - m_{t_1} = \frac{15}{2} \alpha \log_{10}{\frac{t_2}{t_1}}.
\end{equation}

We obtain a date of B-maximum for the supernova component of
$2452606$~JD from our SN~Ia-light curve analysis.
Our fit yields an $\alpha=0.16 \Rightarrow n = 8.5$
for any fixed $\dot{M}$ and $v_w$.  This $n$ is squarely in the range
of values suggested by \citet{chevalier94} as being typical for SN
ejecta.  While \citet{chevalier94} is framed in the context of 
SNe~II, their formalism applies to any SN explosion into a surrounding medium
whose ejecta density profile is described by their analytic model.

The interaction scaling relations presented above are useful for
decomposing the interaction and supernova contributions to the total
light curve of SN~2002ic. This simple, analytic description 
approximates our data reasonably well.
However, more sophisticated theoretical calculations,
which are beyond the scope of this paper,
are necessary
to more quantitatively derive the detailed physical parameters of the
SN ejecta and the CSM (see \citet{chugai04}).

\section{Discussion}
\label{sec:discussion}

\subsection{Inferred CSM structure and Progenitor Mass-Loss History}

\begin{figure}
\plotone{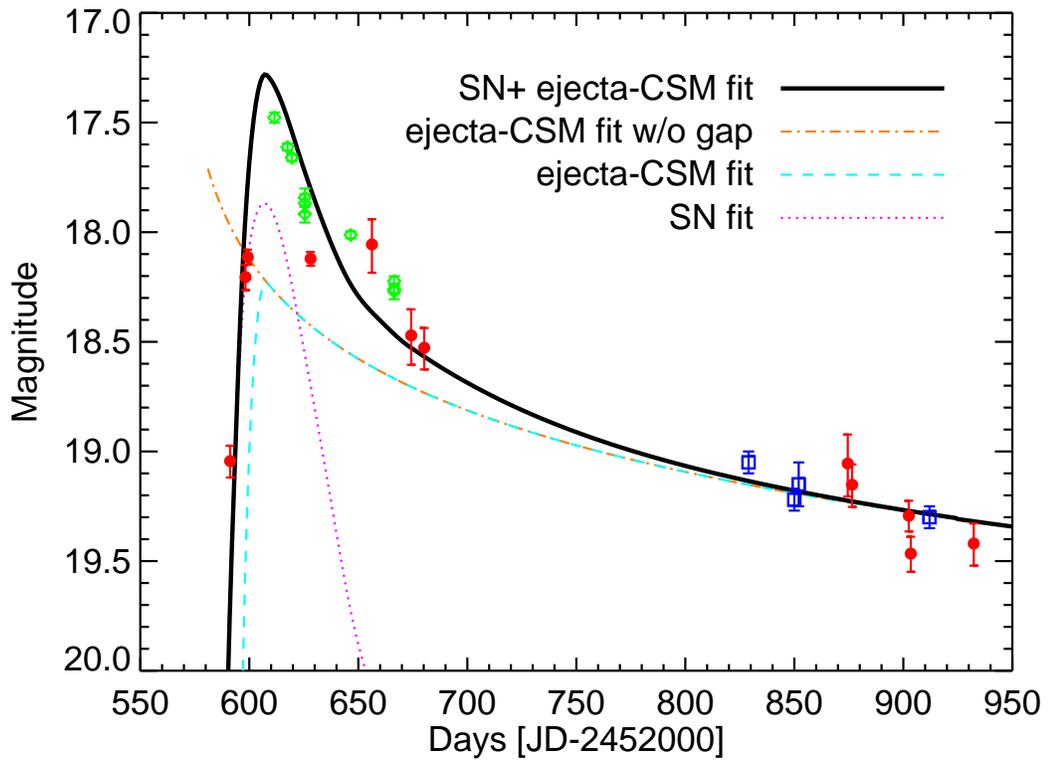}
\caption{
The observed 
photometry compared with
the SN + power-law ejecta-CSM model described in Sec.~\ref{sec:modeling}.
}
\label{fig:2002ic_template_csm_fit}
\end{figure}

We can match the inferred SN ejecta-CSM component of \citet{hamuy03b}
with the interaction model described above and reproduce the light
curve near maximum light by adding the flux from a normal SN~Ia.
Fig.~\ref{fig:2002ic_template_csm_fit} shows our model fit in
comparison with the observed light curve of SN~2002ic.  Note that our
model does not match the observed bump at 40~days after maximum.

\citet{hamuy03b} note a similar disagreement, but the data we present here 
show that this region is clearly a second bump rather than just a very slow
decline.  This discrepancy could be explained by a change in the
density of the circumstellar medium due to a change in the progenitor
mass-loss evolution at that point.  In fact, our simple fit is too bright
before the bump and too dim during the bump, implying more structure
in the underlying CSM than accounted for in our model.  
Any clumpiness in the progenitor wind would have to be on the largest
angular scale to create such a bump and would not explain the new
decline rate shown by our observations to extend out to late time .
We find that our data are consistent with a model comprising three CSM
density regions: (i) an evacuated region out to $20v_s$~days; (ii)
CSM material at a nominal density ($\rho\propto r^{-2}$) out to
$\sim100v_s$~days; and (iii) an increase in CSM density at
$\sim100v_s$~days, with a subsequent $r^{-2}$ fall-off extending
through the $800v_s$~days covered by our observations.
This model agrees well with the light curve of SN~2002ic, but, as it
involves too many parameters to result in a unique fit using only the
photometric data, we do not show it here.


Our data, particularly the pre-maximum observations,
provide key constraints on the nature of the progenitor system
of SN~2002ic.  In the context of our model, a mass-loss gradient of some form
is required by our early data points.
As a computational convenience, our model assumes that the transition
to a nominal circumstellar density is described by
$\sin(\frac{t}{20~\mathrm{days}})$.  If the mass-loss rate had been
constant until just prior to the explosion, then the $t^{-3\alpha}$
model light curve would continue to curve upward and significantly
violate our first data point at the $7\sigma$ level (as shown by the
line extended from the ejecta-CSM component in
Fig.~\ref{fig:2002ic_template_csm_fit}).  If the conversion of kinetic
energy to luminosity is immediate and roughly constant in time, as
assumed in our model, we would conclude that a low-density region must
have existed between the star and \mbox{$20$~days$\cdot v_s$}
out from the star.  For example, as a stellar system transitions from
an AGB star to a proto-planetary nebula (PPN), it changes from
emitting a denser, cooler wind, to a hotter, less dense
wind~\citep{kwok93}.  This hot wind pushes the older wind out farther
and creates a sharp density gradient and possible clumping near the
interface between the cool and hot winds~\citep{young92}.  This
overall structure is similar to that which we infer from our modeling
of SN~2002ic.  Assuming a SN ejecta speed of $v_s=30,000$~km~s$^{-1}$
and a progenitor star hot wind speed of
$v_w=100$~km~s$^{-1}$~\citep{young92,herpin02}, we conclude that the
hot wind must have begun just $\sim15$~years prior to the SN
explosion.  Alternatively, there is also the possibility that the
conversion from kinetic energy to optical luminosity is for some
reason significantly less efficient at very early times.

It is interesting to note that the observed light curve decline rate
of SN~2002ic after $40$~days past maximum light is apparently constant
during these observations.  Spectroscopic study
\citep{wang04} shows the highest observed velocity of the ejecta
to be around $11000$~km~s$^{-1}$ at day~$200$ after maximum light.
If we assume a constant expansion rate, these observations of continuing
emission through $\sim320$~days after maximum provide a lower limit of
$\sim3\times10^{16}$~cm for the spatial extent of the CSM.  Compared to a
nominal pre-explosion stellar wind speed of $10$~km~s$^{-1}$, the ejecta is
moving $\sim1000$ times more rapidly and thus has overtaken the progenitor
wind from the past $\sim800$~years.  The overall smoothness of the
late-time light curve shows the radial density profile of the CSM to
be similarly smooth and thus implies a fairly uniform mass-loss rate
between $100$--$800$ years prior to the SN explosion.

We take the lack of enhanced flux at early times and the bump after
maximum light as evidence for a gap between the SN progenitor and the
dense CSM as well as a significant further change in the mass-loss of
the progenitor system $\sim100$~years prior to the SN explosion.

\subsection{Reinterpretation of Past SNe~IIn}

These new results prompt a reexamination of supernovae previously
classified as Type~IIn, specifically SN~1988Z \citep{iauc4691,stathakis91},
SN~1997cy \citep{iauc6706,turatto00,germany00}, and
SN~1999E \citep{iauc7091,siloti00,rigon03}. These supernovae bear
striking similarities in their light curves and their late-time
spectra to SN~2002ic.  However, SN~2002ic is the only one of these
supernovae to have been observed early in its evolution.  If SN~2002ic
had been observed at the later times typical of the observations of
these Type IIn SNe, it would not have been identified as a Type~Ia.
It is interesting to note that \citet{chugai94} found from models of
light curves of SN~1988Z that the mass of the SN~1988Z supernova
ejecta is on the order of 1~$M_\sun$, which is consistent with a
SN~Ia.



\begin{figure}
\plotone{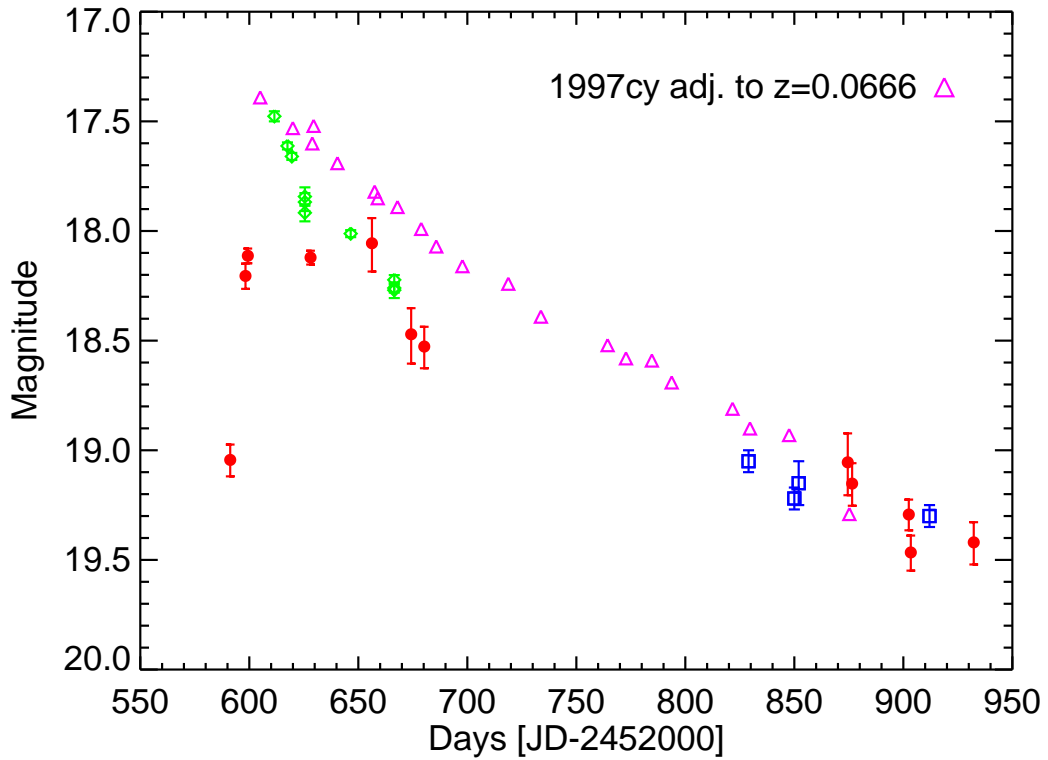}
\caption{The NEAT unfiltered and Hamuy V-band observations of SN~2002ic
compared to the K-corrected V-band observations of SN~1997cy from
\citet{germany00}.  No date of maximum or magnitude uncertainties are
available for SN~1997cy.  Here the maximum observed magnitude for SN~1997cy
has been adjusted to the redshift of 2002ic, z=0.0666~\citep{hamuy03b}, 
and the date of the first light curve point of SN~1997cy has been set to the 
date of maximum for SN~2002ic from our V-band fit.}
\label{fig:2002ic_1997cy}
\end{figure}

\begin{figure}
\plotone{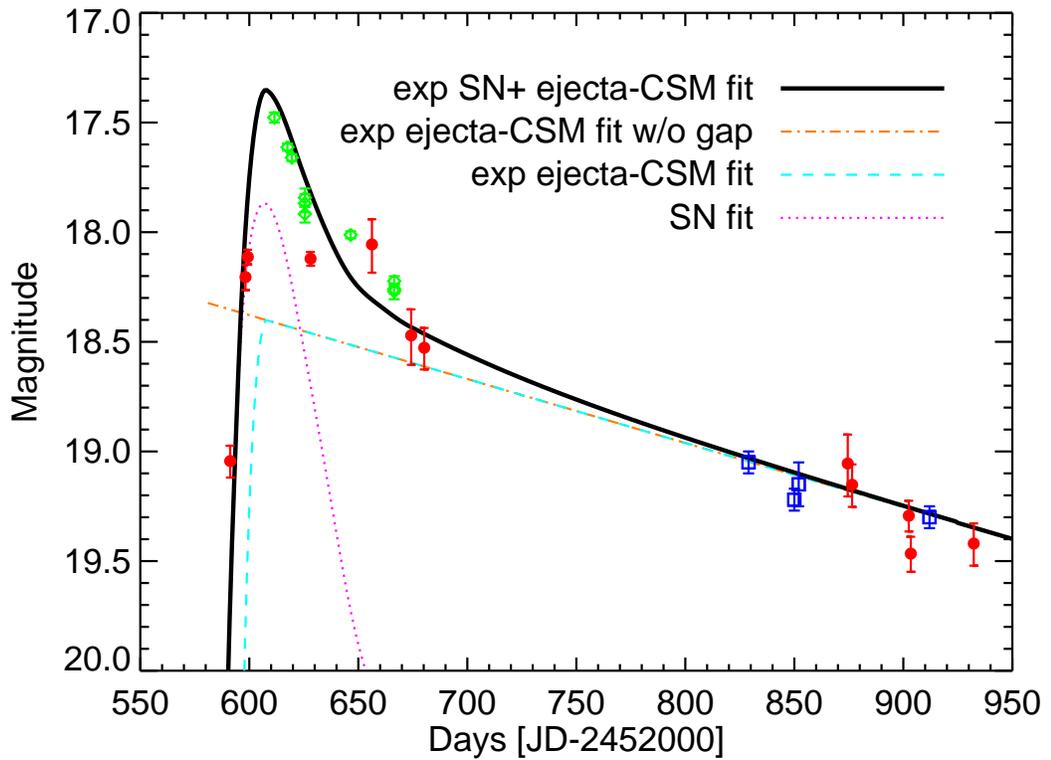}
\caption{
The observed 
photometry compared with
the SN + exponential ejecta-CSM model described in Sec.~\ref{sec:discussion}.
}
\label{fig:2002ic_template_csm_exp_fit}
\end{figure}

\begin{figure}
\plotone{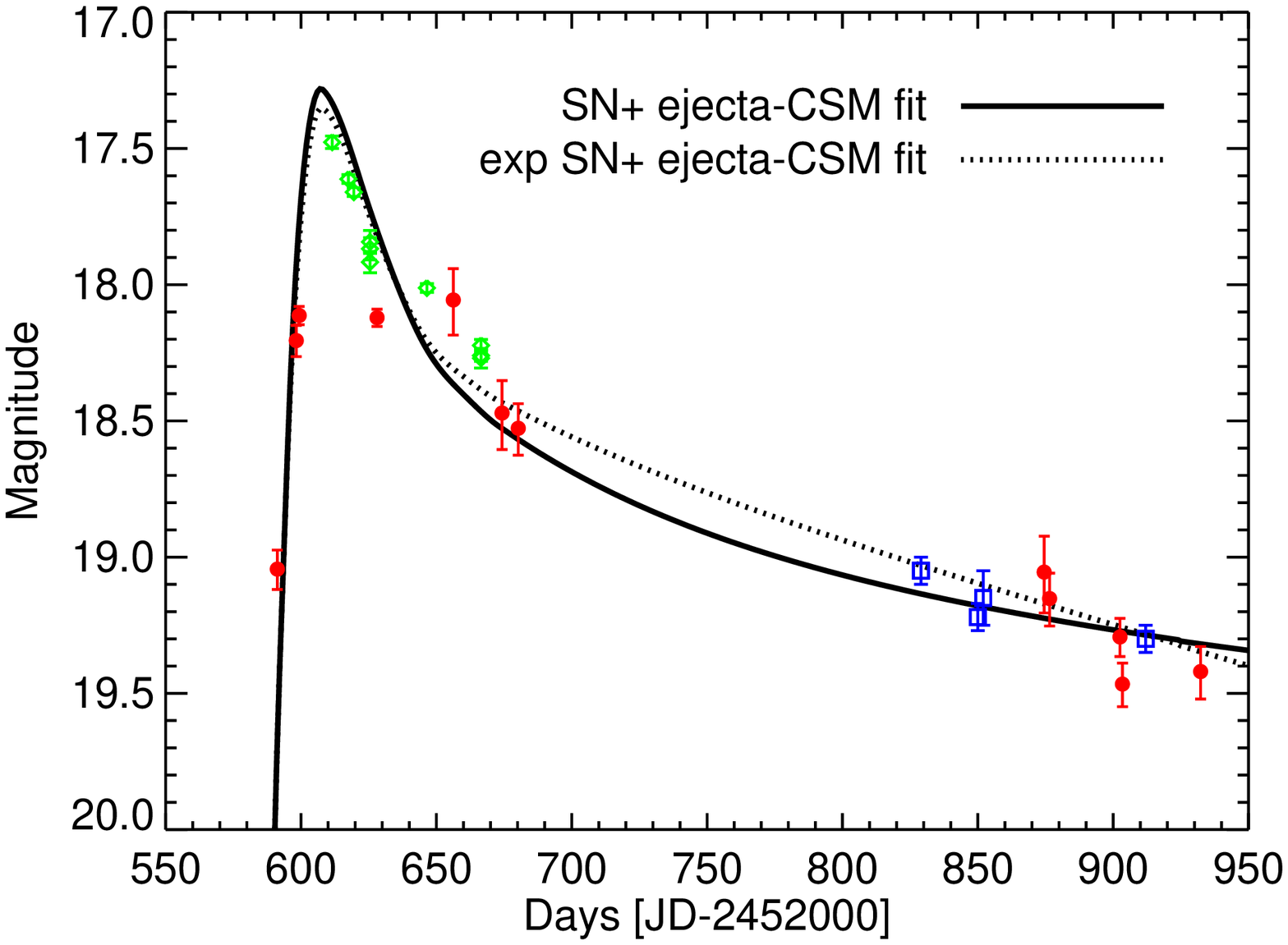}
\caption{
A comparison of fits with power-law (solid) and exponential (dotted) SN ejecta density profiles.
}
\label{fig:2002ic_template_csm_both_fit}
\end{figure}

We next explore the possibility that SN~1997cy and SN~1999E (a close
parallel to SN~1997cy) may have been systems like SN~2002ic.
\citet{hamuy03b} found that available spectra of SN~1997cy were 
very similar to post-maximum spectra of SN~2002ic.  We complement this
spectroscopic similarity with a comparison of the photometric behavior
of SN~1997cy and SN~2002ic.  As shown in Fig.~\ref{fig:2002ic_1997cy},
the late-time behavior of both SNe appear remarkably similar with both
SNe fading by $\sim2.5$~magnitudes 8 months after their respective
discoveries.  The luminosity decay rate of the ejecta-CSM interaction
is directly related to the assumed functional form for the ejecta
density and the mass-loss rate (Eq.~\ref{eq:ejecta_density}).  The
observed late-time light curves of SN~1997cy and SN~1999E clearly
follow a linear magnitude decay with time, which implies an
exponential flux vs. time dependence: $m \propto t \Rightarrow
\mathrm{flux} \propto e^{Ct}$.  If the ejecta density followed an
exponential rather than a power-law decay, the magnitude would
similarly follow a linear magnitude-time decay.
Fig.~\ref{fig:2002ic_template_csm_exp_fit} shows a fit to the light
curve of SN~2002ic using the framework of Sec.~\ref{sec:modeling} 
but using an exponential SN-ejecta density profile.
\citet{chevalier01} suggest that SNe~Ia follow exponential
ejecta profiles~\citep{dwarkadas98} while core-collapse SNe follow
power-law decays~\citep{chevalier89,matzner99}.  Thus, if SN~1997cy
and SN~1999E had been core-collapse events, they would have been
expected to show power-law declines.  Instead, their decline behavior
lends further credence to the idea that they were SN~Ia events rather
than core-collapse SNe.  Although we modeled the light curve of
SN~2002ic using a power-law ejecta profile, SN~2002ic was not observed
between 100 and 200 days after explosion, so its decay behavior during
that time is not well constrained.  Its late-time light curve is
consistent with the linear magnitude behavior of SN~1997cy.  We fit
such a profile to our data (see
Fig.~\ref{fig:2002ic_template_csm_exp_fit}~\&~\ref{fig:2002ic_template_csm_both_fit})
and arrive at an exponential fit to the flux of the form $e^{-0.003
t}$ where $t$ is measured in days.  As the solution for the SN-ejecta
interaction is not analytic, we cannot immediately relate the
exponential decay parameter to any particular property of the SN
ejecta.  Taken together in the context of the \citet{chevalier94}
model, SN~2002ic, SN~1997cy, and SN~1999E lend support to numerical
simulations of the density profiles of SNe~Ia explosions.

If we take the time of maximum for SN~1997cy to be the earliest light curve
point from \citet{germany00} and shift the magnitudes from the
redshift of SN~1997cy, $z=0.063$~\citep{germany00}, to the redshift of
SN~2002ic ($z=0.0666$), we find that the luminosity of both SNe agree
remarkably well.  This further supports that hypothesis that SN~1997cy
and SN~2002ic are related events.  However, we note that the explosion
date of SN~1997cy is uncertain and may have been 2--3 months prior to
the discovery date~\citep{germany00,turatto00}.

Fig.~\ref{fig:2002ic_template_csm_fit}~\&~\ref{fig:2002ic_template_csm_exp_fit}
show that neither a power-law nor an exponential model allow for a
significant ejecta-CSM contribution before maximum light.  In each
figure, the ``[exp] ejecta-CSM fit w/o gap'' line shows how the SN
ejecta-CSM interaction would continue if the density profile remained
the same.  Both lines significantly disagree with the earliest light
curve point.  This is consistent with our earlier conclusion that the light
curve is dominated by the SN until near maximum light.




\subsection{Relation to Proto-Planetary Nebulae?}

The massive CSM and spatial extent inferred for SN~2002ic
are surprisingly similar to certain
PPNe and the atmospheres of very late red giant
stars evolving to PPNe. Such structures are normally short-lived (less
than or on the order of $1000$~years).  The polarization seen by
\citet{wang04} suggests the presence of a disk-like structure
surrounding SN~2002ic.  Furthermore, the H$\alpha$ luminosity and mass
and size estimates suggest a clumpy medium.  Combined with the
evidence presented here for a possible transition region between a
slow and fast wind, we are left with an object very similar to
observed PPNe.  We encourage more detailed radiative hydrodynamic
modeling of SNe~Ia in a surrounding medium as our data provide
valuable constraints on this important early-time phase.

Of particular interest are bi-polar PPNe where a
WD companion emits a fast wind that shapes the AGB star wind while
simultaneously accreting~\citep{soker00} mass from the AGB star.

Typical thermonuclear supernovae are believed to have accretion 
time scales of $10^{7}$~years, yet several SNe Ia (SN~2002ic and
possibly SN~1988Z, SN~1997cy, and SN~1999E) out of several hundred
have been observed to show evidence for significant CSM.  If the
presence of a detectable CSM is taken as evidence that these SNe
exploded within a particular $\sim1000$~year-period in their
respective evolution, such as the PPN phase, this coincidence would
imply a factor of $\sim100$ enhancement 
\linebreak
($10^7$~years~$ /
1000$~years~$/100$) in the supernova explosion rate during this
period.  Thus we suggest that it is not a coincidence that the
supernova explosion is triggered during this phase.

\section{Conclusions}
\label{sec:conclusions}

The supernova 2002ic exhibits the light curve behavior and hydrogen
emission of a Type~IIn supernova after maximum but was
spectroscopically identified as a Type~Ia supernova near maximum
light.  The additional emission is attributed to a contribution from
surrounding CSM.  This emission remains quite significant
$\sim$11~months after the explosion.  The discovery of dense CSM 
surrounding a Type~Ia supernova strongly
favors the binary nature of Type~Ia progenitor systems to explain the
simultaneous presence of at least one degenerate object and
substantial material presumably ejected by a significant stellar wind.
However, it is as yet unclear whether the available data for SN~2002ic 
can prove or disprove either the single- or the double-degenerate scenario,
although the inferred resemblance to PPN systems is suggestive.
The early-time light curve data presented in this paper strongly
suggest the existence of a
$\sim15~(\frac{v_w}{10~\mathrm{km/s}})^{-1}$~year gap between the
exploding object and the surrounding CSM.
Our discovery and early- through
late-time photometric followup of SN~2002ic suggests a
reinterpretation of some Type~IIn events as Type~Ia thermonuclear
explosions shrouded by a substantial layer of circumstellar material.

\section{Acknowledgments}
\label{sec:acknowledgments}

We would like to acknowledge our fruitful collaboration with the NEAT
group at the Jet Propulsion Laboratory, operated under contract NAS7-030001 with the National Aeronautics and Space Administration, which provided the images for our supernova work.  
We thank Mario Hamuy for sharing his BVI photometry of stars in the
field of SN~2002ic. 
We are grateful to Nikolai Chugai for helpful comments.
This research used resources of the National Energy Research Scientific
Computing Center, which is supported by the Office of Science of the U.S.
Department of Energy under Contract No. DE-AC03-76SF00098. We would like
to thank them for making available the significant computational 
resources required for our supernova search.  In addition, we thank
NERSC for a generous allocation of computing time on the IBM SP
RS/6000 used by Peter Nugent and Rollin Thomas to
reconstruct the response curve for the NEAT detector on our behalf.
HPWREN is operated by the University of California, San Diego under
NSF Grant Number ANI-0087344.
This work was supported in part by the Director, Office of Science,
Office of High Energy and Nuclear Physics, of the US Department of
Energy under Contract DE-AC03-76SF000098.
WMWV was supported in part by a National Science Foundation Graduate
Research Fellowship.

We thank the anonymous referee for helpful and detailed comments
that improved the scientific clarity of this manuscript.

\clearpage



\begin{deluxetable}{lrllr}
\tablewidth{0pc}
\tablecaption{The unfiltered magnitudes for SN~2002ic as observed by the
NEAT telescopes and shown in Fig.~\ref{fig:2002ic_VLT}.  The left
brackets ([) denote limiting magnitudes at a signal-to-noise of 3.  A
systematic uncertainty of $0.4$~magnitudes in the overall calibration
is not included in the tabulated uncertainties.  (See
Sec.~\ref{sec:processing} for further discussion of our calibration). }
\ssp
\tablehead{
\colhead{JD - 2452000} & \colhead{E Mag} & \multicolumn{2}{c}{E Mag} & \colhead{Telescope} \\
\colhead{}             & \colhead{}      & \multicolumn{2}{c}{Uncertainty} & \colhead{}    
}
\startdata
    195.4999 &  [  20.52  &            &            &        Palomar 1.2-m \\
    224.2479 &  [  20.44  &            &            &        Palomar 1.2-m \\
    250.2492 &  [  21.01  &            &            &        Palomar 1.2-m \\
    577.4982 &  [  20.29  &            &            &      Haleakala 1.2-m \\
    591.2465 &     19.04  &   $-$  0.07  &   $+$  0.07  &        Palomar 1.2-m \\
    598.2519 &     18.20  &   $-$  0.06  &   $+$  0.06  &        Palomar 1.2-m \\
    599.3306 &     18.11  &   $-$  0.03  &   $+$  0.03  &        Palomar 1.2-m \\
    628.0956 &     18.12  &   $-$  0.03  &   $+$  0.03  &        Palomar 1.2-m \\
    656.2508 &     18.06  &   $-$  0.13  &   $+$  0.12  &      Haleakala 1.2-m \\
    674.2524 &     18.47  &   $-$  0.13  &   $+$  0.12  &      Haleakala 1.2-m \\
    680.2519 &     18.53  &   $-$  0.10  &   $+$  0.09  &      Haleakala 1.2-m \\
    849.5003 &  [  18.88  &            &            &      Haleakala 1.2-m \\
    853.4994 &  [  18.54  &            &            &      Haleakala 1.2-m \\
    855.4963 &  [  19.32  &            &            &      Haleakala 1.2-m \\
    858.4986 &  [  19.23  &            &            &      Haleakala 1.2-m \\
    860.4992 &  [  18.74  &            &            &      Haleakala 1.2-m \\
    864.5017 &  [  17.17  &            &            &      Haleakala 1.2-m \\
    874.4982 &     19.05  &   $-$  0.15  &   $+$  0.13  &      Haleakala 1.2-m \\
    876.4998 &     19.15  &   $-$  0.10  &   $+$  0.09  &      Haleakala 1.2-m \\
    902.4989 &     19.29  &   $-$  0.07  &   $+$  0.07  &        Palomar 1.2-m \\
    903.4138 &     19.47  &   $-$  0.08  &   $+$  0.08  &        Palomar 1.2-m \\
    932.2942 &     19.42  &   $-$  0.10  &   $+$  0.09  &        Palomar 1.2-m \\
\enddata
\label{tab:2002ic_lightcurve}
\dsp
\end{deluxetable}

\begin{deluxetable}{lll}
\tablewidth{0pc}
\tablecaption{The V-band magnitudes for SN~2002ic as synthesized from the VLT spectrophotometry of \citet{wang04} and shown in Fig.~\ref{fig:2002ic_template}.}
\ssp
\tablehead{
\colhead{JD - 2452000} & \colhead{V Mag} & \colhead{V Mag}       \\
\colhead{}             & \colhead{}      & \colhead{Uncertainty} 
}
\startdata
     829 &     19.05  &   $\pm$  0.05  \\
     850 &     19.22  &   $\pm$  0.05  \\
     852 &     19.15  &   $\pm$  0.10  \\
     912 &     19.30  &   $\pm$  0.05  \\
\enddata
\label{tab:2002ic_VLT}
\dsp
\end{deluxetable}



\clearpage

\clearpage


\part{Summary}
\label{part:summary}

\chapter{Conclusions and Future Directions}
\label{chp:conclusions}

The Nearby Supernova Factory is poised to change the way we look at
Type~Ia supernovae.  The search pipeline that was developed as the focus
of this dissertation
will allow for the discovery and subsequent follow-up of $100$~SNe~Ia
per year out to a redshift of $z=0.1$.  At the end of three years, the SNfactory will have studied
$300$~SNe~Ia in unprecedented detail and will be able to answer many of
the questions regarding unknown systematics and variation of SNe~Ia
maximum magnitudes.  In addition, the SNfactory data set will provide an ideal
resource to search for improved luminosity distance indicators from SNe~Ia.

The pilot test of the SNfactory search pipeline discovered 83 new
supernovae and found a total 99 supernovae (Chapter~\ref{chp:supernovae_found}).  The automated search
pipeline currently reduces $30,000$ images a night to less than $100$
subtractions that need to be scanned by a trained scientist.  
The SNfactory search pipeline has already proven its ability to search
on three different detector-telescope
combinations~(Chapter~\ref{chp:search_pipeline}).  Additional searching
resources could be easily integrated into the pipeline to expand or,
if necessary, replace the current data stream.
Close integration of the SNfactory discovery and follow-up
pipelines will allow for automated discovery and observation of SNe~Ia
discovered soon after explosion.
The SNIFS instrument, custom-designed and built for the SNfactory project, has been commissioned and has begun
to observe supernovae found using the SNfactory search pipeline.

The successful prototype run of the SNfactory search pipeline demonstrates that 
a large automated supernova search can be constructed to find the
100 nearby SNe~Ia per year required by the SNfactory science goals.
Improvements in star masking, detection of detector artifacts, and
image quality standards were all required to reduce the number of
subtractions that had to be examined by human scanners.  Problematic issues in
processing such large amounts of data (60~GB/night) and coordinating
large number of jobs on computing clusters were identified and
addressed.  A salient technical lesson learned was the difficulty in
simultaneously accessing a large number of images stored on
centralized disk vaults.  The storage of images, both locally and on
archival storage quickly became a limiting bottleneck.  Data
management issues such as these will impact future high-volume
supernova experiments such as LSST and SNAP.


By their nature, supernova searches are sensitive to contrast against
the host galaxy.  The SNfactory prototype search and subsequent
calculations of rates from the search reveals the significant effect a
minimum percent increase requirement has in reducing the number
supernovae found (Chapter~\ref{chp:rates}).  
As the percent increase cut is not applied in the higher-redshift work
of the SCP, this effect only applies to the SNfactory work presented here.
In retrospect, the magnitude of this effect is
apparent, but this experience offers a general lesson on the
importance of simulations in predicting and checking the performance of
large-scale, automated programs.

The subtraction code used by the SNfactory should be improved to yield fewer false
candidates.  Explorations of more sophisticated convolution kernels,
such as that of \citet{alard98}, are one promising route toward reducing the
number of non-astrophysical detections in the SNfactory subtractions.
An exploration of additional candidate quality metrics and
multi-dimensional score cuts should be pursued to reduce the number of
non-supernova candidates currently passed by the automated search
scanning software.

The automated, large-sky survey of the SNfactory search has provided a
well-defined basis for a calculation of SN~Ia rates (Chapter~\ref{chp:rates}).  The analysis
presented here is the first to measure the efficiency and control time
of a multi-epoch supernova search in a truly self-consistent manner.  A study of a
sample of the tens of thousands of square degrees searched during the
fall of 2002 and spring of 2003 yielded a 90\% confidence interval for
the SN~Ia rate of $\snrate$~SNe~Ia/yr/Mpc$^3$, or $\snrateSNu$~SNu,
from a sample of $\sneiaused$ effective SNe~Ia,
\footnote{See Sec.~\ref{sec:snrate} for a discussion of the fractional typing used here to determine the number of SNe~Ia from the incompetely typed search sample.} 
in general agreement with previous
studies of nearby SNe~Ia.  This analysis highlighted the importance of
a full understanding of the detection efficiency of the search
pipeline.  In addition, the effect of the underlying galaxy population
light on the search selection criteria 
was realized and accounted for using a comprehensive sample of
galaxies from the SDSS.  Supernova searches at higher redshifts will
require comparable galaxy catalogs to properly measure the search
sensitivities and progenitor populations.  The techniques outlined in
this dissertation should establish a new standard for calculating SN
rates from supernova searches.

In the first year of operation, the SNfactory will easily double the
number of supernovae found using its well-characterized search
pipeline.  The supernovae discovered in the full operational phase of the
SNfactory will allow for an improved measurement of supernova rates because
all promising candidates will have spectroscopic observations to
confirm or reject the new object as a supernova.  This extended sample
will provide better coverage of supernovae in low-surface-brightness
galaxies as well as statistics on supernova types as a function of
host galaxy type.

The method of calculating SN rates described in Chapter~\ref{chp:rates}
should be extended to cover the full SNfactory sample.  Additionally,
an investigation of the SN rate of non-Type Ia SNe would provide a perfect
complement to the SN~Ia rate from the SNfactory.  Core-collapse SN
rates are desirable as they are a direct tracer of the star formation
history of galaxies.  The time delay between stellar formation and
SN~Ia explosions can be measured by comparing the relative rates of
SNe~Ia and core-collapse SNe.  This delay can offer further hints
toward the nature of the progenitor systems of SNe~Ia.

In several years, when the SNfactory search data will provide QSO time
variability baselines from a week to almost 8 years, the QSO
variability investigation presented in
Appendix~\ref{apx:qso_microlensing} should be revisited in
collaboration with those currently working with QSOs as part of the
Palomar Consortium.  The expanded range of time baselines from further
years of the survey will allow for a well-constrained integration of
the observed QSO power spectrum with existing QSO variability studies.

Several unusual supernovae that challenge conventional classification
were found by the SNfactory prototype search.  One of these,
SN~2002cx, is a SN~Ia that exhibits SN~1991T-like spectral features but is
under-luminous like SN~1991bg~\citep{li03}.  Studying 
category-spanning SNe~Ia like SN~2002cx is important in understanding the true
diversity of SNe~Ia.  Perhaps the most interesting supernova found in
the initial search was SN~2002ic.  This supernova was the first SN~Ia
to exhibit hydrogen emission lines~\citep{hamuy03b}.  This emission,
together with a highly elevated light curve for at least a year after
explosion, was interpreted as evidence for circumstellar material
surrounding the progenitor system of SN~2002ic.  This material has a
density structure consistent with a post-AGB star that has just ended
its final mass-loss phase.  Several other previously observed
supernovae were proposed to be possibly related to SN~2002ic.  This
identification of a specific period in the evolution of the stellar
system of SN~2002ic offers a unique insight into the progenitors of
SNe~Ia.  Both SN~2002cx and SN~2002ic highlight the importance of
surveying the entire population of SNe~Ia to understand the
behavior and origin of these invaluable cosmological tools.

The final comprehensive SNfactory spectrophotometric data set on
300~SNe~Ia will allow for a wide range of studies of SN~Ia
phenomenology and physics.  In particular, investigations into
correlations between spectral features and luminosity distance
indicators should yield improvements in cosmological measurements
using these important standardizable candles.  This analysis should be
done at yearly intervals and eventually lead to a final analysis after
the last final reference spectrum is taken.  The wealth of data to be
accumulated by the SNfactory should occupy future graduate students
for years to come.




\ssp
\part{References}

\bibliographystyle{astroads}
\bibliography{apj-jour,thesis_refs}

\begin{thebibliography}{183}
\expandafter\ifx\csname natexlab\endcsname\relax\def\natexlab#1{#1}\fi
\expandafter\ifx\csname href\endcsname\relax
  \def\href#1#2{}\fi
\expandafter\ifx\csname urllinklabel\endcsname\relax
  \def\urllinklabel{[LINK]}\fi
\expandafter\ifx\csname adsurllinklabel\endcsname\relax
  \def\adsurllinklabel{[ADS]}\fi

\bibitem[{{Alard}(2000)}]{alard00}
{Alard}, C. 2000, \aaps, 144, 363


\bibitem[{{Alard} \& {Lupton}(1998)}]{alard98}
{Alard}, C. \& {Lupton}, R.~H. 1998, \apj, 503, 325


\bibitem[{{Aldering} {et~al.}(2002{\natexlab{a}}){Aldering}, {Adam},
  {Antilogus}, {Astier}, {Bacon}, {Bongard}, {Bonnaud}, {Copin}, {Hardin},
  {Henault}, {Howell}, {Lemonnier}, {Levy}, {Loken}, {Nugent}, {Pain},
  {Pecontal}, {Pecontal}, {Perlmutter}, {Quimby}, {Schahmaneche}, {Smadja}, \&
  {Wood-Vasey}}]{aldering02b}
{Aldering}, G., {Adam}, G., {Antilogus}, P., {Astier}, P., {Bacon}, R.,
  {Bongard}, S., {Bonnaud}, C., {Copin}, Y., {Hardin}, D., {Henault}, F.,
  {Howell}, D.~A., {Lemonnier}, J., {Levy}, J., {Loken}, S.~C., {Nugent},
  P.~E., {Pain}, R., {Pecontal}, A., {Pecontal}, E., {Perlmutter}, S.,
  {Quimby}, R.~M., {Schahmaneche}, K., {Smadja}, G., \& {Wood-Vasey}, W.~M.
  2002{\natexlab{a}}, in Survey and Other Telescope Technologies and
  Discoveries. Edited by Tyson, J. Anthony; Wolff, Sidney. Proceedings of the
  SPIE, Volume 4836, pp. 61-72 (2002)., 61--72
 \href{http://adsabs.harvard.edu/cgi-bin/nph-bib_query?bibcode=2002SPIE.4836..%
.61A&amp;db_key=AST}{\adsurllinklabel}

\bibitem[{{Aldering} {et~al.}(2002{\natexlab{b}}){Aldering}, {Akerlof},
  {Amanullah}, {Astier}, {Barrelet}, {Bebek}, {Bergstrom}, {Bercovitz},
  {Bernstein}, {Bester}, {Bonissent}, {Bower}, {Carithers}, {Commins}, {Day},
  {Deustua}, {DiGennaro}, {Ealet}, {Ellis}, {Eriksson}, {Fruchter}, {Genat},
  {Goldhaber}, {Goobar}, {Groom}, {Harris}, {Harvey}, {Heetderks}, {Holland},
  {Huterer}, {Karcher}, {Kim}, {Kolbe}, {Krieger}, {Lafever}, {Lamoreux},
  {Lampton}, {Levi}, {Levin}, {Linder}, {Loken}, {Malina}, {Massey}, {McKay},
  {McKee}, {Miquel}, {Moertsell}, {Mostek}, {Mufson}, {Musser}, {Nugent},
  {Oluseyi}, {Pain}, {Palaio}, {Pankow}, {Perlmutter}, {Pratt}, {Prieto},
  {Refregier}, {Rhodes}, {Robinson}, {Roe}, {Sholl}, {Schubnell}, {Smadja},
  {Smoot}, {Spadafora}, {Tarle}, {Tomasch}, {von der Lippe}, {Vincent},
  {Walder}, \& {Wang}}]{aldering02a}
{Aldering}, G., {Akerlof}, C.~W., {Amanullah}, R., {Astier}, P., {Barrelet},
  E., {Bebek}, C., {Bergstrom}, L., {Bercovitz}, J., {Bernstein}, G.~M.,
  {Bester}, M., {Bonissent}, A., {Bower}, C., {Carithers}, W.~C., {Commins},
  E.~D., {Day}, C., {Deustua}, S.~E., {DiGennaro}, R.~S., {Ealet}, A., {Ellis},
  R.~S., {Eriksson}, M., {Fruchter}, A., {Genat}, J., {Goldhaber}, G.,
  {Goobar}, A., {Groom}, D.~E., {Harris}, S.~E., {Harvey}, P.~R., {Heetderks},
  H.~D., {Holland}, S.~E., {Huterer}, D., {Karcher}, A., {Kim}, A.~G., {Kolbe},
  W.~F., {Krieger}, B., {Lafever}, R., {Lamoreux}, J.~C., {Lampton}, M.~L.,
  {Levi}, M.~E., {Levin}, D.~S., {Linder}, E.~V., {Loken}, S.~C., {Malina}, R.,
  {Massey}, R., {McKay}, T., {McKee}, S.~P., {Miquel}, R., {Moertsell}, E.,
  {Mostek}, N., {Mufson}, S., {Musser}, J.~A., {Nugent}, P.~E., {Oluseyi},
  H.~M., {Pain}, R., {Palaio}, N.~P., {Pankow}, D.~H., {Perlmutter}, S.,
  {Pratt}, R., {Prieto}, E., {Refregier}, A., {Rhodes}, J., {Robinson}, K.~E.,
  {Roe}, N., {Sholl}, M., {Schubnell}, M.~S., {Smadja}, G., {Smoot}, G.~F.,
  {Spadafora}, A., {Tarle}, G., {Tomasch}, A.~D., {von der Lippe}, H.,
  {Vincent}, D., {Walder}, J.-P., \& {Wang}, G. 2002{\natexlab{b}}, in Future
  Research Direction and Visions for Astronomy. Edited by Dressler, Alan M.
  Proceedings of the SPIE, Volume 4835, pp. 146-157 (2002)., 146--157
 \href{http://adsabs.harvard.edu/cgi-bin/nph-bib_query?bibcode=2002SPIE.4835..%
146A&amp;db_key=AST}{\adsurllinklabel}

\bibitem[{{Aldering} {et~al.}(2004){Aldering}, {Althouse}, {Amanullah},
  {Annis}, {Astier}, {Baltay}, {Barrelet}, {Basa}, {Bebek}, {Bergstrom},
  {Bernstein}, {Bester}, {Bigelow}, {Blandford}, {Bohlin}, {Bonissent},
  {Bower}, {Brown}, {Campbell}, {Carithers}, {Commins}, {Craig}, {Day},
  {DeJongh}, {Deustua}, {Diehl}, {Dodelson}, {Ealet}, {Ellis}, {Emmet},
  {Fouchez}, {Frieman}, {Fruchter}, {Gerdes}, {Gladney}, {Goldhaber}, {Goobar},
  {Groom}, {Heetderks}, {Hoff}, {Holland}, {Huffer}, {Hui}, {Huterer}, {Jain},
  {Jelinsky}, {Karcher}, {Kent}, {Kahn}, {Kim}, {Kolbe}, {Krieger}, {Kushner},
  {Kuznetsova}, {Lafever}, {Lamoureux}, {Lampton}, {Fevre}, {Levi}, {Limon},
  {Lin}, {Linder}, {Loken}, {Lorenzon}, {Malina}, {Marriner}, {Marshall},
  {Massey}, {Mazure}, {McKay}, {McKee}, {Miquel}, {Morgan}, {Mortsell},
  {Mostek}, {Mufson}, {Musser}, {Nugent}, {Oluseyi}, {Pain}, {Palaio},
  {Pankow}, {Peoples}, {Perlmutter}, {Prieto}, {Rabinowitz}, {Refregier},
  {Rhodes}, {Roe}, {Rusin}, {Scarpine}, {Schubnell}, {Sholl}, {Smadja},
  {Smith}, {Smoot}, {Snyder}, {Spadafora}, {Stebbins}, {Stoughton},
  {Szymkowiak}, {Tarle}, {Taylor}, {Tilquin}, {Tomasch}, {Tucker}, {Vincent},
  {von der Lippe}, {Walder}, {Wang}, \& {Wester}}]{snap04}
{Aldering}, S.~C.~G., {Althouse}, W., {Amanullah}, R., {Annis}, J., {Astier},
  P., {Baltay}, C., {Barrelet}, E., {Basa}, S., {Bebek}, C., {Bergstrom}, L.,
  {Bernstein}, G., {Bester}, M., {Bigelow}, B., {Blandford}, R., {Bohlin}, R.,
  {Bonissent}, A., {Bower}, C., {Brown}, M., {Campbell}, M., {Carithers}, W.,
  {Commins}, E., {Craig}, W., {Day}, C., {DeJongh}, F., {Deustua}, S., {Diehl},
  T., {Dodelson}, S., {Ealet}, A., {Ellis}, R., {Emmet}, W., {Fouchez}, D.,
  {Frieman}, J., {Fruchter}, A., {Gerdes}, D., {Gladney}, L., {Goldhaber}, G.,
  {Goobar}, A., {Groom}, D., {Heetderks}, H., {Hoff}, M., {Holland}, S.,
  {Huffer}, M., {Hui}, L., {Huterer}, D., {Jain}, B., {Jelinsky}, P.,
  {Karcher}, A., {Kent}, S., {Kahn}, S., {Kim}, A., {Kolbe}, W., {Krieger}, B.,
  {Kushner}, G., {Kuznetsova}, N., {Lafever}, R., {Lamoureux}, J., {Lampton},
  M., {Fevre}, O.~L., {Levi}, M., {Limon}, P., {Lin}, H., {Linder}, E.,
  {Loken}, S., {Lorenzon}, W., {Malina}, R., {Marriner}, J., {Marshall}, P.,
  {Massey}, R., {Mazure}, A., {McKay}, T., {McKee}, S., {Miquel}, R., {Morgan},
  N., {Mortsell}, E., {Mostek}, N., {Mufson}, S., {Musser}, J., {Nugent}, P.,
  {Oluseyi}, H., {Pain}, R., {Palaio}, N., {Pankow}, D., {Peoples}, J.,
  {Perlmutter}, S., {Prieto}, E., {Rabinowitz}, D., {Refregier}, A., {Rhodes},
  J., {Roe}, N., {Rusin}, D., {Scarpine}, V., {Schubnell}, M., {Sholl}, M.,
  {Smadja}, G., {Smith}, R.~M., {Smoot}, G., {Snyder}, J., {Spadafora}, A.,
  {Stebbins}, A., {Stoughton}, C., {Szymkowiak}, A., {Tarle}, G., {Taylor}, K.,
  {Tilquin}, A., {Tomasch}, A., {Tucker}, D., {Vincent}, D., {von der Lippe},
  H., {Walder}, J., {Wang}, G., \& {Wester}, W. 2004, ArXiv Astrophysics
  e-prints


\bibitem[{{Allen} {et~al.}(2002){Allen}, {Schmidt}, \& {Fabian}}]{allen02}
{Allen}, S.~W., {Schmidt}, R.~W., \& {Fabian}, A.~C. 2002, \mnras, 334, L11
 \href{http://adsabs.harvard.edu/cgi-bin/nph-bib_query?bibcode=2002MNRAS.334L.%
.11A&amp;db_key=AST}{\adsurllinklabel}

\bibitem[{{Antonucci}(1993)}]{antonucci93}
{Antonucci}, R. 1993, \araa, 31, 473
 \href{http://adsabs.harvard.edu/cgi-bin/nph-bib_query?bibcode=1993ARA%26A..31%
..473A&db_key=AST}{\adsurllinklabel}

\bibitem[{{Bahcall} {et~al.}(2003){Bahcall}, {Dong}, {Bode}, {Kim}, {Annis},
  {McKay}, {Hansen}, {Schroeder}, {Gunn}, {Ostriker}, {Postman}, {Nichol},
  {Miller}, {Goto}, {Brinkmann}, {Knapp}, {Lamb}, {Schneider}, {Vogeley}, \&
  {York}}]{bahcall03}
{Bahcall}, N.~A., {Dong}, F., {Bode}, P., {Kim}, R., {Annis}, J., {McKay},
  T.~A., {Hansen}, S., {Schroeder}, J., {Gunn}, J., {Ostriker}, J.~P.,
  {Postman}, M., {Nichol}, R.~C., {Miller}, C., {Goto}, T., {Brinkmann}, J.,
  {Knapp}, G.~R., {Lamb}, D.~O., {Schneider}, D.~P., {Vogeley}, M.~S., \&
  {York}, D.~G. 2003, \apj, 585, 182
 \href{http://adsabs.harvard.edu/cgi-bin/nph-bib_query?bibcode=2003ApJ...585..%
182B&amp;db_key=AST}{\adsurllinklabel}

\bibitem[{{Bennett} {et~al.}(2003){Bennett}, {Halpern}, {Hinshaw}, {Jarosik},
  {Kogut}, {Limon}, {Meyer}, {Page}, {Spergel}, {Tucker}, {Wollack}, {Wright},
  {Barnes}, {Greason}, {Hill}, {Komatsu}, {Nolta}, {Odegard}, {Peiris},
  {Verde}, \& {Weiland}}]{bennett03}
{Bennett}, C.~L., {Halpern}, M., {Hinshaw}, G., {Jarosik}, N., {Kogut}, A.,
  {Limon}, M., {Meyer}, S.~S., {Page}, L., {Spergel}, D.~N., {Tucker}, G.~S.,
  {Wollack}, E., {Wright}, E.~L., {Barnes}, C., {Greason}, M.~R., {Hill},
  R.~S., {Komatsu}, E., {Nolta}, M.~R., {Odegard}, N., {Peiris}, H.~V.,
  {Verde}, L., \& {Weiland}, J.~L. 2003, \apjs, 148, 1
 \href{http://adsabs.harvard.edu/cgi-bin/nph-bib_query?bibcode=2003ApJS..148..%
..1B&amp;db_key=AST}{\adsurllinklabel}

\bibitem[{{Bes Optics}(2003)}]{schott03}
{Bes Optics}. 2003, Schott Optical Filters, \\ 
  http://www.besoptics.com/html/body\_schott\_rg610\_filter\_glass.html
 \href{http://www.besoptics.com/html/body_schott_rg610_filter_glass.html}{\urllinklabel}

\bibitem[{Blanc(2002)}]{blanc02}
Blanc, G. 2002, Master's thesis, Saclay


\bibitem[{{Blanc} {et~al.}(2004){Blanc}, {Afonso}, {Alard}, {Albert},
  {Aldering}, {Amadon}, {Andersen}, {Ansari}, {Aubourg}, {Balland}, {Bareyre},
  {Beaulieu}, {Charlot}, {Conley}, {Coutures}, {Dahlen}, {Derue}, {Fan},
  {Ferlet}, {Folatelli}, {Fouque}, {Garavini}, {Glicenstein}, {Goldman},
  {Goobar}, {Gould}, {Graff}, {Gros}, {Haissinski}, {Hamadache}, {Hardin},
  {Hook}, {deKat}, {Kent}, {Kim}, {Lasserre}, {LeGuillou}, {Lesquoy}, {Loup},
  {Magneville}, {Marquette}, {Maurice}, {Maury}, {Milsztajn}, {Moniez},
  {Mouchet}, {Newberg}, {Nobili}, {Palanque-Delabrouille}, {Perdereau},
  {Prevot}, {Rahal}, {Regnault}, {Rich}, {Ruiz-Lapuente}, {Spiro}, {Tisserand},
  {Vidal-Madjar}, {Vigroux}, {Walton}, \& {Zylberajch}}]{blanc04}
{Blanc}, G., {Afonso}, C., {Alard}, C., {Albert}, J.~N., {Aldering}, G.,
  {Amadon}, A., {Andersen}, J., {Ansari}, R., {Aubourg}, E., {Balland}, C.,
  {Bareyre}, P., {Beaulieu}, J.~P., {Charlot}, X., {Conley}, A., {Coutures},
  C., {Dahlen}, T., {Derue}, F., {Fan}, X., {Ferlet}, R., {Folatelli}, G.,
  {Fouque}, P., {Garavini}, G., {Glicenstein}, J.~F., {Goldman}, B., {Goobar},
  A., {Gould}, A., {Graff}, D., {Gros}, M., {Haissinski}, J., {Hamadache}, C.,
  {Hardin}, D., {Hook}, I.~M., {deKat}, J., {Kent}, S., {Kim}, A., {Lasserre},
  T., {LeGuillou}, L., {Lesquoy}, E., {Loup}, C., {Magneville}, C.,
  {Marquette}, J.~B., {Maurice}, E., {Maury}, A., {Milsztajn}, A., {Moniez},
  M., {Mouchet}, M., {Newberg}, H., {Nobili}, S., {Palanque-Delabrouille}, N.,
  {Perdereau}, O., {Prevot}, L., {Rahal}, Y.~R., {Regnault}, N., {Rich}, J.,
  {Ruiz-Lapuente}, P., {Spiro}, M., {Tisserand}, P., {Vidal-Madjar}, A.,
  {Vigroux}, L., {Walton}, N.~A., \& {Zylberajch}, S. 2004, ArXiv Astrophysics
  e-prints


\bibitem[{{Blanton} {et~al.}(2003{\natexlab{a}}){Blanton}, {Hogg}, {Bahcall},
  {Baldry}, {Brinkmann}, {Csabai}, {Eisenstein}, {Fukugita}, {Gunn}, {Ivezi{\'
  c}}, {Lamb}, {Lupton}, {Loveday}, {Munn}, {Nichol}, {Okamura}, {Schlegel},
  {Shimasaku}, {Strauss}, {Vogeley}, \& {Weinberg}}]{blanton03a}
{Blanton}, M.~R., {Hogg}, D.~W., {Bahcall}, N.~A., {Baldry}, I.~K.,
  {Brinkmann}, J., {Csabai}, I., {Eisenstein}, D., {Fukugita}, M., {Gunn},
  J.~E., {Ivezi{\' c}}, {\v Z}., {Lamb}, D.~Q., {Lupton}, R.~H., {Loveday}, J.,
  {Munn}, J.~A., {Nichol}, R.~C., {Okamura}, S., {Schlegel}, D.~J.,
  {Shimasaku}, K., {Strauss}, M.~A., {Vogeley}, M.~S., \& {Weinberg}, D.~H.
  2003{\natexlab{a}}, \apj, 594, 186
 \href{http://adsabs.harvard.edu/cgi-bin/nph-bib_query?bibcode=2003ApJ...594..%
186B&db_key=AST}{\adsurllinklabel}

\bibitem[{{Blanton} {et~al.}(2003{\natexlab{b}}){Blanton}, {Hogg}, {Bahcall},
  {Brinkmann}, {Britton}, {Connolly}, {Csabai}, {Fukugita}, {Loveday},
  {Meiksin}, {Munn}, {Nichol}, {Okamura}, {Quinn}, {Schneider}, {Shimasaku},
  {Strauss}, {Tegmark}, {Vogeley}, \& {Weinberg}}]{blanton03b}
{Blanton}, M.~R., {Hogg}, D.~W., {Bahcall}, N.~A., {Brinkmann}, J., {Britton},
  M., {Connolly}, A.~J., {Csabai}, I., {Fukugita}, M., {Loveday}, J.,
  {Meiksin}, A., {Munn}, J.~A., {Nichol}, R.~C., {Okamura}, S., {Quinn}, T.,
  {Schneider}, D.~P., {Shimasaku}, K., {Strauss}, M.~A., {Tegmark}, M.,
  {Vogeley}, M.~S., \& {Weinberg}, D.~H. 2003{\natexlab{b}}, \apj, 592, 819
 \href{http://adsabs.harvard.edu/cgi-bin/nph-bib_query?bibcode=2003ApJ...592..%
819B&db_key=AST}{\adsurllinklabel}

\bibitem[{{Branch} {et~al.}(1995){Branch}, {Livio}, {Yungelson}, {Boffi}, \&
  {Baron}}]{branch95}
{Branch}, D., {Livio}, M., {Yungelson}, L.~R., {Boffi}, F.~R., \& {Baron}, E.
  1995, \pasp, 107, 1019
 \href{http://adsabs.harvard.edu/cgi-bin/nph-bib_query?bibcode=1995PASP..107.1%
019B&db_key=AST}{\adsurllinklabel}

\bibitem[{{Branch} \& {Miller}(1993)}]{branch93}
{Branch}, D. \& {Miller}, D.~L. 1993, \apjl, 405, L5


\bibitem[{{Branch} \& {Tammann}(1992)}]{branch92}
{Branch}, D. \& {Tammann}, G.~A. 1992, \araa, 30, 359


\bibitem[{Braun(2003)}]{hpwren}
Braun, H.-W. 2003, {High Performance Wireless Research and Education Network},
  {\em http://hpwren.ucsd.edu/}
 \href{http://hpwren.ucsd.edu/}{\urllinklabel}

\bibitem[{{Burbidge}(1967)}]{burbidge67}
{Burbidge}, E.~M. 1967, \araa, 5, 399
 \href{http://adsabs.harvard.edu/cgi-bin/nph-bib_query?bibcode=1967ARA%26A...5%
..399B&db_key=AST}{\adsurllinklabel}

\bibitem[{{Cappellaro} {et~al.}(1999{\natexlab{a}}){Cappellaro}, {Evans}, \&
  {Turatto}}]{cappellaro99}
{Cappellaro}, E., {Evans}, R., \& {Turatto}, M. 1999{\natexlab{a}}, \aap, 351,
  459
 \href{http://adsabs.harvard.edu/cgi-bin/nph-bib_query?bibcode=1999A%26A...351%
..459C&db_key=AST}{\adsurllinklabel}

\bibitem[{{Cappellaro} \& {Turatto}(1988)}]{cappellaro88}
{Cappellaro}, E. \& {Turatto}, M. 1988, \aap, 190, 10
 \href{http://adsabs.harvard.edu/cgi-bin/nph-bib_query?bibcode=1988A%26A...190%
...10C&db_key=AST}{\adsurllinklabel}

\bibitem[{{Cappellaro} {et~al.}(1999{\natexlab{b}}){Cappellaro}, {Turatto}, \&
  {Mazzali}}]{iauc7091}
{Cappellaro}, E., {Turatto}, M., \& {Mazzali}, P. 1999{\natexlab{b}}, \iaucirc,
  7091, 1
 \href{http://adsabs.harvard.edu/cgi-bin/nph-bib_query?bibcode=1999IAUC.7091..%
..1C&db_key=AST}{\adsurllinklabel}

\bibitem[{{Cappellaro} {et~al.}(1997){Cappellaro}, {Turatto}, {Tsvetkov},
  {Bartunov}, {Pollas}, {Evans}, \& {Hamuy}}]{cappellaro97}
{Cappellaro}, E., {Turatto}, M., {Tsvetkov}, D.~Y., {Bartunov}, O.~S.,
  {Pollas}, C., {Evans}, R., \& {Hamuy}, M. 1997, \aap, 322, 431
 \href{http://adsabs.harvard.edu/cgi-bin/nph-bib_query?bibcode=1997A%26A...322%
..431C&db_key=AST}{\adsurllinklabel}

\bibitem[{{Cardelli} {et~al.}(1989){Cardelli}, {Clayton}, \&
  {Mathis}}]{cardelli89}
{Cardelli}, J.~A., {Clayton}, G.~C., \& {Mathis}, J.~S. 1989, \apj, 345, 245


\bibitem[{{Chevalier} \& {Fransson}(1994)}]{chevalier94}
{Chevalier}, R.~A. \& {Fransson}, C. 1994, \apj, 420, 268
 \href{http://adsabs.harvard.edu/cgi-bin/nph-bib_query?bibcode=1994ApJ...420..%
268C&db_key=AST}{\adsurllinklabel}

\bibitem[{{Chevalier} \& {Fransson}(2001)}]{chevalier01}
---. 2001, ArXiv Astrophysics e-prints, 0110060
 \href{http://www.arxiv.org/abs/astro-ph/0110060}{\urllinklabel}

\bibitem[{{Chevalier} \& {Soker}(1989)}]{chevalier89}
{Chevalier}, R.~A. \& {Soker}, N. 1989, \apj, 341, 867
 \href{http://adsabs.harvard.edu/cgi-bin/nph-bib_query?bibcode=1989ApJ...341..%
867C&amp;db_key=AST}{\adsurllinklabel}

\bibitem[{{Chugai} {et~al.}(2002){Chugai}, {Blinnikov}, {Fassia}, {Lundqvist},
  {Meikle}, \& {Sorokina}}]{chugai02}
{Chugai}, N.~N., {Blinnikov}, S.~I., {Fassia}, A., {Lundqvist}, P., {Meikle},
  W.~P.~S., \& {Sorokina}, E.~I. 2002, \mnras, 330, 473
 \href{http://adsabs.harvard.edu/cgi-bin/nph-bib_query?bibcode=2002MNRAS.330..%
473C&amp;db_key=AST}{\adsurllinklabel}

\bibitem[{{Chugai} \& {Danziger}(1994)}]{chugai94}
{Chugai}, N.~N. \& {Danziger}, I.~J. 1994, \mnras, 268, 173
 \href{http://adsabs.harvard.edu/cgi-bin/nph-bib_query?bibcode=1994MNRAS.268..%
173C&amp;db_key=AST}{\adsurllinklabel}

\bibitem[{{Chugai} \& {Yungelson}(2004)}]{chugai04}
{Chugai}, N.~N. \& {Yungelson}, L.~R. 2004, Astron. Letters, 30, 65


\bibitem[{{Commins}(2004)}]{commins04}
{Commins}, E.~D. 2004, New Astronomy Review, 48, 567


\bibitem[{{Cumming} {et~al.}(1996){Cumming}, {Lundqvist}, {Smith}, {Pettini},
  \& {King}}]{cumming96}
{Cumming}, R.~J., {Lundqvist}, P., {Smith}, L.~J., {Pettini}, M., \& {King},
  D.~L. 1996, \mnras, 283, 1355
 \href{http://adsabs.harvard.edu/cgi-bin/nph-bib_query?bibcode=1996MNRAS.283.1%
355C&db_key=AST}{\adsurllinklabel}

\bibitem[{{Dahlen} {et~al.}(2004){Dahlen}, {Strolger}, {Riess}, {Mobasher},
  {Chary}, {Conselice}, {Ferguson}, {Fruchter}, {Giavalisco}, {Livio}, {Madau},
  {Panagia}, \& {Tonry}}]{dahlen04}
{Dahlen}, T., {Strolger}, L., {Riess}, A.~G., {Mobasher}, B., {Chary}, R.,
  {Conselice}, C.~J., {Ferguson}, H.~C., {Fruchter}, A.~S., {Giavalisco}, M.,
  {Livio}, M., {Madau}, P., {Panagia}, N., \& {Tonry}, J.~L. 2004, ArXiv
  Astrophysics e-prints


\bibitem[{{Deng} {et~al.}(2004){Deng}, {Kawabata}, {Ohyama}, {Nomoto},
  {Mazzali}, {Wang}, {Jeffery}, {Iye}, {Tomita}, \& {Yoshii}}]{deng04}
{Deng}, J., {Kawabata}, K.~S., {Ohyama}, Y., {Nomoto}, K., {Mazzali}, P.~A.,
  {Wang}, L., {Jeffery}, D.~J., {Iye}, M., {Tomita}, H., \& {Yoshii}, Y. 2004,
  \apjl, 605, L37


\bibitem[{{Dwarkadas} \& {Chevalier}(1998)}]{dwarkadas98}
{Dwarkadas}, V.~V. \& {Chevalier}, R.~A. 1998, \apj, 497, 807
 \href{http://adsabs.harvard.edu/cgi-bin/nph-bib_query?bibcode=1998ApJ...497..%
807D&amp;db_key=AST}{\adsurllinklabel}

\bibitem[{{Evans}(1997)}]{evans97}
{Evans}, R. 1997, Publications of the Astronomical Society of Australia, 14,
  204


\bibitem[{{Filippenko}(1997)}]{filippenko97}
{Filippenko}, A.~V. 1997, \araa, 35, 309
 \href{http://adsabs.harvard.edu/cgi-bin/nph-bib_query?bibcode=1997ARA%26A..35%
..309F&amp;db_key=AST}{\adsurllinklabel}

\bibitem[{{Filippenko} {et~al.}(1992){Filippenko}, {Richmond}, {Branch},
  {Gaskell}, {Herbst}, {Ford}, {Treffers}, {Matheson}, {Ho}, {Dey}, {Sargent},
  {Small}, \& {van Breugel}}]{filippenko92}
{Filippenko}, A.~V., {Richmond}, M.~W., {Branch}, D., {Gaskell}, M., {Herbst},
  W., {Ford}, C.~H., {Treffers}, R.~R., {Matheson}, T., {Ho}, L.~C., {Dey}, A.,
  {Sargent}, W.~L.~W., {Small}, T.~A., \& {van Breugel}, W.~J.~M. 1992, \aj,
  104, 1543
 \href{http://adsabs.harvard.edu/cgi-bin/nph-bib_query?bibcode=1992AJ....104.1%
543F&db_key=AST}{\adsurllinklabel}

\bibitem[{{Freedman} {et~al.}(2001){Freedman}, {Madore}, {Gibson}, {Ferrarese},
  {Kelson}, {Sakai}, {Mould}, {Kennicutt}, {Ford}, {Graham}, {Huchra},
  {Hughes}, {Illingworth}, {Macri}, \& {Stetson}}]{freedman01}
{Freedman}, W.~L., {Madore}, B.~F., {Gibson}, B.~K., {Ferrarese}, L., {Kelson},
  D.~D., {Sakai}, S., {Mould}, J.~R., {Kennicutt}, R.~C., {Ford}, H.~C.,
  {Graham}, J.~A., {Huchra}, J.~P., {Hughes}, S.~M.~G., {Illingworth}, G.~D.,
  {Macri}, L.~M., \& {Stetson}, P.~B. 2001, \apj, 553, 47


\bibitem[{{Frei} \& {Gunn}(1994)}]{frei94}
{Frei}, Z. \& {Gunn}, J.~E. 1994, \aj, 108, 1476
 \href{http://adsabs.harvard.edu/cgi-bin/nph-bib_query?bibcode=1994AJ....108.1%
476F&db_key=AST}{\adsurllinklabel}

\bibitem[{{Gal-Yam} \& {Maoz}(2004)}]{gal-yam04}
{Gal-Yam}, A. \& {Maoz}, D. 2004, \mnras, 347, 942


\bibitem[{{Garnavich} {et~al.}(2002){Garnavich}, {Holland}, {Schmidt},
  {Krisciunas}, {Smith}, {Suntzeff}, {Becker}, {Miceli}, {Miknaitis}, {Rest},
  {Stubbs}, {Filippenko}, {Jha}, {Li}, {Challis}, {Kirshner}, {Matheson},
  {Barris}, {Tonry}, {Riess}, {Leibundgut}, {Sollerman}, {Spyromilio},
  {Clocchiatti}, {Pompea}, \& {High-Z Supernova Search Team}}]{garnavich02}
{Garnavich}, P.~M., {Holland}, S.~T., {Schmidt}, B.~P., {Krisciunas}, K.,
  {Smith}, R.~C., {Suntzeff}, N.~B., {Becker}, A., {Miceli}, A., {Miknaitis},
  G., {Rest}, A., {Stubbs}, C., {Filippenko}, A.~V., {Jha}, S., {Li}, W.,
  {Challis}, P., {Kirshner}, R.~P., {Matheson}, T., {Barris}, B., {Tonry},
  J.~L., {Riess}, A.~G., {Leibundgut}, B., {Sollerman}, J., {Spyromilio}, J.,
  {Clocchiatti}, A., {Pompea}, S., \& {High-Z Supernova Search Team}. 2002,
  \baas, 34, 1233
 \href{http://adsabs.harvard.edu/cgi-bin/nph-bib_query?bibcode=2002AAS...201.7%
809G&amp;db_key=AST}{\adsurllinklabel}

\bibitem[{{Garnavich} {et~al.}(1998){Garnavich}, {Kirshner}, {Challis},
  {Tonry}, {Gilliland}, {Smith}, {Clocchiatti}, {Diercks}, {Filippenko},
  {Hamuy}, {Hogan}, {Leibundgut}, {Phillips}, {Reiss}, {Riess}, {Schmidt},
  {Schommer}, {Spyromilio}, {Stubbs}, {Suntzeff}, \& {Wells}}]{garnavich98a}
{Garnavich}, P.~M., {Kirshner}, R.~P., {Challis}, P., {Tonry}, J., {Gilliland},
  R.~L., {Smith}, R.~C., {Clocchiatti}, A., {Diercks}, A., {Filippenko}, A.~V.,
  {Hamuy}, M., {Hogan}, C.~J., {Leibundgut}, B., {Phillips}, M.~M., {Reiss},
  D., {Riess}, A.~G., {Schmidt}, B.~P., {Schommer}, R.~A., {Spyromilio}, J.,
  {Stubbs}, C., {Suntzeff}, N.~B., \& {Wells}, L. 1998, \apjl, 493, L53+


\bibitem[{{Gerardy} {et~al.}(2004){Gerardy}, {H{\"o}flich}, {Quimby}, {Wang},
  {Wheeler}, {Fesen}, {Marion}, {Nomoto}, \& {Schaefer}}]{gerardy04}
{Gerardy}, C.~L., {H{\"o}flich}, P., {Quimby}, R., {Wang}, L., {Wheeler},
  J.~C., {Fesen}, R.~A., {Marion}, G.~H., {Nomoto}, K., \& {Schaefer}, B.~E.
  2004, \apj, {\em in press}
 \href{http://www.arxiv.org/abs/astro-ph/0309639}{\urllinklabel}

\bibitem[{{Germany} {et~al.}(2000){Germany}, {Reiss}, {Sadler}, {Schmidt}, \&
  {Stubbs}}]{germany00}
{Germany}, L.~M., {Reiss}, D.~J., {Sadler}, E.~M., {Schmidt}, B.~P., \&
  {Stubbs}, C.~W. 2000, \apj, 533, 320
 \href{http://adsabs.harvard.edu/cgi-bin/nph-bib_query?bibcode=2000ApJ...533..%
320G&db_key=AST}{\adsurllinklabel}

\bibitem[{{Goldhaber} {et~al.}(2001){Goldhaber}, {Groom}, {Kim}, {Aldering},
  {Astier}, {Conley}, {Deustua}, {Ellis}, {Fabbro}, {Fruchter}, {Goobar},
  {Hook}, {Irwin}, {Kim}, {Knop}, {Lidman}, {McMahon}, {Nugent}, {Pain},
  {Panagia}, {Pennypacker}, {Perlmutter}, {Ruiz-Lapuente}, {Schaefer},
  {Walton}, \& {York}}]{goldhaber01}
{Goldhaber}, G., {Groom}, D.~E., {Kim}, A., {Aldering}, G., {Astier}, P.,
  {Conley}, A., {Deustua}, S.~E., {Ellis}, R., {Fabbro}, S., {Fruchter}, A.~S.,
  {Goobar}, A., {Hook}, I., {Irwin}, M., {Kim}, M., {Knop}, R.~A., {Lidman},
  C., {McMahon}, R., {Nugent}, P.~E., {Pain}, R., {Panagia}, N., {Pennypacker},
  C.~R., {Perlmutter}, S., {Ruiz-Lapuente}, P., {Schaefer}, B., {Walton},
  N.~A., \& {York}, T. 2001, \apj, 558, 359
 \href{http://adsabs.harvard.edu/cgi-bin/nph-bib_query?bibcode=2001ApJ...558..%
359G&db_key=AST}{\adsurllinklabel}

\bibitem[{{Gutierrez} {et~al.}(1996){Gutierrez}, {Garcia-Berro}, {Iben},
  {Isern}, {Labay}, \& {Canal}}]{gutierrez95}
{Gutierrez}, J., {Garcia-Berro}, E., {Iben}, I.~J., {Isern}, J., {Labay}, J.,
  \& {Canal}, R. 1996, \apj, 459, 701
 \href{http://adsabs.harvard.edu/cgi-bin/nph-bib_query?bibcode=1996ApJ...459..%
701G&db_key=AST}{\adsurllinklabel}

\bibitem[{{Hamuy} {et~al.}(2002){Hamuy}, {Maza}, \& {Phillips}}]{iauc8028}
{Hamuy}, M., {Maza}, J., \& {Phillips}, M. 2002, \iaucirc, 8028, 2
 \href{http://adsabs.harvard.edu/cgi-bin/nph-bib_query?bibcode=2002IAUC.8028..%
..2H&db_key=AST}{\adsurllinklabel}

\bibitem[{{Hamuy} {et~al.}(2003){Hamuy}, {Phillips}, {Suntzeff}, \&
  {Maza}}]{iauc8151}
{Hamuy}, M., {Phillips}, M., {Suntzeff}, N., \& {Maza}, J. 2003, \iaucirc,
  8151, 2
 \href{http://adsabs.harvard.edu/cgi-bin/nph-bib_query?bibcode=2003IAUC.8151..%
..2H&db_key=AST}{\adsurllinklabel}

\bibitem[{{Hamuy} {et~al.}(1995){Hamuy}, {Phillips}, {Maza}, {Suntzeff},
  {Schommer}, \& {Aviles}}]{hamuy95}
{Hamuy}, M., {Phillips}, M.~M., {Maza}, J., {Suntzeff}, N.~B., {Schommer},
  R.~A., \& {Aviles}, R. 1995, \aj, 109, 1
 \href{http://adsabs.harvard.edu/cgi-bin/nph-bib_query?bibcode=1995AJ....109..%
..1H&db_key=AST}{\adsurllinklabel}

\bibitem[{Hamuy {et~al.}(2003)Hamuy, Phillips, Suntzeff, Maza, {Gonz\'alez},
  Roth, Krisciunas, Morrell, Green, Persson, \& McCarthy}]{hamuy03b}
Hamuy, M., Phillips, M.~M., Suntzeff, N.~B., Maza, J., {Gonz\'alez}, L.~E.,
  Roth, M., Krisciunas, K., Morrell, N., Green, E.~M., Persson, S.~E., \&
  McCarthy, P.~J. 2003, Nature, 424, 651
 \href{http://www.ociw.edu/~mhamuy/SNe/}{\urllinklabel}

\bibitem[{{Hamuy} {et~al.}(1996){Hamuy}, {Phillips}, {Suntzeff}, {Schommer},
  {Maza}, \& {Aviles}}]{hamuy96}
{Hamuy}, M., {Phillips}, M.~M., {Suntzeff}, N.~B., {Schommer}, R.~A., {Maza},
  J., \& {Aviles}, R. 1996, \aj, 112, 2391
 \href{http://adsabs.harvard.edu/cgi-bin/nph-bib_query?bibcode=1996AJ....112.2%
391H&db_key=AST}{\adsurllinklabel}

\bibitem[{{Hamuy} {et~al.}(1993){Hamuy}, {Phillips}, {Wells}, \&
  {Maza}}]{hamuy93}
{Hamuy}, M., {Phillips}, M.~M., {Wells}, L.~A., \& {Maza}, J. 1993, \pasp, 105,
  787


\bibitem[{{Hardin} {et~al.}(2000){Hardin}, {Afonso}, {Alard}, {Albert},
  {Amadon}, {Andersen}, {Ansari}, {Aubourg}, {Bareyre}, {Bauer}, {Beaulieu},
  {Blanc}, {Bouquet}, {Char}, {Charlot}, {Couchot}, {Coutures}, {Derue},
  {Ferlet}, {Glicenstein}, {Goldman}, {Gould}, {Graff}, {Gros}, {Haissinski},
  {Hamilton}, {de Kat}, {Kim}, {Lasserre}, {Lesquoy}, {Loup}, {Magneville},
  {Mansoux}, {Marquette}, {Maurice}, {Milsztajn}, {Moniez},
  {Palanque-Delabrouille}, {Perdereau}, {Pr{\' e}vot}, {Regnault}, {Rich},
  {Spiro}, {Vidal-Madjar}, {Vigroux}, {Zylberajch}, \& {The EROS
  Collaboration}}]{hardin00}
{Hardin}, D., {Afonso}, C., {Alard}, C., {Albert}, J.~N., {Amadon}, A.,
  {Andersen}, J., {Ansari}, R., {Aubourg}, {\' E}., {Bareyre}, P., {Bauer}, F.,
  {Beaulieu}, J.~P., {Blanc}, G., {Bouquet}, A., {Char}, S., {Charlot}, X.,
  {Couchot}, F., {Coutures}, C., {Derue}, F., {Ferlet}, R., {Glicenstein},
  J.~F., {Goldman}, B., {Gould}, A., {Graff}, D., {Gros}, M., {Haissinski}, J.,
  {Hamilton}, J.~C., {de Kat}, J., {Kim}, A., {Lasserre}, T., {Lesquoy}, {\'
  E}., {Loup}, C., {Magneville}, C., {Mansoux}, B., {Marquette}, J.~B.,
  {Maurice}, {\' E}., {Milsztajn}, A., {Moniez}, M., {Palanque-Delabrouille},
  N., {Perdereau}, O., {Pr{\' e}vot}, L., {Regnault}, N., {Rich}, J., {Spiro},
  M., {Vidal-Madjar}, A., {Vigroux}, L., {Zylberajch}, S., \& {The EROS
  Collaboration}. 2000, \aap, 362, 419


\bibitem[{{Hatano} {et~al.}(2000){Hatano}, {Branch}, {Lentz}, {Baron},
  {Filippenko}, \& {Garnavich}}]{hatano00}
{Hatano}, K., {Branch}, D., {Lentz}, E.~J., {Baron}, E., {Filippenko}, A.~V.,
  \& {Garnavich}, P.~M. 2000, \apjl, 543, L49
 \href{http://adsabs.harvard.edu/cgi-bin/nph-bib_query?bibcode=2000ApJ...543L.%
.49H&db_key=AST}{\adsurllinklabel}

\bibitem[{{Hawkins}(2000)}]{hawkins00}
{Hawkins}, M.~R.~S. 2000, \aaps, 143, 465
 \href{http://adsabs.harvard.edu/cgi-bin/nph-bib_query?bibcode=2000A%26AS..143%
..465H&db_key=AST}{\adsurllinklabel}

\bibitem[{{Hawkins}(2002)}]{hawkins02}
---. 2002, \mnras, 329, 76
 \href{http://adsabs.harvard.edu/cgi-bin/nph-bib_query?bibcode=2002MNRAS.329..%
.76H&db_key=AST}{\adsurllinklabel}

\bibitem[{{Herpin} {et~al.}(2002){Herpin}, {Goicoechea}, {Pardo}, \&
  {Cernicharo}}]{herpin02}
{Herpin}, F., {Goicoechea}, J.~R., {Pardo}, J.~R., \& {Cernicharo}, J. 2002,
  \apj, 577, 961
 \href{http://adsabs.harvard.edu/cgi-bin/nph-bib_query?bibcode=2002ApJ...577..%
961H&amp;db_key=AST}{\adsurllinklabel}

\bibitem[{{Hillebrandt} \& {Niemeyer}(2000)}]{hillebrandt00}
{Hillebrandt}, W. \& {Niemeyer}, J.~C. 2000, \araa, 38, 191
 \href{http://adsabs.harvard.edu/cgi-bin/nph-bib_query?bibcode=2000ARA%26A..38%
..191H&db_key=AST}{\adsurllinklabel}

\bibitem[{{Hoeflich} \& {Khokhlov}(1996)}]{hoeflich96}
{Hoeflich}, P. \& {Khokhlov}, A. 1996, \apj, 457, 500
 \href{http://adsabs.harvard.edu/cgi-bin/nph-bib_query?bibcode=1996ApJ...457..%
500H&db_key=AST}{\adsurllinklabel}

\bibitem[{{Iben} \& {Tutukov}(1984)}]{iben84}
{Iben}, I. \& {Tutukov}, A.~V. 1984, \apjs, 54, 335
 \href{http://adsabs.harvard.edu/cgi-bin/nph-bib_query?bibcode=1984ApJS...54..%
335I&db_key=AST}{\adsurllinklabel}

\bibitem[{{Iwamoto} {et~al.}(1999){Iwamoto}, {Brachwitz}, {Nomoto},
  {Kishimoto}, {Umeda}, {Hix}, \& {Thielemann}}]{iwamoto99}
{Iwamoto}, K., {Brachwitz}, F., {Nomoto}, K., {Kishimoto}, N., {Umeda}, H.,
  {Hix}, W.~R., \& {Thielemann}, F. 1999, \apjs, 125, 439
 \href{http://adsabs.harvard.edu/cgi-bin/nph-bib_query?bibcode=1999ApJS..125..%
439I&db_key=AST}{\adsurllinklabel}

\bibitem[{{Jaffe} {et~al.}(2001){Jaffe}, {Ade}, {Balbi}, {Bock}, {Bond},
  {Borrill}, {Boscaleri}, {Coble}, {Crill}, {de Bernardis}, {Farese},
  {Ferreira}, {Ganga}, {Giacometti}, {Hanany}, {Hivon}, {Hristov},
  {Iacoangeli}, {Lange}, {Lee}, {Martinis}, {Masi}, {Mauskopf}, {Melchiorri},
  {Montroy}, {Netterfield}, {Oh}, {Pascale}, {Piacentini}, {Pogosyan},
  {Prunet}, {Rabii}, {Rao}, {Richards}, {Romeo}, {Ruhl}, {Scaramuzzi},
  {Sforna}, {Smoot}, {Stompor}, {Winant}, \& {Wu}}]{jaffe01}
{Jaffe}, A.~H., {Ade}, P.~A., {Balbi}, A., {Bock}, J.~J., {Bond}, J.~R.,
  {Borrill}, J., {Boscaleri}, A., {Coble}, K., {Crill}, B.~P., {de Bernardis},
  P., {Farese}, P., {Ferreira}, P.~G., {Ganga}, K., {Giacometti}, M., {Hanany},
  S., {Hivon}, E., {Hristov}, V.~V., {Iacoangeli}, A., {Lange}, A.~E., {Lee},
  A.~T., {Martinis}, L., {Masi}, S., {Mauskopf}, P.~D., {Melchiorri}, A.,
  {Montroy}, T., {Netterfield}, C.~B., {Oh}, S., {Pascale}, E., {Piacentini},
  F., {Pogosyan}, D., {Prunet}, S., {Rabii}, B., {Rao}, S., {Richards}, P.~L.,
  {Romeo}, G., {Ruhl}, J.~E., {Scaramuzzi}, F., {Sforna}, D., {Smoot}, G.~F.,
  {Stompor}, R., {Winant}, C.~D., \& {Wu}, J.~H. 2001, Physical Review Letters,
  86, 3475
 \href{http://adsabs.harvard.edu/cgi-bin/nph-bib_query?bibcode=2001PhRvL..86.3%
475J&amp;db_key=AST}{\adsurllinklabel}

\bibitem[{{Kessler}(2002)}]{kessler02}
{Kessler}, R. 2002, private communication


\bibitem[{{Khokhlov}(1991{\natexlab{a}})}]{khokhlov91b}
{Khokhlov}, A.~M. 1991{\natexlab{a}}, \aap, 245, 114
 \href{http://adsabs.harvard.edu/cgi-bin/nph-bib_query?bibcode=1991A%26A...245%
..114K&db_key=AST}{\adsurllinklabel}

\bibitem[{{Khokhlov}(1991{\natexlab{b}})}]{khokhlov91a}
---. 1991{\natexlab{b}}, \aap, 245, L25
 \href{http://adsabs.harvard.edu/cgi-bin/nph-bib_query?bibcode=1991A%26A...245%
L..25K&db_key=AST}{\adsurllinklabel}

\bibitem[{{Kim} {et~al.}(1996){Kim}, {Goobar}, \& {Perlmutter}}]{kim96}
{Kim}, A., {Goobar}, A., \& {Perlmutter}, S. 1996, \pasp, 108, 190


\bibitem[{{Kim}(1999)}]{kim99}
{Kim}, M.~Y. 1999, Ph.D.~Thesis


\bibitem[{{Knop} {et~al.}(2003){Knop}, {Aldering}, {Amanullah}, {Astier},
  {Blanc}, {Burns}, {Conley}, {Deustua}, {Doi}, {Ellis}, {Fabbro}, {Folatelli},
  {Fruchter}, {Garavini}, {Garmond}, {Garton}, {Gibbons}, {Goldhaber},
  {Goobar}, {Groom}, {Hardin}, {Hook}, {Howell}, {Kim}, {Lee}, {Lidman},
  {Mendez}, {Nobili}, {Nugent}, {Pain}, {Panagia}, {Pennypacker}, {Perlmutter},
  {Quimby}, {Raux}, {Regnault}, {Ruiz-Lapuente}, {Sainton}, {Schaefer},
  {Schahmaneche}, {Smith}, {Spadafora}, {Stanishev}, {Sullivan}, {Walton},
  {Wang}, {Wood-Vasey}, \& {Yasuda}}]{knop03}
{Knop}, R.~A., {Aldering}, G., {Amanullah}, R., {Astier}, P., {Blanc}, G.,
  {Burns}, M.~S., {Conley}, A., {Deustua}, S.~E., {Doi}, M., {Ellis}, R.,
  {Fabbro}, S., {Folatelli}, G., {Fruchter}, A.~S., {Garavini}, G., {Garmond},
  S., {Garton}, K., {Gibbons}, R., {Goldhaber}, G., {Goobar}, A., {Groom},
  D.~E., {Hardin}, D., {Hook}, I., {Howell}, D.~A., {Kim}, A.~G., {Lee}, B.~C.,
  {Lidman}, C., {Mendez}, J., {Nobili}, S., {Nugent}, P.~E., {Pain}, R.,
  {Panagia}, N., {Pennypacker}, C.~R., {Perlmutter}, S., {Quimby}, R., {Raux},
  J., {Regnault}, N., {Ruiz-Lapuente}, P., {Sainton}, G., {Schaefer}, B.,
  {Schahmaneche}, K., {Smith}, E., {Spadafora}, A.~L., {Stanishev}, V.,
  {Sullivan}, M., {Walton}, N.~A., {Wang}, L., {Wood-Vasey}, W.~M., \&
  {Yasuda}, N. 2003, \apj, 598, 102


\bibitem[{{Kriessler} {et~al.}(1998){Kriessler}, {Humphreys}, {Cabanela},
  {Rees}, {Ngo}, \& {Srivastava}}]{apscatalog98}
{Kriessler}, J.~R., {Humphreys}, R.~M., {Cabanela}, J.~E., {Rees}, R.~F.,
  {Ngo}, H., \& {Srivastava}, J. 1998, \baas, 30, 900
 \href{http://adsabs.harvard.edu/cgi-bin/nph-bib_query?bibcode=1998AAS...192.5%
509K&db_key=AST}{\adsurllinklabel}

\bibitem[{{Kwok}(1993)}]{kwok93}
{Kwok}, S. 1993, \araa, 31, 63
 \href{http://adsabs.harvard.edu/cgi-bin/nph-bib_query?bibcode=1993ARA%26A..31%
...63K&amp;db_key=AST}{\adsurllinklabel}

\bibitem[{{Landolt}(1992)}]{landolt92}
{Landolt}, A.~U. 1992, \aj, 104, 340
 \href{http://adsabs.harvard.edu/cgi-bin/nph-bib_query?bibcode=1992AJ....104..%
340L&amp;db_key=AST}{\adsurllinklabel}

\bibitem[{{Landy}(2002)}]{landy02}
{Landy}, S.~D. 2002, \apjl, 567, L1


\bibitem[{{Lantz}(2003)}]{lantz03}
{Lantz}, B.~{\em et~al.}. 2003, in {Proceedings of the SPIE}, 146


\bibitem[{{Leibundgut}(1990)}]{leibundgut90}
{Leibundgut}, B. 1990, \aap, 229, 1


\bibitem[{{Li} {et~al.}(2003){Li}, {Filippenko}, {Chornock}, {Berger},
  {Berlind}, {Calkins}, {Challis}, {Fassnacht}, {Jha}, {Kirshner}, {Matheson},
  {Sargent}, {Simcoe}, {Smith}, \& {Squires}}]{li03}
{Li}, W., {Filippenko}, A.~V., {Chornock}, R., {Berger}, E., {Berlind}, P.,
  {Calkins}, M.~L., {Challis}, P., {Fassnacht}, C., {Jha}, S., {Kirshner},
  R.~P., {Matheson}, T., {Sargent}, W.~L.~W., {Simcoe}, R.~A., {Smith}, G.~H.,
  \& {Squires}, G. 2003, \pasp, 115, 453
 \href{http://adsabs.harvard.edu/cgi-bin/nph-bib_query?bibcode=2003PASP..115..%
453L&amp;db_key=AST}{\adsurllinklabel}

\bibitem[{{Linder}(2003)}]{linder03a}
{Linder}, E.~V. 2003, Physical Review Letters, 90, 91301
 \href{http://adsabs.harvard.edu/cgi-bin/nph-bib_query?bibcode=2003PhRvL..90i1%
301L&amp;db_key=AST}{\adsurllinklabel}

\bibitem[{{Linder} \& {Jenkins}(2003)}]{linder03b}
{Linder}, E.~V. \& {Jenkins}, A. 2003, \mnras, 346, 573


\bibitem[{{Livio}(2001)}]{livio01}
{Livio}, M. 2001, "Supernovae and Gamma-Ray Bursts: The Greatest Explosions
  since the Big Bang" (Cambridge University Press), 334


\bibitem[{{Livio} \& {Riess}(2003)}]{livio03}
{Livio}, M. \& {Riess}, A.~G. 2003, \apjl, 594, L93


\bibitem[{{Madau} {et~al.}(1998){Madau}, {della Valle}, \&
  {Panagia}}]{madau98b}
{Madau}, P., {della Valle}, M., \& {Panagia}, N. 1998, \mnras, 297, L17+


\bibitem[{Malmquist(1924)}]{malmquist24}
Malmquist, K.~G. 1924, Medd. Lund Astron. Obs. Ser. II, 32, 64


\bibitem[{Malmquist(1936)}]{malmquist36}
---. 1936, Stockholm Observatory Medd.


\bibitem[{{Maoz} \& {Gal-Yam}(2004)}]{maoz04}
{Maoz}, D. \& {Gal-Yam}, A. 2004, \mnras, 347, 951


\bibitem[{{Matzner} \& {McKee}(1999)}]{matzner99}
{Matzner}, C.~D. \& {McKee}, C.~F. 1999, \apj, 510, 379
 \href{http://adsabs.harvard.edu/cgi-bin/nph-bib_query?bibcode=1999ApJ...510..%
379M&amp;db_key=AST}{\adsurllinklabel}

\bibitem[{{Mazzali} {et~al.}(1998){Mazzali}, {Cappellaro}, {Danziger},
  {Turatto}, \& {Benetti}}]{mazzali98}
{Mazzali}, P.~A., {Cappellaro}, E., {Danziger}, I.~J., {Turatto}, M., \&
  {Benetti}, S. 1998, \apjl, 499, L49+
 \href{http://adsabs.harvard.edu/cgi-bin/nph-bib_query?bibcode=1998ApJ...499L.%
.49M&amp;db_key=AST}{\adsurllinklabel}

\bibitem[{{Mazzali} {et~al.}(2001){Mazzali}, {Nomoto}, {Cappellaro},
  {Nakamura}, {Umeda}, \& {Iwamoto}}]{mazzali01}
{Mazzali}, P.~A., {Nomoto}, K., {Cappellaro}, E., {Nakamura}, T., {Umeda}, H.,
  \& {Iwamoto}, K. 2001, \apj, 547, 988
 \href{http://adsabs.harvard.edu/cgi-bin/nph-bib_query?bibcode=2001ApJ...547..%
988M&db_key=AST}{\adsurllinklabel}

\bibitem[{{Modjaz} \& {Li}(2001)}]{iauc7645}
{Modjaz}, M. \& {Li}, W.~D. 2001, \iaucirc, 7645, 1
 \href{http://adsabs.harvard.edu/cgi-bin/nph-bib_query?bibcode=2001IAUC.7645..%
..1M&db_key=AST}{\adsurllinklabel}

\bibitem[{Monet {et~al.}(1996)Monet, Bird, Canzian, Harris, Reid, Rhodes, Sell,
  Ables, Dahn, Guetter, Henden, Leggett, Levison, Luginbuhl, Martini, Monet,
  Pier, Riepe, Stone, Vrba, \& Walker}]{usnoa1}
Monet, D., Bird, A., Canzian, B., Harris, H., Reid, N., Rhodes, A., Sell, S.,
  Ables, H., Dahn, C., Guetter, H., Henden, A., Leggett, S., Levison, H.,
  Luginbuhl, C., Martini, J., Monet, A., Pier, J., Riepe, B., Stone, R., Vrba,
  F., \& Walker, R. 1996, USNO-SA1.0 Catalog (U.S. Naval Observatory,
  Washington DC)


\bibitem[{{Muller} {et~al.}(1992){Muller}, {Newberg}, {Pennypacker},
  {Perlmutter}, {Sasseen}, \& {Smith}}]{muller92}
{Muller}, R.~A., {Newberg}, H.~J.~M., {Pennypacker}, C.~R., {Perlmutter}, S.,
  {Sasseen}, T.~P., \& {Smith}, C.~K. 1992, \apjl, 384, L9


\bibitem[{{Nielsen} \& {Odewahn}(1994)}]{nielsen94}
{Nielsen}, M.~L. \& {Odewahn}, S.~C. 1994, \baas, 26, 1498
 \href{http://adsabs.harvard.edu/cgi-bin/nph-bib_query?bibcode=1994AAS...18510%
709N&db_key=AST}{\adsurllinklabel}

\bibitem[{{Nomoto}(1982{\natexlab{a}})}]{nomoto82b}
{Nomoto}, K. 1982{\natexlab{a}}, \apj, 257, 780
 \href{http://adsabs.harvard.edu/cgi-bin/nph-bib_query?bibcode=1982ApJ...257..%
780N&db_key=AST}{\adsurllinklabel}

\bibitem[{{Nomoto}(1982{\natexlab{b}})}]{nomoto82a}
---. 1982{\natexlab{b}}, \apj, 253, 798
 \href{http://adsabs.harvard.edu/cgi-bin/nph-bib_query?bibcode=1982ApJ...253..%
798N&db_key=AST}{\adsurllinklabel}

\bibitem[{{Nomoto} {et~al.}(1997){Nomoto}, {Iwamoto}, \&
  {Kishimoto}}]{nomoto97}
{Nomoto}, K., {Iwamoto}, K., \& {Kishimoto}, N. 1997, Science, 276, 1378
 \href{http://adsabs.harvard.edu/cgi-bin/nph-bib_query?bibcode=1997Sci...276.1%
378N&db_key=AST}{\adsurllinklabel}

\bibitem[{{Nomoto} \& {Sugimoto}(1977)}]{nomoto77}
{Nomoto}, K. \& {Sugimoto}, D. 1977, \pasj, 29, 765
 \href{http://adsabs.harvard.edu/cgi-bin/nph-bib_query?bibcode=1977PASJ...29..%
765N&db_key=AST}{\adsurllinklabel}

\bibitem[{{Nugent} {et~al.}(2002){Nugent}, {Kim}, \& {Perlmutter}}]{nugent02}
{Nugent}, P., {Kim}, A., \& {Perlmutter}, S. 2002, \pasp, 114, 803
 \href{http://adsabs.harvard.edu/cgi-bin/nph-bib_query?bibcode=2002PASP..114..%
803N&db_key=AST}{\adsurllinklabel}

\bibitem[{{Nugent} {et~al.}(1995){Nugent}, {Phillips}, {Baron}, {Branch}, \&
  {Hauschildt}}]{nugent95}
{Nugent}, P., {Phillips}, M., {Baron}, E., {Branch}, D., \& {Hauschildt}, P.
  1995, \apjl, 455, L147+
 \href{http://adsabs.harvard.edu/cgi-bin/nph-bib_query?bibcode=1995ApJ...455L.%
147N&amp;db_key=AST}{\adsurllinklabel}

\bibitem[{{Odewahn}(1995)}]{odewahn95}
{Odewahn}, S.~C. 1995, \pasp, 107, 770
 \href{http://adsabs.harvard.edu/cgi-bin/nph-bib_query?bibcode=1995PASP..107..%
770O&db_key=AST}{\adsurllinklabel}

\bibitem[{{Odewahn} {et~al.}(1993){Odewahn}, {Humphreys}, {Aldering}, \&
  {Thurmes}}]{odewahn93}
{Odewahn}, S.~C., {Humphreys}, R.~M., {Aldering}, G., \& {Thurmes}, P. 1993,
  \pasp, 105, 1354
 \href{http://adsabs.harvard.edu/cgi-bin/nph-bib_query?bibcode=1993PASP..105.1%
354O&db_key=AST}{\adsurllinklabel}

\bibitem[{{Odewahn} {et~al.}(1992){Odewahn}, {Stockwell}, {Pennington},
  {Humphreys}, \& {Zumach}}]{odewahn92}
{Odewahn}, S.~C., {Stockwell}, E.~B., {Pennington}, R.~L., {Humphreys}, R.~M.,
  \& {Zumach}, W.~A. 1992, \aj, 103, 318
 \href{http://adsabs.harvard.edu/cgi-bin/nph-bib_query?bibcode=1992AJ....103..%
318O&db_key=AST}{\adsurllinklabel}

\bibitem[{{O'Donnell}(1994)}]{odonnell94}
{O'Donnell}, J.~E. 1994, \apj, 422, 158


\bibitem[{{P{\' e}contal} {et~al.}(2003){P{\' e}contal}, {Aldering}, {Adam},
  {Antilogus}, {Astier}, {Copin}, {H{\' e}nault}, {Lemonnier}, {Nugent},
  {Pain}, {P{\' e}contal}, {Perlmutter}, {Quimby}, {Smadja}, \&
  {Wood-Vasey}}]{pecontal03}
{P{\' e}contal}, E., {Aldering}, G., {Adam}, G., {Antilogus}, P., {Astier}, P.,
  {Copin}, Y., {H{\' e}nault}, F., {Lemonnier}, J.-P., {Nugent}, P., {Pain},
  R., {P{\' e}contal}, A., {Perlmutter}, S., {Quimby}, R., {Smadja}, G., \&
  {Wood-Vasey}, M. 2003, in From Twilight to Highlight: The Physics of
  Supernovae. Proceedings of the ESO/MPA/MPE Workshop held in Garching,
  Germany, 29-31 July 2002, p. 404., 404--+
 \href{http://adsabs.harvard.edu/cgi-bin/nph-bib_query?bibcode=2003fthp.conf..%
404P&amp;db_key=AST}{\adsurllinklabel}

\bibitem[{{Pain} {et~al.}(2002){Pain}, {Fabbro}, {Sullivan}, {Ellis},
  {Aldering}, {Astier}, {Deustua}, {Fruchter}, {Goldhaber}, {Goobar}, {Groom},
  {Hardin}, {Hook}, {Howell}, {Irwin}, {Kim}, {Kim}, {Knop}, {Lee}, {Lidman},
  {McMahon}, {Nugent}, {Panagia}, {Pennypacker}, {Perlmutter}, {Ruiz-Lapuente},
  {Schahmaneche}, {Schaefer}, \& {Walton}}]{pain02}
{Pain}, R., {Fabbro}, S., {Sullivan}, M., {Ellis}, R.~S., {Aldering}, G.,
  {Astier}, P., {Deustua}, S.~E., {Fruchter}, A.~S., {Goldhaber}, G., {Goobar},
  A., {Groom}, D.~E., {Hardin}, D., {Hook}, I.~M., {Howell}, D.~A., {Irwin},
  M.~J., {Kim}, A.~G., {Kim}, M.~Y., {Knop}, R.~A., {Lee}, J.~C., {Lidman}, C.,
  {McMahon}, R.~G., {Nugent}, P.~E., {Panagia}, N., {Pennypacker}, C.~R.,
  {Perlmutter}, S., {Ruiz-Lapuente}, P., {Schahmaneche}, K., {Schaefer}, B., \&
  {Walton}, N.~A. 2002, \apj, 577, 120
 \href{http://adsabs.harvard.edu/cgi-bin/nph-bib_query?bibcode=2002ApJ...577..%
120P&db_key=AST}{\adsurllinklabel}

\bibitem[{{Pain} {et~al.}(1996){Pain}, {Hook}, {Deustua}, {Gabi}, {Goldhaber},
  {Groom}, {Kim}, {Kim}, {Lee}, {Pennypacker}, {Perlmutter}, {Small}, {Goobar},
  {Ellis}, {McMahon}, {Glazebrook}, {Boyle}, {Bunclark}, {Carter}, \&
  {Irwin}}]{pain96}
{Pain}, R., {Hook}, I.~M., {Deustua}, S., {Gabi}, S., {Goldhaber}, G., {Groom},
  D., {Kim}, A.~G., {Kim}, M.~Y., {Lee}, J.~C., {Pennypacker}, C.~R.,
  {Perlmutter}, S., {Small}, I.~A., {Goobar}, A., {Ellis}, R.~S., {McMahon},
  R.~G., {Glazebrook}, K., {Boyle}, B.~J., {Bunclark}, P.~S., {Carter}, D., \&
  {Irwin}, M.~J. 1996, \apj, 473, 356
 \href{http://adsabs.harvard.edu/cgi-bin/nph-bib_query?bibcode=1996ApJ...473..%
356P&db_key=AST}{\adsurllinklabel}

\bibitem[{{Pain} \& {The {SNLS} Collaboration}(2003)}]{pain03}
{Pain}, R. \& {The {SNLS} Collaboration}. 2003, in From Twilight to Highlight:
  The Physics of Supernovae. Proceedings of the ESO/MPA/MPE Workshop held in
  Garching, Germany, 29-31 July 2002, p. 408., 408--+
 \href{http://adsabs.harvard.edu/cgi-bin/nph-bib_query?bibcode=2003fthp.conf..%
408P&amp;db_key=AST}{\adsurllinklabel}

\bibitem[{{Pence}(2003)}]{fits}
{Pence}, W.~D. 2003, Flexible Image Transport System
 \href{http://fits.gsfc.nasa.gov}{\urllinklabel}

\bibitem[{{Perlmutter} {et~al.}(1998){Perlmutter}, {Aldering}, {della Valle},
  {Deustua}, {Ellis}, {Fabbro}, {Fruchter}, {Goldhaber}, {Groom}, {Hook},
  {Kim}, {Kim}, {Knop}, {Lidman}, {McMahon}, {Nugent}, {Pain}, {Panagia},
  {Pennypacker}, {Ruiz-Lapuente}, {Schaefer}, \& {Walton}}]{perlmutter98a}
{Perlmutter}, S., {Aldering}, G., {della Valle}, M., {Deustua}, S., {Ellis},
  R.~S., {Fabbro}, S., {Fruchter}, A., {Goldhaber}, G., {Groom}, D.~E., {Hook},
  I.~M., {Kim}, A.~G., {Kim}, M.~Y., {Knop}, R.~A., {Lidman}, C., {McMahon},
  R.~G., {Nugent}, P., {Pain}, R., {Panagia}, N., {Pennypacker}, C.~R.,
  {Ruiz-Lapuente}, P., {Schaefer}, B., \& {Walton}, N. 1998, \nat, 391, 51
 \href{http://adsabs.harvard.edu/cgi-bin/nph-bib_query?bibcode=1998Natur.391..%
.51P&amp;db_key=AST}{\adsurllinklabel}

\bibitem[{{Perlmutter} {et~al.}(1999){Perlmutter}, {Aldering}, {Goldhaber},
  {Knop}, {Nugent}, {Castro}, {Deustua}, {Fabbro}, {Goobar}, {Groom}, {Hook},
  {Kim}, {Kim}, {Lee}, {Nunes}, {Pain}, {Pennypacker}, {Quimby}, {Lidman},
  {Ellis}, {Irwin}, {McMahon}, {Ruiz-Lapuente}, {Walton}, {Schaefer}, {Boyle},
  {Filippenko}, {Matheson}, {Fruchter}, {Panagia}, {Newberg}, {Couch}, \& {The
  Supernova Cosmology Project}}]{perlmutter99}
{Perlmutter}, S., {Aldering}, G., {Goldhaber}, G., {Knop}, R.~A., {Nugent}, P.,
  {Castro}, P.~G., {Deustua}, S., {Fabbro}, S., {Goobar}, A., {Groom}, D.~E.,
  {Hook}, I.~M., {Kim}, A.~G., {Kim}, M.~Y., {Lee}, J.~C., {Nunes}, N.~J.,
  {Pain}, R., {Pennypacker}, C.~R., {Quimby}, R., {Lidman}, C., {Ellis}, R.~S.,
  {Irwin}, M., {McMahon}, R.~G., {Ruiz-Lapuente}, P., {Walton}, N., {Schaefer},
  B., {Boyle}, B.~J., {Filippenko}, A.~V., {Matheson}, T., {Fruchter}, A.~S.,
  {Panagia}, N., {Newberg}, H.~J.~M., {Couch}, W.~J., \& {The Supernova
  Cosmology Project}. 1999, \apj, 517, 565
 \href{http://adsabs.harvard.edu/cgi-bin/nph-bib_query?bibcode=1999ApJ...517..%
565P&db_key=AST}{\adsurllinklabel}

\bibitem[{{Perlmutter} {et~al.}(1997){Perlmutter}, {Gabi}, {Goldhaber},
  {Goobar}, {Groom}, {Hook}, {Kim}, {Kim}, {Lee}, {Pain}, {Pennypacker},
  {Small}, {Ellis}, {McMahon}, {Boyle}, {Bunclark}, {Carter}, {Irwin},
  {Glazebrook}, {Newberg}, {Filippenko}, {Matheson}, {Dopita}, {Couch}, \& {The
  Supernova Cosmology Project}}]{perlmutter97b}
{Perlmutter}, S., {Gabi}, S., {Goldhaber}, G., {Goobar}, A., {Groom}, D.~E.,
  {Hook}, I.~M., {Kim}, A.~G., {Kim}, M.~Y., {Lee}, J.~C., {Pain}, R.,
  {Pennypacker}, C.~R., {Small}, I.~A., {Ellis}, R.~S., {McMahon}, R.~G.,
  {Boyle}, B.~J., {Bunclark}, P.~S., {Carter}, D., {Irwin}, M.~J.,
  {Glazebrook}, K., {Newberg}, H.~J.~M., {Filippenko}, A.~V., {Matheson}, T.,
  {Dopita}, M., {Couch}, W.~J., \& {The Supernova Cosmology Project}. 1997,
  \apj, 483, 565
 \href{http://adsabs.harvard.edu/cgi-bin/nph-bib_query?bibcode=1997ApJ...483..%
565P&db_key=AST}{\adsurllinklabel}

\bibitem[{{Perlmutter}(1997)}]{perlmutter97a}
{Perlmutter}, S.~{\em et al.}. 1997, in NATO ASIC Proc. 486: Thermonuclear
  Supernovae, 749--+
 \href{http://adsabs.harvard.edu/cgi-bin/nph-bib_query?bibcode=1997thsu.conf..%
749P&amp;db_key=AST}{\adsurllinklabel}

\bibitem[{{Phillips}(1993)}]{phillips93}
{Phillips}, M.~M. 1993, \apjl, 413, L105
 \href{http://adsabs.harvard.edu/cgi-bin/nph-bib_query?bibcode=1993ApJ...413L.%
105P&db_key=AST}{\adsurllinklabel}

\bibitem[{{Phillips} {et~al.}(1999){Phillips}, {Lira}, {Suntzeff}, {Schommer},
  {Hamuy}, \& {Maza}}]{phillips99}
{Phillips}, M.~M., {Lira}, P., {Suntzeff}, N.~B., {Schommer}, R.~A., {Hamuy},
  M., \& {Maza}, J. 1999, \aj, 118, 1766


\bibitem[{{Pinto} \& {Eastman}(2000{\natexlab{a}})}]{pinto00a}
{Pinto}, P.~A. \& {Eastman}, R.~G. 2000{\natexlab{a}}, \apj, 530, 744
 \href{http://adsabs.harvard.edu/cgi-bin/nph-bib_query?bibcode=2000ApJ...530..%
744P&db_key=AST}{\adsurllinklabel}

\bibitem[{{Pinto} \& {Eastman}(2000{\natexlab{b}})}]{pinto00b}
---. 2000{\natexlab{b}}, \apj, 530, 757
 \href{http://adsabs.harvard.edu/cgi-bin/nph-bib_query?bibcode=2000ApJ...530..%
757P&db_key=AST}{\adsurllinklabel}

\bibitem[{{Pollas} {et~al.}(1988){Pollas}, {Cappellaro}, {Turatto}, \&
  {Candeo}}]{iauc4691}
{Pollas}, C., {Cappellaro}, E., {Turatto}, M., \& {Candeo}, G. 1988, \iaucirc,
  4691, 1
 \href{http://adsabs.harvard.edu/cgi-bin/nph-bib_query?bibcode=1988IAUC.4691..%
..1P&db_key=AST}{\adsurllinklabel}

\bibitem[{{Pravdo} {et~al.}(1999){Pravdo}, {Rabinowitz}, {Helin}, {Lawrence},
  {Bambery}, {Clark}, {Groom}, {Levin}, {Lorre}, {Shaklan}, {Kervin},
  {Africano}, {Sydney}, \& {Soohoo}}]{pravdo99}
{Pravdo}, S.~H., {Rabinowitz}, D.~L., {Helin}, E.~F., {Lawrence}, K.~J.,
  {Bambery}, R.~J., {Clark}, C.~C., {Groom}, S.~L., {Levin}, S., {Lorre}, J.,
  {Shaklan}, S.~B., {Kervin}, P., {Africano}, J.~A., {Sydney}, P., \& {Soohoo},
  V. 1999, \aj, 117, 1616
 \href{http://adsabs.harvard.edu/cgi-bin/nph-bib_query?bibcode=1999AJ....117.1%
616P&db_key=AST}{\adsurllinklabel}

\bibitem[{{Pritchet} \& {Collaboration}(2004)}]{pritchet04}
{Pritchet}, C.~J. \& {Collaboration}, S. 2004, ArXiv Astrophysics e-prints


\bibitem[{{Pskovskii}(1977)}]{pskovskii77b}
{Pskovskii}, I.~P. 1977, Soviet Astronomy, 21, 675


\bibitem[{{Pskovskii}(1970)}]{pskovskii70}
{Pskovskii}, Y.~P. 1970, Soviet Astronomy, 14, 798


\bibitem[{{Richmond} {et~al.}(1998){Richmond}, {Filippenko}, \&
  {Galisky}}]{richmond98}
{Richmond}, M.~W., {Filippenko}, A.~V., \& {Galisky}, J. 1998, \pasp, 110, 553
 \href{http://adsabs.harvard.edu/cgi-bin/nph-bib_query?bibcode=1998PASP..110..%
553R&db_key=AST}{\adsurllinklabel}

\bibitem[{{Riess} {et~al.}(1998{\natexlab{a}}){Riess}, {Filippenko}, {Challis},
  {Clocchiatti}, {Diercks}, {Garnavich}, {Gilliland}, {Hogan}, {Jha},
  {Kirshner}, {Leibundgut}, {Phillips}, {Reiss}, {Schmidt}, {Schommer},
  {Smith}, {Spyromilio}, {Stubbs}, {Suntzeff}, \& {Tonry}}]{riess98a}
{Riess}, A.~G., {Filippenko}, A.~V., {Challis}, P., {Clocchiatti}, A.,
  {Diercks}, A., {Garnavich}, P.~M., {Gilliland}, R.~L., {Hogan}, C.~J., {Jha},
  S., {Kirshner}, R.~P., {Leibundgut}, B., {Phillips}, M.~M., {Reiss}, D.,
  {Schmidt}, B.~P., {Schommer}, R.~A., {Smith}, R.~C., {Spyromilio}, J.,
  {Stubbs}, C., {Suntzeff}, N.~B., \& {Tonry}, J. 1998{\natexlab{a}}, \aj, 116,
  1009
 \href{http://adsabs.harvard.edu/cgi-bin/nph-bib_query?bibcode=1998AJ....116.1%
009R&amp;db_key=AST}{\adsurllinklabel}

\bibitem[{{Riess} {et~al.}(1998{\natexlab{b}}){Riess}, {Nugent}, {Filippenko},
  {Kirshner}, \& {Perlmutter}}]{riess98b}
{Riess}, A.~G., {Nugent}, P., {Filippenko}, A.~V., {Kirshner}, R.~P., \&
  {Perlmutter}, S. 1998{\natexlab{b}}, \apj, 504, 935


\bibitem[{{Riess} {et~al.}(1995){Riess}, {Press}, \& {Kirshner}}]{riess95}
{Riess}, A.~G., {Press}, W.~H., \& {Kirshner}, R.~P. 1995, \apjl, 438, L17
 \href{http://adsabs.harvard.edu/cgi-bin/nph-bib_query?bibcode=1995ApJ...438L.%
.17R&db_key=AST}{\adsurllinklabel}

\bibitem[{{Riess} {et~al.}(1996){Riess}, {Press}, \& {Kirshner}}]{riess96}
---. 1996, \apj, 473, 88
 \href{http://adsabs.harvard.edu/cgi-bin/nph-bib_query?bibcode=1996ApJ...473..%
.88R&db_key=AST}{\adsurllinklabel}

\bibitem[{{Riess} {et~al.}(2004){Riess}, {Strolger}, {Tonry}, {Casertano},
  {Ferguson}, {Mobasher}, {Challis}, {Filippenko}, {Jha}, {Li}, {Chornock},
  {Kirshner}, {Leibundgut}, {Dickinson}, {Livio}, {Giavalisco}, {Steidel},
  {Benitez}, \& {Tsvetanov}}]{riess04b}
{Riess}, A.~G., {Strolger}, L., {Tonry}, J., {Casertano}, S., {Ferguson},
  H.~C., {Mobasher}, B., {Challis}, P., {Filippenko}, A.~V., {Jha}, S., {Li},
  W., {Chornock}, R., {Kirshner}, R.~P., {Leibundgut}, B., {Dickinson}, M.,
  {Livio}, M., {Giavalisco}, M., {Steidel}, C.~C., {Benitez}, N., \&
  {Tsvetanov}, Z. 2004, ArXiv Astrophysics e-prints


\bibitem[{{Rigon} {et~al.}(2003){Rigon}, {Turatto}, {Benetti}, {Pastorello},
  {Cappellaro}, {Aretxaga}, {Vega}, {Chavushyan}, {Patat}, {Danziger}, \&
  {Salvo}}]{rigon03}
{Rigon}, L., {Turatto}, M., {Benetti}, S., {Pastorello}, A., {Cappellaro}, E.,
  {Aretxaga}, I., {Vega}, O., {Chavushyan}, V., {Patat}, F., {Danziger}, I.~J.,
  \& {Salvo}, M. 2003, \mnras, 340, 191
 \href{http://adsabs.harvard.edu/cgi-bin/nph-bib_query?bibcode=2003MNRAS.340..%
191R&db_key=AST}{\adsurllinklabel}

\bibitem[{{Sabine} {et~al.}(1997){Sabine}, {Baines}, \& {Howard}}]{iauc6706}
{Sabine}, S., {Baines}, D., \& {Howard}, J. 1997, \iaucirc, 6706, 1
 \href{http://adsabs.harvard.edu/cgi-bin/nph-bib_query?bibcode=1997IAUC.6706..%
..1S&db_key=AST}{\adsurllinklabel}

\bibitem[{{Sandage}(1961)}]{sandage61}
{Sandage}, A.~R. 1961, Sky and Telescope, 21


\bibitem[{{Schlegel} {et~al.}(1998){Schlegel}, {Finkbeiner}, \&
  {Davis}}]{schlegel98}
{Schlegel}, D.~J., {Finkbeiner}, D.~P., \& {Davis}, M. 1998, \apj, 500, 525
 \href{http://adsabs.harvard.edu/cgi-bin/nph-bib_query?bibcode=1998ApJ...500..%
525S&db_key=AST}{\adsurllinklabel}

\bibitem[{{Schmidt}(1969)}]{schmidt69}
{Schmidt}, M. 1969, \araa, 7, 527
 \href{http://adsabs.harvard.edu/cgi-bin/nph-bib_query?bibcode=1969ARA%26A...7%
..527S&db_key=AST}{\adsurllinklabel}

\bibitem[{{Schneider} {et~al.}(2003){Schneider}, {Fan}, {Hall}, {Jester},
  {Richards}, {Stoughton}, {Strauss}, {SubbaRao}, {Vanden Berk}, {Anderson},
  {Brandt}, {Gunn}, {Gray}, {Trump}, {Voges}, {Yanny}, {Bahcall}, {Blanton},
  {Boroski}, {Brinkmann}, {Brunner}, {Burles}, {Castander}, {Doi},
  {Eisenstein}, {Frieman}, {Fukugita}, {Heckman}, {Hennessy}, {Ivezi{\' c}},
  {Kent}, {Knapp}, {Lamb}, {Lee}, {Loveday}, {Lupton}, {Margon}, {Meiksin},
  {Munn}, {Newberg}, {Nichol}, {Niederste-Ostholt}, {Pier}, {Richmond},
  {Rockosi}, {Saxe}, {Schlegel}, {Szalay}, {Thakar}, {Uomoto}, \&
  {York}}]{schneider03}
{Schneider}, D.~P., {Fan}, X., {Hall}, P.~B., {Jester}, S., {Richards}, G.~T.,
  {Stoughton}, C., {Strauss}, M.~A., {SubbaRao}, M., {Vanden Berk}, D.~E.,
  {Anderson}, S.~F., {Brandt}, W.~N., {Gunn}, J.~E., {Gray}, J., {Trump},
  J.~R., {Voges}, W., {Yanny}, B., {Bahcall}, N.~A., {Blanton}, M.~R.,
  {Boroski}, W.~N., {Brinkmann}, J., {Brunner}, R., {Burles}, S., {Castander},
  F.~J., {Doi}, M., {Eisenstein}, D., {Frieman}, J.~A., {Fukugita}, M.,
  {Heckman}, T.~M., {Hennessy}, G.~S., {Ivezi{\' c}}, {\v Z}., {Kent}, S.,
  {Knapp}, G.~R., {Lamb}, D.~Q., {Lee}, B.~C., {Loveday}, J., {Lupton}, R.~H.,
  {Margon}, B., {Meiksin}, A., {Munn}, J.~A., {Newberg}, H.~J., {Nichol},
  R.~C., {Niederste-Ostholt}, M., {Pier}, J.~R., {Richmond}, M.~W., {Rockosi},
  C.~M., {Saxe}, D.~H., {Schlegel}, D.~J., {Szalay}, A.~S., {Thakar}, A.~R.,
  {Uomoto}, A., \& {York}, D.~G. 2003, \aj, 126, 2579
 \href{http://adsabs.harvard.edu/cgi-bin/nph-bib_query?bibcode=2003AJ....126.2%
579S&amp;db_key=AST}{\adsurllinklabel}

\bibitem[{{Schneider}(1993)}]{schneider93}
{Schneider}, P. 1993, \aap, 279, 1
 \href{http://adsabs.harvard.edu/cgi-bin/nph-bib_query?bibcode=1993A%26A...279%
....1S&db_key=AST}{\adsurllinklabel}

\bibitem[{{Schwartz} \& {Li}(2002)}]{iauc8014}
{Schwartz}, M. \& {Li}, W. 2002, in \iaucirc, 1--+


\bibitem[{{Shaw}(1979)}]{shaw79}
{Shaw}, R.~L. 1979, \aap, 76, 188
 \href{http://adsabs.harvard.edu/cgi-bin/nph-bib_query?bibcode=1979A%26A....76%
..188S&db_key=AST}{\adsurllinklabel}

\bibitem[{{Shen} {et~al.}(2003){Shen}, {Mo}, {White}, {Blanton}, {Kauffmann},
  {Voges}, {Brinkmann}, \& {Csabai}}]{shen03}
{Shen}, S., {Mo}, H.~J., {White}, S.~D.~M., {Blanton}, M.~R., {Kauffmann}, G.,
  {Voges}, W., {Brinkmann}, J., \& {Csabai}, I. 2003, \mnras, 343, 978
 \href{http://adsabs.harvard.edu/cgi-bin/nph-bib_query?bibcode=2003MNRAS.343..%
978S&db_key=AST}{\adsurllinklabel}

\bibitem[{{Siloti} {et~al.}(2000){Siloti}, {Schlegel}, {Challis}, {Jha},
  {Kirshner}, \& {Garnavich}}]{siloti00}
{Siloti}, S.~Z., {Schlegel}, E.~M., {Challis}, P., {Jha}, S., {Kirshner},
  R.~P., \& {Garnavich}, P. 2000, \baas, 32, 1538
 \href{http://adsabs.harvard.edu/cgi-bin/nph-bib_query?bibcode=2000AAS...197.8%
107S&db_key=AST}{\adsurllinklabel}

\bibitem[{{Sloan Digital Sky Survey}(2003)}]{sdss_dr1}
{Sloan Digital Sky Survey}. 2003, {Sloan Digital Sky Survey Data Relase 1},
  {\em http://www.sdss.org/dr1/}, last accessed: 2003 July 22
 \href{http://www.sdss.org/dr1/}{\urllinklabel}

\bibitem[{{Sloan Digital Sky Survey}(2004)}]{sdss_dr2}
---. 2004, {Sloan Digital Sky Survey Data Relase 2}, {\em
  http://www.sdss.org/dr2/}, last accessed: 2004 March 28
 \href{http://www.sdss.org/dr2/}{\urllinklabel}

\bibitem[{{Soker} \& {Rappaport}(2000)}]{soker00}
{Soker}, N. \& {Rappaport}, S. 2000, \apj, 538, 241
 \href{http://adsabs.harvard.edu/cgi-bin/nph-bib_query?bibcode=2000ApJ...538..%
241S&amp;db_key=AST}{\adsurllinklabel}

\bibitem[{{Space Telescope Science Institute}(2003)}]{dss}
{Space Telescope Science Institute}. 2003, {Digitized Sky Survey}
 \href{http://www-gsss.stsci.edu/DSS/dss\_home.htm}{\urllinklabel}

\bibitem[{{Spergel} {et~al.}(2003){Spergel}, {Verde}, {Peiris}, {Komatsu},
  {Nolta}, {Bennett}, {Halpern}, {Hinshaw}, {Jarosik}, {Kogut}, {Limon},
  {Meyer}, {Page}, {Tucker}, {Weiland}, {Wollack}, \& {Wright}}]{spergel03}
{Spergel}, D.~N., {Verde}, L., {Peiris}, H.~V., {Komatsu}, E., {Nolta}, M.~R.,
  {Bennett}, C.~L., {Halpern}, M., {Hinshaw}, G., {Jarosik}, N., {Kogut}, A.,
  {Limon}, M., {Meyer}, S.~S., {Page}, L., {Tucker}, G.~S., {Weiland}, J.~L.,
  {Wollack}, E., \& {Wright}, E.~L. 2003, \apjs, 148, 175
 \href{http://adsabs.harvard.edu/cgi-bin/nph-bib_query?bibcode=2003ApJS..148..%
175S&amp;db_key=AST}{\adsurllinklabel}

\bibitem[{{Spyromilio} {et~al.}(1992){Spyromilio}, {Meikle}, {Allen}, \&
  {Graham}}]{spyromilio92}
{Spyromilio}, J., {Meikle}, W.~P.~S., {Allen}, D.~A., \& {Graham}, J.~R. 1992,
  \mnras, 258, 53P
 \href{http://adsabs.harvard.edu/cgi-bin/nph-bib_query?bibcode=1992MNRAS.258P.%
.53S&db_key=AST}{\adsurllinklabel}

\bibitem[{{Stathakis} \& {Sadler}(1991)}]{stathakis91}
{Stathakis}, R.~A. \& {Sadler}, E.~M. 1991, \mnras, 250, 786
 \href{http://adsabs.harvard.edu/cgi-bin/nph-bib_query?bibcode=1991MNRAS.250..%
786S&db_key=AST}{\adsurllinklabel}

\bibitem[{Stetson(2004)}]{stetson04}
Stetson, P. 2004, {Stetson Photometric Standard Fields}, {\em
  http://cadcwww.dao.nrc.ca/standards/}, last accessed: March 2004
 \href{{http://cadcwww.dao.nrc.ca/standards/}}{\urllinklabel}

\bibitem[{{Strauss} {et~al.}(2002){Strauss}, {Weinberg}, {Lupton}, {Narayanan},
  {Annis}, {Bernardi}, {Blanton}, {Burles}, {Connolly}, {Dalcanton}, {Doi},
  {Eisenstein}, {Frieman}, {Fukugita}, {Gunn}, {Ivezi{\' c}}, {Kent}, {Kim},
  {Knapp}, {Kron}, {Munn}, {Newberg}, {Nichol}, {Okamura}, {Quinn}, {Richmond},
  {Schlegel}, {Shimasaku}, {SubbaRao}, {Szalay}, {Vanden Berk}, {Vogeley},
  {Yanny}, {Yasuda}, {York}, \& {Zehavi}}]{strauss02}
{Strauss}, M.~A., {Weinberg}, D.~H., {Lupton}, R.~H., {Narayanan}, V.~K.,
  {Annis}, J., {Bernardi}, M., {Blanton}, M., {Burles}, S., {Connolly}, A.~J.,
  {Dalcanton}, J., {Doi}, M., {Eisenstein}, D., {Frieman}, J.~A., {Fukugita},
  M., {Gunn}, J.~E., {Ivezi{\' c}}, {\v Z}., {Kent}, S., {Kim}, R.~S.~J.,
  {Knapp}, G.~R., {Kron}, R.~G., {Munn}, J.~A., {Newberg}, H.~J., {Nichol},
  R.~C., {Okamura}, S., {Quinn}, T.~R., {Richmond}, M.~W., {Schlegel}, D.~J.,
  {Shimasaku}, K., {SubbaRao}, M., {Szalay}, A.~S., {Vanden Berk}, D.,
  {Vogeley}, M.~S., {Yanny}, B., {Yasuda}, N., {York}, D.~G., \& {Zehavi}, I.
  2002, \aj, 124, 1810
 \href{http://adsabs.harvard.edu/cgi-bin/nph-bib_query?bibcode=2002AJ....124.1%
810S&amp;db_key=AST}{\adsurllinklabel}

\bibitem[{{Strolger} {et~al.}(2004){Strolger}, {Riess}, {Dahlen}, {Livio},
  {Panagia}, {Challis}, {Tonry}, {Filippenko}, {Chornock}, {Ferguson},
  {Koekemoer}, {Mobasher}, {Dickinson}, {Giavalisco}, {Casertano}, {Hook},
  {Blondin}, {Leibundgut}, {Nonino}, {Rosati}, {Spinrad}, {Steidel}, {Stern},
  {Garnavich}, {Matheson}, {Grogin}, {Hornschemeier}, {Kretchmer}, {Laidler},
  {Lee}, {Lucas}, {de Mello}, {Moustakas}, {Ravindranath}, {Richardson}, \&
  {Taylor}}]{strolger04}
{Strolger}, L.~., {Riess}, A.~G., {Dahlen}, T., {Livio}, M., {Panagia}, N.,
  {Challis}, P., {Tonry}, J.~L., {Filippenko}, A.~V., {Chornock}, R.,
  {Ferguson}, H., {Koekemoer}, A., {Mobasher}, B., {Dickinson}, M.,
  {Giavalisco}, M., {Casertano}, S., {Hook}, R., {Blondin}, S., {Leibundgut},
  B., {Nonino}, M., {Rosati}, P., {Spinrad}, H., {Steidel}, C.~C., {Stern}, D.,
  {Garnavich}, P.~M., {Matheson}, T., {Grogin}, N., {Hornschemeier}, A.,
  {Kretchmer}, C., {Laidler}, V.~G., {Lee}, K., {Lucas}, R., {de Mello}, D.,
  {Moustakas}, L.~A., {Ravindranath}, S., {Richardson}, M., \& {Taylor}, E.
  2004, ArXiv Astrophysics e-prints


\bibitem[{{Tammann}(1970)}]{tammann70}
{Tammann}, G.~A. 1970, \aap, 8, 458
 \href{http://adsabs.harvard.edu/cgi-bin/nph-bib_query?bibcode=1970A%26A.....8%
..458T&db_key=AST}{\adsurllinklabel}

\bibitem[{{Tegmark} {et~al.}(2003){Tegmark}, {Strauss}, {Blanton}, {Abazajian},
  {Dodelson}, {Sandvik}, {Wang}, {Weinberg}, {Zehavi}, {Bahcall}, {Hoyle},
  {Schlegel}, {Scoccimarro}, {Vogeley}, {Berlind}, {Budavari}, {Connolly},
  {Eisenstein}, {Finkbeiner}, {Frieman}, {Gunn}, {Hui}, {Jain}, {Johnston},
  {Kent}, {Lin}, {Nakajima}, {Nichol}, {Ostriker}, {Pope}, {Scranton},
  {Seljak}, {Sheth}, {Stebbins}, {Szalay}, {Szapudi}, {Xu}, \&
  {others}}]{tegmark03}
{Tegmark}, M., {Strauss}, M., {Blanton}, M., {Abazajian}, K., {Dodelson}, S.,
  {Sandvik}, H., {Wang}, X., {Weinberg}, D., {Zehavi}, I., {Bahcall}, N.,
  {Hoyle}, F., {Schlegel}, D., {Scoccimarro}, R., {Vogeley}, M., {Berlind}, A.,
  {Budavari}, T., {Connolly}, A., {Eisenstein}, D., {Finkbeiner}, D.,
  {Frieman}, J., {Gunn}, J., {Hui}, L., {Jain}, B., {Johnston}, D., {Kent}, S.,
  {Lin}, H., {Nakajima}, R., {Nichol}, R., {Ostriker}, J., {Pope}, A.,
  {Scranton}, R., {Seljak}, U., {Sheth}, R., {Stebbins}, A., {Szalay}, A.,
  {Szapudi}, I., {Xu}, Y., \& {others}, . 2003, ArXiv Astrophysics e-prints
 \href{http://adsabs.harvard.edu/cgi-bin/nph-bib_query?bibcode=2003astro.ph.10%
723T&amp;db_key=PRE}{\adsurllinklabel}

\bibitem[{{The CFHTLS Supernova Program}(2004)}]{cfhls}
{The CFHTLS Supernova Program}. 2004, Canada-France-Hawaii Legacy Survey
 \href{http://cfht.hawaii.edu/Science/CFHLS/}{\urllinklabel}

\bibitem[{{The ESSENCE Team}(2003)}]{essence}
{The ESSENCE Team}. 2003, {Equation of State: SupErNovae trace Cosmic
  Expansion}
 \href{http://www.ctio.noao.edu/essence/}{\urllinklabel}

\bibitem[{{The QUEST Collaboration}(2003)}]{quest}
{The QUEST Collaboration}. 2003, The Palomar-QUEST Variability Survey
 \href{http://hepwww.physics.yale.edu/quest/palomar.html}{\urllinklabel}

\bibitem[{{Turatto} {et~al.}(2000){Turatto}, {Suzuki}, {Mazzali}, {Benetti},
  {Cappellaro}, {Danziger}, {Nomoto}, {Nakamura}, {Young}, \&
  {Patat}}]{turatto00}
{Turatto}, M., {Suzuki}, T., {Mazzali}, P.~A., {Benetti}, S., {Cappellaro}, E.,
  {Danziger}, I.~J., {Nomoto}, K., {Nakamura}, T., {Young}, T.~R., \& {Patat},
  F. 2000, \apjl, 534, L57
 \href{http://adsabs.harvard.edu/cgi-bin/nph-bib_query?bibcode=2000ApJ...534L.%
.57T&db_key=AST}{\adsurllinklabel}

\bibitem[{{Turner}(2001)}]{turner01}
{Turner}, M.~S. 2001, \pasp, 113, 653
 \href{http://adsabs.harvard.edu/cgi-bin/nph-bib_query?bibcode=2001PASP..113..%
653T&amp;db_key=AST}{\adsurllinklabel}

\bibitem[{{Ulrich} {et~al.}(1997){Ulrich}, {Maraschi}, \& {Urry}}]{ulrich97}
{Ulrich}, M., {Maraschi}, L., \& {Urry}, C.~M. 1997, \araa, 35, 445
 \href{http://adsabs.harvard.edu/cgi-bin/nph-bib_query?bibcode=1997ARA%26A..35%
..445U&db_key=AST}{\adsurllinklabel}

\bibitem[{{U.S. Department of Energy}(2004)}]{esnet}
{U.S. Department of Energy}. 2004, {Energy Sciences Network}, {\em
  http://www.es.net/}
 \href{http://www.es.net/}{\urllinklabel}

\bibitem[{{V\'eron-Cetty} \& {V\'eron}(2001)}]{veron-cetty01}
{V\'eron-Cetty}, M.-P. \& {V\'eron}, P. 2001, {A Catalog of Quasars and Active
  Nuclei (10th Edition)}, {\em
  http://www.obs-hp.fr/www/catalogues/veron2\_10/veron2\_10.html}, last
  accessed: July 2003
 \href{{http://www.obs-hp.fr/www/catalogues/veron2_10/veron2_10.html}}{\urllinklabel}

\bibitem[{{Wagner} \& {Witzel}(1995)}]{wagner95}
{Wagner}, S.~J. \& {Witzel}, A. 1995, \araa, 33, 163
 \href{http://adsabs.harvard.edu/cgi-bin/nph-bib_query?bibcode=1995ARA%26A..33%
..163W&db_key=AST}{\adsurllinklabel}

\bibitem[{{Wang} {et~al.}(2003{\natexlab{a}}){Wang}, {Baade}, {H{\" o}flich},
  {Khokhlov}, {Wheeler}, {Kasen}, {Nugent}, {Perlmutter}, {Fransson}, \&
  {Lundqvist}}]{wang03a}
{Wang}, L., {Baade}, D., {H{\" o}flich}, P., {Khokhlov}, A., {Wheeler}, J.~C.,
  {Kasen}, D., {Nugent}, P.~E., {Perlmutter}, S., {Fransson}, C., \&
  {Lundqvist}, P. 2003{\natexlab{a}}, \apj, 591, 1110
 \href{http://adsabs.harvard.edu/cgi-bin/nph-bib_query?bibcode=2003ApJ...591.1%
110W&db_key=AST}{\adsurllinklabel}

\bibitem[{{Wang} {et~al.}(2004){Wang}, {Baade}, {H{\" o}flich}, {Wheeler},
  {Kawabata}, \& {Nomoto}}]{wang04}
{Wang}, L., {Baade}, D., {H{\" o}flich}, P., {Wheeler}, J.~C., {Kawabata}, K.,
  \& {Nomoto}, K. 2004, \apjl, 604, L53


\bibitem[{{Wang} {et~al.}(2003{\natexlab{b}}){Wang}, {Goldhaber}, {Aldering},
  \& {Perlmutter}}]{wang03}
{Wang}, L., {Goldhaber}, G., {Aldering}, G., \& {Perlmutter}, S.
  2003{\natexlab{b}}, \apj, 590, 944
 \href{http://adsabs.harvard.edu/cgi-bin/nph-bib_query?bibcode=2003ApJ...590..%
944W&amp;db_key=AST}{\adsurllinklabel}

\bibitem[{{Wang} {et~al.}(1997){Wang}, {Hoeflich}, \& {Wheeler}}]{wang97}
{Wang}, L., {Hoeflich}, P., \& {Wheeler}, J.~C. 1997, \apjl, 483, L29+
 \href{http://adsabs.harvard.edu/cgi-bin/nph-bib_query?bibcode=1997ApJ...483L.%
.29W&db_key=AST}{\adsurllinklabel}

\bibitem[{{Webbink}(1984)}]{webbink84}
{Webbink}, R.~F. 1984, \apj, 277, 355
 \href{http://adsabs.harvard.edu/cgi-bin/nph-bib_query?bibcode=1984ApJ...277..%
355W&db_key=AST}{\adsurllinklabel}

\bibitem[{{Weller} \& {Albrecht}(2002)}]{weller02}
{Weller}, J. \& {Albrecht}, A. 2002, \prd, 65, 103512


\bibitem[{{Wheeler}(2003)}]{wheeler02}
{Wheeler}, J.~C. 2003, American Journal of Physics, 71, 11


\bibitem[{{Wheeler} \& {Harkness}(1990)}]{wheeler90}
{Wheeler}, J.~C. \& {Harkness}, R.~P. 1990, Reports of Progress in Physics, 53,
  1467
 \href{http://adsabs.harvard.edu/cgi-bin/nph-bib_query?bibcode=1990RPPh...53.1%
467W&db_key=AST}{\adsurllinklabel}

\bibitem[{{Windhorst} {et~al.}(1991){Windhorst}, {Burstein}, {Mathis},
  {Neuschaefer}, {Bertola}, {Buson}, {Koo}, {Matthews}, {Barthel}, \&
  {Chambers}}]{windhorst91}
{Windhorst}, R.~A., {Burstein}, D., {Mathis}, D.~F., {Neuschaefer}, L.~W.,
  {Bertola}, F., {Buson}, L.~M., {Koo}, D.~C., {Matthews}, K., {Barthel},
  P.~D., \& {Chambers}, K.~C. 1991, \apj, 380, 362
 \href{http://adsabs.harvard.edu/cgi-bin/nph-bib_query?bibcode=1991ApJ...380..%
362W&db_key=AST}{\adsurllinklabel}

\bibitem[{{Wood-Vasey}(2002)}]{iauc8019}
{Wood-Vasey}, W.~M. 2002, \iaucirc, 8019, 1
 \href{http://adsabs.harvard.edu/cgi-bin/nph-bib_query?bibcode=2002IAUC.8019..%
..1W&db_key=AST}{\adsurllinklabel}

\bibitem[{{Wood-Vasey} {et~al.}(2002{\natexlab{a}}){Wood-Vasey}, {Aldering},
  {Lee}, {Loken}, {Nugent}, {Perlmutter}, {Quimby}, {Siegrist}, {Wang}, {Knop},
  {Antilogus}, {Astier}, {Hardin}, {Pain}, {Copin}, {Smadja}, {Adam}, {Bacon},
  {Lemmonier}, {Pecontal}, {Kessler}, \& {Nearby Supernova Factory
  Collaboration}}]{wood-vasey03aas1}
{Wood-Vasey}, W.~M., {Aldering}, G., {Lee}, B.~C., {Loken}, S., {Nugent}, P.,
  {Perlmutter}, S., {Quimby}, R., {Siegrist}, J., {Wang}, L., {Knop}, R.~A.,
  {Antilogus}, P., {Astier}, P., {Hardin}, D., {Pain}, R., {Copin}, Y.,
  {Smadja}, G., {Adam}, G., {Bacon}, R., {Lemmonier}, J., {Pecontal}, E.,
  {Kessler}, R., \& {Nearby Supernova Factory Collaboration}.
  2002{\natexlab{a}}, American Astronomical Society Meeting, 201, 0
 \href{http://adsabs.harvard.edu/cgi-bin/nph-bib_query?bibcode=2002AAS...201.5%
608W&db_key=AST}{\adsurllinklabel}

\bibitem[{{Wood-Vasey} {et~al.}(2004{\natexlab{a}}){Wood-Vasey}, {Aldering},
  {Lee}, {Loken}, {Nugent}, {Perlmutter}, {Siegrist}, {Wang}, {Antilogus},
  {Astier}, {Hardin}, {Pain}, {Copin}, {Smadja}, {Gangler}, {Castera}, {Adam},
  {Bacon}, {Lemonnier}, {P{\' e}contal}, {P{\' e}contal}, \&
  {Kessler}}]{wood-vasey04a}
{Wood-Vasey}, W.~M., {Aldering}, G., {Lee}, B.~C., {Loken}, S., {Nugent}, P.,
  {Perlmutter}, S., {Siegrist}, J., {Wang}, L., {Antilogus}, P., {Astier}, P.,
  {Hardin}, D., {Pain}, R., {Copin}, Y., {Smadja}, G., {Gangler}, E.,
  {Castera}, A., {Adam}, G., {Bacon}, R., {Lemonnier}, J.-P., {P{\' e}contal},
  A., {P{\' e}contal}, E., \& {Kessler}, R. 2004{\natexlab{a}}, New Astronomy
  Review, 48, 637


\bibitem[{{Wood-Vasey} {et~al.}(2003{\natexlab{a}}){Wood-Vasey}, {Aldering}, \&
  {Nugent}}]{iauc8149}
{Wood-Vasey}, W.~M., {Aldering}, G., \& {Nugent}, P. 2003{\natexlab{a}},
  \iaucirc, 8149, 2
 \href{http://adsabs.harvard.edu/cgi-bin/nph-bib_query?bibcode=2003IAUC.8149..%
..2W&db_key=AST}{\adsurllinklabel}

\bibitem[{{Wood-Vasey} {et~al.}(2003{\natexlab{b}}){Wood-Vasey}, {Aldering}, \&
  {Nugent}}]{iauc8082}
---. 2003{\natexlab{b}}, \iaucirc, 8082, 1
 \href{http://adsabs.harvard.edu/cgi-bin/nph-bib_query?bibcode=2003IAUC.8082..%
..1W&db_key=AST}{\adsurllinklabel}

\bibitem[{{Wood-Vasey} {et~al.}(2003{\natexlab{c}}){Wood-Vasey}, {Aldering}, \&
  {Nugent}}]{iauc8089}
---. 2003{\natexlab{c}}, \iaucirc, 8089, 2
 \href{http://adsabs.harvard.edu/cgi-bin/nph-bib_query?bibcode=2003IAUC.8089..%
..2W&db_key=AST}{\adsurllinklabel}

\bibitem[{{Wood-Vasey} {et~al.}(2003{\natexlab{d}}){Wood-Vasey}, {Aldering}, \&
  {Nugent}}]{iauc8141}
---. 2003{\natexlab{d}}, \iaucirc, 8141, 1
 \href{http://adsabs.harvard.edu/cgi-bin/nph-bib_query?bibcode=2003IAUC.8141..%
..1S&db_key=AST}{\adsurllinklabel}

\bibitem[{{Wood-Vasey} {et~al.}(2002{\natexlab{b}}){Wood-Vasey}, {Aldering},
  {Nugent}, {Helin}, {Pravdo}, {Hicks}, \& {Lawrence}}]{iauc7902}
{Wood-Vasey}, W.~M., {Aldering}, G., {Nugent}, P., {Helin}, E.~F., {Pravdo},
  S., {Hicks}, M., \& {Lawrence}, K. 2002{\natexlab{b}}, \iaucirc, 7902, 3
 \href{http://adsabs.harvard.edu/cgi-bin/nph-bib_query?bibcode=2002IAUC.7902..%
..3W&db_key=AST}{\adsurllinklabel}

\bibitem[{{Wood-Vasey} {et~al.}(2003{\natexlab{e}}){Wood-Vasey}, {Aldering},
  {Nugent}, \& {Li}}]{iauc8053}
{Wood-Vasey}, W.~M., {Aldering}, G., {Nugent}, P., \& {Li}, K.
  2003{\natexlab{e}}, \iaucirc, 8053, 1
 \href{http://adsabs.harvard.edu/cgi-bin/nph-bib_query?bibcode=2003IAUC.8053..%
..1W&db_key=AST}{\adsurllinklabel}

\bibitem[{{Wood-Vasey} {et~al.}(2003{\natexlab{f}}){Wood-Vasey}, {Aldering},
  {Nugent}, {Mulchaey}, \& {Phillips}}]{iauc8088}
{Wood-Vasey}, W.~M., {Aldering}, G., {Nugent}, P., {Mulchaey}, J., \&
  {Phillips}, M. 2003{\natexlab{f}}, \iaucirc, 8088, 2
 \href{http://adsabs.harvard.edu/cgi-bin/nph-bib_query?bibcode=2003IAUC.8088..%
..2W&db_key=AST}{\adsurllinklabel}

\bibitem[{{Wood-Vasey} {et~al.}(2004{\natexlab{b}}){Wood-Vasey}, {Wang}, \&
  {Aldering}}]{wood-vasey04b}
{Wood-Vasey}, W.~M., {Wang}, L., \& {Aldering}, G. 2004{\natexlab{b}}, ArXiv
  Astrophysics e-prints


\bibitem[{{Young} {et~al.}(1992){Young}, {Serabyn}, {Phillips}, {Knapp},
  {Guesten}, \& {Schulz}}]{young92}
{Young}, K., {Serabyn}, G., {Phillips}, T.~G., {Knapp}, G.~R., {Guesten}, R.,
  \& {Schulz}, A. 1992, \apj, 385, 265
 \href{http://adsabs.harvard.edu/cgi-bin/nph-bib_query?bibcode=1992ApJ...385..%
265Y&amp;db_key=AST}{\adsurllinklabel}

\bibitem[{{Young} {et~al.}(1995){Young}, {Baron}, \& {Branch}}]{young95}
{Young}, T.~R., {Baron}, E., \& {Branch}, D. 1995, \apjl, 449, L51+
 \href{http://adsabs.harvard.edu/cgi-bin/nph-bib_query?bibcode=1995ApJ...449L.%
.51Y&db_key=AST}{\adsurllinklabel}

\bibitem[{{Zackrisson} \& {Bergvall}(2003)}]{zackrisson03a}
{Zackrisson}, E. \& {Bergvall}, N. 2003, \aap, 399, 23
 \href{http://adsabs.harvard.edu/cgi-bin/nph-bib_query?bibcode=2003A%26A...399%
...23Z&db_key=AST}{\adsurllinklabel}

\bibitem[{{Zackrisson} {et~al.}(2003){Zackrisson}, {Bergvall}, {Marquart}, \&
  {Helbig}}]{zackrisson03b}
{Zackrisson}, E., {Bergvall}, N., {Marquart}, T., \& {Helbig}, P. 2003, \aap,
  408, 17
 \href{http://adsabs.harvard.edu/cgi-bin/nph-bib_query?bibcode=2003A%26A...408%
...17Z&amp;db_key=AST}{\adsurllinklabel}

\bibitem[{{Zwicky}(1938)}]{zwicky38}
{Zwicky}, F. 1938, \apj, 88, 529
 \href{http://adsabs.harvard.edu/cgi-bin/nph-bib_query?bibcode=1938ApJ....88..%
529Z&db_key=AST}{\adsurllinklabel}

\bibitem[{{Zwicky}(1942)}]{zwicky42}
---. 1942, \apj, 96, 28
 \href{http://adsabs.harvard.edu/cgi-bin/nph-bib_query?bibcode=1942ApJ....96..%
.28Z&db_key=AST}{\adsurllinklabel}

\end{thebibliography}

\dsp


\newpage
\appendix
\part{Appendices}
\label{part:appendices}

\chapter{SNfactory Summary}
\label{apx:snfactory_summary}

\noindent
{\bf [The following appeared in New Astronomy Review as \citet{wood-vasey04a}.]}

\section{Abstract}The Nearby Supernova Factory (SNfactory) is an ambitious project to
find and study in detail approximately 300 nearby Type Ia supernovae (SNe~Ia)
at redshifts \linebreak
\mbox{$0.03<z<0.08$}.  This program will provide an
exceptional data set of well-studied SNe in the nearby smooth
Hubble flow that can be used as calibration for the current and
future programs designed to use SNe to measure the cosmological
parameters.
The first key ingredient for this program is a reliable supply of
Hubble-flow SNe systematically discovered in unprecedented
numbers using the same techniques as those used in distant SNe
searches.  In 2002, 35 SNe were found using our test-bed
pipeline for automated SN search and discovery. The pipeline
uses images from the asteroid search conducted by the Near Earth
Asteroid Tracking group at JPL.  Improvements in our subtraction
techniques and analysis have allowed us to increase our effective
SN discovery rate to $\sim$12 SNe/month in 2003.


\section{Introduction}

Type Ia supernovae (SNe~Ia) have proven extremely useful as
standardizable candles to explore the expansion history of the
Universe~\citep{perlmutter97a,perlmutter98a,garnavich98a,riess98a,perlmutter99}.
Ambitious follow-on experiments
are just starting, SNLS~\citep{pain03}, ESSENCE~\citep{garnavich02},
or have been planned, SNAP~\citep{aldering02a},
to extend the revolutionary result that the Universe is accelerating
to precise statements about the constituents and history of the
Universe.  However, a key assumption underlying these experiments is
that the current observed diversity in SNe~Ia will be well-behaved and
calibrated to allow for the desired precision measurements.
This assumption can be tested in large part using nearby SNe~Ia.
Moreover, there is the possibility for such nearby studies to uncover
new relationships that will make SNe~Ia even better
standard candles, much as the width-luminosity relation has brought
SNe~Ia to their already impressive level of standardization.

The Nearby Supernova Factory (SNfactory) 
project is designed to bring this improved understanding of
SNe~Ia~\citep{aldering02b, pecontal03,lantz03}.  Over the course of
three years, it will study 300~SNe~Ia in the nearby smooth Hubble
flow~\ref{fig:sweet_spot}.  These SNe will be observed with a dedicated instrument, the
SuperNova Integral Field Spectrograph (SNIFS), which is currently in
the final stages of construction.  SNIFS will provide simultaneous
spectrophotometric coverage of both the SN and the host galaxy at
3--5~day intervals during the rise and fall of each SN.  This
unprecedented dataset will provide a wealth of information on SNe Ia
and allow for an improved calibration of SNe~Ia for use in cosmology.

\section{The Supernova Search Dataset}

The SNfactory searches for SNe using wide-field images obtained in
collaboration with the Near Earth Asteroid Tracking (NEAT)
group~\citep{pravdo99} at the Jet Propulsion
Laboratory (JPL).  In their quest for asteroids, the NEAT
group observes hundreds of fields each night by taking three images
of each field, spaced fifteen to thirty minutes apart, and searching for
objects that move by more than a couple of arcseconds over this
period.  The SNfactory uses this temporal spacing to
eliminate asteroids and reduce cosmic-ray contamination.
The NEAT observing program covers the observable sky from $-40$ to
$+40$ degrees in declination every one to two weeks.

Since 2001, the Palomar 1.2-m. Oschin telescope has been used 
by
the SNfactory-NEAT collaboration.  The NEAT group outfitted this
telescope with an automated system for control and observations and
added a 3-chip, 3~\sqdeg~field-of-view CCD camera (NEAT12GEN2) at the
spherical focal plane of this Schmidt reflector.  We spent 2001
designing, testing, and verifying the search pipeline with
the large data stream from this telescope.  Full search operations
began in the fall of 2002 and continued through April of 2003 when the
NEAT12GEN2 camera was replaced by the Yale QUEST group~\citep{quest}
with a 112-chip, 9~\sqdeg~field-of-view camera (QUESTII) capable of
both drift-scan and point-and-track observations.  QUESTII became
operational in August 2003 and is now providing data for the SNfactory search.  The NEAT group also uses the Haleakala
1.2-m MSSS telescope, but as this telescope has a smaller field of
view and poor image quality, the SNfactory has focused its search
efforts on the images from the Palomar Oschin telescope.

\section{Data Processing}

The SNfactory, in collaboration with the High Performance Wireless
Research and Education Network~\citep{hpwren}, has
established a high-speed, 6~Megabyte-per-second (MBps) radio internet
link to the San Diego Supercomputer Center (SDSC) from the Palomar
observatory.  Images are transmitted from the telescope and stored at
the National Energy Research Supercomputing Center (NERSC) High
Performance Storage System (HPSS).  The bandwidth from SDSC to
NERSC allows for near real-time transfer of 20--50~GB per night. 
 From HPSS the data are transferred to the 200-node NERSC Parallel
Distributed Systems Facility (PDSF) 
and
are submitted for simultaneous processing on the PDSF cluster
in groups based on the dark frame taken
closest in time.
The processed images are then registered with our
image database, renamed to match our canonical name format,
and saved to HPSS.

The SNfactory searches these data for SNe using 
image subtraction.  
The computers use a
sophisticated suite of image tools, but there
remains a significant, although ever-decreasing, amount of human interpretation needed to discriminate
the good SN candidates from the bad.
The same software used by the SNfactory to search for nearby (z $\le
0.1$) SNe is used by the Supernova Cosmology Project (SCP) to search
for distant SNe.
Our subtraction software begins by spatially registering all of the
images of a given field to a common reference system.
The images to be used as a reference are
stacked into a single coadded image,
while the images to be searched are coadded into
two separate images to allow for later checks 
for asteroids and cosmic rays. 
To account for differences in the effective point-spread-function from
variations in atmospheric and telescope conditions, we
calculate a convolution kernel to match each coadded image
to the worst-seeing coadded image of the set. 
An automated scanning program
looks for objects in the convolution-matched, subtracted
image and applies a variety of selection criteria 
to eliminate cosmic rays, asteroids, and subtraction and detector artifacts.
This program compiles a list of interesting candidates to be looked at in more
detail by a human scanner.
Every day, human scanners consult the list of interesting candidates
and decide whether or not a computer-flagged candidate appears to be a real, variable object.
Once we have a promising candidate, we submit it to the
target list for the next night of observation with NEAT.  
After obtaining confirmation images of a candidate that
reveal it to exhibit the appropriate behavior for a SN, we announce
the apparent SN in the IAU Circulars.  A confirmation spectrum is
desirable but currently not always possible.  When the SNIFS
instrument is installed on the Hawaii 2.2-m telescope, it will be
automatically scheduled to confirm and follow SNe.  Spectra will be
taken of each SN at 3--5~day intervals over a span of roughly 60~days.

\section{Results}

\begin{figure}
\plotone{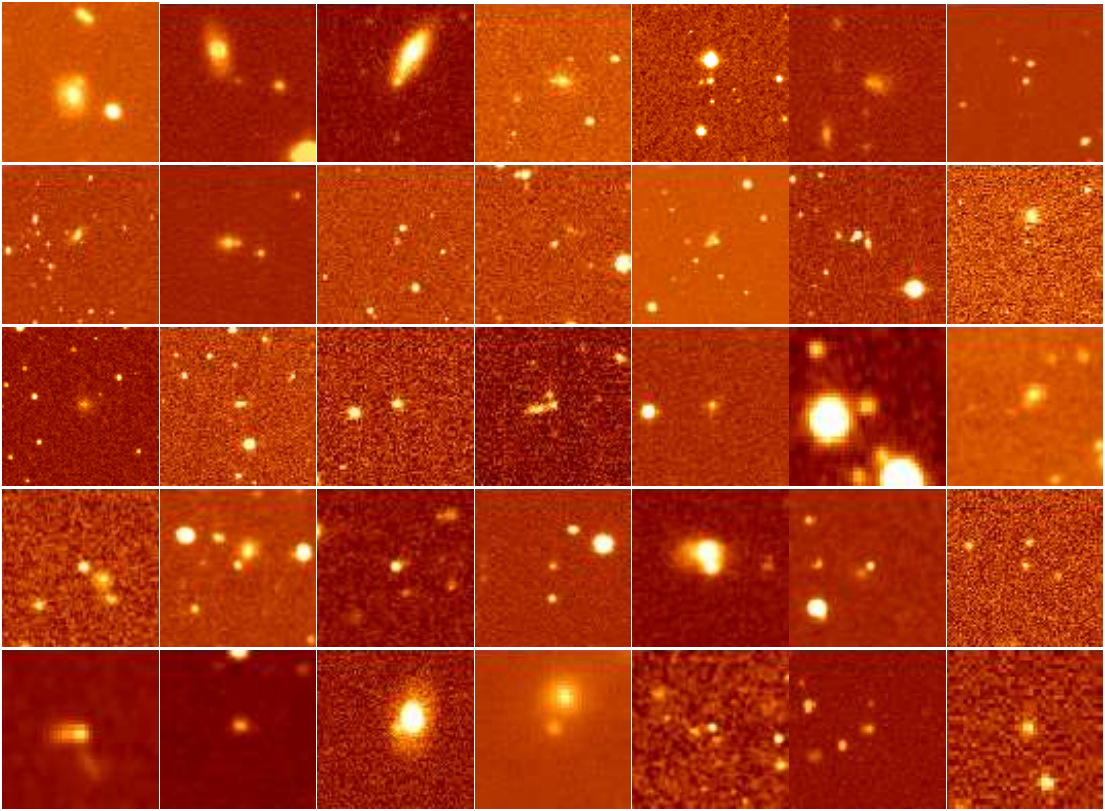}
\caption{
A mosaic of the 35 supernovae found by the Nearby Supernova
Factory pipeline in 2002.  Each image is centered on the respective supernova.
}
\label{fig:sne_mosaic}
\end{figure}
\begin{figure}
\plotone{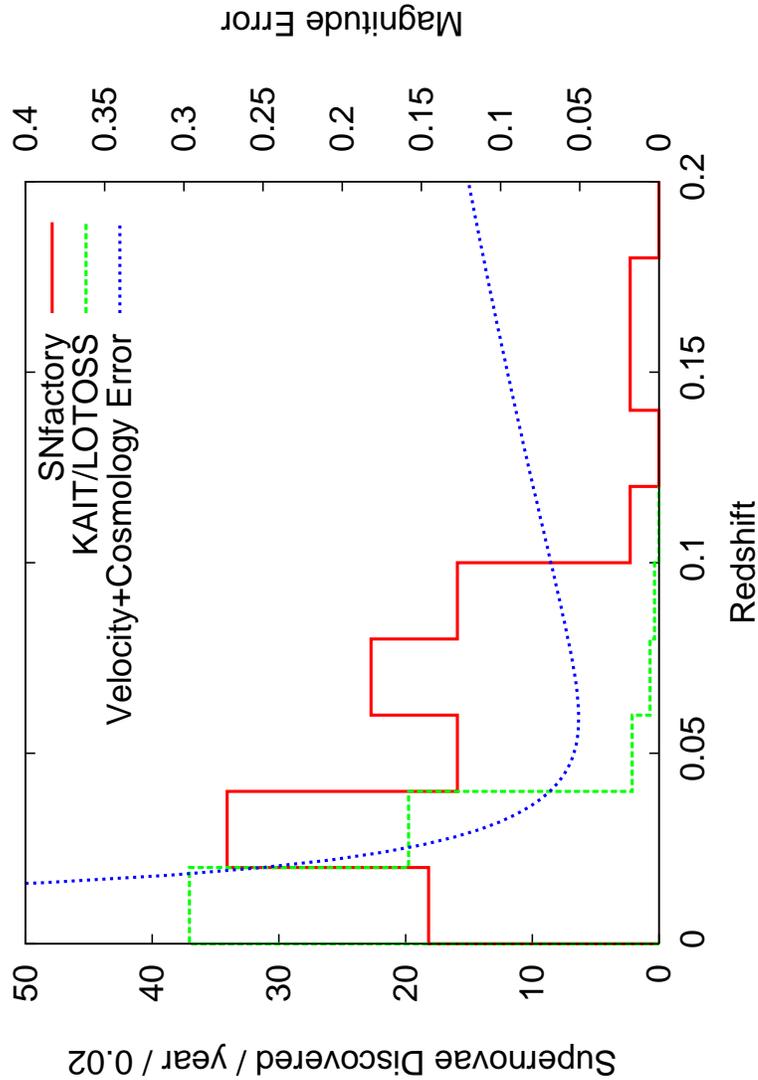}
\caption{
The SNfactory is operating in the ``sweet spot'' redshift range
between peculiar-velocity noise and cosmological uncertainty.  The
SNfactory curve is the redshift distribution of supernovae found and
spectroscopically confirmed in our search to date scaled up to 100
SNe/year.  The velocity error is for an assumed $300$~km/s velocity
dispersion.  The cosmology error is modeled as the difference between
an Einstein-de Sitter cosmology and a Universe with $\Omega_M=0.3$ and
$\Omega_\Lambda=0.7$.}
\label{fig:sweet_spot}
\end{figure}

Eighty-three SNe have been found using the techniques described
above and have been accepted by the International Astronomical Union.
Fig.~\ref{fig:sne_mosaic} shows the 35 SNe we discovered in
2002.  In the first five months of 2003, we found and reported an
additional 48 SNe.  In addition, our search identified
another 17 SNe that had already been reported by other groups.
We are currently running with a very conservative set of selection
criteria and need human eyes to scan $\sim5$\% of the successful
subtractions for each night.  However, many of the SNe discovered in
2003 could have been found using very restrictive criteria that
would have required human scanning for $<1$\% of the subtractions.  
To further reduce our scanning
burden, efforts are ongoing to understand the parameter space of our
candidate scores to uniquely identify the SNe.

In order to understand our search results, we have developed a simple search
simulator.  We use a V-band lightcurve template from
\citet{goldhaber01} to model the rise and fall of SNe~Ia.\footnote{We
have found that the NEAT unfiltered magnitudes track V-band SN
lightcurves quite well.}  The simulator calculates the amount of time
a supernova would be visible at a given redshift assuming a normal
(stretch~$=1$) SN~Ia. We considered a redshift range from $0<z<0.2$,
a limiting unfiltered magnitude of 19.5, and 
several different repeat coverage cadences for our simulations.  We
included the typical NEAT sky coverage rate, $S_D = 500~\sqdeg$/night,
and assumed $S_O=$10,000$~\sqdeg$ of usable sky a night.  There is a
maximum effective cadence, $C_\mathrm{max}$, beyond which one is just
idling the telescope: $C_\mathrm{max} = S_O / ( S_D -
\frac{S_0}{365~\mathrm{days}})$.  
For the NEAT observation program, this 
cadence is approximately 20 days.  

\begin{figure}
\plotone{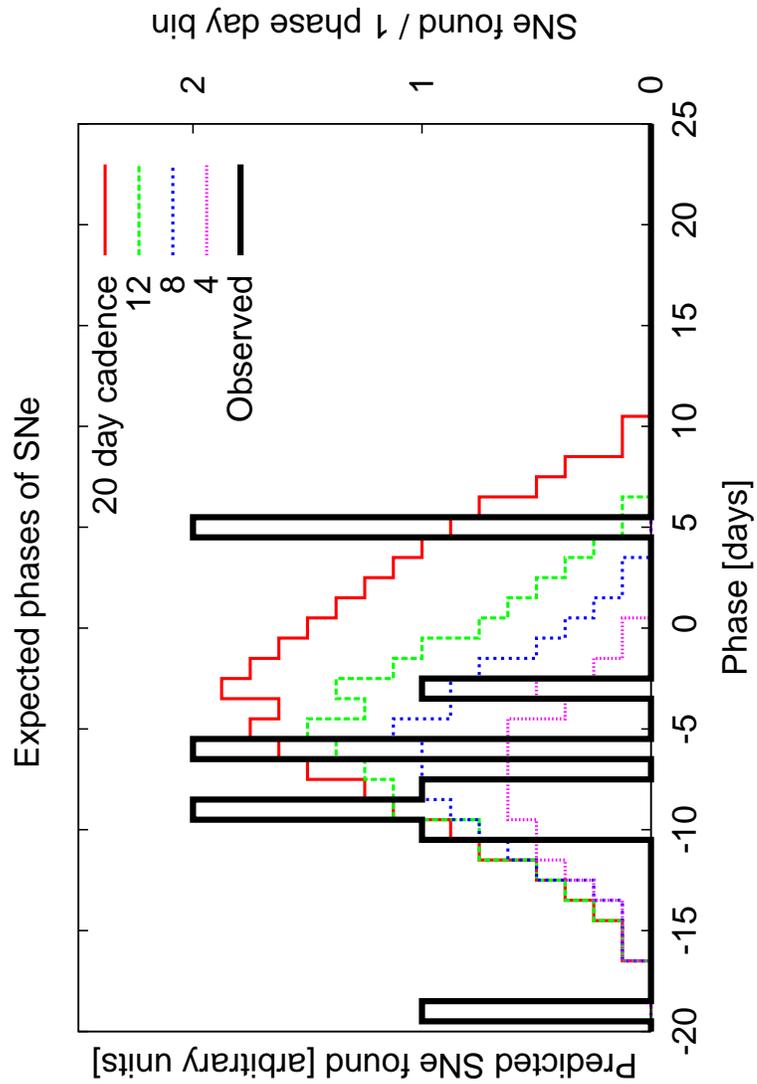}
\caption{
The distribution of discovery epoch for the SNe~Ia with determined dates of maximum found in the
SNfactory dataset compared with the
simulations described in the text.  The model curves are not calculated in
absolute units and have been scaled for comparison with the
observations.  As the model curves show, a shorter cadence gives fewer
supernovae but better constraints on the epoch at discovery.  }
\label{fig:compare_phase}
\end{figure}

An important goal of the SNfactory is to discover SNe early enough
after explosion to follow them through their rise and fall.
Fig.~\ref{fig:compare_phase} shows a comparison of our simulations of
the SNe discovery phase with
the actual discovery phase for SNe from the SNfactory search having
well-known dates of maximum.  For shorter sky coverage cadences
less sky can be covered and so fewer SNe are discovered overall,
although the epoch of each SN can be better constrained.
Fig.~\ref{fig:compare_phase} clearly shows that the SNfactory search
pipeline is successfully finding SNe~Ia early in their lightcurves.


\section{Conclusion}

The Nearby Supernova Factory search pipeline is operational and has
proven the ability to discover $\sim12$~SNe/month $\Rightarrow$
$\sim$150~SNe/year.  As $\frac{2}{3}$ of the supernovae discovered in
our search have been SNe~Ia, we expect to discover $\sim100$~SNe~Ia/year.
Most of these supernovae have been discovered sufficiently early to enable
detailed study starting before maximum light.  This extensive study will enable
improvements in the use of SNe~Ia for cosmological measurements,
and provide a wealth of information on the supernovae themselves.


\chapter{QSOs and Microlensing}
\label{apx:qso_microlensing}

\newcommand{\numQSOpoints}{20309}
\newcommand{\numQSOdiffs}{49489}
\newcommand{\numQSOs}{982}
\newcommand{\numQSOmagRADec}{3131}
\newcommand{\QSOdev}{0.57}
\newcommand{\QSOshortdev}{0.48}
\newcommand{\QSOyeardev}{0.52}
\newcommand{\QSOlongdev}{0.65}

\section{Introduction}

Quasi-stellar objects (QSOs) were discovered by Sandage and
Mathews over forty years ago~\citep{sandage61,burbidge67,schmidt69} and were eventually
understood as a class of active galactic nuclei
(AGN)~\citep{antonucci93}.  Due to historical convention,
there are a variety of names for these AGNs according to 
the different observed properties: 
quasars (radio-loud and radio-quiet), X-ray galaxies, Seyfert
galaxies, low ionization nuclear emission region (Liner) galaxies, and
broad absorption line objects (BALs).  The ``quasar'' designation has
generally been used for unresolved, point-like objects, while the
various galaxy classifications have been assigned when there is a resolved
object.  But all of these classifications share the common model of a
massive central engine that powers energetic radiation that is subject to
variability over both short and long time scales.  I will refer to the
general class of these objects as QSOs in this study.

QSOs have been known to exhibit variability in their flux since 
soon after their discovery~\citep{burbidge67,schmidt69,wagner95,ulrich97}.  This variability
could be due to causes either intrinsic or extrinsic to the active
galactic nucleus itself.  One possible candidate for a source of
extrinsic variability is microlensing by compact objects between the
QSO and the observer~\citep{schneider93}.
\citet{schneider93} presented work constraining the composition of
possible compact objects given the observed statistical variations in
the lightcurves of QSOs.  \citet{zackrisson03a,zackrisson03b}
(hereafter referred to as ZBZ) found that a microlensing model
could explain the observed average structure function of QSOs, but 
that it did
not explain  the
scale of variability for nearby QSOs ($z<1$)
or the high variability ($> 0.35$~mag) at any redshift.  
They concluded that
while microlensing may be a possible explanation for some part of the
observed QSO variability, there must be another mechanism, possibly
intrinsic, at work to explain all of the observed fluctuations.
ZBZ focused on QSO variability on the
scale of one-to-ten years and averaged over smaller, intra-year
variability.  The NEAT dataset provides a new opportunity to investigate
variations on the time scale of a week to a few years.  This chapter
details the results of an investigation of this relatively unexplored
region of the QSO time-variability power spectrum.


\section{The NEAT data set}

The NEAT data was originally analyzed for QSO variability
by Roland Rudas for his senior thesis in Physics at UC Berkeley
(2003).  He analyzed $229$~QSOs visible in the NEAT survey and
developed initial numbers on the $\Delta t$ vs. $\Delta m$
distribution.  I have built on that work by extending the QSO sample to
\numQSOs~objects, computing the time-variability power spectrum from the light curves, calculating the variation amplitude probability density function and 
investigating the effect of a median versus mean algorithm for
computing the power spectrum and other measures of variability.

\section{QSO Catalogs}

The QSO catalogs of \citet{veron-cetty01}~(VC01) 
and \citet{schneider03}~(S03)
were used as the source of QSOs for this work.
The $40,473$ QSOs presented in these catalogs ($23,760$ from VC01 plus $16,713$ from S03) were
checked against the NEAT dataset for QSO variability.
Out of those $40,473$~QSOs, \numQSOmagRADec~were covered by images from the NEAT survey and were brighter than $V<20$ (VC01) or $g<20$ (S03).
Requiring a signal-to-noise $\ge 5$ for each light curve points (see below), 
left a final total of \numQSOs~QSOs for this analysis.

\section{QSO Photometry}

For each QSO matching the criteria above, all of the NEAT images 
covering its coordinates were retrieved from 
long-term storage on the NERSC HPSS system (see Chapter~\ref{chp:search_pipeline}) and
and calibrated against the
{USNO-A1.0} catalog~\citep{usnoa1} ``R''-band magnitudes.  While
the {USNO-A1.0} catalog suffers from plate-to-plate uncertainties
of $0.2$ magnitudes (see Appendix~\ref{apx:calibration} for more details on the NEAT calibration), this should not affect the results
presented here as this is a differential analysis and all of the 
NEAT images of a given QSO substantially, if not completely, overlap with the same
plate---each POSS plate is $6.5\deg\times6.5\deg$ while the NEAT images are at most $0.25$\deg on a side---so the relative calibration for a given QSO is almost always derived from within the same POSS plate.
Each image was photometered using aperture
photometry scaled to the observed x- and y-FWHM measured for the image.
No extinction corrections were performed.
The magnitudes measured from all of the images for a given QSO were saved in
a light curve file for the QSO for later analysis.

\begin{figure}
\includegraphics[angle=270]{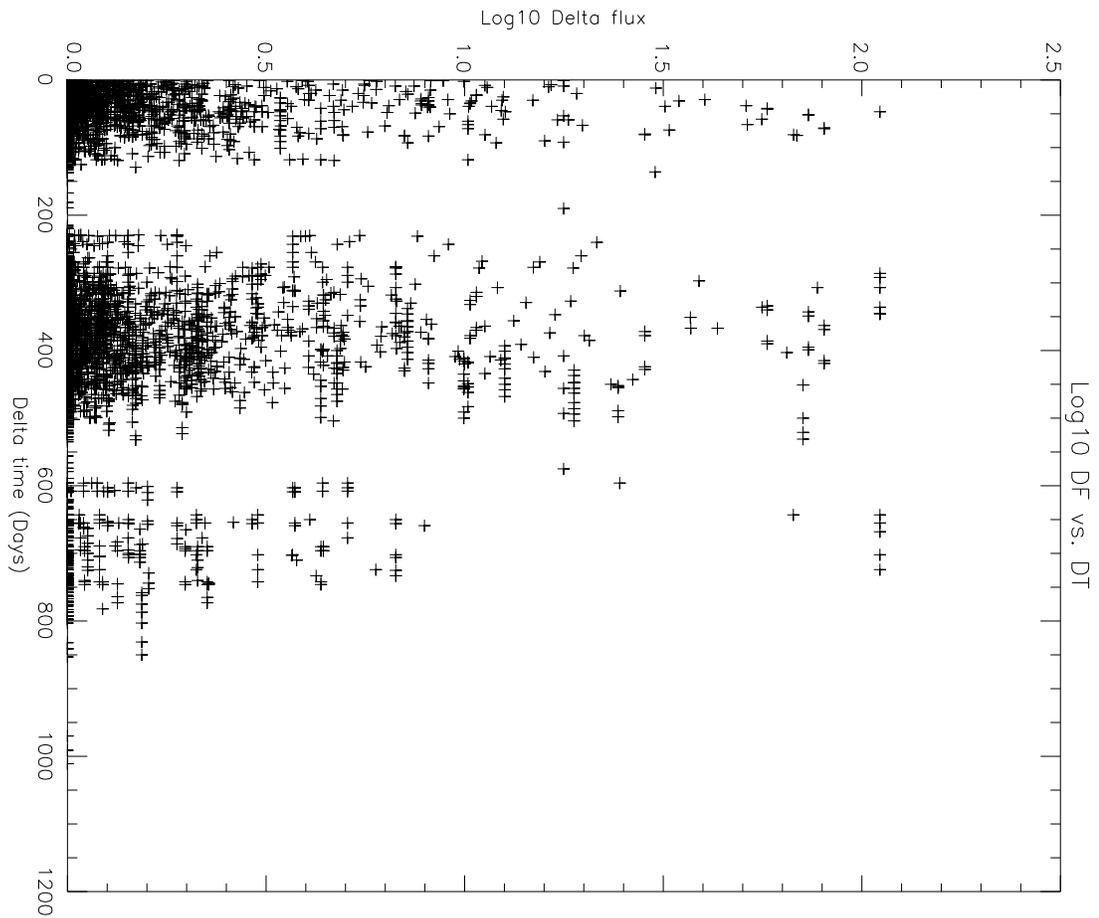}
\caption{A scatter plot of $\log_{10}{\Delta f}$ versus $\Delta t$ from the QSOs covered in the NEAT dataset.}
\label{fig:qso_df_dt}
\end{figure}

\begin{figure}
\plotone{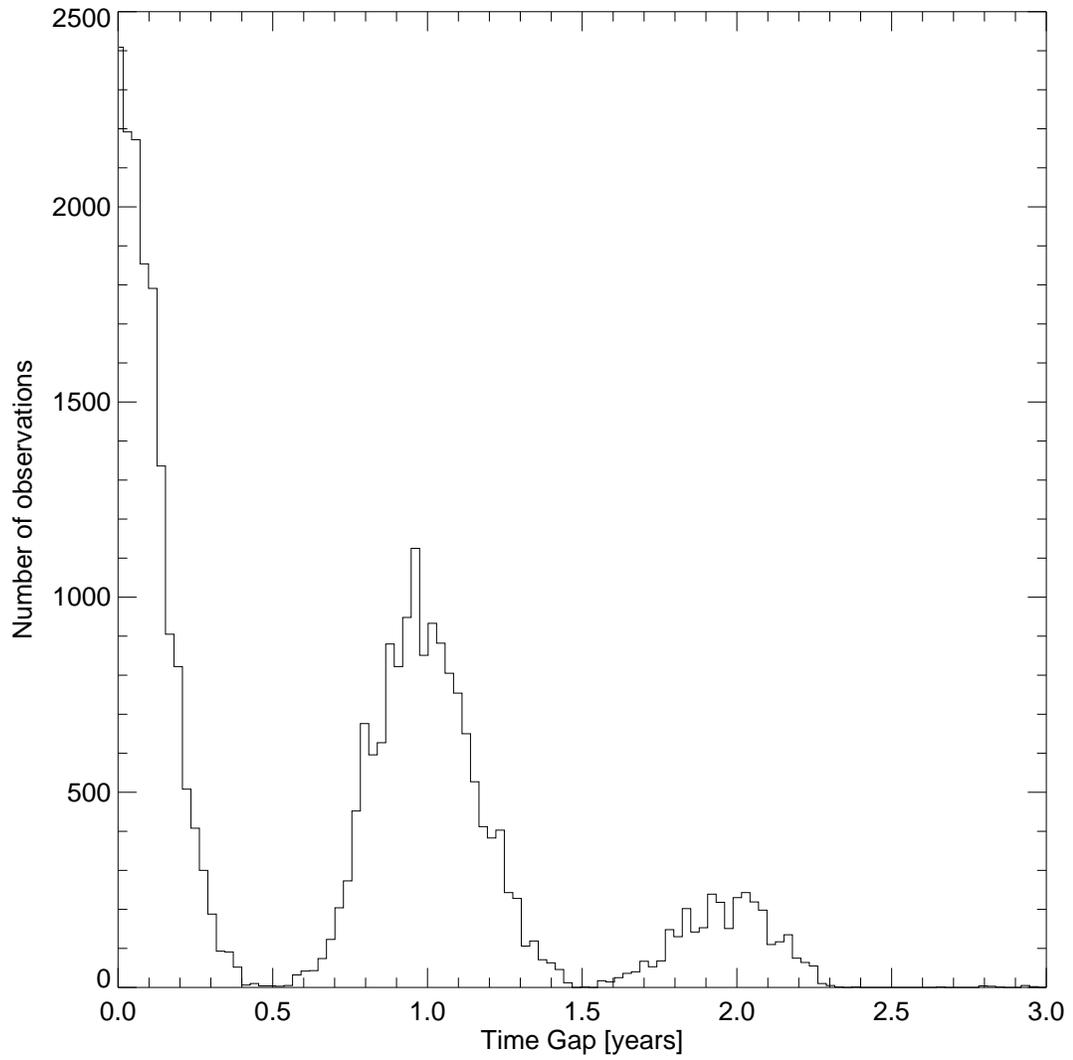}
\caption{Histogram of observed time baselines, $\Delta t$.}
\label{fig:hist_of_dt}
\end{figure}

\begin{figure}
\plotone{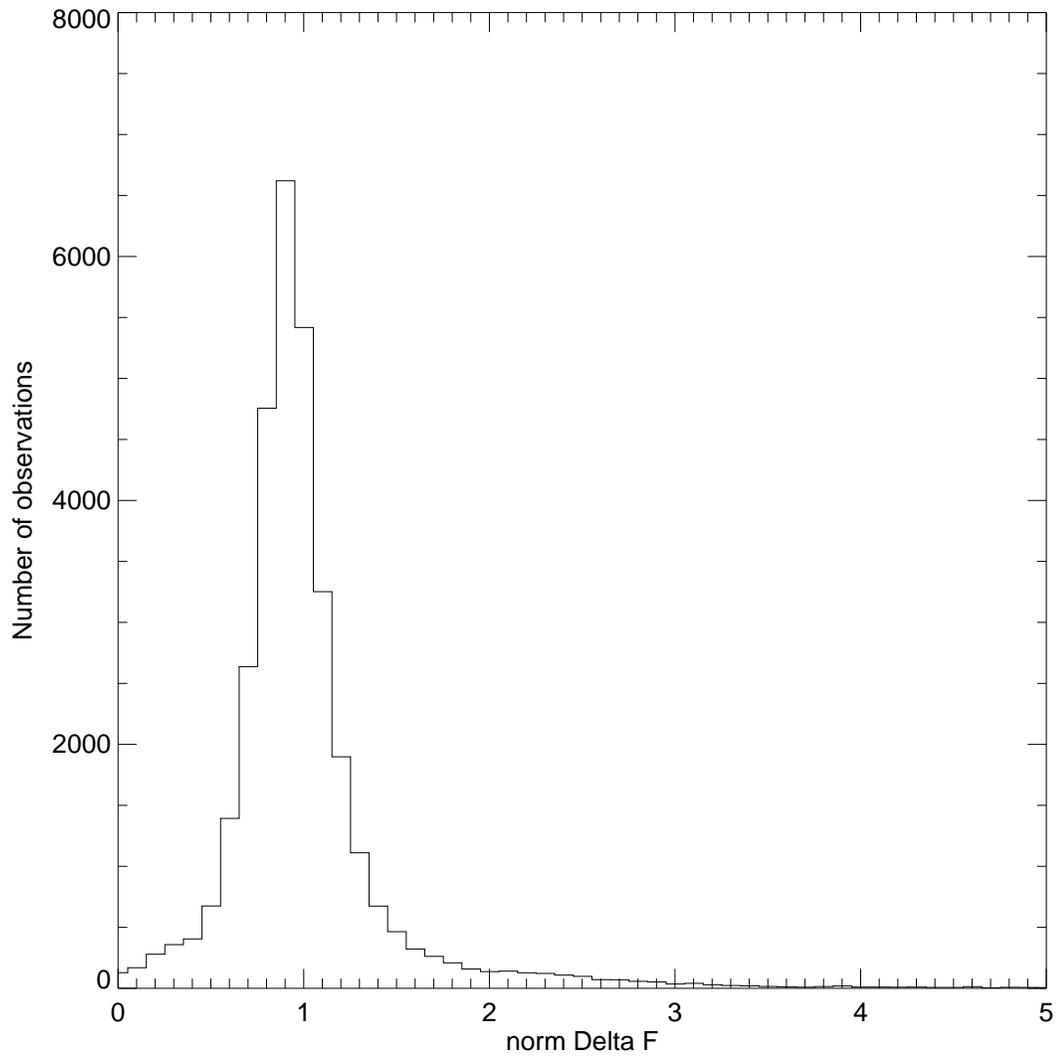}
\caption{Histogram of observed relative change in flux, $\Delta f = \frac{f_j}{f_i}$.}
\label{fig:hist_of_df}
\end{figure}

In the analysis of the light curve properties,
data from the same night was medianed together.
This median step was served to eliminate outliers derived from bad images.
As most observations were from triplets of images during a given night,
a good data point was obtained for most nights of observation.
Fig.~\ref{fig:qso_df_dt} shows a scatter plot of $\Delta f$ versus
$\Delta t$. 
Figs.~\ref{fig:hist_of_dt}~\&~\ref{fig:hist_of_df} show the one-dimensional histograms from Fig.~\ref{fig:qso_df_dt} for $\Delta t$ and $\Delta f$, respectively.
Figs.~\ref{fig:qso_structure_function_avg}, \ref{fig:qso_structure_function_med}, \ref{fig:qso_structure_function_avg_log}, and \ref{fig:qso_structure_function_med_log} show the power spectrum
histogram for $\Delta f$.
See Fig~\ref{fig:qso_dm2_mean_med} for a histogram of the magnitude fluctuations observed in my sample.

\section{Power Spectrum of QSO Variability}

A generalized analysis of variability naturally leads to a power-spectrum analysis.
The flux of the
variability at each time scale is often a predicted characteristic
of modeling of the phenomenon.  As QSO variability is poorly understood,
a power spectrum analysis is a useful way of comparing different 
physical models whose effect on the power-spectrum can be calculated theoretically.  ZBZ present their results in the context of a 
power spectrum.
Fig.~\ref{fig:qso_structure_function_avg} and
Fig.~\ref{fig:qso_structure_function_med} show the QSO structure
function as calculated from the NEAT light curve using Eq.~2 of ZBZ 
(rewritten as Eq.~\ref{eq:qso_structure_function_avg} in this text).
The QSO structure function, $S(\tau)$, is defined as
\begin{equation}
S(\tau) = \sqrt{ \mathrm{mean}\left( \Sigma_{i<j} \left( m_j - m_i \right)^2 \right) }
\label{eq:qso_structure_function_avg}
\end{equation}
\begin{equation}
S(\tau) = \sqrt{ \mathrm{median}\left( \Sigma_{i<j} \left( m_j - m_i \right)^2 \right) }
\label{eq:qso_structure_function_med}
\end{equation}

ZBZ
calculated the average structure 
function by taking the average of Eq.~\ref{eq:qso_structure_function_avg}
over each QSO.  This approach suffers the disadvantage of weighting
all of the QSOs used in the analysis equally even if they have
significantly different numbers of data points in their lightcurves.
For the ZBZ analysis, their coverage relatively homogeneous and so this was less of an issue.
In the analysis of the NEAT data, the number of data points contributed by each QSO
varied significantly so I chose to consider all of the
$\Delta t$ and $\Delta f$ values together and calculated my
statistics treating each of those pairings equally.
Since I am testing the hypothesis that QSO variability 
is caused by microlensing events, which are caused by
physical conditions uncorrelated from the QSOs, this treatment is valid.

Statistical comparisons of $\Delta f$ versus $\Delta t$ and
$m$ versus $\Delta t$ were made for each QSO where
\begin{eqnarray}
\Delta f & = & \frac{f_j}{f_i} \\
\Delta m & = & -2.5 \log_{10}{\Delta f} \\
\Delta t & = & t_j - t_i \\
& & \forall~i < j.
\end{eqnarray}
A total of of \numQSOpoints~measurements and \numQSOdiffs~differences were obtained
from this process.

\section{Discussion}

\begin{figure}
\plotone{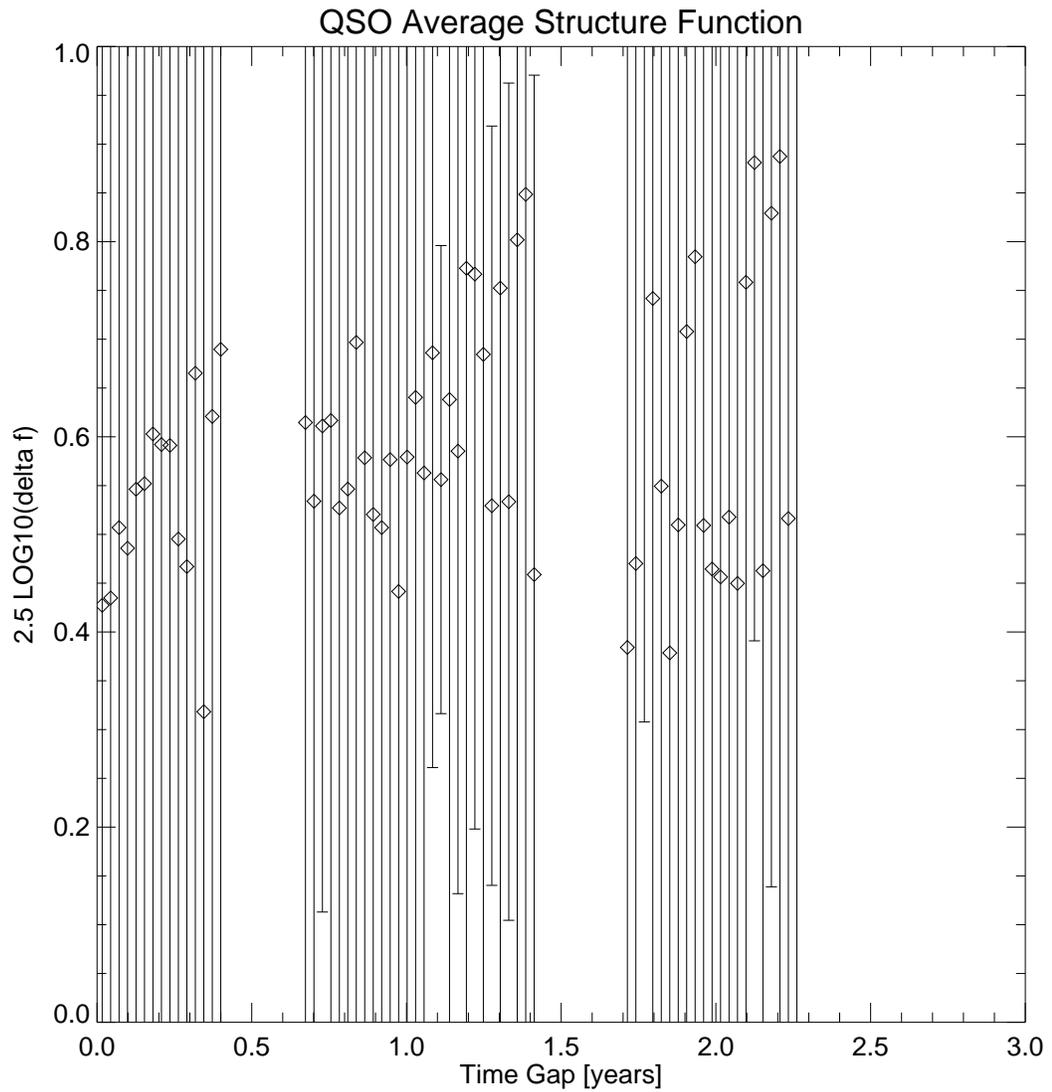}
\caption{The QSO average structure function (as defined in \citet{zackrisson03b}) derived from an analysis of the QSOs in the NEAT dataset.  The error bars are computed as the standard deviation from the mean of the $\Delta f$ distribution.}
\label{fig:qso_structure_function_avg}
\end{figure}

\begin{figure}
\plotone{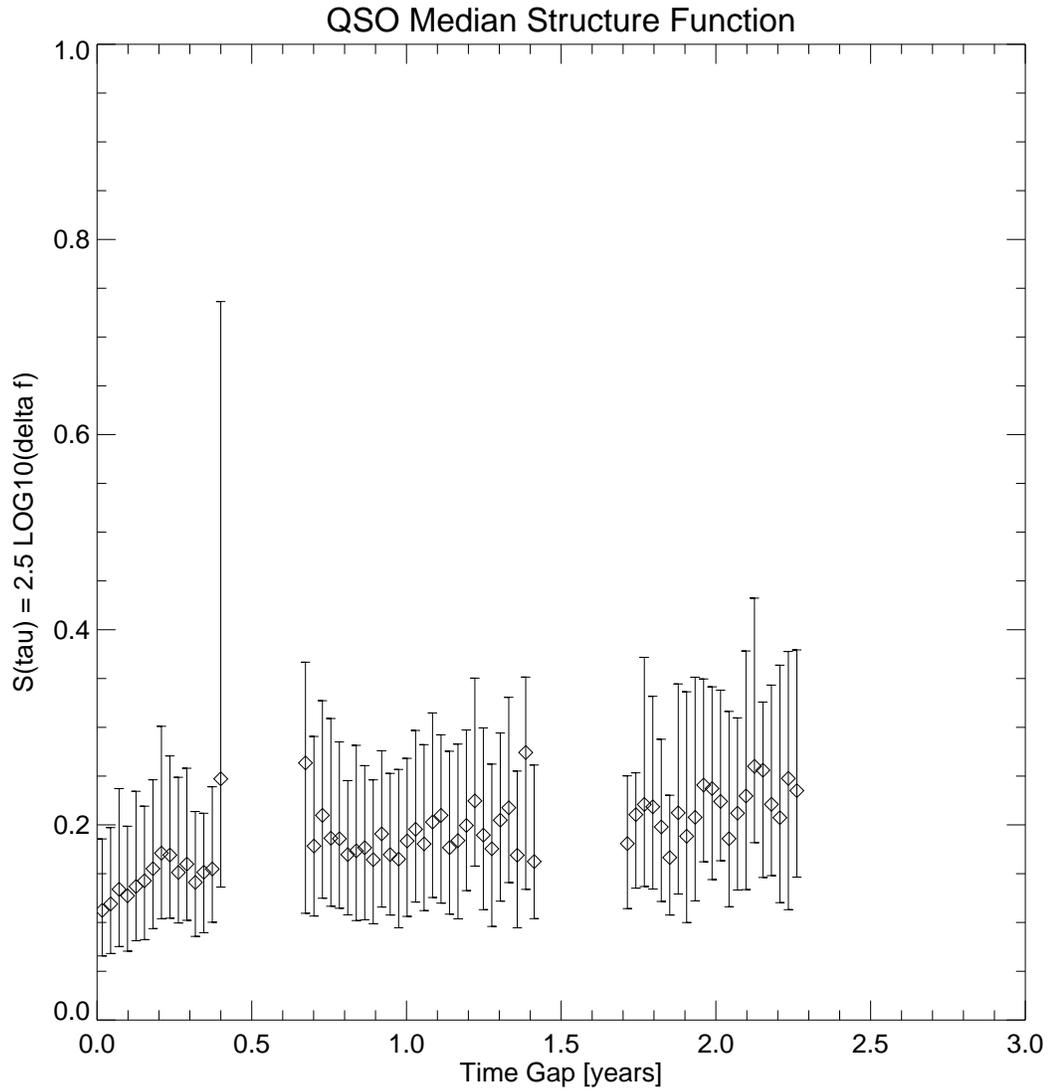}
\caption{The QSO median structure function (as in Fig.~\ref{fig:qso_structure_function_avg} but using a median instead of average on the change in magnitude) derived from an analysis of the QSOs in the NEAT dataset.  The error bars are computed based on including $\pm$34\% of the area around the median.}
\label{fig:qso_structure_function_med}
\end{figure}

\begin{figure}
\plotone{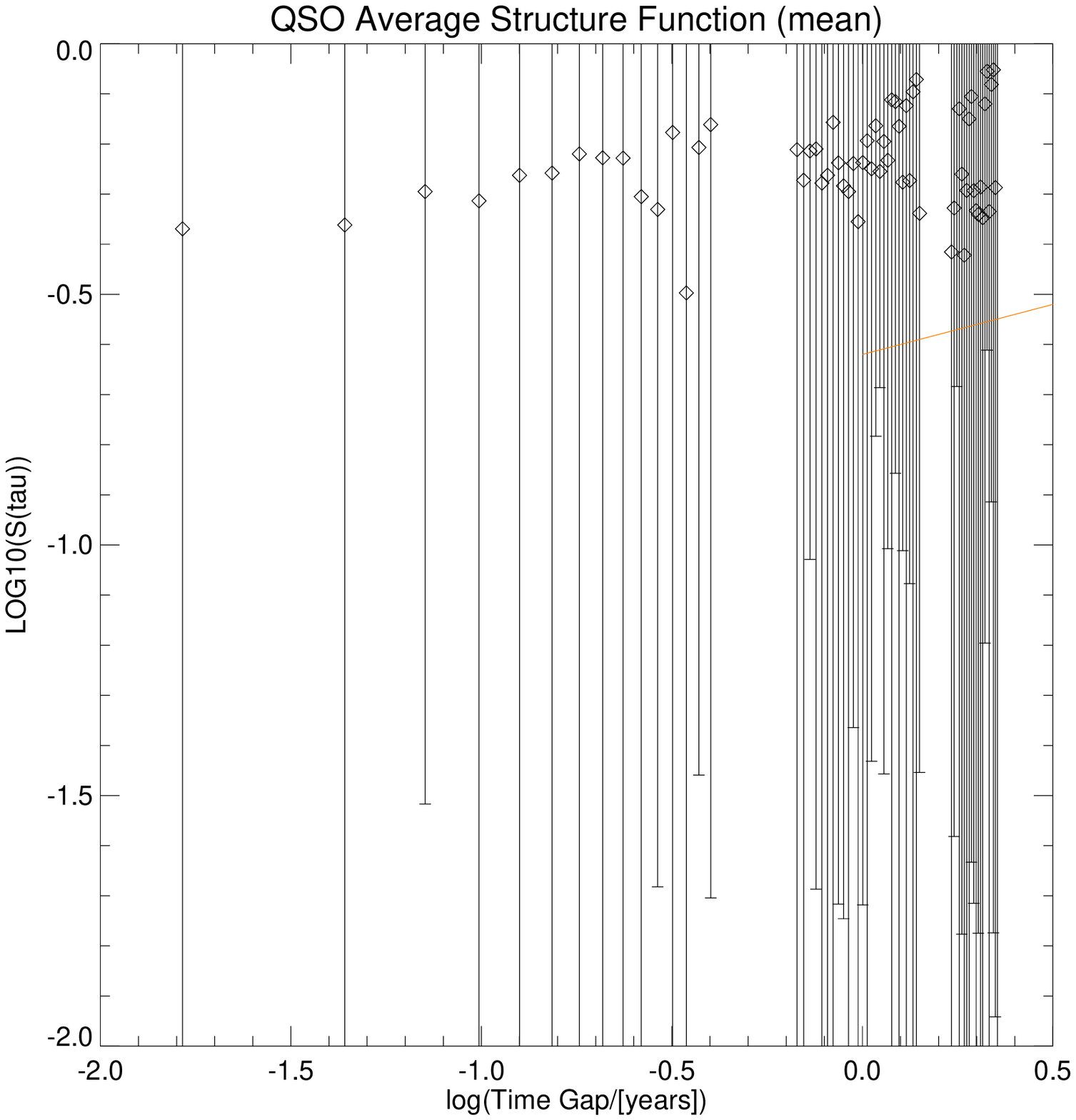}
\caption{Fig.~\ref{fig:qso_structure_function_avg} plotted in log space.  
The line in orange is the power-spectrum slope of ZBZ.}
\label{fig:qso_structure_function_avg_log}
\end{figure}

My work presented here shows a QSO power spectrum consistent with
some of the microlensing models analyzed in ZBZ
 to time scales down to a week on a slope that matches 
that found by ZBZ for time scales of 1--20 years.
Figs.~\ref{fig:qso_structure_function_avg}~\&~\ref{fig:qso_structure_function_med} show the results of my power-spectrum analysis for the mean and median amplitudes. 
The difference between taking the mean (Eq.~\ref{eq:qso_structure_function_avg}) versus median value (Eq.~\ref{eq:qso_structure_function_med}) in the 
calculation of the structure function is clear.
Fig.~\ref{fig:qso_dm2_mean_med} show a histogram of the distribution of
$\Delta m^2$ along with the mean and median of the distribution.  The
difference between these numbers for the entire sample accounts for
the difference in the two methods of calculating the structure
function.  Fig.~\ref{fig:qso_structure_function_mean_vs_med} shows the
ratio of Fig.~\ref{fig:qso_structure_function_avg} and
Fig.~\ref{fig:qso_structure_function_med} and the expected offset from
a Gaussian distribution (green line) and the actual distribution of
$\Delta m^2$ (blue line).  To compare with the results of
\citet{zackrisson03b}, I show the mean structure function in Figs.~\ref{fig:qso_structure_function_avg}~\&~\ref{fig:qso_structure_function_avg_log}, but I wish to point out the significant weight that
large variations have on the average magnitude fluctuation and the difficulty in assigning a meaningful error bar to the mean given the distribution shown in Fig.~\ref{fig:qso_dm2_mean_med}.  Thus, for the rest of the analysis, the median value will be used.

\begin{figure}
\plotone{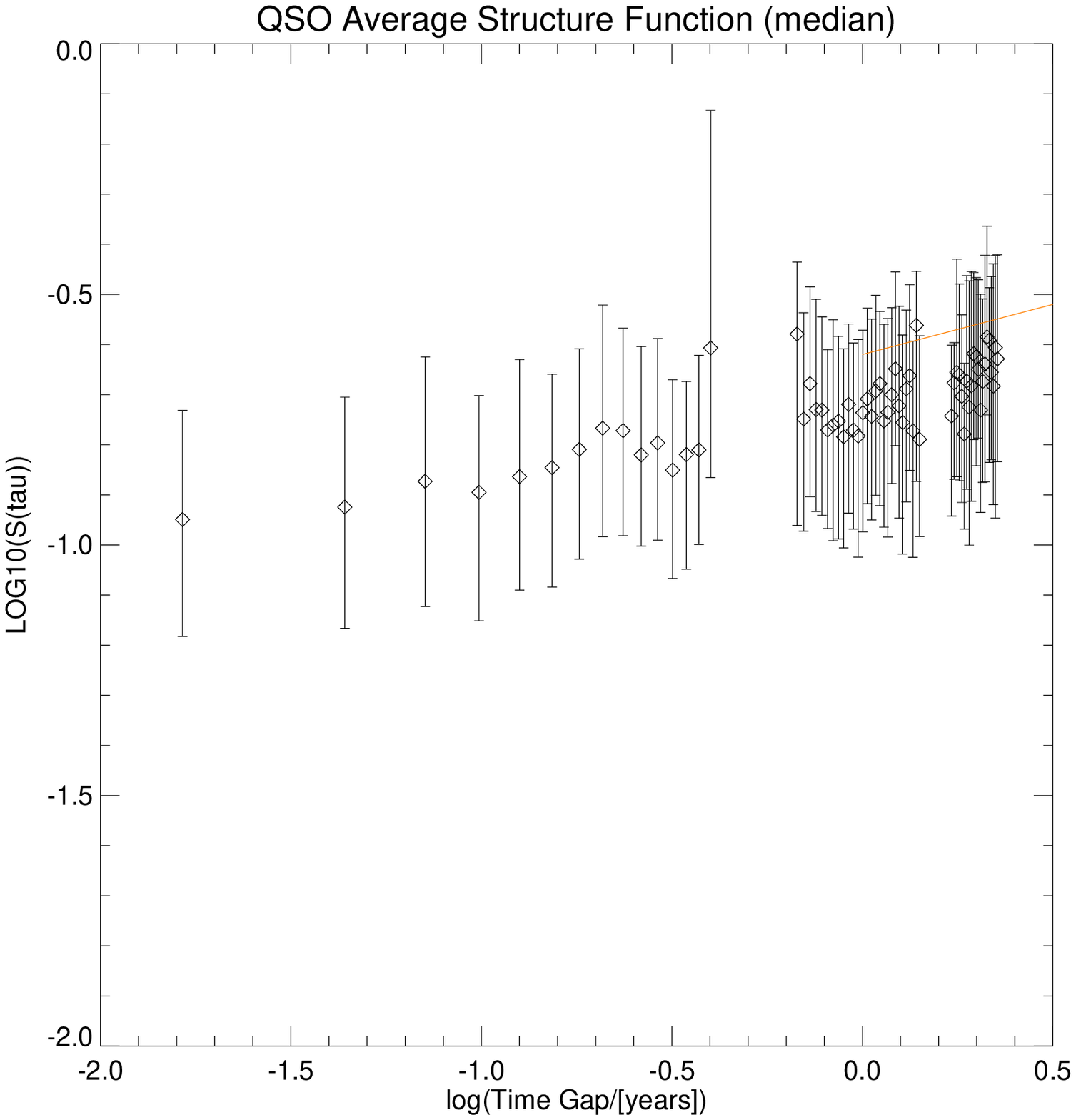}
\caption{Fig.~\ref{fig:qso_structure_function_med} plotted in log space.
The line in orange is the power-spectrum slope of ZBZ.}
\label{fig:qso_structure_function_med_log}
\end{figure}
\begin{figure}
\plotone{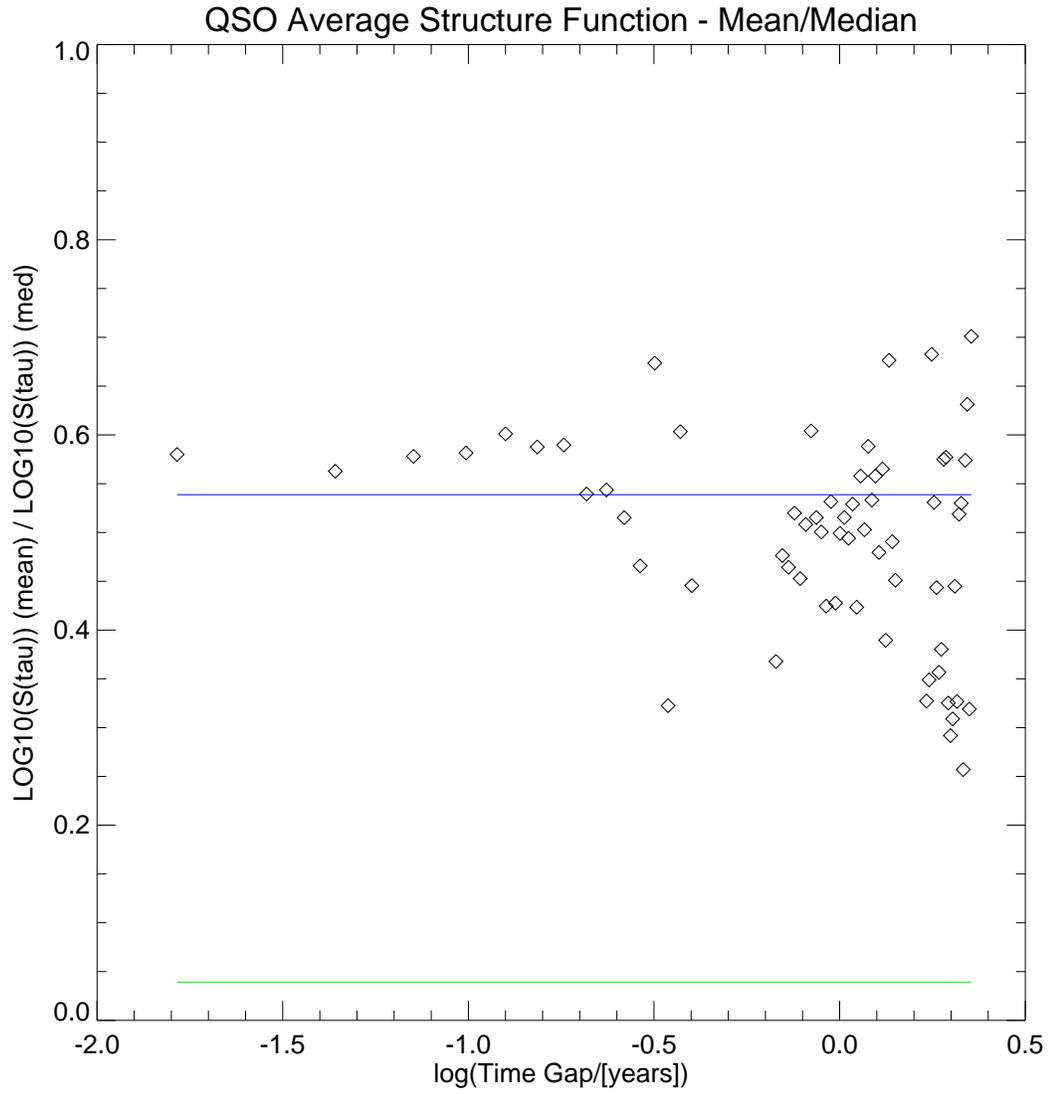}
\caption{The ratio of the QSO structure function as calculated using the mean versus median of flux change values.}
\label{fig:qso_structure_function_mean_vs_med}
\end{figure}
\begin{figure}
\plotone{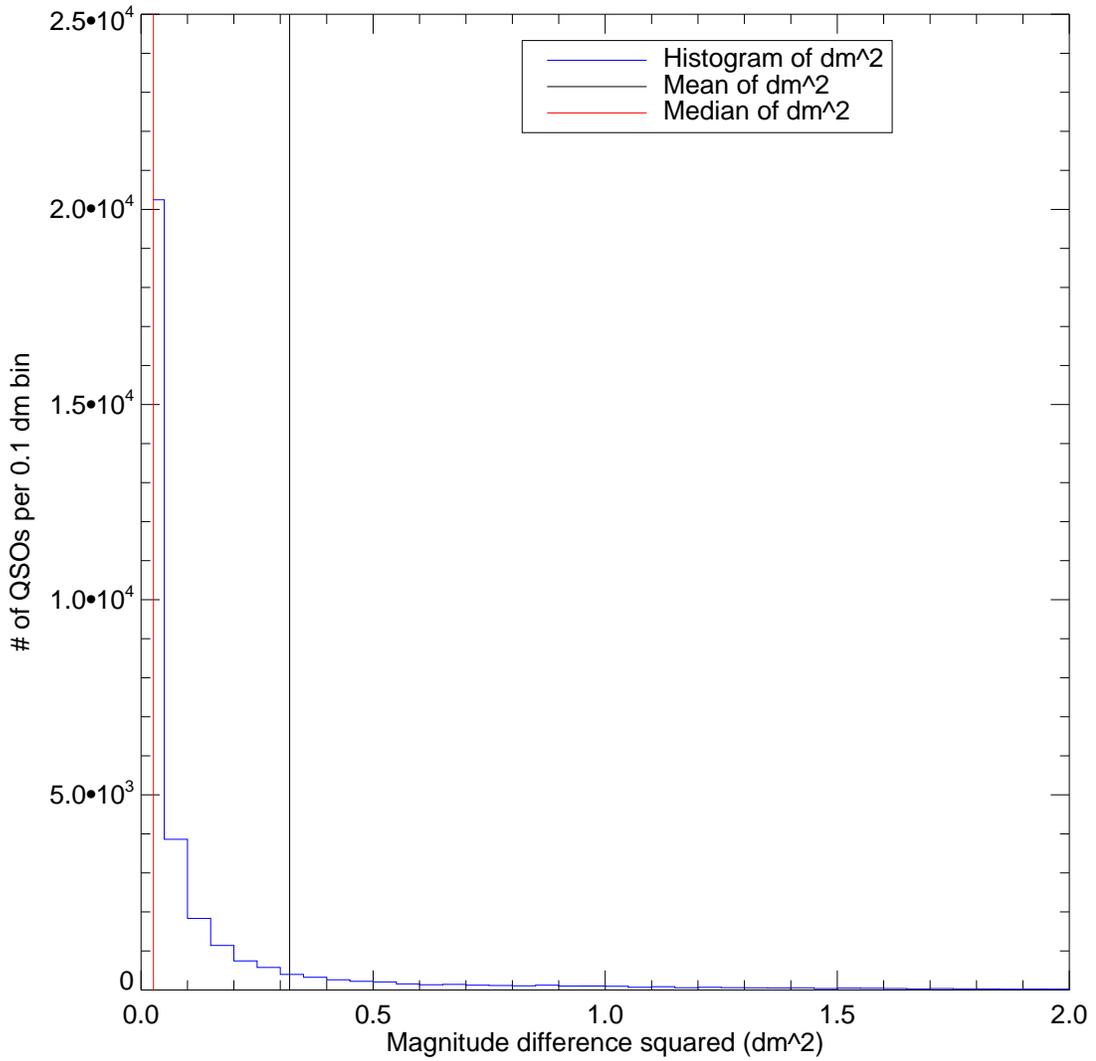}
\caption{The distribution of magnitude fluctuations for all time
scales.  Note the significant difference between the mean (black line)
and median (red line) of the distribution.}
\label{fig:qso_dm2_mean_med}
\end{figure}

The power-spectrum slope of \citep{zackrisson03b} is indicated in
Figs.~\ref{fig:qso_structure_function_avg_log}~\&~\ref{fig:qso_structure_function_med_log}.  
Note that the solution from ZBZ has a slope completely consistent
with the analysis presented here.  This slope is a major factor
in determining the source of QSO variability and its continuation to
smaller time scales  is consistent with a range of microlensing
models considered by ZBZ.
These
studies do not entirely overlap as my survey has a smaller baseline
and fewer data points at larger time gaps.  This different time baseline coverage was a
large part of the motivation for undertaking this study.  However, my larger and
differently biased QSO sample may also be part of the explanation for
the difference observed in the region from 1 to 3 years.

\citet{zackrisson03b} stated that Hawkins observed a
dispersion of $0.13$~magnitudes from the intra-year plates and thus
they estimated that the contribution on these time scales was an
uncorrelated $\sigma_m = 0.065$.  Assuming the use of the term ``dispersion''
means standard deviation of the distribution without any relative uncertainty weighting, I observe a
significantly higher standard deviation of $\QSOyeardev$~magnitudes for my
sample of time baselines less than one year.  This is not too
different from the $\QSOlongdev$~magnitudes seen in my sample of time baselines
greater than one year and is comparable to the $\QSOdev$~magnitudes seen for
my overall sample.

The smaller time-gap region also shows a flatter slope than that found
by \citep{hawkins02} or ZBZ.
As
\citet{zackrisson03b} note, this is consistent 
with microlensing from objects of mass $M_\mathrm{compact} <
10^{-5}~M_\sun$.  ZBZ argued against this possibility as 
such a small slope and thus relatively higher power at small scales
was inconsistent with their observed structure function.  However, the
structure function I present here shows more power on small time scales and is
consistent with the small slope expected from microlensing.


\begin{figure}
\plotone{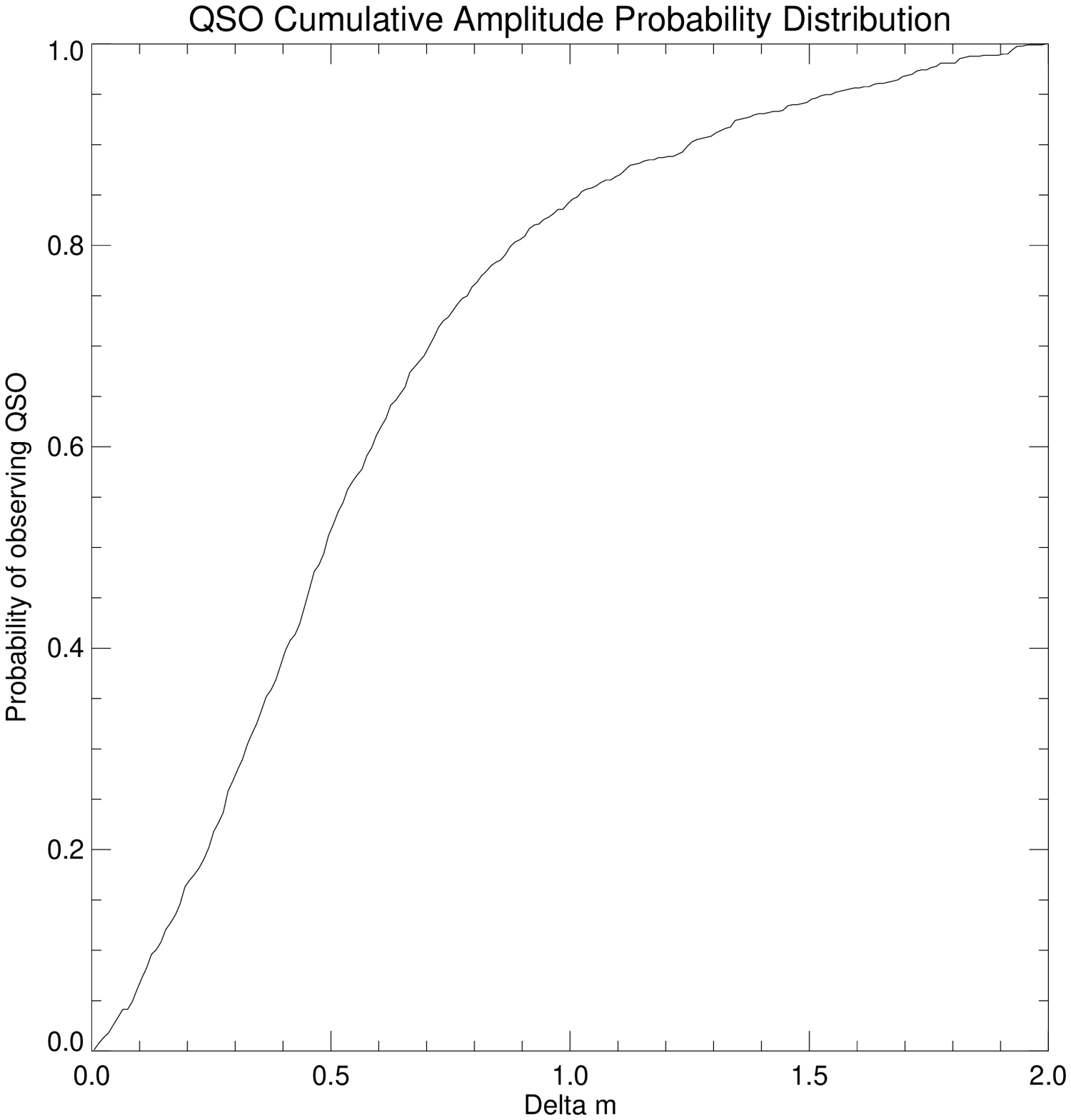}
\caption{The cumulative probability amplitude distribution for my QSO
sample.  This plot varies significantly from that of
\citet{zackrisson03b} and is more consistent with the microlensing
models presented therein.}
\label{fig:qso_cum_amp}
\end{figure}

\subsection{Cumulative Probability Amplitude}
\label{subsec:cum_amp}

\citet{zackrisson03b} also compare the cumulative probability amplitude 
distribution of their sample with models.  I obtain a very different
distribution than ZBZ at both small and large time scales.
Fig.~\ref{fig:qso_cum_amp} shows a smoothly increasing concave curve
without any sharp cutoff or significant inflection point as seen in
Figs.~3~\&~4 of \citet{zackrisson03b}.  There is no large cutoff at
$0.35$.  Rather, 40\% of objects have a variability less than
$\delta m = 0.35$.  This allows a range of models from Fig.~5 of
\citet{zackrisson03b}.  Note that the modeling of observational uncertainties 
can be important 
in considering small-scale variability.  \citet{zackrisson03b} include
the error of the measurements in their models.  This makes it difficult to
truly compare my sample with different error bars to the models they show
in their paper.  However, the power observed on small scales in this
analysis is much larger than the observational uncertainties.


\subsection{QSO amplitude redshift dependence}
\label{subsec:amp_z}

Fig.~\ref{fig:qso_amp_z} shows the distribution of QSO mean amplitude,
\begin{equation}
\delta m = \mathrm{max}(m(t))-\mathrm{min}(m(t))
\label{eq:meanamp}
\end{equation}
versus redshift.
The error bars represent the standard deviation of the sample in each
redshift bin.  
There is a large degree in uncertainty in these values
and the distribution itself is highly non-Gaussian around the
calculated mean as can be seen in Fig.~\ref{fig:qso_hist_amp}.
Fig.~\ref{fig:qso_amp_z} is the analogous plot to Fig.~6 of
\citet{zackrisson03b} where they note their error bars as
``the uncertainty in the position of the observed mean,'' which we
take to mean that they were using the error in the calculation of the
mean and not the standard deviation of the distribution.  For a
distribution such as that shown in Fig.~\ref{fig:qso_hist_amp}, the
quoted error bars in Fig.~6 of \citet{zackrisson03b} would seem misleading
in my analysis. 

In addition, for my sample, the $\delta\,m$ statistic of
Eq.~\ref{eq:meanamp} is not a good statistic as my light curve
is inhomogeneous and thus QSOs with more observations are more likely to
have larger values of $\delta\,m$.  In general $\delta\,m$ is 
a quantity dependent on the observational program.  I prefer an
analysis of the median difference for each QSO, as shown in 
Fig.~\ref{fig:qso_amp_med_z}.  If this statistic can be
compared with Fig.~6 of \citet{zackrisson03b} then my results
are consistent with the microlensing models with characteristic
variability amplitudes of $0.2$ and are not a strong function of
redshift.
This flat dependence of variation amplitude with redshift is
consistent with the findings of \citet{hawkins00}.



\begin{figure}
\plotone{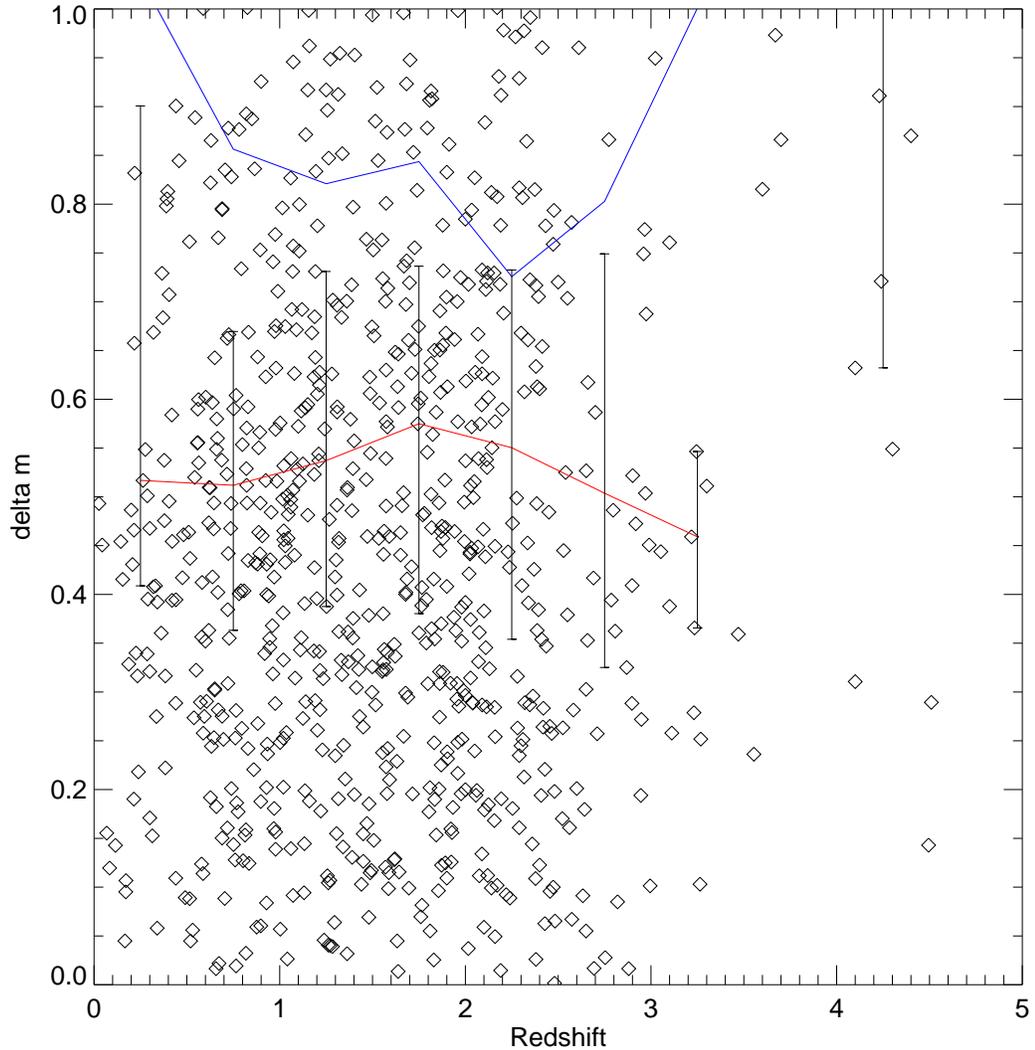}
\caption{A scatter plot of mean amplitude (Eq.~\ref{eq:meanamp}) 
versus redshift for the
QSOs in this analysis.  Binned version of the data is shown by the
over-plotted lines.  The blue line is the mean, while the red line is
the median of the distribution.  The standard deviation of the points
in each redshift bin are shown as errors on the mean line.  A similar
uncertainty in the median line can be postulated, but note that the
distribution of mean amplitudes is not symmetric about the mean value
but instead is piled up to zero.  \citet{zackrisson03b} finds a
maximum mean amplitude of approximately 0.4 mag at a redshift of
$z=0.5$.  This is easily consistent with the error bars implied by the
distribution of values in the plot.}
\label{fig:qso_mean_amp_z}
\end{figure}

\begin{figure}
\plotone{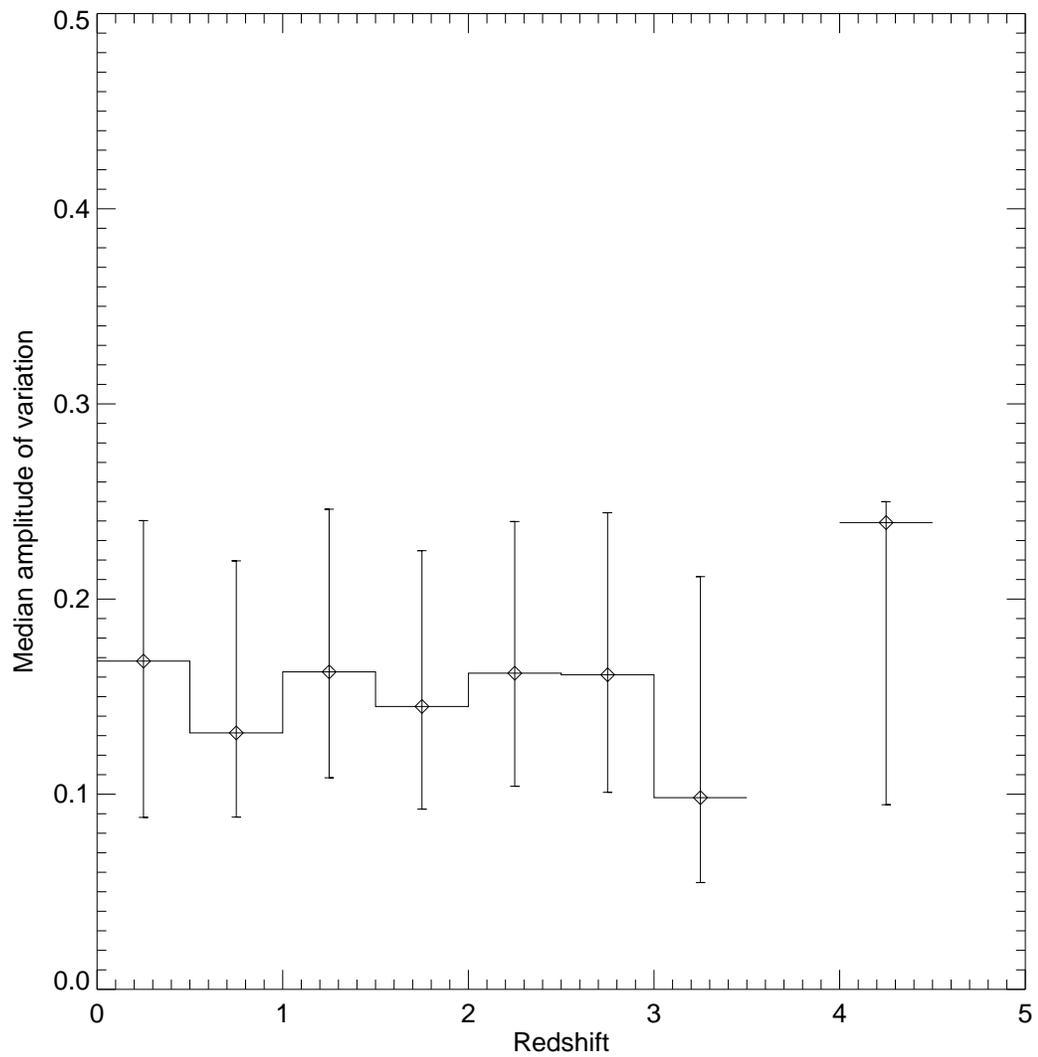}
\caption{
The median values from Fig.~\ref{fig:qso_mean_amp_z}.  The error bars are
from a 68\% inclusion of points around the median value.
}
\label{fig:qso_amp_med_z}
\end{figure}

\begin{figure}
\plotone{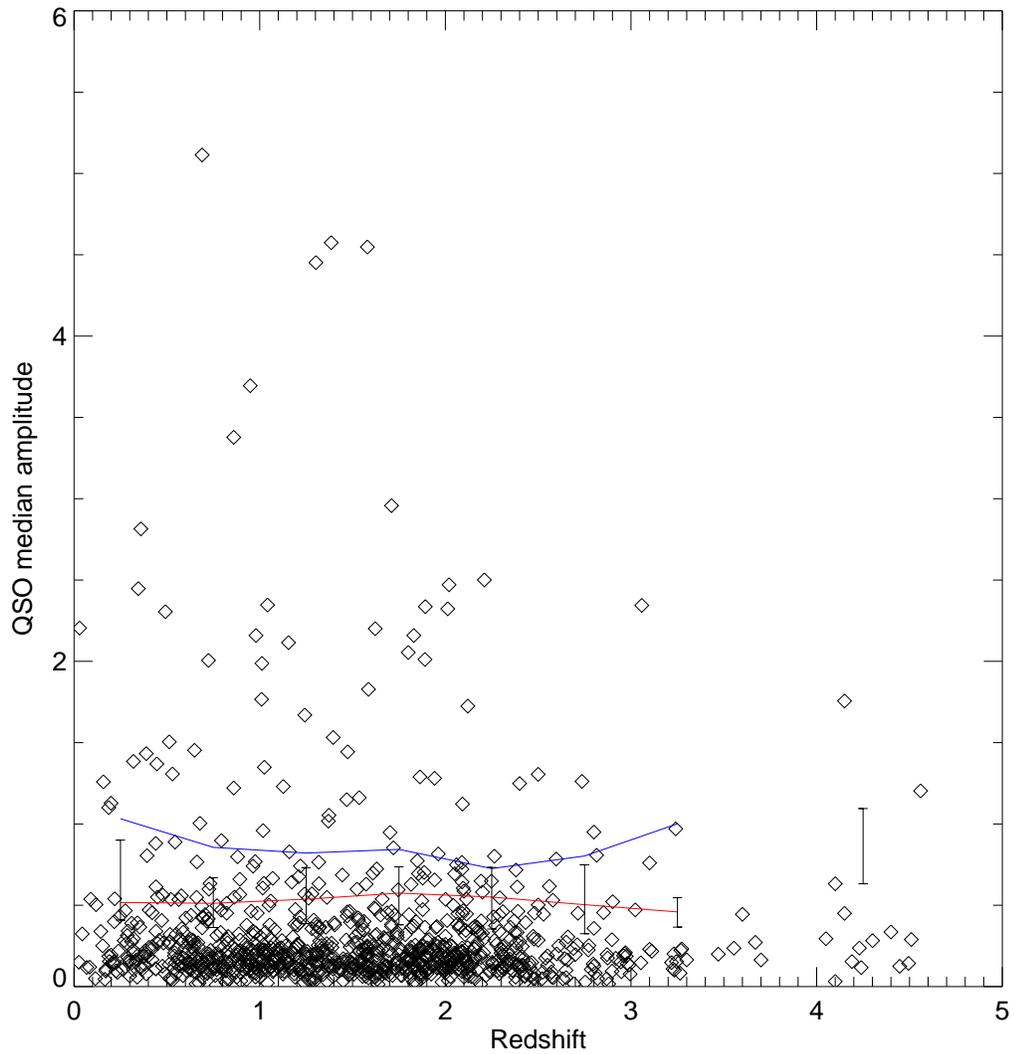}
\caption{A scatter plot of median amplitude 
versus redshift for the
QSOs in this analysis.  Similar to Fig.~\ref{fig:qso_mean_amp_z} except
uses the median of all of the variability amplitudes for each QSO.
}
\label{fig:qso_amp_z}
\end{figure}

\begin{figure}
\plotone{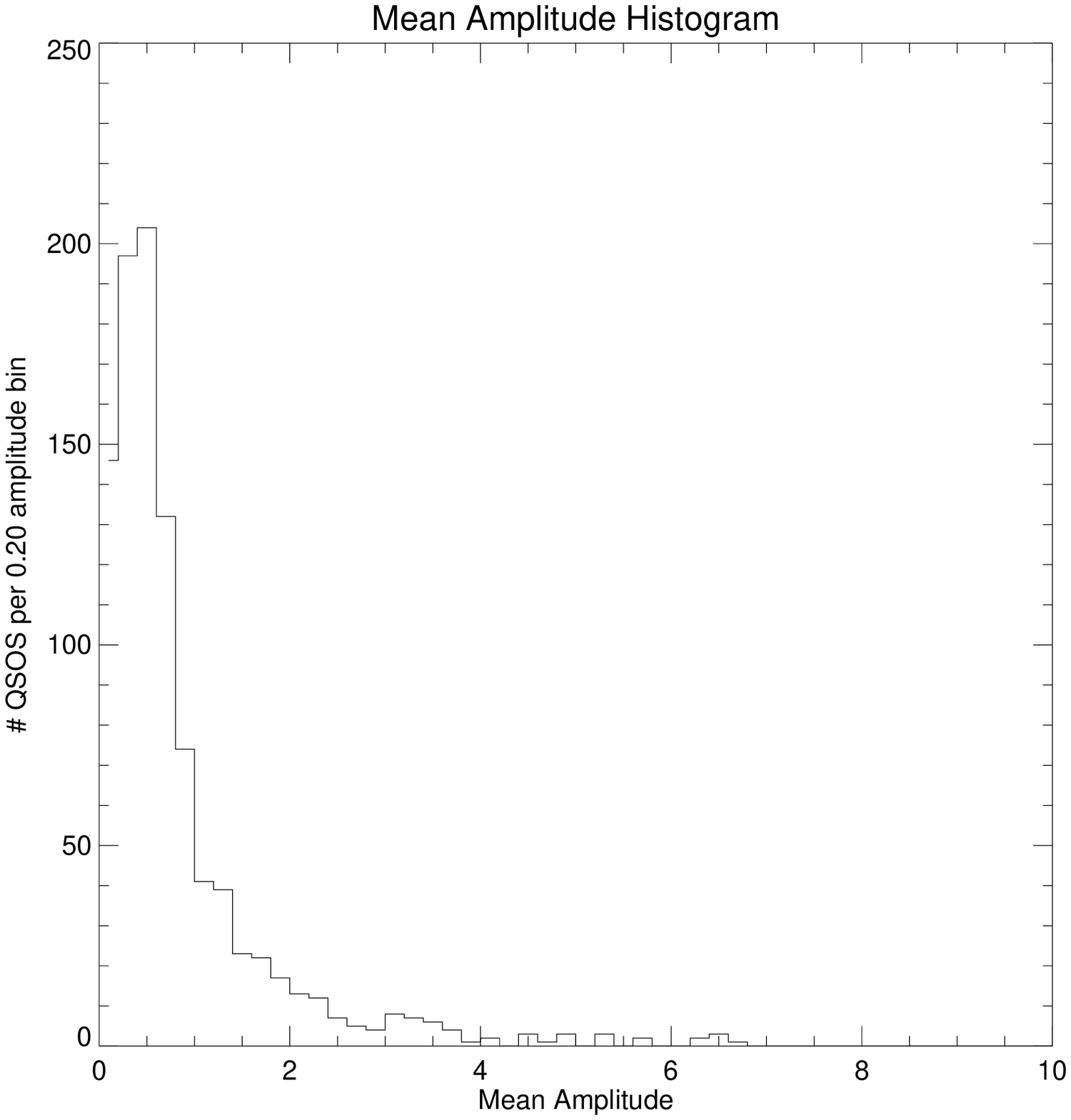}
\caption{The histogram of mean amplitudes for my QSO sample.  Note
the concentration near zero.  This makes the statistics of
Fig.~\ref{fig:qso_amp_z} difficult to interpret as a Gaussian
distribution.}
\label{fig:qso_hist_amp}
\end{figure}

\section{Conclusions}

The NEAT QSO sample exhibits an average structure function consistent
with that of \citet{hawkins02} and ZBZ, but I find significantly
different properties in the cumulative probability distribution, the
fraction of low-variability objects.  My amplitude-redshift relation
agrees with \citet{hawkins02} but not with ZBZ.  These differences
generally ease the constraints on the microlensing models used by ZBZ
and my results are consistent with a range of QSO microlensing models
presented in ZBZ.

Further investigations using an expanded dataset from the NEAT
searches will allow a revisiting of these effects and the eventual use
of a QSO sample that is an order of magnitude larger than that
of~\citet{hawkins00} and is selected in a different and less-biased
way.  More sophisticated understanding of the model predictions with
regards to the distribution of QSO variability fluctuations around the
mean and median distributions is encouraged to standardize the analysis
of different QSO-variability datasets.

I conclude that the different selection biases in the two samples
and the use of averaging to analyze variability 
lead to the different results.  I am particularly concerned with the
low-variability numbers of the \citet{hawkins00} sample and find that
intra-year variations are significant and comparable to the
longer-term variations focused on in ZBZ.
Small, $10^{-5}$~$M_\sun$,
compact objects can provide the observed variability and small slope
observed in my sample.  Thus I elaborate on the of conclusion
\citet{zackrisson03b} to state that my results are consistent with
microlensing as an explanation for the observed variability of QSOs in
my sample and allow that if there is another mechanism for
variability that it is a longer-scale process that leads to
variability on the time scale of years instead of months.  This
division by time scale of the contributions of different variability
mechanisms satisfies some of the more difficult light-travel and
energy-transport time problems in many central-engine-based
explanations of short-term, achromatic luminosity variations in QSOs,
but allows that 
central-engine variability could dominate QSO variability on longer
time scales.


\section{Acknowledgments}

I would like the acknowledge the significant contributions of Roland
Rudas who did the initial study of QSO variability in the NEAT data
as part of his senior thesis in Physics at UC Berkeley.



\chapter{Calibration of the NEAT System Response}
\label{apx:calibration}

\section{Pre-Introduction}

This chapter documents attempts to calibrate the NEAT photometric systems.
These attempts were largely unsuccessful and are documented here as an
aid to future efforts in this direction.  The calibration of the 
NEAT detector systems is 
discussed in Chapter~\ref{chp:2002ic} for the specific case of SN~2002ic.
In general, no color calibration was found that yielded improved
results over the na\"ive approach of calibrating NEAT magnitudes off of
stars from the ``red'' USNO~A1.0 catalog~\citep{usnoa1}.

\section{Introduction} 

The first two NEAT detectors on Haleakala and Palomar were used in an
open, unfiltered mode.  This choice was made by the NEAT group
because asteroids shine in reflected sunlight and an open filter
maximizes the amount of light detected from an asteroid.  However, it can also
let in more light from other sources, most notably the night sky
background.  In the optical regime, the night sky brightness is mostly
due to scattered light and so is blue.  Thus an $R$ filter can help
reduce the contamination of night sky light and in bright time there
is a gain in detection sensitivity.  However, in dark time, when scattered
light is less of an issue, there is a
loss in detection sensitivity due to cutting out more light from the
object.  These two factors are roughly comparable~\citep{kessler02}.

\begin{figure}
\includegraphics{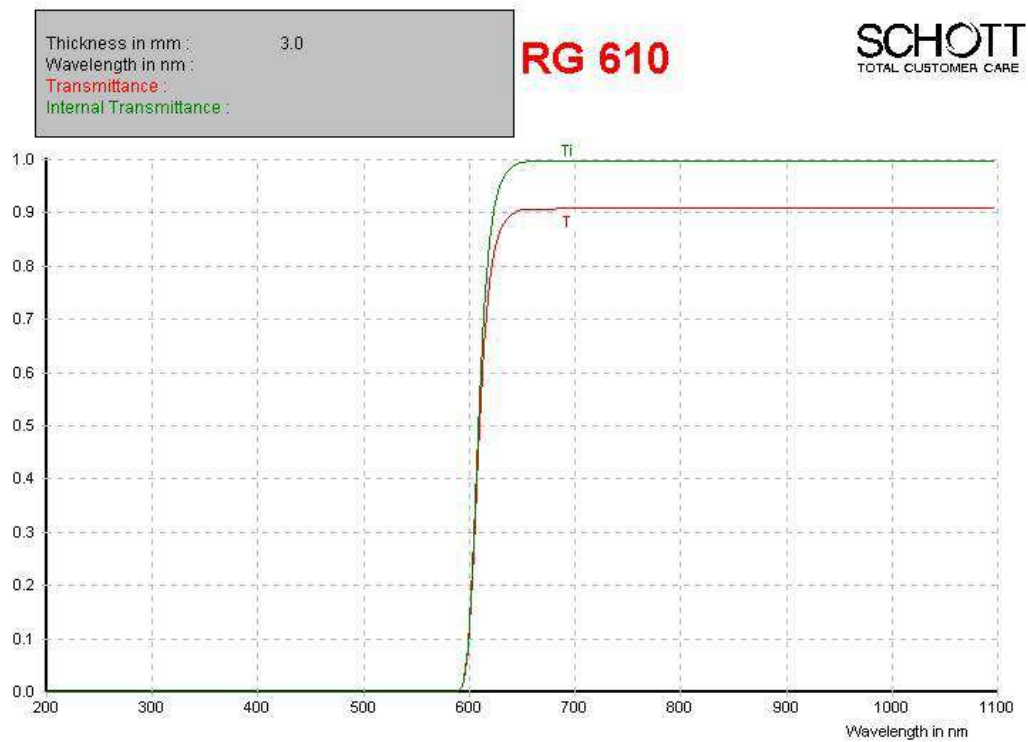}
\caption{The filter used by the NEAT asteroid search on all four fingers of the QUESTII camera \citep{schott03}.}
\label{fig:rg610}
\end{figure}

After some discussions with the NEAT group, they agreed to observe
with a filter when using the new QUESTII camera installed on the
Palomar 1.2-m telescope.  This could have been a definite advantage
for the SNfactory as it would have allowed for calibration of supernova observations
against standard filters.  Unfortunately, the NEAT group chose to use an RG-610
filter~(Fig. \ref{fig:rg610}).  While a known filter shape in combination
with the measured QE curves for the QUESTII camera will allow
for a well-constrained understanding of the effective response
function, it won't be possible to cleanly compare these magnitudes
with other observations in more standard filters such as $UBVRI$
or $ugriz$.  The filter choice for the NEAT QUESTII observations
is currently being reconsidered for summer 2004 operations and beyond.

Leaving aside the filter for the new QUESTII camera for the moment, there
are years of unfiltered data with one hundred supernovae 
observed with NEAT detectors and it would be valuable to place this
on a known photometric system.
This goal requires the calibration of historical NEAT detectors (the Haleakala
NEAT4GEN2 and the Palomar NEAT12GEN4) to determine a total effective
response of the system as a function of wavelength.  This response 
function depends on a number of factors
including the quantum efficiency of the CCD, the optical properties of
the telescope and the atmospheric transmission.

\section{Calibration Techniques}

Two major opportunities for calibration presented themselves: Landolt standard
fields~\citep{landolt92} and the Sloan Digital Sky Survey (SDSS)~\citep{sdss_dr1}.  Over the course of
the past few years of NEAT observations, thousands of images have been taken of a
variety of Landolt standards.  This extensive and repeated coverage provides the opportunity to determine effective color terms between the Landolt
standard stars in $UBVRI$ and the unfiltered NEAT observations.  The
supplementary work of \citet{stetson04} is also available for 
a more thorough calibration of the Landolt standard fields.
The
SDSS survey provides five-band $ugriz$ colors for millions of stars.
Using the overlap between the SDSS Data Release 1~\citep{sdss_dr1} (SDSS Data Release 2 was not yet released when the work described in this chapter was done), we
can thus similarly calculate color terms between $ugriz$ and the
unfiltered NEAT observations.  Calibration against the Landolt fields
has the advantage of directly providing color terms for the $UBVRI$
filters commonly used in supernova work.  Calibrating against the SDSS
survey is better in all other ways due to the increased sky coverage
and far greater number of stars.  Some combination of the two would be
the best solution.
 
A general caveat is that color terms between filters can only be
defined with respect to an assumed spectral shape.  Thus the color
terms that can be derived from the comparisons mentioned above are
strictly only valid for an archetypical star representing the aggregate
of all the stars used in the calibration and are not entirely valid
for calibration of supernova lightcurves.  This is related to
the general problem of S and K corrections so vital to precision
photometry but is beyond the precision we can achieve here.

\section{SDSS Calibration}

The SDSS DR1 provides 88~million objects in $3324$\sq\deg across
the sky~\citep{sdss_dr2} (see Fig.~\ref{fig:sdss_coverage}).  These
stripes overlap with the NEAT coverage pattern (compare Fig.~\ref{fig:sdss_coverage} and Fig.~\ref{fig:neat_sky_coverage}) in a number of regions
and some SDSS field is normally covered during the course of each night's
observations.  This is in contrast to the Landolt fields,
which are observed more rarely.

\begin{figure}
\plotone{sdss_dr1_coverage}
\caption{The RA and Dec sky coverage of the SDSS DR1~\citep{sdss_dr1}.}
\label{fig:sdss_coverage}
\end{figure}

\begin{figure}
\includegraphics[angle=270]{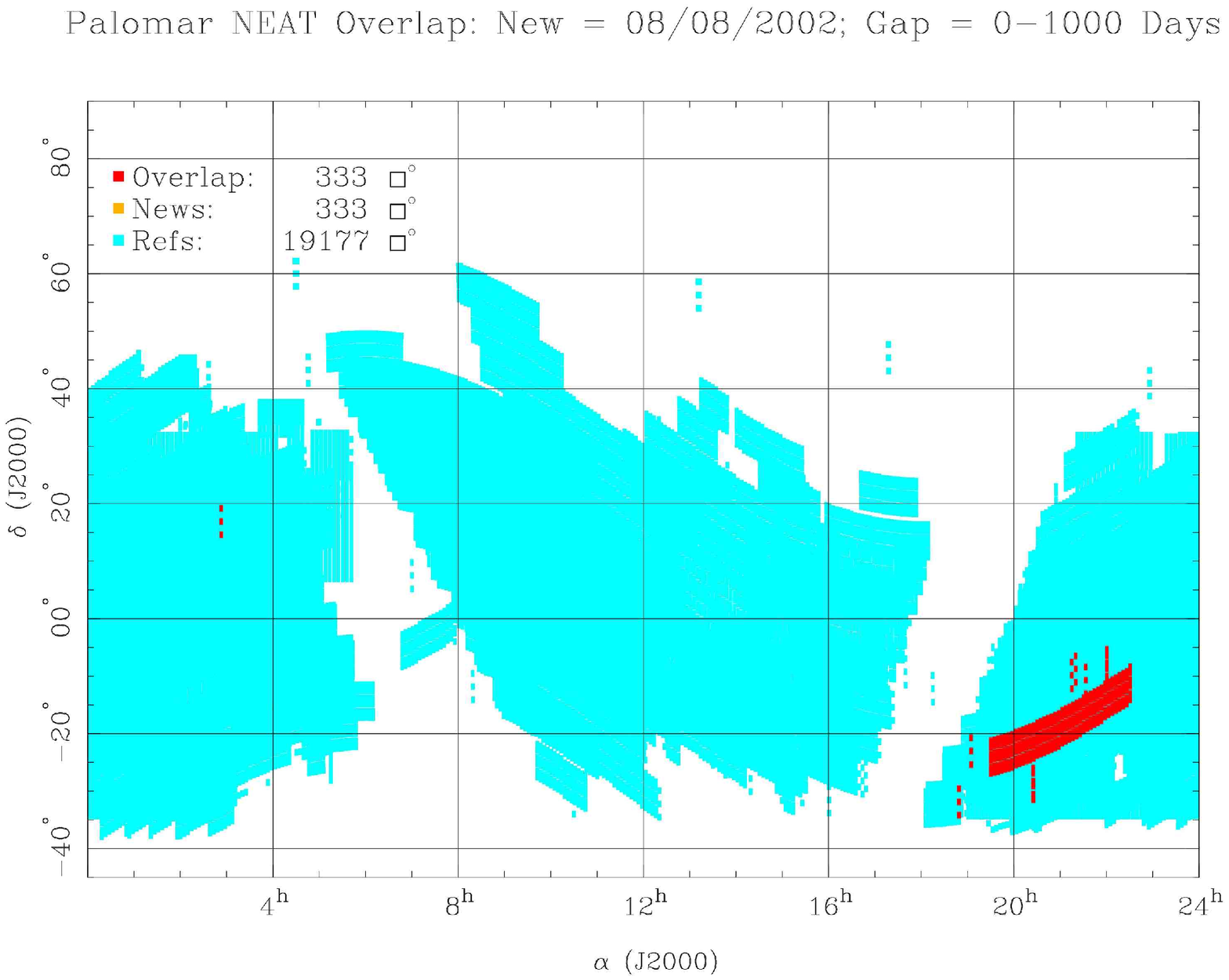}
\caption{The RA and Dec coverage of the NEAT survey (light coloring) with
a sample night of search coverage (dark coloring).  This figure is for
search images taken the night of 8 August 2002.  The nightly sky
coverage has since doubled with the new QUESTII camera.}
\label{fig:neat_sky_coverage}
\end{figure}

The SDSS color bands are defined in terms of the specific Gunn filters
mounted on the Apache Point 2.5-m telescope.  The Gunn system is an AB
magnitude system where there is no filter-by-filter zeropoint adjustment to
match a stellar spectrum.  Only one color zeropoint is set and the
rest just come from the $\log{\mathrm{flux}_{\mathrm{filter}}}$ relative to the nominal $\log{\mathrm{flux}}$
value.  By contrast, in the Vega system a model of the star Vega is defined to have
a color of 0 in each filter $UBVRI$.  This is clearly different from what one would
get from just taking the $\log$ of the flux as flux increases and
decreases with wavelength over the bands considered depending on the
temperature of the star.  The AB and Vega system are matched to each
other near the center of the Gunn $g$ filter, and thus the correction for $g$ is
relatively small.

The SDSS colors were converted to the Bessel V-band by using
the Gunn$\rightarrow$Bessel corrections of \citet{frei94} (AB to Vega
magnitude correction, see Eqs.~\ref{eq:ab_vega_r}~\&~\ref{eq:ab_vega_g}) and
\citet{windhorst91} (\mbox{Gunn Vega$\rightarrow$Bessel Vega}, see
Eq.~\ref{eq:gunn_bessel}).
%
%
%
%
%
%
%
\begin{eqnarray}
g_\mathrm{Vega} &=& g_\mathrm{AB} + 0.013 \pm 0.002 \\
\label{eq:ab_vega_g}
r_\mathrm{Vega} &=& r_\mathrm{AB} + 0.226 \pm 0.003 \\
\label{eq:ab_vega_r}
V_\mathrm{Vega} &=& g_\mathrm{Vega}-0.03-0.42~(g_\mathrm{Vega} - r_\mathrm{Vega})
\label{eq:gunn_bessel}
\end{eqnarray}

\begin{figure}
\plotone{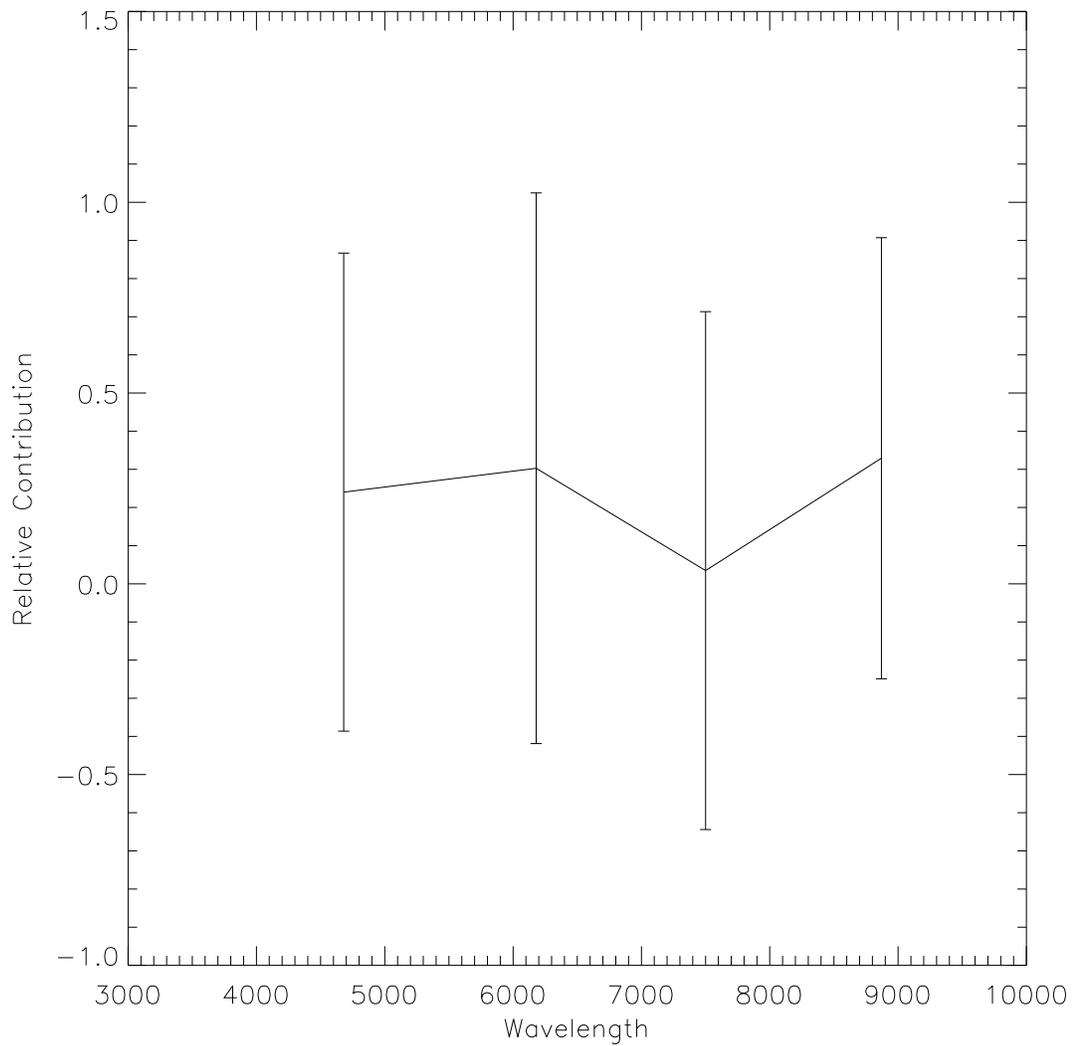}
\caption{The smooth continuum profiles of many stars over the SDSS
bands leads to degeneracies in fitting for the contribution of
individual filters to the NEAT unfiltered magnitudes.}
\label{fig:sdss_degeneracy}
\end{figure}

\begin{figure}
\plotone{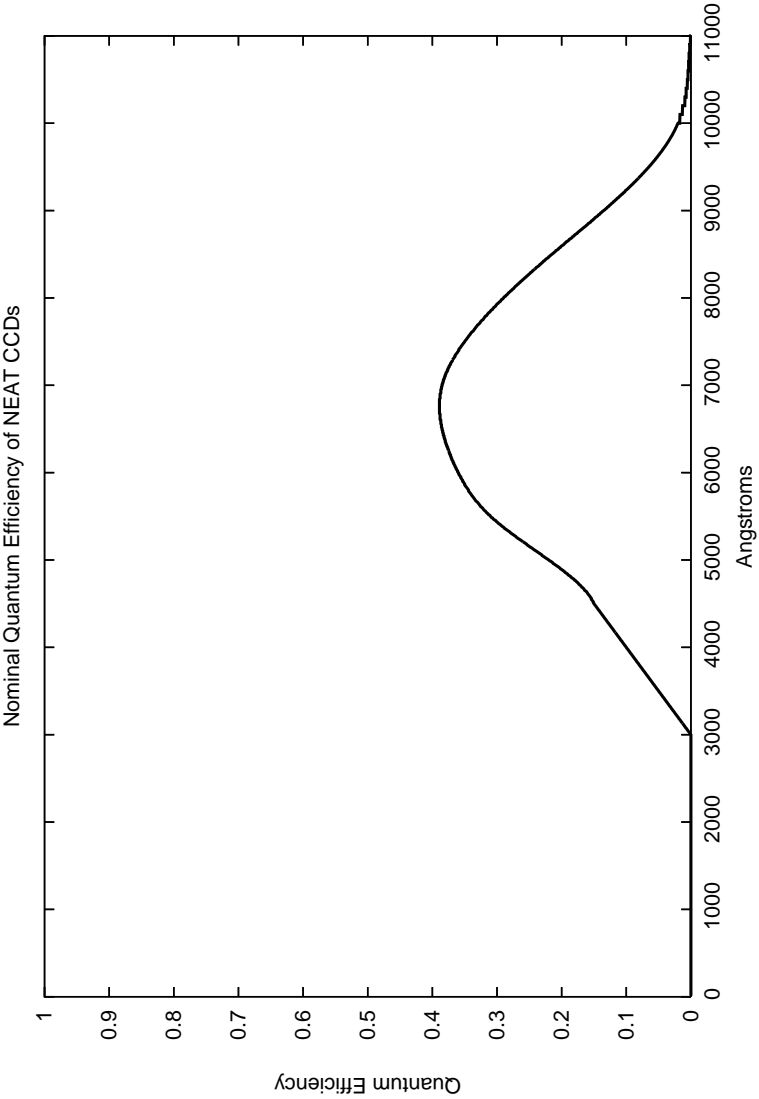}
\caption{The response curve of a CCD similar to the ones used by the
NEAT cameras.  The response curve has been modified from the source
template by assuming a linear decrease from 4500~\AA\ to 3000~\AA\ to
remove the effect of the UV-fluorescent coating present on the CCD
analyzed for this measurement.}
\label{fig:neat_ccd_qe_ref}
\end{figure}

\begin{figure}
\plotone{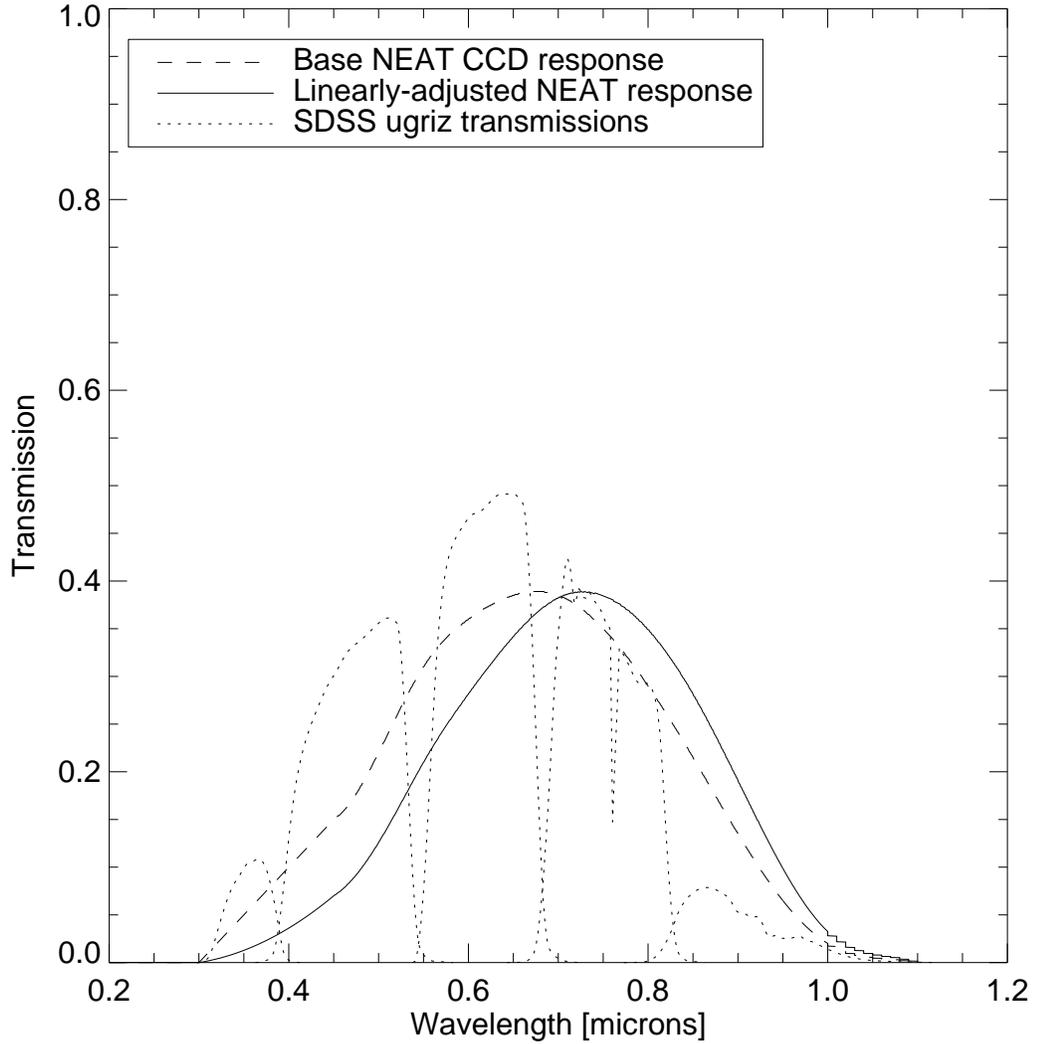}
\caption{The modulated response curve of
Fig.~\ref{fig:neat_ccd_qe_ref} as fit for the NEAT Haleakala MSSS
camera.  The functional form of of the modification is
$\mathrm{QE}(\lambda) = \mathrm{QE_\mathrm{ref_mod}}(\lambda) \times
(a +b \lambda)$.  The fit parameters here are $a = -2.84159E-01\pm4.59E-03$, $b =1.24385E-04\pm6.4348661e-07$ with a reduced $\chi^2 = 6.2$.
}
\label{fig:neat_ccd_qe_msss}
\end{figure}

A number of approaches to deriving a color term for the NEAT
observations were tried.  First, an individual filter-by-filter
comparison was done to derive a color offset term between each filter
and the NEAT magnitude.  Next, the contribution of each filter to the
NEAT magnitudes was fit.  Unfortunately, there was too much degeneracy
between different filter bands to obtain a good overall fit.
Fig.~\ref{fig:sdss_degeneracy} shows the wide range of fit values
derived from this method.  Finally, a constrained template was
assumed, corresponding to the quantum efficiency (QE) curve of a CCD
similar to the NEAT CCD used at Palomar but with a UV-sensitive
fluorescent coating.  To remove the effect of the UV-sensitive coating
from this response curve, the QE was assumed to linearly decrease from
the listed value at $4500$~\AA\ to zero at $3000$~\AA\ (see
Fig.~\ref{fig:neat_ccd_qe_ref}).  In the attempted fit, the response
curve was modulated by a first-order polynomial, whose coefficients
were fitted using the data from all of the SDSS stars observed for a
given night.  This assured a continuity and smoothness of the
contribution from each SDSS filter to the response function of the
NEAT CCD.  The same method was used to fit independently for the
appropriate coefficients for both the Palomar NEAT12GEN2 and Haleakala
NEAT4GEN2 CCDs.  Unfortunately, this linear fit was not sufficient to
reproduce good agreement with the SDSS magnitudes.  Again, the
degeneracies limited the resolution of the fit.  While the fit
parameters (see Fig.~\ref{fig:neat_ccd_qe_msss}) are very precise,
they do not actually represent a good fit to the data as evidence by
the large $\chi^2$.

\section{Peter Nugent's NEAT Response Curve}

\begin{figure}
\plotone{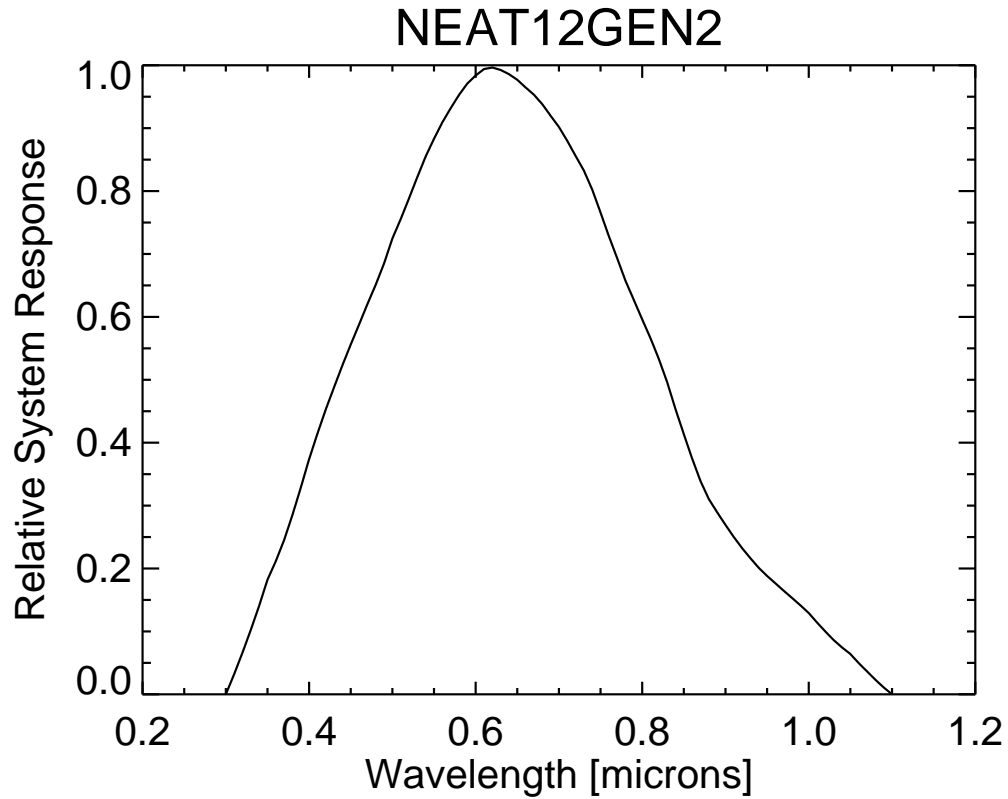}
\caption{The system response curve of the NEAT12GEN2 detector as
calculated by Peter Nugent using a tens of thousands of SDSS stars in
combination with matching stellar spectra.}
\label{fig:nugent_response_curve_NEAT12GEN2}
\end{figure}

As these attempts at calibration were not meeting with success,
Peter Nugent graciously spent some time using the observations of tens
of thousands of SDSS stars by the NEAT telescope to reconstruct the
response curve of the telescope-detector systems.  He used the $ugriz$
information for the stars together with stellar spectra matched to the
colors of each star (including the effects of atmospheric extinction)
to fit for the response curve (see
Fig.~\ref{fig:nugent_response_curve_NEAT12GEN2}).
An undergraduate summer student, Marc Rafelski, set up
our initial interface code with the SDSS server.

This fit did yield good agreement with the data with
a reduced $\chi^2$ near 1.
However, it did not agree with the external calibration
of SN~2002ic.

\subsection{Calibration of the SN~2002ic lightcurve}

SN~2002ic was a very unusual supernovae that is covered in more
detail in Chapter~\ref{chp:2002ic}.  It was the first supernova
that was compared with published photometric observations.
This supernova was one of the immediate motivations 
to try to understand the color calibration of the NEAT
detectors.
The calibration efforts started with a focus on the NEAT images of SN~2002ic
(see Sec.~\ref{chp:2002ic}).  All of the images from the light curve
were considered and other images from the corresponding nights were
compared against the SDSS catalog.  Any overlapping images were
calibrated against the catalog and a color term to the SDSS $ugriz$
system was calculated.

This color term was used for the night and was extended to
non-overlapping images, namely the images of SN~2002ic, which doesn't
lie in a region covered by the SDSS catalog.  


\begin{figure}
\plotone{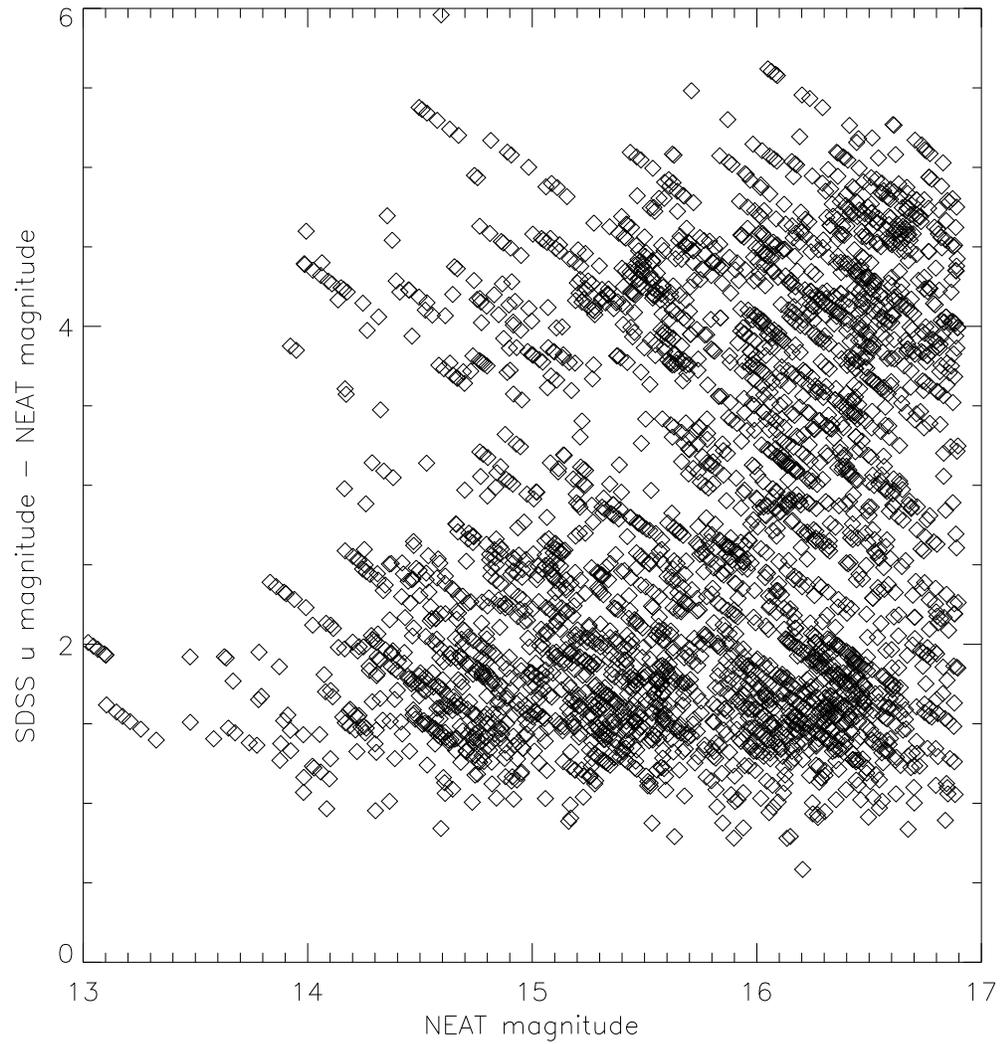}
\caption{The SDSS $u$-band magnitudes $-$ NEAT magnitudes of roughly $1,000$ stars as a function of NEAT magnitude.  No extinction correction has been performed.  The lines of points are from single stars observed multiple times with different derived NEAT magnitudes.}
\label{fig:neat_mag_sdss_u}
\end{figure}

\begin{figure}
\plotone{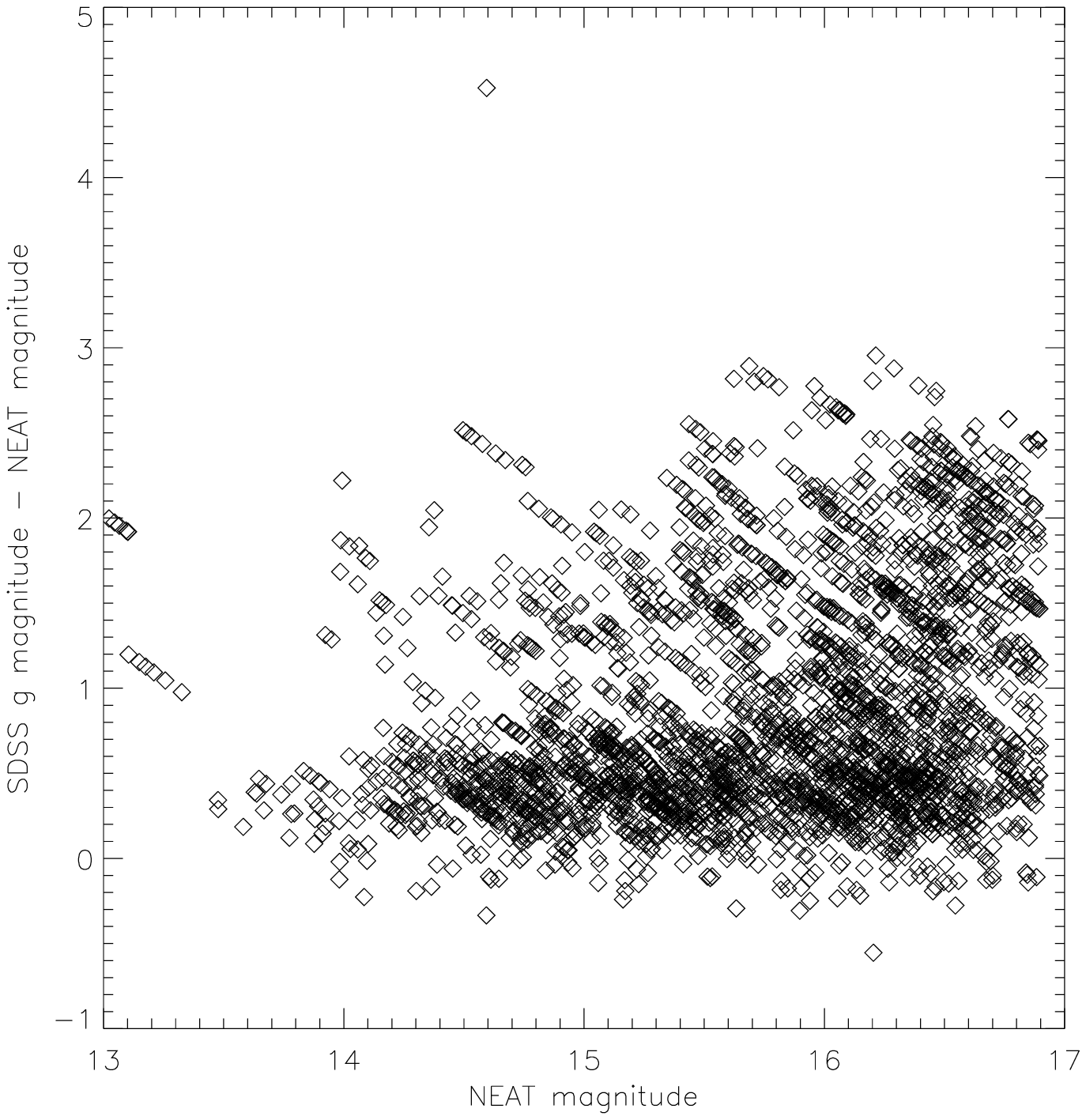}
\caption{As Fig.~\ref{fig:neat_mag_sdss_g} but for SDSS $g$' band.}
\label{fig:neat_mag_sdss_g}
\end{figure}

\begin{figure}
\plotone{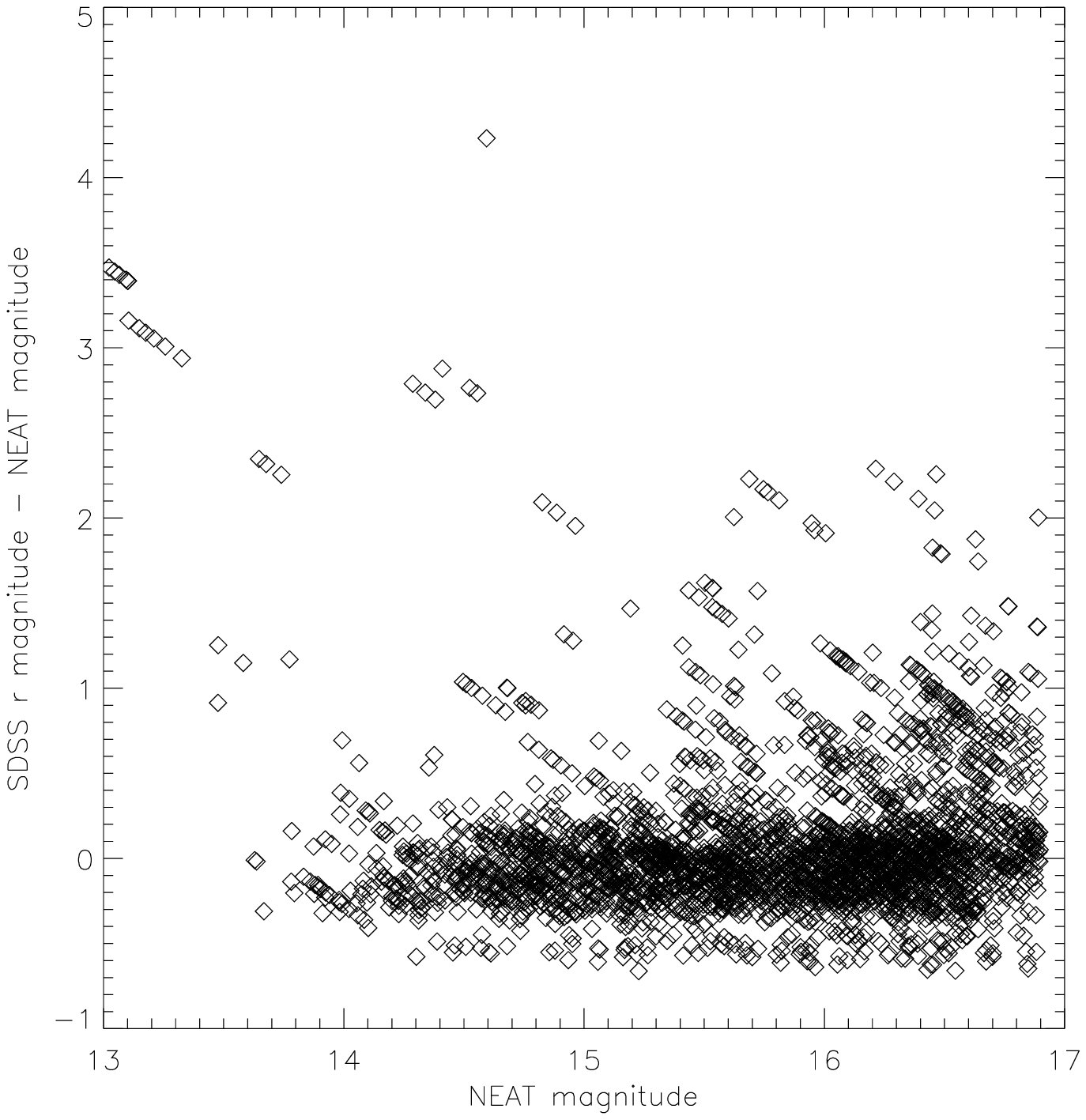}
\caption{As Fig.~\ref{fig:neat_mag_sdss_r} but for SDSS $r$' band.}
\label{fig:neat_mag_sdss_r}
\end{figure}

\begin{figure}
\plotone{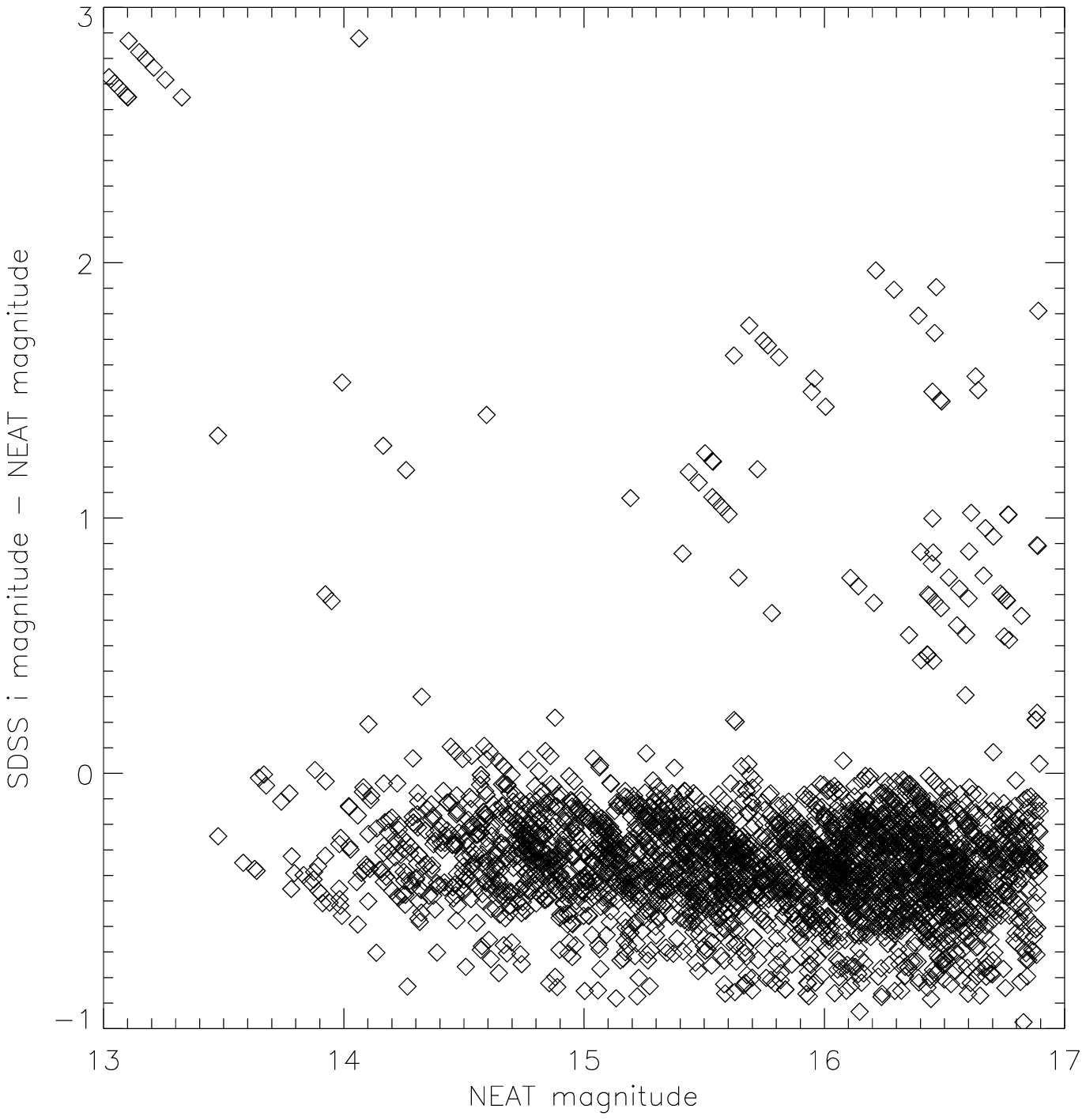}
\caption{As Fig.~\ref{fig:neat_mag_sdss_u} but for SDSS $i$' band.}
\label{fig:neat_mag_sdss_i}
\end{figure}

\begin{figure}
\plotone{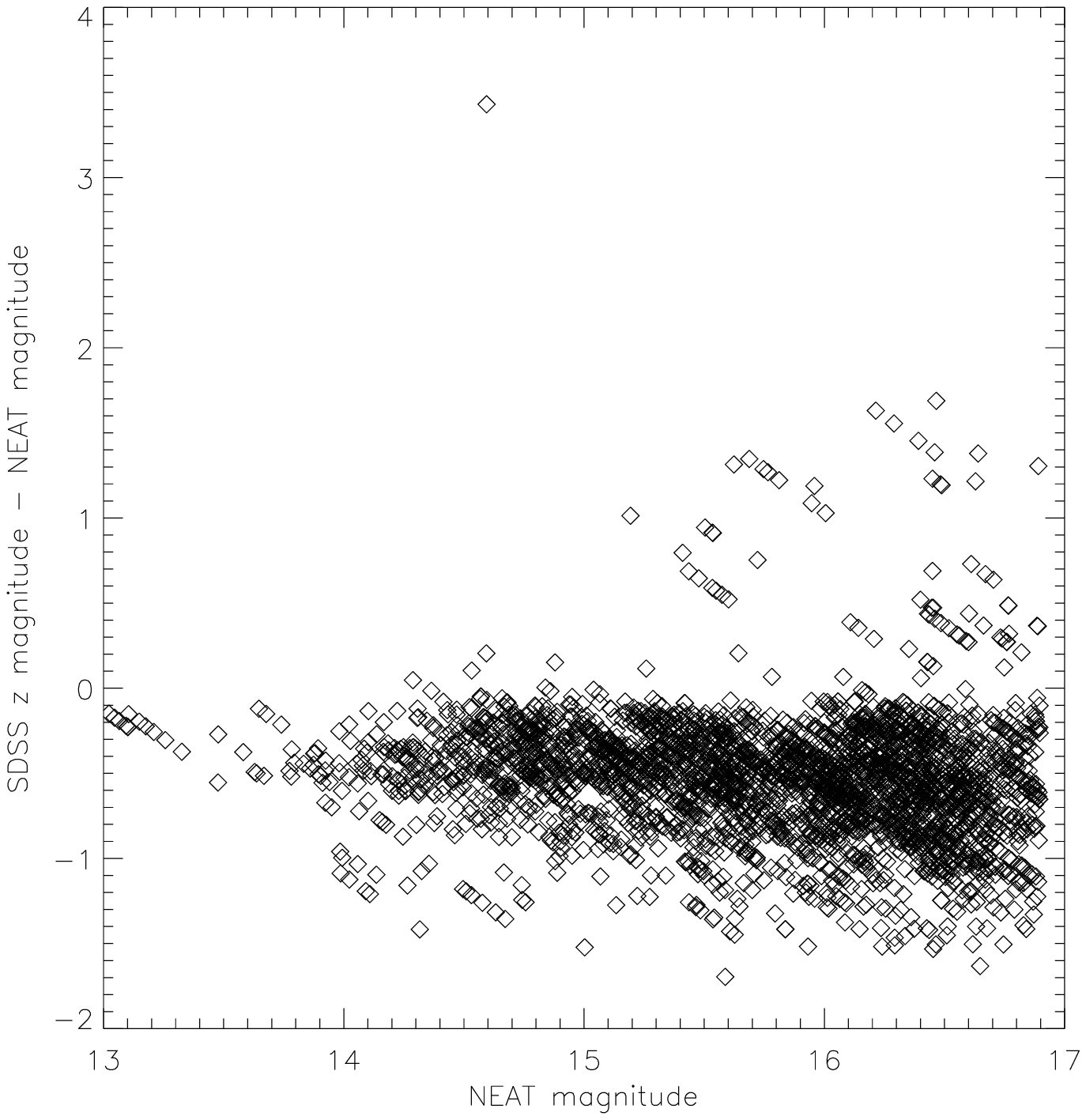}
\caption{As Fig.~\ref{fig:neat_mag_sdss_u} but for SDSS $z$' band.}
\label{fig:neat_mag_sdss_z}
\end{figure}


The availability of flux-calibrated spectroscopic observations for
2002ic would make it possible to do some more refined color corrections with
a spectrum of a supernova instead of a star, but these spectra are
only available at later times for SN~2002ic. 

\section{Landolt and Stetson Calibration}

Peter Stetson has been compiling a list of secondary standard
stars~\citep{stetson04} in the fields of Landolt ~\citep{landolt92}.
These Stetson standard stars were used to calibrate the color response
and zeropoint of the NEAT detectors.  The color range of these
standards varied by field but was nominally in the range $0.3 < B-V <
1.5$.  This allowed for a calculation of the color term
for the NEAT detectors versus the standard $UBVRI$ system.

The magnitude offset is still something that needs to be set for an individual
field as the zeropoint used for the NEAT magnitudes is the zeropoint
from the USNO~A1.0 solution.  This zeropoint can vary by $0.2$
magnitudes from plate-to-plate from the original POSS survey~\citep{usnoa1}.

Guillaume Blanc compiled an astrometrically-corrected Landolt catalog
that was useful in undertaking this calibration program.

\section{Hamuy Secondary Standard Star Calibration}

Mario Hamuy was kind enough to share his $BVI$ photometry of stars in
the field of SN~2002ic.  This allowed for direct calculation of the 
offset between the USNO~A1.0 zeropoint for the field and the offset to
use in Eq.~\label{eq:NEATcolor_quad}.  Unfortunately, this offset gave
magnitude corrections of $\sim0.4$~magnitudes for the USNO-A1.0-calibrated
magnitudes, which again was in conflict
with the better agreement found with the \citet{hamuy03b} and \citet{deng04}
V-band data.


\section{Summary}

Despite some relatively in-depth efforts to determine color terms
for the NEAT magnitudes calibrated to the USNO-A1.0 POSS-E/R system,
 an appropriate correction
that actually reduces the uncertainty in the magnitude
could not be determined.
Since
the USNO-A1.0 fields may vary in relative calibration by as much
as $0.2$~magnitudes, it may be possible that the field of
SN~2002ic was coincidentally calibrated in such a way as to
yield magnitudes comparable to V-band for SN~Ia objects.  
The SNfactory will continue treating the USNO-calibrated
magnitudes as the best understanding of the NEAT 
system until a better calibration is obtained.

\chapter{Always dither like your parental unit told you}
\label{apx:dither}

This is an account of why you should always dither in a pattern that
does not repeat CCD column or row values.

\section{Introduction}

When using a CCD for optical astronomical imaging, it is good
practice to take several exposures of the target and to dither the
exact pointing of the telescope for each exposure.  This dithering is
to use different pixels on the CCD to observe a given object in the
field.  CCDs often exhibit cosmetic defects, bad columns, and other
problems, and it is important to select a variety of positions so
that a given problem does not affect all of a night's observation.  

Sources external to the detector, such as cosmic rays, also tend to
prompt the taking of multiple dithered images.
Although dithering for cosmic ray rejection is not strictly necessary,
it is helpful.

When dithering, it is unfortunately more common than it should be to
simple do offsets along one axis and then the other.  Thus, if one has a
triplet of images, as is the case for the NEAT observations, one might
end up with a dither pattern like that shown in
Fig.~\ref{fig:dither_simple}.  However, this pattern does not maximize
the advantages of dithering.  One should sample a different row and a
different column with each dither.  Fig.~\ref{fig:dither_better} shows
such an improved dither pattern.

\section{The NEAT dither pattern}

\begin{figure}
\plotone{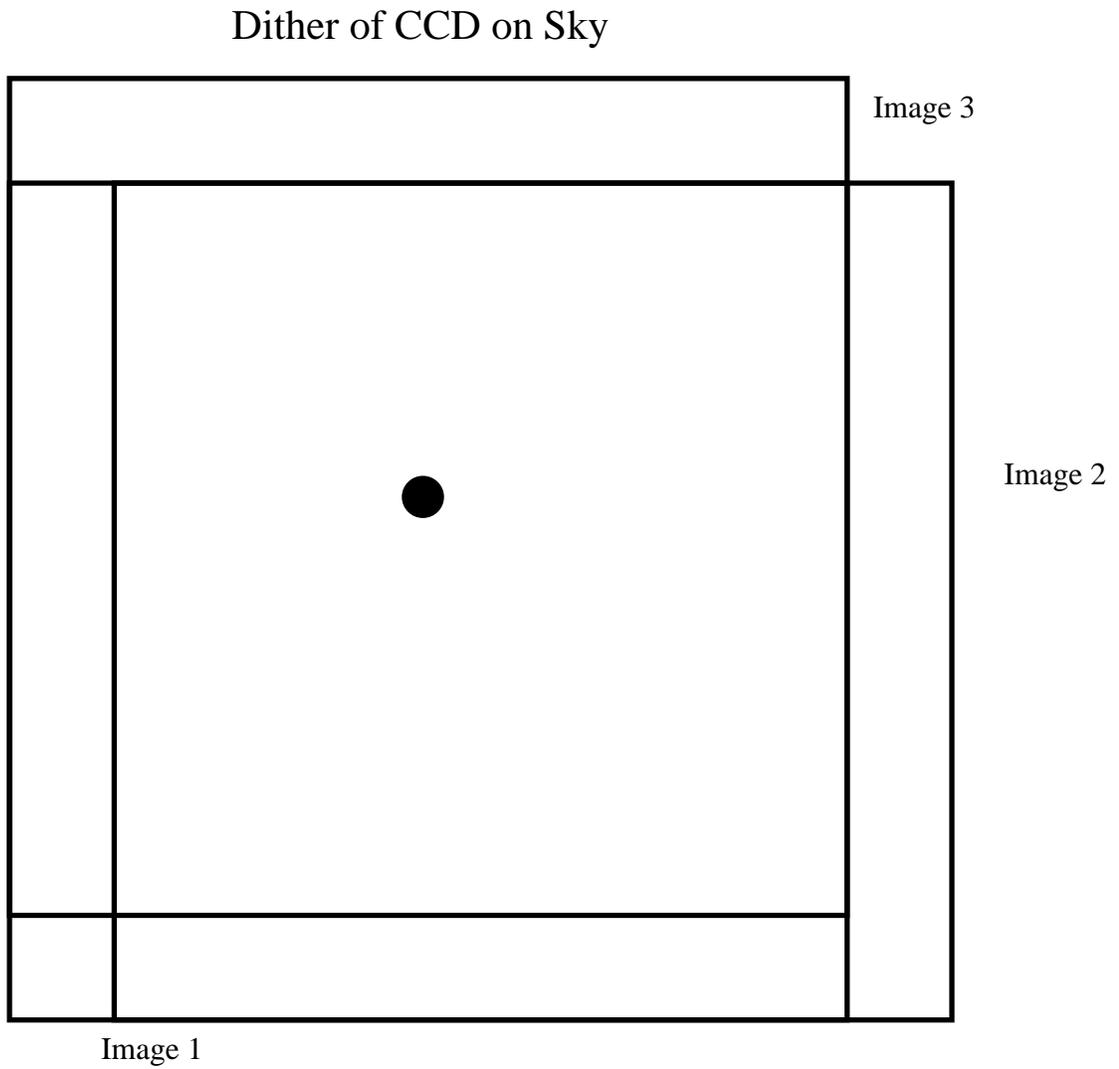}
\caption{A sub-optimal dither pattern on the sky.}
\label{fig:dither_simple}
\end{figure}

\begin{figure}
\plotone{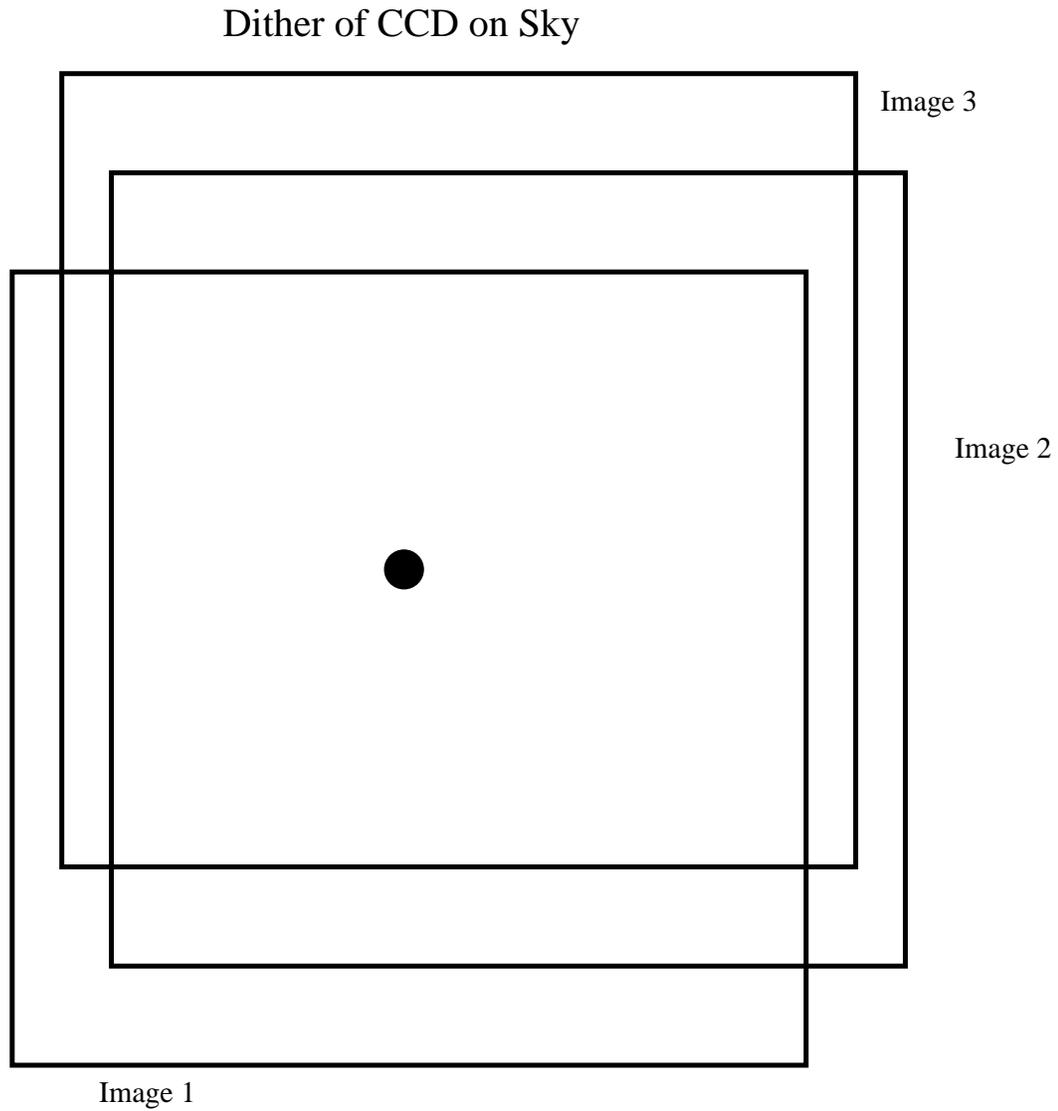}
\caption{A better dither pattern on the sky.  Note that each object is
now in a different row and column.}
\label{fig:dither_better}
\end{figure}

The NEAT group has used the dither pattern shown in
Fig.~\ref{fig:dither_simple} on both the Palomar and Haleakala
telescopes since the inception of their programs.

In May of 2003 it was determined that this sub-optimal dither
pattern was leading to a problem.  The Haleakala NEAT4GEN2 detector
suffers from electronic ghosts where bright stars imaged onto the CCD
can affect other amplifiers during readout.  This electronic ghosting
has the property that it tracks the dither in one direction but moves
opposite to it along the other axis.

\section{CCD Ghosts}

\begin{figure}
\plotone{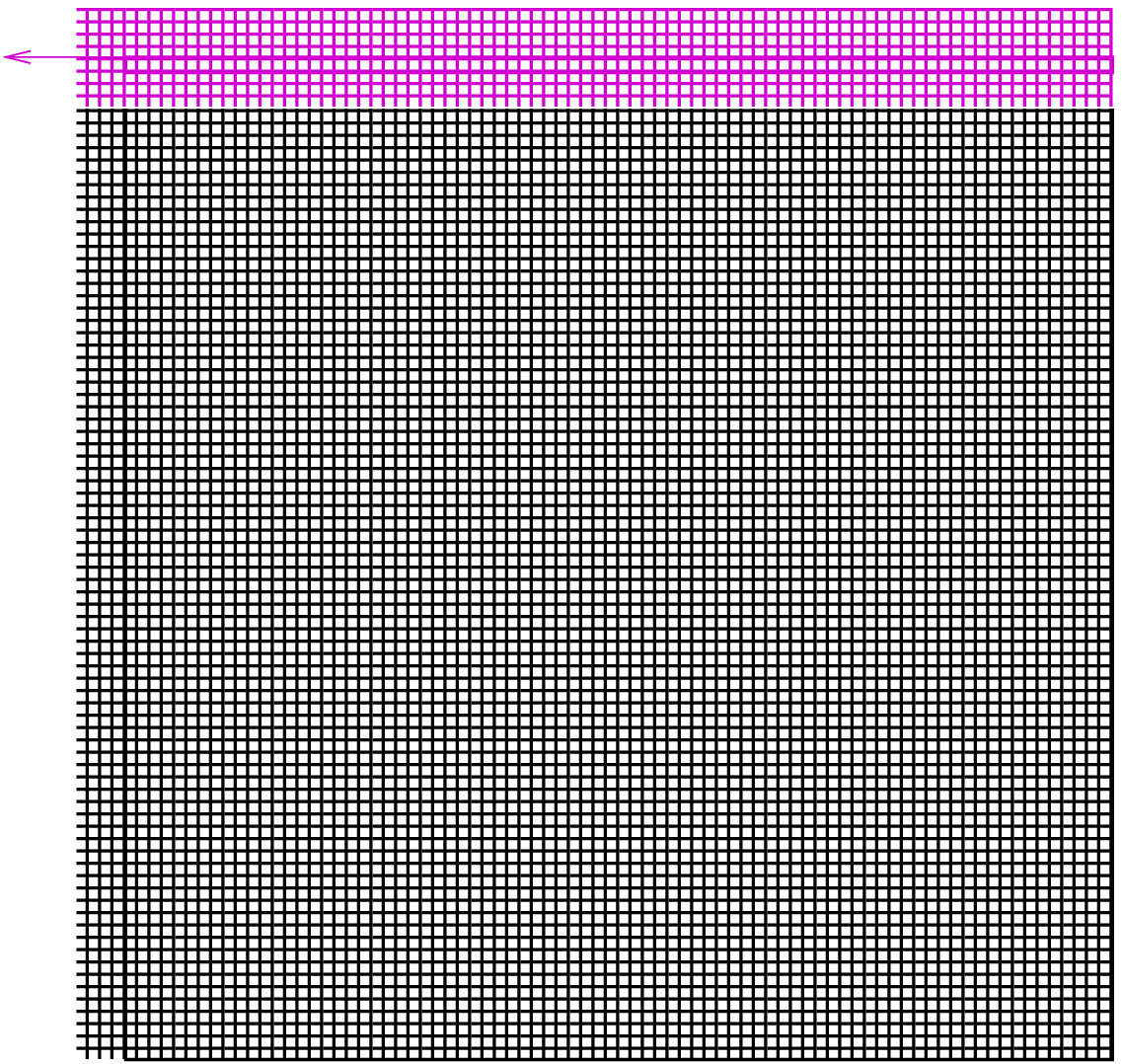}
\caption{A CCD is readout by an amplifier on one side of the CCD.  The
charge in each pixel is shifted row-by-row to the serial readout which
is then readout.  It is important to understand the nature of CCDs
when understand problems that can occur when using them.}
\label{fig:ccd_readout}
\end{figure}

Fig.~\ref{fig:ccd_readout} gives a schematic representation of how a
CCD is read out.  An amplifier reads a CCD out row-by-row.  The row at
the edge of the CCD is read out serially and then all of the rows on
the CCD are shifted down one row toward the edge.  Then the next row
is read, and so on until all of the rows on the CCD have been read.

\begin{figure}
\plotone{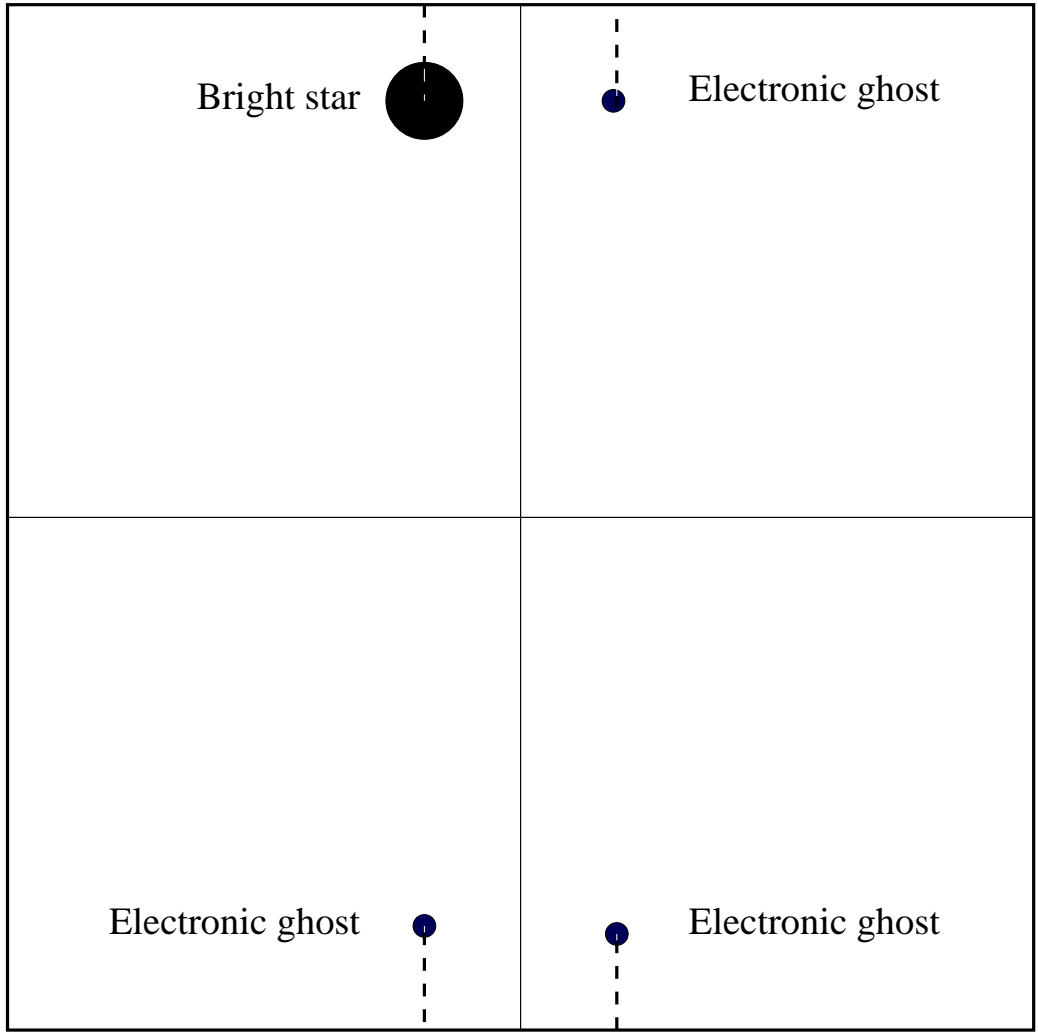}
\caption{Electronic ghosts with multiple amplifiers.  In a CCD read
out with multiple amplifiers, all of the electronics share a common
ground and temporary signal spikes from one amplifier can leak to
other amplifiers and produce electronic ghosts.}
\label{fig:electronic_ghosts}
\end{figure}

When a particularly bright pixel is read out, a mild surge can affect
the electronics of the amplifier.  If there are multiple amplifiers in
the detector, the other amplifiers can be affected by this surge and
an electronic ghost of the bright pixel can be registered on the other
amplifiers.  This can result in a very dim image of a bright star
being imposed on a different amplifier.  The location of this ghost
depends on the direction of the read out of the CCD.  See
Fig.~\ref{fig:electronic_ghosts} for an illustration of this effect.

\section{Electronic Ghosts Meet Sub-Optimal Dithering}

This all conspired to create a problem for the SNfactory supernova search pipeline.

\begin{figure}
\plotone{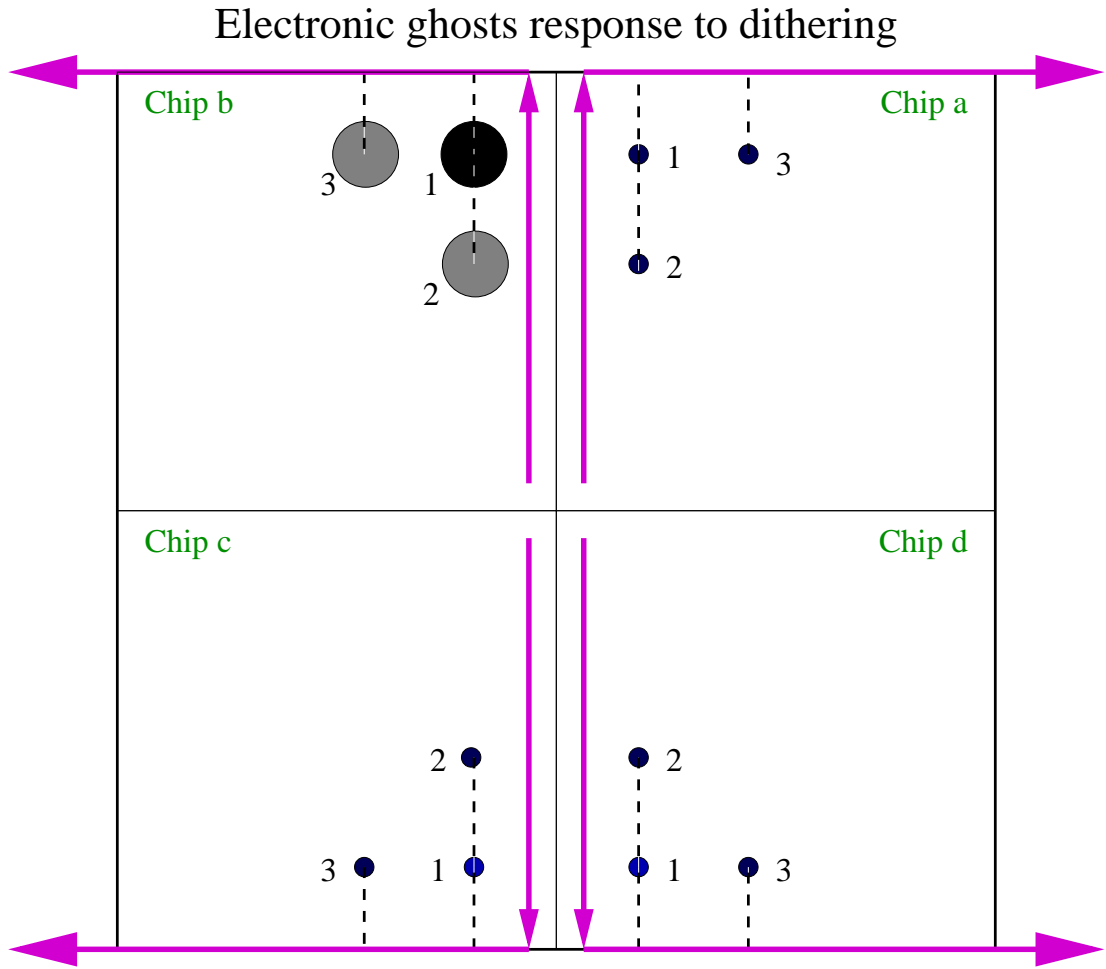}
\caption{The electronic ghosts will move with or opposite to objects
on the sky depending on the direction of sky motion relative to the
direction of CCD readout (indicated by arrows for each quadrant).  
In the example shown here, the electronic ghost moves with the sky field
on Chip {\bf a} between dither positions 1 and 2, and on Chip {\bf c}
between dither positions 1 and 3.}
\label{fig:dither_ghosts}
\end{figure}

The problem was that an electronic ghost would only move in response to
a dither in the x-direction on the CCD and didn't respond to a dither
in the y-direction.  This meant that given an image dither pattern of
(+13,0), (0,+13) relative to the initial image, the electronic ghost
would be at the same location in images 1 and 3.  In image 2 it would
be offset by twice the dither amount since it moves in opposition to
dithers in the x-direction.  See Fig.~\ref{fig:dither_ghosts}

For a few months it was believed that such objects were asteroids that 
were caught just in their retrograde motion with respect to Earth.
An undergraduate was working on asteroid orbit calculations in the
spring of 2003, but those calculations  hadn't yet incorporated to
check so whether relative retrograde asteroid motion made sense as an
explanation for this events.  For example, a sample candidate that was improperly
saved by mistake is 'S2003-180'.  This candidate was mistakenly submitted 
for follow up and its source was revealed.

After a few weeks the images of this candidate from the follow up
observations were observed and the candidate appeared to move.
After some quick calculations, it was clear that this object was not moving
in a way consistent with a solar system object (asteroid, KBO, etc.).
It was moving too slowly.  Upon a more careful analysis it became clear
that the object was not moving in a linear fashion over time but rather moved
around from night to night in a relatively scattered way.

It was eventually realized that this was a ghost in the CCD image.
Upon further checking with Steve Pravdo, he explained that the MSSS
detector did exhibit electronic ghosts.  Because of the nature of
electronic ghosts, the NEAT dither pattern (see
Fig.~\ref{fig:dither_ghosts}) placed the electronic ghosts in
the same places in some of the dithers.

\begin{figure}
\plotone{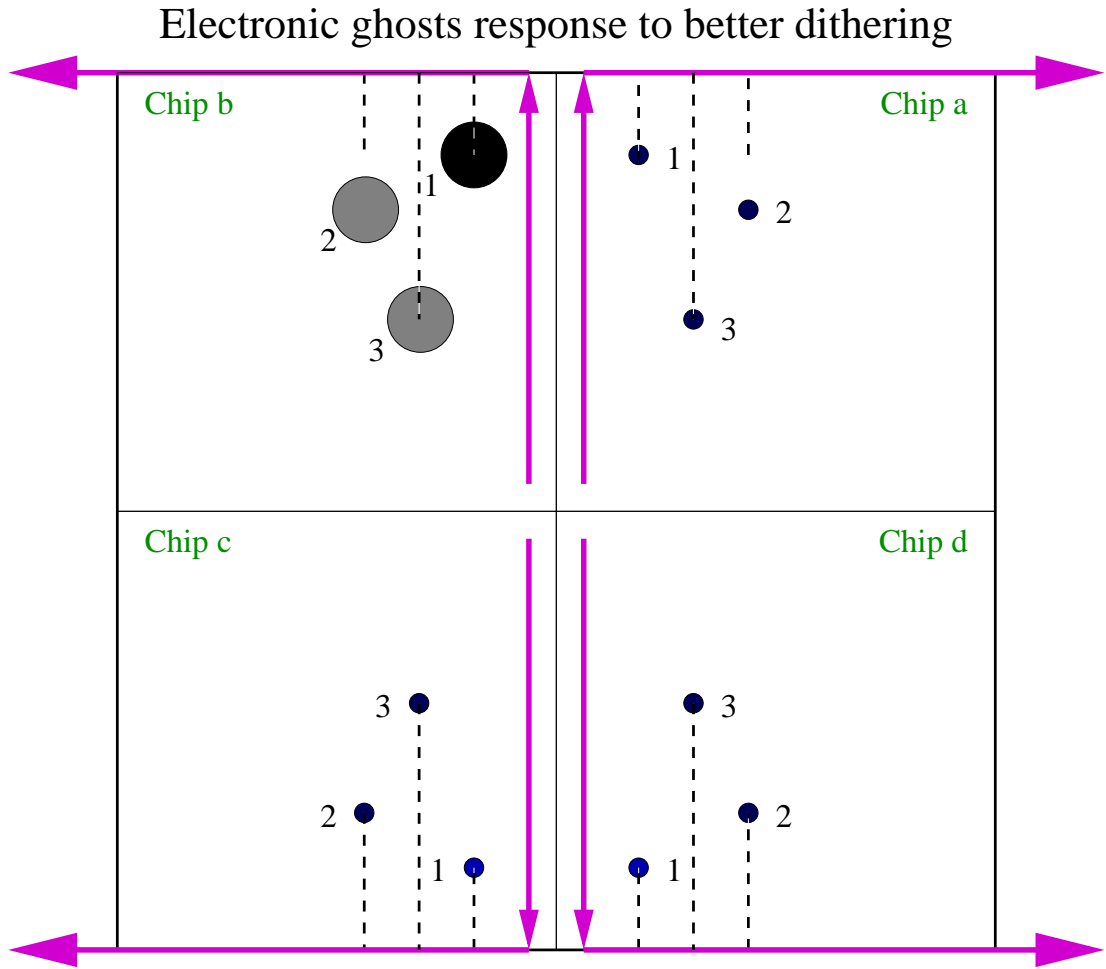}
\caption{A new dither pattern that results in no ghosts being in the
same position in the different images.}
\label{fig:dither_ghosts_better}
\end{figure}

This realization led to a recommendation for a dither pattern of the
form of that shown in Fig.~\ref{fig:dither_ghosts_better}.  This
dither pattern results in electronic ghosts being placed in different
locations with respect to surrounding stars.  Thus, when the images are
registered and moved to the same reference frame the ghosts do not
cause false candidates.

It is estimated that improvement reduced the number of subtractions that
had to be examined by hand by 10 percentage points for images taken
with the NEAT4GEN2 and NEAT12GEN2 cameras.

\chapter{Compressed FITS Performance Benchmarking for SNfactory}
\label{apx:compressed_fits}

\section{Summary}

An investigation was performed to determine whether the FITS files
should be processed compressed or uncompressed in the SNfactory
pipeline.  It was found that using compressed FITS files was in fact
slower than using non-compressed FITS files with the Deeplib software
used by the SNfactory.

Briefly, this resulted because the search pipeline was not actually
I/O limited for the individual image manipulation operations.  There
was a timing log as part of the processing steps that would allow for a
more precise analysis to try to reduce image processing time.  It is
possible that the way the Deeplib software manipulates files (by using
pixel streams) is not optimal when dealing with compressed files, but
this is the unlike to be the central issue with using compressed 
versus uncompressed FITS file.

The results of this study resulted in a decision to use uncompressed
rather that compressed FITS files for the search pipeline image
processing.

\section{Timing Tests}

After a few simple tests, the initial assumption that operating on
compressed files would be faster given that the search pipeline would be be I/O bound appears
not to hold:
\begin{Verbatim}[fontsize=\scriptsize]
pdsflx004 81% time sky `findimage feb112002oschinab84551.fts`
Image sigma: 20.2457
1.150u 0.350s 0:02.14 70.0%     0+0k 0+0io 923pf+0w
\end{Verbatim}

This file had not been accessed recently and there was no 
reason it should be in the NFS or disk cache of the computer or
diskvault.  The \verb'1.150u' refers to 1.15 user CPU seconds, while the
\verb'0.350s' refers to the 0.35 seconds of system time, the \verb'0:02.14'
indicates total wallclock time, and the \verb'70.0%' is the CPU
percentage.

A sample subtraction with the QUESTII camera in point-and-track mode
took 9 minutes, 30 seconds with a CPU percentage of 43.8\%.  This
was from a combination of I/O and database access times:

\begin{Verbatim}[fontsize=\scriptsize]
pdsflx006 15% time idl `whichdir 20030926 neat`\
/subdir/idlsubsep262003palomaag101307_360_240_6_i_aux
198.630u 51.750s 9:31.40 43.8%  0+0k 0+0io 716764pf+0w
\end{Verbatim}

A \code{time} command was added to the processing of images to get
up-to-date statistics on the processing of the images. 
Testing indicated that the compression imposes a significant penalty:

\begin{Verbatim}[fontsize=\scriptsize]
pdsflx004 157% time imarith sep22003questaaa32.fts.gz x sep22003questdaa75.fts.gz \
   test2.fts.gz
mode : wb
5.150u 0.590s 0:06.12 93.7%     0+0k 0+0io 626pf+0w
pdsflx004 158% time imarith sep22003questaaa32.fts.gz x sep22003questdaa75.fts.gz \ 
   test2.fts.gz
mode : wb
5.010u 0.720s 0:07.68 74.6%     0+0k 0+0io 626pf+0w
pdsflx004 159% time imarith sep22003questaaa32.fts.gz x sep22003questdaa75.fts.gz \ 
   test2.fts.gz
mode : wb
5.410u 0.790s 0:06.80 91.1%     0+0k 0+0io 626pf+0w
pdsflx004 160% time imarith sep52003questaaa96.fts x sep52003questaaa59.fts \ 
   test.fts
mode : wb0
1.950u 0.440s 0:03.79 63.0%     0+0k 0+0io 625pf+0w
pdsflx004 161% time imarith sep52003questaaa96.fts x sep52003questaaa59.fts \ 
   test.fts
mode : wb0
1.950u 0.350s 0:02.59 88.8%     0+0k 0+0io 625pf+0w
pdsflx004 162% time imarith sep52003questaaa96.fts x sep52003questaaa59.fts \ 
   test.fts
mode : wb0
2.000u 0.300s 0:02.39 96.2%     0+0k 0+0io 625pf+0w
pdsflx004 166% time imarith sep22003questaaa32.fts.gz x sep22003questdaa75.fts.gz \ 
   test2.fts
mode : wb0
3.600u 0.290s 0:04.04 96.2%     0+0k 0+0io 626pf+0w
pdsflx004 167% time imarith sep22003questaaa32.fts.gz x sep22003questdaa75.fts.gz \ 
   test2.fts
mode : wb0
3.550u 0.250s 0:03.86 98.4%     0+0k 0+0io 626pf+0w
pdsflx004 168% time imarith sep22003questaaa32.fts.gz x sep22003questdaa75.fts.gz \ 
   test2.fts
mode : wb0
3.560u 0.370s 0:04.02 97.7%     0+0k 0+0io 626pf+0w
pdsflx004 169% time imarith sep52003questaaa96.fts x sep52003questaaa59.fts \ 
   test.fts.gz
mode : wb
3.120u 0.430s 0:03.60 98.6%     0+0k 0+0io 625pf+0w
pdsflx004 170% time imarith sep52003questaaa96.fts x sep52003questaaa59.fts \ 
   test.fts.gz
mode : wb
3.170u 0.340s 0:03.54 99.1%     0+0k 0+0io 625pf+0w
pdsflx004 171% time imarith sep52003questaaa96.fts x sep52003questaaa59.fts \ 
   test.fts.gz
mode : wb
3.120u 0.280s 0:03.49 97.4%     0+0k 0+0io 625pf+0w
\end{Verbatim}

The first three tests were from compressed input to compressed output files.
The second three tests were from non-compressed input to non-compressed output files.
The third three tests were from compressed input to non-compressed output files.
The fourth three tests were from non-compressed input to compressed output files.

The fastest image arithmetic tests were the non-compressed to non-compressed runs, with an
average of $2.9$~seconds total elapsed time.  The slowest were the
compressed to compressed runs, with an average of $7.2$~seconds.  The
mixed runs were in between these two extremes with an average of $4.0$~seconds 
for the compressed to non-compressed run and $3.5$~seconds for
the non-compressed to compressed run.

As doubling the available disk space for the SNfactory was far cheaper
than doubling our available processors, it was clear that using
non-compressed files was the better approach.

\end{document}